\documentclass[a4paper,11pt,twoside]{ThesisStyle}

\usepackage{subcaption}

\usepackage{amsmath,amssymb}             
\renewcommand{\leq}{\leqslant}
\renewcommand{\geq}{\geqslant}

\usepackage[mono=false]{libertine}
\usepackage[left=1.01in,right=1.01in,top=2cm,bottom=2cm,includefoot,includehead,headheight=13.6pt]{geometry}

\usepackage{amscd,amsthm,yhmath,bbm}
\usepackage{tikz}
\usetikzlibrary{through,angles}
\usetikzlibrary{matrix,arrows}
\usepackage{graphicx,stmaryrd}
\usepackage{color}
\usepackage{mathrsfs}
\usepackage{subcaption}  
\usepackage{mathtools}
\usepackage{enumitem}
\usepackage{relsize}
\usepackage{changepage}

\newenvironment{blockfigure}{\vspace{-0.3\baselineskip}\begin{center}}{\end{center}\vspace{-0.3\baselineskip}}

\newcommand{\blockcaption}[2][\textwidth]{\par\centering{\parbox{#1}{\captionof{figure}{#2}}}}

\newtheorem{theorem}{Theorem}[section]
\newtheorem{example}[theorem]{Example}
\newtheorem{proposition}[theorem]{Proposition}
\newtheorem{lemma}[theorem]{Lemma}
\newtheorem{corollary}[theorem]{Corollary}
\newtheorem{conjecture}[theorem]{Conjecture}
\theoremstyle{definition}
\newtheorem{definition}[theorem]{Definition}
\theoremstyle{remark}
\newtheorem{remark}[theorem]{Remark}
\theoremstyle{algo}
\newtheorem{algo}[theorem]{Algorithm}
\theoremstyle{notation}
\newtheorem{notation}[theorem]{Notation}

\setlist{itemsep=0pt}

\newcommand{\dd}{\mathrm{d}}
\newcommand{\beq}{\begin{equation}}
\newcommand{\eeq}{\end{equation}}
\newcommand{\bea}{\begin{eqnarray}}
\newcommand{\eea}{\end{eqnarray}}
\newcommand{\bs}[1]{\ensuremath{\boldsymbol{#1}}}

\newcommand{\C}{\mathbb C}

\newcommand{\Q}{\mathbb Q}

\newcommand{\N}{\mathbb N}

\newcommand{\Res}{\mathop{\,\rm Res\,}}
\newcommand*{\bigcdot}{\raisebox{-0.25ex}{\scalebox{1.2}{$\cdot$}}}
\DeclareMathOperator\Tr{Tr}

\makeatletter
\def\bign#1{\mathclose{\hbox{$\left#1\vbox to8.5\p@{}\right.\n@space$}}\mathopen{}}
\def\Bign#1{\mathclose{\hbox{$\left#1\vbox to11.5\p@{}\right.\n@space$}}\mathopen{}}
\def\biggn#1{\mathclose{\hbox{$\left#1\vbox to14.5\p@{}\right.\n@space$}}\mathopen{}}
\def\Biggn#1{\mathclose{\hbox{$\left#1\vbox to17.5\p@{}\right.\n@space$}}\mathopen{}}
\makeatother

\usepackage[nottoc, notlof, notlot]{tocbibind}
\usepackage{minitoc}
\usepackage{aecompl}

\usepackage[intoc]{nomencl}

\makenomenclature

\usepackage{ifpdf}

\ifpdf
  \usepackage[a4paper,pagebackref,hyperindex=true]{hyperref}
\else
  \usepackage{graphicx}
  \DeclareGraphicsExtensions{.ps,.eps}
  \usepackage[a4paper,dvipdfm,pagebackref,hyperindex=true]{hyperref}
\fi

\graphicspath{{.}{images/}}

\renewcommand*{\backref}[1]{}
\renewcommand*{\backrefalt}[4]{%
\ifcase #1 %
(Not cited.)%
\or
(Cited on page~#2.)%
\else
(Cited on pages~#2.)%
\fi}

\usepackage{color}
\definecolor{linkcol}{rgb}{0,0,0.4} 
\definecolor{citecol}{rgb}{0.5,0,0} 

\hypersetup
{
bookmarksopen=true,
pdfmenubar=true, 
pdfhighlight=/O, 
colorlinks=true, 
pdfpagemode=None, 
pdfpagelayout=SinglePage, 
pdffitwindow=true, 
linkcolor=linkcol, 
citecolor=citecol, 
urlcolor=linkcol 
}

\usepackage{rotating}                    
\usepackage{fancyhdr}                    

\let\headruleORIG\headrule
\renewcommand{\headrule}{\color{black} \headruleORIG}

\usepackage{colortbl}
\arrayrulecolor{black}

\fancypagestyle{plain}{
  \fancyhead{}
  \fancyfoot{}
  
}

\usepackage{algorithm}
\usepackage[noend]{algorithmic}

\makeatletter

\def\cleardoublepage{\clearpage\if@twoside \ifodd\c@page\else%
  \hbox{}%
  \thispagestyle{empty}
  \newpage%
  \if@twocolumn\hbox{}\newpage\fi\fi\fi}

\makeatother
 
{%

\hrulefill
\vspace*{0.5cm}%
\end{minipage}
}

\let\minitocORIG\minitoc
\renewcommand{\minitoc}{\minitocORIG \vspace{1.5em}}

\usepackage{multirow}
\usepackage{slashbox}

{ \begin{list}%
	{$\bullet$}%
	{\setlength{\labelwidth}{25pt}%
	 \setlength{\leftmargin}{30pt}%
	 \setlength{\itemsep}{\parsep}}}%
{ \end{list} }

\begin{document}

\begin{titlepage}
\begin{center}

\vspace*{.06\textheight}

{\Huge \bfseries On discrete surfaces\par}\vspace{0.6cm}
{\LARGE \bfseries Enumerative geometry, matrix models \\ and universality classes   via topological recursion}\vspace{2cm} 
 

\Large \textbf{Dissertation}\\[0.3cm] 
\Large \textbf{zur}\\[0.3cm]
\Large \textbf{Erlangung des Doktorgrades (Dr. rer. nat.)}\\[0.3cm]
\Large \textbf{der}\\[0.3cm]
\Large \textbf{Mathematisch-Naturwissenschaftlichen Fakult\"{a}t}\\[0.3cm]
\Large \textbf{der}\\[0.3cm]
\Large \textbf{Rheinischen Friedrich-Wilhelms-Universit\"{a}t Bonn}\\[2.4cm]
\Large \textbf{vorgelegt von}\\[0.4cm]
\huge \textbf{Elba Garcia-Failde}\\[0.3cm]
\Large \textbf{aus}\\[0.3cm]
\Large \textbf{Barcelona, Spanien}\\[0.8cm]

\vfill

{\large Bonn, 2018}\\[4cm] 

\vfill
\end{center}
\newpage
\pagestyle{empty}
\begin{center}
\vfill
Angefertigt mit Genehmigung der Mathematisch-Naturwissenschaftlichen Fakult\"{a}t der Rheinischen Friedrich-Wilhelms-Universit\"{a}t Bonn
\end{center}
\vspace{17.5cm}
1. Gutachter: Prof. Dr. Don Bernard Zagier\\[0.3cm]
2. Gutachter: Dr. Ga\"{e}tan Borot\\[0.3cm]
Tag der Promotion: 11.~Oktober 2018\\[0.3cm]
Erscheinungsjahr: 2019\\[0.3cm]

\end{titlepage}

\pagenumbering{roman}

\cleardoublepage
\thispagestyle{empty}
\section*{Zusammenfassung}

The main objects under consideration in this thesis are called maps, a certain class of graphs embedded on surfaces. We approach our study of these objects from different perspectives, namely bijective combinatorics, matrix models and analysis of critical behaviors. Our problems have a powerful relatively recent tool in common, which is the so-called topological recursion introduced by Chekhov, Eynard and Orantin around 2007. Further understanding general properties of this procedure also constitutes a motivation for us.

We introduce the notion of fully simple maps, which are maps with non self-intersecting disjoint boundaries. In contrast, maps where such a restriction is not imposed are called ordinary.  We study in detail the combinatorial relation between fully simple and ordinary maps with topology of a disk or a cylinder. We show that the generating series of simple disks is given by the functional inversion of the generating series of ordinary disks. We also obtain an elegant formula for cylinders. These relations reproduce the relation between (first and second order) correlation moments and free cumulants established by Collins et al~\cite{Secondorderfreeness} in the setting of free probability, and implement the exchange transformation $x \leftrightarrow y$ on the spectral curve in the context of topological recursion.  
These interesting features motivated us to investigate fully simple maps, which turned out to be interesting combinatorial objects by themselves.
We then propose a combinatorial interpretation of the still not well understood exchange symplectic transformation of the topological recursion.

We provide a matrix model interpretation for fully simple maps, via the formal hermitian matrix model with external field.
We also deduce a universal relation between generating series of fully simple maps and of ordinary maps, which involves double monotone Hurwitz numbers. In particular, (ordinary) maps without internal faces -- which are generated by the Gaussian Unitary Ensemble -- and with boundary perimeters $(\lambda_1,\ldots,\lambda_n)$ are strictly monotone double Hurwitz numbers with ramifications $\lambda$ above $\infty$ and $(2,\ldots,2)$ above $0$. Combining with a recent result of Dubrovin et al.~\cite{DiYang}, this implies an ELSV-like formula for these Hurwitz numbers.

Later, we consider ordinary maps endowed with a so-called $O(\mathsf{n})$ loop model, which is a classical model in statistical physics. We consider a probability measure on these objects, thus providing a notion of randomness, and our goal is to determine which shapes are more likely to occur regarding the nesting properties of the loops decorating the maps. In this context, we call volume the number of vertices of the map and we want to study the limiting objects when the volume becomes arbitrarily large, which can be done by studying the generating series at dominant singularities. An important motivation comes from the conjecture that the geometry of large random maps is universal.

We pursue the analysis of nesting statistics in the $O(\mathsf{n})$ loop model on random maps of arbitrary topologies in the presence of large and small boundaries, which was initiated for maps with the topology of disks and cylinders in~\cite{BBD}. For this purpose we rely on the topological recursion results of~\cite{BEOn,BEO} for the enumeration of maps in the $O(\mathsf{n})$ model. We characterize the generating series of maps of genus $g$ with $k$ boundaries and~$k'$ marked points which realize a fixed nesting graph, which is associated to every map endowed with loops and encodes the information regarding non-separating loops, which are the non-contractible ones on the complement of the marked elements. These generating series are amenable to explicit computations in the so-called loop model with bending energy on triangulations, and we characterize their behavior at criticality in the dense and in the dilute phases, which are the two universality classes characteristic of the $O(\mathsf{n})$ loop model. We extract interesting qualitative conclusions, e.g., which nesting graphs are more probable to occur.

We also argue how this analysis can be generalized to other problems in enumerative geometry satisfying the topological recursion, and apply our method to study the fully simple maps introduced in the first part of the thesis.



\cleardoublepage
\thispagestyle{empty}
\section*{Acknowledgments}

First and foremost, I would like to thank my two supervisors. I am very grateful to Ga\"etan Borot, who has been my de facto supervisor, for being always morally present, extremely generous with his time and helpful with his advice. He was always willing to share, discuss and help. Many thanks also for all the patience listening and answering to all my vague ideas and doubts. I would also like to express my sincere gratitude to my other supervisor Don Zagier, for all the inspiring and fun discussions, especially for being so great at transmitting his passion for mathematics, which was often really encouraging.

Moreover, I would like to thank Werner Ballmann, for being my second official supervisor and part of my committee, and Albrecht Klemm for also being a member of the jury. Equally, I am thankful to the Max Planck Institute for Mathematics for providing me with economic support and with a great working environment. Many thanks also to the non-scientific staff, who were extremely cooperative.

I am also indebted to Sergey Shadrin for a great welcome in the University of Amsterdam during a two-month research stay. It was amazing to experience his teamwork style. Many thanks to Reinier Kramer, Danilo Lewa\'nski and Sergey Shadrin for all the insteresting discussions.

I would also like to thank the scientific community around my research topics. I am grateful for all the conferences I had the opportunity to attend, which were certainly a source of inspiration. I want to especially thank Rapha\"{e}l Belliard, Vincent Bouchard, Norman Do and Danilo Lewa\'nski for fruitful discussions regarding my thesis. I would also like to mention Olivia Dumitrescu, Bertrand Eynard, Olivier Marchal and Motohico Mulase for the encouragement and advice. I also thank Alessandro, Anupam, Axel, Danilo, Dima, Marvin, Max, Norm, Petya, Paul, Priya, Reinier, Rapha\"{e}l, Sasha and Vincent for all the fun in the conferences.

I especially thank Alessandro Giacchetto for many nice discussions, for his help in the last months and for some pictures of this thesis.

I am also grateful to Roland Speicher and Felix Leid for interesting discussions on some aspects of my thesis. I also thank Timothy Budd for his interest and for bringing useful references to our attention.

I thank many coworkers for useful discussions and motivation to learn things together, especially Gabriele Bogo, John Alexander Cruz Morales, N\'{e}stor Le\'on Delgado, Ana Ros Camacho, Julia Semikina, Alessandro Valentino, Di Yang and Federico Zerbini.

Finally, I would also like to express my gratitude to the people who were really supportive on the personal side. 

Muchas gracias a mis padres, a mi tutata y a mi cu\~{n}ado, que es como mi hermano, por todo el apoyo, por sobrellevar todos los inconvenientes de la distancia y por venir a verme con tanta ilusi\'on. Als meus petitons Arnau i Paula, que ja s'han fet molt menys petitons, per ajudar-me, amb tanta facilitat, a recordar les coses m\'{e}s importants de la vida.

Tamb\'e voldria agrair tot el suport de persones que s'han comportat com la meva fam\'ilia i que, per tant, considero com a tal: al Joan, a la Montse, a l'Albert, a la Mom, a la Isabel i al Manel.

To Bonniatos, for being there for everything, especially for parties. Also to Planckers and MensaMensaMensa for all the nice moments together.
Tengo que mencionar especialmente a Ana (per totes les sessions de pipes), a Elena (por todos los bailes, todo el esperanto y toda la ayuda durante mi ceguera), a N\'{e}stor (por las risas y las confesiones interminables) y a Fede (sencillamente por todo lo vivido). I also want to thank Julia (for sharing the last tough moments) and Pietro (for being there for last minute technical issues).

A las muchachas: Geo, Jessi, Raquel y Rosario. Por todo lo que hemos compartido y lo que hemos vivido creciendo juntas, por estar incre\'{i}blemente ah\'{i} estando tan lejos, porque cuando volvemos a vernos no hay distancia entre nosotras, a\'{u}n siendo tan diferentes. Por el documental tan genial que me dedicasteis y que tantas veces he mirado cuando ten\'{i}a ataques de morri\~{n}a. Y a ti Xario, por ser la mejor mejor amiga que se puede tener.

Als meus companys d'universitat, amb qui sovint vaig descobrir que no era l'\'unica en moltes coses. Per\`{o} sobretot, al David, la Celia, la Laia i la Marta, amb qui vaig compartir tantes bogeries matem\`{a}tiques que ens van unir molt\'{i}ssim. Tot i que cadasc\'{u} sigui a una punta del m\'{o}n, us continuo sentint molt a prop dia a dia.

E infine a te, Andrea, perch\'{e} trovarsi \`{e} stato incredibile.


\tableofcontents

\mainmatter

\chapter{Introduction}
\label{chap:intro}








In this chapter we give the necessary background on six different, but quite related, topics which are especially relevant for this thesis. We also include some new notions specific to this thesis and new points of view adapted to our subsequent work. The main purpose of this chapter is to make the thesis as self-contained as possible, focusing at the same time on brevity and on giving an overview of the bigger picture where the results of the thesis fit in. This of course very often implies providing references for more details whenever we consider it necessary or interesting.

Another goal is achieved in the seventh section, where we provide an outline of the results of the thesis with the support of the objects and techniques previously introduced.

\section{Maps}\label{maps}

Maps can intuitively be thought as graphs drawn on surfaces or discrete surfaces obtained from gluing polygons, and they receive different names in the literature: \emph{embedded graphs}, \emph{ribbon graphs}, \emph{discrete surfaces}, \emph{fat graphs}... Enumeration of maps by combinatorial methods has been intensively studied since the pioneering work of Tutte in his series of ``Census'' articles \cite{Tutte1,Tutte2,Tutte3,Tutte4} in the 60s. His initial motivation was to prove the famous Four-color theorem. Even if he did not achieve his goal, he managed to enumerate triangulations and quadrangulations, and in general his research stimulated important development in the theory of generating series. He found a recursive decomposition of maps by deleting an edge in every step, which led to equations satisfied by the generating functions of maps: Tutte's equations \cite{TutteQ}.

Nowadays, maps can be seen from different points of view, from which they have natural generalizations: they can be encoded by a sequence of $3$ permutations ($\sigma,\alpha,\varphi$) or interpreted as maps from a Riemann surface to the Riemann sphere. Actually, this a consequence of many beautiful, unexpected relations to other fields, as far as Galois theory, quantum field theory and string theory.

For maps, the permutation $\alpha$ has to be an involution. When we take $\alpha$ to be arbitrary, the sequence of three permutations encode a more general object called \emph{hypermap}, which can be seen to be equivalent to a map whose underlying graph is bipartite together with a two-coloring of the vertices. Grothendieck coined the term \emph{dessins d'enfants} in 1984 \cite{Esquisse} for hypermaps in the context of Galois theory.
By Bely\u\i \ theorem \cite{BelyiThm}, Riemann surfaces that can be defined as algebraic curves over $\overline{\Q}$ are precisely the ones that can be described by hypermaps (identifying them with certain coverings of the sphere with three branch points). This allows us to translate the action of the absolute Galois group ${\rm Aut}(\overline{\Q}\,\vert\,\Q)$ on curves defined over $\overline{\Q}$ to an action on dessins d'enfants. As a consequence of the theorem, this action is faithful, what suggests that very elementary objects \footnote{ ``... of which any child's drawing scrawled on a bit of paper (at least if the drawing is made without lifting the pencil) gives a perfectly explicit example'', Alexander Grothendieck in his {\it Esquisse d'un programme} \cite{Esquisse}.} such as dessins d'enfants can help understanding a very intricate object such as the absolute Galois group. This discovery made a very strong impression on Grothendieck and inspired further development in connection with the field of Galois theory (see e.g. the collection of papers \cite{NumberFields}).


In physics, summing over maps is a well-defined discrete replacement for the non-obviously defined path integral over all possible metrics on a given surface which underpin two-dimensional quantum gravity. The observation by t'Hooft \cite{tHooft} that maps are Feynman diagrams for the large rank expansion of gauge theories led Br\'ezin-Itzykson-Parisi-Zuber \cite{BIPZ} to the discovery that hermitian matrix integrals are generating series of maps. The rich mathematical structure of matrix models -- integrability, representation theory of $U(\infty)$, Schwinger-Dyson equations, etc. -- led to further insights into the enumeration of maps, \textit{e.g.} \cite{ACM,dFZJ}. It also inspired further developments, putting the problem of counting maps into the more general context of enumerative geometry of surfaces, together with geometry on the moduli space of curves \cite{HarerZagier,Witten,Kontsevich}, volumes of the moduli space \cite{Mirzakhani}, Gromov-Witten theory \cite{Witten,BKMP}, Hurwitz theory \cite{EMS,EynardHarnad}, etc. and unveiling a common structure of ``topological recursion'' \cite{EORev,Eynardbook}.


The idea of this section is to briefly make the intuitive idea of maps precise. Further details about the classical theory are very well exposed in \cite{Eynardbook,LandoZvonkin,BouttierChapter}. We also introduce a new type of maps that we call simple and fully simple maps, and will constitute a crucial object of study in this thesis. To emphasize that classical maps do not necessarily satisfy any condition of simplicity, we call them \emph{ordinary}. We also introduce here the combinatorial side of the $O(\mathsf{n})$ loop model, whose objects of study are maps decorated with loops. This is one of the most famous models in statistical physics whose study is also an important motivation for this thesis. We will describe its natural context on random maps in a subsequent section. We also define stuffed maps in general, which are even more general than maps endowed with loops, and we call classical maps \emph{usual} to make the distinction clear in this context.

\subsection{Ordinary, usual maps}\label{OrdMapsIntro}

\begin{definition}\label{usual_map}
An \emph{embedded graph} of genus $g$ is a connected graph $\Gamma$ embedded into a connected orientable surface $X$ of genus $g$ such that $X \setminus \Gamma$ is a disjoint union of connected components, called ${\it faces}$, each of them homeomorphic to an open disk. 

Every edge belongs to two faces (which may be the same) and we call {\it length} of a face the number of edges belonging to it.

We say that an embedded graph has $n$ {\it boundaries}, when it has $n$ marked faces, labeled $1,\ldots,n$, which we require to contain a marked edge, called {\it root}, represented by an arrow following the convention that the marked face sits on the left side of the root. A face which is not marked receives the name of {\it inner face}.

Two embedded graphs $\Gamma_i\subset X_i$, $i=1,2$, are {\it isomorphic} if there exists an orientation preserving homeomorphism $\varphi: X_1 \rightarrow X_2$ such that  $\left.\varphi\right|_{\Gamma_1}$ is a graph isomorphism between $\Gamma_1$ and $\Gamma_2$, and the restriction of $\varphi$ to the marked edges is the identity.

A \emph{map} is an isomorphism class of embedded graphs.
\end{definition}

\begin{remark}
Since the faces of a map have a length given by the graph structure, we will often call them \emph{polygons}. In our definition, maps are thought as graphs on a surface, but they can equivalently be viewed as surfaces built from gluing two types of polygons\footnote{The term map in this context actually comes from thinking the building polygons as countries and seas in a world which is not necessarily a sphere, but an orientable surface of any topology.}: marked and unmarked ones. This idea will be precise after introducing the permutational model to encode maps with a sequence of three permutations that will prescribe the lengths of the building polygons and how to glue them.
\end{remark}

\begin{blockfigure}
		\includegraphics[width=.22\textwidth]{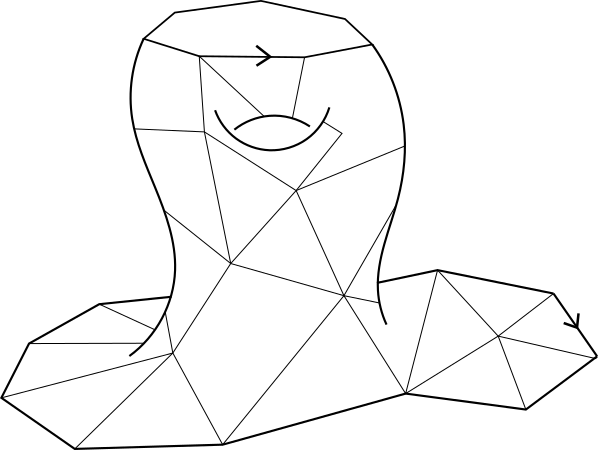}
		\hspace{1.5cm}
		\includegraphics[width=.37\textwidth]{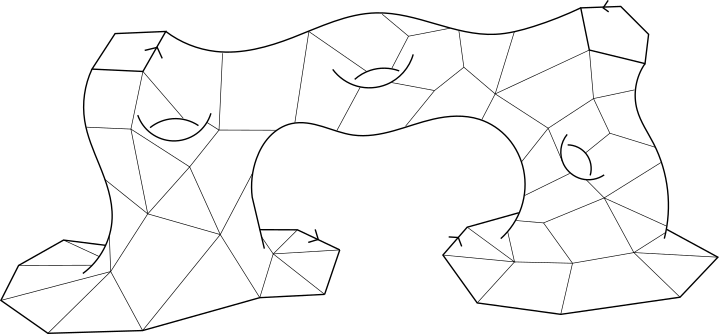}
		\vspace{0.3cm}
		\blockcaption{Two maps: genus $1$, $2$ boundaries (left), and genus $3$, $4$ boundaries (right).}
	\end{blockfigure}

Note that we include the connectedness condition in the definition of map. We will also work with disjoint unions of maps and we will specify we are dealing with a non-connected map whenever it is necessary.
Observe that $\sum_{f\in F} {\rm length}(f) = 2|E|$, where the sum is taken over the set of faces $F$ of the map and $E$ denotes the set of edges.

We call \emph{volume} the number of vertices of a map.

The vertices, edges and faces of a map are its \emph{cells}. Two of them are \emph{incident} if one is contained in the boundary of the other. Any edge doubly incident with a face is called an \emph{isthmus}\footnote{This term is justified by the Jordan curve theorem that implies that an edge contained in a cycle of the graph $\Gamma$ (i.e. in a simple closed curve in embedded graph $\Gamma\subset X$) is necessarily incident with two different faces.}.

We call {\it planar} a map of genus $0$. We call {\it map of topology} $(g,n)$ a map of genus $g$ with $n$ boundaries. For the cases $(0,1)$ and $(0,2)$ we use the special names: {\it disks} and {\it cylinders}, respectively.


To each map $\mathcal{M}$ we may associate its \emph{dual map} $\mathcal{M}^*$, which is obtained,
roughly speaking, by exchanging the roles of vertices and
faces. More precisely, the dual map has a vertex in the interior of every face of $\mathcal{M}$. Moreover, to every edge $e$ of $\mathcal{M}$ we associate a dual edge crossing $e$ and incident to the vertices of $\mathcal{M}^*$ in the two faces of $\mathcal{M}$ which are incident to $e$. Observe, that if $e$ is an isthmus, the dual edge is a self-loop, i.e. an edge that connects a vertex to itself. The operation of passing to the dual is an involution.

A map in which all faces are of length $k$ is called a \emph{$k$-angulation}; in particular, for $k=3,4$ we talk about \emph{triangulations} and \emph{quadrangulations}, respectively. Dually, a map in which all vertices are of degree $k$ is called \emph{$k$-valent}; in particular, \emph{trivalent} and \emph{quadrivalent} for the cases $k=3,4$.

A map is called \emph{bipartite} if all its vertices can be colored using two colors in such a way that no edge links two vertices of the same color. Equivalently, it is a map that does not contain any odd-length cycles. 

It is worth mentioning the following bijection due to Tutte, which shows bipartite quadrangulations are particularly interesting maps:
\begin{proposition}
There exists an explicit bijection between bipartite quadrangulations of genus $g$ and $f$ faces, and maps of the same genus $g$ with $f$ edges.
\end{proposition}

\begin{remark}
Observe that planar quadrangulations are bipartite, but such a statement is not true in positive genus.
\end{remark}

It is often easier to work with maps with bounded face degrees. That is what makes this elementary bijection so useful, since it turns maps with arbitrary face degrees into maps with bounded degrees.

\subsubsection{Permutational model}\label{permModel}

The embedding of the graph into an oriented surface provides the extra information of a cyclic order of the edges incident to a vertex. More precisely, we consider half-edges, each of them incident to exactly one vertex. Let $H$ be the set of half-edges and observe that $|H|=2|E|$. We label the half-edges by $1,\ldots, 2|E|$ in an arbitrary way.

\begin{figure}[h!]
 \begin{center}
        \def\svgwidth{\columnwidth}
        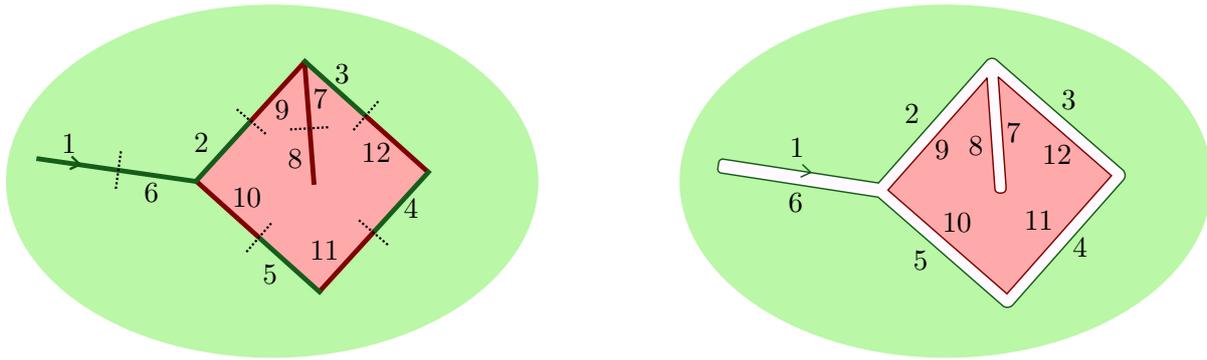
        	\caption{Two different ways of representing half-edges: in both cases by numbered segments, but on the left the two half-edges forming an edge are drawn consecutively, while on the right two consecutive half-edges belong to the same face and two half-edges forming an edge are drawn parallel. When maps are depicted as on the right, they are usually called ribbon graphs. We will sometimes use this representation because it can be a bit clearer, but will usually use the simpler representation on the left. On the left, the half-edges incident to a vertex are clearly the ones touching the vertex in the drawing; on the right, with our conventions, the half-edges incident to a vertex are the ones on the left viewed from the vertex in question.}\label{permutations}
\end{center}
\end{figure}

Every map with labeled half-edges can be encoded by a so-called {\it combinatorial map}, which consists of a pair of permutations $(\sigma,\alpha)$ acting on $H$ such that all cycles of $\alpha$ have length $2$. Given a half-edge $h\in H$, let $\sigma(h)$ be the the half-edge after $h$ when turning around its vertex according to the orientation fixed for the underlying surface (by convention counterclockwise). On the other hand, let $\alpha(h)$ be the other half-edge of the edge to which $h$ belongs. The information that $\alpha$ provides is encoded in the graph structure of the map, while $\sigma$ characterizes the additional data of a map given by the embedding of the graph in the surface.
\begin{itemize}
\item[$\bullet$] A cycle of $\sigma$ corresponds to a vertex in the map.
\item[$\bullet$] Every cycle of $\alpha$ corresponds to an edge of the map.
\item[$\bullet$] The faces may be represented by cycles of a permutation, called $\varphi$, of $H$.
\end{itemize}
Observe that with the convention that the face orientation is also counterclockwise, we obtain
\beq
\label{abc1}\sigma\circ\alpha\circ\varphi = {\rm id},
\eeq
and hence $\varphi$ can be determined by $\sigma^{-1}\alpha^{-1}$.

Rooting an edge in a face amounts to marking the associated label in the corresponding cycle of $\varphi$. Such cycles containing a root will be ordered and correspond to boundaries of the map.

\begin{example}{\rm
In Figure \ref{permutations}, we have
$$
\begin{array}{rcl}
\sigma & = & (5\ 11)(4\ 12)(3\  9\  7)(2\  6\  10), \\
\alpha & = & (1\  6)(2\  9)(7\  8)(3\  12)(4\  11)(5\  10), \\
\varphi & = & \sigma^{-1}\circ\alpha^{-1} = (1\  2\  3\  4\  5\  6)(7\  8\  9\  10\  11\  12), 
\end{array}
$$
where the root is the half-edge labeled $1$.
}
\end{example}

The lengths of the cycles of $\sigma$ and $\varphi$ correspond to the degrees of vertices and faces, respectively. The Euler characteristic is given by
$$
\chi(\sigma,\alpha)= |\mathcal{C}(\sigma)| - |\mathcal{C}(\alpha)| + |\mathcal{C}(\varphi)| - n,
$$
where $\mathcal{C}(\cdot)$ denotes the set of cycles of a permutation and $n$ is the number of cycles of $\varphi$ containing a root.

$G = \langle \sigma,\alpha \rangle$ is the {\it cartographic group}. Its orbits on the set of half-edges determine the connected components of the map. If the action of $G$ on $H$ is transitive, the map is connected, and its genus $g$ is given by the formula:
$$
2 - 2g - n= \chi(\sigma,\alpha),
$$
where $n$ is the number of boundaries. If all orbits contain a root, the map is called {\it $\partial$-connected}.

\subsubsection{Automorphisms}\label{aut}
Let us consider the decomposition $H= H^{u}\sqcup H^{\partial}$, where $H^{u}$ is the set of half-edges belonging to unmarked faces and $H^{\partial}$ is the set of half-edges belonging to boundaries.

Observe that from a combinatorial map one can retrieve all the information of the original map. So there is a one-to-one correspondence between maps with labeled half-edges and combinatorial maps. There is a canonical way of labeling half-edges in boundaries: assigning the first label to the root, continuing by cyclic order of the boundary and taking into account that boundaries are ordered. However, we can label half-edges of unmarked faces in many different ways. To obtain a bijective correspondence with unlabeled maps, we have to identify configurations which differ by a relabeling of $H^u$, i.e. $(\sigma,\alpha)\sim(\gamma\sigma\gamma^{-1},\gamma\alpha\gamma^{-1})$ with $\gamma$ any permutation acting on $H$ such that $\left.\gamma\right|_{H^{\partial}}={\rm Id}_{H^{\partial}}$. We call such an equivalence class {\it unlabeled combinatorial map} and we denote it by $[(\sigma,\alpha)]$. Note that unlabeled combinatorial maps are in bijection with the unlabeled maps we defined at the beginning of this section.

\begin{definition}
Given a combinatorial map $(\sigma, \alpha)$ acting on $H$, we call $\gamma$ an {\it automorphism} if it is a permutation acting on $H$ such that $\left.\gamma\right|_{H^{\partial}}={\rm Id}_{H^{\partial}}$ and
$$
\sigma = \gamma\sigma\gamma^{-1}, \ \ \ \ \ \ \alpha=\gamma\alpha\gamma^{-1}.
$$
\end{definition}
Observe that for connected maps with $n\geq 1$ boundaries, the only automorphism is the identity. Note also that these special relabelings that commute with $\sigma$ and $\alpha$, and we call automorphisms, exist because of a symmetry of the (unlabeled) map. The symmetry factor $\left|{\rm Aut}(\sigma,\alpha)\right|$ of a map is its number of automorphisms. 

We denote $\text{Gl}(\sigma,\alpha)$ the number of elements in the class $[(\sigma,\alpha)]$ and $\text{Rel}(\sigma,\alpha)$ the total number of relabelings of $H^u$, which is $|H^u|!$, if we consider completely arbitrary labels for the half-edges. By the orbit-stabilizer theorem, we have $\text{Gl}(\sigma,\alpha)=\frac{\text{Rel}(\sigma,\alpha)}{\left|{\rm Aut}(\sigma,\alpha)\right|}$.

\subsection{Simple and fully simple maps}\label{fsmaps}

We will say that a face $f$ of a map is \emph{simple} when at most two edges of $f$ are incident to every vertex in $f$. In the definition of maps, polygons may be glued along edges without restrictions, in particular faces may not be simple. This leads to singular situations, somehow at odds with the intuition of what a neat discretization of a surface should look like. This definition is the one naturally prescribed by the Feynman diagram expansion of hermitian matrix models. It is also one for which powerful combinatorial (generalized Tutte's recursion, Schaeffer bijection, etc.) and analytic/geometric (matrix models, integrability, topological recursion, etc.) methods can be applied. Within such methods, it is possible to count maps with restrictions of a global nature (topology, number of vertices, number of polygons of degree $k$), and on the number of boundaries and their perimeters. Combined with probabilistic techniques, they helped in the development of a large corpus of knowledge about the geometric properties of random maps. 

Tutte introduced in \cite{Tutte4} the notion of planar non-separable map, in which faces must all be simple. Some of the combinatorial methods aforementioned have been extended to handle non-separable maps -- see \textit{e.g.} \cite{Nonsep1,Nonsep2} --, but the analytic methods have not been explored and the probabilistic aspects not as much.

Brown, a student of Tutte, studied non-separable maps of arbitrary genus \cite{Nonsep}, which were refined later in \cite{Nonsep0}, distinguishing between the notions of graph-separability and map-separability, which only coincide for planar maps. Maps with only simple faces are still non-separable for arbitrary genus. However, our notion of simplicity is much stronger for non-planar maps than both notions of non-separability. The non-simplicity of separable maps was the easiest type for us to handle, since it could be treated in an analogous way as non-simplicity for planar maps; it is the non-simplicity of non-separable maps the one that reflects the much more intricate structure for higher topologies.

\begin{blockfigure}
\vspace{0.15cm}
		\includegraphics[width=.3\textwidth]{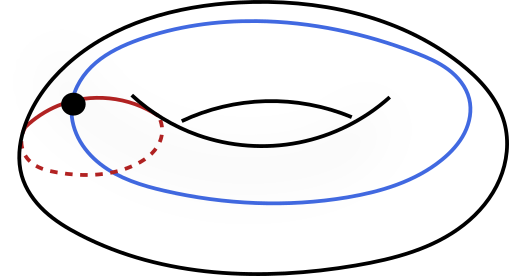}
		\hspace{1.5cm}
		\includegraphics[width=.3\textwidth]{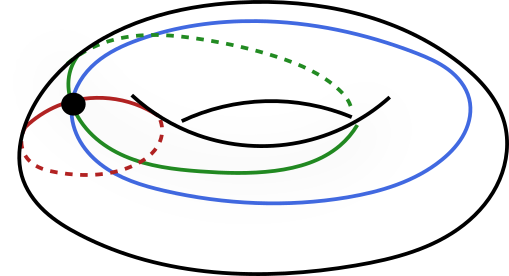}
		\vspace{0.15cm}
		\blockcaption{The two unmarked maps of genus $1$ with one vertex which can be thought as the gluing of a square (left) and the gluing of two triangles (right), which have $|{\rm Aut}| = 4,6$, respectively. The map in Figure \ref{permutations} is separable and the faces are not simple. For genus $0$, a map is non-separable if and only if all faces are simple. This is not true for higher genus. For example, the two maps on this Figure are non-separable, but their faces are not simple. This type of non-simplicity in which removing the problematic vertex does not disconnect the map appears only for higher genus and is more complicated to deal with than the one in separable maps, such as Figure \ref{permutations}, which is the only one that appears for planar maps.}
	\end{blockfigure}

Regarding the planar case, we consider an intermediate problem, \textit{i.e.} the enumeration of maps where only the boundaries are imposed to be simple. This more refined problem for genus $0$, and much more refined for higher genus, is interesting by itself. Actually, when there are several boundaries, we are led to distinguish maps in which each boundary is simple from an even more restrictive kind of maps that we will call fully simple maps. Moreover, we find remarkable combinatorial and algebraic properties that also justify the relevance of this problem a posteriori.

We discovered later\footnote{We thank Timothy Budd for bringing these references to our attention.} that Krikun \cite{Krikun} enumerated planar fully simple triangulations (which he called ``maps with holes''), using a combinatorial identity due to Tutte, and Bernardi and Fusy recently recovered Krikun's formula and also provided an expression for the number of planar fully simple quadrangulations (which they call simply ``maps with boundaries''), via a bijective procedure \cite{BernardiFusy}. The general enumeration problem of fully simple maps that we consider in this thesis regards arbitrary bounded degrees of the internal faces and any genus. We are not aware of other references where some instances of this problem were studied.

\begin{definition}
We call a boundary $B$ \emph{simple} if no more than two edges belonging to $B$ are incident to a vertex. We say that a map is {\it simple} if all boundaries are simple.
\end{definition}

\begin{figure}[h!]
 \begin{center}
        \def\svgwidth{\columnwidth}
        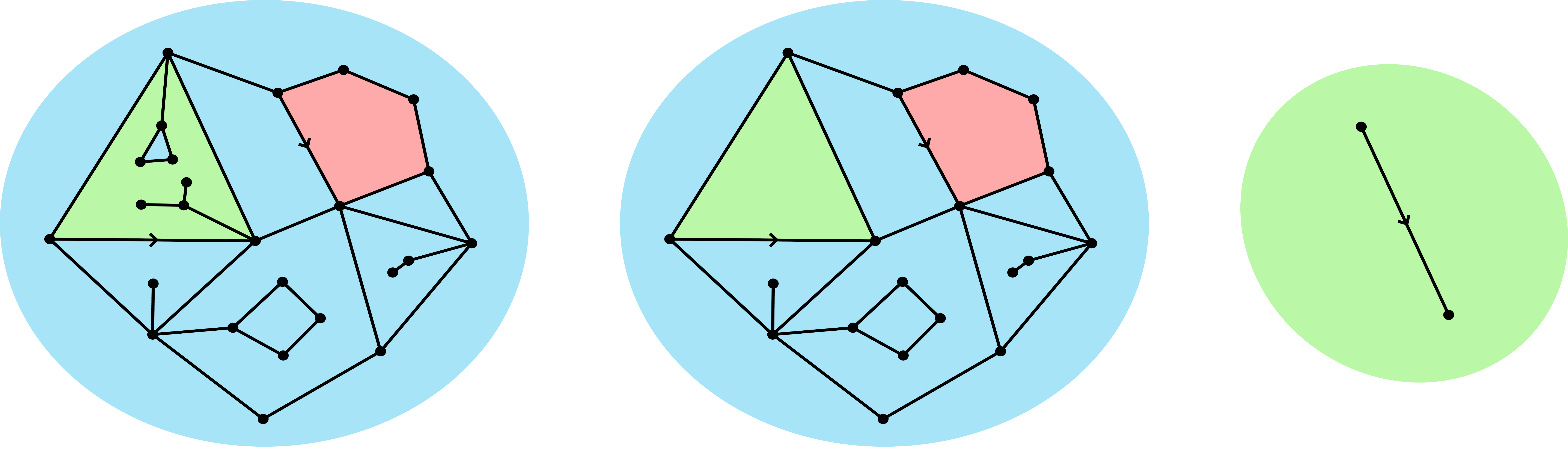
        	\caption{(a) is an ordinary cylinder where the green boundary is non-simple and (b) is a simple cylinder. (c) is the only simple map where two edges in the boundary are identified.}\label{degenerate simple map}
\end{center}
\end{figure}

To acquire an intuition about what this concept means, observe that the condition for a boundary to be simple is equivalent to not allowing its edges to be identified, except for the degenerate case of a boundary of length $2$ where the two edges are identified but is actually considered to be a simple map (see Figure \ref{degenerate simple map}.(c)).

We will call {\it ordinary} the maps introduced in the previous section to emphasize that they are not necessarily simple.

\begin{definition}
We say a boundary $B$ is {\it fully simple} if no more than two edges belonging to any boundary are incident to a vertex of $B$.
We say that a map is {\it fully simple} if all boundaries are fully simple.
\end{definition}

Again, one can visualize the concept of a fully simple boundary as a simple boundary which moreover does not share any vertex with any other boundary.

\begin{figure}[h!]
 \begin{center}
        \def\svgwidth{\columnwidth}
        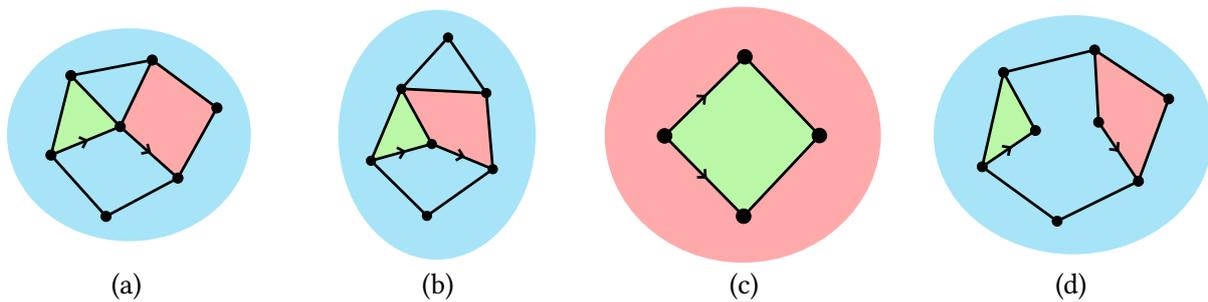
        	\caption{Four simple maps: (a), (b) and (c) are non-fully simple, and (d) is fully-simple. In (a) the two boundaries share a vertex, in (b) an edge, and in (c) they are completely glued to each other.}\label{totally glued}
\end{center}
\end{figure}
\subsection{Generating series}

We introduce now the notations and conventions for the generating series of ordinary, usual maps, and simple and fully simple maps.

Let $\mathbb{M}^{(g,n)}(v)$ be the set of maps of genus $g$ and $n$ boundaries with internal faces of degrees $\geq 3$ and $\leq d <\infty$, and with $v$ vertices.
We take the convention that $\mathbb{M}^{(0,1)}(1)$ contains only one map which consists of a single vertex and no edges; it is the map of genus $0$ with $1$ boundary of length $0$. Apart from this degenerate case, we always consider that boundaries have length $\geq 1$. 

We define the generating series of maps of genus $g$ and $n$ boundaries of fixed lengths $l_1,\ldots,l_n$ as follows:
\beq\label{GenFctF}
F_{l_1,\ldots,l_n}^{[g]}\coloneqq \sum_{v\geq 1} u^v \sum_{\mathcal{M}\in\mathbb{M}^{(g,n)}(v)}\ \  \frac{\prod_{j\geq 3}t_j^{n_j(\mathcal{M})}}{|{\rm Aut}\, \mathcal{M}|}\prod_{i=1}^n \delta_{l_i,\ell_i(\mathcal{M})},
\eeq
where $n_j(\mathcal{M})$ denotes the number of unmarked faces of length $j$ of $\mathcal{M}$ and $\ell_i(\mathcal{M})$ the length of the $i$-th boundary of $\mathcal{M}$. We note that by convention, $F_0^{[0]}=1$. For $n=0$, we denote $F^{[g]}$ the generating series of closed maps of genus $g$. We remark that the dependence on the weights $u,t_j$ will be omitted.


It can be easily checked that $\mathbb{M}^{(g,n)}(v)$ is a finite set, that is the number of maps is finite after fixing the topology $(g,n)$ and the number of vertices $v$. Thus, the generating series $F^{[g]}$ and $F_{l_1,\ldots,l_n}^{[g]}$ are formal power series in $u$ whose coefficients are rational polynomials of the $t_j$'s:
$$
F^{[g]}, F_{l_1,\ldots,l_n}^{[g]} \in \mathbb{Q}\left[t_3,\ldots, t_d\right][[u]].
$$

Summing over all possible lengths, we define the generating series of maps of genus $g$ and $n$ boundaries as follows:
\beq\label{correlators}
W_n^{[g]}(x_1,\ldots, x_n)\coloneqq \sum_{l_1,\ldots,l_n \geq 0} \frac{F^{[g]}_{l_1,\ldots,l_n}}{x_1^{1+l_1}\cdots x_n^{1+l_n}}.
\eeq
We have that $W_n^{[g]}(x_1,\ldots, x_n)\in \mathbb{Q}\left[\frac{1}{x_1},\ldots, \frac{1}{x_n},t_3,t_4,\ldots,t_d\right][[u]]$ and observe that
$$
F_{l_1,\ldots,l_n}^{[g]} = (-1)^n \underset{x_1\rightarrow\infty}{{\rm Res}}\cdots \underset{x_n\rightarrow\infty}{{\rm Res}} x_1^{l_1}\cdots x_n^{l_n} W_n^{[g]}(x_1,\ldots,x_n) \dd x_1\cdots \dd x_n.
$$

We denote $H^{[g]}_{k_1,\ldots,k_n}$ the analogous generating series for fully simple maps of genus $g$ and $n$ boundaries of fixed lengths $k_1,\ldots,k_n$ and we introduce the following more convenient generating series for fully simple maps with boundaries of all possible lengths:
\beq\label{FScorrelators}
X_{n}^{[g]}(w_1,\ldots,w_n) \coloneqq \sum_{k_1,\ldots,k_n \geq 0} H^{[g]}_{k_1,\ldots,k_n} w_1^{k_1 - 1} \ldots w_n^{k_n - 1}.
\eeq

Finally, we denote $G^{[g]}_{k_1,\ldots,k_m | l_1, \ldots, l_n}$ the generating series of maps with $m$ simple boundaries of lengths $k_1,\ldots, k_m$ and $n$ ordinary boundaries of lengths $l_1, \ldots, l_n$. We write
\beq\label{Scorrelators}
Y_{m|n}^{[g]}(w_1,\ldots,w_m\mid x_1,\ldots,x_n) = \sum_{(\mathbf{k},\mathbf{l}) \in \mathbb{N}^{m}\times\mathbb{N}^n} \frac{w_1^{k_1 - 1}\cdots w_m^{k_m - 1}}{x_1^{l_1 + 1}\cdots x_n^{l_n + 1}}\,G_{\mathbf{k}|\mathbf{l}}^{[g]}.
\eeq

We use the following simplification for maps with only simple boundaries: $G^{[g]}_{k_1,\ldots,k_m}$ for $G^{[g]}_{k_1,\ldots,k_m |}$, and $Y_{m}^{[g]}$ for $Y_{m|}^{[g]}$.

Observe that for maps with only one boundary the concepts of simple and fully simple coincide. Therefore $G^{[g]}_{k}=H^{[g]}_{k}$ and $Y_{1}^{[g]}=X_{1}^{[g]}$.

For all the generating series introduced we allow to omit the information about the genus in the case of $g=0$. We also use the simplification of removing the information about the number of boundaries if $n=1$. In this way, $W$ and $X$ stand for $W_1^{[0]}$ and $X_1^{[0]}$.

\subsubsection{Alternative definitions}

We consider now $\mathbb{M}_{l_1,\dots,l_n}^{(g,n)}$ to be the set of maps of genus $g$ with $n$ boundaries of fixed lengths $l_1,\ldots,l_n$.

One can notice that the parameter $u$ keeping track of the number of vertices is redundant, since fixing the genus and boundary lengths, the volume $v$ can be deduced from the number of internal faces $n_j$ of every possible length $j\geq 1$. Therefore, we could choose to define the generating series of maps with $u=1$ as follows:
\beq\label{u=1}
F_{l_1,\ldots,l_n}^{[g]}\coloneqq \sum_{\mathcal{M}\in\mathbb{M}_{l_1,\dots,l_n}^{(g,n)}}\ \  \frac{\prod_{j\geq 1}t_j^{n_j(\mathcal{M})}}{|{\rm Aut}\, \mathcal{M}|}.
\eeq
We remark that $\mathbb{M}_{l_1,\dots,l_n}^{(g,n)}$ is not finite, but we still have
\beq
F^{[g]}, F_{l_1,\ldots,l_n}^{[g]} \in \mathbb{Q}\left[[t_1, t_2,\ldots\right]],
\eeq
that is the number of maps is finite after fixing the topology $(g,n)$ and the number of internal faces $n_j$ of every possible length $j\geq 1$.

In this case, we also get $W_n^{[g]}(x_1,\ldots, x_n)\in \mathbb{Q}\left[\frac{1}{x_1},\ldots, \frac{1}{x_n}\right][[t_1, t_2,\ldots]]$.

Observe that with this alternative definition we allow internal faces to have any degree. If we set $t_1, t_2=0$ in \eqref{u=1} and $u=1$ in \eqref{GenFctF}, both definitions give the same generating series.

However, very often it is more convenient to work with formal series in just one formal variable $u$ than with multiple formal variables. So we will keep the original definition in general and specify in which cases we work with this other definition for convenience.

Furthermore, despite having required unmarked faces to have length $\geq 3$ and $\leq d <\infty$, this restriction will not be necessary for many manipulations. In some cases, we may even be forced to remove it. Again, we shall keep it in general and indicate the special cases in which we will drop it. In those cases, we will work with the definition introduced here \eqref{u=1}, possibly allowing $t_1, t_2\neq 0$.

This alternative definition and all these remarks work analogously for all the other generating series that we have introduced.

\subsection{Stuffed maps}\label{stuffed}

We introduce stuffed maps as in \cite{Bstuff}, which encompass usual maps since by substitution one may consider stuffed maps as maps whose elementary cells are themselves maps. 
\begin{definition}
An {\it elementary 2-cell} of genus $h$ and $k$ boundaries of lengths $m_1,\ldots, m_k$ is a connected orientable surface of genus $h$ with boundaries $B_1, \ldots, B_k$ endowed with a set $V_i\subset B_i$ of $m_i\geq 1$ vertices. The connected components of $B_i\setminus V_i$ are called {\it edges}. We require that each boundary has a marked edge, called the root, and by following the cyclic order, the rooting induces a labeling of the edges of the boundaries. We say that such an elementary $2$-cell is {\it of topology} $(h,k)$.

A {\it stuffed map} of genus $g$ and $n$ boundaries of lengths $l_1,\ldots, l_n$ is the object obtained from gluing $n$ labeled elementary 2-cells of topology $(0,1)$ with boundaries of lengths $l_1,\ldots,l_n$, and a finite collection of unlabeled elementary 2-cells by identifying edges of opposite orientation and with the same label in such a way that the resulting surface has genus $g$. The labeled cells are considered as {\it boundaries} of the stuffed map, and the marked edges which do not belong to the boundary are forgotten after gluing. The unlabelled cells are again referred to as \emph{internal faces}.
\end{definition}

\begin{remark}
A map in the \emph{usual} sense is a stuffed map composed only of elementary 2-cells with the topology of a disk. Therefore, stuffed maps could alternatively be defined with the same kind of definition we gave for usual maps in Definition \ref{usual_map} as embedded graphs, just waiving the condition that faces are homeomorphic to disks. 

Stuffed maps are also considered up to cellular continuous deformation preserving the roots of the boundaries.
\end{remark}

We denote $\widehat{\mathbb{M}}^{(g,n)}(v)$ the set of stuffed maps of genus $g$ and $n$ boundaries of fixed number of vertices $v$ (and with the internal faces of the topology of a disk of degrees $3\geq j\geq d$ for some $d<\infty$).

To every stuffed map $\mathcal{M}$ we assign a Boltzmann weight, which we denote $w(\mathcal{M})$ with the following factors: 
\begin{itemize}
\item a symmetry factor $\left|{\rm Aut}(\mathcal{M})\right|^{-1}$ as previously for maps,
\item a weight $t_{m_1,\ldots,m_k}^h$ per unlabeled elementary 2-cell of genus $h$ and $k$ boundaries, depending symmetrically on the lengths $\bold{m}=(m_1,\ldots,m_k)$.
\end{itemize}

Slightly extending the notation for usual maps, we add a hat to the previous symbols to denote the generating series of stuffed maps of topology $(g,n)$:
\beq\label{Stuffedcorrelators}
\widehat{W}_n^{[g]}(x_1,\ldots,x_n)= \sum_{v\geq 1} u^v\sum_{\mathcal{M}\in\widehat{\mathbb{M}}^{(g,n)}(v)} \frac{w(\mathcal{M})}{x_1^{\ell_1(\mathcal{M})+1}\cdots x_n^{\ell_n(\mathcal{M})+1}}.
\eeq
We have that $\widehat{W}_n^{[g]}(x_1,\ldots,x_n)\in \mathbb{Q}\left[(x_j^{-1})_j, (t^h_{\bold{m}})_{\bold{m},h}\right][[u]]$.

A slight generalization of the permutational model for maps also works for stuffed maps. Let $F$ be the number of unlabeled elementary $2$-cells, considered as the internal faces of the stuffed map. A {\it combinatorial stuffed map} $((\sigma,\alpha),\bigsqcup^F_{p=1} f_p, (h_p)_{p=1}^F)$ consists of the following data:
\begin{itemize}
\item As for maps, a pair of permutations $(\sigma,\alpha)$ on the set of half-edges $H = H^u\sqcup H^{\partial}$, where $\alpha$ is a fixed-point free involution whose cycles represent the edges of the stuffed map, and $\mathcal{C}(\sigma)$ corresponds to the set of vertices. 
The cycles of $\varphi \coloneqq (\sigma\circ\alpha)^{-1}$ are associated to the boundaries of elementary $2$-cells.
\item A partition $\bigsqcup^F_{p=1} f_p$ of $\mathcal{C}(\left.\varphi\right|_{H^u})$, where every part $f_p$ corresponds to an unlabeled elementary $2$-cell with boundaries given by the cycles in $f_p$.
\item A sequence of non-negative integers $(h_p)_{p=1}^F$, where every $h_p$ is the genus of the unlabeled elementary $2$-cell $f_p$.
\end{itemize}

To define the notion of connectedness for stuffed maps, we consider the following equivalence relation on the set of half-edges:
\begin{itemize}
\item[$(i)$] $h \sim \sigma(h)$ and $h \sim \alpha(h)$,
\item[$(ii)$] if $h,h'$ are in two cycles $c,c' \in f_{p}$ for some $p$, then $h \sim h'$.
\end{itemize}
Each equivalence class on $H$ corresponds to a connected component of the stuffed map. We say that a stuffed map  is $\partial$-{\it connected} if each equivalence class has a non-empty intersection with $H^{\partial}$. Observe that the notion of connectedness for maps relies on the equivalence class generated only by $(i)$.

Since the concepts of simplicity and fully simplicity that we introduced for maps in Subsection \ref{fsmaps} only refer to properties of the boundaries, they clearly extend to stuffed maps because the elementary 2-cells corresponding to boundaries in stuffed maps are imposed to be of the topology of a disk, as for maps.

\subsection{The combinatorial $O(\mathsf{n})$ loop model}\label{CombOnModel}
The $O(\mathsf{n})$ model admits a famous representation in terms of loops \cite{DoMuNiSc81,NienhuisCG} with $\mathsf{n}$ the fugacity per loop.

In this context we may also consider maps with $k'$ marked points, and by convention we do not assume that the $k'$ marked points necessarily sit on pairwise distinct vertices. We call \emph{marked element} (or \emph{mark} for short) either a marked point or a marked face (boundary).

\begin{figure}[htpb]
  \centering
  \includegraphics[width=.3\textwidth]{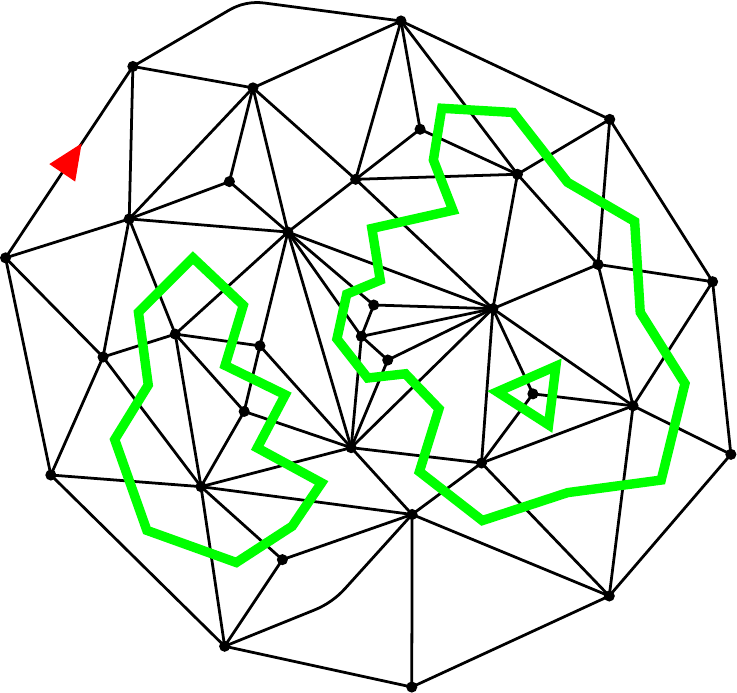}
  \caption{A planar triangulation with a boundary of length $10$ (with
    root in red, the distinguished face being the outer face), endowed
    with a loop configuration (drawn in green).}
  \label{fig:myloopconfig_all}
\end{figure}

\begin{definition}
A \emph{loop} is an undirected simple closed path on the dual map (i.e. it covers edges and vertices of the dual map, and
hence visits faces and crosses edges of the original map) which does not visit any boundary. A \emph{loop configuration} is a collection of disjoint loops. A
\emph{configuration} of the $O(\mathsf{n})$ loop model is a map endowed with a loop configuration.
\end{definition}
Our notion of loop is not to be confused with the graph-theoretical notion of loop (edge incident twice to the same vertex which we previously called self-loop to avoid confusion). A loop configuration can be viewed alternatively as a collection of \emph{crossed} edges such that every face of the map is incident to either $0$ or $2$ crossed edges. See Figure~\ref{fig:myloopconfig_all} for an example of a configuration of the $O(\mathsf{n})$ loop model.

We also employ the name \emph{usual map} here for a map without a loop configuration.
Usual maps carrying self-avoiding loop configurations are equivalent to stuffed maps for which we allow unlabeled elementary 2-cells to have the topology of a disk (usual faces) or of a cylinder (rings of faces carrying the loops). By equivalence, we mean here an equality of generating series after a suitable change of formal variables.

\begin{remark}
  In the original formulation of \cite{GaudinKostov,KOn,KSOn,EKOn},
  the loops cover vertices and edges of the map itself. Our motivation
  for drawing them on the dual map is that it makes our combinatorial
  decompositions easier to visualize.
\end{remark}

\subsubsection{Permutational model and automorphisms}
To extend the permutational model that we defined for usual maps to configurations of the $O(\mathsf{n})$ loop model, we need to include also the information of the loop configuration. We consider the loop configuration as a collection of crossed edges. For every loop, we pick one of the two possible sides of the loop as the ``inside''. Our construction will of course be independent of which side we choose. We want to transform the collection of crossed edges forming the loop into a subset of half-edges, since the permutations act on the set of half-edges $H$. 

We choose by convention that the half-edges forming the loop will be the half-edges which belong to the crossed edges of the loop and sit on the left seen from inside (here we are imagining the map depicted as a ribbon graph). All the half-edges corresponding to a loop configuration will be collected in a set denoted $H^o$. Note that interchanging the outsides and the insides, this set transforms to $\alpha(H^o)$ and that by definition of loop configurations, $\alpha(H^o)\subset H^u$. Now we are ready to encode the loop configuration with a fourth permutation $\theta$ acting on $H^o\subset H$. Given $h\in H^o$, we define $\theta(h)$ as the next half-edge in the loop following a fixed orientation from the inside (by convention counterclockwise). Observe that, by construction of the loops, there exists $l_1\geq 0$ such that 
\beq\label{theta}
\theta(h)=\alpha \varphi^{l_1+1}(h).
\eeq
The cycles of $\theta$ correspond to loops in the configuration. 

To make the construction independent of the choices, we identify $(\theta,H^o)\sim (\alpha\theta^{-1}\alpha^{-1},\alpha(H^o))$. A \emph{combinatorial configuration} $((\sigma,\alpha),[(\theta,H^o)])$ consists of a combinatorial map $(\sigma,\alpha)$ and an equivalence class $[(\theta,H^o)]$ encoding the information about loops as we just described. As before, we need to identify configurations that differ by a relabeling of $H^u$, i.e., 
$$
((\sigma,\alpha),[(\theta,H^o)])\sim((\gamma\sigma\gamma^{-1},\gamma\alpha\gamma^{-1}),[(\gamma\theta\gamma^{-1},\gamma H^o)])
$$ 
with $\gamma$ any permutation acting on $H$ such that $\left.\gamma\right|_{H^{\partial}}={\rm Id}_{H^{\partial}}$.
We call such an equivalence class {\it unlabeled combinatorial configuration} and we denote it by $[(\sigma,\alpha),(\theta,H^o)]$. Note that unlabeled combinatorial configurations are in bijection with the (unlabeled) configurations we defined at the beginning of this section.

\begin{definition}\label{autLoopModel}
Given a combinatorial configuration $((\sigma,\alpha),[(\theta,H^o)])$ acting on $H$, we call $\gamma$ an {\it automorphism} if it is a permutation acting on $H$ such that $\left.\gamma\right|_{H^{\partial}}={\rm Id}_{H^{\partial}}$ and
$$
\sigma = \gamma\sigma\gamma^{-1}, \ \ \ \ \ \ \alpha=\gamma\alpha\gamma^{-1}, \ \ \ \ \ \ \theta=\gamma\theta\gamma^{-1}.
$$
\end{definition}
Observe that because of \eqref{theta}, if $\gamma\in{\rm Aut}(\sigma,\alpha)$ and for all $h\in H^o$, we have $\gamma\theta\gamma^{-1}(h)\in H^o$, then we automatically have that $\theta=\gamma\theta\gamma^{-1}$. So the condition $\theta=\gamma\theta\gamma^{-1}$ can be relaxed to: $\gamma\theta\gamma^{-1}(H^o)=H^o$.

Again, for connected maps with $k\geq 1$ boundaries, the only automorphism is the identity.
As before, these special relabelings that commute with $\sigma$ and $\alpha$, and preserve the loops, which we called automorphisms, exist because of a symmetry of the (unlabeled) configuration $C=[(\sigma,\alpha),(\theta,H^o)]$, i.e. of the underlying map together with the loop configuration. The symmetry factor $\left|{\rm Aut}(C)\right|$ of a configuration is its number of automorphisms. 

Finally, let us also comment how to include in this formalism configurations with $k^{\prime}\geq 1$ marked points. Let $H^{\bullet}_i$ be the subset of half-edges incident to the $i$-th marked point, for $i=1,\ldots,k^{\prime}$. By definition of maps with marked points, some of these subsets can coincide. We consider 
$$
((\sigma,\alpha),[(\theta,H^o)], (H^{\bullet}_i)_{i=1}^{k^{\prime}})
$$ 
a \emph{combinatorial configuration with $k^{\prime}$ marked points}. Including the extra subsets $H^{\bullet}_i$ is equivalent to marking the corresponding cycles in $\sigma$: 
$$
\sigma_i^{\bullet}\coloneqq \left.\sigma\right|_{H^{\bullet}_i}, \text{ for } i=1,\dots,k^{\prime}. 
$$
We call $\gamma$ an \emph{automorphism} of $((\sigma,\alpha),[(\theta,H^o)], (H^{\bullet}_i)_{i=1}^{k^{\prime}})$ if we require that $\gamma\sigma_i^{\bullet}\gamma^{-1}=\sigma_i^{\bullet}$, for $i=1,\ldots,k^{\prime}$, additionally to the conditions imposed in Definition \ref{autLoopModel}. Since $\sigma = \gamma\sigma\gamma^{-1}$, the restriction  $\gamma\sigma_i^{\bullet}\gamma^{-1}=\sigma_i^{\bullet}$ can also be relaxed to $\gamma\sigma_i^{\bullet}\gamma^{-1}(H_i^{\bullet})=H_i^{\bullet}$.

\subsubsection{The nesting graphs}
\label{Markm}

%
%

Given a configuration $\mathcal{C}$ of the $O(\mathsf{n})$ loop model on a map $\mathcal{M}$ of genus $g$, we may cut the underlying surface
along every loop, which splits it into several connected components
$c_1,\ldots,c_N$. Let $\Gamma_0$ be the graph on the vertex set
$\{c_1,\ldots,c_N\}$ in which there is an edge between $c_i$ and $c_j$ if and only if they have a common boundary, \textit{i.e.}\ they touch each other
along a loop (thus the edges of $\Gamma_0$ correspond to the loops of $\mathcal{C}$). We assign to each vertex $\mathsf{v}$ the genus $h(\mathsf{v})$ of the corresponding connected component and for each marked element in $\mathcal{M}$ belonging to 
a connected component $c_i$, we put a mark on the  corresponding vertex of $\Gamma_0$. If the map is planar, $\Gamma_0$ is a tree and all its vertices carry genus $0$. We call $\Gamma_0$ the \emph{primary nesting graph} of $\mathcal{M}$.


We define the \emph{nesting graph} $\Gamma$ from $\Gamma_0$ by repeatedly performing the following two steps until they leave the graph unchanged:
\begin{itemize}
\item[$(i)$] erasing all vertices that correspond to connected components which, in the complement of all loops in $\mathcal{M}$, are homeomorphic to disks, and the edge incident to each of them; 
\item[$(ii)$] replacing any maximal simple path of the form $v_0 - v_1 - \cdots  - v_{P}$ with $P\geq 2$, where the vertices $(v_i)_{i = 1}^{P-1}$ represent connected components homeomorphic to cylinders, by a single edge
$$
v_0 \mathop{-}^{P} v_{P}
$$
carrying a length $P$. By convention, edges which are not obtained in this way carry a length $P = 1$. We call $P$ the \emph{depth} of the edge or, alternatively, the \emph{arm length} of the arm in the configuration corresponding to this edge in $\Gamma$.
\end{itemize}
The outcome is $(\Gamma,\star,\mathbf{P})$, where $\Gamma$ is the \emph{nesting graph}, which is connected and has vertices labeled by genera such that
$$  
g = b_1(\Gamma) + \sum_{\mathsf{v} \in V(\Gamma)} h(\mathsf{v}).
$$


\begin{figure}[h!]
\begin{center}
\def\svgwidth{0.9\columnwidth}
 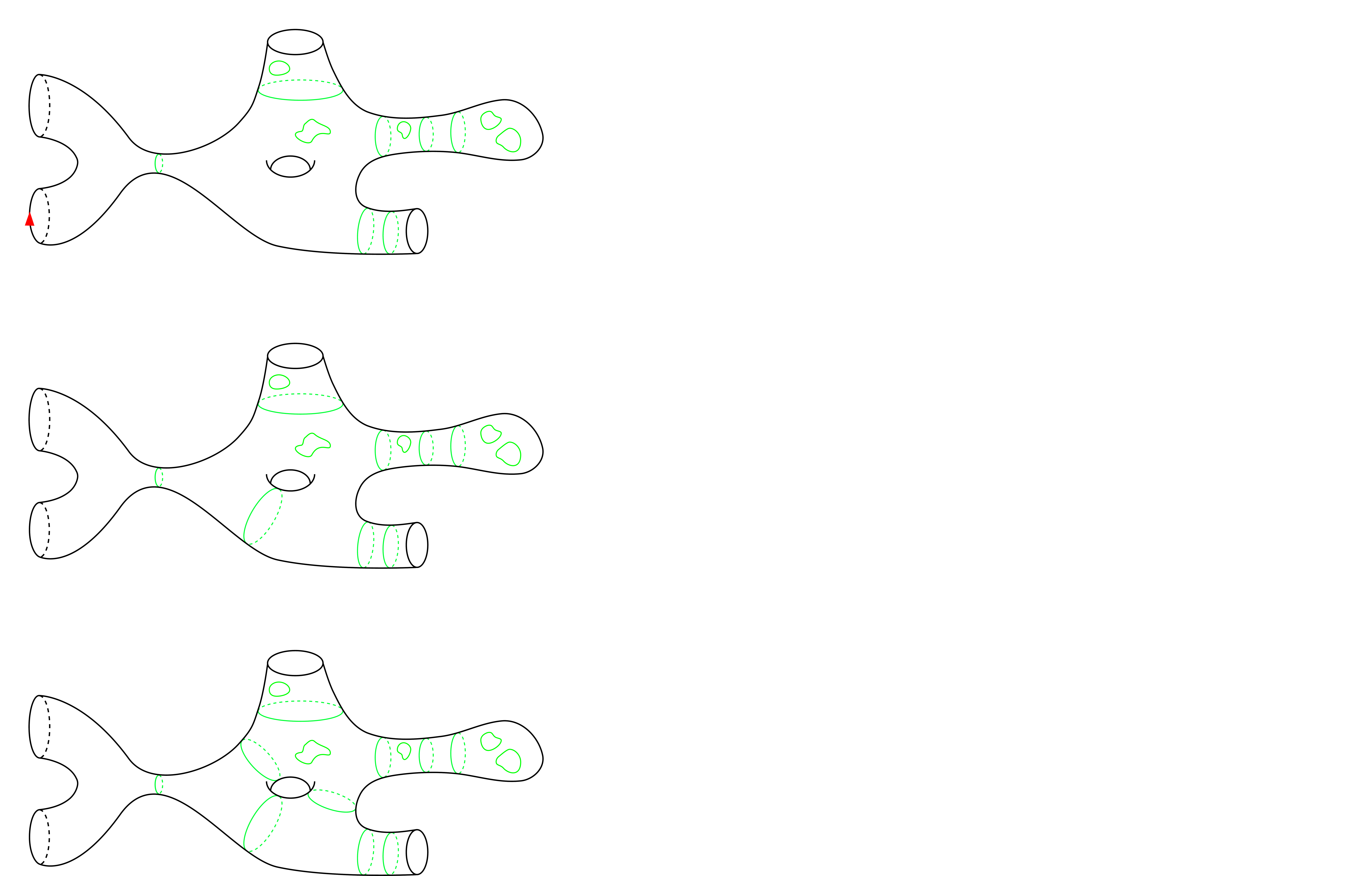
 \caption{  \label{fig:nestinggraph} Left: schematic representations of loop configurations on a
    map of genus $1$ with $4$ boundaries. Center: associated primary nesting graphs, where every
    red vertex carries the marks of the boundaries which belong to the corresponding
    connected component in the map. Right: associated nesting graphs,
    where every edge is labelled with its depth. All vertices carry genus $0$, except
    $\mathsf{v}$ in the first case which has $h(\mathsf{v})=1$.}
    \end{center}
\end{figure}

\begin{definition}\label{separating}
In a configuration $\mathcal{C}$ with a non empty set of marked elements $\mathcal{E}$, a loop is \emph{separating} if it is not contractible in $\mathcal{C}\setminus \mathcal{E}$.
\end{definition}

The sequence of depths $\mathbf{P}$ records the number of consecutive separating loops for each edge. By construction, given the total genus $g$ and a finite set of marked elements, one can only obtain finitely many inequivalent nesting graphs. Finally, $\star$ is the assignment of the marked elements of $\mathcal{M}$ to the vertices of $\Gamma$. The valency $d(\mathsf{v})$ of vertices $\mathsf{v}$ of the nesting graph $\Gamma$ with no boundaries must satisfy:
$$
2h(\mathsf{v}) - 2 + d(\mathsf{v}) > 0.
$$
Observe that every edge of the primary nesting graph $\Gamma_0$ represents a loop from the configuration, while every edge of the nesting graph $\Gamma$ corresponds to a sequence of separating loops.

\subsubsection{Boltzmann weights}
\label{bendI}

In the $O(\mathsf{n})$ loop model we introduce a weight $\mathsf{n}$ to keep track of the number of loops.
In addition to this ``nonlocal'' parameter, we need also some
``local'' parameters, controlling in particular the size of the maps
and of the loops. Precise instances of the model can be defined in
various ways.

The origin of the $O(\mathsf{n})$ loop model is in statistical physics and even if here we are defining a deterministic model, we refer to instances on maps with some notion of randomness that we will make precise later.

The simplest instance is the \emph{$O(\mathsf{n})$ loop model on random
  triangulations} \cite{GaudinKostov,KOn,KSOn,EKOn}: here we require
the underlying map to be a triangulation, possibly with boundaries and marked points.
There are two local parameters $\mathsf{g}$ and $\mathsf{h}$, which are the weights per
face (triangle) distinct from a boundary and which is, respectively, not visited and visited
by a loop. The Boltzmann weight attached to a configuration $\mathcal{C}$ with $k\geq 1$ boundaries is
thus $w(\mathcal{C})=\mathsf{n}^{\mathcal{L}} \mathsf{g}^{T} \mathsf{h}^{T'}$, with $\mathcal{L}$ the number of loops of $\mathcal{C}$,
$T$ its number of unvisited triangles and $T'$ its number of
visited triangles.

\begin{figure}[h!]
\begin{center}
\def\svgwidth{0.5\columnwidth}
 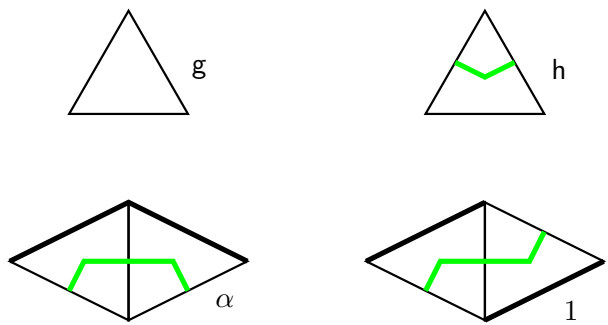
  \caption{Top row: local weights for the $O(\mathsf{n})$ loop model on random
    triangulations. Bottom row: in the bending energy model, an extra
    weight $\alpha$ is attached to each segment of a loop between two
    successive turns in the same direction.}
  \label{fig:curv}
  \end{center}
\end{figure}

A slight generalization of this model is the \emph{bending energy
  model} \cite{BBG12b}, where we incorporate in the Boltzmann weight
$w(\mathcal{C})$ an extra factor $\alpha^B$, where $B$ is the number of pairs of
successive loop turns in the same direction, see
Figure~\ref{fig:curv}. Another variant is the $O(\mathsf{n})$ loop model on
random quadrangulations considered in \cite{BBG12a} (and its ``rigid''
specialization).

In the general $O(\mathsf{n})$ loop model, the Boltzmann weight of a configuration is:
\beq\label{Bweight}
w(\mathcal{C}) = \frac{1}{|{\rm Aut}\,C|}\,\mathsf{n}^{\mathcal{L}(\mathcal{C})} \prod_{l \geq 3} t_{l}^{N_{l}(\mathcal{C})} \prod_{\substack{\{l_1,l_2\} \\ l_1 + l_2 \geq 1}} t_{l_1,l_2}^{N_{l_1,l_2}(\mathcal{C})},
\eeq
where $N_l$ is the number of unvisited faces of degree $l$, and $N_{l_1,l_2}$ is the number of visited faces of degree $(l_1 + l_2 + 2)$ whose boundary consists, in cyclic order with an arbitrary orientation, of $l_1$ uncrossed edges, $1$ crossed edge, $l_2$ uncrossed edges and $1$ crossed edge. As the loops are not oriented here, $N_{l_1,l_2} = N_{l_2,l_1}$ and we also assume $t_{l_1,l_2} = t_{l_2,l_1}$. 


\subsubsection{Generating series}

We now define the basic generating series of interest. Fixing three integers $k,k^{\prime} \geq 0$ and $g \geq 0$, we consider the ensemble of allowed configurations of the $O(\mathsf{n})$ model where the underlying map is a connected surface of genus $g$, with $k$ boundaries of respective lengths $\ell_1,\ell_2,\ldots,\ell_k \geq 1$ (called perimeters) and $k'$ marked points. The corresponding generating series is then the sum of the Boltzmann weights $w(\mathcal{C})$ of all such
configurations. We find convenient to add an auxiliary weight $u$ per
vertex, and define
\begin{equation}
  \label{eq:Fdef}
  \mathcal{F}^{[g,\bullet k^{\prime}]}_{\ell_1,\ldots,\ell_k} \coloneqq \delta_{k,1}\delta_{\ell_1,0}\,u + \sum_{\mathcal{C}} u^{|\mathcal{V}(\mathcal{C})|} w(\mathcal{C}),
\end{equation}
where the sum runs over all desired configurations $\mathcal{C}$, here over configurations of genus $g$, $k^{\prime}$ marked points and $k$ boundaries of fixed lengths $\ell_1,\ldots,\ell_k$, and $|\mathcal{V}(\mathcal{C})|$ denotes the number of vertices of the underlying map of $\mathcal{C}$, also called \emph{volume}. Sometimes we use the more complete notation $\mathcal{F}^{[g,k,\bullet k^{\prime}]}_{\ell_1,\ldots,\ell_k}$ indicating also the number of boundaries, even if that information is already contained in the number of fixed lengths we give. We simply write $\mathcal{F}^{[g]}_{\ell_1,\ldots,\ell_k}$ ($\mathcal{F}_{\ell_1,\ldots,\ell_k}$ for $g=0$) when there are no marked points, and $\mathcal{F}^{[g,\bullet]}_{\ell_1,\ldots,\ell_k}$ ($\mathcal{F}^{\bullet}_{\ell_1,\ldots,\ell_k}$ when $g=0$) when there is one marked point.
We also call \emph{cylinders} the planar maps with $k = 2$ boundaries, and \emph{disks} the planar maps with $k = 1$ boundary endowed with $O(\mathsf{n})$ loop models. The first term in \eqref{eq:Fdef} accounts for the map consisting of a single vertex in the sphere.

In the course of studying the $O(\mathsf{n})$ loop model, we will also need the generating series of usual maps, in this context with internal faces of length $l\geq 1$. The Boltzmann weight of a configuration in this case is chosen to be
\beq
\label{usualmap} w(\mathcal{M}) = \frac{1}{|{\rm Aut}\,\mathcal{M}|}\,\prod_{l \geq 1} t_l^{N_{l}(\mathcal{M})},
\eeq
and the generating series $F^{[g,\bullet k']}_{\ell_1,\ldots,\ell_k}$ is defined as previously where this time we sum over configurations without loops, i.e. over usual maps. Observe that for $k'=0$ this is the same generating series introduced for usual maps in \eqref{GenFctF}, but here we have added a weight $u$ per vertex. It will be clear from the context when we consider this weight specialized to $1$ and do not keep track of the number of vertices, and when make use of this more refined version. We employ the same name for simplicity.

\section{Hurwitz numbers}\label{IntroHNs}

The goal of this section is not only to introduce the different enumeration problems that Hurwitz numbers solve, but also to give an idea of the relations among them, which give the possibility to involve different intuitions to study them, making the subject very rich and interesting. For an elementary, nice introduction to these different points of view, see for example \cite{LandoZvonkin}. For an also elementary, but much more complete, introduction to standard Hurwitz theory (including the enumeration of coverings of a surface of genus greater than $0$), see for example \cite{CavalieriMiles}.

Let us first fix some notations regarding partitions of an integer. If $\lambda$ is a partition of an integer $L>0$, we denote $|\lambda|$ the sum of its elements (size) and $\ell(\lambda)$ the number of elements in $\lambda$ (length). Associated to a permutation $\beta\in\mathfrak{S}_L$, with conjugacy class $[\beta]$, we can form a partition $\lambda=\lambda_{[\beta]}$ -- by collecting the lengths of cycles in $\beta$ -- for which we have $|\mathcal{C}(\beta)| = \ell(\lambda_{[\beta]})$. We also denote $t(\lambda_{[\beta]})=t(\beta) = t([\beta]) = L - \ell(\lambda_{[\beta]})$, which can be checked to be the minimal number of transpositions in a factorization of $\beta$. Recall that actually the set of conjugacy classes in $\mathfrak{S}_{L}$ is in bijection with the set of partitions of $L$. We use the notation $C_{\lambda}$ for the conjugacy class in $\mathfrak{S}_{L}$ described by the partition $\lambda$. We denote
$$
|{\rm Aut}\,\lambda |:= \frac{L!}{|C_{\lambda}|},\qquad L = |\lambda|.
$$
For a permutation $\beta$, one also defines $|{\rm Aut}\,\beta|\coloneqq |{\rm Aut}\,\lambda_{[\beta]}|$. For a transposition $\tau$ of $a$ and $b$, we denote ${\rm max}\,\tau = {\rm max}\,(a,b)$.

\subsection{Topological definition}

The most natural context for ramified coverings is complex analysis, where one has much more structure. For this reason, we consider $X$ to be a compact Riemann surface. Actually, non-constant maps between Riemann surfaces can be identified with ramified coverings. However, one can see that ramified coverings encode purely topological information.
\begin{definition}
Consider $y_1,\ldots,y_{m}\in\C P^1$ pairwise distinct points and let $\lambda_1,\ldots,\lambda_m$ be partitions of a non-negative integer $L$. The {\it Hurwitz number} $h^{\circ,g}_{\lambda_1,\ldots,\lambda_m}$ is the weighted number of isomorphism classes of ramified coverings $\pi: X \to \C P^1$:
\beq 
h^{\circ,g}_{\lambda_1,\ldots,\lambda_m} = \sum_{[\pi]} \frac{1}{|{\rm Aut}\, \pi|},
\eeq
where
\begin{enumerate}[itemsep=0\baselineskip]
\item X is connected and has genus $g$,
\item the branch locus of $\pi$ is $Y=\{y_1,\ldots,y_{m}\}$,
\item the ramification profile of $\pi$ at $y_j$ is $\lambda_j$.
\end{enumerate}
\end{definition}
The Riemann existence theorem ensures that this enumeration is independent of the configuration of the branch points, and hence Hurwitz numbers are well-defined and symmetric on the $\lambda_j$'s.
The Euler characteristic of the total space is given by the Riemann-Hurwitz formula:
$$
2-2g=\chi(X) = 2L - \sum_{j = 1}^{m} t(\lambda_j).
$$
If $X$ has $s$ connected components of genera $g_1, \ldots, g_s$, we define its genus as
$$
g(X)\coloneqq g_1 + \ldots + g_s + 1 -s.
$$
If we allow $X$ to be disconnected in the previous definition, we denote the corresponding Hurwitz number by $h^{\bullet,g}_{\lambda_1,\ldots,\lambda_m}$. With the appropriate definition of genus we considered, the Riemann-Hurwitz formula remains unchanged for disconnected covers.

\subsection{Monodromy representations}

Choose a base-point $p\in \C P^1 \setminus Y$ and a bijection 
$$
\varphi: \pi^{-1}(p) \to \{1,\ldots,L\}
$$ 
labeling its preimages. We call a pair $(\pi, \varphi)$ $p$\text{\it -labeled} covering. An isomorphism $f: (\pi_1,\varphi_1)\to (\pi_2,\varphi_2)$ of $p$-labeled coverings should additionally satisfy: $\varphi_2\circ f=\varphi_1$. Choosing a labelling allows us to encode the monodromy action of $\gamma\in\pi_1( \C P^1\setminus Y, p)$ on $\pi^{-1}(p)$ with a permutation $\sigma_{\gamma}\in\mathfrak{S}_{L}$.

Loops $\gamma_j$ around every special point $y_j$ in $Y$ give a presentation of the fundamental group:
$$
\pi_1(\C P^1 \setminus Y) = \langle\gamma_1,\ldots,\gamma_{m} \mid  \gamma_1\cdots\gamma_{m}\rangle.
$$

The correspondence $\rho: \gamma  \mapsto \sigma_{\gamma}$ is a well-defined group homomorphism, which we call {\it monodromy representation of type} $(\lambda_1,\ldots,\lambda_m)$ if $\rho([\gamma_j]) \in C_{\lambda_j}$ for every $j$. 

Observe that monodromy representations are purely topological constructions.

The monodromy representation associated to a $p$-labeled ramified covering with ramification profile over $y_j$ given by $\lambda_j$ is of type $(\lambda_1,\ldots,\lambda_m)$ by construction, and the $\beta_j\coloneqq\rho(\gamma_j)$ satisfy $\beta_m \circ \cdots \circ \beta_{1} = {\rm id}$.

Moreover, one can check that two isomorphic $p$-labeled ramified coverings give rise to the same monodromy representations.

This important correspondence is actually a bijection, which allows carrying the Hurwitz problem of counting ramified coverings of the sphere into a group-theoretic setting.
$$
\begin{array}{ccc}
\left\{ \begin{array}{c}p\text{-labeled ramified coverings} \\  \pi: X \to \C P^1 \text{ with branch locus } Y  \\ \text{and ramification profile } \lambda_j \text{ over } y_j \end{array}\right\}_{\biggn/{\sim}}  & \leftrightarrow & \left\{ \begin{array}{c}\rho: \pi_1( \C P^1\setminus Y, p) \to \mathfrak{S}_{L} \\ \text{ monodromy representations } \\ \text{of type } (\lambda_1,\ldots,\lambda_m) \end{array} \right\}.
\end{array}
$$

We sketch the idea behind the inverse construction. Given a monodromy representation $\rho$, consider the action of $G\coloneqq {\rm Im} \, \rho$ on $\llbracket 1,L \rrbracket$. For a point $x\in\llbracket 1,L \rrbracket$, let $G_x$ be the stabilizer of $x$ in $G$. The subgroup $\rho^{-1}(G_x)\subset \pi_1(\C P^1\setminus Y,p)$ has index $L$ and, by the classification of unramified coverings\footnote{The classification theorem that establishes a bijection between conjugacy classes of subgroups of index $L$ of the fundamental group of a space and its $L$-sheeted unramified coverings is often called Galois correpondence of covering theory.}, determines an $L$-sheeted unramified covering of $\C P^1\setminus Y$. Stabilizers corresponding to distinct points in $\llbracket 1,L \rrbracket$ are conjugate in $\mathfrak{S}_L$ and conjugate subgroups of the fundamental group give rise to isomorphic unramified coverings. 

We finally complete our unramified covering of a punctured sphere to a ramified covering of the sphere whose associated monodromy representation will be exactly $\rho$. First, we compactify our base space by adding to it the missing points in $Y$ and it becomes $\C P^1$. Then we need to add the preimages of $Y$ to the total space. For every $y_j \in Y$, we denote $y_j^{(i)} \in \pi^{-1}(y_j)$ its preimage in the $i$-th sheet (the sheet containing the point in $\pi^{-1}(p)$ labeled $i$). We identify the points $y_j^{(i)}$ and $y_j^{(\beta_j(i))}$  for every $y_j \in Y$. In this way, we get $|\pi^{-1}(y_j)|=|\mathcal{C}(\beta_j)|=\lambda_j$. It can be showed that relabeling of $\pi^{-1}(p)$ amounts to changing the elements $\beta_j$ within the same conjugacy classes.

Let $MR_{\lambda_1,\ldots,\lambda_m}$ be the set of monodromy representations of type $(\lambda_1,\ldots,\lambda_m)$. Let us relate the cardinality of this set with Hurwitz numbers. Given a ramified covering $\pi$, there are $L!$ ways of labeling the preimages $\pi^{-1}(p)$. An automorphism of $\pi$ is an isomorphism of $p$-labeled coverings where $\pi$ remains the same, but the labeling changes. By the orbit-stabilizer theorem, the number of distinct monodromy representations associated to the same $\pi$ with different labelings is
$$
m_{\pi} = \frac{L!}{|{\rm Aut}\, \pi|},
$$
which is also the number of isomorphism classes of $p$-labeled coverings for the given $\pi$.

Therefore,
\beq
h^{\bullet,g}_{\lambda_1,\ldots,\lambda_m} = \sum_{[\pi]} \frac{1}{|{\rm Aut}\, \pi|} = \sum_{[\pi]} \frac{m_{\pi}}{L!} = \frac{|MR_{\lambda_1,\ldots,\lambda_m}|}{L!}\,.
\eeq

The covering space $X$ is connected if and only if $\beta_1,\ldots,\beta_{m}$ act transitively on $\llbracket 1,L \rrbracket$, in which case we call the monodromy representation {\it connected}. All the analogous results follow in the connected setting imposing this extra transitivity condition.

This characterization of Hurwitz numbers in terms of monodromy is very useful. For example, we can deduce that, since $\pi_1(\C P^1\setminus Y)$ is finitely generated and $\mathfrak{S}_L$ is finite, Hurwitz numbers are also finite.

\subsection{The group algebra of the symmetric group}

The center of the group algebra $Z(\mathbb{C}[\mathfrak{S}_{L}])$ of the symmetric group $\mathfrak{S}_{L}$ has two interesting basis labeled by partitions. The most obvious one is formed by
$$
\hat{C}_{\lambda} = \sum_{\gamma \in C_{\lambda}} \gamma, \ \ \lambda\vdash L.
$$
The second one is the basis of orthogonal idempotents, which can be related to the first one via the characters of the symmetric group as follows:
\beq\label{changebasis}
\hat{\Pi}_{\lambda} = \frac{\chi_{\lambda}({\rm id})}{L!} \sum_{\mu \vdash L}\chi_{\lambda}(C_{\mu})\,\hat{C}_{\mu} \ \  \text{ and } \ \  \hat{C}_{\mu} = \frac{1}{|{\rm Aut}\,\mu |} \sum_{\lambda \vdash L} \frac{L!}{\chi_{\lambda}({\rm id})}\chi_{\lambda}(C_{\mu})\hat{\Pi}_{\lambda}.
\eeq
The orthogonality of the characters of $\mathfrak{S}_{L}$ implies that
$$
\hat{\Pi}_{\lambda}\hat{\Pi}_{\mu} = \delta_{\lambda,\mu} \hat{\Pi}_{\lambda}.
$$
Now we can make yet another translation of the Hurwitz problem, immediately realizing that counting the monodromy represetations of type $(\lambda_1,\ldots,\lambda_m)$ can be viewed as choosing the appropriate coefficient in the following product of these basis elements:
\beq
h^{\bullet,g}_{\lambda_1,\ldots,\lambda_m} = \frac{1}{L!}[\hat{C}_{(1^L)}] \hat{C}_{\lambda_m}\cdots \hat{C}_{\lambda_2} \hat{C}_{\lambda_1},
\eeq
where $(1^L)$ denotes the partition $(1,\ldots,1)$ of length and size $L$. In a more elementary language, this is simply
\beq
h^{\bullet,g}_{\lambda_1,\ldots,\lambda_m} = \frac{1}{L!}|\{(\sigma_1,\ldots,\sigma_m)\mid \sigma_i\in C_{\lambda_i}, \sigma_m\cdots\sigma_1 = {\rm id}\}|.
\eeq

\begin{remark}
With this last interpretation of Hurwitz numbers and the permutational model for maps given in \ref{permModel}, one can observe that taking $m=3$ and $\lambda_2=(2,\ldots,2)$,
$$
|{\rm Aut}\,\lambda_3|\, h^{\circ,g}_{\lambda_1,(2,\ldots,2),\lambda_3}
$$
enumerates maps of genus $g$ with $\ell(\lambda_3)$ boundaries of lengths given by $\lambda_3$ and no internal faces.

In the same way, the Hurwitz problem with $m=3$, but for a general $\lambda_2$, is related to the enumeration of hypermaps that we mentioned in the introduction of Section \ref{maps}.
\end{remark}

\subsection{Various types of double Hurwitz numbers}\label{doubleHNsIntro}

We call {\it ordinary double Hurwitz numbers} the Hurwitz numbers $h^{\bullet,g}_{\lambda_1,\ldots,\lambda_m}$ with $\lambda_2=\cdots=\lambda_{m-1} = (2,1^{L-2})$. We will now denote them $[P_{1,k}]_{\lambda,\mu}$, with $k=m-2$ the number of simple ramifications, $\lambda = \lambda_1$ and $\mu = \lambda_m$. Our goal is to study the Hurwitz numbers counting $L$-sheeted coverings of the sphere with ramification profile $\lambda$ over $0$, $\mu$ over $\infty$ and other ramifications encoded by any given element $B\in Z(\mathbb{C}[\mathfrak{S}_{L}])$:
\beq
\frac{1}{L!}[\hat{C}_{(1^L)}]\, \hat{C}_{\mu}\, B \, \hat{C}_{\lambda}. \nonumber
\eeq

The Jucys-Murphy elements of $\mathbb{C}[\mathfrak{S}_{L}]$ are defined (see \cite{Jucys,Murphy}) by
$$
\hat{J}_{1}=0, \ \  \hat{J}_{k} = \sum_{i = 1}^{k-1} (i \ k), \ \  k=2,\ldots, L.
$$
Their key property is that the symmetric polynomials in the elements $(\hat{J}_{k})_{k = 2}^L$ span $Z(\mathbb{C}[\mathfrak{S}_{L}])$, see \textit{e.g.} \cite{Meliot}.

Let $r$ be a symmetric polynomial in infinitely many variables. We define 
$$
r(\hat{J})\coloneqq r((\hat{J}_k)_{k=2}^L, 0,0,\ldots).
$$
The operator of multiplication by $r(\hat{J})$ in $\mathbb{C}[\mathfrak{S}_{L}]$ acts diagonally on the basis $\hat{\Pi}_{\lambda}$ of idempotents, with eigenvalues equal to the evaluation on the content of the partition $\lambda$:
$$
r(\hat{J})\hat{\Pi}_{\lambda} = r({\rm cont}(\lambda),0,0,\ldots)\hat{\Pi}_{\lambda},\qquad {\rm cont}(\lambda) \coloneqq (j-i)_{(i,j) \in \mathbb{Y}_{\lambda}},
$$
where $\mathbb{Y}_{\lambda}$ is the Young diagram associated to the partition $\lambda$.
A function of the form $\lambda \mapsto r({\rm cont}(\lambda))$ is called {\it content function}. We denote $r({\rm cont}(\lambda),0,0,\ldots)$ by just $r({\rm cont}(\lambda))$.

In the conjugacy class basis, the action of multiplication by $r(\hat{J})$ has a combinatorial meaning \cite{HarnadGuay}. We define the {\it double Hurwitz numbers} associated with $r$ by the formula
\beq
R_{\lambda,\mu} \coloneqq \frac{1}{L!}[\hat{C}_{(1^L)}]\,\hat{C}_{\mu}\,r(\hat{J})\,\hat{C}_{\lambda}.
\eeq
The following expression is well-known to be equivalent to our definition:
\beq\label{characters_def}
R_{\mu,\lambda} = \frac{1}{|{\rm Aut\, \lambda}||{\rm Aut\, \mu}|}\sum_{\nu \vdash L} \chi_{\nu}(C_{\mu})\, r(\text{cont}\, \nu) \,\chi_{\nu}(C_{\lambda}).
\eeq
We can also characterize double Hurwitz numbers via the following decomposition:
\beq
\label{rjC}r(\hat{J})\hat{C}_{\mu} = \sum_{\lambda \vdash L} |{\rm Aut}\,\lambda|\,R_{\mu,\lambda}\,\hat{C}_{\lambda},
\eeq
which can be checked to be equivalent to \eqref{characters_def} using the formulas \eqref{changebasis} to change between the conjugacy class basis and the idempotent basis. 
The definition of the Jucys-Murphy elements implies that $|{\rm Aut}\,\lambda|\,R_{\mu,\lambda}$ is a weighted number of paths in the Cayley graph of $\mathfrak{S}_{L}$ generated by transpositions, starting at an arbitrary permutation with cycle type $\mu$ and ending at an (arbitrary but) fixed permutation with cycle type $\lambda$. We can hence define several variations of the Hurwitz numbers using the standard bases of symmetric polynomials evaluated at the Jucys-Murphy elements. We mention here the most relevant for this thesis:

\vspace{0.2cm}

\noindent {\it\textbf{Ordinary.}} $p_1(\hat{J})$ is the sum of all transpositions. Therefore
$$
p_1(\hat{J})^{k} = \hat{C}_{(2,1^{L-2})}^k= \sum_{\tau_1,\ldots,\tau_k} \tau_1 \cdots \tau_k.
$$
This is the reason why we denoted $[P_{1,k}]_{\mu,\lambda}$ the ordinary double Hurwitz numbers.

\vspace{0.2cm}

\noindent {\it\textbf{Strictly monotone.}} For the elementary symmetric polynomial $e_k$, we have
$$
e_k(\hat{J}) = \sum_{\substack{\tau_1,\ldots,\tau_k \\ (\max \tau_i)_{i = 1}^k\, \text{ strictly increasing}}} \tau_1 \cdots \tau_k.
$$
Then, the strictly monotone Hurwitz numbers enumerate
$$[E_k]_{\lambda,\mu} = \frac{1}{L!}|\{(\alpha,\tau_1,\ldots,\tau_k)\mid \alpha\in C_{\lambda}, \tau_k\cdots\tau_1\alpha\in C_{\mu}, \tau_i\in C_{(2,1^{L-1})}, {\rm max}\, \tau_i<{\rm max}\, \tau_{i+1}, \forall i\}|.$$

\vspace{0.1cm}
\noindent {\it\textbf{Weakly monotone.}} For the complete symmetric polynomials $h_k$,
$$
h_k(\hat{J}) = \sum_{\substack{\tau_1,\ldots,\tau_k \\ (\max \tau_i)_{i = 1}^k\,\,{\rm weakly}\,\,{\rm increasing}}} \tau_1 \cdots \tau_k.
$$
Then, the weakly monotone Hurwitz numbers enumerate
$$[H_k]_{\lambda,\mu} = \frac{1}{L!}|\{(\alpha,\tau_1,\ldots,\tau_k)\mid \alpha\in C_{\lambda}, \tau_k\cdots\tau_1\alpha\in C_{\mu}, \tau_i\in C_{(2,1^{L-1})}, {\rm max}\, \tau_i\leq{\rm max}\, \tau_{i+1}, \forall i\}|.$$

\vspace{0.1cm}

We refer to either of the two last cases as {\it monotone} Hurwitz numbers.


\section{Matrix models}\label{IntroMMs}


This section is a summary of the results concerning the method of matrix models in map enumeration which are the most relevant for this thesis. The reader interested in a more complete exposition can consult the same references suggested for maps. 


We already commented that maps became a popular tool in theoretical physics after the influential discoveries of the 70s, which gave a powerful incentive for developments in random matrices, quantum gravity, string theory and enumerative geometry.
In 1974, while studying quantum chromodynamics, t'Hooft \cite{tHooft} noticed that the leading contribution in the Feynman diagram expansion is given by planar diagrams, which can be identified with planar maps, and higher genus contributions correspond to correction terms. In 1978, the physicists Br\'{e}zin, Itzykson, Parisi and Zuber \cite{BIPZ} turned this idea of a topological expansion into a general relation between random matrices and map enumeration. 


The method of matrix models was reinvented in 1986 by the mathematicians Harer and Zagier \cite{HarerZagier} in order to solve an enumerative problem that arose for them as a step in the computation of the orbifold Euler characteristic of the moduli space of curves. This paper attracted much attention in the mathematical community and stimulated new interesting research that culminated in Kontsevich proof \cite{Kontsevich} of Witten conjecture.



For us, matrix integrals constitute a very useful formal tool to deal with generating series of maps. For example, Tutte's equations became equivalent to loop equations (related to integration by parts). It is important to note that these are sometimes called formal matrix integrals to distinguish them from convergent matrix integrals (see e.g. \cite{RMT}), which instead are studied within analysis and probability. In this context of asymptotic analysis in which one cares about the convergence of the integrals, finding the asymptotic expansion of the integrals constitutes a challenging issue and it is instead map enumeration which sometimes contributes towards its solution.

\subsection{Gaussian model and Wick formula}


We introduce the ordinary volume form in the space $\mathcal{H}_N$ of $N \times N$ Hermitian matrices
$$
\dd M = \prod_{i = 1}^N \dd M_{i,i} \prod_{i < j} \dd {\rm Re}\,M_{i,j} \  \dd {\rm Im}\,M_{i,j}.
$$
The most studied random matrix ensembles are the Gaussian ensembles. The Gaussian Unitary Ensemble (GUE$(N)$) is described by the Gaussian probability measure on $\mathcal{H}_N$:
\beq\label{GUE}
\dd\nu_{{\rm GUE}}(M) = \frac{\dd M}{Z_{{\rm GUE}}}\,e^{-N\mathrm{Tr}\,\frac{M^2}{2}},\qquad Z_{{\rm GUE}} = \int_{\mathcal{H}_{N}} \dd M\,e^{-N\mathrm{Tr}\,\frac{M^2}{2}} = 2^{\frac{N}{2}}\left(\frac{\pi}{N}\right)^{\frac{N^2}{2}}.
\eeq
This measure is unitary invariant, which allows us to diagonalize Hermitian matrices conjugating by a unitary matrix whenever this facilitates a certain computation.

The expansion in Feynman diagrams of a matrix integral consists of performing a formal power series expansion with respect to some parameters around a Gaussian point.
We denote $\langle \mathcal{\cdot} \rangle_{{\rm GUE}}$ the expectation value with respect to the Gaussian measure $\dd\mu_{{\rm GUE}}$. The matrix elements have covariance:
\beq\label{contrac} 
\langle M_{a,b}M_{c,d} \rangle_{{\rm GUE}} = \frac{1}{N}\delta_{a,d}\delta_{b,c}.
\eeq
The following theorem reduces the integration of any polynomial of Gaussian random variables of even degree to products of covariances (sometimes called \emph{propagators} by the physical community):
\begin{theorem}[Wick formula \cite{Wick}]
The expectation value of a product of an odd number of Gaussian random variables is $0$ and the product of an even number is given by
\beq
\langle X_{i_1}\cdots X_{i_{2m}} \rangle_{{\rm GUE}} = \sum_{\pi\in\mathcal{P}_2(2m)}\prod_{(r,s)\in\pi} \langle X_{i_r} X_{i_{s}} \rangle_{{\rm GUE}},
\eeq 
where $\mathcal{P}_2(2m)$ is the set of all pairings of the set $[2m]$.
\end{theorem}

\subsection{Generating series of maps and the one-matrix model}\label{MMOrdMapsIntro}

\subsubsection{Partition function  and closed maps}
Considering the following formal power series
$$
V(x)\coloneqq\sum_{k \geq 1} \frac{t_k\,x^k}{k},
$$
which we call \emph{potential}, the formal expression $Z\coloneqq\langle e^{N{\rm Tr} V(M)}\rangle_{{\rm GUE}}$
can be viewed as the \emph{partition funtion} of the \emph{Hermitian one-matrix model}, where $V$ plays the role of a non-Gaussian perturbation, with $V=0$ corresponding to the particular case of a Gaussian model. We should keep in mind the dependence of the partition function on $N$ and $V$, even if we omit it in the notation.

Let ${\rm IF}(\mathcal{M})$ denote the set of internal faces of a map $\mathcal{M}$, which is the set of all faces, if the map is closed, because the set of boundaries $\partial(\mathcal{M})$ is empty. Using Wick formula, one can show that $Z$ can be interpreted as the generating function of disconnected closed maps:
\beq\label{partitionfct}
Z =  \sum_{\substack{\text{Maps } \mathcal{M}  \\ \text{with } \partial(\mathcal{M})=\emptyset}} \frac{N^{\chi(\mathcal{M})}}{|{\rm Aut}(\mathcal{M})|}\,\prod_{f \in\, {\rm IF}(\mathcal{M})}t_{\ell(f)}.
\eeq
We define the so-called \emph{free energy} by
\beq\label{logexp}
F\coloneqq  \log Z.
\eeq
Recall that we denoted $F^{[g]}$ the generating series of closed maps of genus $g$, whose dependence on $V$ (in the context of maps seen as the collection of weights of internal faces) was omitted for simplicity.
When we have a generating series counting disconnected objects, it is a well-known trick in combinatorics (see e.g. \cite{gfcts}) that taking the logarithm produces the generating series counting only the connected objects:
\begin{proposition}[\cite{BIPZ}]
\beq
F = \sum_{g \geq 0} N^{2-2g} F^{[g]}.
\eeq
\end{proposition}
This statement is just a formalization of the ideas present in \cite{BIPZ}. We remark that the power of $N$ sorts maps by their Euler characteristic, which was already t'Hooft's discovery in \cite{tHooft}. 

In Section~\ref{MMs}, we will derive this expansion in terms of maps in some detail. This type of calculations are usually done in dual terms, i.e. gluing stars instead of directly gluing polygons, which is a more traditional language in physics. In the end of those computations, invoking the dual construction is how one can see maps clearly appearing in the expansion. We did the computation directly in terms of gluing polygons as building pieces for maps to illustrate a different derivation (but completely equivalent), which will also make clearer the refinement that we will need later, when we introduce boundaries, to derive a similar relation between matrix integrals and the generating series of fully simple maps.

In our derivation, we represent the factors appearing in the expansion of $Z$ graphically as certain elements in a map illustrated in Figure \ref{GraphRep}. The arrows in the end of half-edges in this picture do not represent roots (which are always depicted with arrows in the middle of half-edges), they remind us of certain conventions we took for maps, especially when we represent them like here as ribbon graphs:
\begin{itemize}
\item Faces are taken with counterclockwise orientation.
\item Half-edges incident to a vertex are the ones on the left, viewed from the vertex, which correspond to the outgoing ones. 
\item Vertices are also considered with counterclockwise orientation.
\end{itemize}
\begin{figure}[h!]
 \begin{center}
        \def\svgwidth{\columnwidth}
       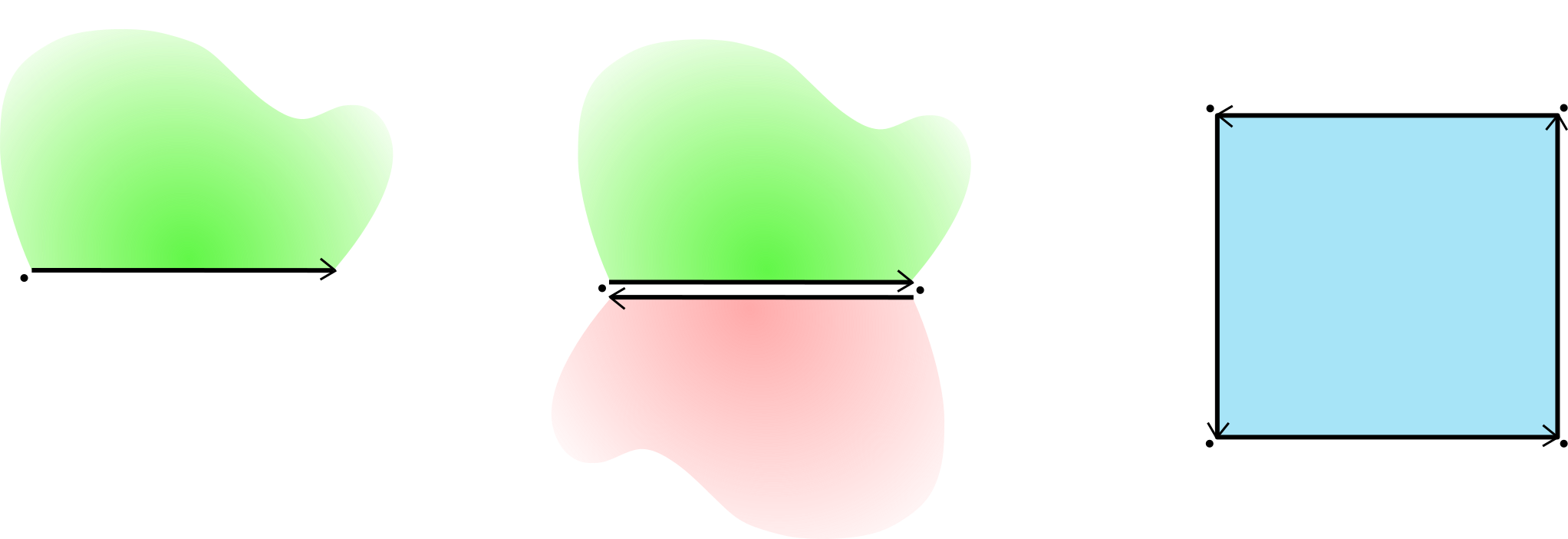
    	\caption{(a) An element of the matrix corresponds to a half-edge with its two legs labelled by the indices. (b) Pairing two matrix entries is represented as two half-edges forming an edge.  (c) Products of elements of that form are the ones appearing in ${\rm Tr}\, M^4$ and are represented by quadrangles.}\label{GraphRep}
\end{center}
\end{figure}

Every term in ${\rm Tr}\, M^k$, which is a product of $k$ matrix elements, represents $k$ half-edges cyclically ordered forming a face so that the last index of one is the same as the first index of the next one. In the picture, we can see the case $k=4$.
Wick's theorem, applied to matrix Gaussian random variables, says that the expectation value of products of matrix elements can be computed pairing them in all possible ways and taking the corresponding product of propagators. These pairings are represented by edges of the map and the precise condition that the graphical representation suggests that the indices of the legs of the half-edges should coincide depicts precisely the condition for the propagators \eqref{contrac} to be non-zero. The idea is that a pairing that does not satisfy this restriction and does not produce a valid edge in a map will not contribute to the expectation value.

\subsubsection{Maps with boundaries} 

\textbf{Cumulants.}
Let $(X_j)_{j\geq1}$ be a sequence of random variables. The generating series of their cumulants is defined in terms of the logarithm of the generating series of expectation values as follows:
\beq
\log \left\langle e^{\sum_{j\geq 1} t_j X_j}\right\rangle = \sum_{n\geq 1} \frac{1}{n!} \sum_{j_1,\ldots,j_n \geq 1} \kappa_n(X_{j_1},\ldots,X_{j_n}) \prod_{i=1}^n t_{j_i}.
\eeq
This implies the following moment-cumulant relation
\beq\label{easyMomCum}
\langle X_1 \cdots X_n\rangle = \sum_{\pi\in \mathcal{P}(n)} \prod_{B\in\pi} \kappa_{|B|}(X_j \mid j\in B),
\eeq
where $\mathcal{P}(n)$ denotes the set of partitions of $[n]$. This relation can be inverted, giving the cumulants in terms of the moments:
\beq\label{inverseMomCum}
\kappa_n(X_1, \ldots, X_n) = \sum_{\pi\in \mathcal{P}(n)} (|\pi|-1)! (-1)^{|\pi|-1} \prod_{B\in\pi} \Big\langle \prod_{j\in B} X_j\Big\rangle.
\eeq
 where $|\pi|$ denotes the number of blocks of $\pi\in\mathcal{P}(n)$. \\

\noindent{\textbf{Correlators and open maps.}}
These ideas will become clearer in Section~\ref{MMs}, where we will actually extend the computation of the partition function \eqref{partitionfct} in terms of closed maps to some expectation values $\langle \cdot\rangle$ with respect to the measure
\begin{equation}\label{ordinaryMeasure}
\dd\nu(M) = \frac{\dd M}{Z}\,\exp\left\{N\,{\rm Tr}\Big(-\frac{M^2}{2} + V(M)\Big)\right\},
\end{equation}
which can be interpreted as the generating series:
\beq\label{TracesGSMAps}
\Big\langle \prod_{i = 1}^n {\rm Tr}\,M^{l_i} \Big\rangle = \sum_{\substack{\partial \text{-connected}\  \mathcal{M}  \\ \text{with } \partial \text{ lengths}\,\,(l_i)_{i = 1}^n}} \frac{N^{\chi(\mathcal{M})}}{|{\rm Aut}(\mathcal{M})|}\,\prod_{f \in\, {\rm IF}(\mathcal{M})}t_{\ell(f)},
\eeq
where $\partial$\emph{-connected} means that every connected component contains at least one boundary.

For  open maps with $n\geq 1$ boundaries, we need to keep track of the number of boundaries of every connected component in order to enumerate disconnected maps in terms of the number of connected maps. For this reason, the exponential formula \eqref{logexp} that we employed before has to be refined. A standard argument shows that taking cumulants for maps with boundaries, we obtain the generating series of connected maps. So for open maps, cumulants play the role that logarithm played for closed maps. It is not difficult to convince oneself of this fact by understanding the moment-cumulant relation \eqref{easyMomCum}, which has the simple combinatorial meaning that we described, and then inverting it to obtain \eqref{inverseMomCum}.

We recall that $F^{[g]}_{l_1,\ldots,l_n}$ was the generating series of maps of genus $g$ with $n$ boundaries of lengths given by $l_1,\ldots,l_n$ introduced in \eqref{GenFctF} in Section \ref{maps}.
\begin{proposition}[\cite{Eynardbook}]
$$
\kappa_n({\rm Tr}\,M^{l_1},\ldots,{\rm Tr}\,M^{l_n}) = \sum_{g \geq 0} N^{2-2g-n} F^{[g]}_{l_1,\ldots,l_n}.
$$
\end{proposition}

Continuing with the terminology coming from physics, we can also consider the so-called \emph{correlators}
\beq
W_n(x_1,\ldots,x_n) \coloneqq \sum_{l_1,\ldots,l_n\geq 0} \frac{\kappa_n({\rm Tr}\,M^{l_1},\ldots,{\rm Tr}\,M^{l_n})}{x_1^{l_1+1}\cdots x_n^{l_n+1}}.
\eeq
This is sometimes written in the physicist's literature with the following shorter expression:
$$
\kappa_n\left({\rm Tr}\,\frac{1}{x_1-M},\ldots,{\rm Tr}\,\frac{1}{x_n-M}\right),
$$
which holds in a formal sense. Finally, we also have a topological expansion
$$
W_n(x_1,\ldots,x_n) = \sum_{g\geq 0} N^{2-2g-n}W_n^{[g]}(x_1,\ldots,x_n),
$$
where $W_n^{[g]}(x_1,\ldots,x_n)$ are the generating series of maps of topology $(g,n)$ introduced in \eqref{correlators} in Section \ref{maps}.

Considering the following operator
\beq\label{create}
\frac{\partial}{\partial V(x)} = \sum_{k\geq 1}\frac{k}{x^{k+1}}\frac{\partial}{\partial t_k},
\eeq
we can view, as usual, the correlators as derivatives of the partition function:
$$
W_n(x_1,\ldots,x_n) = \frac{\partial}{\partial V(x_n)} \cdots \frac{\partial}{\partial V(x_1)} Z,
$$
which makes a lot of sense combinatorially, since the operator \eqref{create} creates a boundary of length $k$ weighted by $x^{-(k+1)}$.

\subsection{Tutte equations, loop equations, Virasoro constraints}\label{VirasoroIntro}

In 1963, Tutte discovered a recursive equation for counting disks, by finding bijections between an initial disk and the resulting disks after having removed an edge. This can be translated into a recursive equation satisfied by the generating series of disks of fixed lengths and eventually to an equation for the generating series collecting all lengths $W(x)$. Using the same technique, one can find recursive equations for the generating series of maps of higher topologies. The only added complication is that more terms will appear due to the different possibilities that can occur after deleting an edge because of the higher genus and number of boundaries.

Tutte's equations can also be derived in the context of matrix integrals, where they are called loop equations. Loop equations can be deduced simply by integration by parts or, alternatively, by a change of variables\footnote{Exploiting the invariance of an integral under change of variables gives rise in a more general context in theoretical physics to Schwinger-Dyson equations, which are quantum analogues of the Euler-Lagrangian equations, since they are the equations of motion for the Green's functions in quantum field theories.}. It is faster to integrate a matrix integral by parts than to find bijections between sets of maps, and hence it is simpler to derive loop equations with matrix model techniques.

In the context of string theory and conformal field theory, it is known that partition functions must satisfy Virasoro constraints. Loop equations for generating functions of maps can also be rewritten as differential operators annihilating the partition function and, in this form, they also receive the name of Virasoro constraints.

These equations and their derivations in all their forms, and their solutions can be found in \cite{Eynardbook}. 



\subsection{Motivations coming from physics}

Finally, we give a rough idea of the connection of these methods to physics, where they have their origin, because it is commonly difficult for mathematicians who lack a physical background to make these inspirational connections, which are often a source of interesting mathematical problems. 
However, for a more complete exposition, the curious reader who is not familiar with these theories is referred to specialized literature, such as \cite{QFT-String}. 

%

\subsubsection{Matrix models as perturbative quantum field theories}
In classical mechanics, a particle follows a path $\phi$ that minimizes the functional of the action $S(\phi)$ or, more generally, is stationary. The Feynman path integral formulation of quantum mechanics replaces the notion of a classical trajectory by an average over all the possible trajectories, which measures the contribution of the quantum trajectory, and is called  probability \emph{amplitude} (or correlator). The simplest amplitude is of the form:
$$
\int_{\phi\in\Phi} e^{-\frac{i}{\hbar} S(\phi)} d\phi,
$$
where $\Phi$ is the (inifinite-dimensional) space of paths. Typically, this quantity depends on the integration procedure and hence is not well-defined. As a solution, Feynman proposed to discretize the paths by the so-called \emph{Feynman diagrams}, over which the integral can be computed. Therefore, one can find an expansion of the integral around $\hbar=0$, in which contributions of Feynman diagrams are organized by the number of cycles in the graph. This approach receives the name of \emph{perturbative quantum field theory} and it allowed to make numerous predictions which were experimentally confirmed, even if one would like to have a non-perturbative theory eventually. 


Here one can already see the relation to our section, in which we also found 
expansions of integrals in terms of ``Feynman diagrams'', in our case around a 
Gaussian matrix. In the case of matrix models, the spacetime, in which the 
fields are defined, is just a point, and our fields take values in the space 
of Hermitian matrices $\mathcal{H}_N$. Our potential $V$ plays the role of an 
interaction term which depends on some parameters, and the goal is to study 
the dependence of the partition function on these parameters. When there is no 
interaction, the measure becomes Gaussian and it is called \emph{free field}. 
This is the reason why matrix models can be viewed as $0$-dimensional quantum 
field theories. 

\subsubsection{Maps as discrete approximations of surfaces}\label{WittenKontsevich}

While in quantum field theory, particles are thought as points and their evolution in time can be represented by Feynman diagrams,
in string theory, particles are modeled by closed\footnote{One can also talk about open strings, but here we restrict to closed string theory for simplicity.} simple curves, called strings, and as they evolve in time they create a surface which plays the role of the trajectory.  By analogy with quantum field theory, amplitudes of $n$ strings are computed by expansions over Feynman diagrams, which correspond now to surfaces with $n$ boundaries sorted by their genera. We may suppose, for instance, that the time segment is finite and hence, the surfaces become compact.

The path integrals are now replaced by integrals over the space of surfaces, which may seem worse than before, but in some aspects it has great advantages. There are two main strategies in order to reduce these integrals to finite dimensional ones. The first one is to classify these surfaces by their conformal structure, which leads directly to integrate over the compactification of the moduli space of Riemann surfaces of genus $g$ with $n$ boundaries $\overline{\mathcal{M}}_{g,n}$, which is well-known to have dimension $3g-3+n$. The other approach considers a discrete approximation of a surface by a map, which also serves as a motivation for the next section of this thesis. When both methods can be used, they produce consistent results, which motivated the type of beautiful mathematics coming from dualities in physics.

Actually, such a physical argument led Witten \cite{Witten}, in 1991, to formulate his conjecture that the generating series of $\psi$-class intersection numbers on $\overline{\mathcal{M}}_{g,n}$ is a tau-function of the KdV hierarchy\footnote{This can be rephrased as follows: Let $F$ be the generating series of $\psi$-class intersection numbers on $\overline{\mathcal{M}}_{g,n}$. Then, the partition function $e^F$ is a solution of the KdV hierarchy. This was showed to be equivalent to the Virasoro constraints associated to the KdV hierarchy \cite{DVV}, which consist of a system of linear PDEs, while the KdV hierarchy is a system of non-linear PDEs.}. He was studying a particular model of two-dimensional quantum gravity with infinite-dimensional integrals over the space of Riemannian metrics on a surface. Again, the first technique to reduce such a calculation to an integral over a finite dimensional space consists in classifying metrics by their conformal class, and the second approach consists in triangulating the surfaces, which produces a very natural way of discretizing a metric. As the number of triangles tends to infinity, these discrete metrics start approximating random metrics and the infinite-dimensional integrals are equivalent to summations over the set of all triangulations, hence reducing them to asymptotic enumerations of such triangulations, which are known to be governed by the KdV hierarchy. 

In 1992, Kontsevich \cite{Kontsevich} gave the first proof of Witten conjecture, using precisely matrix model techniques to enumerate a certain type of maps. He gave a cellular decomposition of the moduli space $\overline{\mathcal{M}}_{g,n}$, using maps of genus $g$ with $n$ faces and vertices of degree greater than $2$ to label the cells, where the cells of top dimension are the ones labelled by trivalent maps. In this way, there is no loss of information and there is no need to take any limit. He also used a Hermitian one-matrix model\footnote{Different from the most classical one that we described in this section.} to enumerate them.

\section{Large random (decorated) maps and universality}\label{largeRandom}

In physics, it is very common to use mathematical models to study nature. We may refer to a specific model in statistical physics, which depends on some parameters, as the \emph{microscopic} setup. A model which is too malleable, i.e. that leads to different conclusions after small variations, is of course considered a bad model that will never capture the details of nature. In other words, the only important properties deduced from a statistical physics model are those which are \emph{universal}, that is those that do not depend on the microscopic details of the model, which are called \emph{macroscopic}. An important motivation to study large random maps comes from the generally accepted conjecture that their geometry is universal, i.e. that there should exist ensembles of random metric spaces depending on a small set of data (like the central charge and a symmetry group attached to the problem) which describe the continuum limit of random maps.

As we explained in the previous section, an important physical motivation to study random maps is to use them as a method of discretization of random surfaces, since computing transition amplitudes in gauge theory or string theory requires averaging over random surfaces, which replaces the traditional Feynman path integrals from quantum field theory. Two-dimensional quantum gravity aims at the description of these random continuum limits of random maps and physical processes on them, and the universal theory which should underly them is \emph{Liouville quantum gravity} (LQG), possibly coupled to a conformal field theory (CFT) \cite{Ginsparg-Moore,Ginsparg}. This theory models the quantum fluctuation of the gravitational field by a random metric over a two dimensional space-time. The interaction between the gravitational field and matter is modeled by an external field, often coming from CFT. It is generally accepted, and proved in some cases \cite{Smirnov01,ChelkakSmirnov12}, that certain random decorations over fixed lattices, as in the classical Ising and $O(\mathsf{n})$ statistical models, have scaling limits described by a field from CFT. Therefore a natural way of constructing LQG is to define a discretization of the random metric of the space-time, using random undecorated maps, and afterwards to couple it to some decorations, which represent the coupling to matter.

Theoretical physics predicts the celebrated KPZ relation \cite{KPZ}, which links the critical exponents of classical models over fixed lattices to those over a dynamic randomly chose lattice, i.e. over random maps. Another motivation to analyze models of random decorated maps is to verify if this prediction is always valid and gaining understanding of the mechanism behind it.

It is widely believed that after a Riemann conformal map to a given planar domain, the proper conformal structure for the continuum limit of random planar maps  weighted by the partition functions of various statistical models is described by LQG (see, {\it e.g.}, the reviews \cite{Ginsparg-Moore,Ginsparg,MR2073993} and  \cite{DuplantierICM2014,LeGallICM2014}). In the particular case of random planar maps without decorations, called \emph{pure gravity}, the universal metric structure of the Brownian map \cite{Gall1,MierMapP,LeGallICM2014} has very recently been identified with that directly constructed from LQG \cite{map_making,qlebm,quantum_spheres0,MS2,MS3,MSQL}. Nevertheless, little is known on the metric properties of large random maps weighted by an $O(\mathsf{n})$ model, even from a physical point of view. Moreover, most of the works described above are restricted to planar maps. So another motivation is to study the critical exponents characteristic of some models to be able to compare them to those of continuum objects that are their conjectured scaling limits \cite{sheffield2009,zbMATH06121652,LQGRS,Sheffield16}.

There are several important challenges in this branch of mathematical physics. The first one is to explore the macroscopic properties of the universality classes through solvable models, in which we can explicitly compute the partition function, for us the generating series of certain type of maps. And the second one is to verify if universality really exists for a certain type of models comparing the results of different microscopic models which belong to the same universality classes, according to the physical predictions. In this thesis, we will focus on task one for the $O(\mathsf{n})$ loop model on random maps of arbitrary topologies. 

Apart from the physical motivations, decorated maps are also very interesting from the mathematical point of view because of their rich combinatorial and probabilistic structure. Finally, understanding rigorously the emergent fractal geometry of such limit objects constitutes nowadays another major problem in mathematical physics, and to establish the convergence of random maps towards such limit objects is in general a difficult problem. Solving various problems of map enumeration is often instrumental in this program, as it provides useful probabilistic estimates.

We do not elaborate much on the scaling limits of random maps. For a complete introduction to these aspects in the planar and non-decorated cases, see \cite{miermont2014aspects}. In \cite{miermont2009random}, a special emphasis is put on the fractal properties of the random metric spaces involved, for example providing a detailed calculation of the Hausdorff dimension of the scaling limits of random planar quadrangulations.


\subsection{Random Boltzmann maps}

In Section \ref{maps} we introduced the combinatorial $O(\mathsf{n})$ loop model, whose configurations consist of maps decorated with loops. All the parameters we used as weights were formal and the generating series were formal power series, which are very useful for combinatorics. In this section, we want to use those weights to define a probability measure in order to formalize the intuitive concept of random map. For this purpose we want to choose the parameters to be real positive numbers so that the models are \emph{well-defined}, i.e. the weights induce a probability distribution over the set of configurations. This is the natural setting for the $O(\mathsf{n})$ loop model, viewed as a statistical ensemble of configurations in which $\mathsf{n}$ plays the role of a fugacity per loop. We remind the reader that to make the distinction explicit, maps without a statistical physics model will be called usual maps.
 
\begin{definition}
In the context of usual maps, we say that the weight sequence $(u,(t_l))$ of nonnegative real numbers is \emph{admissible} if $F_{\ell}^{\bullet}$ (evaluated at this weight sequence) is finite, for any $\ell\geq 1$. Analogously, for $O(\mathsf{n})$-configurations, we say that the weight sequence $(u, \mathsf{n},(t_l),(t_{l_1,l_2}))$ ($(\mathsf{n},\mathsf{g},\mathsf{h},\alpha)$ for the bending energy model) of nonnegative real numbers is \emph{admissible} if the corresponding disk generating series from \eqref{eq:Fdef} $\mathcal{F}_{\ell}^{\bullet}$ (evaluated at this weight sequence) is finite, for any $\ell\geq 1$.
\end{definition}

\begin{remark}
One can check that the assumption of finiteness of the non-pointed generating series $F_{\ell}$ or $\mathcal{F}_{\ell}$ is equivalent to admissibility. Is can also be shown that, for admissible weight sequences, we have 
$$
F_{\ell_1,\ldots,\ell_k}^{[g,\bullet k^{\prime}]}<\infty \ \text{  and  } \  \mathcal{F}_{\ell_1,\ldots,\ell_k}^{[g,\bullet k^{\prime}]}<\infty,
$$
for all $g, k$ and all $\ell_1,\ldots,\ell_k$.
\end{remark}

For usual maps, if we record all the possible boundary perimeters at the same time as in \eqref{correlators}, we have that for admissible vertex and face weights, 
$$
W_k^{[g,\bullet k']}(x_1,\ldots,x_k) \in \mathbb{Q}[[x_1^{-1},\ldots,x_k^{-1}]].
$$
For admissible weight sequences, we analogously define for the $O(\mathsf{n})$ loop model:
\beq\label{Oncorrelators}
\mathcal{W}_k^{[g,\bullet k^{\prime}]}(x_1,\ldots,x_k) = \sum_{\ell_1,\ldots, \ell_k \geq 0} \frac{\mathcal{F}_{\ell_1,\ldots,\ell_k}^{[g,\bullet k^{\prime}]}}{x_1^{\ell_1 + 1}\cdots x_k^{\ell_k + 1}} \in \mathbb{Q}[[x_1^{-1},\ldots,x_k^{-1}]].
\eeq

From now on, we denote by $\mathbf{t}$ any admissible weight sequence. Now we can define a measure associated to a specific weight sequence $\mathbf{t}$ on the set of configurations, evaluating the Boltzmann weights $w$ defined in \eqref{Bweight} at $\mathbf{t}$:
\beq\label{BoltMeasure}
\mathfrak{W}_{\mathbf{t}}(\mathcal{C})\coloneqq u^{|\mathcal{V}(\mathcal{C})|} w(C),
\eeq
for every configuration $\mathcal{C}$.

We endow a configuration $\mathcal{C}_0$ of genus $g$ with $k$ boundaries of fixed lengths $\ell_1,\ldots,\ell_k$ and $k'$ marked points with the following \emph{$\mathbf{t}$-Boltzmann law}:
\beq\label{probMeasure}
\mathbb{P}^{[g,\bullet k^{\prime}]}_{\ell_1,\ldots,\ell_k}(\mathcal{\mathcal{C}} = \mathcal{C}_0) \coloneqq \frac{\mathfrak{W}_{\mathbf{t}}(\mathcal{C})}{\mathcal{F}_{\ell_1,\ldots,\ell_k}^{[g,\bullet k^{\prime}]}}.
\eeq
We call such a pair $(\mathcal{C}_0,\mathbf{t})$ a $\mathbf{t}$-Boltzmann distributed \emph{random configuration}. If there is an admissible weight sequence $\mathbf{t}_u$ such that $\mathfrak{W}_{\mathbf{t}_u}(\mathcal{C})=1$, for all configurations $\mathcal{C}$ in an ensemble $\mathbb{U}$, we call $\{(\mathcal{C},\mathbf{t}_u)\mid \mathcal{C}\in\mathbb{U}\}$ the set of \emph{uniform random configurations}.

\begin{remark}
Specializing this construction to configurations without loops, we also define $\mathbf{t}$-Boltzmann probability laws on usual maps. We can also consider models where we fix the underlying map $\mathcal{M}$ and only the loop configuration is random. We refer to such models as $O(\mathsf{n})$ loop models over a fixed lattice which is deterministic.
\end{remark}

\subsection{Critical points}
Let $\mathbf{t}=(u,\tilde{\mathbf{t}})$ be a weight sequence where we fix the non-vertex weights sequence $\tilde{\mathbf{t}}$ and let the vertex weight $u$ vary in such a way that $\mathbf{t}$ is admissible. There exists a critical value $u_c$ above which the model for usual maps (or for $O(\mathsf{n})$ configurations) ceases to be well-defined:
$$
u_{c} \coloneqq \sup\{u \geq 0\mid F_{\ell}^{\bullet} < \infty \  (\mathcal{F}_{\ell}^{\bullet} < \infty) \text{ for any } \ell\geq 1\}.
$$
If $u_{c} = 1$ (resp. $u_{c} < 1$, $u_{c} > 1$), we say that the model is at a \emph{critical} (resp. \emph{subcritical}, \emph{supercritical}) point. 

\subsubsection{Large size asymptotics and singularities}
It is a general principle that studying the large order behavior of a sequence is equivalent to analyzing the behavior of the generating series close to its singularities. The rigorous results that relate the asymptotic expansion of a function near its dominant singularity to the asymptotic expansion of its coefficients are called \emph{transfer theorems} (see \cite[Chapter VI]{Flajolet}). More precisely, assuming that $A(u)$ is analytic in a so called $\Delta$-domain\footnote{For a fixed $u_c\in \mathbb{R}$, $0<\phi<\frac{\pi}{2}$ and $R>1$, we define
$$
\Delta(\phi,R)\coloneqq \{ z\in\mathbb{C}\mid |z|<R, z\neq u_c, |{\rm arg}(z-u_c)|>\phi\}.
$$
We say a domain is a \emph{$\Delta$-domain} at $u_c$ is it is of the form $\Delta(\phi,R)$ for some $R$ and $\phi$. See \cite[Chapter VI]{Flajolet} for some pictures and examples.} at some $u_c$, we obtain the following equivalence of asymptotic behaviors, for $r\in\mathbb{R}\setminus \mathbb{Z}_{\leq 0}$:
\beq
A(u)\coloneqq \sum_{v\geq 0} A_v u^v \underset{u\rightarrow u_c}{\sim} \mathsf{a}\left(1-\frac{u}{u_c}\right)^{-r} = \mathsf{a}\sum_{v \geq 0} \frac{\Gamma(v + r)}{v!\,\Gamma(r)}\left(\frac{u}{u_c}\right)^{v} \ \Leftrightarrow \ A_{v}  \underset{v\rightarrow \infty}{\sim} \mathsf{a}\, u_c^{-v} \frac{v^{r-1}}{\Gamma(r)},
\eeq
for some constant $\mathsf{a}>0$.

Going back to our models, at a critical point, the generating series $W(x)$ (or $\mathcal{W}(x)$) has a singularity when $u \rightarrow 1^{-}$, and the nature (universality class) of this singularity is characterized by some critical exponents. It is known that transfer theorems are applicable to our models.

\subsubsection{Example: usual quadrangulations}
In order to consider quadrangulations, we only keep the weight associated to quadrangles, i.e. $t_k=\delta_{k,4}t_4$. In this special case, $F_{\ell}$ counts quadrangulations with $v$ vertices, $n_4$ quadrangles and a boundary of size $\ell$. Observe that we have $V=1+n_4+\ell/2$, so $V$ and $n_4$ are not independent and $\ell$ must be even. One can compute (see for example \cite[Chapter~3]{Eynardbook}):
\beq\label{F2l}
F_{2l}=\gamma^{2l}\frac{(2l)!}{l!(l+2)!}((2l+2)u-l\gamma^2), \ \ \ \ \ F_{2l+1}=0,
\eeq
where $\gamma^2=\frac{1-\theta}{6t_4}$, with $\theta=\sqrt{1-12ut_4}$. The generating series $F_{2l}$ is finite if and only if the series $\theta=\theta(u,t_4)$ converges, i.e. for $ut_4\leq \frac{1}{12}$. In that case, we say that $(u,t_4)$ is an admissible weight sequence. If we fix $t_4=\frac{1}{12}$, the model is at a critical point since $u_c=1$.

If we keep any $u=1$, we fix $t_4$ to any value such that $u_c>1$, and we consider the boundary length to become large, we obtain the following asymptotic behavior
\beq\label{perimeterExpSub}
F_{2l}\underset{l\rightarrow \infty}{\sim}\frac{2-\gamma^2}{\sqrt{\pi}}(2\gamma)^{2l} l^{-\frac{3}{2}}.
\eeq
On the other hand, if we consider $t_4=\frac{1}{12}$ and $u=u_c=1$, then $\gamma^2=2$ and we have the following critical behavior
\beq\label{perimeterExpCrit}
F_{2l}=\frac{2^{l+1}(2l)!}{l!(l+2)!}\underset{l\rightarrow \infty}{\sim}\frac{2}{\sqrt{\pi}}\, 8^{l}\, l^{-\frac{5}{2}}.
\eeq
Expanding \eqref{F2l} into powers of $u$ and $t_4$, we find that the number of rooted planar quadrangulations with $n_4$ quadrangles is:
$$
3^{n_4}\frac{(2l)!}{l!(l-1)!}\frac{(2n_4+l-1)!}{(l+n_4+1)!n_4!}.
$$
In particular, for $l=2$, we recover the number of rooted quadrangulations with $f$ faces, including the boundary, which was already computed by Tutte:
$$
[t_4^{f-1}u^{f+2}]F_{4}=\frac{2\cdot 3^f(2f)!}{f!(f+2)!}.
$$
Taking into account that the number of vertices is $V=f+2$, we find the following asymptotic behavior when we send the volume $V$ to infinity:
$$
[t_4^{V-3}u^{V}]F_4\underset{V\rightarrow \infty}{\sim}\frac{2}{\sqrt{\pi}}\, 12^{V}\, V^{-\frac{5}{2}}.
$$
The exponent $-5/2$ implies that the series $F_4$ is indeed convergent for $t_4=1/12$ and for all $u\leq u_c=1$. Finally, evaluating $F_4$ at a fixed quadrangle weight $t_4=\frac{1}{12}$, we obtain the following behavior of the generating series of rooted quadrangulations with fixed large volume:
\beq\label{volumeExp}
[u^{V}]F_4\underset{V\rightarrow \infty}{\sim}\frac{2\cdot 12^{3}}{\sqrt{\pi}}\,  V^{-\frac{5}{2}} = \frac{2\cdot 4\cdot 12^{3}}{3\,\Gamma(-3/2)}\,  V^{-\frac{3}{2}-1} \ \Rightarrow \  \left.F_4\right|_{{\rm sing}} \underset{u\rightarrow 1}{\sim} 4608 (1-u)^{3/2},
\eeq
where $\left.F_4\right|_{{\rm sing}}$ denotes the leading singular part in the asymptotic expansion of $F_{4}$ around $u=1$.
\subsection{The Brownian universality class: pure gravity, usual maps}

The geometry of large random planar maps with faces of bounded degrees (\textit{e.g.}, quadrangulations) is fairly well understood. From the point of view of statistical physics, all types of $\mathbf{q}$-Boltzmann distributed random maps without decorations (and with faces of bounded degrees) are different microscopic descriptions of a same macroscopic system. Therefore, one expects the appearance of the same universal object in the limit, i.e. when the volume of our maps tends to infinity. It is in general a very difficult problem to establish rigorously the convergence to this continuum object, but in this case it has been proved for many families of maps that the scaling limit is a random compact metric space called the Brownian map \cite{MarckertMokkadem,GallMap,MierMapP,Gall1}. The complete proof of convergence in the Gromov-Hausdorff sense was only obtained in \cite{MierMapP,Gall1} for uniform quadrangulations and triangulations. After this, the same limit was found in many more cases, see for example \cite{marzouk2016scaling} for planar maps with given degree faces and \cite{Bettinelli10} for maps of all genera. 



\begin{figure}[htpb]
  \centering
  \includegraphics[width=.65\textwidth]{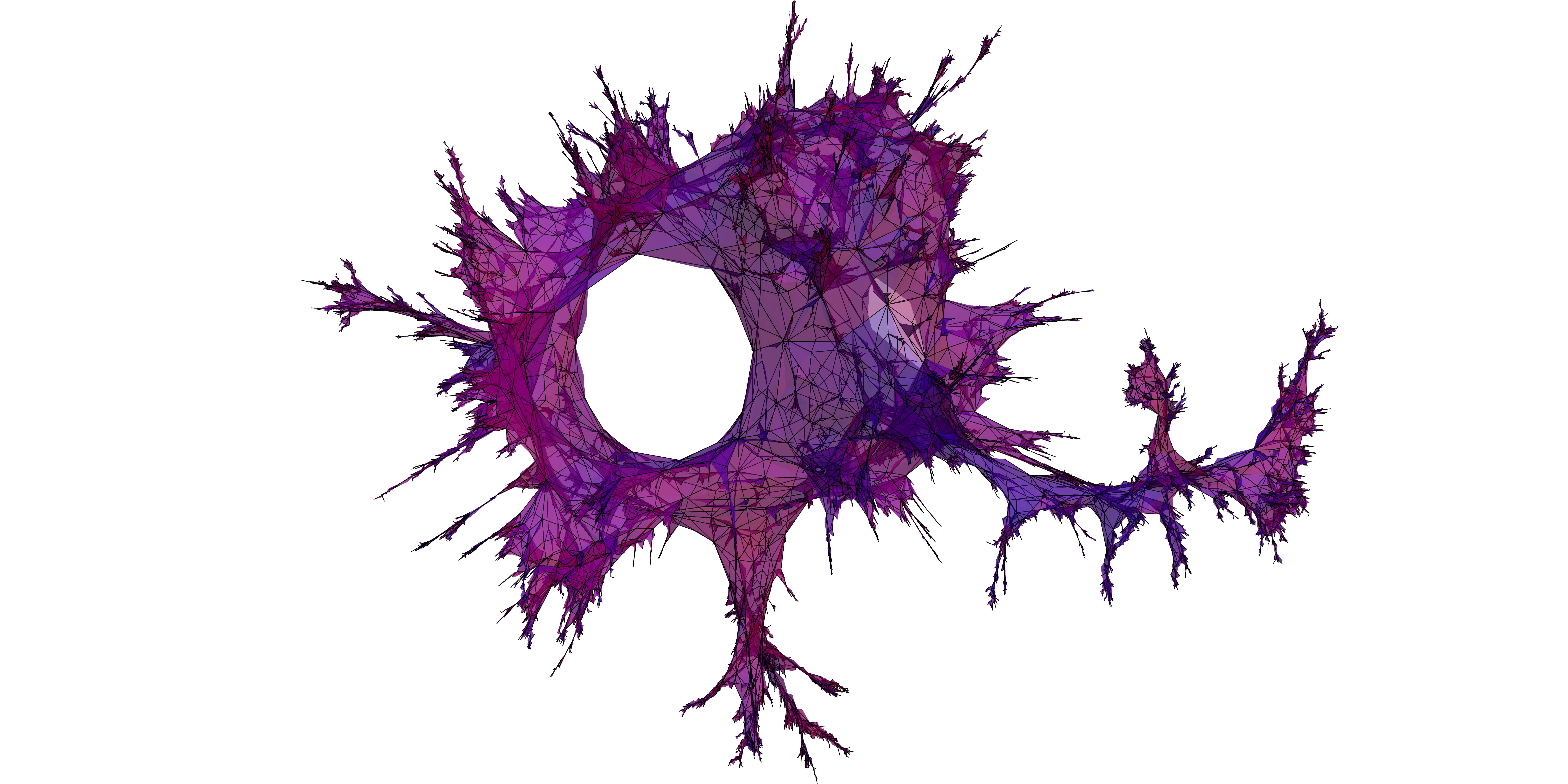}
  \caption[Caption for LOF]{Brownian torus.\protect\footnotemark}
\end{figure}
\footnotetext{Many thanks to J\'{e}r\'{e}mie Bettinelli for the nice picture. More related computer simulations can be found on his \href{http://www.normalesup.org/~bettinel/simul.html}{website}.}

This universality class is often referred to as \emph{pure gravity}  in physics.

Recent progress generalized part of this understanding to planar maps containing faces whose degrees are drawn from a heavy tail distribution. In particular, the limiting object is the so-called $\alpha$-stable map, which can be coded in terms of stable processes whose parameter $\alpha$ is related to the power law decay of the degree distribution \cite{MierGall}.

\subsubsection{Volume and perimeter exponents}

The exponent $-5/2$ from \eqref{volumeExp}, called \emph{volume exponent}, and $-3/2$ and $-5/2$ from \eqref{perimeterExpSub} and \eqref{perimeterExpCrit}, called \emph{perimeter exponents}, are common to all the maps in the pure gravity universality class and, hence, we say that they are also universal and characteristic of the Brownian universality class. There are many features related to this universality class, but these exponents are particularly interesting since they can also be computed for decorated maps, which is interesting to compare the distinct universality classes.

\subsection{The $O(\mathsf{n})$ loop model}\label{ProbOnMod}

The next class of interesting models concerns random maps equipped with a statistical physics model, combinatorially maps endowed with some decorations, like percolation \cite{Kazakov88}, the Ising model \cite{Kazakov86,Boulatov-Kazakov}, or the $Q$-Potts model \cite{Daul,BonnetEynard,PZinn}. It is well-known, at least on fixed lattices \cite{FKcluster,Boulatov-Kazakov,Truong,PerkWu,NienhuisCG}, that the $Q$-state Potts model can be reformulated as a fully packed loop model with a fugacity $\sqrt{Q}$ per loop. For random maps this equivalence is explained in detail in \cite{BBG12c}. In this thesis, we will be especially interested in the $O(\mathsf{n})$ loop model in general.

A remarkable feature of the $O(\mathsf{n})$ model is that it gives rise to two new universality classes which depend continuously on $\mathsf{n}$, called \emph{dense} or \emph{dilute} in respect to the behavior of macroscopic loops, as can be detected at the level of critical exponents \cite{1982PhRvL..49.1062N,1984JSP....34..731N,NienhuisCG,1988PhRvL..61.1433D,GaudinKostov,KOn,1990NuPhB.340..491D,KSOn,PagesjaunesOn}. The famous KPZ relations \cite{KPZ} (see also \cite{MR981529,MR1005268}) relate, at least from the physics point of view, the critical exponents of these models on a fixed regular lattice, with their corresponding critical exponent on random planar maps, as was repeatedly checked for a series of models \cite{KPZ,1988PhRvL..61.1433D,1990NuPhB.340..491D,KOn,MR2112128,BBD}. 

\begin{center}
\begin{figure}[h!]
\begin{center}
\def\svgwidth{0.6\columnwidth}
\scalebox{1}{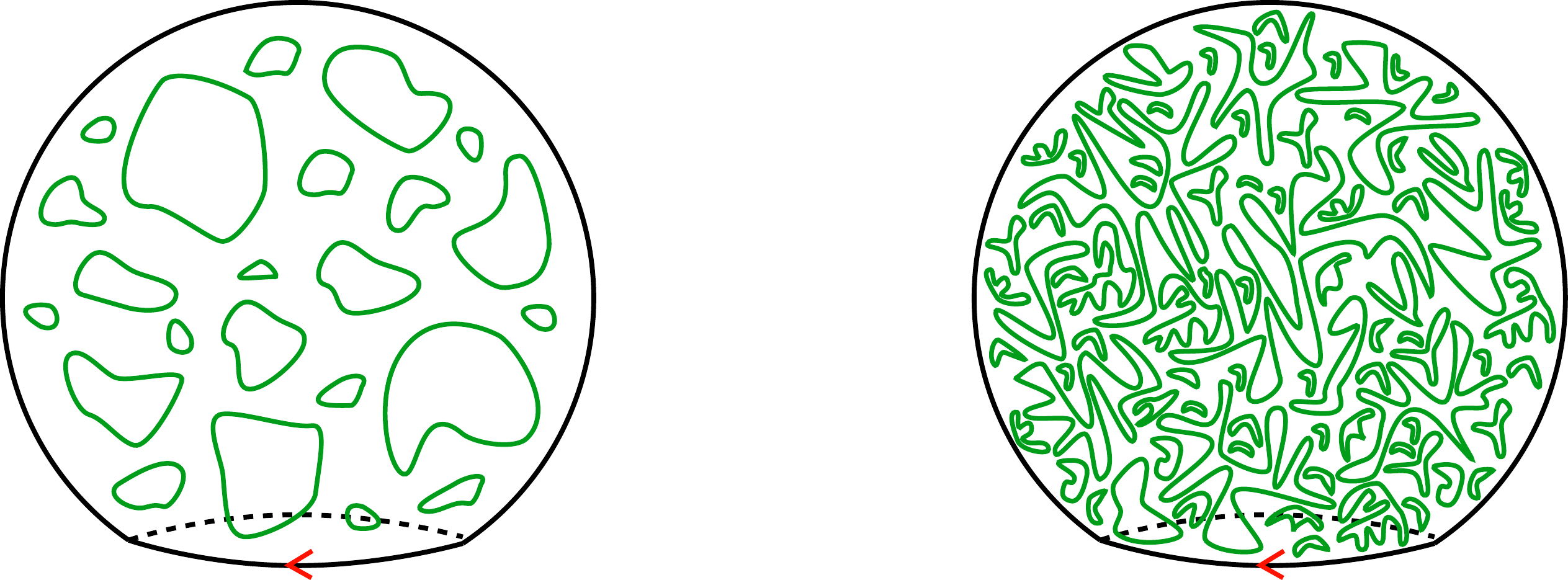}
\caption{\label{DenseDilute}Schematic illustration of dilute and dense phases (left and right, respectively).}\label{decomp}
\end{center}
\end{figure}
\end{center}

The \emph{gasket} of a disk (concept introduced in \cite{BBG12a}) is obtained by removing the interior of the outermost loops. In the dense phase, the loops on the gasket are believed to touch themselves and each other in the scaling limit, while in the dilute phase are believed to be simple and dispersed, avoiding each other, as is illustrated in Figure~\ref{DenseDilute}.

\subsubsection{Phase diagram}\label{IntroPhase}

In this thesis we will mainly focus on the bending energy model, since the generating series will be amenable to computations in this model, which still contains the two universality classes believed to be characteristic of the general $O(\mathsf{n})$ loop model.

\begin{figure}[h!]
\begin{center}
\def\svgwidth{0.5\columnwidth}
 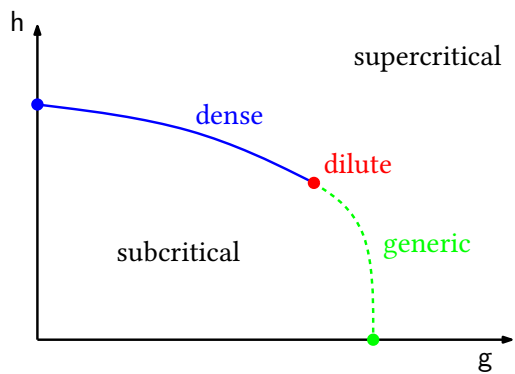
 \caption{\label{Qualiphas} The phase diagram of the model with bending energy is qualitatively insensitive to the value of $\mathsf{n} \in (0,2)$ and $\alpha$ not too large.}
    \end{center}
\end{figure}

The phase diagram of the model with bending energy was rigorously determined in \cite{BBG12b,BBD}, and is plotted qualitatively in Figure~\ref{Qualiphas}, see also the early works \cite{KOn,GaudinKostov} for $\alpha = 1$. The same universality classes and qualitatively the same phase diagram were obtained for the rigid $O(\mathsf{n})$ loop model on quadrangulations \cite{BBG12a}, and are expected for more general loops models, where $\mathsf{g}$ and $\mathsf{h}$ should be thought as a weight per unvisited and visited faces, respectively. 

For $\mathsf{n}\in(0,2)$, there are three universality classes in the model with bending energy: \emph{generic}, non-generic \emph{dilute} and non-generic \emph{dense}. For $\mathsf{n} > 0$, we find a dense critical line, which ends with a dilute critical point, and continues as a generic critical line. For $\mathsf{n} = 0$, only the generic critical line remains, which is the universality class of pure gravity, already present in maps without loops. The non-generic universality class will be the most interesting for us, since it is specific to the loop model, and it corresponds to a regime where macroscopic loops continue to exist in maps of volume $V \rightarrow \infty$ \cite{KOn,BEThese}. If $(\mathsf{g},\mathsf{h})$ is chosen on the non-generic critical line but we keep the vertex weight $u < 1$, the model remains off-critical, and the distance to criticality is governed by $(1 - u) \rightarrow 0$.

If $\mathsf{g}=0$, the model is called \emph{fully packed}, i.e. all internal faces are visited by loops.

\subsubsection{Critical exponents} 
\begin{notation}
We agree that $F \stackrel{\bigcdot}{\sim} G$ means there exists a constant $C > 0$ such that $F \sim CG$ in the asymptotic regime under study.
\end{notation}
We will now describe the critical exponents that characterize the universality classes of the $O(\mathsf{n})$ loop model. In the case of the dense and dilute universality classes, the exponents will be rational functions of the following paramenter:
\beq
b=\frac{1}{\pi}\arccos\left(\frac{\mathsf{n}}{2}\right),
\eeq
which decreases from $\frac{1}{2}$ to $0$ as $\mathsf{n}$ increases from $0$ to $2$.
Actually, the four phases of the model can be defined by the universal perimeter exponent $a\in\left[\frac{3}{2},\frac{5}{2}\right]$, given by the asymptotic behavior as we keep $u=1$ but take the boundary to be of large perimeter:
$$
\mathcal{F}_{\ell}\underset{\ell\rightarrow\infty}{\stackrel{\bigcdot}{\sim}}\gamma_{+} \ell^{-a},\ \ \ \  \ell\rightarrow\infty,
$$
where $\gamma_{+}$ is a non-universal constant and with the following correspondence to the critical phases:
$$
a=\begin{cases}
3/2, & \text{ subcritical}, \\
2-b, & \text{ critical, non-generic and dense}, \\
2+b, & \text{ critical, non-generic and dilute}, \\
5/2, & \text{ critical generic}.
\end{cases}
$$

To find the volume exponent, we keep the model at a critical point, i.e. $u_{c}=1$. The type of singularity of the generating series of disks for fixed perimeter $\ell$ around a critical point is encoded by the so-called \emph{string susceptibility exponent} $\gamma_{{\rm str}}$:
\beq
\left.\mathcal{F}_{\ell}\right|_{{\rm sing}}  \stackrel{\bigcdot}{\sim} (1-u)^{1-\gamma_{{\rm str}}}, \ \ \ u \rightarrow 1.
\eeq
Since $u$ is coupled to the volume, the generating series of disks of fixed length with fixed large volume behaves as:
\beq
[u^{V}]\mathcal{F}_{\ell}\stackrel{\bigcdot}{\sim}V^{\gamma_{{\rm str}}-2}, \ \ \ V\rightarrow \infty.
\eeq 
For the $O(\mathsf{n})$ loop model, the string susceptibility exponent $\gamma_{{\rm str}}\in[-1,0]$ may take the following values:
$$
\gamma_{{\rm str}}=\begin{cases}
-\frac{1}{2}, & \text{ generic}, \\
-\frac{b}{1-b}, & \text{ dense}, \\
-b, & \text{ dilute}.
\end{cases}
$$
Recall that the KPZ relation relates the string susceptibility exponent to the central charge $\mathfrak{c}$ of conformal field theory: 
\beq
\gamma_{{\rm str}}(\mathfrak{c})=\frac{\mathfrak{c}-1-\sqrt{(1-\mathfrak{c})(25-\mathfrak{c})}}{12}.
\eeq
The case of pure gravity corresponds to the trivial CFT $\mathfrak{c}=0$, which gives $\gamma_{{\rm str}}(0)=-1/2$. The critical point of the Ising model gives $\mathfrak{c}=1/2$, and hence $\gamma_{{\rm str}}(0)=-1/3$, as predicted by the exact solution over random maps \cite{Kazakov86,Boulatov-Kazakov}. 
These and some other exponents for random planar maps are summarized in \cite[Figure 4]{BBD}.
\subsubsection{Relation to other models}

We finally remind the reader of some important instances of the $O(\mathsf{n})$ loop model:
\begin{itemize}
\item$\mathsf{n}=0$ in the dilute phase: pure gravity.
\item$\mathsf{n}=1$ in the dense phase: critical percolation.
\item$\mathsf{n}=1$ in the dilute phase: Ising model (spin clusters).
\item$\mathsf{n}=\sqrt{2}$ in the dense phase: Ising model (FK clusters).
\item$\mathsf{n}=2$ (dilute and dense exponents coincide): Kosterlitz-Thouless.
\item$\mathsf{n}=\sqrt{Q}$ in the dense phase: $Q$-Potts and its FK cluster boundaries.
\end{itemize}

\section{Topological recursion (TR)}\label{TRIntro}

Topological recursion (often abbreviated as TR) is an ubiquitous procedure developed in the last decade which uses basically residue computations on a Riemann surface. It was initially discovered by B. Eynard, N. Orantin and L. Chekhov around 2004 in the context of large size asymptotic expantions in random matrix theory \cite{AMM,CEO06, E1MM,CE061,CE062} and established as an independent universal theory around 2007 \cite{EOFg}. 

A characteristic feature of this recursion formula is that it is related to many different fields as enumerative geometry, volumes of moduli spaces, Gromov-Witten invariants, integrable systems, geometric quantization, mirror symmetry, matrix models, knot theory and string theory, which has always been surprising and at the same time exciting. Moreover, many aspects of this theory still remain a mystery. 

Nowadays TR has already several robust generalizations which helped giving it more structure and placing it in different contexts, as well as opening many new interesting questions. TR plays a central role in this thesis: it serves both as a powerful tool and as an important motivation. In this introduction, we focus on giving a brief overview of the original formulation, and just mention other related topics, recent developments and generalizations.

\subsection{The recursive formula}

The method of topological recursion associates to a so-called spectral curve $\mathcal{S}$, which consists of a Riemann surface with some extra data, a doubly indexed family of meromorphic multi-differentials $\omega_{g,n}$ on $\Sigma^n$:
\begin{center}
TR:\ \ Spectral curve $\mathcal{S}$ $\leadsto$ Invariants $\omega_{g,n}$ ($\mathfrak{F}_g = \omega_{0}^{[g]}$).
\end{center}

\begin{definition}[Input] A {\it spectral curve} $\mathcal{S}=(\Sigma, x, \omega_{0,1}, \omega_{0,2})$ is defined by the following data:
\begin{itemize}
\item $\Sigma$ is a Riemann surface,
\item $x: \Sigma \rightarrow \mathbb{C}$ is a meromorphic function with finitely many and simple critical points (denoted ${\rm Cr}(x) \coloneqq\{a\in \Sigma\mid  x^{\prime}(a)=0\} $), which can be thought as a ramified covering with ${\rm Cr}(x)$ the set of ramification points,
\item $\omega_{0,1}$ is a meromorphic $1$-form on $\Sigma$, often written $\omega_{0,1}=y\,\dd x$ with $y: \Sigma \rightarrow\mathbb{C}$ holomorphic on a neighborhood of every $a \in {\rm Cr}(x)$ and $dy(a) \neq 0$ (curves satisfying these conditions for $y$ are called \emph{regular}),
\item $\omega_{0,2}$ is a symmetric bi-differential on $\Sigma \times \Sigma$ with double poles along the diagonal and vanishing residues, that is locally
$$
\omega_{0,2}(z_1,z_2) = \frac{\dd z_1 \dd z_2}{(z_1-z_2)^2} + \overbrace{h(z_1,z_2)}^{\text{holomorphic}}.
$$
\end{itemize}
\end{definition}
Let $\hat{g}$ be the genus of $\Sigma$. One may also include in the input of TR the extra information of a Torelli marking, which is a choice of a symplectic basis $\{\{\mathcal{A}_i\}_{i=1}^{\hat{g}},\{\mathcal{B}_i\}_{i=1}^{\hat{g}}\}$ of $H_1(\Sigma,\mathbb{Z})$. If we consider a bi-differential $B(z_1,z_2)$ on $\Sigma\times\Sigma$ satisfying the properties imposed for $\omega_{0,2}$, the non-holomorphic part is fixed and has the form we gave in the definition. Additionally imposing the following normalization on the $\mathcal{A}$-cycles of $H_1(\Sigma,\mathbb{Z})$ for $B(z_1,z_2)$:
$$
\oint_{\mathcal{A}_i}B(z_1,\cdot)=0
$$
fixes also its holomorphic part. Thus, $B(z_1,z_2)$ is the unique bi-differential with those properties. Such a bi-differential has a natural construction in algebraic geometry and is called the normalized \emph{fundamental differential of the second kind}\footnote{It often receives the misleading name of \emph{Bergman kernel} in the community.} on $\Sigma$. In case the spectral curve is of genus $0$, i.e. $\Sigma = \mathbb{C}P^1$, it is known that
$$
B(z_1,z_2)=\frac{\dd z_1\dd z_2}{(z_1-z_2)^2}.
$$
So far, almost all cases where the meaning of the TR invariants is understood have spectral curves of genus $0$.

If $\Sigma$ is compact, and $x$ and $y$ are meromorphic, they must be algebraically dependent, i.e. $E(x,y)=0$, with $E$ some polynomial in two variables. Then, the functions $x$ and $y$ provide a parametric representation of a plane curve with $\Sigma$ as parameter space:
$$
\{(x(z),y(z))\mid z\in\Sigma\}=\{(x,y)\mid E(x,y)=0\}.
$$

Since the critical points of $x$ are simple, there exist neighborhoods $U_a$ around each $a\in {\rm Cr}(x)$ and a holomorphic map $\sigma_a: U_a\rightarrow U_a$ such that $x(z)=x(\sigma_a(z))$ for all $z\in U_a$, $\sigma_a \neq {\rm id}$, $\sigma_a^2 = {\rm id}$ and $\sigma_a(a)=a$.
\begin{definition}[Output]
Given a spectral curve $\mathcal{S}$, we define the \emph{TR amplitudes} (also known as \emph{TR correlators} or \emph{TR invariants}), for all $g \geq 0$, $n\geq 1$ with $2g-2+n >0$, :
\begin{align}
\omega_{g,n}(z_1,\ldots,z_n) & \coloneqq\sum_{a\in {\rm Cr}(x)} \underset{z=a}{{\rm Res}} \, K_a(z_1,z)\Big( \omega_{g-1,n+1}(z,\sigma_a(z),z_2,\ldots,z_n) +\nonumber \\
\label{TRampl} & +  \sum_{\mathclap{\substack{h =0,\ldots, g \\ I \sqcup J =\{2,\ldots,n\}} }}{}^{'} \omega_{h, |I|+1}(z,z_I)\, \omega_{g-h,|J|+1}(\sigma_a(z),z_J)\Big),
\end{align}
with $\sum{}^{'}$ meaning we omit $(h,I)=(0,\emptyset)$ and $(h,J)=(g,\{2,\ldots,n\})$, i.e. we omit the terms involving $\omega_{0,1}$. The \emph{recursion kernel} is defined as follows: 
$$
K_a(z_1,z)\coloneqq\frac{\int_{z'=\sigma_a(z)}^z \omega_{0,2}(z_1,z')}{2(\omega_{0,1}(z)-\omega_{0,1}(\sigma_a(z))}.
$$

Let $\Phi(z)$ be a primitive of $\omega_{0,1}(z)$, that is $\dd\Phi(z)=\omega_{0,1}(z)$. For $g\geq 2$ and $n=0$, we define
\beq\label{TRsympInv}
\mathfrak{F}_g\coloneqq\omega_{g,0}=\frac{1}{2-2g}\sum_{a\in {\rm Cr}(x)} \underset{z=a}{{\rm Res}}  (\omega_{g,1}(z)\Phi(z)),
\eeq
which we remark is independent of the choice of primitive $\Phi$.
\end{definition}
 
We called the invariants $\omega_{g,n}$ multi-differentials on $\Sigma^n$. More precisely, the $\omega_{g,n}$'s are
meromorphic sections of the bundle $\bigotimes_{i=1}^n\pi_i^* T^*\Sigma \rightarrow \Sigma^n$, where $\pi_i:\Sigma^n\rightarrow \Sigma$ denotes the projection onto the $i$th factor.

The $\mathfrak{F}_g$'s are complex numbers\footnote{In general, they are believed to belong to the field over which the spectral curve $\mathcal{S}$ was defined, but this has not been justified yet.}.
We do not give here the definition of $\mathfrak{F}_0$ and $\mathfrak{F}_1$, since they are more involved and we will not use them in this thesis.

\begin{figure}[h!]
    \begin{center}
        \centering
        \def\svgwidth{\columnwidth}
        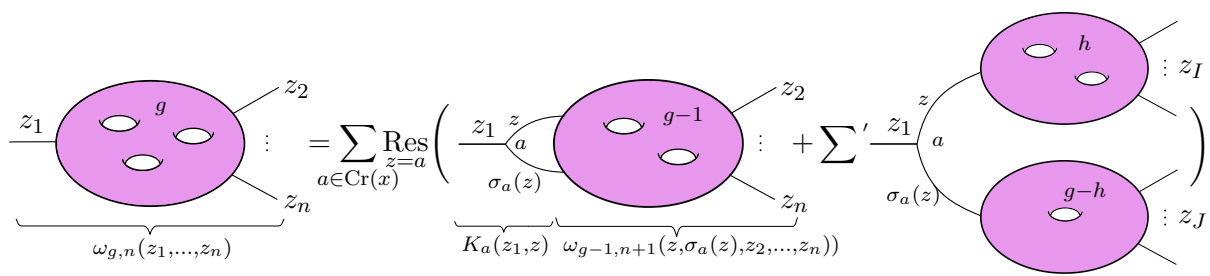
    \end{center}
\caption{\label{TR} Graphical representation of the recursive formula of TR. With this graphical representation, the method of topological recursion can be thought as a process of first solving the problem for the simplest topologies: disks $(0,1)$ and cylinders $(0,2)$, and afterwards understanding the problem for higher topologies $(g,n)$ by gluing pairs of pants (surfaces of topology $(0,3)$) and hence increasing the minus Euler characteristic $-\chi = 2g-2+n$.}
\end{figure}

The topological recursion has a graphical representation, which is often useful to illustrate computations. We depict every amplitude $\omega_{g,n}$ with a surface of genus $g$ with $n$ marked points (or $n$ boundaries) and the recursion kernel with a surface with $3$ marked points (or a pair of pants), which can be drawn for simplicity as a trivalent vertex like in the Figure \ref{TR}.

\begin{remark}
The topological recursion receives its name because it is a recursion on the minus Euler characteristic $-\chi_{g,n} = 2g-2+n$ of the surfaces associated to the correlators $\omega_{g,n}$.
\end{remark}

The table in Figure \ref{tableEx} shows that the TR amplitudes often encode interesting geometric or combinatorial data. It is a great advantage to prove that a certain problem satisfies TR because it provides a way of calculating the correlators $\omega_{g,n}$ and, even if there may have already been formulas to calculate those correlators and in some cases even more efficiently, the problem still gains a lot of structure for free, just for being part of this universal theory, like for example the relation to integrable hierarchies or quantum curves.

The first spectral curve $x=\frac{1}{2}y^2$ of our table is called the Airy curve. It is a fundamental result that the Virasoro constraints for the $\psi$-class intersection numbers (Witten-Kontsevich theorem we mentioned in \ref{WittenKontsevich}) is equivalent to TR for the Airy curve. 

An important feature of TR is that the correlators $\omega_{g,n}$ can be written in general in terms of intersection numbers of tautological classes. Any regular spectral curve locally resembles the Airy curve, and hence the $\omega_{g,n}$'s are related to $\psi$-class intersection numbers, defined as integrals over $\overline{\mathcal{M}}_{g,n}$. Later, Eynard \cite{INTR0,INTR} associated a tautological class in a moduli space of decorated stable Riemann surfaces to any spectral curve in such a way that the integral of the class over the moduli space coincides with the TR amplitudes of that spectral curve. Finally, Dunin-Barkowski, Orantin, Shadrin and Spitz \cite{DBOSS} showed that TR computes the correlators of any semi-simple Cohomological Field Theory translating Givental data into local spectral curve data. 




\renewcommand{\arraystretch}{1.5}

\begin{figure}[h!]
\begin{center}
\begin{tabular}{ |p{5.5cm}|p{7.5cm}| }
 \hline
 Spectral curve $\mathcal{S}$ & TR amplitudes $\omega_{g,n}$ \\
 \hline
 $x=\frac{1}{2}y^2$ \ \cite{EOFg}  & $\psi$-class intersection numbers on $\overline{\mathcal{M}}_{g,n}$  \\
 $x=ye^{-y}$ \ \cite{BEMS,EMS,DLN,BSLM} & Simple (and orbifold) Hurwitz numbers \\
 $x=2\cosh y$ \ \cite{NS,DBOSS} & Gromov-Witten invariants of $\mathbb{C}P^1$ \\
  $x=\frac{1}{y(1-y)}$ \ \cite{DN1} & Dessins d'enfants \\
 $y=\frac{-\sin(2\pi\sqrt{x})}{2\pi}$ \ \cite{EOMirzakhani}  & Weil Petersson volumes of $\mathcal{M}_{g,n}(L_1,\ldots,L_n)$  \\
 Mirror curve of a toric CY $3$-fold \cite{BKMP,EO-BKMP,FLZ}  & Gromov-Witten total ancestor potential for the $3$-fold  \\
 $A$-polynomial of a knot \ \cite{DFM,BEApol} & Conjecturally, wave function associated to the correlators $\leadsto$  colored Jones polynomial  \\
 \hline
\end{tabular}
\caption{\label{tableEx} Table with examples of TR problems.}
\end{center}
\end{figure}

\subsection{Main properties}

Topological recursion has many remarkable properties, and here we describe briefly the main ones:
\vspace{0.1cm}

$\bullet$ {\it Symmetry:} The TR correlators $\omega_{g,n}(z_1,\ldots,z_n)$ are symmetric under permutation of the variables $z_1,\ldots,z_n$. We remark that this property is a priori not apparent since $z_1$ plays a special role in the definition of the recursion.
\vspace{0.1cm}

$\bullet$ {\it Pole structure:} For stable topologies, i.e. for $2g-2+n>0$, the TR correlators $\omega_{g,n}(z_1,\ldots,z_n)$ are meromorphic, with poles only at critical points $a\in{\rm Cr}(x)$ and vanishing residues.
\vspace{0.1cm}

$\bullet$ {\it Homogeneity:} The TR correlators $\omega_{g,n}$ are homogeneous of degree $2-2g-n$, i.e. under the transformation $\omega_{0,1}\mapsto\lambda\omega_{0,1}$ for some $\lambda\in\mathbb{C}^*$, they transform as $\omega_{g,n}\mapsto\lambda^{2-2g-n}\omega_{g,n}$.
\vspace{0.1cm}

$\bullet$ {\it Dilaton equation:} For any $(g,n)$ such that $2g-2+n>0$, we have
$$
\sum_{a\in {\rm Cr}(x)} \underset{z=a}{{\rm Res}}\ \omega_{g,n+1}(z_1,\ldots,z_n,z)\Phi(z) =(2-2g-n)\, \omega_{g,n}(z_1,\ldots,z_n),
$$
where $\Phi$ is such that $\dd \Phi = y\, \dd x =\omega_{0,1}$. This property was used to define the $\mathfrak{F}_g$'s for $g\geq 2$.
\vspace{0.1cm}

The TR correlators have many more properties that we do not elaborate on here, such as: they satisfy deformation equations (such as Theorem~\ref{deformationsTR}), have some modular properties~\cite{EMhol}, behave well under taking singular limits of families of spectral curves~\cite[Theorem 5.3.2]{Eynardbook} and are believed to satisfy Hirota-like equations~\cite{BEInt}.

Finally, a very deep and still not well understood property that constituted a motivation for this thesis:

\subsubsection{Symplectic invariance}\label{sympinv} If two spectral curves $\mathcal{S}$ and $\mathcal{S'}$ are symplectically equivalent, that is $\dd x\wedge \dd y =\dd x'\wedge \dd y'$, then a relation between $\mathfrak{F}_g[\mathcal{S}]$ and $\mathfrak{F}_g[\mathcal{S'}]$ is expected. More concretely, they are proved to be equal for any symplectic transformation which does not imply exchanging $x$ and $y$ (see \cite{EOFg}). For this reason, the $\mathfrak{F}_g$'s are called \emph{symplectic invariants}. The exchanging transformation $(x,y)\mapsto(-y,x)$ is then considered to be the most mysterious and interesting one. These invariants are also expected to be equal after exchanging $x$ and $y$ up to some correction terms whose exact form is still under scrutiny (see \cite{EO2MM,EOxy} for some progress towards the relation for algebraic compact curves). One of the motivations for some results in this thesis was also to gain some understanding of this property in this complicated case.

\subsection{TR for ordinary maps}\label{TRMaps}

In the context of maps, $\Sigma$ is the curve on which the generating series of disks can be maximally analytically continued with respect to its parameter $x$ coupled to the boundary perimeter, and $\Sigma$ has a distinguished point $[\infty]$ corresponding to $x \rightarrow \infty$.

\begin{theorem}\cite{E1MM}\label{ordmaps} Let $x(z)\coloneqq\alpha + \gamma \left( z+\frac{1}{z}\right)$, with $\alpha$ and $\gamma$ parameters determined in terms of the weights $(u, t_3, t_4,\ldots)$. Then, up to a correction term, and written as a bi-differential form, the generating series of ordinary, usual cylinders is the fundamental differential of the second kind for rational spectral curves:
\beq 
W_2^{[0]}(x(z_1), x(z_2)) \dd x(z_1) \dd x(z_2) + \frac{\dd x(z_1) \dd x(z_2)}{(x(z_1)-x(z_2))^2} = B(z_1,z_2). \nonumber
\eeq
In general, we rewrite the generating series of maps in terms of the variables $z_i$ and as multi-differential forms in $\mathbb{C}P^1$:
\beq
\label{WngTR}\omega_{g,n}(z_1,\ldots,z_n)=W_n^{[g]}(x(z_1),\ldots, x(z_n)) \dd x(z_1)\cdots \dd x(z_n) + \delta_{g,0}\delta_{n,2}\frac{\dd x(z_1) \dd x(z_2)}{(x(z_1)-x(z_2))^2}. 
\eeq
Then, for $2g-2+n>0$, the multi-differential forms $\omega_{g,n}(z_1,\ldots,z_n)$ satisfy the topological recursion applied to the following spectral curve:
$$
\left(\mathbb{C}P^1, x, W_1^{(0)}(x(z))\dd x(z), B(z_1,z_2)\right).
$$ 
\end{theorem}
Here $z_i$ is a generic name for points in $\Sigma$, and \eqref{WngTR} means the equality of Laurent expansion near $z_i \rightarrow [\infty]$.

\subsubsection{TR for $O(\mathsf{n})$ configurations}\label{TRConf}

The generating series $\mathcal{W}^{[g]}_k$ of $O(\mathsf{n})$ configurations also satisfy the topological recursion \cite{BEOn,BEO}. We will give details on the base cases in Section \ref{ONsection} and, even more explicitly for the particular case of the bending energy model in Section \ref{S4} and Appendix \ref{App1}, including the special parametrization of the spectral curve, but we already comment that the spectral curve is non-algebraic in general and $\Sigma$ is realized naturally as the universal cover of a torus.


\subsection{Generalizations}\label{generalizationsTR}

We finish with a brief overview of the generalizations of TR:
\vspace{0.1cm}

$\bullet$ {\it Higher order ramifications:} TR can be generalized to curves with non-simple ramification points, i.e. points where the order of ramification is higher than $2$ \cite{BouchardEynard,HigherRam}.
\vspace{0.1cm}

$\bullet$ {\it Irregular curves:} The outcome of TR for irregular spectral curves is explored in \cite{DN1,DN2}. For instance, in the table of examples we included the spectral curve for the enumeration problem of dessins d'enfants, which is an irregular curve.
\vspace{0.1cm}

$\bullet$ {\it Blobbed topological recursion (BTR):} TR provides solutions of loop equations \cite{EOFg,BEO}. The set of all solutions to the loop equations is given by a generalization of TR called blobbed TR \cite{BSblob}. The initial data of TR $(\omega_{0,1},\omega_{0,2})$ is here enriched by the so-called blobs, which are symmetric holomorphic forms. We remark that multitrace matrix models from the matrix model point of view and stuffed maps from the combinatorial point of view, which generalize the one-matrix model and usual maps, satisfy the blobbed TR.
\vspace{0.1cm}

$\bullet$ {\it Quantum Airy structures (KS-TR):} In 2017, Kontsevich and Soibelman \cite{KontsevichSoibelman} reformulated TR seeing it as a quantization of quadratic Lagrangians in the symplectic vector space $T^*V$, for some vector space $V$. This procedure takes as input a quantum Airy structure, which consists of a particular family of at most quadratic differential operators on $V$ which form a Lie algebra and whose coefficients are encoded in four tensors $(A,B,C,D)$. This approach allows a more algebraic and slightly more general presentation of the possible initial data for TR, and was further studied in \cite{ABCD}. The output produced consists of a formal series of functions on $V$, annihilated by the operators that conform the input. If the Lie algebra formed by the operators is a direct sum of copies of the Borel subalgebra of the Virasoro algebra, this formalism is equivalent to the TR presented in this section.
\vspace{0.1cm}

$\bullet$ {\it Geometric recursion (GR):} This very recent theory \cite{GR} takes as input a functor $E$ from a category of surfaces to a category of topological vector spaces, together with gluing data, and produces functorial assignments $\Sigma \mapsto \Omega_{\Sigma}\in E(\Sigma)$.  Any initial data for TR can be lifted to an input data for GR corresponding to continuous functions over Teichm\"{u}ller spaces\footnote{Actually, we are omitting the subtlety that in general one gets distributions over Teichm\"{u}ller spaces. This is circumvented by introducing a regularization parameter which deforms the spectral curve in such a way that one obtains continuous functions as desired. When the parameter is sent to $0$, the deformed spectral curve approaches the original one in a non-singular way.}, valued in a Frobenius algebra, in such a way that integrating GR amplitudes over the moduli space, one recovers the TR amplitudes after a Laplace transform on boundary lengths. This is exactly the procedure that relates Mirzakhani's recursion for the Weil-Petersson volumes, which was found using hyperbolic geometry, to TR in the fifth example of our table. This construction generalizes all previous versions of TR and aims at producing quantities associated to surfaces which are invariant under mapping class group transformations.
\vspace{0.1cm}

\begin{remark}
The initial data of KS-TR, given by $(A,B,C,D)$, corresponds exactly to the coefficients $\mathcal{C}_{0,3}\ (A)$, $\mathcal{C}_{1,1}\ (D)$, $\tilde{\mathcal{K}}\ (B)$ and $\mathcal{K}\ (C)$ that we used in Section \ref{TRtr} to analyze the critical behavior of TR amplitudes. We decided to use these building blocks for TR before the formulation of KS-TR was discovered.
\end{remark}

\begin{figure}[h!]
    \begin{center}
        \centering
        \def\svgwidth{\columnwidth}
        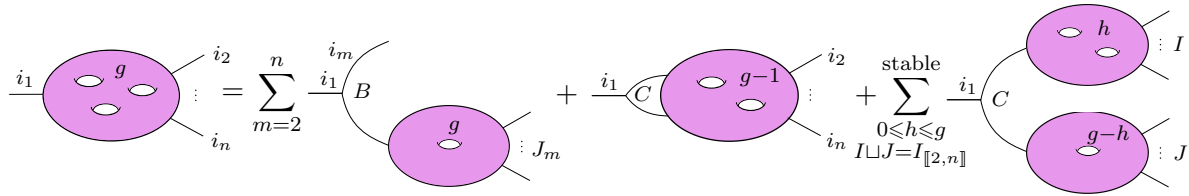
    \end{center}
\caption{\label{KSTR} Graphical representation of the recursive formula of KS-TR. We have used the notations $I_{\llbracket 2,n\rrbracket}\coloneqq\{i_2,\ldots,i_n\}$, $J_m\coloneqq I_{\llbracket 2,n\rrbracket}\setminus \{i_m\}$ and ``stable'' to indicate that we exclude the terms involving disks and cylinders, i.e. $(h,|I|+1),(g-h,|J|+1)\not\in \{(0,1),(0,2)\}$. $B$ and $C$ correspond to our $\tilde{\mathcal{K}}$ and $\mathcal{K}$, respectively (see Proposition \ref{cocor1}).}
\end{figure}

\section{Free probability}\label{IntroFP}

Dan-Virgil Voiculescu introduced free probability in the 1980's \cite{Voiculescu1} to address the free group factors isomorphism problem, an open problem in the theory of von Neumann algebras. Whereas Voiculescu's original approach \cite{VDN} is quite analytical and operator algebraic in nature, we will focus on the combinatorial aspects of free probability, which were initiated by Roland Speicher \cite{Speicher}, who introduced the concept of free cumulants using the lattice of non-crossing partitions. The combinatorial approach to (first order) free probability is nicely and exhaustively introduced in \cite{NicaSpeicher}, which is a good source, together with many other surveys \cite{SpeicherSurvey1,SpeicherSurvey2,NovakFP} for the curious reader who wants to complete the details of this very compact summary. 

\subsection{Free independence}

The first main feature of free probability is that we will allow the algebras of random variables to be non-commutative.
\begin{definition}
A {\it non-commutative probability space} $(\mathcal{A},\varphi)$ consists of a unital algebra $\mathcal{A}$ over $\C$, whose elements are called {\it (non-commutative) random variables}, and a unital linear functional $\varphi \colon \mathcal{A}\to \C$. We say that $(\mathcal{A},\varphi)$ is {\it tracial} if we additionally impose that $\varphi(ab)=\varphi(ba)$ for all $a,b\in\mathcal{A}$.
\end{definition}

Following the terminology from classical probability theory, we call $\varphi(a)$ the {\it expectation value} of $a\in\mathcal{A}$ and, in general, $\varphi(a^k)$, $k\geq 0$, are called the {\it moments}.

Free independence, which was modeled on the free product of groups, constitutes the basic notion which turns non-commutative probability into free probability.
\begin{definition}
Two random variables $a,b$ in a non-commutative probability space $(\mathcal{A},\varphi)$ are defined to be {\it freely independent} or {\it free} if $\varphi(P_1(a)Q_1(b)\cdots P_k(a)Q_k(b))=0$, whenever $P_i,Q_i$ are polynomials such that $\varphi(P_i(a))=\varphi(Q_i(b))=0$ for all $i=1,\ldots,k$.
\end{definition}

This should be thought as an analog of the classical notion of independence. Classical independence on this more general setting of non-commutative probability corresponds to the notion of tensor product:
\begin{definition}
Two random variables $a,b$ in a non-commutative probability space $(\mathcal{A},\varphi)$ are said to be {\it classically} (or {\it tensor}) {\it independent} if they commute: $ab=ba$, and $\varphi(P(a)Q(b))=0$, whenever $P,Q$ are polynomials such that $\varphi(P(a))=\varphi(Q(b))=0$.
\end{definition}

As in the case of classical independence, but not as obviously, free independence provides (and is determined by) a special rule to calculate joint moments of free independent variables using just moments of the single variables.

It is important to note that freeness is not a generalization of classical independence. As we have seen, both notions can be formulated within the framework of non-commutative probability. However, while classical independence requires commutativity, free independence becomes rather trivial if commutativity is imposed\footnote{One can show that commuting random variables can only be freely independent if at least one of them has vanishing variance, which implies that is also almost surely constant. It can also be checked that constant random variables are freely independent from everything.}. 

One can actually prove that these are the only two natural notions of independence possible.

\subsubsection{Examples}

\hspace{\parindent} {\it \textbf{Classical.}} Let $(\Omega, \mathcal{F}, P)$ be a probability space in the classical sense. Consider the algebra of genuine random variables $\mathcal{A}= L^{\infty -}(\Omega, P) \coloneqq  \bigcap_{p=1}^{\infty} L^p(\Omega, P)$ and $\varphi$ given by the classical expectation value:
$$
\varphi(a)\coloneqq \langle a\rangle = \int_{\Omega} a(\omega)\, d P(\omega).
$$
Classical probability spaces are always commutative.
\vspace{0.1cm}

{\it \textbf{Matrix spaces.}} Let $(\mathcal{A},\varphi)\coloneqq (\mathcal{M}_N(\C), {\rm tr})$, with $\mathcal{M}_N(\C)$ the algebra of $N\times N$ complex matrices and ${\rm tr}\colon \mathcal{M}_N(\C)\to \C$ the trace normalized such that ${\rm tr}\,{\rm Id}=1$. This probability space is commutative only when $N=1$.
\vspace{0.1cm}

{\it \textbf{Random matrices.}} Combining the previous deterministic example with the first one, which is genuinely random but commutative, provides an important model for free probability, namely the algebra of random matrices $\mathcal{A}\coloneqq\mathcal{M}_N(L^{\infty -}(\Omega, P))$ with $\varphi(M)\coloneqq \langle {\rm tr}\, M\rangle$.
\vspace{0.1cm}

{\it \textbf{Group algebras.}} Consider $(\C G, \tau_G)$, where $\C G$ is the group algebra of a group $G$ and the functional $\tau_G(\alpha)$ selects the coefficient of the identity for every $\alpha\in\C G$.
\vspace{0.1cm}

\begin{remark}
The last example provides a purely algebraic model for freely independent random variables, which actually was the one that motivated the concept of ``free'' independence: the subgroups $G_i$, $i\in I$, of $G$ are free if and only if the subalgebras $\C G_i$ of $\C G$ are freely independent in the non-commutative probability space $(\C G,\tau_G)$. 
\end{remark}

\subsection{Free cumulants}\label{1stOrderFCsSection}

We denote $\mathcal{P}(n)$ the set of partitions of $[n]$ and $|\pi|$ the number of blocks of $\pi\in\mathcal{P}(n)$. A partition is called {\it non-crossing} if no two blocks ``cross'' each other, i.e.~if labeling the vertices of a regular $n$-gon from $1$ to $n$, the convex hulls of different blocks of the partition are pairwise disjoint. We denote $NC(n)$ the set of non-crossing partitions of $[n]$. A partition where each block consists of exactly two elements, is called {\it pairing}. We denote $\mathcal{P}_2(n)$ and $NC_2(n)$ the sets of pairings and non-crossing pairings, respectively.

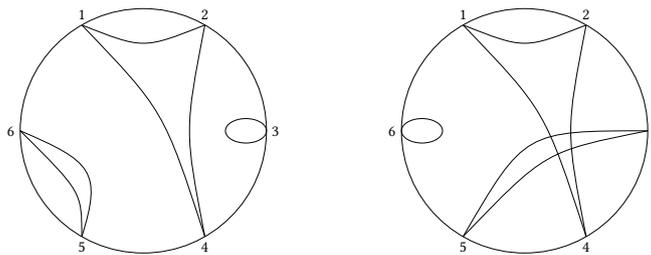
\begin{figure}[h!]
\begin{center}
\scalebox{0.54}{
\begin{tikzpicture}
\draw (60:3) node[anchor=south]{2} .. controls (0,2) .. (120:3) node[anchor=south]{1};
\draw (60:3) .. controls (1,0) .. (-60:3) node[anchor=north]{4};
\draw (120:3) .. controls (0.5,0.5) .. (-60:3);
\draw (2.5,0) ellipse (0.5cm and 0.3cm);
\draw (3,0) node[anchor=west]{3};
\draw (180:3) node[anchor=east]{6} .. controls (-1.5,-1.5) .. (240:3) node[anchor=north]{5};
\draw (180:3) .. controls (-1.25,-0.75) and (-1,-1) .. (240:3);
\draw (0,0) circle (3cm);
\end{tikzpicture}
\qquad\qquad\qquad
\begin{tikzpicture}
\draw (60:3) node[anchor=south]{2} .. controls (0,2) .. (120:3) node[anchor=south]{1};
\draw (60:3) .. controls (1,0) .. (-60:3) node[anchor=north]{4};
\draw (120:3) .. controls (0.5,0.5) .. (-60:3);
\draw (-2.5,0) ellipse (0.5cm and 0.3cm);
\draw (-3,0) node[anchor=east]{6};
\draw (0:3) node[anchor=west]{3} .. controls (0,0) .. (240:3) node[anchor=north]{5};
\draw (0:3) .. controls (0.5,-0.5) .. (240:3);
\draw (0,0) circle (3cm);
\end{tikzpicture}}
\caption{Left: Illustration of a non-crossing partition. Right: Illustration of a partition which is not non-crossing.}
\end{center}
\end{figure}

We will introduce now very useful objects, called free cumulants, which will allow to characterize free independence in a much simpler way than on the level of moments.

We first motivate their definition using non-crossing partitions by giving the idea behind the free version of the Central Limit Theorem. Let $a_1, a_2,\ldots \in\mathcal{A}$ identically distributed random variables which are either tensor or free independent with $\varphi(a_i)=0$ and $\sigma^2\coloneqq\varphi(a_i^2)$.

The CLT gives information about the behavior of 
$$
S_N\coloneqq \frac{a_1+\cdots+a_N}{\sqrt{N}}, \text{ when } N\to \infty .
$$
We denote $\ker (i_1,\ldots,i_n)$ the partition $\pi\in\mathcal{P}(n)$ whose blocks correspond to the different values of the indices, i.e. $r$ and $s$ belong to the same block of $\ker i$ if and only if $i_r=i_s$. Because of independence and all $a_i$ have the same distribution, we have $\varphi(a_{i_1}\cdots a_{i_n})=\varphi(a_{j_1}\cdots a_{j_n})$ whenever $\ker i = \ker j$. For every $\pi\in\mathcal{P}(n)$, let us denote $c_{\pi}$ the common value  of $\varphi(a_{i_1},\ldots,a_{i_n})$ for all $i$ with $\ker i =\pi$. Therefore,
$$
\varphi(S_N^n)=\frac{1}{N^{n/2}}\sum_{1\leq i(1),\ldots,i(n)\leq N} \varphi(a_{i(1)}\cdots a_{i(n)})=\frac{1}{N^{n/2}}\sum_{\pi\in\mathcal{P}(n)} c_{\pi}\cdot \vert\{i\colon [n]\to[N]\mid \ker i =\pi \}\vert.
$$
One can compute that in the limit:
$$
\lim_{N\to\infty} \varphi(S_N^n)=\sum_{\pi\in\mathcal{P}_2(n)} c_{\pi},
$$
which clearly vanishes, if $n$ is odd. For $n$ even and tensor independence we get
$$
\lim_{N\to\infty} \varphi(S_N^n)=\sum_{\pi\in\mathcal{P}_2(n)} \sigma^n = \sigma^n(n-1)!!.
$$
For free independence, however, partitions with the property that consecutive indices will coincide successively, i.e. non-crossing partitions, are the only ones that will contribute
$$
\lim_{N\to\infty} \varphi(S_N^{2k})=\sum_{\pi\in NC_2(2k)} \sigma^n =\sigma^{2k}C_{k}.
$$
It is well-known that Catalan numbers $C_k$ enumerate $NC_2(2k)$.

Let $\kappa_{\pi}[a_1,\ldots,a_n]\coloneqq\prod_{B\in\pi}\kappa_{|B|}(a_{i_1},\ldots,a_{i_s})$, where $i_1,\ldots,i_s\in B$, denote a product of classical cumulants. We recall the moment-cumulant relation in the classical setting
$$
\varphi(a_1\cdots a_n) = \sum_{\pi\in\mathcal{P}(n)} \kappa_{\pi}[a_1,\ldots,a_n].
$$
Motivated by our free version of the CLT, we define:
\begin{definition}
The {\it free cumulants} $k_{\pi}$ are given by the non-crossing moment-cumulant relation:
\beq\label{1stOrderFCs}
\varphi(a_1\cdots a_n) = \sum_{\pi\in NC(n)} k_{\pi}[a_1,\ldots,a_n].
\eeq
\end{definition}
It can be proved that this equation can be inverted by M\"{o}bius inversion:
$$
k_n(a_1,\ldots, a_n) = \sum_{\pi\in NC(n)} \varphi_{\pi}[a_1,\ldots,a_n]\mu(\pi,1_n),
$$
where $\varphi_{\pi}$ is a product of moments according to the block structure of $\pi$ and $\mu$ is the so-called M\"{o}bius function on $NC(n)$.

As classical cumulants characterize tensor independence by the vanishing of mixed classical cumulants \cite{Rota}, free cumulants constitute important objects which characterize free independence in an analogous way. As a consequence, and again analogously as in the classical world, we also get that free cumulants linearize the problem of adding free variables:
$$
k_n(x+y,\ldots,x+y)=k_n(x,\ldots,x) + k_n(y,\ldots,y), \text{ if } x \text{ and } y \text{ are free}.
$$
This property gives one of the main reasons to call such objects cumulants.

Apart from the analogy between classical and free cumulants in a probability context, another beautiful analogy was introduced \cite{NovakSniady} in the context of combinatorics as a different approach to introduce free cumulants: coupled to a notion called geometric connectedness as classical cumulants are related to usual connectedness.

\subsection{Higher order free probability}

The connection of free probability with random matrix theory \cite{MingoSpeicherBook} has been very enriching for both fields and is quite relevant for us, especially the introduction of second order freeness in \cite{MingoSpeicher,MMS07}. Finally, higher order freeness was introduced in \cite{Secondorderfreeness}, but has not been understood much further for the moment. Since it constitutes both an important motivation for us and a possible source of applications, we also provide a compact introduction to its main objects. Basically, while usual free cumulants are defined using non-crossing partitions, second order free cumulants use annular non-crossing permutations and increasing the order will involve more complicated non-crossing permutations which will be handled with general objects called partitioned permutations.

\subsubsection{Second order freeness}

\begin{definition} Let $(\mathcal{A},\varphi)$ be a tracial non-commutative probability space. Consider additionally a bilinear functional $\varphi_2\colon\mathcal{A}\times\mathcal{A} \to \C$ which is symmetric and tracial in both arguments and satisfies that $\varphi_2(1,a)=0$, for all $a\in \mathcal{A}$. We say that $(\mathcal{A},\varphi,\varphi_2)$ is a {\it second order non-commutative probability space}.
\end{definition}
Let $(\mathcal{A}_i)_{i\in I}$ be a family of unital subalgebras of $\mathcal{A}$. We say that $a_i\in\mathcal{A}_{j_i}$, for $i=1,\ldots,n$, are {\it cyclically alternating} if $j_1\neq j_2 \neq \cdots \neq j_n\neq j_1$.
\begin{definition}
We say that $(\mathcal{A}_i)_{i\in I}$ are {\it free of second order} if they are free with respect to $\varphi$ and, given two centered and cyclically alternating tuples $a_1,\ldots,a_m$ and $b_1,\ldots,b_n$, we have that, for $(m,n)\neq (1,1)$,
$$
\varphi_2(a_1\cdots a_m,b_n\cdots b_1)=\delta_{mn} \sum_{k=0}^{n-1}\prod_{i=1}^n \varphi(a_i b_{i+k}),
$$
where the indices of $b_i$ are considered modulo $n$, and for $(m,n)=(1,1)$, we have $\varphi_2(a,b)=0$, if $a\in \mathcal{A}_i$ and $b\in \mathcal{A}_j$, with $i\neq j$.
\end{definition}

\subsubsection{Preliminaries on partitioned permutations}\label{partitionedPermutations}

Let $\mathcal{U}=\{U_1,\ldots,U_k\}$ and $\mathcal{V}=\{V_1,\ldots,V_l\}$ be partitions of the same set. We say that $\mathcal{U}\leq\mathcal{V}$ if for every block $U_i$ there is some block $V_j$ in $\mathcal{V}$ that contains it: $U_i\subset V_j$. We denote $\mathcal{U}\vee \mathcal{V}$ the smallest partition $\mathcal{W}$ such that $\mathcal{U}\leq \mathcal{W}$ and $\mathcal{V}\leq \mathcal{W}$, by $1_B$ the biggest partition $\{B\}$ of the partitions of a set $B$ and by $0_B$ the smallest. More concretely, we denote by $1_n$ the biggest partition $\{[n]\}\in\mathcal{P}(n)$ and by $0_n$ the smallest partition consisting of $n$ blocks with one element. 

Given a permutation $\beta\in\mathfrak{S}_n$, we can associate $0_{\beta}\in\mathcal{P}(n)$ to it by forgetting the order of the cycles.

Let us denote by $\gamma_n\in\mathfrak{S}_n$ the cycle $(1 \ 2 \ \ldots \ n)$. For all $\beta\in\mathfrak{S}_n$, it can be checked that $t(\beta)+t(\gamma_n\beta^{-1})\leq n-1$. If we have equality, we call $\beta$ a {\it non-crossing permutation}. We denote the set of non-crossing permutations by $\mathfrak{S}_{n}^{NC}$. Observe that a non-crossing permutation $\beta$ can be identified with its partition $0_{\beta}$ because there exists only one possible order of the blocks of $\beta$ making it into a non-crossing permutation.

Let $L = {\ell_1} +\cdots +\ell_k$. We denote by $\gamma_{{\ell_1},\ldots,\ell_k}$ the product of $k$ cycles:
$$
(1 \ 2 \ \ldots \ \ell_1)(\ell_1+1 \ \ell_1+2 \ \ldots \ \ell_1+\ell_2) \cdots (\ell_1+\cdots +\ell_{k-1}+1 \ \ldots \ L)\in\mathfrak{S}_{L}.
$$

We call {\it partitioned permutation} a pair $(\mathcal{V},\beta)$, $\beta\in\mathfrak{S}_n$ and $\mathcal{V}\in\mathcal{P}(n)$, with $\mathcal{V}\geq 0_{\beta}$. We denote $\mathcal{P}\mathfrak{S}_n$ the set of partitioned permutations of $n$ elements and also $\mathcal{P}\mathfrak{S}\coloneqq\bigcup_{n\in\N}\mathcal{P}\mathfrak{S}_n$. We define the length of a partitioned permutation by $|(\mathcal{U}, \alpha)|\coloneqq 2(n-|\mathcal{U}|) - t(\alpha)$ and the product on $\mathcal{P}\mathfrak{S}_n$ as follows:
$$
(\mathcal{U},\alpha)\cdot (\mathcal{V},\beta) \coloneqq \begin{cases} (\mathcal{U}\vee \mathcal{V}, \alpha\beta), & \text{ if } |(\mathcal{U},\alpha)|+|(\mathcal{V},\beta)| = |(\mathcal{U}\vee\mathcal{V},\alpha\beta)|,\\
0, & \text{ otherwise.}
\end{cases}
$$

For two functions $f,g\colon \mathcal{P}\mathfrak{S}\to\C$, we define their convolution $f*g \colon \mathcal{P}\mathfrak{S}\to\C$ by
$$
(f*g)(\mathcal{U},\alpha) \coloneqq \sum_{\substack{(\mathcal{V},\beta),(\mathcal{W},\gamma)\in\mathcal{P}\mathfrak{S}(n) \\ (\mathcal{V},\beta)\cdot (\mathcal{W},\gamma)=(\mathcal{U},\alpha)}} f(\mathcal{V},\beta) \cdot g(\mathcal{W},\gamma), \ \ \forall (\mathcal{U},\alpha)\in\mathcal{P}\mathfrak{S}_n.
$$
A function $f\colon \mathcal{P}\mathfrak{S}\to\C$ is {\it multiplicative} if $f(1_n,\beta)$ depends only on $[\beta]$ and 
$$
f(\mathcal{V},\beta)=\prod_{V\in\mathcal{V}} f(1_V,\left.\beta\right|_{V}).
$$ 
Consider the multiplicative function on $\mathcal{P}\mathfrak{S}_n$ given by
$$
\delta(\mathcal{V},\beta)\coloneqq \begin{cases} 1, & \text{ if } (\mathcal{V},\beta)=(0_n, {\rm id}_{\mathfrak{S}_n}), \\ 0, & \text{ otherwise.}\end{cases}
$$
The convolution of multiplicative functions on $\mathcal{P}\mathfrak{S}$ is commutative and has $\delta$ as the unit element.

For a fixed $(\mathcal{U},\alpha)\in\mathcal{P}\mathfrak{S}$, we say that $(\mathcal{V},\beta)\in\mathcal{P}\mathfrak{S}$ is $(\mathcal{U},\alpha)-${\it non-crossing} if
$$
(\mathcal{V},\beta)\cdot (0_{\beta^{-1}\alpha},\beta^{-1}\alpha)=(\mathcal{U},\alpha).
$$
We denote $\mathcal{P}\mathfrak{S}^{NC}(\mathcal{U},\alpha)$ the set of $(\mathcal{U},\alpha)-$non-crossing partitioned permutations. This terminology comes from the fact that $(1_n,\gamma_n)-$non-crossing partitioned permutations can be identified with non-crossing permutations and hence with non-crossing partitions:
$$
\mathcal{P}\mathfrak{S}^{NC}(1_n,\gamma_n) = \{(0_{\beta},\beta)\mid \beta\in NC(n)\}.
$$

\begin{figure}[h!]
\begin{center}
\centering
\scalebox{0.5}{
\begin{tikzpicture}
\draw (60:3) node[anchor=south]{1};
\draw (0:3) node[anchor=west]{2};
\draw (-60:3) node[anchor=north]{3};
\draw (-120:3) node[anchor=north]{4};
\draw (-180:3) node[anchor=east]{5};
\draw (-240:3) node[anchor=south]{6};
\draw (90:1) node[anchor=north]{7};
\draw (162:1) node[anchor=west]{8};
\draw (234:1) node[anchor=south]{9};
\draw (306:1) node[above left]{10};
\draw (378:1) node[anchor=east]{11};
\draw (60:3) .. controls (0,1.5) .. (-240:3);
\draw (60:3) .. controls (0,1) .. (-240:3);
\draw (-2.5,0) ellipse (0.5cm and 0.3cm);
\draw (-120:3) .. controls (0,-1.5) .. (306:1);
\draw (-60:3) .. controls (2,-1) .. (378:1);
\draw (-60:3) .. controls (0,-2) .. (-120:3);
\draw (306:1) .. controls (1.5,-0.5) .. (378:1);
\draw (162:1) .. controls (-1.5,-0.5) .. (234:1);
\draw (162:1) .. controls (-1.25,-0.5) .. (234:1);
\draw (90:1) .. controls (1,1.75) and (2,0) .. (0:3);
\draw (90:1) .. controls (1,1) and (2,0) .. (0:3);
\draw (0,0) circle (3cm);
\draw (0,0) circle (1cm);
\path[step=1.0,black,thin,xshift=0.5cm,yshift=0.5cm] (-3,-5) grid (5,3);
\end{tikzpicture}
\qquad
\begin{tikzpicture}
\path[step=1.0,black,thin,xshift=0.5cm,yshift=0.5cm] (-3,-5) grid (5,3);
\draw (-6,0) ++ (90:3) node[anchor=south]{1} coordinate (1);
\draw (-6,0) ++ (30:3) node[above right]{2} coordinate (2);
\draw (-6,0) ++ (-30:3) node[anchor=west]{3} coordinate (3);
\draw (-6,0) ++ (-90:3) node[anchor=north]{4} coordinate (4);
\draw (-6,0) ++ (-150:3) node[anchor=east]{5} coordinate (5);
\draw (-6,0) ++ (-210:3) node[above left]{6} coordinate (6);
\draw (-6,0) ++ (0,-3.3) ellipse (0.15cm and 0.3cm);
\draw (-6,0) ++ (-120:0.875) coordinate (X);
\pic [draw, angle radius=3.125cm] {angle=6--X--3};
\draw (1) .. controls (-3,3.5) .. (2);
\draw (1) .. controls (-3.5,3) .. (2);
\draw (3) .. controls (-5,-4.75) and (-7.5,-4) .. (5);
\draw (5) .. controls (-9.5,0) .. (6);
\draw (3,0) ++ (90:3) node[anchor=south]{7} coordinate (7);
\draw (3,0) ++ (18:3) node[anchor=west]{8} coordinate (8);
\draw (3,0) ++ (-54:3) node[anchor=north]{9} coordinate (9);
\draw (3,0) ++ (-126:3) node[below left]{10} coordinate (10);
\draw (3,0) ++ (-198:3) node[anchor=east]{11} coordinate (11);
\draw (11) .. controls (1,1) .. (7);
\draw (10) .. controls (1.5,-1) .. (7);
\draw (11) .. controls (1,-1) .. (10);
\draw (8) .. controls (7,-1.5) .. (9);
\draw (8) .. controls (7.5,-2) .. (9);
\draw [dashed] (7) .. controls (-2.5,4) .. (1);
\draw [dashed] (2) .. controls (-2.5,1) and (-0.5,-1) .. (10);
\draw (-6,0) circle (3cm);
\draw (3,0) circle (3cm);
\end{tikzpicture}}
\caption{Illustrations of the two different types of $(1_{6+5},\gamma_{6,5})-$non-crossing partitioned permutations. They correspond to $(\mathcal{V}_1,\beta_1)$ (left) and $(\mathcal{V}_2,\beta_2)$ (right), with $\beta_1=(1\ 6)(2\ 7)(8\ 9)(5)(3\ 4\ 10 \ 11)\in S_{6+5}$, $\mathcal{V}_1=0_{\beta_1}$, and $\beta_2=(1\ 2)(3\ 5\ 6)(4)(7\ 10 \ 11)(8\ 9)$, $\mathcal{V}_2=\{\{3,5,6\},\{4\},\{1,2,7,10,11\},\{8,9\}\}$.}
\end{center}
\end{figure}
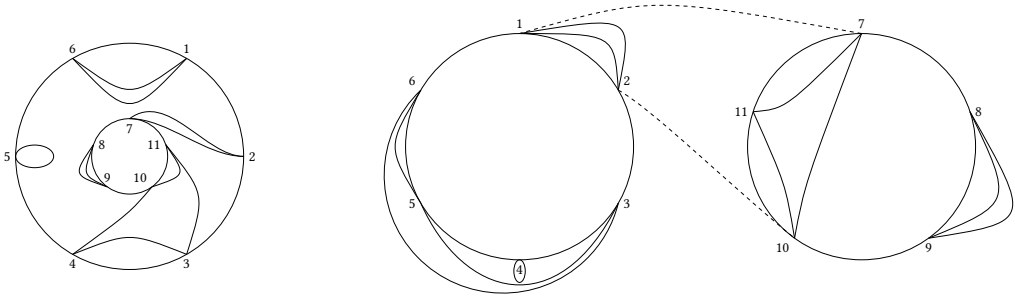

We define the {\it Zeta function} $\zeta$ on $\mathcal{P}\mathfrak{S}$ by
$$
\zeta(\mathcal{V},\beta)\coloneqq \begin{cases} 1, & \text{ if } \mathcal{V}=0_{\beta}, \\
0, & \text{ if } \mathcal{V}>0_{\beta}.\end{cases}
$$
It can be showed that there exists a unique inverse of $\zeta$ under convolution: the {\it M\"{o}bius function} $\mu$, i.e., $\zeta*\mu =\delta=\mu *\zeta$.

Observe that with these notations, we have
$$
(f*\zeta)(\mathcal{U},\alpha)=\sum_{(\mathcal{V},\beta)\in\mathcal{P}\mathfrak{S}^{NC}(\mathcal{U},\alpha)} f(\mathcal{V},\beta).
$$

\subsubsection{Higher order freeness and cumulants}

\begin{definition}
A {\it higher-order (non-commutative) probability space} (HOPS) consists of a unital algebra $\mathcal{A}$ and a collection of multilinear functionals $\varphi_n \colon \mathcal{A}^n \to \C$, $n\in\N$, which are tracial in each argument, symmetric, and satisfy that $\varphi_1(1)=1$ and $\varphi_n(1,a_2,\ldots,a_n)=0$ for all $n\geq 2$ and all $a_2,\ldots,a_n\in\mathcal{A}$.
\end{definition}

Let $\mathcal{P}\mathfrak{S}(\mathcal{A})$ denote $\bigcup_{n\in\N}(\mathcal{P}\mathfrak{S}\times\mathcal{A}^n)$. For a function
$$
\begin{array}{rrcl}
f \colon & \mathcal{P}\mathfrak{S}(\mathcal{A}) & \to & \C \\
& (\mathcal{V},\beta)\times (a_1,\ldots,a_n) & \mapsto & f(\mathcal{V},\beta)[a_1,\ldots,a_n]
\end{array}
$$
and a function $g\colon \mathcal{P}\mathfrak{S}\to\C$, we define their convolution $f*g \colon \mathcal{P}\mathfrak{S}(\mathcal{A})\to\C$ by
$$
(f*g)(\mathcal{U},\alpha)[a_1,\ldots,a_n] \coloneqq \sum_{\substack{(\mathcal{V},\beta),(\mathcal{W},\gamma)\in\mathcal{P}\mathfrak{S}(n) \\ (\mathcal{V},\beta)\cdot (\mathcal{W},\gamma)=(\mathcal{U},\alpha)}} f(\mathcal{V},\beta)[a_1,\ldots,a_n] \cdot g(\mathcal{W},\gamma).
$$

We say a function $f\colon\mathcal{P}\mathfrak{S}(\mathcal{A}) \to \C$ is {\it multiplicative} if
$$
f(\mathcal{V},\beta)[a_1,\ldots,a_n] = \prod_{B\in\mathcal{V}} f(1_B,\left.\beta\right|_{B})[(a_1,\ldots,a_n)_B]
$$
and $f(1_n,\sigma^{-1}\beta\sigma)[a_{\sigma(1)},\ldots,a_{\sigma(n)}]=f(1_n,\beta)[a_1,\ldots,a_n]$, for all $a_1,\ldots,a_n\in\mathcal{A}, \beta,\sigma\in \mathfrak{S}_n.$

Note that with this extended setting on $\mathcal{P}\mathfrak{S}(\mathcal{A})$, $\delta$ is still the unit and $f=g*\zeta$ is still equivalent to $g=f*\mu$ for multiplicative functions $f,g\in \mathcal{P}\mathfrak{S}(\mathcal{A})$, after inserting the formal variables $a_1,\ldots,a_n$ at the right positions.

Given a HOPS, we will use the $\varphi_n$ to produce a multiplicative moment function, using the machinery of partitioned permutations:
$$
\varphi(1_L,\gamma_{\ell_1,\ldots,\ell_n})[a_1,\ldots,a_L] \coloneqq \varphi_{n}(a_1\cdots a_{\ell_1},\ldots, a_{\ell_1+\cdots+\ell_{k-1}+1}\cdots a_L), \text{ where } \ell_1+\cdots + \ell_n =L,
$$
and extend this by multiplicativity to the whole $\mathcal{P}\mathfrak{S}(\mathcal{A})$. For example: 
$$
\varphi(\{1,2,5\}\{3,4,6\},(1)(2)(3)(5)(4\ 6))[a_1,\ldots,a_6]=\varphi_3(a_1,a_2,a_5)\varphi_2(a_3,a_4a_6).
$$
\begin{definition}\label{defFreeCumulants}
The {\it (higher order) free cumulants} are functions on $\mathcal{P}\mathfrak{S}(\mathcal{A})$ given by $k\coloneqq \varphi * \mu$.
\end{definition}
As in the definition of first order free cumulants, this is equivalent to $\varphi = k*\zeta$, i.e.,
$$
\varphi(\mathcal{U},\alpha)[a_1,\ldots,a_n] = \sum_{(\mathcal{V},\beta)\in \mathcal{P}\mathfrak{S}^{NC}(\mathcal{U},\alpha)} k(\mathcal{V},\beta)[a_1,\ldots,a_n], \ \ \forall (\mathcal{U},\alpha)[a_1,\ldots,a_n]\in\mathcal{P}\mathfrak{S}(\mathcal{A}).
$$
The definition of higher order freeness is in terms of the vanishing of mixed higher order free cumulants:
\begin{definition}
Two random variables $x,y\in\mathcal{A}$ are {\it free of all orders} if for all $n\geq 2$ and all $a_i\in \{x,y\}$, $1\leq i\leq n$, satisfying that there are at least two indices $i,j$ such that $a_i=x$ and $a_j=y$, we have that $k(1_n,\beta)[a_1,\ldots,a_n]=0$, for all $\beta\in\mathfrak{S}_n$.
\end{definition}
The number of cycles $\vert\mathcal{C}(\beta)\vert$ determines the order of the free cumulant $k(1_n,\beta)$. 
The vanishing of mixed cumulants of first order, and of first and second order, respectively, is equivalent to the definitions of first and second order freeness we gave in terms of moments. An explicit characterization of freeness in terms of moments was not found for higher order.

Let $\lambda = (\lambda_1,\ldots,\lambda_l)$ be a partition of a number. We introduce some special notations for when we consider the distribution of just one random variable $a\in\mathcal{A}$: $\varphi^a_{\lambda_1,\ldots,\lambda_l}=\varphi^a(\lambda)\coloneqq \varphi(1_{|\lambda|},\beta)[a,\ldots,a]=\varphi_{l}(a^{\lambda_1},\ldots,a^{\lambda_{l}})$ and $k^a_{\lambda_1,\ldots,\lambda_l}=k^a(\lambda)\coloneqq k(1_{|\lambda|},\beta)[a,\ldots,a]$, where $\beta\in C_{\lambda}$.

The vanishing of mixed cumulants implies again that, for $a, b\in\mathcal{A}$ free of all orders, we have $k^{a+b}(\lambda)=k^a(\lambda)+k^b(\lambda)$, for all partitions $\lambda$.

\subsection{$R$-transform formulas for first and second orders}

Consider the following generating series of first and second order moments and free cumulants: the {\it Cauchy transforms}
$$
G(x)=\frac{1}{x}+\sum_{\ell\geq 1} \frac{\varphi^a_{\ell}}{x^{\ell+1}} \ \text{ and } \ G(x_1,x_2)=\sum_{\ell_1,\ell_2\geq 1} \frac{\varphi^a_{\ell_1,\ell_2}}{x_1^{\ell_1+1}x_2^{\ell_2+1}},
$$
and the {\it R-transforms}
$$
\mathcal{R}(w)=\sum_{\ell\geq 1} k^a_{\ell} w^{\ell-1} \ \text{ and } \ \mathcal{R}(w_1,w_2)=\sum_{\ell_1,\ell_2\geq 1} k^a_{\ell_1,\ell_2} w_1^{\ell_1-1}w_2^{\ell_2-1}.
$$
\begin{theorem}\label{R-transform} The moment-cumulant relations for first and second order are equivalent to the functional relations:
\beq\label{voiculescu}
\frac{1}{G(x)}+\mathcal{R}(G(x))=x,
\eeq
\beq\label{speicher}
G(x_1,x_2)=G^{\prime}(x_1)G^{\prime}(x_2)\left(\mathcal{R}(G(x_1),G(x_2))+\frac{1}{(G(x_1)-G(x_2))^2}\right)-\frac{1}{(x_1-x_2)^2}.
\eeq
\end{theorem}
The first formula \eqref{voiculescu} is well-known in free probability, was given by Voiculescu in \cite{Voiculescu2} and is sometimes referred to as the $R$-transform machinery. In \cite{Secondorderfreeness}, they introduced higher order cumulants and freeness, but they were able to find such important formulas relating moments and free cumulants only for second order \eqref{speicher}, and already in a quite intricate way. Even if the conceptual framework is the same for any order, the complexity of the combinatorial objects involved makes the computations in higher orders too complicated.
\subsection{Asymptotic freeness for random matrices}\label{FreenessRMs}

Let $(\mathcal{A}_N,\varphi^{(N)})$ be HOPSs. We say that $X_N\in\mathcal{A}_N$ has a limit distribution of all orders if there exists a HOPS $(\mathcal{A},\varphi)$ such that $X_N\to x$, $N\to\infty$, for some $x\in\mathcal{A}$, i.e., every $\varphi_n^{(N)}$ of polynomials evaluated in the $X_N$ converges to the $\varphi_n$ of the same polynomials evaluated on $x$.

\begin{definition}
We say that $X_N$ and $Y_N$ in $\mathcal{A}_N$ are {\it asymptotically free of all orders} if $X_N\to x$ and $Y_N\to y$, when $N\to\infty$, and $x$ and $y$ are free of all orders in a HOPs $(\mathcal{A},\varphi)$.
\end{definition}

We saw that random matrices provide a good model of non-commutative probability spaces and, actually, Voiculescu already discovered that random matrices constitute an important asymptotic model of free random variables.

Let $A=(A_N)_{N\in\N}$ be a complex random matrix ensemble. We define our $n$-th order correlation moments as the scaled limits of classical cumulants of $n$ traces of powers of our matrices:
\beq\label{correlationMoments}
\varphi_{l_1,\ldots,l_n}^A\coloneqq \lim_{N\to \infty} N^{n-2}\kappa_n({\rm Tr}\, A_N^{l_1},\ldots,{\rm Tr}\, A_N^{l_n}),
\eeq
which constitute the limiting distribution of all orders of $A$.

\begin{theorem}
Let $A=(A_N)_{N\in\N}$ and $B=(B_N)_{N\in\N}$ be random matrix ensembles with limit distributions of all orders. If $A_N$ and $B_N$ are independent and at least one of them is unitarily invariant, then $A_N$ and $B_N$ are asymptotically free of all orders.
\end{theorem}
This result (\cite[Theorem 8.2]{Secondorderfreeness}) generalizes Voiculescu's analogous one for first order \cite{Voiculescu3}, which showed that freeness arises in a natural way in the important world of random matrix theory. The original motivation to introduce higher order freeness was to study the asymptotic behavior of random matrices in general, and in particular, second order to understand the global fluctuations of the eigenvalues.

Another reason for the name free cumulants is that for a unitarily invariant random matrix ensemble, they can also be expressed as special limits of classical cumulants of the entries of the matrices (included in \cite[Theorem 4.4]{Secondorderfreeness}):

\begin{theorem}\label{freeCum}
Consider $l_1,\ldots,l_n\geq 1$ with $\sum_{i=1}^n l_i=L$. If $A$ is a unitarily invariant random matrix ensemble with $A_N=(a_{ij}^{(N)})_{i,j=1}^N$, its $n$-th order free cumulants can be written as: 
\beq
k_{l_1,\ldots,l_n}^A = \lim_{N\to\infty} N^{n-2+L}\kappa_L(a_{i_{1,1}, i_{1,2}}^{(N)}, a_{i_{1,2}, i_{1,3}}^{(N)},\ldots, a_{i_{1,l_1}, i_{1,1}}^{(N)}, \ldots, a_{i_{n,1}, i_{n,2}}^{(N)}, a_{i_{n,2}, i_{n,3}}^{(N)},\ldots, a_{i_{n,l_n}, i_{n,1}}^{(N)}),
\eeq
for any choice of pairwise disjoint cycles: $\gamma_j=(i_{j,1} \ i_{j,2} \ldots i_{j,l_j})$ in $\mathfrak{S}_L$ for $j=1,\ldots,n$.
\end{theorem}


\section{Outline of results}

This thesis is mainly based on the results appearing in the articles: \cite{BGF17} (Part~1) and \cite{BGF16} (Part~2), which are both joint work with G.~Borot.
Before entering into the details regarding each part, we list in Figures~\ref{Notations1}-\ref{Notations2} the notations used throughout the whole thesis with the purpose of avoiding confusion. We also remark that when we talk about some kind of maps, we mean by default \emph{connected} maps, unless otherwise stated.

\begin{figure}[h!]
\begin{center}
\begin{tabular}{ |p{3.2cm}|p{7.5cm}| }
 \hline
 \textbf{Notation} & \textbf{Generating series of some type of maps} \\
 \hline
 $F, W$  \eqref{GenFctF}, \eqref{correlators}  & Usual, ordinary maps  \\
 $G, Y$  \eqref{Scorrelators}  & Simple maps  \\
 $H, X$  \eqref{FScorrelators}  & Fully simple maps  \\
 $\widehat{F}, \widehat{W}$  \eqref{Stuffedcorrelators}  & Stuffed maps  \\
 $\mathcal{F}, \mathcal{W}$ \eqref{eq:Fdef}, \eqref{Oncorrelators}  & $O(\mathsf{n})$ configurations  \\
 $\mathsf{F}, \mathsf{W}$  \eqref{RenormalizedFs}, \eqref{renFmarked}  & Usual maps with renormalized faces, i.e., $O(\mathsf{n})$ configurations only with non-separating loops  \\
 $\mathscr{F}, \mathscr{W}$ \eqref{magi}, \eqref{FixedNGcorrelators}  & $O(\mathsf{n})$  configurations with a fixed nesting graph  \\
 \hline
 \end{tabular}
\caption{\label{Notations1} Summary of generating series of maps.}
\end{center}
\end{figure}

\begin{figure}[h!]
 \begin{center}
\begin{tabular}{ |p{3.2cm}|p{7.5cm}| }
 \hline
 \textbf{Notation} & \textbf{Objects} \\
 \hline
 $\omega_{g,n}$ \eqref{TRampl} & TR amplitudes \\
 $\mathfrak{F_g}$  \eqref{TRsympInv} & TR $n=0$ invariants \\
 $\mathfrak{W}_{\mathbf{t}}$ \eqref{BoltMeasure} & Boltzmann probability measure \\
  $\hat{F}_A, \hat{W}_A$  \eqref{HMMEFcorrelators} & Free energy and correlators of the Hermitian matrix model with external field $A$ \\
 \hline
\end{tabular}
\caption{\label{Notations2} Summary of notations.}
\end{center}
\end{figure}

\subsection{Fully simple maps, Hurwitz numbers and topological recursion}
\ \\

Our main objects of study in this part will be fully simple maps, which we introduced in Section~\ref{fsmaps}. The vocabulary we adopt to refer to the distinct classes of maps is summarized in the tables below, where $(B_i)_{i = 1}^n$ denote the boundary faces.


\begin{center}
\begin{figure}[h!]
\begin{center}
\begin{tabular}{|c|c|}
\hline
\textbf{Boundary type} & \textbf{Description} \\
\hline
Ordinary & No restriction \\
 Simple & No vertex in $B_i$ with more than $2$ incoming half-edges from $B_i$ \\
Fully simple & No vertex in $B_i$ with more than $2$ incoming half-edges from $\bigcup_{j} B_j$ \\
\hline
\end{tabular}
\caption{\label{Fig1} Classes of maps with respect to the restriction imposed to the boundaries.}
\end{center}
\end{figure}
\end{center}

We study in detail the combinatorial relation between fully simple and ordinary disks and cylinders. We show that the generating series of simple disks is given by the functional inversion of the generating series of ordinary disks. We also obtain an elegant formula for cylinders. These relations reproduce the relation between (first and second order) correlation moments and free cumulants that we stated in Theorem \ref{R-transform}, and implement the exchange transformation $x \leftrightarrow y$ on the spectral curve in the context of topological recursion. 

These interesting features constituted our main motivation to study fully simple maps. We then propose a combinatorial interpretation of the still not well understood exchange symplectic transformation of the topological recursion that we commented on in Section \ref{sympinv}. We explain all the ideas towards a proof of this interpretation for usual maps and state precisely what remains to be checked. The starting point is a matrix model interpretation of fully simple maps, via the formal hermitian matrix model with external field. 

We also deduce a universal relation between generating series of fully simple maps and of ordinary maps which involves double monotone Hurwitz numbers. In particular, (ordinary) maps without internal faces -- which are generated by the Gaussian Unitary Ensemble -- and with boundary perimeters $(\lambda_1,\ldots,\lambda_n)$ are strictly monotone double Hurwitz numbers with ramifications $\lambda$ above $\infty$ and $(2,\ldots,2)$ above $0$. Combining with a recent result of Dubrovin et al. \cite{DiYang}, this implies an ELSV-type formula for these Hurwitz numbers.

\vspace{0.2cm}
\begin{center}
\begin{figure}
\begin{center}
\begin{tabular}{|c|c|c|}
\hline
\textbf{Type of maps} & \textbf{Topology of inner faces} & \textbf{Matrix model \eqref{mesmap}-\eqref{mesmapgeneral}} \\
\hline
Usual & Disks & $t_{d}$ \\
With loops \cite{BEO} & Disks and cylinders & $t_{0;d_1}$ and $t_{0;d_1,d_2}$ \\
Stuffed \cite{Bstuff} & Arbitrary & All $t_{h;d_1,\ldots,d_k}$ \\
\hline
\end{tabular}
\caption{\label{Fig2} Classes of maps with respect to the topology imposed to the internal faces.}
\end{center}
\end{figure}
\end{center}

\subsubsection{Disks and cylinders via combinatorics}

For planar maps with one (disks) or two (cylinders) boundaries, we give in Sections~\ref{Section3}-\ref{Section4} a bijective algorithm which reconstructs ordinary maps from fully simple maps. This algorithm is not sensitive to the assumption -- included in the definition of usual maps -- that faces must be homeomorphic to disks. Therefore, it applies to all types of maps described in Figure~\ref{Fig2}.

We deduce two remarkable formulas for the corresponding generating series. We denoted $F_{\ell}$ (resp. $H_{\ell}$) the generating series of ordinary (resp. fully simple) disks with perimeter $\ell$, and
$$
W(x) = \frac{1}{x} + \sum_{\ell \geq 1} \frac{F_{\ell}}{x^{\ell + 1}},\qquad X(w) = \frac{1}{w} + \sum_{\ell \geq 1} H_{\ell}w^{\ell - 1}\,.
$$
\begin{proposition}
\label{P10} For all types of maps in Figure~\ref{Fig2}, $X(W(x)) = x$.
\end{proposition}

Let $F_{\ell_1,\ell_2}$ (resp. $H_{\ell_1,\ell_2})$ be the generating series of ordinary (resp. fully simple) cylinders with perimeters $(\ell_1,\ell_2)$, and
$$
W_{2}^{[0]}(x_1,x_2) = \sum_{\ell_1,\ell_2 \geq 1} \frac{F_{\ell_1,\ell_2}}{x_1^{\ell_1 + 1}x_2^{\ell_2 + 1}},\qquad X_{2}^{[0]}(w_1,w_2) = \sum_{\ell_1,\ell_2 \geq 1} H_{\ell_1,\ell_2}\,w_{1}^{\ell_1 - 1}w_{2}^{\ell_2 - 1}\,.
$$
\begin{theorem}
\label{P20}
For all types of maps in Figure~\ref{Fig2}, if one sets $x_i = X(w_i)$ or equivalently $w_i = W(x_i)$, 
$$
\bigg(W_{2}^{[0]}(x_1,x_2) + \frac{1}{(x_1 - x_2)^2}\bigg)\dd x_1\dd x_2 = \bigg(X_{2}^{[0]}(w_1,w_2) + \frac{1}{(w_1 - w_2)^2}\bigg)\dd w_1\dd w_2.
$$
\end{theorem}
The identities of both results are equalities of formal series in $x_i \rightarrow \infty$ and $w_i \rightarrow 0$.

\subsubsection{Matrix model interpretation}\label{MMInterpretationIntro}

In the sections about the matrix model interpretation, when we talk about some type of maps, we mean by default \emph{not necessarily connected} maps, since it is more natural in this setting.

As we explained in Section~\ref{MMOrdMapsIntro}, the generating series of ordinary maps with prescribed boundary perimeters $(\ell_i)_{i = 1}^n$ are computed as the moments $\langle {\rm Tr}\,M^{\ell_1} \cdots {\rm Tr}\,M^{\ell_n} \rangle$ in the formal hermitian matrix model with
\beq
\label{mesmap} \dd\rho(M) = \dd M\,\exp[-N\,{\rm Tr}\,V(M)],\qquad V(x) = \frac{x^2}{2} - \sum_{d \geq 1} \frac{t_dx^d}{d}\,,
\eeq
where $t_d$ is the weight per $d$-gon, and the weight of a map of Euler characteristic $\chi$ is proportional to $N^{\chi}$, as we saw in \eqref{TracesGSMAps}.  Restricting to connected maps amounts to considering the cumulant expectation values $\kappa_n({\rm Tr}\,M^{\ell_1},\ldots,{\rm Tr}\,M^{\ell_n})$ instead of the moments. More generally, the measure
\beq
\label{mesmapgeneral} \dd\rho(M) = \dd M\,\exp\bigg(- N\,{\rm Tr}\,\frac{M^2}{2} + \sum_{h \geq 0} \sum_{k \geq 1} \sum_{d_1,\ldots,d_k \geq 1} \frac{N^{2 - 2h - k}}{k!}\,\frac{t_{h;d_1,\ldots,d_k}}{d_1\cdots d_k} {\rm Tr}\,M^{d_1} \cdots {\rm Tr}\,M^{d_k}\bigg)
\eeq
generates maps with loops or stuffed maps.

We show in Section~\ref{Matcomb} that the generating series of fully simple maps with prescribed boundary perimeters $(\ell_i)_{i = 1}^n$ in these models are computed as $\langle \mathcal{P}_{\gamma_1}(M) \cdots \mathcal{P}_{\gamma_n}(M)\rangle$, where $\gamma$ is a permutation of $\{1,\ldots,L\}$ with $n$ disjoint cycles $(\gamma_i)_{i = 1}^n$ of respective lengths $(\ell_i)_{i = 1}^n$, $L = \sum_{i = 1}^n \ell_i$, and $\mathcal{P}_{\gamma_i}(M) = \prod_{j} M_{j,\gamma_i(j)}$. We can write $\langle \mathcal{P}^{(\ell_1)}(M) \cdots \mathcal{P}^{(\ell_n)}(M)\rangle=\langle\mathcal{P}_{\lambda}(M)\rangle$, since this quantity does not depend on the permutation $\gamma$, but only on the lengths $(\ell_i)_{i = 1}^n$, which are encoded into a partition $\lambda$. Again, the cumulants
$$
\kappa_{n}\big(\mathcal{P}_{\gamma_1}(M),\ldots,\mathcal{P}_{\gamma_n}(M)\big) = \kappa_{n}\big(\mathcal{P}^{(\ell_1)}(M), \ldots, \mathcal{P}^{(\ell_n)}(M)\big)
$$
generate only connected maps. 

\subsubsection{From fully simple to ordinary via Hurwitz theory}

The expression $\prod_{i} \mathcal{P}_{\gamma_i}(M)$ is not invariant under $U_N$-conjugation. Yet, as the measure $\rho$ is $U_N$-invariant, its expectation value must be expressible in terms of $U_N$-invariant observables, i.e. as a linear combination of $\langle \prod_{i} {\rm Tr}\,M^{m_i} \rangle$. In other words, we can express the fully simple generating series in terms of the ordinary generating series. The precise formula is derived via Weingarten calculus in Section~\ref{HNsWeingarten}.
\begin{theorem}\label{EHH}
If $\rho$ is a unitarily invariant measure on $\mathcal{H}_N$, in particular for the measures \eqref{mesmap}-\eqref{mesmapgeneral} generating any type of map in Figure~\ref{Fig2},
\bea
 \frac{\langle\mathcal{P}_{\lambda}(M)\rangle}{|{\rm Aut}\,\lambda|}  & = & \sum_{\mu \vdash |\lambda|} N^{-|\mu|} \bigg(\sum_{k \geq 0} (-N)^{-k} [H_{k}]_{\lambda,\mu}\bigg) \Big\langle \prod_{i = 1}^{\ell(\mu)} {\rm Tr}\,M^{\mu_i} \Big\rangle\,, \\
\label{subfree}\frac{\Big\langle \prod_{i = 1}^{\ell(\mu)} {\rm Tr}\,M^{\mu_i} \Big\rangle}{|{\rm Aut}\,\mu|} & = & \sum_{\lambda \vdash |\mu|} N^{|\lambda|} \bigg( \sum_{k \geq 0} N^{-k}\,[E_{k}]_{\mu,\lambda}\bigg) \langle \mathcal{P}_{\lambda}(M) \rangle\,,
\eea
where $[H_{k}]_{\lambda,\mu}$ are double weakly monotone Hurwitz numbers and $[E_k]_{\lambda,\mu}$ are double strictly monotone Hurwitz numbers.
\end{theorem}
Formula \eqref{subfree} is appealing as it is subtraction-free, and suggests the existence of a bijection describing ordinary maps as gluings of a fully simple map ``along'' a strictly monotone ramified covering, i.e., a hypermap or dessin d'enfant. We postpone such a bijective proof of \eqref{subfree} to a future work.

\subsubsection{Combinatorial interpretation of the matrix model with external field}

As a by-product, we show in Section~\ref{HMMEF} that the partition function of the formal hermitian matrix model with external field $A \in \mathcal{H}_{N}$
$$
\check{Z}(A) = \int_{\mathcal{H}_{N}} \dd\mu(M)\,\exp[N{\rm Tr}(MA)]
$$
is a generating series of (not necessarily connected) fully simple maps in the following sense:
\begin{proposition}\label{HMMEFfsMaps} If $\mu$ is a unitarily invariant measure on $\mathcal{H}_{N}$ -- in particular for all types of maps in Figure~\ref{Fig2},
\beq\label{HMMEFfsMapsEq}
\hat{\mathcal{Z}}(A)\coloneqq\frac{\check{Z}(A)}{\check{Z}(0)} = \sum_{\lambda} \frac{|C_{\lambda}|}{|\lambda|!}\,N^{|\lambda|}\langle \mathcal{P}_{\lambda}(M) \rangle \prod_{i = 1}^{\ell(\lambda)} {\rm Tr}\,A^{\lambda_i}\,.
\eeq 
\end{proposition}
 
\subsubsection{Application: an ELSV-type formula}

If we have a model in which the generating series of fully simple maps are completely known, \eqref{subfree} can be used to compute a certain class of monotone Hurwitz numbers in terms of generating series of maps. This is the case for the Gaussian Unitary Ensemble, i.e., $t_d = 0$ for all $d$ in \eqref{mesmap}. As the matrix entries are independent,
$$
\langle\mathcal{P}_{\lambda}(M)\rangle_{{\rm GUE}} = \prod_{i = 1}^{\ell(\lambda)} \frac{\delta_{\lambda_i,2}}{N}\,.
$$ 
Combinatorially, this formula is also straightforward: as the maps generated by the GUE have no internal faces, the only connected fully simple map is the disk of perimeter 2. Dubrovin et al.~\cite{DiYang} recently proved a formula relating the GUE moments with all $\ell_i$ even to cubic Hodge integrals. Combining their result with our \eqref{subfree} specialized to the GUE, we deduce in Section~\ref{Section11} an ELSV-like formula for the so-called $2$-orbifold strictly monotone Hurwitz numbers.

\begin{proposition}
Let $[E^{\circ}_{g}]_{\lambda,\mu}$ denote the connected, strictly monotone Hurwitz numbers. 
For $g \geq 0$ and $n \geq 1$ such that $2g - 2 + n > 0$, and $\mu=(2m_1,\ldots,2m_n)$, we have
$$
|{\rm Aut}\,\mu|\,[E^{\circ}_{g}]_{\mu,(2,\ldots,2)} = 2^{g} \int_{\overline{\mathcal{M}}_{g,n}} [\Delta] \cap \Lambda(-1)\Lambda(-1)\Lambda(\tfrac{1}{2})\exp\Big(-\sum_{k \geq 1} \frac{\kappa_{k}}{k}\Big) \prod_{i = 1}^n \frac{m_i\,{2m_i \choose m_i}}{1 - m_i\psi_i}\,,
$$
where 
$$
[\Delta] = \sum_{h = 0}^{g} \frac{[\Delta_h]}{2^{3h}(2h)!},
$$
and $[\Delta_h]$ is the class of $\overline{\mathcal{M}}_{g - h,n + 2h}$ included in $\overline{\mathcal{M}}_{g,n}$ by identifying the $2h$ last punctures pairwise.
\end{proposition}

The recent work of Dubrovin, Yang and Zagier~\cite{DubrovinYangZagier} is also related to GUE moments and cubic Hodge integrals, and other classical combinatorial problems, such as Hurwitz numbers, which may some interesting connections to our results.

\subsubsection{Topological recursion interpretation}


As we explained in Section~\ref{TRMaps} of the introduction, it was proved in \cite{E1MM,C05,Eynardbook} for maps, and \cite{BEOn,BEO} for maps with loops, that the generating series of ordinary maps $W_{n}^{[g]}$ satisfies the topological recursion (hereafter TR) that we introduced in Section~\ref{TRIntro}. 


In the context of maps, $\Sigma$ is the curve on which the generating series of disks can be maximally analytically continued with respect to its parameter $x$ coupled to the boundary perimeter, and $\Sigma$ has a distinguished point $[\infty]$ corresponding to $x \rightarrow \infty$.
\begin{theorem}\label{TRRRR} The TR amplitudes for the initial data
\beq
\label{inidataordmap} \left\{\begin{array}{lll} p = x \\ \lambda = w = W_{1}^{[0]}(x) \\ B(z_1,z_2) = \Big(W_{2}^{[0]}(x(z_1),x(z_2)) + \frac{1}{(x(z_1) - x(z_2))^2}\Big)\dd x(z_1)\dd x(z_2) \end{array}\right.
\eeq
compute the generating series of usual maps or maps with loops, through
\beq
\label{WngTR2}W_{n}^{[g]}(x(z_1),\ldots,x(z_n)) = \frac{\omega_{g,n}(z_1,\ldots,z_n)}{\dd x(z_1)\cdots \dd x(z_n)},\qquad 2g - 2 + n > 0.
\eeq
Here $z_i$ is a generic name for points in $\Sigma$, and \eqref{WngTR2} means the equality of Laurent expansion near $z_i \rightarrow [\infty]$.
\end{theorem}

\noindent{\textbf{Symplectic invariance.}}
The most remarkable, and still mysterious feature of the topological recursion is the expected symplectic invariance of the $n=0$ correlators $\mathfrak{F}_g$. We explained in Section \ref{sympinv} the current status of this property. Basically, the $\mathfrak{F}_g$'s remain invariant under any symplectic change of variables $(x,y)\mapsto \phi(x,y)$ on the spectral curves that does not involve the exchange transformation $(x,y)\mapsto (-y,x)$. In some simple cases, the $\mathfrak{F}_g$'s are also invariant under the exchange transformation (this was believed to be always the case in \cite{EO2MM}); in many other cases, they are invariant up to some determined correction terms (given in \cite{EOxy}), and in general they are believed to be invariant up to some corrections terms which are still under investigation. 

In the case of topological strings on toric Calabi-Yau threefolds, symplectic invariance amounts to the framing independence of the closed sector, albeit involving curves given by a polynomial relation between $e^{p}$ and~$e^{\lambda}$. However, this is one of the few instances where the reason behind symplectic invariance is understood.

Symplectic invariance and the mysterious exchange transformation constituted an important motivation for us to study fully simple maps, since our generating series of ordinary and fully simple maps obviously agree for $n=0$ boundaries:  $F^{[g]} = W_{0}^{[g]}=X_{0}^{[g]}=H^{[g]}$, as the condition for maps to be fully simple affects only the boundaries\footnote{The generating series of closed maps $F^{[g]}$ is related to the symplectic invariants of TR applied to the spectral curve given by \eqref{inidataordmap}, which we denote $\mathfrak{F}_g[p,\lambda]$, as follows:
$$
F^{[g]}=\frac{B_{2g}u^{2-2g}}{2g(2-2g)}+\mathfrak{F}_g[p,\lambda],
$$
where $B_{2g}$ is the $2g$th Bernoulli number. The proof of this fact \cite{Eynardbook} shows that $F^{[g]}-\mathfrak{F}_g$ is a constant independent of the $t_k$'s, which can be computed at $t_k=0$ using the Gaussian matrix model. We believe there should be a combinatorial proof of this relation viewing the constant term as the correction eliminating or adding the extra maps counted with $\mathfrak{F}_g[p,\lambda]$ and that there can be a similar, but for sure more complicated, relation for fully simple maps, i.e., between $H^{[g]}$ and $\mathfrak{F}_g[-\lambda,p]$, in light of our~Conjecture \ref{MainConj}.}. 

Moreover, Propositions~\ref{P10}-\ref{P20} tell us that swapping $\lambda$ and $p$ in the initial data \eqref{inidataordmap} amounts to replacing the generating series of ordinary disks and cylinders with their fully simple version. We see it as the planar tip of an iceberg:

\begin{conjecture}
\label{MainConj} For usual maps or maps with loops, let $\check{\omega}_{n}^{[g]}$ be the TR amplitudes for the initial data \eqref{inidataordmap} after the exchange of $x$ and $w$. We have
\beq
\label{XngTR}X_{n}^{[g]}(w(z_1),\ldots,w(z_n)) = \frac{\check{\omega}_{n}^{[g]}(z_1,\ldots,z_n)}{\dd w(z_1)\cdots \dd w(z_n)},\qquad \text{for } 2g - 2 + n > 0.
\eeq
This is an equality of formal Laurent series when $z_i \rightarrow [\infty]$.
\end{conjecture}
The validity of this conjecture would give a combinatorial interpretation to the symplectic invariance, which we hope could shed some light on this deep enigmatic feature.  

\vspace{0.2cm}

\noindent\textbf{Progress towards a proof and supporting data for quadrangulations.}
In Section~\ref{Section9} of the thesis we give all the ideas towards a possible proof of Conjecture~\ref{MainConj} for usual maps. We manage to reduce the problem to a technical condition concerning a milder version of symplectic invariance for the \mbox{$1$-hermitian} matrix model with external field. We confirmed experimentally that the condition is satisfied for some particular cases, but do not have an argument to give a general proof at the moment.

Even if we could make the technical condition work to complete our proof of the conjecture through the study of the formal $1$-hermitian matrix model with external field, it would still be desirable to find a combinatorial proof, as it may give an independent proof of symplectic invariance for the initial data related to maps -- i.e., a large class of curves of genus $0$ --, it would help understand symplectic invariance for the fundamental instance of maps and the enumeration problem of fully simple maps explicitly, and it may be naturally generalizable for all types of maps in the Figure~\ref{Fig2}.

Apart from the general proof for disks and cylinders, and the ideas towards the full proof for usual maps, we gathered some combinatorial evidence supporting our conjecture in Section~\ref{Quadrangulations}. In fact, there is no a priori reason for the coefficients of expansion of $\check{\omega}_{n}^{[g]}$ to be positive integers. Besides, for the same given perimeters, there should be less fully simple maps than maps. For the initial data corresponding to quadrangulations, we have checked in Sections~\ref{ToriSection} and \ref{PairPantsSection} that, for the topologies $(1,1)$ and $(0,3)$, positivity and the expected inequalities hold for the coefficients of~$\check{\omega}_{1,1}$ and $\check{\omega}_{0,3}$, obtained after fixing the number of internal quadrangles, for boundaries up to length~$14$.

For the pair of pants case (topology $(0,3)$) the evidence is much stronger. In 2017, O.~Bernardi and \'{E}.~Fusy gave a formula for the number of fully simple planar quadrangulations with boundaries of prescribed even lengths in \cite{BernardiFusy}. We computed the outcome of their formula and they perfectly match our conjectural numbers for the cases of even lengths. Their formula also agrees with our enumeration for disks and cylinders. From our results for cylinders and our conjectural numbers for pairs of pants, one can observe that an analogous formula seems to be true also in presence of some odd boundaries\footnote{We thank Timothy Budd for bringing this reference to our attention and pointing out that our data suggests an analogous formula hold when some boundaries are of odd length.}.

If our conjecture is true, it would solve theoretically the problem of enumeration of fully simple maps in full generality. Moreover, the algorithm of TR allows to solve explicitly the first cases of the iteration. So our conjecture would produce another proof of the formula of \cite{BernardiFusy} for cylinders and pairs of pants with even lengths, would allow to prove the cases in presence of odd lengths and would produce the first explicit formulas for numbers of fully simple maps of positive genus. We give explicit formulas for ordinary and conjecturally fully simple maps of genus $1$ with $1$ boundary in Section~\ref{ToriSection}.

Furthermore, the fact that the conjecture produces the right numbers for pairs of pants seems to indicate that our technical condition should hold in general; otherwise, we believe the presence of non-zero correction terms in our technical condition should be manifested from the beginning of the recursion. Finally, even if the conjecture was not true and there were some non-zero correction terms modifying our technical condition, the data suggests there is a combinatorial problem behind since we obtain positive integers, so the correction terms may give rise to a simpler combinatorial problem complementing the number of fully simple maps.


\subsubsection{Relations to free probability}

Let $M=(M_N)_{N\in\N}$ be a unitarily invariant hermitian random matrix ensemble. In the important setting of random matrices discussed in Section~\ref{FreenessRMs}, we give an interpretation of the $n$th order free cumulants (Definition~\ref{defFreeCumulants}) in terms of the connected fully simple observables (which are classical cumulants of the $\mathcal{P}$'s) that we defined in Section~\ref{MMInterpretationIntro} of this outline:
$$
k_{\ell_1,\ldots,\ell_n}^M = \lim_{N \rightarrow \infty} N^{n - 2 + \sum_{i} \ell_i} \kappa_n\big(\mathcal{P}_{\gamma_1}(M),\ldots,\mathcal{P}_{\gamma_n}(M)\big),
$$
where $\gamma_j := (i_{j,1},\ldots,i_{j,\ell_j})_{j = 1}^n$ are pairwise disjoint cycles of respective lengths $\ell_j$.

In the case of the measure \eqref{mesmap}, this will give a combinatorial interpretation of higher order free cumulants via planar fully simple maps $k_{\ell_1,\ldots,\ell_n}^M =H_{\ell_1,\ldots,\ell_n}$, as the correlation moments defined in \eqref{correlationMoments} are in this setting generating series of planar ordinary maps $\varphi_{\ell_1,\ldots,\ell_n}^M =F_{\ell_1,\ldots,\ell_n}$.

The results of Proposition~\ref{P10} for simple disks and Theorem~\ref{P20} for fully simple cylinders coincide with the formulas found for generating series of the first and second order free cumulants in \cite{Secondorderfreeness}, given in Theorem~\ref{R-transform} of the introduction. We proved these formulas via combinatorics of maps, independently of \cite{Secondorderfreeness}, and also explained that they are natural in light of the topological recursion. The restriction of our Conjecture~\ref{MainConj} to genus $0$ would give in principle a recursive algorithm to compute the higher order free cumulants of the matrix $M$ sampled from the large $N$ limit of the measure \eqref{mesmap}. This is interesting as the relation at the level of generating series between $n$th order free cumulants and $n$th correlation moments, called $R$-transform machinery, is not otherwise known for $n \geq 3$ as of writing, thus imposing to work with their involved combinatorial definition via partitioned permutations.

We explain in Section~\ref{SFree} that a possible generalization of Conjecture~\ref{MainConj} to stuffed maps -- whose generating series in the ordinary case are governed by blobbed TR (see \ref{generalizationsTR} for some comments on this generalization of TR) -- would shed light on computation of generating series of higher order cumulants in the full generality of \cite{Secondorderfreeness}. Given the universality of the TR structure, one may also wonder if a universal theory of approximate higher order free cumulants can be formulated taking into account the higher genus amplitudes.

\subsubsection{Virasoro constraints}
Section \ref{virasoro} is based on joint work in progress with G.~Borot and D.~Lewa\'nski. Our goal is to deduce explicit Virasoro constraints -- in the sense explained in Section \ref{VirasoroIntro} -- for fully simple maps.

The partition function $Z(\mathbf{p})$ of the formal $1$-hermitian matrix model is known to be a tau function of the KP hierarchy. It can be proved by standard techniques that the partition function $\hat{\mathcal{Z}}(A)$ of the matrix model with external field is also a tau function of the KP hierarchy. Using our combinatorial interpretation of the matrix model with external field \eqref{HMMEFfsMapsEq}, we can show that the transition element from $Z(\mathbf{p})$ to $\hat{\mathcal{Z}}(A)$ is the universal convolution operator $\mathcal{O}$ associated with double weakly monotone Hurwitz numbers. Conjugating the Virasoro constraints known for $Z(\mathbf{p})$ by the operator $\mathcal{O}$ gives Virasoro constraints for $\hat{\mathcal{Z}}(A)$, hence for fully simple maps. Our aim is to deduce Tutte's equations associated to the Virasoro constraints, which seem otherwise too complicated to obtain from bijective combinatorial methods.

We briefly introduce the semi-infinite wedge formalism which is used to compute the Virasoro constraints explicitly. We provide an explicit derivation for the case of usual disks, which serves as a toy model for more complicated topologies and also as a check of our technique since in this case we are able to produce the same result using the definition of (first order) free cumulants via non-crossing partitions \eqref{1stOrderFCs}. We also aim at giving explicit Virasoro constraints for at least the topologies: $(0,2)$, $(0,3)$ and $(1,1)$. Apart from obtaining explicit results for low topologies, our motivations are: again to give some insight in the computation of $R$-transform machinery formulas in the context of free probability for $n\geq 3$, and to aid the understanding of both the combinatorial problem of fully simple maps -- especially in relation to ordinary maps -- and the complicated objects used to define higher order free cumulants called partitioned permutations, introduced in \ref{partitionedPermutations}. 

%
%


\subsection{Large random maps with small and large boundaries, and loop nesting}

We pursue the analysis of nesting statistics in the $O(\mathsf{n})$ loop model on random maps, initiated for maps with the topology of disks and cylinders in \cite{BBD}, here for arbitrary topologies. For this purpose we rely on the topological recursion results of \cite{BEOn,BEO} for the enumeration of maps in the $O(\mathsf{n})$ model. We characterize the generating series of configurations of genus $g$ with $k$ boundaries and $k'$ marked points which realize a fixed nesting graph. These generating series are amenable to explicit computations in the loop model with bending energy on triangulations, and we characterize their behavior at criticality in the dense and in dilute phases that we introduced in Section~\ref{IntroPhase}. The method we develop to analyze the critical behavior for configurations of higher topologies is actually a general procedure to study criticality for any enumerative problem satisfying the topological recursion.

\subsubsection{Combinatorics of $O(\mathsf{n})$ configurations and their nesting} 

We introduced in detail the $O(\mathsf{n})$ loop model from the combinatorial point of view in Section \ref{CombOnModel} and from the probabilistic point of view in Section \ref{ProbOnMod}. The nesting graphs from Section \ref{Markm} are a crucial tool for us to study the nesting properties of $O(\mathsf{n})$ configurations. Actually, the main goal of this part of the thesis is to study the generating series of configurations realizing a fixed nesting graph, which we denote with script letters: $\mathscr{W}$.
We will also encounter generating series of $O(\mathsf{n})$ configurations not keeping track of nestings, which we defined in \eqref{Oncorrelators}, and denoted $\mathcal{W}$. We review the substitution approach of \cite{BBG12a} in Section~\ref{SSub}, describing disks with an $O(\mathsf{n})$ loop model as usual maps whose faces can also be disks with an $O(\mathsf{n})$ loop model. Generically, generating series of usual maps whose faces can also be disks with an $O(\mathsf{n})$ loop model -- i.e., of configurations carrying only non-separating loops -- will be denoted $\mathsf{W}$. We may impose geometric constraints on the maps under consideration, by fixing the genus $g$, the number $k'$ of marked points, the number of boundaries $k$ and their respective perimeters $(\ell_i)_{i = 1}^k$, the volume (i.e., the total number of vertices) $V$, and maybe the arm lengths $(P(\mathsf{e}))_{\mathsf{e}}$. This is conveniently handled at the level of generating series by including extra Boltzmann weights, respectively $\prod_{i} x_i^{-(\ell_i + 1)}$, $u^{V}$, and $\prod_{\mathsf{e}} s(\mathsf{e})^{P(\mathsf{e})}$. 


In Section~\ref{CombDecompConf}, we give a combinatorial decomposition of configurations in terms of their associated nesting graph, which permits to study the critical behavior of the whole configurations via the analysis of the critical behavior of every type of piece, resulting in the basic formula for the generating series of maps with a fixed nesting graph in Proposition~\ref{magi}.


In Section~\ref{Sanal}, we review the analytic properties of these generating series, i.e., in which sense the Boltzmann weights can be considered as nonnegative real-valued parameters instead of formal parameters, and their characterization by functional equations already known in the literature. We study how the topological recursion for $\mathcal{W}^{[g,k]}$ and $\mathsf{W}^{[g,k]}$ commented in \ref{TRConf}, which will be our main tool to handle configurations or arbitrary topologies, can be used in practice. We also explain in Section~\ref{addinm} how the addition of extra marked points can easily be handled at the level of generating series. These results are valid in the general $O(\mathsf{n})$ loop model, where loops are allowed to cross faces of any degree.

We specialize them in Section~\ref{S4} to the $O(\mathsf{n})$ loop model on triangulations with bending energy~$\alpha$, which we recall also depended on the parameter $\mathsf{h}$ per triangle visited by a loop, and $\mathsf{g}$ per empty triangle. This model is the simplest one which is amenable to an explicit solution (in terms of theta functions), and still contains the dense and dilute universality classes which are specific to loop models. At this point, it is convenient to introduce the parameter $b \in (0,\tfrac{1}{2})$ such that
$$
\mathsf{n} = 2\cos(\pi b).
$$
We review the expression for the generating series of disks and cylinders (Section~\ref{elparam}), which constitute the non-trivial initial data allowing to reach higher topologies. We also transform in Section~\ref{TRtr} the topological recursion formula for $\mathcal{W}^{(g,k)}$ into a more explicit sum over trivalent graphs, which will be suited for later analysis.

\subsubsection{Critical behavior of loop nesting}
\label{Sintrocrit} We reviewed in Section~\ref{IntroPhase} the phase diagram of this bending energy model. The properties of the special functions, and some details necessary to obtain this phase diagram as well as for later use, are collected in Appendix~\ref{AppA}-\ref{AppD} which are mostly taken from \cite{BBD}. For fixed $\mathsf{n} \in (0,2)$, $\alpha$ not too large and vertex weight $u = 1$, it features in the $(\mathsf{g},\mathsf{h})$ plane a \emph{non-generic} critical line, beyond which the generating series of pointed disks are divergent. As is well-known, the radius of convergence is actually the same for generating series of maps of any topology. The critical exponents in the interior (resp. at the tip) of the non-generic critical line pertain to the dense (resp. dilute) universality class. Beyond this point, the critical line continues to a \emph{generic} line, which corresponds to the universality class of pure gravity. We focus on the non-generic critical line as it is specific to the loop models.
If $(\mathsf{g},\mathsf{h})$ is chosen on the non-generic critical line but we keep the vertex weight $u < 1$, the model remains off-critical. The distance to criticality is governed by $(1 - u) \rightarrow 0$. At the level of the explicit solution in terms of theta functions, approaching criticality corresponds to a trigonometric limit with a modulus scaling like
\beq
\label{ququ} q \sim [(1 - u)/q_*]^{c},
\eeq
with an exponent distinguishing between the dense and the dilute phase
$$
c = \left\{\begin{array}{lll} \tfrac{1}{1 - b} & & {\rm dense}, \\ 1 & & {\rm dilute}. \end{array}\right.
$$
It is related to the famous string susceptibility exponent $\gamma_{{\rm str}}$ by $c = -\gamma_{{\rm str}}b$. All other exponents can be expressed in terms of $b$ and $c$, and we will give expressions valid for both universality classes using
$$
\mathfrak{d} = \left\{\begin{array}{lll} 1 & & {\rm dense}, \\ -1 & & {\rm dilute}. \end{array} \right.
$$

The main contribution of this part of the thesis is the analysis of singularities of the generating series under consideration for $(\mathsf{g},\mathsf{h})$ on the non-generic critical line, in the limit $u \rightarrow 1$, here conveniently traded for $q \rightarrow 0$ according to \eqref{ququ}. This is done in several steps in Section~\ref{critit}-\ref{S6} summarized below. We then perform in Section~\ref{FixedV} a saddle point analysis to extract the asymptotics of the desired generating series of maps with fixed volume $V \rightarrow \infty$. The analysis reveals two interesting regimes for boundary perimeters: either we impose the boundary to be
\begin{itemize}
\item[$\bullet$] microscopic (``small''), when $\ell_i$ is kept finite,
\item[$\bullet$] or macroscopic (``large''), here corresponding to $\ell_i\,V^{c/2}$ for fixed $\ell_i$.
\end{itemize}
We argue in Section~\ref{crititmarked} that, as far as critical exponents for asymptotics are concerned, marked points behave like small boundaries. So, we choose to present here our results in a simpler form in absence of marked points.

\begin{center}
\begin{figure}[h!]
\begin{center}
\includegraphics[width=0.85\textwidth]{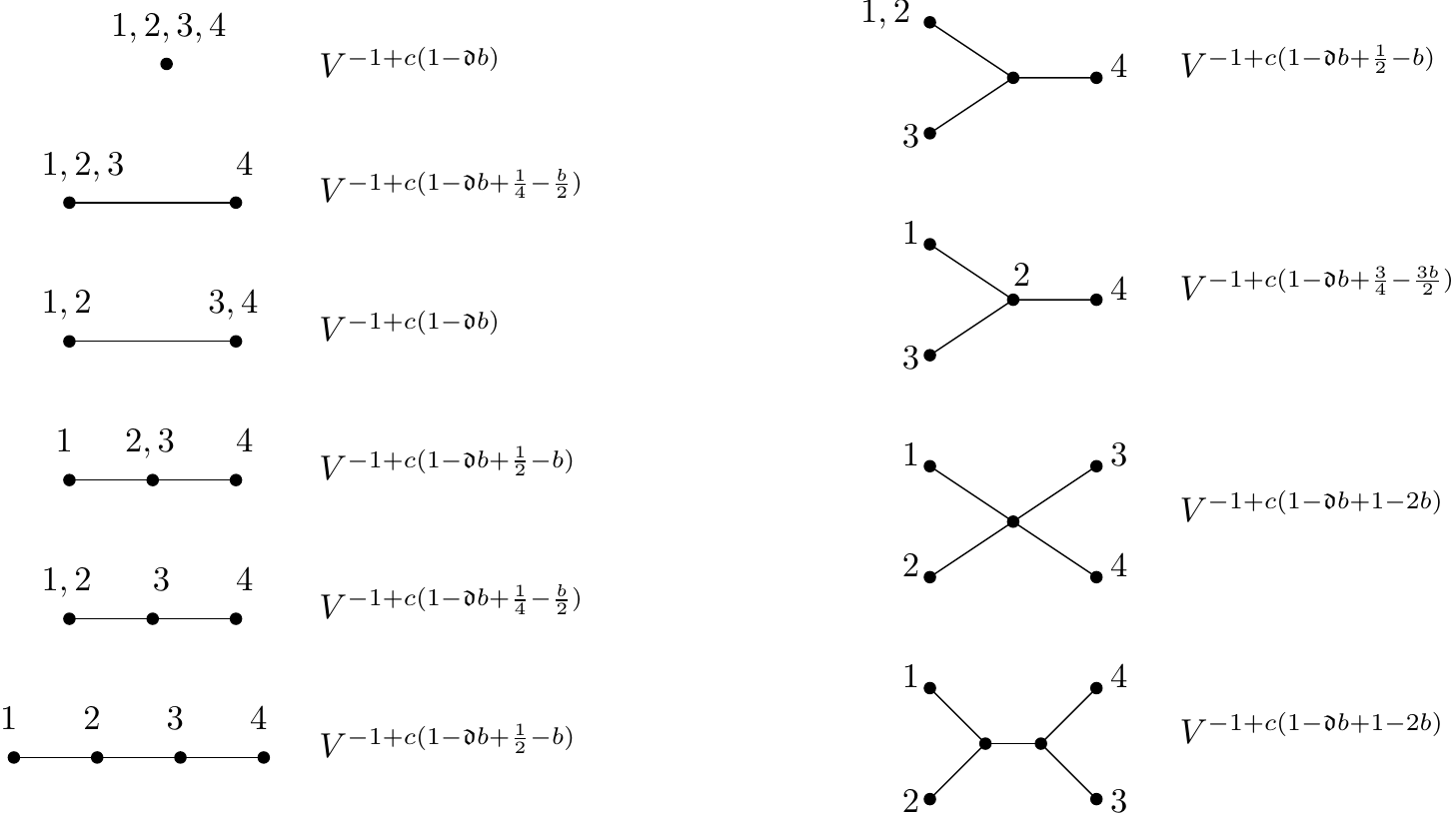}
\caption{\label{04graph} The possible nesting graphs for planar maps with  $4$ microscopic boundaries labeled $1,2,3,4$ (up to permutations of the labels), and the order of magnitude of the number of maps realizing them for large volume $V$. For $\mathsf{n} \in (0,2)$, i.e. $b \in (0,\tfrac{1}{2})$, the greatest order of magnitude is achieved for the two last graphs in the right column.}
\end{center}
\end{figure}
\end{center}

Our first main result (Theorem~\ref{LAPA}) concerning the generating series of maps with fixed nesting graph reads:

\begin{theorem}\label{thm1nesting}
Assume $2 g - 2 + k > 0$. Let $k_{{\rm L}}$ be the number of macroscopic boundaries, $k_{{\rm S}}$ the number of microscopic boundaries, and $k = k_{{\rm L}} + k_{{\rm S}}$. Let also $k_{{\rm S}}^{(0,2)}$ be the number of microscopic boundaries that belong to a genus $0$ connected component of the complement of all loops which does not contain any other boundary and was adjacent to exactly one separating loop (before cutting). Consider $(\Gamma,\star)$ the data of a nesting graph after having forgotten the information on the arm lengths $\mathbf{P}$. The generating series of configurations of genus $g$ realizing $(\Gamma,\star)$ behaves as
$$ 
\Big[ u^{V} \prod_{i = 1}^{k_{{\rm L}}} x_i^{-(\ell_iV^{c/2}+1)} \prod_{i = k_{{\rm L}} + 1}^{k_{{\rm L}} + k_{{\rm S}}} x_i^{-(\ell_i + 1)}\Big] \mathscr{W}^{[g,k]}_{\Gamma,\star}\;\, \mathop{\sim}^{\bigcdot}\;\, V^{-1\,+\,c[(2g - 2 + k)(1 - \mathfrak{d}\frac{b}{2}) - \frac{1}{4}k_{{\rm S}} + (\frac{1}{4} - \frac{b}{2})k_{{\rm S}}^{(0,2)}]},
$$
when $V\rightarrow \infty$.
\end{theorem}
As $b \in (0,\tfrac{1}{2})$, we see that the nesting graphs most likely to occur are those in which each microscopic mark -- either a marked point or a microscopic boundary -- belongs to a genus $0$ univalent vertex which does not carry any other mark. This is exemplified in Figure~\ref{04graph} for maps of genus $0$ with $4$ microscopic marks. The analog statement for cylinders can easily be extracted from \cite{BBD} and is rederived here as Theorem~\ref{LAPB}.

Our second main result (Theorem~\ref{igfsgb} in Section~\ref{Fixedas}) concerning nesting properties describes the large deviation function of (large) arm lengths in configurations realizing a given nesting graph. It is instructive to first review the result for cylinders obtained in \cite{BBD}, which is expressed in terms of the function
\bea
J(p)\,=\,J_{\mathsf{n}}(p) & = & \sup_{s \in [0,2/\mathsf{n}]} \big\{p\ln(s) + {\arccos}(\mathsf{n}s/2) - {\rm arccos}(\mathsf{n}/2)\big\} \nonumber \\
\label{Jdef222} & = & p\ln\Big(\frac{2}{\mathsf{n}}\,\frac{p}{\sqrt{1 + p^2}}\Big) + {\rm arccot}(p) - {\rm arccos}(\mathsf{n}/2).
\eea
plotted in Figure~\ref{Jplot}. It has the following properties:
\begin{itemize}
\item[$\bullet$] $J(p) \geq 0$ for positive $p$, and achieves its minimum value $0$ at $p_{{\rm opt}} = \frac{\mathsf{n}}{\sqrt{4 - \mathsf{n}^2}}$ given below.
\item[$\bullet$] $J(p)$ is strictly convex, and $J''(p) = \frac{1}{p(p^2 + 1)}$.
\item[$\bullet$] $J(p)$ has a slope $\ln(2/\mathsf{n})$ when $p \rightarrow \infty$.
\item[$\bullet$] When $p \rightarrow 0$, we have $J(p) = {\rm arcsin}(\mathsf{n}/2) + p\ln(p) + O(p)$.
\end{itemize}


\begin{theorem}
Fix positive variables $(\ell_1,\ell_2)$ independent of $V$, and $p$ positive such that $p \ll \ln V\!$.
The probability that, in a cylinder with volume $V\rightarrow\infty$, the two boundaries of perimeters $(L_1,L_2)$ are separated by $P$ loops admits the following asymptotics when $V \rightarrow \infty$ 
\bea
\mathbb{P}\Big[P = \frac{c \ln V}{\pi}\,p\,\Big|\,V,\,L_1 = \ell_1,\,L_2 = \ell_2\Big] & \stackrel{\bigcdot}{\sim} & (\ln V)^{-\frac{1}{2}}\,V^{-\frac{c}{\pi}\,J(p)}, \nonumber \\
\mathbb{P}\Big[P = \frac{c\ln V}{2\pi}\,p\,\Big|\,V,\,L_1 = \ell_1,\,L_2 = \ell_2 V^{\frac{c}{2}}\Big] &\stackrel{\bigcdot}{\sim}  & (\ln V)^{-\frac{1}{2}}\,V^{-\frac{c}{2\pi}\,J(p)}, \nonumber \\
\mathbb{P}\Big[P = \frac{c \ln V}{\pi}\,p\,\Big|\,V,\,L_1 = \ell_1 V^{\frac{c}{2}},\,L_2 = \ell_2 V^{\frac{c}{2}}\Big] & \stackrel{\bigcdot}{\sim} & V^{-\frac{c}{\pi}\,p(\mathsf{e})\ln \frac{2}{\mathsf{n}}}. \nonumber  
\eea
\end{theorem} 
We observe that the typical order of magnitude of the number of separating loops between 
a small boundary and any other type of boundary is $\ln V$. More precisely, $\frac{\pi \jmath P}{c\ln V}$ is almost surely equal to the value $p_{{\rm opt}}$, at which the large deviation reaches its minimum value zero, and the fluctuations of $P$ are Gaussian of order $\sqrt{\ln V}$ due to the quadratic behavior of $J(p)$ near $p = p_{{\rm opt}}$. Here $\jmath$ is a normalization constant which is equal to $1$ for two microscopic boundaries and to $2$ for one microscopic and one macroscopic boundary. On the other hand, the arms with both boundaries large will typically contain finitely many separating loops.

For maps of any topology and any nesting graph, we show in this thesis (Theorem~\ref{igfsgb} in Section~\ref{Fixedasarms}) that individual arms have exactly the same behavior when $V\rightarrow \infty$: arm lengths are asymptotically independent from one another, where
\begin{itemize}
\item arms corresponding to edges incident to a genus $0$ univalent vertex carrying as only mark one microscopic boundary typically contain infinitely many loops with depth of order $\ln V$ and large deviation function given by \eqref{Jdef222};
\item the other arms will typically contain finitely many separating loops.
\end{itemize}


\begin{theorem}\label{thm2nesting}
\label{Main2}Assume $2g - 2 + k > 0$, fix a nesting graph $(\Gamma,\star)$, and choose which boundaries are microsopic or macroscopic. Let $E(\Gamma)$ the set of edges of $\,\Gamma$, $E_{0,2}^{{\rm S}}(\Gamma)$, the set of edges incident to a genus $0$ univalent vertex carrying as only mark one microscopic boundary. Consider the regime
\beq
P(\mathsf{e}) = \frac{c\ln V\,p(\mathsf{e})}{\jmath(\mathsf{e})\pi},\qquad \jmath(\mathsf{e}) = \left\{\begin{array}{lll} 2, & & \text{ if}\,\,\mathsf{e} \in E_{0,2}^{{\rm S}}(\Gamma), \\ 1, & & \text{ otherwise}, \end{array}\right.
\eeq
where $p(\mathsf{e})$ may depend on $V$ but remains bounded away from $0$, and negligible in front of $\ln V$. The probability to have arm lengths $\mathbf{P}$ in maps realizing $(\Gamma,\star)$, of volume $V\rightarrow\infty$ with boundary perimeters $L_i = \ell_i$ for the microscopic ones, and $L_i = \ell_i\,V^{\frac{c}{2}}$ for the macroscopic ones, with fixed $\ell_i > 0$, behaves as
$$
\mathbb{P}^{[g,k]}\big[\mathbf{P}|\Gamma,\star,V,\mathbf{L}\big]\; \stackrel{\bigcdot}{\sim} \prod_{\mathsf{e} \in E(\Gamma)\setminus E_{0,2}^{{\rm S}}(\Gamma)} V^{-\frac{c}{\pi}\,p(\mathsf{e})\ln \frac{2}{\mathsf{n}}} \prod_{\mathsf{e} \in E_{0,2}^{{\rm S}}(\Gamma)} \frac{V^{-\frac{c}{2\pi}\,J[p(\mathsf{e})]}}{\sqrt{\ln V}}.
$$
\end{theorem}

For arms corresponding to $\mathsf{e} \in E_{0,2}^{{\rm S}}(\Gamma)$, the Gaussian fluctuations of depths at order $\sqrt{\ln V}$ around $\frac{cp_{{\rm opt}}}{2\pi}\,\ln V$ are precisely described in Corollary~\ref{Gaussflu}. 

In \cite{BBD}, Borot, Bouttier and Duplantier showed that the nesting properties of loops on disks and cylinders weighted by an $O(\mathsf{n})$ model are in perfect agreement with the known nesting properties of ${\rm CLE}_{\kappa}$ \cite{MWW}, after taking into account a suitable version of the KPZ relations \cite{KPZDS}. Our Theorem~\ref{Main2} for arbitrary topologies could in principle be converted into a prediction of extreme nestings of any topology for ${\rm CLE}_{\kappa}$ using the same techniques. However, one faces two additional difficulties here. First, ${\rm CLE}_{\kappa}$ is not properly defined on Riemann surfaces of any topology. And second, even if it were, to be able to compare to our results regarding random loop configurations on random underlying maps, we would need the underlying Riemann surface to be randomly chosen on the moduli space for a fixed topology $(g,n)$. Thus one would need to suitably average out against that randomness.


\subsubsection{Large random maps with small and large boundaries}

The task of Section~\ref{S5} is to derive, for $2g - 2 + k > 0$, the non-generic critical behavior of:
\begin{itemize}
\item[$\bullet$] the generating series $\mathcal{W}^{(g,k)}$ of $O(\mathsf{n})$ configurations, and
\item[$\bullet$] the generating series $\mathsf{W}^{(g,k)}$ of $O(\mathsf{n})$ configurations carrying only non-separating loops,
\end{itemize}
in presence of an arbitrary fixed number of microscopic and macroscopic boundaries (Theorem~\ref{ouqusf}). Here we work in the so-called canonical ensemble, i.e., considering the generating series depending on Boltzmann weights $u$ for vertices and $x_i$ for boundary perimeters. When all boundaries are macroscopic, the result easily follows from the property ``commuting with singular limits'' of the topological recursion, see e.g. \cite[Theorem 5.3.2]{Eynardbook}. The situation is much more involved in the presence of microscopic boundaries, and our analysis in this case is new. 

Our analysis of the critical behavior of the topological recursion amplitudes, which is outlined after Proposition~\ref{2g2mr} in Section~\ref{TRtr}, is in fact more general than the $O(\mathsf{n})$ model, and it may be of use for other problems in enumerative geometry satisfying TR, such as investigating degenerations of semi-simple cohomological field theories. Concretely, we start from the sum over colored trivalent graphs for $\mathsf{W}^{(g,k)}$ and $\mathcal{W}^{(g,k)}$ described in Section~\ref{diagTR}. We analyze the critical behavior of the weights of vertices and of edges in Appendix~\ref{AppBB}, and collect the result in Section~\ref{Sbuildi}. The difference between $\mathsf{W}$ and $\mathcal{W}$ comes only from the edge weights, so both cases can be treated in parallel. Then, we determine in Section~\ref{NextS} for fixed genus $g$, fixed number of boundaries $k$, and \emph{fixed coloring of the $k$ legs}, which graphs give the leading contribution in the critical regime. This is the most technical part, since the formula for the critical exponent of this leading contribution in Lemma~\ref{Cbehavior} is quite intrincate, but we should remember that it does not a priori have a combinatorial meaning. 

The quantities which do have a meaning are $\mathcal{W}$ and $\mathsf{W}$, and they are obtained by summing all these contributions over the colorings of the legs. We find that the final result for the critical behavior of $\mathcal{W}$ and $\mathsf{W}$ in Theorem~\ref{ouqusf} is much simpler. This result does not concern nesting but is interesting per se. Moreover, the critical behavior of $\mathsf{W}$ is a key step to prove the two previously announced theorems describing criticality after fixing a nesting graph: Theorem~\ref{thm1nesting} and Theorem~\ref{thm2nesting}. The outcome clearly displays the affine dependence of the critical exponents on the Euler characteristic of the maps. The generating series of disks $\mathcal{W}(x)$ (which is equal in this case to $\mathsf{W}(x)$) are known to be holomorphic for $x\in\mathbb{C}\setminus [\gamma_-,\gamma_+]$.
\begin{theorem}\label{TROnAsymp}
\textit{The generating series of $\,O(\mathsf{n})$-configurations of genus $g$ with $k$ boundaries, of which $k_S$ are microscopic, has the following critical behavior in the critical regime $q \rightarrow 0$:
$$
\mathcal{W}^{[g,k]}(x_1,\ldots,x_k)\, \stackrel{\bigcdot}{\sim}\, q^{(2 g - 2 + k)(\mathfrak{d}\frac{b}{2}-1)-\frac{k}{2} + \frac{b + 1}{2}k_{S}},
$$
and for the generating series of configurations with only non-separating loops:
$$
\mathsf{W}^{[g,k]}(x_1,\ldots,x_k)\, \stackrel{\bigcdot}{\sim}\, q^{(2 g - 2 + k)(\mathfrak{d}\frac{b}{2}-1)-\frac{k}{2} + \frac{3}{4}k_{S}},
$$
where the errors are uniform, for $\tilde{x}_i = q^{-1/2}(x_i - \gamma_+)$ in any compact, if the $i$th boundary is large, and for $x_i$ in any compact away from $[\gamma_-,\gamma_+]$, if the $i$th boundary is small.}
\end{theorem}

We devote Section~\ref{GeneralizationCriticality} to comment on the generalization of this procedure to any problem satisfying TR under some mild conditions and specialize it to obtain the analog of Theorem \ref{TROnAsymp} in the generic critical line, which is the universality class corresponding to pure gravity, i.e., for usual maps without loops. We also apply our result to the generating series of fully simple maps in presence of large and small boundaries assuming our Conjecture~\ref{MainConj} is true, i.e., to study the critical behavior of the main objects of the first part of the thesis.

We proceed in Section~\ref{S6} to examine the dominant contribution to the critical behavior of $\mathscr{W}_{\Gamma,\star}$, the generating series for fixed nesting graph $(\Gamma,\star)$. The starting point is the combinatorial formula of Proposition~\ref{P212}, which is an appropriate gluing along the given nesting graph $\Gamma$ of vertex weight and edge weights. The vertex weights are the $\mathsf{W}$'s for which we have already obtained the critical behavior in Theorem~\ref{ouqusf}. The edge weights are the generating series $\mathcal{W}_{s}^{(2)}$ for cylinders remembering the number of separating loops between two boundaries, and some variants of these obtained by attaching a loop around one ($\hat{\mathcal{W}}_{s}^{(2)}$) or both ($\tilde{\mathcal{W}}_{s}^{(2)}$) of their boundaries which are defined in Section~\ref{Srefff}; we determine their critical behavior in Section~\ref{armsS}, thanks to the explicit formula for $\mathcal{W}_{s}^{(2)}$ from Proposition~\ref{p15}. We deduce the critical behavior of $\mathscr{W}_{\Gamma,\star}$'s by a saddle point analysis in Section~\ref{CoscrFs} (Theorem~\ref{CoscrF}). This is then converted, as explained in Section~\ref{Sintrocrit}, into asymptotics in the so-called microcanonical ensemble, i.e., for fixed and large volume, fixed large and small boundary perimeters, and finally large and fixed arm lengths in Section~\ref{FixedV}.

A word of caution concerning the canonical ensemble: the dominant contributions depend on the set of variables for which one wishes to study the singularities. If one is only interested later on in fixing the volume and boundary perimeters, one should study singularities with respect to $u$ -- via the variable $q$ -- and $x_i$'s. 
In Section~\ref{Fixedasarms}, we refine the analysis, studying terms containing different types of singularities with respect to the collection of Boltzmann weights $\mathbf{s}$ for the separating loops in order to determine the dominant behavior if we also fix arm lengths. 

The saddle point analysis here is facilitated as similar handlings already appeared for cylinder generating series in \cite{BBD}, and the technical aspects of this part of the thesis rather focus on the combinatorics of maps of higher topology and, more generally, on the analysis of the critical behavior of TR amplitudes for any configuration $(k_L,k_S)$ for $0\leq k_S,\,k_L\leq k_L+k_S=k$.

\part[Fully simple maps]{Fully simple maps \\ \ \\ \large{New combinatorial interpretations \\ of the matrix model with external field and of higher order free cumulants}}


\chapter{Enumeration}
\label{chap:intro}

This part of the thesis is based on joint work with G.~Borot \cite{BGF17}. Our main objects of study are the fully simple maps that we introduced in Section~\ref{fsmaps}, especially in comparison to the classical ordinary maps that were reviewed in Section~\ref{OrdMapsIntro}, where we also set all the notations up tailored for our use in this part of the thesis. We first study combinatorially the cases of disks and cylinders, and later we explain how we expect the problem to relate to the topological recursion setting. In the next Chapter we analyze a natural generalization of this problem from the matrix model point of view. We gave an introduction to the classical $1$-hermitian matrix model and how it relates to the problem of enumeration of ordinary maps in Section~\ref{IntroMMs}. In particular, we will give a matrix model interpretation to the enumeration of fully simple maps, which will allow us to relate this problem to the classical ordinary map enumeration through double monotone Hurwitz numbers, which were introduced in Section~\ref{IntroHNs}. Finally, in Chapter~\ref{chap:apps}, we provide some applications of our results: to free probability, to obtain a new ELSV formula and towards finding Virasoro constraints for fully simple maps.

\section{Base cases: disks and cylinders}

We start by giving a combinatorial decomposition of ordinary disks into a simple disk and some ordinary disks with smaller boundaries, follow by relating in a similar way ordinary cylinders to simple cylinders and finish by giving an algorithm to decompose simple cylinders into fully simple ones, simple disks and other known pieces. We translate our bijections into interesting formulas relating the different generating series.

\subsection{Simple disks from ordinary disks}\label{disks}
\label{Section3}

We can decompose an ordinary disk $\mathcal{M}$ with boundary of length $\ell >0$ into a simple disk $\mathcal{M}^s$ with boundary of length $1\leq \ell ' \leq \ell$ and ordinary disks of lengths $\ell_i < \ell$, using the following procedure:
\begin{algo}(from ordinary to simple)\label{alg1}
Set $\mathcal{M}^s:=\mathcal{M}$. We run over all edges of $\mathcal{M}$, starting at the root edge $e_1$ and following the cyclic order around the boundary. When we arrive at a vertex $v_i$ from an edge $e_i$, we create two vertices out of it: the first remains on the connected component containing $e_i$, while the second one glues together the remaining connected components, giving a map $\mathcal{M}_i$. We then update $\mathcal{M}^s$ to be the first connected component and proceed to the next edge on it. Every $\mathcal{M}_i$, for $i=1, \ldots, \ell '$, is an ordinary map consisting of
\begin{itemize}
\item a single vertex whenever $v_i$ was simple, or
\item a map with a boundary of positive length with the marked edge being the edge in $\mathcal{M}_i$ following $e_i$ in $\mathcal{M}$.
\end{itemize}
\end{algo}

\begin{example} \emph{Consider a non-simple map with a boundary of length $11$ (non-simple vertices are circled). Applying the algorithm we obtain the simple map $\mathcal{M}^s$ of length $3$, and $3$ ordinary maps $\mathcal{M}_1$, $\mathcal{M}_2$, $\mathcal{M}_3$.
    \begin{center}
        \def\svgwidth{\columnwidth}
        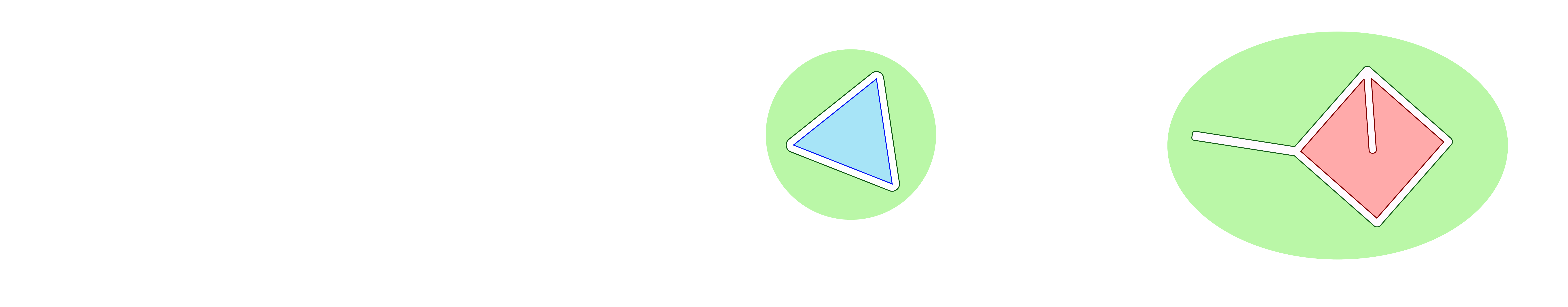
    \end{center}
    \vspace{0.3cm}
The maps should be regarded as drawn on the sphere and the outer face is in all cases the boundary.}
\end{example}

Using the decomposition given by the algorithm, we find that $X$ and $W$ are reciprocal functions:
\begin{proposition}\label{0,1}
\beq
\label{relmaster1} x = X(W(x)).
\eeq
\end{proposition}
\noindent \textbf{Proof.}
Since the algorithm establishes a bijection, we find that
\beq
\forall \ell \geq 1,\qquad F_{\ell} = \sum_{\ell' = 1}^{\ell} H_{\ell'} \sum_{\substack{\ell_1,\ldots,\ell_{\ell'} \geq 0 \\ \sum_i (\ell_i + 1) = \ell}} \prod_{i = 1}^{\ell'} F_{\ell_i},
\eeq
which implies, at the level of resolvents:
\begin{align*}
W(x) & = \sum_{\ell \geq 0} \frac{F_{\ell}}{x^{\ell+1}} = \frac{F_0}{x} + \sum_{\ell\geq 1} \frac{F_\ell}{x^{\ell+1}} = \frac{H_0}{x} + \sum_{\ell \geq 1} \frac{1}{x^{\ell+1}} \sum_{\ell' = 1}^{\ell} H_{\ell'} \sum_{\substack{\ell_1,\ldots,\ell_{\ell'} \geq 0 \\ \sum_i (\ell_i + 1) = \ell}} \prod_{i = 1}^{\ell'} F_{\ell_i}  \\
 & = \frac{1}{x}\sum_{\ell' \geq 0} H_{\ell'}(W(x))^{\ell'} = \frac{W(x)}{x} X(W(x)). 
\end{align*}
We remind the reader that $F_0=H_0=1$ by convention.
\hfill $\Box$

\subsection{Ordinary, simple and fully simple cylinders}\label{cyl}
\label{Section4}
\subsubsection{Replacing an ordinary boundary by a simple boundary}
Let us consider a planar map $\mathcal{M}$ with one ordinary boundary of length $\ell$, and one simple boundary of length $k$. We apply the procedure described in Algorithm \ref{alg1} to the ordinary boundary. We have to distinguish two cases depending on the nature of $\mathcal{M}^s$:
\begin{itemize}
\item either $\mathcal{M}^s$ is a planar map with one simple boundary of some length $\ell'$, and another simple boundary of length $k$ (which we did not touch). Then, the rest of the pieces $\mathcal{M}_1, \ldots, \mathcal{M}_{\ell'}$ are planar maps with one ordinary boundary of lengths $\ell_1,\ldots,\ell_{\ell'}$.
\item or $\mathcal{M}^s$ is a planar map with one simple boundary of some length $\ell'$. And the rest consists of a disjoint union of:
\begin{itemize}
\item a planar map with the simple boundary of length $k$ which bordered $\mathcal{M}$ initially, and another ordinary boundary with some length $\ell_1$, 
\item and  $\ell' - 1$ planar maps with one ordinary boundary of lengths $\ell_2,\ldots,\ell_{\ell'}$.
\end{itemize}
\end{itemize}
This decomposition is again a bijection, and hence
\beq
G_{k|\ell} = \sum_{\ell' = 1}^{\ell}  \sum_{\substack{\ell_1,\ldots,\ell_{\ell'} \geq 0 \\ \sum_i (\ell_i + 1) = \ell}} \left(G_{k,\ell'}\prod_{i = 1}^{\ell'} F_{\ell_i}+ \ell'\,G_{\ell'} G_{k|\ell_1}\prod_{i = 2}^{\ell'} F_{\ell_i}\right).
\eeq
We deduce, at the level of resolvents, that
\beq
Y_{1|1}(w\mid x) = \frac{W(x)}{x} Y_2(w,W(x))+\frac{Y_{1|1}(w\mid x)}{x} (\partial_w(w X(w)))_{w=W(x)}.
\eeq
Isolating $Y_2$, we obtain:
\beq
\label{2simple} Y_{2}(w,W(x)) = -Y_{1|1}(w\mid x) (\partial_{w}X(w))_{w=W(x)} .
\eeq

\subsubsection{From ordinary cylinders to simple cylinders}

We consider the following operator
\beq
\frac{\partial}{\partial V(x)} = \sum_{k \geq 0} \frac{k}{x^{k + 1}} \frac{\partial}{\partial t_{k}}\,,
\eeq
which creates an ordinary boundary of length $k$ weighted by $x^{-(k + 1)}$. Therefore, we have
\begin{align*}
W_{n}^{[g]}(x_1,\ldots,x_n) & = \frac{\partial}{\partial V(x_2)}\cdots\frac{\partial}{\partial V(x_n)}\,W_{1}^{[g]}(x_1),  \\
Y_{1|n}^{[g]}(w \mid x_1,\ldots,x_n) & = \frac{\partial}{\partial V(x_1)}\cdots \frac{\partial}{\partial V(x_n)}\,Y_{1}^{[g]}(w). 
\end{align*}
Applying $\frac{\partial}{\partial V(x_1)}$ to equation (\ref{relmaster1}), we obtain
\beq
0 =  (\partial_w X(w))_{w=W(x_1)} \frac{\partial}{\partial V(x_1)}W(x_1) + Y_{1|1}(W(x_1)\mid x_2) \nonumber
\eeq
and hence
\beq
\label{ge1}Y_{1 | 1}(W(x_1) \mid x_2) = - W_{2}(x_1,x_2)\,(\partial_w X(w))_{w = W(x_1)}.
\eeq
Finally, combining equations (\ref{2simple}) and (\ref{ge1}), we obtain the following relation between ordinary and simple cylinders:
\beq\label{simpleOrdCyl}
Y_2(W(x_1),W(x_2))= W_2(x_1,x_2) (\partial_w X(w))_{w = W(x_1)} (\partial_w X(w))_{w = W(x_2)}.
\eeq

\subsubsection{From simple cylinders to fully simple cylinders}

We describe an algorithm which expresses a planar map $M$ with two simple boundaries in terms of planar fully simple maps. The idea is to merge simple boundaries that touch each other. By definition, a simple boundary which is not fully simple shares at least one vertex with another boundary. By convention, whenever we refer to cyclic order in the process, we mean cyclic order of the first boundary. Let $B_1$ and $B_2$ denote the first and the second boundaries respectively.
\begin{definition}
A {\it pre-shared piece of length $m >0$} is a sequence of $m$ consecutive edges in $B_1$ which are shared with $B_2$. We define a {\it pre-shared piece of length $m =0$} to be a vertex which both boundaries $B_1$ and $B_2$ have in common. 

The first vertex $sv_1$ of the first edge and the second vertex $sv_2$ of the last edge in a pre-shared piece are called the {\it endpoints}.
If $m=0$, the endpoints coincide by convention with the only vertex of the pre-shared piece: $sv_1=sv_2$.

We say that a  pre-shared piece of length $m \geq 0$ is a {\it shared piece of length $m \geq 0$} if the edge in $B_1$ that arrives to $sv_1$ and the edge in $B_1$ outgoing from $sv_2$ are not shared with $B_2$.

We define the {\it interior} of a shared piece to be the shared piece minus the two endpoints.
The interior of a shared vertex is empty.
\end{definition}

Before describing the decomposition algorithm, we describe a special case which corresponds to maps as in Figure \ref{totally glued}.(c), which we will exclude. Consider a map whose two only faces are the two simple boundaries. The only possibility is that they have the same length and are completely glued to each other. We will count this kind of maps apart.
\begin{algo}\label{AlgCyls}(From simple cylinders to fully simple disks or cylinders)
\begin{enumerate}
\item Save the position of the marked edge on each boundary.

\item Denote $r$ the number of shared pieces. If $r=0$, we already have a fully simple cylinder and we stop the algorithm. Otherwise, denote the shared pieces by $S_0, \ldots, S_{r-1}$. Save their lengths $m_0, \ldots, m_{r-1}$, labeled in cyclic order, and shrink their interiors so that only shared vertices remain.

Since we have removed all common edges and boundaries are simple,  every shared vertex has two non-identified incident edges from $B_1$ and two from $B_2$.

\item Create two vertices $v_1, v_2$ out of each shared vertex $v$ in such a way that each $v_j$ has exactly one incident edge from $B_1$ and one from $B_2$, which were consecutive edges for the cyclic order at $v$ in the initial map.

In this way, we got rid of all shared pieces, and we obtain a graph drawn on the sphere formed by $r$ connected components which are homeomorphic to a disk. We consider each connected component separately, and we glue to their boundary a face homeomorphic to a disk.

\item For $i = 0,\ldots , r-1$, call $\mathcal{M}_i$ the connected component which was sharing a vertex with $S_i$ and $S_{i + 1\,({\rm mod}\, r)}$. Mark the edge in $\mathcal{M}_i$ which belonged to the first boundary and was outgoing from $sv_2^i$ in $\mathcal{M}_i$. Then, $\mathcal{M}_i$ becomes a simple disk. Denote by $\ell_i'$ (resp. $\ell_i''$) the number of edges of the boundary of $\mathcal{M}_i$ previously belonging to $B_1$ (resp. $B_2$). Then, the boundary of $M_i$  has perimeter $\ell_i'+\ell_i''$.
\end{enumerate}
\end{algo}

\begin{figure}[h!]
 \begin{center}
        \def\svgwidth{\columnwidth}
        \input{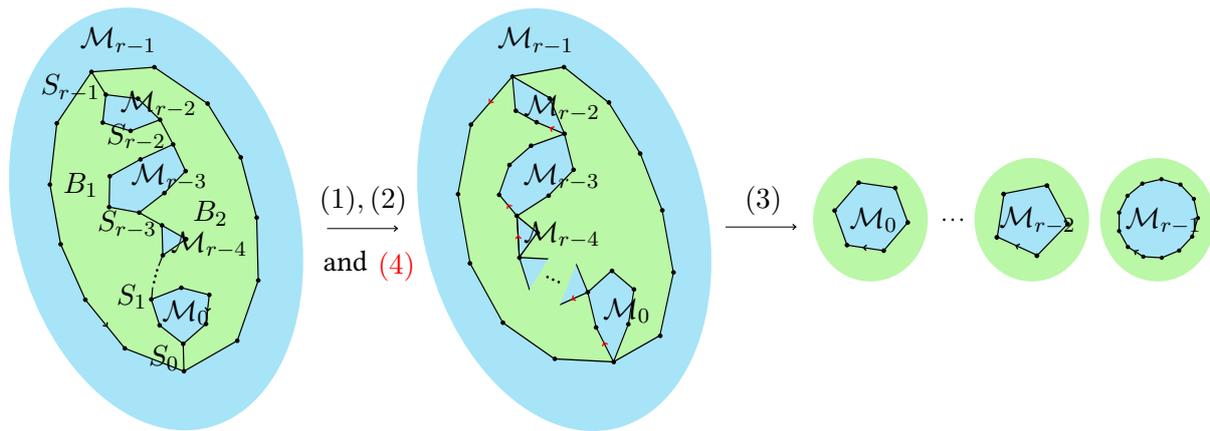}
        	\caption{Schematic representation of Algorithm \ref{AlgCyls}. The two green faces are the two simple, but non-fully simple, boundaries and the blue part represents the inner faces. The shared pieces $S_0,\ldots,S_{r-1}$ are drawn schematically as shared pieces of length $1$. $\mathcal{M}_{r-1}$ is drawn as the outer face, but it does not play a special role.}
\end{center}
\end{figure}

Observe that by construction, $\ell_i', \ell_i'' \geq 1$ and $m_i\geq 0$. Moreover, note that the only map of length $\ell_i'+\ell_i''\geq 2$ and considered simple which is not allowed as $\mathcal{M}_i$ is the map with one boundary of length $2$ where the two edges are identified as in Figure \ref{degenerate simple map}.(c), since this would correspond to a shared piece of length $1$ and it would have been previously removed.

This decomposition is a bijection, since we can retrieve the original map from all the saved information and the obtained fully simple maps. To show this, we describe the inverse algorithm:

\begin{algo}(From fully simple disks or cylinders to simple cylinders)
\begin{enumerate}
\item Let $r$ be the number of given (fully) simple discs. If $r=0$, we already had a fully simple cylinder and the algorithms become trivial. Otherwise, observe that every $\mathcal{M}_i$, for $i = 0,\ldots , r-1$, is a planar disk with two distinguished vertices $v_1^i$ and $v_2^i$. The first one $v_1^i$ is the starting vertex of the root edge $e_1^{i}$ and the second one $v_2^i$ is the ending vertex of the edge $e_{\ell_i'}^{i}$, where the edges are labeled according to the cyclic order of the boundary.

\item For $i = 0,\ldots , r-1$, consider shared pieces $S_i$ of lengths $m_i$.

\item Glue $sv_1^{i}$ of a shared piece $S_{i}$  to $v_2^{i-1\,({\rm mod}\, r)}$ in $\mathcal{M}_{i - 1\,({\rm mod}\, r)}$ and $sv_2^{i}$ of $S_{i}$ to $v_1^{i}$ in $\mathcal{M}_{i}$. 

All the marked edges in the $\mathcal{M}_i$'s should belong to the same simple face, which we call $B_1$. We call $B_2$ the other face, which is bordered by following the edges from $v_2^i$ to $v_1^i$ in every $\mathcal{M}_i$, and the shared piece $S_i$ from $sv_2^i$ to $sv_1^i$, for $i = 1,\ldots,r$.

\item Remove the $r$ markings in $B_1$ and recover the roots in $B_1$ and $B_2$, which are now part of our data.

We have glued the $r$ simple disks and shared pieces into a map with two simple (not fully simple) boundaries $B_1$ and $B_2$.
\end{enumerate}
\end{algo}

This bijection translates into the following relation between generating series of simple and fully simple cylinders:
\begin{proposition}\label{0,2}
\beq
Y_2(w_1,w_2) =  X_2(w_1,w_2) + \partial_{w_1}\partial_{w_2}\ln\left(\frac{w_1 - w_2}{X(w_1) - X(w_2)}\right).
\eeq
\end{proposition}
\noindent \textbf{Proof.}
Let us introduce:
$$
\tilde{X}(w) = X(w) - w^{-1} - w = \sum_{\ell \geq 1} \tilde{H}_{\ell}\,w^{\ell - 1},
$$
the generating series of (fully) simple disks, excluding the disk with boundary of length $0$ which consists of a single vertex, and the simple disk with boundary of length $2$ in which the two edges of the boundary are identified, as in Figure \ref{degenerate simple map}.(c).

Then, using the bijection we established, we obtain that
\beq
G_{L_1,L_2} = H_{L_1,L_2} + \delta_{L_1,L_2}\,L_1  + \sum_{r \geq 1} \sum_{\substack{\ell_i',\,\,\ell_i'' > 0,\,\,m_i \geq 0 \\ \sum_{i=0}^{r-1} \ell_i' + \sum_{i} m_i = L_1 \\ \sum_{i=0}^{r-1}  \ell_i'' + \sum_{i} m_i = L_2}} \frac{L_1L_2}{r} \prod_{i = 1}^r \tilde{H}_{\ell_i' + \ell_i''}, \nonumber 
\eeq
where the first term of the right hand side counts the case $r=0$ in which the simple cylinders were already fully simple and the second term counts the degenerate case we excluded from the algorithm. We already observed that this degenerate case can only occur if $L_1=L_2$ and there are $L_1^2$ possibilities for the two roots, but we also divide by $L_1$ because of the cyclic symmetry of this type of cylinders.

Summing over lengths $L_1, L_2 \geq 1$ with a weight $w_1^{L_1 - 1}w_2^{L_2 - 1}$, we get:
\bea
Y_2(w_1,w_2) & = & X_2(w_1,w_2) - \partial_{w_1}\partial_{w_2}\ln(1 - w_1w_2) \nonumber \\
& & + \partial_{w_1}\partial_{w_2}\left(\sum_{r \geq 1} \frac{1}{r}\bigg(\sum_{m \geq 0} (w_1w_2)^m\bigg)^r\bigg(\sum_{\ell' ,\ell' > 0} \tilde{H}_{\ell' + \ell''} w_1^{\ell'}w_2^{\ell''}\bigg)^{r}\right). \nonumber
\eea
Let us remark that
$$
\sum_{\substack{\ell',\ell'' > 0 \\ \ell' + \ell'' = \ell  }} w_1^{\ell'}w_2^{\ell''} = \frac{w_{1}^{\ell+ 1} - w_2^{\ell + 1}}{w_1 - w_2} - w_1^{\ell} - w_2^{\ell}.
$$
Therefore,
\bea
\sum_{\ell',\ell'' \geq 1} w_1^{\ell'}w_2^{\ell''}\,\tilde{H}_{\ell' + \ell''} & = & \frac{w_1^2\tilde{X}(w_1) - w_2^2\tilde{X}(w_2)}{w_1 - w_2} - w_1\tilde{X}(w_1) - w_2\tilde{X}(w_2) \nonumber \\
& = & \frac{w_1w_2(\tilde{X}(w_1) - \tilde{X}(w_2))}{w_1 - w_2} \nonumber \\
& = & w_1w_2\,\frac{(X(w_1) - X(w_2))}{w_1 - w_2} - (1 - w_1w_2). \nonumber
\eea
And finally, 
\bea
Y_2(w_1,w_2) & = & X_2(w_1,w_2) - \partial_{w_1}\partial_{w_2}\ln(1 - w_1w_2) \nonumber \\
& & - \partial_{w_1}\partial_{w_2}\ln\left[1 - \frac{1}{1 - w_1w_2}\left(w_1w_2\,\frac{X(w_1) - X(w_2)}{w_1 - w_2} - (1 - w_1w_2)\right)\right] \nonumber \\
& = & X_2(w_1,w_2) - \partial_{w_1}\partial_{w_2}\ln\left[-w_1w_2\,\frac{X(w_1) - X(w_2)}{w_1 - w_2}\right] \nonumber \\
& = & X_2(w_1,w_2) + \partial_{w_1}\partial_{w_2}\ln\left(\frac{w_1 - w_2}{X(w_1) - X(w_2)}\right). \nonumber
\eea
\hfill $\Box$


\section{Towards a combinatorial interpretation of symplectic invariance}\label{symplinv}
\label{Section5}

We explain how the formulas obtained in the previous section fit naturally in the universal setting of topological recursion, state a precise conjecture on how this would generalize for higher topologies, and give some numerical evidence and illustration in the particular case of quadrangulations.

\subsection{TR for fully simple maps}

We remind the reader that the generating series of ordinary maps satisfy the topological recursion, as we stated precisely in Theorem \ref{ordmaps}.
Moreover, the important, and still mysterious, property of symplectic invariance can be reviewed in Section \ref{sympinv}.

Let $w(z)\coloneqq W_1^{(0)}(x(z))\dd x(z)$. The spectral curve for ordinary maps was:
\beq\label{inidataordmap}
\mathcal{S}\coloneqq\left(\mathbb{C}P^1, x, W_1^{(0)}(x(z))\dd x(z), B(z_1,z_2)\right).
\eeq 

Let us apply the exchange transformation:
$$
(x,w) \to (-w,x),
$$
that is we consider the spectral curve $\check{\mathcal{S}}$ given by $(w,x)$ with extra initial data 
$$
(\check{\omega}_{0,1}(z_1), \check{\omega}_{0,2}(z_1,z_2)) \coloneqq  (x(z) \dd w(z), B(z_1,z_2)).
$$
For $2g-2+n >0$, we call $\check{\omega}_{g,n}$ the TR amplitudes for this spectral curve.

It is natural to wonder whether the $\check{\omega}_{g,n}$ also solve some enumerative problem and, in that case, which kind of objects they are counting. We propose an answer which also offers a combinatorial interpretation of the important property of symplectic invariance:
\begin{conjecture}\label{conj}
The invariants $\check{\omega}_{g,n}$ enumerate fully simple maps of genus $g$ with $n$ boundaries, in the following sense:
\beq
\check{\omega}_{g,n}(z_1,\ldots,z_n)=X_n^{[g]}(w(z_1),\ldots, w(z_n)) \dd w(z_1)\cdots \dd w(z_n) + \delta_{g,0}\delta_{n,2}\frac{\dd w(z_1) \dd w(z_2)}{(w(z_1)-w(z_2))^2}. \nonumber
\eeq
\end{conjecture}

Using non-combinatorial techniques, we give in Section~\ref{PRof} the path to a possible proof of this conjecture. We manage to reduce the problem to a technical condition regarding the symplectic invariance of the so-called hermitian matrix model with external field.

Observe that if $n=0$, that is we consider maps without boundaries, we have in a very natural way that $W_0^{[g]}=X_0^{[g]}$. It would be interesting to investigate the relation between the TR $n=0$ invariants $\mathfrak{F}_g[\mathcal{S}]$ and $\mathfrak{F}_g[\mathcal{\check{S}}]$. We believe there should be a combinatorial justification for their relation which could be explored with a combinatorial proof of our conjecture, but this is beyond the scope of this thesis.

We dedicate the rest of this chapter to prove some first cases of this conjecture in a combinatorial way, to give some evidence for the conjecture in general and to comment some possible generalizations.


Using our formulas relating the generating series of fully simple disks and cylinders with the ordinary ones, we obtain a combinatorial proof of the first two base cases of the conjecture:
\begin{theorem}
\label{simplTh} Conjecture~\ref{conj} is true for the two base cases $(g,n)= (0,1)$ and $(0,2)$.
\end{theorem}
\noindent \textbf{Proof.}
By Proposition \ref{0,1}, we obtain
$$
\check{\omega}_{0,1}(z) \coloneqq x(z) \dd w(z) = X(W(x(z))) \dd w(z),
$$
which is equal to $X(w(z)) \dd w(z)$ by definition. This proves the theorem for $(g,n)= (0,1)$.

For cylinders, we have, by definition: $\check{\omega}_{0,2}(z_1,z_2)=B(z_1,z_2)$. Substituting the expression for the generating series of ordinary cylinders in terms of the one for simple cylinders given by formula (\ref{simpleOrdCyl}) in the equation for the fundamental differential of the second kind from Theorem \ref{ordmaps}, we obtain
\bea
\omega_{0,2}(z_1,z_2) = B(z_1,z_2)  = & W_2(x(z_1), x(z_2)) \dd x(z_1) \dd x(z_2) + \frac{\dd x(z_1) \dd x(z_2)}{(x(z_1)-x(z_2))^2} \nonumber \\
 = & Y_2(w(z_1), w(z_2)) \dd w(z_1) \dd w(z_2) + \frac{\dd x(z_1) \dd x(z_2)}{(x(z_1)-x(z_2))^2}. \nonumber
\eea
Finally, using Proposition \ref{0,2}, we get the theorem for $(g,n)= (0,2)$:
\beq
\check{\omega}_{0,2}(z_1,z_2)=X_2(w(z_1), w(z_2)) \dd w(z_1) \dd w(z_2) + \frac{\dd w(z_1) \dd w(z_2)}{(w(z_1)-w(z_2))^2}. \nonumber
\eeq
\hfill $\Box$


\subsection{Supporting data for quadrangulations}\label{Quadrangulations}

In this section we compare the number of fully simple, simple and ordinary disks and cylinders in the case in which all the internal faces are quadrangulations, which allows us to make computations explicitly using our results for the base topologies: $(0,1)$ and $(0,2)$. We also compare the conjectural number of fully simple quadrangulations to the number of ordinary ones for topologies $(1,1)$ and $(0,3)$, whose outcomes are given by the first iteration of the algorithm of topological recursion. The reasonable outcomes support our conjecture that, after the exchange transformation, TR counts some more restrictive kind of maps. 

The $(1,1)$ topology is especially interesting since it is the first case with genus $g>0$. For topology $(1,1)$, we provide explicit general formulas for the number of ordinary maps and for the conjectural number of fully simple maps, which we extract from TR. We also give a combinatorial argument that indeed shows that the conjecture provides the right numbers for the first possible length: $\ell = 2$. 

The $(0,3)$ topology is also particularly relevant since in that case we find substantial evidence for our conjecture, using the formulas proved in \cite{BernardiFusy} for fully simple planar quadrangulations with even boundary lengths. Moreover, this is one of the most relevant cases for one of our motivations coming from free probability, since TR for fully simple maps of topology $(0,3)$ would provide new interesting formulas relating the third order free cumulants to the third order correlation moments that we introduced in Section~\ref{IntroFP}. We will elaborate on this application in Section~\ref{SFree}.

We consider maps whose internal faces are all quadrangles \cite{Tutte2}, that is $t_{j} = t \delta_{j,4}$ where we denote here $t$ the weight per internal quadrangle. The spectral curve is given by
$$
x(z) = c\Big(z + \frac{1}{z}\Big),\qquad w(z) = \frac{1}{cz} - \frac{tc^3}{z^3},\qquad 
$$
with
\beq\label{cSeries}
c = \sqrt{\frac{1 - \sqrt{1 - 12t}}{6t}} = 1 + \frac{3t}{2} + \frac{63}{8}t^2 + \frac{891}{16}t^3 + \frac{57915}{128}t^4 + O(t^4).
\eeq
The zeroes of $\dd x$ are located at $z = \pm 1$, and the deck transformation is $\iota(z) = \frac{1}{z}$.  The zeroes of $\dd w$ are located at $z = \pm c^2 \sqrt{3t}$, and the deck transformation is
$$
\check{\iota}(z) = \frac{c^2z(c^2t + \sqrt{4tz^2 - 3c^4t^2})}{2(z^2 - tc^4)}.
$$
Consider the multidifferentials $\omega_{g,n}$ and $\check{\omega}_{g,n}$ on $\mathbb{P}^1$ as at the beginning of the section but with initial data specialized for quadrangulations.
We define
\bea
F_{\ell_1,\ldots,\ell_n}^{[g]} & \!\!\!\! = \!\!\!\! & (-1)^n \Res_{z_1 \rightarrow \infty} (x(z_1))^{\ell_1} \cdots \Res_{z_n \rightarrow \infty} (x(z_n))^{\ell_n} \bigg(\omega_{g,n}(z_1,\ldots,z_n) - \delta_{g,0}\delta_{n,2} \frac{\dd x(z_1)\dd x(z_2)}{(x(z_1) - x(z_2))^2}\bigg), \nonumber \\
\check{F}_{k_1,\ldots,k_n}^{[g]} & \!\!\!\! = \!\!\!\! & \Res_{z_1 \rightarrow \infty} (w(z_1))^{-k_1} \cdots \Res_{z_n \rightarrow \infty} (w(z_n))^{-k_n}\bigg(\check{\omega}_{g,n}(z_1,\ldots,z_n) - \delta_{g,0}\delta_{n,2}\frac{\dd w(z_1)\dd w(z_2)}{(w(z_1) - w(z_2))^2}\bigg). \nonumber
\eea
We know that $\check{F}_{k_1}=H_{k_1}$ and $\check{F}_{k_1,k_2}=H_{k_1,k_2}$, and we conjecture $\check{F}_{k_1,\ldots,k_n}^{[g]}=H_{k_1,\ldots,k_n}^{[g]}$ in general.

\subsubsection{Disks}
We explore two tables to compare the coefficients $[t^{Q}]\,F_{\ell}$ and $[t^{Q}]\,\check{F}_{k}$. It is remarkable that all the $[t^{Q}]\,\check{F}_{k}$ are nonnegative integers, which already suggested a priori that they may be counting some objects. Theorem \ref{simplTh} identifies $[t^{Q}]\,\check{F}_{k}$ with the number of (fully) simple disks $[t^{Q}]\,H_{k}$.

If the length of the boundary is odd, the number of disks is obviously $0$.

Observe that if we consider a boundary of length $\ell=2$, the number of ordinary disks is equal to the number of (fully) simple disks because the only two possible boundaries of length $2$ are simple in genus $0$. If the two vertices get identified in the non-degenerate case, either the genus is increased or an internal face of length $1$ appears, which is not possible because we are counting quadrangulations.

\begin{center}
\begin{figure}[h!]
\begin{center}
{\scriptsize \begin{tabular}{|c||c|c|c|c|c|c|c|c|c|}
\hline
$\ell$ & $Q = 0$ & $1$ & $2$ & $3$ & $4$ & $5$ & $6$ & $7$ & $8$ \\
\hline\hline
$\mathbf{2}$ & $1$ & $2$ & $9$ & $54$ & $378$ & $2916$ & $24057$ & $208494$ & $1876446$ \\
\hline
$\mathbf{4}$ & $2$ & $9$ & $54$ & $378$ & $2916$ & $24057$ & $208494$ & $1876446$ & $17399772$ \\
\hline
$\mathbf{6}$ & $5$ & $36$ & $270$ & $2160$ & $18225$ & $160380$ & $1459458$ & $13646880$ & $130489290$ \\
\hline
$\mathbf{8}$ & $14$ & $140$ & $1260$ & $11340$ & $103950$ & $972972$ & $9287460$ & $90221040$ & $890065260$ \\
\hline
\end{tabular}}
\caption{Number of ordinary disks with boundary of length $\ell$ and $Q$ quadrangles.}
\end{center}
\end{figure}
\end{center}

We also remark that the $[t^{Q}]\,H_{k}$ in the second table are (much) smaller than the corresponding $[t^{Q}]\,F_{k}$, and for small number of quadrangles some of them are $0$, which makes sense due to the strong geometric constraints to form maps with simple boundaries and a small number of internal faces.

\begin{center}
\begin{figure}[h!]
\begin{center}
{\scriptsize \begin{tabular}{|c||c|c|c|c|c|c|c|c|c|}
\hline
$k$ & $Q = 0$ & $1$ & $2$ & $3$ & $4$ & $5$ & $6$ & $7$ & $8$ \\
\hline\hline
$\mathbf{2}$ & $1$ & $2$ & $9$ & $54$ & $378$ & $2916$ & $24057$ & $208494$ & $1876446$ \\
\hline
$\mathbf{4}$ & $0$ & $1$ & $10$ & $90$ & $810$ & $7425$ & $69498$ & $663390$ & $6444360$ \\
\hline
$\mathbf{6}$ & $0$ & $0$ & $3$ & $56$ & $756$ & $9072$ & $103194$ & $1143072$ & $12492144$ \\
\hline
$\mathbf{8}$ & $0$ & $0$ & $0$ & $12$ & $330$ & $5940$ & $89100$ & $1211760$ & $15540822$ \\
\hline
\end{tabular}}
\caption{\label{FSdisks} Number of simple disks with boundary of length $k$ and $Q$ quadrangles.}
\end{center}
\end{figure}
\end{center}

\subsubsection{Cylinders}

We explore now the number of cylinders imposing different constraints to the boundaries: $[t^{Q}]\,F_{\ell_1,\ell_2}$ (ordinary) and $[t^{Q}]\,H_{k_1,k_2}$ (fully simple), and also $[t^{Q}]\,G_{k_1\mid\ell_1}$ (one simple boundary, one ordinary boundary) and $[t^{Q}]\,G_{k_1,k_2\mid}$ (simple) given by the formulas (\ref{ge1}) and (\ref{simpleOrdCyl}) respectively.

Since we know how to convert an unmarked quadrangle into an ordinary boundary of length $4$, we can relate the outcomes for cylinders with at least one of the boundaries being ordinary of length $4$ to the previous results for disks as follows:
\bea
\bullet & 4\frac{\partial}{\partial t} F_{\ell_1} = F_{\ell_1,4} & \Rightarrow \ \    4Q[t^Q]\,F_{\ell_1} = [t^{Q-1}]\, F_{\ell_1,4}, \nonumber \\
\bullet & 4\frac{\partial}{\partial t} G_{k_1} = G_{k_1\mid 4} & \Rightarrow \ \   4Q [t^Q]\,G_{k_1}= [t^{Q-1}]\, G_{k_1\mid 4}. \nonumber
\eea
If the sum of the lengths of the two boundaries is odd, the number of quadrangulations is obviously $0$.

Observe that the results also satisfy the following inequalities:
$$
[t^{Q}]\,F_{l_1,l_2} \geq [t^{Q}]\,G_{l_1\mid l_2}\geq [t^{Q}]\,G_{l_1,l_2} \geq [t^{Q}]\,H_{l_1,l_2},
$$
which are compatible with the combinatorial interpretation that Theorem \ref{simplTh} offers, since we are imposing further constraints whenever we force a boundary to be simple or, even more, fully simple.

We also obtain more and more zeroes for small number of quadrangles as we impose stronger conditions on the boundaries.
 
\begin{center}
\begin{figure}[h!]
\begin{center}
{\scriptsize 
\scalebox{0.95}{\begin{tabular}{|c||c|c|c|c|c|c|c|c|c|}
\hline
$(\ell_1,\ell_2)$ & $Q = 0$ & $1$ & $2$ & $3$ & $4$ & $5$ & $6$ & $7$ & $8$ \\
\hline\hline
$\mathbf{(1,1)}$ & $1$ & $3$ & $18$ & $135$ & $1134$ & $10206$ & $96228$ & $938223$ & $9382230$ \\
\hline
$\mathbf{(3,1)}$ & $3$ & $18$ & $135$ & $1134$ & $10206$ & $96228$ & $938223$ & $9382230$ & $95698746$\\
\hline
$\mathbf{(5,1)}$ & $10$ & $90$ & $810$ & $7560$ & $72900$ & $721710$ & $7297290$ & $75057840$ & $782989740$ \\
\hline
$\mathbf{(7,1)}$ & $35$ & $420$ & $4410$ & $45360$ & $467775$ & $4864860$ & $51081030$ & $541326240$ & $5785424190$ \\
\hline
$\mathbf{(9,1)}$ & $126$ & $1890$ & $22680$ & $255150$ & $2806650$ & $30648618$ & $334348560$ & $3653952120$ & $40052936700$ \\
\hline\hline
$\mathbf{(2,2)}$ & $2$ & $12$ & $90$ & $756$ & $6804$ & $64152$ & $625482$ & $6254820$ & $63799164$ \\
\hline
$\mathbf{(4,2)}$ & $8$ & $72$ & $648$ & $6048$ & $58320$ & $577368$ & $5837832$ & $60046272$ & $626391792$ \\
\hline
$\mathbf{(6,2)}$ & $30$ & $360$ & $3780$ & $38880$ & $400950$ & $4169880$ & $43783740$ & $463993920$ & $4958935020$ \\
\hline
$\mathbf{(8,2)}$ & $112$ & $1680$ & $20160$ & $226800$ & $2494800$ & $27243216$ & $297198720$ & $3247957440$ & $335602610400$ \\
\hline\hline
$\mathbf{(3,3)}$ & $12$ & $108$ & $972$ & $9072$ & $87480$ & $866052$ & $8756748$ & $90069408$ & $939587688$ \\
\hline
$\mathbf{(5,3)}$ & $45$ & $540$ & $5670$ & $58320$ & $601425$ & $6254820$ & $65675610$ & $695990880$ & $7438402530$ \\
\hline
$\mathbf{(7,3)}$ & $168$ & $2520$ & $30240$ & $340200$ & $3742200$ & $40864824$ & $445798080$ & $4871936160$ & $53403915600$ \\
\hline
$\mathbf{(9,3)}$ & $630$ & $11340$ & $153090$ & $1871100$ & $21891870$ & $250761420$ & $2841962760$ & $32042349360$ & $360476430300$ \\
\hline\hline
$\mathbf{(4,4)}$ & $36$ & $432$ & $4536$ & $46656$ & $481140$ & $5003856$ & $52540488$ & $556792704$ & $5950722024$ \\
\hline
$\mathbf{(6,4)}$ & $144$ & $2160$ & $25920$ & $291600$ & $3207600$ & $35026992$ & $382112640$ & $4175945280$ & $45774784800$ \\
\hline
$\mathbf{(8,4)}$ & $560$ & $10080$ & $136080$ & $1663200$ & $19459440$ & $222899040$ & $2526189120$ & $28482088320$ & $320423493600$ \\
\hline
\end{tabular}}}
\caption{Number of ordinary cylinders with boundaries of lengths $(\ell_1,\ell_2)$ and $Q$ quadrangles: $[t^{Q}]\,F_{\ell_1,\ell_2}$.}
\end{center}
\end{figure}
\end{center}


\begin{center}
\begin{figure}[h!]
\begin{center}
\scalebox{0.95}{{\scriptsize \begin{tabular}{|c||c|c|c|c|c|c|c|c|c|}
\hline
$(k_1,\ell_1)$ & $Q = 0$ & $1$ & $2$ & $3$ & $4$ & $5$ & $6$ & $7$ & $8$ \\
\hline\hline
$\mathbf{(1,1)}$ & $1$ & $3$ & $18$ & $135$ & $1134$ & $10206$ & $96228$ & $938223$ & $9382230$ \\
\hline
$\mathbf{(3,1)}$ & $0$ & $3$ & $36$ & $378$ & $3888$ & $40095$ & $416988$ & $4378374$ & $46399392$\\
\hline
$\mathbf{(5,1)}$ & $0$ & $0$ & $15$ & $315$ & $4725$ & $62370$ & $773955$ & $9287460$ & $109306260$ \\
\hline
$\mathbf{(7,1)}$ & $0$ & $0$ & $0$ & $84$ & $2520$ & $49140$ & $793800$ & $11566800$ & $158233824$ \\
\hline
$\mathbf{(9,1)}$ & $0$ & $0$ & $0$ & $0$ & $495$ & $19305$ & $463320$ & $8860995$ & $148551975$ \\
\hline\hline
$\mathbf{(2,2)}$ & $2$ & $12$ & $90$ & $756$ & $6804$ & $64152$ & $625482$ & $6254820$ & $63799164$ \\
\hline
$\mathbf{(4,2)}$ & $0$ & $8$ & $120$ & $1440$ & $16200$ & $178200$ & $1945944$ & $21228480$ & $231996960$ \\
\hline
$\mathbf{(6,2)}$ & $0$ & $0$ & $42$ & $1008$ & $16632$ & $235872$ & $3095820$ & $38864448$ & $474701472$ \\
\hline
$\mathbf{(8,2)}$ & $0$ & $0$ & $0$ & $240$ & $7920$ & $166320$ & $2851200$ & $43623360$ & $621632880$ \\
\hline\hline
$\mathbf{(1,3)}$ & $3$ & $18$ & $135$ & $1134$ & $10206$ & $96228$ & $938223$ & $9382230$ & $95698746$\\
\hline
$\mathbf{(3,3)}$ & $3$ & $36$ & $378$ & $3888$ & $40095$ & $416988$ & $4378374$ & $46399392$ & $495893502$\\
\hline
$\mathbf{(5,3)}$ & $0$ & $15$ & $315$ & $4725$ & $62370$ & $773955$ & $9287460$ & $109306260$ & $1271521800$ \\
\hline
$\mathbf{(7,3)}$ & $0$ & $0$ & $84$ & $2520$ & $49140$ & $793800$ & $11566800$ & $158233824$ & $2076818940$ \\
\hline
$\mathbf{(9,3)}$ & $0$ & $0$ & $0$ & $495$ & $19305$ & $463320$ & $8860995$ & $148551975$ & $2287700415$ \\
\hline\hline
$\mathbf{(2,4)}$ & $8$ & $72$ & $648$ & $6048$ & $58320$ & $577368$ & $5837832$ & $60046272$ & $626391792$ \\
\hline
$\mathbf{(4,4)}$ & $4$ & $80$ & $1080$ & $12960$ & $148500$ & $1667952$ & $18574920$ & $206219520$ & $2288739240$ \\
\hline
$\mathbf{(6,4)}$ & $0$ & $24$ & $672$ & $12096$ & $181440$ & $2476656$ & $32006016$ & $399748608$ & $4882643712$ \\
\hline
$\mathbf{(8,4)}$ & $0$ & $0$ & $144$ & $5280$ & $118800$ & $2138400$ & $33929280$ & $497306304$ & $6911094960$ \\
\hline
\end{tabular}}}
\caption{Number of cylinders with the first boundary simple of length $k_1$ and the second boundary ordinary of length $\ell_1$: $[t^{Q}]\,G_{k_1\mid\ell_1}$.}
\end{center}
\end{figure}
\end{center}

\begin{center}
\begin{figure}[h!]
\begin{center}
\scalebox{0.95}{{\scriptsize \begin{tabular}{|c||c|c|c|c|c|c|c|c|c|}
\hline
$(k_1,k_2)$ & $Q = 0$ & $1$ & $2$ & $3$ & $4$ & $5$ & $6$ & $7$ & $8$ \\
\hline\hline
$\mathbf{(1,1)}$ & $1$ & $3$ & $18$ & $135$ & $1134$ & $10206$ & $96228$ & $938223$ & $9382230$ \\
\hline
$\mathbf{(1,3)}$ & $0$ & $3$ & $36$ & $378$ & $3888$ & $40095$ & $416988$ & $4378374$ & $46399392$\\
\hline
$\mathbf{(1,5)}$ & $0$ & $0$ & $15$ & $315$ & $4725$ & $62370$ & $773955$ & $9287460$ & $109306260$ \\
\hline
$\mathbf{(1,7)}$ & $0$ & $0$ & $0$ & $84$ & $2520$ & $49140$ & $793800$ & $11566800$ & $158233824$ \\
\hline
$\mathbf{(1,9)}$ & $0$ & $0$ & $0$ & $0$ & $495$ & $19305$ & $463320$ & $8860995$ & $148551975$ \\
\hline\hline
$\mathbf{(2,2)}$ & $2$ & $12$ & $90$ & $756$ & $6804$ & $64152$ & $625482$ & $6254820$ & $63799164$ \\
\hline
$\mathbf{(2,4)}$ & $0$ & $8$ & $120$ & $1440$ & $16200$ & $178200$ & $1945944$ & $21228480$ & $231996960$ \\
\hline
$\mathbf{(2,6)}$ & $0$ & $0$ & $42$ & $1008$ & $16632$ & $235872$ & $3095820$ & $38864448$ & $474701472$ \\
\hline
$\mathbf{(2,8)}$ & $0$ & $0$ & $0$ & $240$ & $7920$ & $166320$ & $2851200$ & $43623360$ & $621632880$ \\
\hline\hline
$\mathbf{(3,3)}$ & $3$ & $27$ & $252$ & $2457$ & $24705$ & $253935$ & $2653560$ & $28089828$ & $300480678$\\
\hline
$\mathbf{(3,5)}$ & $0$ & $15$ & $270$ & $3690$ & $45900$ & $547560$ & $6395760$ & $73862280$ & $847681200$ \\
\hline
$\mathbf{(3,7)}$ & $0$ & $0$ & $84$ & $2268$ & $41076$ & $628992$ & $8808912$ & $116940348$ & $1499730876$ \\
\hline
$\mathbf{(3,9)}$ & $0$ & $0$ & $0$ & $495$ & $17820$ & $402435$ & $7341840$ & $118587645$ & $1772680140$ \\
\hline\hline
$\mathbf{(4,4)}$ & $4$ & $48$ & $536$ & $5952$ & $66132$ & $735696$ & $8196552$ & $91476864$ & $1022868648$ \\
\hline
$\mathbf{(4,6)}$ & $0$ & $24$ & $504$ & $7728$ & $105336$ & $1354752$ & $16855776$ & $205426368$ & $2469577896$ \\
\hline
$\mathbf{(4,8)}$ & $0$ & $0$ & $144$ & $4320$ & $85200$ & $1401120$ & $20856960$ & $291942144$ & $3922233840$ \\
\hline
\end{tabular}}}
\caption{Number of simple cylinders with boundaries of lengths $(k_1,k_2)$ and $Q$ quadrangles: $[t^{Q}]\,G_{k_1,k_2}$.}
\end{center}
\end{figure}
\end{center}

\begin{center}
\begin{figure}[h!]
\begin{center}
\scalebox{0.95}{{\scriptsize \begin{tabular}{|c||c|c|c|c|c|c|c|c|c|}
\hline
$(k_1,k_2)$ & $Q = 0$ & $1$ & $2$ & $3$ & $4$ & $5$ & $6$ & $7$ & $8$ \\
\hline\hline
$\mathbf{(1,1)}$ & $0$ & $1$ & $9$ & $81$ & $756$ & $7290$ & $72171$ & $729729$ & $7505784$ \\
\hline
$\mathbf{(1,3)}$ & $0$ & $0$ & $6$ & $108$ & $1458$ & $17820$ & $208494$ & $2388204$ & $27066312$\\
\hline
$\mathbf{(1,5)}$ & $0$ & $0$ & $0$ & $35$ & $945$ & $17010$ & $257985$ & $3572100$ & $46845540$ \\
\hline
$\mathbf{(1,7)}$ & $0$ & $0$ & $0$ & $0$ & $210$ & $7560$ & $170100$ & $3084480$ & $49448070$ \\
\hline
$\mathbf{(1,9)}$ & $0$ & $0$ & $0$ & $0$ & $0$ & $1287$ & $57915$ & $1563705$ & $33011550$ \\
\hline\hline
$\mathbf{(2,2)}$ & $0$ & $0$ & $6$ & $108$ & $1458$ & $17820$ & $208494$ & $2388204$ & $27066312$ \\
\hline
$\mathbf{(2,4)}$ & $0$ & $0$ & $0$ & $40$ & $1080$ & $19440$ & $294840$ & $4082400$ & $53537760$ \\
\hline
$\mathbf{(2,6)}$ & $0$ & $0$ & $0$ & $0$ & $252$ & $9072$ & $204120$ & $3701376$ & $59337684$ \\
\hline
$\mathbf{(2,8)}$ & $0$ & $0$ & $0$ & $0$ & $0$ & $1584$ & $71280$ & $1924560$ & $40629600$ \\
\hline\hline
$\mathbf{(3,3)}$ & $0$ & $0$ & $0$ & $48$ & $1296$ & $23328$ & $353808$ & $4898880$ & $64245312$\\
\hline
$\mathbf{(3,5)}$ & $0$ & $0$ & $0$ & $0$ & $315$ & $11340$ & $255150$ & $4626720$ & $74172105$ \\
\hline
$\mathbf{(3,7)}$ & $0$ & $0$ & $0$ & $0$ & $0$ & $2016$ & $90720$ & $2449440$ & $51710400$ \\
\hline
$\mathbf{(3,9)}$ & $0$ & $0$ & $0$ & $0$ & $0$ & $0$ & $12870$ & $694980$ & $21891870$ \\
\hline\hline
$\mathbf{(4,4)}$ & $0$ & $0$ & $0$ & $0$ & $300$ & $10800$ & $243000$ & $4406400$ & $70640100$ \\
\hline
$\mathbf{(4,6)}$ & $0$ & $0$ & $0$ & $0$ & $0$ & $2016$ & $90720$ & $2449440$ & $51710400$ \\
\hline
$\mathbf{(4,8)}$ & $0$ & $0$ & $0$ & $0$ & $0$ & $0$ & $13200$ & $712800$ & $22453200$ \\
\hline
\end{tabular}}}
\caption{\label{FScyl} Number of fully simple cylinders with boundaries of lengths $(k_1,k_2)$ and $Q$ quadrangles: $[t^{Q}]\,H_{k_1,k_2}$.}\label{fig:fscyl}
\end{center}
\end{figure}
\end{center}

\vspace{-0.5cm}

Observe that forcing a boundary of length $1$ or $2$ to be simple does not have any effect in the planar case and therefore the corresponding rows in the first three tables coincide. 
However, imposing that the cylinder is fully simple is much stronger, so in the last table (Figure \ref{fig:fscyl}) all the entries are (much) smaller.

\subsubsection{Tori with 1 boundary}\label{ToriSection}

We compute
\bea\label{tori1}
\omega_{1,1}(z) & = & \frac{z^3(tc^4z^4 + z^2(1 - 5tc^4) + tc^4)}{c(z^2 - 1)^5(1 - 3tc^4)^2}\,\dd z,  \\
\label{tori2}\check{\omega}_{1,1}(z) & = & \frac{3t^2c^9z^5[(3tc^4-2)z^4 + 3tc^4(9tc^4-1)z^2 - 27t^3c^{12}]}{(3tc^4-z^2)^5(1 - 3tc^4)^2}\,\dd z. 
\eea

We present in Figure~\ref{ordinaryTori} the number of tori with $1$ ordinary boundary of perimeter $\ell$ and $Q$ internal quadrangles, as given by Theorem~\ref{ordmaps}.
\begin{center}
\begin{figure}[h!]
\begin{center}
\scalebox{0.88}{{\scriptsize \begin{tabular}{|c||c|c|c|c|c|c|c|c|c|c|}
\hline
$\ell$ & $Q = 0$ & $1$ & $2$ & $3$ & $4$ & $5$ & $6$ & $7$ & $8$ \\
\hline\hline
$\mathbf{2}$ & $0$ & $1$ & $15$ & $198$ & $2511$ & $31266$ & $385398$ &  $4721004$ & $57590271$ \\
\hline
$\mathbf{4}$ & $1$ & $15$ & $198$ & $2511$ & $31266$ & $385398$ & $4721004$ & $57590271$ & $700465482$  \\
\hline
$\mathbf{6}$ & $10$ & $150$ & $1980$ & $25110$ & $312660$ & $3853980$ & $47210040$ & $575902710$ & $7004654820$  \\
\hline
$\mathbf{8}$ & $70$ & $1190$ & $16590$ & $216720$ & $2748060$ & $34286480$ & $423600030$ & $5199957000$ & $63549802260$  \\
\hline
$\mathbf{10}$ & $420$ & $8190$ & $122850$ & $1678320$ & $21925890$ & $279389250$ & $3505914090$ & $43551655560$ & $537235675200$  \\
\hline
$\mathbf{12}$ & $2310$ & $51282$ & $831600$ & $11962566$ & $162074682$ & $2121490602$ & $27174209832$ & $343061095608$ & $4287091638060$  \\ 
\hline
$\mathbf{14}$ & $12012$ & $300300$ & $5261256$ & $79891812$ & $1126377252$ & $15198795612$ & $199385314128$ & $2565902298960$ & $32572738238040$ \\
\hline
\end{tabular}}}
\caption{\label{ordinaryTori}$[t^{Q}]\,F_{\ell}^{[1]}$.}
\end{center}
\end{figure}
\end{center}

For comparison, we present the coefficients $[t^{Q}]\,\check{F}_{k}^{[1]}$ in the same range. Again, it is remarkable that they are all nonnegative integers; Conjecture~\ref{conj} proposes a combinatorial interpretation for them. We also remark that they are always (much) smaller than the corresponding $[t^{Q}]\,F_{\ell}^{[1]}$, and that some of them for small number of quadrangles are $0$, which indicates as before that they may be counting a subclass of ordinary maps.

Due to the strong geometric constraints to form maps with simple boundaries and few internal faces, our observations support that the $[t^{Q}]\,\check{F}_{k}^{[1]}$ may indeed be counting fully simple tori with $Q$ quadrangles.

Moreover, we can give the following simple combinatorial argument to prove the conjecture provides the right answer for $\ell=2$:

\begin{remark}\label{remarkl2}
$F_{2}^{[1]}=H_{1,1}+H_2^{[1]}$.
\end{remark}
\noindent \textit{Proof.}
Ordinary tori with a boundary of length $2$ can be of the following two types:
\begin{itemize}
\item The boundary is simple and the two edges are not identified. This type of ordinary tori are exactly the fully simple tori, counted by $H_2^{[1]}$.

\item The two edges of the boundary are not identified, but the two vertices are, hence the boundary forms a non-trivial cycle of the torus. This type of ordinary tori are obviously in bijection with fully simple cylinders with boundary lengths $(1,1)$, counted by $H_{1,1}$, since one can just glue the vertices of the two boundaries of the cylinder to recover the torus.
\hfill $\Box$
\end{itemize}

Observing our data, we find that for $Q=0,\ldots,8$, we have $[t^Q](\check{F}_2^{[1]}=F_2^{[1]}- H_{1,1})$. Thus, from the remark, we get $[t^Q]\check{F}_2^{[1]}=[t^Q]H_2^{[1]}$, up to at least $Q=8$ quadrangles. We are going to provide now explicit formulas for genus $1$, which will, in particular, help us prove this for all $Q\geq 0$.

Let $\phi_m=c^{2m}\frac{1+(m-1)\sqrt{1-12t}}{1-12t}$, where we recall from \eqref{cSeries} that $c^2= \frac{1 - \sqrt{1 - 12t}}{6t}.$
Then, 
\bea\label{formulasSeries1}
F_{2(m+1)}^{[1]} & = &\frac{(2m+1)!}{6\,n!^2}\,\phi_m, \ \text{ for } m\geq 0,\\
\label{formulasSeries2}\check{F}_{2m}^{[1]} & = &\frac{(3m)!\, t^{m+1}}{4\,m!(2m-1)!}\,\phi_{3m+1}, \ \text{ for } m\geq 1.
\eea
More explicitly, the number of ordinary (and of conjectural fully simple) tori with one boundary can be computed with the following expansion:
\beq\label{explicitphi}
\phi_m = \sum_{n\geq 0} \left(mr_{m,n} + (1-m)\sum_{i=0}^{n-1}r_{m,i}\frac{2\cdot 3^{n-i}}{n-i}\binom{2(n-i-1)}{n-i-1}\right) (3t)^n,
\eeq
where
$$
\frac{c^{2m}}{1-12t}=\sum_{i\geq 0}r_{m,i}(3t)^i,
$$
with 
\beq\label{explicitCoeff}
r_{m,i} = 2^{m+2i}-\frac{1}{2}\sum_{j=0}^{m/2}(-1)^j\binom{m-j-1}{j}\binom{2(m+i-j)}{m+i-j}.
\eeq
The formulas~\eqref{formulasSeries1}-\eqref{formulasSeries2} can be directly extracted from the expressions obtained from TR: \eqref{tori1}-\eqref{tori2}, and the explicit coefficients \eqref{explicitCoeff} can be computed using Lagrange inversion.

Remarkably, both~\eqref{formulasSeries1} and \eqref{formulasSeries2} are given in terms of the same $\phi_m$ with a shifted index, a shifted power of $t$ and different, but simple, combinatorial prefactors. This suggests that if our conjecture~\ref{conj} is true, there is an equivalent underlying combinatorial problem for ordinary and fully simple rooted tori that is worth investigating.

We can now confirm our conjectural formula for fully simple rooted tori, for $\ell=2$:
\begin{remark}
$\check{F}_2^{[1]}=H_2^{[1]}.$
\end{remark}
\begin{proof}
From our result for cylinders~\ref{simplTh}, we can extract the simple closed formula:
$$
H_{1,1}=c^6t.
$$
Now, using our explicit expressions~\eqref{formulasSeries1} and \eqref{formulasSeries2}, it can be checked that $F_2^{[1]}=H_{1,1}+\check{F}_2^{[1]}$. Hence the claim follows from our previous Remark~\ref{remarkl2}.
\end{proof}

\begin{center}
\begin{figure}[h!]
\begin{center}
{\scriptsize \begin{tabular}{|c||c|c|c|c|c|c|c|c|c|c|}
\hline  $k$ & $Q = 0$ & $1$ & $2$ & $3$ & $4$ & $5$ & $6$ & $7$ & $8$ \\
\hline\hline
 $\mathbf{2}$ & 0 & 0 & 6 & 117 & 1755 & 23976 & 313227 & 3991275 & 50084487 \\
\hline $\mathbf{4}$ & 0 & 0 & 0 & 105 & 2925 & 55215 & 885330 & 13009005 & 181316880 \\
\hline $\mathbf{6}$ & 0 & 0 & 0 & 0 & 1260 & 46116 & 1065960 & 19983348 & 332470656 \\
\hline $\mathbf{8}$ & 0 & 0 & 0 & 0 & 0 & 12870 & 585090 & 16073640 &346928670 \\
\hline $\mathbf{10}$ & 0 & 0 & 0 & 0 & 0 & 0 & 120120 & 6531525 & 208243035 \\
\hline $\mathbf{12}$ & 0 & 0 & 0 & 0 & 0 & 0 & 0 & 1058148 & 66997476 \\
\hline $\mathbf{14}$ & 0 & 0 & 0 & 0 & 0 & 0 & 0 & 0 & 8953560 \\
\hline 
\end{tabular}}
\caption{$[t^{Q}]\,H_{k}^{[1]}$.}
\end{center}
\end{figure}
\end{center}

If our conjecture is true, \eqref{explicitphi} would provide the first formula counting a class of fully simple maps for positive genus $g>0$.

\subsubsection{Pairs of pants: evidence for conjecture for even boundary lengths}\label{PairPantsSection}

Very recently, Bernardi and Fusy \cite{BernardiFusy} were able to count, via a bijective procedure, the number of planar fully simple quadrangulations with boundaries of prescribed even lengths. We write their formula here in terms of our notations:
\begin{theorem} 
Let $Q$ be the number of internal quadrangles and $k_1,\ldots,k_n$ positive even integers with $L=\sum_{i=1}^n k_i$ the total boundary length. If $v=Q-\frac{L}{2}-n+2 \geq 0$, in which case it counts the number of internal vertices, we have that the number of planar fully simple quadrangulations is given by
\beq\label{evenLengths}
[t^Q]H_{k_1,\ldots,k_n} = \alpha(Q,L,n)\prod_{i=1}^n k_i\binom{\frac{3}{2}k_i}{k_i}.
\eeq
where $\alpha(Q,L,n)\coloneqq\frac{3^{Q-\frac{L}{2}}(e-1)!}{v!(L+Q)!}$, with $e=\frac{L}{2}+2Q$ the total number of edges.
\end{theorem}

This formula reproduces the number of fully simple disks $[t^Q]H_{k}$ and cylinders $[t^Q]H_{k_1,k_2}$ in our Figures~\ref{FSdisks} and~\ref{FScyl}, for even boundary lengths $k, k_1, k_2$. Therefore, it can be recovered from our Theorem~\ref{simplTh} for those base topologies.

More importantly, we checked that our conjectural numbers of fully simple pairs of pants give indeed the right numbers for even boundary lengths $1\leq k_1, k_2, k_3 \leq 8$:
$$
[t^Q] \check{F}_{k_1,k_2,k_3} = H_{k_1,k_2,k_3}, \ \ \ \text{ for } 0\leq Q \leq 8,
$$
thus providing solid evidence for our Conjecture~\ref{conj} in the case of quadrangulations of topology $(0,3)$.

From our data for cylinders and pairs of pants, one can propose a similar formula for fully simple quadrangulations with (any) prescribed boundary lengths:
\beq
[t^Q]H_{k_1,\ldots,k_n} = \alpha(Q,L,n)\prod_{i=1}^n \varepsilon(k_i),
\eeq
where
$$
\varepsilon(k)\coloneqq \begin{cases}
\frac{(3l)!}{l!(2l-1)!}, & \text{ if } k=2l, \\
\sqrt{3}\frac{(3l+1)!}{l!(2l)!}, & \text{ if } k=2l+1. \\
\end{cases}
$$
This formula is a conjectural generalization of \eqref{evenLengths} to include the presence of odd boundary lengths. Observe that since the total length $L$ has to be even, the number of odd lengths also has to be even, hence only factors of $3$ will appear for every extra pair of odd lengths.

Again, the case of $[t^Q]H_{k_1,k_2}$ with $k_1$ and $k_2$ odd can be proved from our Theorem~\ref{simplTh}. 

\section{Fully simple pairs of pants}\label{RtransPairPants}
Since we have strong evidence for our Conjecture~\ref{conj} in the case of the topology $(0,3)$, we state here as a consequence a formula relating the generating series of ordinary and fully simple pairs of pants.

Observe that the differential of any meromorphic function $f$ is given by
$$
\dd f(p) = \Res_{z \rightarrow p} B(p,z)\,f(z).
$$
Let us denote collectively the zeroes of $\dd x$ by $\mathbf{a}$, and by $\mathbf{b}$ the zeroes of $\dd w$.
Using the formula in \cite[Theorem 4.1]{EOFg}, we get that
$$
\omega_{3}^{[0]}(z_1,z_2,z_3) = \Res_{z \rightarrow \mathbf{a}} -\frac{B(z,z_1)B(z,z_2)B(z,z_3)}{\dd x(z)\dd w(z)}
$$
and
$$
\check{\omega}_{3}^{[0]}(z_1,z_2,z_3) = \Res_{z \rightarrow \mathbf{b}} -\frac{B(z,z_1)B(z,z_2)B(z,z_3)}{\dd x(z)\dd w(z)}, 
$$
where the residue at a set means the sum over residues at every point in the set.

Using also that for any meromorphic $1$-form $\alpha$ on a compact algebraic curve, which is the case for the spectral curve for maps, we have that
$$
\sum_{p} \Res_{z \rightarrow p} \alpha(z) = 0,
$$
we obtain:
\begin{align}\label{03formula}
& \omega_{3}^{[0]}(z_1,z_2,z_3) + \check{\omega}_{3}^{[0]}(z_1,z_2,z_3) = \Res_{z \rightarrow z_1,z_2,z_3} \frac{B(z,z_1)B(z,z_2)B(z,z_3)}{\dd x(z)\dd y(z)}  \\
& =  \dd_{1}\Big[\frac{B(z_1,z_2)B(z_1,z_3)}{\dd x(z_1)\dd y(z_1)}\Big] + \dd_{2}\Big[\frac{B(z_2,z_1)B(z_2,z_3)}{\dd x(z_2)\dd y(z_2)}\Big] + \dd_3\Big[\frac{B(z_3,z_1)B(z_3,z_2)}{\dd x(z_3)\dd y(z_3)}\Big]. \nonumber
\end{align}

\section{Generalization to stuffed maps}
\label{Section6}

We refer the reader to the section \ref{stuffed} for an introduction to stuffed maps. We recall the intuitive idea that this concept generalizes usual maps in the following sense: stuffed maps are built from the same pieces as usual maps allowing internal faces to have any topology now, in contrast to the condition that all faces have to be of the topology of the disk for usual maps.

Since all the results in Sections \ref{disks} and \ref{cyl} for the generating series of maps only affect the boundaries and do not use the condition that internal faces have the topology of the disk at all, the statements and proofs are still valid in the more general setting of stuffed maps.

\subsection{Conjecture for maps carrying a loop model}

Usual maps carrying self-avoiding loop configurations are equivalent to stuffed maps for which we allow unlabeled elementary 2-cells to have the topology of a disk (usual faces) or of a cylinder (rings of faces carrying the loops). By equivalence, we mean here an equality of generating series after a suitable change of formal variables.


As we mentioned in \ref{TRConf}, the generating series of ordinary maps with loops obey the topological recursion, with initial data $\omega_{0,1}$ and $\omega_{0,2}$ again given by the corresponding generating series of disks and cylinders. As of now, explicit expressions for $\omega_{0,1}$ and $\omega_{0,2}$ are only known for a restricted class of model with loops, \textit{e.g.} those in which loops cross only triangle faces \cite{GK,EKOn,BEOn} maybe taking into account bending \cite{BBG12b}.

The analog of Conjecture \ref{conj} for $O(\mathsf{n})$ configurations reads:
\begin{conjecture}
\label{ConT02} After the exchange transformation $(x,w) \to (-w,x)$ in the initial data of TR for ordinary maps with loops, the TR amplitudes enumerate fully simple maps carrying a loop model.
\end{conjecture}
A proof of this conjecture could be given along the lines of Section~\ref{Section9} if one could first establish that the topological recursion governs the topological expansion in the formal matrix model
\thickmuskip=0mu
$$
\dd\mu(M) = \dd M \exp\bigg(N {\rm Tr}(MA) - N\,{\rm Tr}\,\frac{M^2}{2} + \sum_{h \geq 0} \sum_{k \geq 1} \sum_{d \geq 1} N\,\frac{t_{d}}{d}\,{\rm Tr}\,M^{d} + \sum_{d_1,d_2 \geq 1} \frac{t_{d_1,d_2}}{d_1d_2}\,{\rm Tr}\,M^{d_1}\,{\rm Tr}\,M^{d_2}\bigg) 
$$
which depends on the external hermitian matrix $A$.
\thickmuskip=3mu

According to our previous remark, the conjecture is true for disks and cylinders due to the validity of our Theorem~\ref{simplTh}.

\subsection{Vague conjecture for stuffed maps}\label{vagueConjStuffed}

It was proved in \cite{Bstuff} that the generating series of ordinary stuffed maps satisfy the so-called blobbed topological recursion, which was axiomatized in \cite{BSblob}. In this generalized version of the topological recursion, the invariants $\omega_{n}^{[g]}$ are determined by $\omega_{1}^{[0]}$ and $\omega_{2}^{[0]}$ as before, and additionally by the so-called blobs $\phi_n^{[g]}$ for stable topologies, $2g-2+n >0$.
We conjecture that, after the same change of variables, and a transformation of the blobs \emph{still to be described}, the blobbed topological recursion will enumerate fully simple stuffed maps. Again, according to our previous remark, this conjecture is true for disks and cylinders (whose expression do not involve the blobs).

\chapter{Matrix model interpretation}
\label{chap:intro}




In this chapter we introduce what we will call ordinary and fully simple amplitudes for any unitarily invariant measure in the space $\mathcal{H}_N$ of $N\times N$ hermitian matrices. We will also find that these two types of correlators can be related through monotone Hurwitz numbers, bringing another interesting problem into play. Later, we explain that for particular measures, these can be seen as matrix models for our combinatorial problems from the previous chapter: ordinary and fully simple maps. This interpretation motivated the name we gave to the general correlators. Finally, we give a path to a possible proof of our conjecture for usual maps that after exchanging $x$ and $y$ in the spectral curve for ordinary maps, we obtain a spectral curve whose correlators will enumerate fully simple maps.


\section{Ordinary and fully simple correlators for unitarily invariant matrix models}\label{HNsWeingarten}

We consider an arbitrary measure $\dd \mu(M)$ on the space $\mathcal{H}_N$ of $N\times N$ hermitian matrices which is invariant under conjugation by a unitary matrix. If $\mathcal{O}$ is a polynomial function of the entries of $M$, we denote $\langle \mathcal{O}(M) \rangle$ its expectation value with respect to $\dd \mu(M)$:
$$
\langle \mathcal{O}(M) \rangle = \frac{\int_{\mathcal{H}_N} \dd \mu(M)\,\mathcal{O}(M)}{\int_{\mathcal{H}_N} \dd\mu(M)}.
$$
And, if $\mathcal{O}_1,\ldots,\mathcal{O}_n$ are polynomial functions of the entries of $M$, we denote $\kappa_n(\mathcal{O}_1(M),\ldots,\mathcal{O}_n(M))$ their cumulant with respect to the measure $\dd \mu(M)$.

If $\gamma = (c_1 \,  c_2  \, \ldots \, c_{\ell})$ is a cycle of the symmetric group $\mathfrak{S}_{N}$, we denote
$$
\mathcal{P}_{\gamma}(M) \coloneqq M_{c_1,c_2}\, M_{c_2,c_3}\cdots M_{c_{\ell-1},c_{\ell}}\, M_{c_{\ell},c_{1}} = \prod_{m = 1}^{\ell} M_{c_m,\gamma(c_m)}.
$$
We denote $l(\gamma)$ the length of the cycle $\gamma$.

We will be interested in two types of expectation values:
\beq
\label{Obsdisc} \Big\langle \prod_{i = 1}^n {\rm Tr}\,M^{L_i} \Big\rangle \qquad {\rm and} \qquad \Big\langle \prod_{i = 1}^{n} \mathcal{P}_{\gamma_i}(M) \Big\rangle,
\eeq
where $(L_i)_{i = 1}^n$ is a sequence of nonnegative integers, and $(\gamma_i)_{i = 1}^n$ is a sequence of pairwise disjoint cycles in $\mathfrak{S}_{N}$ with $l(\gamma_i)=L_i$ -- the latter imposes $N$ to be larger than $L = \sum_{i = 1}^n l(\gamma_i)$. The first type of expectation value will be called \textit{ordinary}, and the second one \textit{fully simple}. The terms may be used for the ``disconnected'' version \eqref{Obsdisc}, or for the ``connected'' version obtained by taking the cumulants instead of the expectation values of the product. This terminology will be justified by their combinatorial interpretation in terms of ordinary and fully simple maps in Section~\ref{MMs}.

\begin{remark} The unitary invariance of $\mu$ implies its invariance under conjugation of $M$ by a permutation matrix of size $N$. As a consequence, fully simple expectation values only depend on the conjugacy class of the permutation $\gamma_1\cdots\gamma_n$, thus on the partition $\lambda$ encoding the lengths $\ell_i$ of $\gamma_i$. We can then use without ambiguity the following notations:
\bea
& \Big\langle \prod_{i = 1}^n \mathcal{P}_{\gamma_i}(M) \Big\rangle  =  \big\langle\mathcal{P}_{\lambda}(M)\big\rangle = \Big\langle \prod_{i = 1}^n \mathcal{P}^{(\ell_i)}(M) \Big\rangle, \text{ and} \nonumber \\& \kappa_n(\mathcal{P}_{\gamma_1}(M),\ldots,\mathcal{P}_{\gamma_n}(M)\big)  =  \kappa_n\big(\mathcal{P}^{(\ell_1)}(M),\ldots,\mathcal{P}^{(\ell_n)}(M)\big). \nonumber
\eea
If $N < L$, we convene that these quantities are zero.
\end{remark}

\subsection{Weingarten calculus}

If the (formal) measure on $M$ is invariant under conjugation by a unitary matrix of size $N$, it should be possible to express the fully simple observables in terms of the ordinary ones -- independently of the measure on $M$. This precise relation will be described in Theorem~\ref{Transi}. We first introduce the representation theory framework which proves and explains this result.

\subsubsection{Preliminaries on symmetric functions}\label{SymmetricFunctions}

The character ring of ${\rm GL}_N(\mathbb{C})$ -- i.e. polynomial functions of the entries of $M$, which are invariant by conjugation -- is generated by $p_{l}(M) = {\rm Tr}\,M^{l}$ for $l \geq 0$. It is isomorphic to the ring of symmetric functions in $N$ variables
$$
\mathcal{B}_{N} = \mathbb{Q}[x_1,\ldots,x_N]^{\mathfrak{S}_{N}},
$$
tensored over~$\mathbb{C}$.

Let $\lambda$ be a partition of an integer $L\geq 0$. We will use here all the notations introduced in Section\ref{IntroHNs}. We recall the notation $C_{\lambda}$ for the conjugacy class in $\mathfrak{S}_{L}$ described by the partition $\lambda$. We also denote~$\beta$ an element in $C_{\lambda}$.

The power sum functions $p_{[\beta]}(M) = p_{\lambda}(M) \coloneqq \prod_{i = 1}^{\ell(\lambda)} p_{\lambda_{i}}(M)$  with $\ell(\lambda) \leq N$ form a linear basis of the character ring of ${\rm GL}_N(\mathbb{C})$. Another linear basis is formed by the Schur functions $s_{\lambda}(M)$ with $\ell(\lambda) \leq N$, which have the following expansion in terms of power sum functions:
\beq
\label{SWform} s_{\lambda}(M) = \frac{1}{L!}\,\sum_{\mu \vdash L} |C_{\mu}|\,\chi_{\lambda}(C_{\mu})\,p_{\mu}(M) ,\qquad L = |\lambda|,
\eeq
where $\chi_{\lambda}$ are the characters of $\mathfrak{S}_{L}$. 

The $\mathcal{B}_{N}$ are graded rings, where the grading comes from the total degree of a polynomial. We will work with the graded ring of symmetric polynomials in infinitely many variables, defined as
$$
\mathcal{B} = \mathop{{\rm lim}}_{\infty \leftarrow N} \mathcal{B}_{N}.$$
This is the projective limit using the restriction morphisms $\mathcal{B}_{N + 1} \rightarrow \mathcal{B}_{N}$ sending $p(x_1,\ldots,x_{N + 1})$ to $p(x_1,\ldots,x_{N},0)$. By construction, if $r \in \mathcal{B}$, it determines for any $N \geq 0$ an element $\iota_{N}[r] \in \mathcal{B}_{N}$ by setting
$$
\iota_{N}[r](x_1,\ldots,x_{N}) = r(x_1,\ldots,x_{N},0,0,\ldots).
$$
We often abuse notation and write $r(x_1,\ldots,x_{N})$ for this restriction to $N$ variables. In fact, $\mathcal{B}$ is a free graded ring over $\mathbb{Q}$ with one generator $p_{k}$ in each degree $k \geq 1$. The power sums $p_{\lambda}$ and the Schur elements $s_{\lambda}$ are two homogeneous linear bases for $\mathcal{B}$, abstractly related via \eqref{SWform}. A description of the various bases for $\mathcal{B}$ and their properties in relation to representation theory can be found in \cite{Fultonrep}.

Let $\mathcal{B}^{(d)}$ denote the (finite-dimensional) subspace of homogeneous elements of $\mathcal{B}$ of degree $d$. We later need to consider the tensor product of $\mathcal{B}$ with itself, defined by
$$
\mathcal{B} \hat{\otimes} \mathcal{B} := \bigoplus_{d \geq 0} \Big( \bigoplus_{d_1 + d_2 = d} \mathcal{B}^{(d_1)} \otimes \mathcal{B}^{(d_2)}\Big).
$$

\subsubsection{Moments of the Haar measure}

Unlike $\prod_{i = 1}^n {\rm Tr}\,M^{L_i}$, the expression $\prod_{i = 1}^n \mathcal{P}_{\gamma_i}(M)$ is not unitarily invariant. However, the unitary invariance of the measure implies that
\beq
\label{PMun}\Big\langle \prod_{i = 1}^n \mathcal{P}_{\gamma_i}(M) \Big\rangle  = \Big\langle \int_{\mathcal{U}_N} \dd U\,\prod_{i = 1}^n \mathcal{P}_{\gamma_i}(UMU^{\dagger})\Big\rangle ,
\eeq
where $\dd U$ is the Haar measure on the unitary group. Moments of the entries of a random unitary matrix distributed according to the Haar measure can be computed in terms of representation theory of the symmetric group: this is Weingarten calculus \cite{CollinsWeingarten}. If $N \geq 1$ and $L \geq 0$ are two integers, the Weingarten function is defined as
$$
G_{N,L}(\beta) \coloneqq \frac{1}{L!^2} \sum_{\lambda \vdash L} \frac{\chi_{\lambda}({\rm id})^2 \chi_{\lambda}(\beta)}{s_{\lambda}(1_{N})} ,\ \ \text{ for } \beta \in \mathfrak{S}_{L}.
$$
Note that it only depends on the conjugacy class of $\beta$.

\begin{theorem} \cite{CollinsWeingarten}
\label{UNmoment}
$$
\int_{\mathcal{U}_N} \dd U \bigg(\prod_{l = 1}^L U_{a_l,b_l} U^{\dagger}_{b_l',a_l'}\bigg) = \sum_{\beta,\tau \in \mathfrak{S}_{L}} \bigg(\prod_{l = 1}^L \delta_{a_l,a_{\beta(l)}'} \delta_{b_{l},b_{\tau(l)}'}\bigg)\,G_{N,L}(\beta\tau^{-1}).
$$
\end{theorem}

\subsubsection{From fully simple to ordinary}

We will use this formula to compute \eqref{PMun}. Let $(\gamma_i)_{i = 1}^n$ be pairwise disjoint cycles:
$$
\gamma_i = (j_{i,1} \ \   j_{i,2} \  \ldots \  j_{i,L_i}).
$$
We denote $H^{\partial} = \bigsqcup_{i = 1}^n \{i\} \times (\mathbb{Z}/L_i\mathbb{Z})$,
$$
L = |H^{\partial} | = \sum_{i = 1}^n L_i
$$
and $\varphi^{\partial} \in \mathfrak{S}_{H^{\partial} }$ the product of the cyclic permutations sending $(i,l)$ to $(i,l + 1\,\,{\rm mod}\,\,L_i)$. Our notations here are motivated by the fact that when we take a certain specialization of the measure $\dd\mu$ in Section \ref{MMs}, $H^{\partial}$ will refer to the set of half-edges belonging to boundaries of a map and $\varphi^{\partial}$ will be the permutation whose cycles correspond to the boundaries. 

\begin{proposition}\label{wein}
$$
\Big\langle \prod_{i = 1}^n \mathcal{P}_{\gamma_i}(M) \Big\rangle = \sum_{\mu \vdash L} \tilde{G}_{N,L}(C_{\mu},\varphi^{\partial})\,\big\langle p_{\mu}(M)\big\rangle = \sum_{\lambda \vdash L} \frac{ \chi_{\lambda}(\varphi^{\partial})\chi_{\lambda}({\rm id})}{L!\,s_{\lambda}(1_{N})}\,\big\langle s_{\lambda}(M) \big\rangle,  \nonumber 
$$
with 
$$
\tilde{G}_{N,L}(C,\beta) = \frac{1}{L!^2} \sum_{\lambda \vdash L} |C|\chi_{\lambda}(C)\chi_{\lambda}(\beta)\,\frac{\chi_{\lambda}({\rm id})}{s_{\lambda}(1_N)},\ \ \text{ for } \beta \in \mathfrak{S}_{L}.
$$
\end{proposition}
\noindent \textbf{Proof.}
If $M$ is a hermitian matrix, we denote $\Lambda$ its diagonal matrix of eigenvalues -- defined up to permutation. We then have
$$
\int \dd U\, \prod_{i = 1}^n \mathcal{P}_{\gamma_i}(UMU^{\dagger}) = \sum_{\substack{1 \leq a_{i,l} \leq N \\ (i,l) \in H^{\partial}}} \int_{\mathcal{U}_N} \dd U\,\prod_{(i,l) \in H^{\partial}} \Lambda_{a_{i,l}} U_{j_{i,l},a_{i,l}} U^{\dagger}_{a_{i,l},j_{\varphi^{\partial}(i,l)}},
$$
in which we can substitute Theorem~\ref{UNmoment}. We obtain a sum over $\rho,\tau \in \mathfrak{S}_{H^{\partial}}$ of terms involving 
$$
\sum_{\substack{1 \leq a_{i,l} \leq N \\ (i,l) \in H^{\partial}}} \prod_{(i,l) \in H^{\partial}} \Lambda_{a_{i,l}} \delta_{j_{i,l},j_{\rho(\varphi^{\partial}(i,l))}} \delta_{a_{i,l},a_{\tau(i,l)}} = p_{[\tau]}(\Lambda) \prod_{(i,l) \in H^{\partial}} \delta_{j_{i,l},j_{\rho(\varphi^{\partial}(i,l))}}.
$$
As $p_{[\tau]}(\Lambda)$ is unitarily invariant, it is also equal to $p_{[\tau]}(M)$. Since we assumed the $j_{i,l}$ pairwise disjoint, this is non-zero only if $\rho = (\varphi^{\partial})^{-1}$. Therefore
\bea
\label{eq:34}\Big\langle \prod_{i = 1}^n \mathcal{P}_{\gamma_i}(M) \Big\rangle & = & \sum_{\tau \in \mathfrak{S}_{H^{\partial}}} \big\langle p_{[\tau]}(M) \big\rangle\,G_{N,L}\big((\varphi^{\partial})^{-1}\tau^{-1}\big) = \sum_{\mu \vdash L} \big\langle p_{\mu}(M) \big\rangle\,\tilde{G}_{N,L}(C_{\mu},\varphi^{\partial}),
\eea
with
$$
\tilde{G}_{N,L}(C,\beta) = \sum_{\tau \in C} G_{N,L}(\beta^{-1}\tau),
$$
as $\tau$ and $\tau^{-1}$ are in the same conjugacy class. To go further, we recall the Frobenius formula:
\begin{lemma} See e.g. \cite[Theorem 2]{Zagierapp}. If $C_1,\ldots,C_k$ are conjugacy classes of $\mathfrak{S}_{L}$, the number of permutations $\beta_i \in C_i$ such that
$\beta_1 \circ \cdots \circ \beta_{L} = {\rm id}$ is
$$
\mathcal{N}(C_1,\ldots,C_k) = \frac{1}{L!} \sum_{\lambda \vdash L} \frac{\prod_{i = 1}^k |C_i|\chi_{\lambda}(C_i)}{\chi_{\lambda}({\rm id})^{k - 2}}.
$$
\end{lemma}

Since $G_{N,L}(\beta)$ only depends on the conjugacy class of $\beta$, we compute:
\bea
\tilde{G}_{N,L}(C,\beta) & = & \frac{1}{|[\beta]|} \sum_{\mu \vdash L} \mathcal{N}(C,C_{\mu},[\beta])\,G_{N,L}(C_{\mu}) \nonumber \\ 
& = & \sum_{\mu,\lambda,\lambda' \vdash L} \bigg(\frac{|C|\chi_{\lambda'}(C)\,|C_{\mu}|\chi_{\lambda'}(C_{\mu}) \chi_{\lambda'}(\beta)}{L!\,\chi_{\lambda'}({\rm id})}\bigg)\,\frac{\chi_{\lambda}({\rm id})^2 \chi_{\lambda}(C_{\mu})}{s_{\lambda}(1_N)\,L!^2}. \nonumber
\eea 
The orthogonality of characters of the symmetric group gives
$$
\frac{1}{L!} \sum_{\mu \vdash L} |C_{\mu}|\,\chi_{\lambda'}(C_{\mu})\chi_{\lambda}(C_{\mu}) = \delta_{\lambda,\lambda'}.
$$
Therefore
$$
\tilde{G}_{N,L}(C,\beta) = \frac{1}{L!^2} \sum_{\lambda \vdash L} |C|\chi_{\lambda}(C) \chi_{\lambda}(\beta)\,\frac{\chi_{\lambda}({\rm id})}{s_{\lambda}(1_N)}.
$$
The claim in terms of Schur functions is found by performing the sum over conjugacy classes $C$ in \eqref{eq:34} with the help of \eqref{SWform}. \hfill $\Box$

\subsubsection{Dependence in $N$}

In Theorem~\ref{UNmoment},  the only dependence in the matrix size $N$ comes from the denominator. For a cell  $(i,j)$ in a Young diagram $\mathbb{Y}_{\lambda}$, let ${\rm hook}_{\lambda}(i,j)$ be the hook length at $(i,j)$, where $i = 1,\ldots,\ell(\lambda)$ is the row index and $j = 1,\ldots,\lambda_i$ is the column index. We have the following hook-length formulas, see e.g. \cite{Fultonrep}
\bea
\label{dimS} \chi_{\lambda}({\rm id}) & = & \frac{L!}{\prod_{(i,j) \in \mathbb{Y}_{\lambda}} {\rm hook}_{\lambda}(i,j)}, \\
\label{dimU} s_{\lambda}(1_N) & = & \prod_{(i,j) \in \mathbb{Y}_{\lambda}} \frac{(N + j - i)}{{\rm hook}_{\lambda}(i,j)}.
\eea
Therefore
$$
\tilde{G}_{N,L}(C,\beta) = \frac{1}{L!} \sum_{\lambda \vdash L} \frac{|C|\chi_{\lambda}(C)\chi_{\lambda}(\beta)}{\prod_{(i,j) \in \mathbb{Y}_{\lambda}} (N + j - i)}.
$$

\vspace{0.2cm}

The specialization of formula \eqref{SWform} gives another expression of $s_{\lambda}(1_N)$, and thus of $\tilde{G}_{N,L}(C,\beta)$:
$$
s_{\lambda}(1_N) = \frac{N^{L}\chi_{\lambda}({\rm id})}{L!}\bigg(1 + \sum_{\substack{\mu \vdash L \\ C_{\mu} \neq [1]}} N^{-t(C_{\mu})}\,\frac{|C_{\mu}|\chi_{\lambda}(C_{\mu})}{\chi_{\lambda}({\rm id})}\bigg),
$$
where $t(C) = L - \ell(C)$. We obtain that
$$
\tilde{G}_{N,L}(C,\beta) = \frac{N^{-L}}{L!} \sum_{\lambda \vdash L} \sum_{k \geq 0} (-1)^k \sum_{\substack{\mu_1,\ldots,\mu_k \vdash L \\ C_{\mu_i} \neq [1]}}  \frac{|C|\chi_{\lambda}(C)\chi_{\lambda}(\beta)\prod_{i = 1}^k N^{-t(C_{\mu_i})}|C_{\mu_i}|\chi_{\lambda}(C_{\mu_i})}{\chi_{\lambda}({\rm id})^k}.
$$
If we introduce
$$
A^{(d)}_{L,k}(C,\beta) =\frac{1}{|[\beta]|} \sum_{\substack{\mu_1,\ldots,\mu_k \vdash L \\ \sum_i t(C_{\mu_i}) = d \\ t(C_{\mu_i}) > 0}} \mathcal{N}(C,[\beta],C_{\mu_1},\ldots,C_{\mu_k}),
$$
we can write $\tilde{G}_{N,L}(C,\beta)$ in a compact way
\beq
\label{exptildeG} \tilde{G}_{N,L}(C,\beta) = \sum_{d \geq 0} N^{-(L + d)}\bigg( \sum_{k = 0}^d (-1)^k A^{(d)}_{L,k}(C,\beta)\bigg).
\eeq
Recall that $\varphi^{\partial}$ has $n$ cycles. A priori, $\tilde{G}_{N,L}(C,\varphi^{\partial}) \in O(N^{-L})$, but in fact there are stronger restrictions:
\begin{lemma}
We have a large $N$ expansion of the form:
$$
\tilde{G}_{N,L}(C,\varphi^{\partial}) = \sum_{g \geq 0} N^{-(L + \ell(C) - n + 2g)}\tilde{G}^{(g)}_{L}(C,\varphi^{\partial}).
$$
where $\tilde{G}^{(g)}_{L}$ does not depend on $N$.
\end{lemma}
\noindent\textbf{Proof.} The argument follows \cite{CollinsWeingarten}. $A^{(d)}_{L,k}(C,\varphi^{\partial})$ counts the number of permutations $\tau,\beta_1,\ldots,\beta_k\in \mathfrak{S}_L$ such that $\tau \in C$, $\beta_i \neq {\rm id}$, $\sum_{i = 1}^k t(\beta_i) = d$ and 
\beq
\label{prodperm} \tau \circ \varphi^{\partial}  \circ \beta_1 \circ \cdots \circ \beta_k = {\rm id}.
\eeq
Note that $|t(\sigma) - t(\sigma')| \leq t(\sigma\sigma') \leq t(\sigma) + t(\sigma')$ and thus,
$$
|\ell(C) - n| = |t(\varphi^{\partial}) - t(\tau)| \leq t(\beta_1\cdots \beta_k) \leq \sum_{i = 1}^k t(\beta_i) = d.
$$
Therefore, the coefficient of $N^{-(L + d)}$ in \eqref{exptildeG} is zero unless $d \geq |\ell(C) - n|$. A fortiori we must have $d \geq \ell(C) - n$. Also, computing the signature of \eqref{prodperm} we must have
$$
(-1)^{(L - n) + (L - \ell(C)) + \sum_{i} t(\beta_i)} = 1,
$$
i.e. $n - \ell(C) + d$ is even. We get the claim by calling this even integer $2g$.
\hfill $\Box$

\subsection{Transition matrix via monotone Hurwitz numbers}

We dispose of a general theory relating representation theory, Hurwitz numbers and 2d Toda tau hierarchy, which was pioneered by Okounkov~\cite{Okounkov} and to which many authors contributed. For instance, it is clearly exposed in \cite{WorkOkounkov,HarnadGuay}. It relies on three isomorphic descriptions of the vector space $\mathcal{B}$: as the ring of symmetric functions in infinitely many variables, the direct sum of the centers of the group algebras of the symmetric groups, and the charge~$0$ subspace of the Fock space (aka semi-infinite wedge). After reviewing the aspects of this theory which are relevant for our purposes, we apply it in Section~\ref{SecM} to obtain a nicer form of Proposition~\ref{wein}, namely expressing the transition matrix between ordinary and fully simple observables in terms of monotone Hurwitz numbers.

We refer the reader to the Section \ref{IntroHNs} in the introduction of the thesis for a review on the different characterizations of Hurwitz numbers and their relation to an action of the center of the symmetric group algebra on itself via Jucys-Murphy elements.

\subsubsection{Hypergeometric tau-functions}

We consider Frobenius' characteristic map
$$
\Phi\,:\,\bigoplus_{L \geq 0} Z(\mathbb{Q}[\mathfrak{S}_{L}]) \longrightarrow \mathcal{B}
$$
defined by
$$
\Phi(\hat{C}_{\lambda}) = \frac{p_{\lambda}}{|{\rm Aut}\,\lambda|} = \frac{|C_{\lambda}|\,p_{\lambda}}{L!},\qquad |\lambda| = L.
$$
This map is linear and it is a graded isomorphism -- namely it sends $Z(\mathbb{Q}[\mathfrak{S}_{L}])$ to $\mathcal{B}^{(L)}$. This definition together with the formula \eqref{SWform} and the formula for change of basis \eqref{changebasis} imply that
$$
\Phi(\hat{\Pi}_{\lambda}) = \frac{\chi_{\lambda}({\rm id})}{L!}\,s_{\lambda}.
$$
The action of $Z(\mathbb{Q}[\mathfrak{S}_{L}])$ on itself by multiplication can then be assembled into an action of $\mathcal{B}$ on itself. Concretely, if $r \in \mathcal{B}$ this action is given via Jucys-Murphy elements $\hat{J}$ by
$$
r(\hat{J}) := \Phi \circ\bigg(\bigoplus_{L \geq 0} r\big((\hat{J}_{k})_{k = 2}^L,0,0,\ldots\big)\bigg)\circ \Phi^{-1}.
$$

\begin{definition} 
A {\it hypergeometric tau-function} is an element of $\mathcal{B} \hat{\otimes} \mathcal{B}$ of the form $\sum_{\lambda} A_{\lambda}\,s_{\lambda} \otimes s_{\lambda}$ for some scalar-valued $\lambda \mapsto A_{\lambda}$ function which is a content function.
\end{definition}

\begin{remark} A \emph{2d Toda tau-function} is an element of $\mathcal{B} \hat{\otimes} \mathcal{B}$ which satisfies the Hirota bilinear equations -- these are the analog of Pl\"ucker relations in the Sato Grassmannian. It is known that, if $A$ is a content function, $\sum_{\lambda}  A_{\lambda} s_{\lambda} \otimes s_{\lambda}$ is a 2d Toda tau-function \cite{OrlovSch,Carrell}. We adopt here the name ``hypergeometric'' coined by Harnad and Orlov for those particular 2d Toda tau-functions. Let us mention there exist 2d Toda tau-functions which are diagonal in the Schur basis but with coefficients which are not content functions. 
\end{remark}

We can identify $\mathcal{B}\hat{\otimes} \mathcal{B}$ with the ring of symmetric functions in two infinite sets of variables $\mathbf{z}=(z_1,z_2,\ldots)$ and $\widetilde{\mathbf{z}}=(\widetilde{z}_1,\widetilde{z}_2,\ldots)$.
There is a trivial hypergeometric tau-function:
$$
\mathcal{T}_{\emptyset} := \exp\Big(\sum_{k \geq 1} \frac{1}{k}\,p_{k}(\mathbf{z})\,p_{k}(\widetilde{\mathbf{z}})\Big) = \prod_{i,j = 1}^{\infty} \frac{1}{1 - z_{i}\widetilde{z}_{j}} = \sum_{\lambda} s_{\lambda}(\mathbf{z})\,s_{\lambda}(\widetilde{\mathbf{z}}) = \sum_{\mu} \frac{1}{|{\rm Aut}\,\mu|}\,p_{\mu}(\mathbf{z})\,p_{\mu}(\widetilde{\mathbf{z}}),$$
where the two last sums are over all partitions and for the equality in the middle we have used Cauchy-Littlewood formula \cite[Chapter 1]{Macdonald}. 

An element $r \in \mathcal{B}$ acts on the set of hypergeometric tau-functions by action on the first factor via $r(\hat{J}) \otimes {\rm Id}$.  More concretely, the action on $\mathcal{T}_{\emptyset}$ reads
\beq
\label{TauQ2} \mathcal{T}_{r} = \sum_{\lambda} r({\rm cont}\,\lambda)\,s_{\lambda} \otimes s_{\lambda} = \sum_{L \geq 0} \sum_{|\lambda| = |\mu| = L} R_{\lambda,\mu}\,p_{\lambda}\otimes p_{\mu}\,,
\eeq
where $R_{\lambda,\mu}$ are double Hurwitz numbers, which we introduces in Section~\ref{doubleHNsIntro}.

\subsubsection{Main result}
 \label{SecM}
We prove that the transition matrix from ordinary to fully simple expectation values is given by double weakly monotone Hurwitz numbers $[H_{k}]_{\lambda,\mu}$ (with signs), while the transition matrix from fully simple to ordinary is given by the double strictly monotone Hurwitz numbers $[E_{k}]_{\mu,\lambda}$.

\begin{theorem}
\label{Transi}With respect to any $\mathcal{U}_{N}$-invariant measure on the space $\mathcal{H}_N$ of $N \times N$ hermitian matrices, we obtain
\bea
\frac{\big\langle \mathcal{P}_{\lambda}(M)\big\rangle}{|{\rm Aut}\,\lambda|} & = & \sum_{\mu \vdash |\lambda|} N^{-|\mu|} \Big(\sum_{k \geq 0} (-N)^{-k} [H_{k}]_{\lambda,\mu}\Big)\big\langle p_{\mu}(M) \big\rangle, \nonumber \\
\frac{\big\langle p_{\mu}(M) \big\rangle}{|{\rm Aut}\,\mu|} & = & \sum_{\lambda \vdash |\mu|} N^{|\lambda|} \Big(\sum_{k \geq 0} N^{-k}[E_{k}]_{\mu,\lambda}\Big) \big\langle \mathcal{P}_{\lambda}(M) \big\rangle, \nonumber 
\eea
where $[E_k]_{\mu,\lambda}$ (resp. $[H_k]_{\lambda,\mu}$) are the double Hurwitz numbers related to the elementary symmetric (resp. complete symmetric) polynomials.
\end{theorem}

\noindent \textbf{Proof.} We introduce an auxiliary diagonal matrix $\widetilde{M}$ and deduce from Proposition \ref{wein} and Equation~\eqref{SWform} that
\beq
\label{idgen} \frac{1}{L!} \sum_{\lambda \vdash L} |C_{\lambda}|\, p_{\lambda}(\widetilde{M}) \big\langle \mathcal{P}_{\lambda}(M) \big\rangle = \sum_{\mu \vdash L} \frac{\chi_{\mu}({\rm id})}{L!\,s_{\mu}(1_N)}\,s_{\mu}(\widetilde{M})\,\big\langle s_{\mu}(M)\big\rangle.
\eeq
The formulas \eqref{dimS}-\eqref{dimU} show that $\frac{\chi_{\mu}({\rm id})}{L!\,s_{\mu}(1_N)}$ is a content function coming from the complete symmetric polynomials\footnote{We would like to remark that we discovered a posteriori a result of Novak \cite[Theorem 1.1]{NovakJMelements} phrasing our \eqref{contentfct} in an interesting form, but we keep using our formula here since we find it more elementary.}
\beq\label{contentfct}
\frac{\chi_{\mu}({\rm id})}{L!\,s_{\mu}(1_N)} = \prod_{(i,j) \in \mathbb{Y}_{\mu}} \frac{1}{N + {\rm cont}(i,j)} = N^{-|\mu|} \sum_{k \geq 0} (-N)^{-k}\,h_k({\rm cont}\,\mu).
\eeq
We denote $r_N$ the corresponding element of $\mathcal{B}[[N^{-1}]]$. The identity \eqref{idgen} then translates into:
\beq\label{twistedtau}
 \frac{1}{L!} \sum_{\lambda \vdash L}|C_{\lambda}|\, p_{\lambda}(\widetilde{M})\,\big\langle \mathcal{P}_{\lambda}(M)\big\rangle
=  \sum_{\mu \vdash L}  r_N({\rm cont}\,\,\mu)\,s_{\mu}(\widetilde{M})  \big\langle s_{\mu}(M) \big\rangle.
\eeq
To interpret these expressions as $r_N$ acting on $\mathcal{T}_{\emptyset}$ as in \eqref{TauQ2}, we remind that $\mathcal{T}$ in $\mathcal{B} \hat{\otimes} \mathcal{B}$ can be seen as a function of two sets of infinitely many variables $\Lambda$ and $\widetilde{\Lambda}$. Moreover, we consider $\mathcal{T}$ evaluated at two matrices $M$ and $\widetilde{M}$ of size $N$, by substituting $\Lambda$ (resp. $\widetilde{\Lambda}$) by the set of $N$ eigenvalues of $M$ (resp. $\widetilde{M}$) completed by infinitely many zeros. We then write $\mathcal{T}(M,\widetilde{M})$ to stress that we have a function of two matrices. In this way, we identify the summation of \eqref{twistedtau} over $L \geq 0$ with $\langle \mathcal{T}_{r_N}(M,\widetilde{M})\rangle$, where the expectation value is taken with respect to any unitarily-invariant measure on $M$ -- while $\widetilde{M}$ is a matrix-valued parameter. 

Now comparing with \eqref{TauQ2}, we find that
\beq
\label{2ndequation} \frac{\big\langle \mathcal{P}_{\lambda}(M) \big\rangle}{|{\rm Aut}\,\lambda|} = \sum_{\mu \vdash |\lambda|} (R_N)_{\lambda,\mu}\, \big\langle  p_{\mu}(M) \big\rangle,
\eeq
which yields the first formula we wanted to prove. To obtain the second formula, we observe that
$$
s_N({\rm cont}\,\,\lambda) = \prod_{(i,j) \in \mathbb{Y}_{\lambda}} (N + {\rm cont}(i,j)) = N^{|\lambda|} \sum_{k \geq 0} N^{-k}\,e_k({\rm cont}\,\,\lambda)
$$
defines an element $s_N \in \bigoplus_{d \geq 0} \mathcal{B}^{(d)} \otimes (N^{d}\cdot\mathbb{C}[[N^{-1}]])$, which is inverse to $r_N$ in $\mathcal{B}[N,N^{-1}]]$. We denote $[S_{N}]_{\lambda,\mu}$ the Hurwitz numbers it determines via \eqref{rjC}. 
If we act by $s_N(\hat{J})$ on $\mathcal{T}_{r_N}$, we recover the trivial tau-function:
\beq
\label{T222}\langle \mathcal{T}_{\emptyset}(\widetilde{M},M) \rangle = \sum_{\mu} \frac{p_{\mu}(\widetilde{M})}{|{\rm Aut}\,\mu|}\,\langle p_{\mu}(M) \rangle.
\eeq
On the other hand, representing this action in the power sum basis using \eqref{TauQ2} and \eqref{twistedtau} yields
\beq
\label{T222a}\langle \mathcal{T}_{\emptyset}(\widetilde{M},M) \rangle =\sum_{L \geq 0} \sum_{|\lambda| = |\mu| = L} [S_{N}]_{\mu,\lambda}\,  p_{\mu}(\widetilde{M}) \langle \mathcal{P}_{\lambda}(M) \rangle.
\eeq
Finally, we can identify the coefficients of \eqref{T222} and \eqref{T222a} to obtain the desired formula. \hfill $\Box$

\vspace{0.15cm}

With our proof, we obtain some intermediate formulas that will be useful later. However, for the derivation of Theorem \ref{Transi}, it is not crucial to write our generating series in the form of $\tau$-functions. Using Proposition \ref{wein}, \eqref{contentfct} and the expression \eqref{characters_def} for Hurwitz numbers, Theorem \ref{Transi} is straightforward. Apart from the reason already mentioned, we also consider it is interesting to illustrate the relation with the world of $\tau$-functions.

\subsection{Relation with the matrix model with external field}\label{HMMEF}

The Itzykson-Zuber integral \cite{IZ} is a function of an integer $N$ and two matrices $A$ and $B$ of size $N$ defined by
\beq
\label{IZint} \mathcal{I}_{N}(A,B) := \int_{U(N)} \dd U\,\exp\big[N\,{\rm Tr}(AUBU^{\dagger})\big],
\eeq
where $\dd U$ is the Haar measure on $U(N)$ normalized to have mass $1$. It admits a well-known expansion in terms of characters of the unitary group, i.e. Schur functions:
\begin{theorem} \cite[Eq. (4.6)]{Balentekin}\label{balen}
$$
\mathcal{I}_{N}(A,B) = \sum_{\lambda} \frac{N^{|\lambda|}\chi_{\lambda}({\rm id})}{L!\,s_{\lambda}(1_{N})}\,s_{\lambda}(A)\,s_{\lambda}(B).
$$
\end{theorem}
For any unitarily invariant measure $\mu$ on $\mathcal{H}_{N}$, we define$$
\check{Z}(A) = \int_{\mathcal{H}_{N}} \dd\mu(M)e^{N\,{\rm Tr}(AM)}.
$$
\begin{corollary}
\label{gudn}We denote $\langle \cdot \rangle$ the expectation value and $\kappa_n(\cdot)$ the n-th order cumulant with respect to any unitarily invariant measure $\mu$ on $M\in\mathcal{H}_{N}$.
We have the formulas
\bea  
\frac{\check{Z}(A)}{\check{Z}(0)} & = & 1+ \sum_{\lambda\neq \emptyset} \frac{|C_{\lambda}|}{|\lambda|!}\,N^{|\lambda|} p_{\lambda}(A) \big\langle\mathcal{P}_{\lambda}(M)\big\rangle = \big\langle \mathcal{I}_{N}(A,M) \big\rangle, \nonumber \\
\ln\Big(\frac{\check{Z}(A)}{\check{Z}(0)}\Big) & = & \sum_{n \geq 1}\frac{1}{n!} \sum_{\ell_1,\ldots,\ell_n \geq 1} N^{L}\, \kappa_n
\big(\mathcal{P}^{(\ell_1)}(M),\ldots,\mathcal{P}^{(\ell_n)}(M)\big)\,\prod_{i = 1}^n \frac{p_{\ell_i}(A)}{\ell_i}, \nonumber
\eea  
where we recall that the corresponding expectation value is zero whenever $|\lambda|$ or $L\coloneqq\sum_{i} \ell_i$ exceeds $N$.
\end{corollary}
\noindent \textbf{Proof.} Comparing Theorem \ref{balen} with \eqref{twistedtau} gives the first line. We introduce the factor which allows us to go from (ordered) tuples $(\ell_1,\ldots,\ell_n)$ with $L\coloneqq\sum_{i=1}^n \ell_i$ to (unordered) partitions of $L$:
\beq\label{g}
g_{\lambda}\coloneqq \frac{\ell(\lambda)!}{\prod_{i=1}^L m_i(\lambda) !} = \frac{\ell(\lambda)!\, |C_{\lambda}|\, \prod_{i=1}^{\ell(\lambda)} \lambda_i}{|\lambda| !},
\eeq
where $L = \sum_{i=1}^L i\, m_i(\lambda) = \sum_{i=1}^n \lambda_i$. If we replace now the sum over partitions by the sum over tuples of positive integers multiplying by $\frac{1}{g_{\lambda}}$, we find
\bea
\frac{\check{Z}(A)}{\check{Z}(0)} & = & 1 + \sum_{n \geq 1} \frac{1}{n!} \sum_{\ell_1,\ldots,\ell_n \geq 1}  N^{L}\,\big\langle \mathcal{P}^{(\ell_1)}(M)\cdots \mathcal{P}^{(\ell_n)}(M)\big\rangle\,\prod_{i = 1}^n\frac{p_{\ell_i}(A)}{\ell_i} \nonumber \\ 
& = & \left\langle \exp\left(\sum_{i\geq 1} \frac{N^{\ell_i} p_{\ell_i}(A)}{\ell_i} \mathcal{P}^{(\ell_i)}(M)\right) \right\rangle. \nonumber
\eea
Taking the logarithm gives precisely the cumulant generating series as in the second formula.
\hfill  $\Box$

\vspace{0.2cm}

In other words, the fully simple observables for the matrix model $\mu$ are naturally encoded in the corresponding matrix model with an external field $A$.  Compared to Theorem~\ref{Transi}, this result is in agreement with the combinatorial interpretation of the Itzykson-Zuber integral in terms of double monotone Hurwitz numbers \cite{HCIZ}.

\section{Counting maps again}\label{MMs}\label{Matcomb}

The relation between ordinary and fully simple observables through monotone Hurwitz numbers is universal in the sense that it does not depend on the unitarily invariant measure considered. This section is devoted to the relation between matrix models and the enumeration of maps for a specific unitarily invariant measure. This relation is well-known for ordinary maps, as we explained in the Section \ref{IntroMMs} of the introduction. Here we give a detailed derivation directly gluing polygons, that is working with maps, in contrast to the classical derivation in physics which is in terms of gluing stars, that is working on the dual. As we commented in the introduction, we include this calculation to give a different (but completely equivalent) detailed derivation which will also make clearer the refinement of the argument that we need to provide a matrix model for fully simple maps. This specialization motivated our study of the general ordinary and fully simple observables.

We introduce the Gaussian probability measure on the space $\mathcal{H}_N$ of $N \times N$ hermitian matrices:
$$
\dd\mu_0(M) = \frac{\dd M}{Z_0}\,e^{-N\mathrm{Tr}\,\frac{M^2}{2}},\qquad Z_0 = \int_{\mathcal{H}_{N}} \dd M\,e^{-N\mathrm{Tr}\,\frac{M^2}{2}},$$
and the generating series:
\bea
\tilde{T}_{h,k}(w_1,\ldots,w_k) &=& \sum_{m_1,\ldots,m_k \geq 1} \frac{t^h_{m_1,\ldots,m_k}}{m_1\cdots m_k}\,\prod_{i = 1}^k w_i^{m_i},\nonumber \\
T_{h,k}(w_1,\ldots,w_k) &= &\tilde{T}_{h,k}(w_1,\ldots,w_k) - \delta_{h,0}\delta_{k,1}\frac{w_1^2}{2}. \nonumber
\eea
We consider the formal measure
\beq
\label{eq:mu0}\dd\mu(M) = \frac{\dd M}{Z_0}\,\exp\left(\sum_{h\geq 0,k\geq 1} \frac{N^{2 - 2h - k}}{k!} \mathrm{Tr}\,T_{h,k}\left(M^{(1)}_k,\ldots,M^{(k)}_k\right)\right),
\eeq
where $M^{(i)}_k\coloneqq\bigotimes_{j=1}^{i-1} I_N \otimes M \otimes \bigotimes_{j=i+1}^{k} I_N$, in the sense that the expectation value of any polynomial function of $M$ with respect to this measure is defined as a formal series in the $t$'s.

For a combinatorial map $(\sigma, \alpha)$, we consider here a special structure for the set of half-edges $H= H^u \sqcup H^{\partial}$ which will be convenient for our derivation:
$$
H^{\partial} = \bigsqcup_{i = 1}^n \{i\} \times (\mathbb{Z}/L_i\mathbb{Z}),\qquad H^u = \bigsqcup_{m = 1}^r \{m\} \times (\mathbb{Z}/k_m\mathbb{Z}).
$$
The permutation $\varphi \coloneqq (\sigma\circ\alpha)^{-1}$ acting on $H$, whose cycles correspond to faces of the map, is hence given by $\varphi((i, l))=(i,l+1)$. With this special structure of the set of half-edges, counting the number of relabelings of $H^u$ amounts to choosing an order of the unmarked faces and a root for each of them. Therefore
$$
{\rm Rel}(\sigma,\alpha) = r!\,\prod_{m= 1}^r k_m.
$$

\subsection{Ordinary usual maps}
\label{Maps}
Consider first the case where $T_{h,k} = 0$ for $(h,k) \neq (0,1)$, i.e.
\beq
\label{eq:mes}\dd\mu(M) = \frac{\dd M}{Z_{\rm GUE}}\,\exp\left\{N\,{\rm Tr}\Big(-\frac{M^2}{2} + \sum_{k \geq 1} \frac{t_k\,M^k}{k}\Big)\right\},\qquad Z = \int_{\mathcal{H}_N} \dd\mu(M).
\eeq
We denote $\langle \mathcal{\cdot} \rangle_{{\rm GUE}}$ the expectation value with respect to the Gaussian measure $\dd\mu_0$. The matrix elements have covariance:
\beq
\label{contrac} \langle M_{a,b}M_{c,d} \rangle_{{\rm GUE}} = \frac{1}{N}\delta_{a,d}\delta_{b,c}.
\eeq

Let $(L_i)_{i = 1}^n$ be a sequence of nonnegative integers, and $L = \sum_{i = 1}^n L_i$. The expectation values with respect to $\dd\mu$ are computed, as formal series in $(t_k)_k$:
\beq\label{eq:ordev}
\Big\langle \prod_{i = 1}^n {\rm Tr}\,M^{L_i} \Big\rangle =  \frac{1}{Z} \sum_{r \geq 0} \sum_{k_1,\ldots,k_r \geq 1} \frac{N^r}{r!} \prod_{m = 1}^r \frac{t_{k_m}}{k_m} \sum_{\substack{1 \leq j_{h} \leq N, \\ h \in H}} \Big\langle \prod_{h\in H} M_{j_h,j_{\varphi(h)}}\Big\rangle_{{\rm GUE}}. 
\eeq

With the help of Wick's theorem for the Gaussian measure and (\ref{contrac}), we obtain:
$$
\Big\langle \prod_{h\in H} M_{j_h,j_{\varphi(h)}}\Big\rangle_{{\rm GUE}} =  N^{-\frac{|H|}{2}} \sum_{\alpha \in \mathfrak{I}_H} \prod_{h\in H} \delta_{j_{h},j_{\alpha(\varphi(h))}}.
$$
where $\mathfrak{I}_H\subset \mathfrak{G}_H$ is the set of all fixed-point free involutions, i.e. all pairwise matchings, on $H$.

We observe that the product on the right hand side is $1$ if $h\mapsto j_h$ is constant over the cycles of $\alpha\circ\varphi$, otherwise it is $0$. Therefore,
$$
\sum_{\substack{1 \leq j_{h} \leq N, \\ h \in H}} \Big\langle \prod_{h\in H} M_{j_h,j_{\varphi(h)}}\Big\rangle_{{\rm GUE}} =  N^{-\frac{|H|}{2}} \sum_{\alpha \in \mathfrak{I}_H} N^{|\mathcal{C}(\alpha\circ\varphi)|}.
$$

To recognize \eqref{eq:ordev} as a sum over combinatorial maps, we let $\alpha\in\mathfrak{I}_H$ correspond to the edges of maps whose faces are given by cycles of $\varphi$, and $\sigma\coloneqq (\alpha\circ\varphi)^{-1}$, whose cycles will correspond to the vertices. Observe that $2|\mathcal{C}(\alpha)|=|H|$.

$$
\Big\langle \prod_{i = 1}^n {\rm Tr}\,M^{L_i} \Big\rangle = \frac{1}{Z} \sum_{(\sigma,\alpha)} \frac{N^{|\mathcal{C}(\sigma)| - |\mathcal{C}(\alpha)| + |\mathcal{C}(\varphi)|-n}}{{\rm Rel}\,(\sigma,\alpha)}\,\prod_{f \in \mathcal{C}(\left.\varphi\right|_{H^u})} t_{\ell(f)},
$$
where the sum is taken over non-connected combinatorial maps $(\sigma,\alpha)$ with $n$ boundaries of lengths $L_1,\ldots,L_n$.

\vspace{0.3cm}

\begin{figure}[h!]
 \begin{center}
        \def\svgwidth{\columnwidth}
       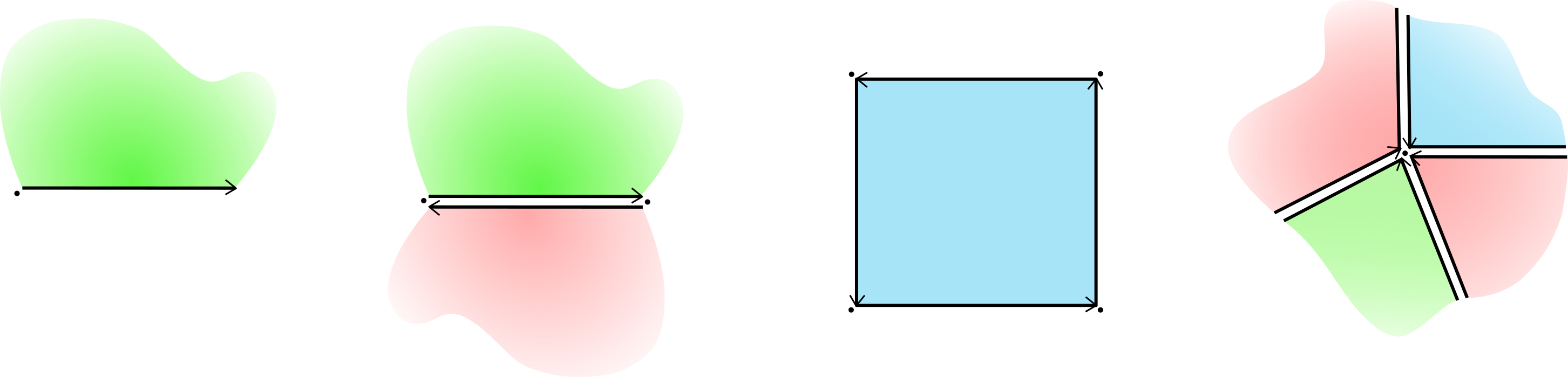
    	\caption{(1) An element of the matrix corresponds to a half-edge with its two legs labelled by the indices. (2) Pairing two matrix entries is represented as two half-edges forming an edge.  (3) Products of elements of that form are the ones appearing in ${\rm Tr}\, M^4$ and are represented by quadrangles. (4) The pairings $\alpha$ that give a non-zero contribution will be the ones that leave the indices invariant in every cycle of $\sigma\coloneqq (\alpha\circ\varphi)^{-1}$, i.e. around every vertex.}\label{GraphRep}
\end{center}
\end{figure}

\vspace{-0.5cm}


To transform this sum over combinatorial maps into a generating series for (unlabeled) maps, we have to multiply by the number of combinatorial maps which give rise to the same unlabeled combinatorial map $[(\sigma,\alpha)]$, which is ${\rm Gl}(\sigma,\alpha)= \frac{{\rm Rel}(\sigma,\alpha)}{|{\rm Aut}(\sigma,\alpha)|}$, as we explained in section \ref{aut}.

A similar computation can be done separately for $Z$, and we find it is the generating series of maps with empty boundary. The contribution of connected components without boundaries factorizes in the numerator and, consequently,
$$
\Big\langle \prod_{i = 1}^n {\rm Tr}\,M^{L_i} \Big\rangle = \sum_{\substack{\partial \text{-connected}\  \mathcal{M} = [(\sigma,\alpha)] \\ \text{with } \partial \text{ lengths}\,\,(L_i)_{i = 1}^n}} \frac{N^{\chi(\sigma,\alpha)}}{|{\rm Aut}(\sigma,\alpha)|}\,\prod_{f \in \mathcal{C}(\left.\varphi\right|_{H^u})}t_{\ell(f)}.
$$
We remark that the power of $N$ sorts maps by their Euler characteristic. 
Finally, a standard argument shows that taking the logarithm for closed maps or the cumulant expectation values for maps with boundaries, we obtain the generating series of connected maps:

\begin{proposition}\cite{Eynardbook}
$$
\ln Z = \sum_{g \geq 0} N^{2-2g} F^{[g]}, \quad \quad
\kappa_n({\rm Tr}\,M^{L_1},\ldots,{\rm Tr}\,M^{L_n}) = \sum_{g \geq 0} N^{2-2g-n} F^{[g]}_{L_1,\ldots,L_n}.
$$
\end{proposition}
This kind of results appeared first for planar maps in \cite{BIPZ}, but for a modern and general exposition see also \cite{Eynardbook}.

\subsection{Ordinary stuffed maps}

For the general formal measure
\bea
\label{eq:mesS}\dd\mu(M) & = & \frac{\dd M \,e^{-N\,{\rm Tr} \frac{M^2}{2}}}{Z_{\rm GUE}}\,\exp\left(\sum_{h\geq 0,k\geq 1} \frac{N^{2 - 2h - k}}{k!} \mathrm{Tr}\,\tilde{T}_{h,k}\left(M^{(1)}_k,\ldots,M^{(k)}_k\right)\right)  \\
& = & \frac{\dd M \,e^{-N\,{\rm Tr} \frac{M^2}{2}}}{Z_{\rm GUE}}\,\exp\left(\sum_{h\geq 0,k\geq 1} \frac{N^{2 - 2h - k}}{k!} N^{2 - 2h - k}\sum_{m_1,\ldots,m_k \geq 1} t^h_{m_1,\ldots,m_k}\,\prod_{i = 1}^k \frac{\mathrm{Tr}\, M^{m_i}}{m_i}\right), \nonumber
\eea
the expectation values are generating series of stuffed maps. We denote here $Z_S$, $\langle \cdot \rangle_S$ and $\kappa_n(\cdot)_S$ the partition function, the expectation value and the n-th order cumulant expectation values with respect to this general measure to distinguish these more general expressions from the previous ones.

A generalization of the technique reviewed in \S~\ref{Maps} shows that
$$
\big\langle {\rm Tr}\,M^{L_1} \cdots {\rm Tr}\,M^{L_n} \big\rangle_S = \sum_{\substack{\partial\text{-connected stuffed map } \mathcal{M}  \\ {\rm with}\, \partial \,  {\rm lengths} \,(L_i)_{i = 1}^n}} \frac{N^{\chi(\mathcal{M})}}{|{\rm Aut}(\mathcal{M})|}\,\prod_{p = 1}^F t^{h_p}_{(\ell(c))_{c\in f_p}}.
$$
As in the previous subsection, taking the logarithm or the cumulant expectation values in absence or presence of boundaries respectively, we obtain the generating series of connected stuffed maps:
\begin{proposition}\cite{Bstuff}\label{propStuffed}
$$
\ln Z_S = \sum_{g \geq 0} N^{2-2g} \widehat{F}^{[g]}, \quad \quad
\kappa_n({\rm Tr}\,M^{L_1},\ldots,{\rm Tr}\,M^{L_n})_S = \sum_{g \geq 0} N^{2-2g-n} \widehat{F}^{[g]}_{L_1,\ldots,L_n}.
$$
\end{proposition}

\subsection{Fully simple maps}

Let $(\gamma_i)_{i = 1}^n$ be pairwise disjoint cycles:
\beq
\label{gammapre}\gamma_i = (j_{i,1} \rightarrow j_{i,2} \rightarrow \cdots \rightarrow j_{i,L_i})
\eeq
and $L = \sum_{i = 1}^n L_i$.
We want to compute $\big\langle \prod_{i = 1}^n \mathcal{P}_{\gamma_i}(M) \big\rangle$ and the idea is that only fully simple maps will make a non-zero contribution, so we will be able to express it as a generating series for fully simple maps. Let us describe this expression for the measure \eqref{eq:mes} in terms of maps. Repeating the steps of \S~\ref{Maps}, we obtain 
\bea
\Big\langle \prod_{i = 1}^n \mathcal{P}_{\gamma_i}(M)  \Big\rangle & = &  \frac{1}{Z} \sum_{r \geq 0} \sum_{k_1,\ldots,k_r \geq 1} \frac{N^r}{r!} \prod_{m = 1}^r \frac{t_{k_m}}{k_m} \sum_{\substack{1 \leq j_{h} \leq N, \\ h \in H^u}} \Big\langle \prod_{h\in H} M_{j_h,j_{\varphi(h)}}\Big\rangle_{0} \nonumber\\
& = &  \frac{1}{Z} \sum_{r \geq 0} \sum_{k_1,\ldots,k_r \geq 1} \frac{N^r}{r!} \prod_{m = 1}^r \frac{t_{k_m}}{k_m} N^{\frac{-|H|}{2}} \sum_{\substack{1 \leq j_{h} \leq N, \\ h \in H^u}}  \sum_{\alpha \in \mathfrak{I}_H} \prod_{h\in H} \delta_{j_{h},j_{\alpha(\varphi(h))}}. \nonumber 
\eea

We consider as before the permutation $\sigma \coloneqq (\alpha\circ\varphi)^{-1}$ which will correspond to the vertices of the maps. The difference with \eqref{eq:ordev} lies in the summation over indices $j_{h}$ between $1$ and $N$ only for $h \in H^u$, while $j_{h}$ for $h = (i,l) \in H^{\partial}$ is prescribed by \eqref{gammapre}. As $j_{i,l}$ are pairwise distinct, the only non-zero contributions to the sum will come from maps for which $(i,l) \in H^{\partial}$ belong to pairwise distinct cycles of $\sigma$ and this is the characterization for fully simple maps in the permutational model setting. The function $h \mapsto j_h$ must be constant along the cycles of $\sigma$, and its value for every $h \in H^{\partial}$ is prescribed by \eqref{gammapre}. So, the number of independent indices of summation among $(j_{h})_{h \in H^u}$ is
$$
|\mathcal{C}(\sigma)| - L.
$$

\vspace{-1.2cm}

\begin{figure}[h!]
\centering
\begin{subfigure}{.5\textwidth}
  \centering
\def\svgwidth{0.63\columnwidth}
       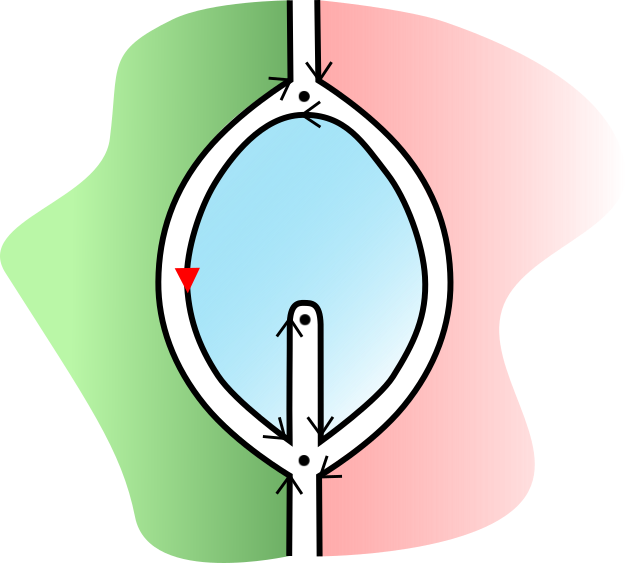
\end{subfigure}%
\begin{subfigure}{.5\textwidth}
  \centering
\def\svgwidth{0.63\columnwidth}
       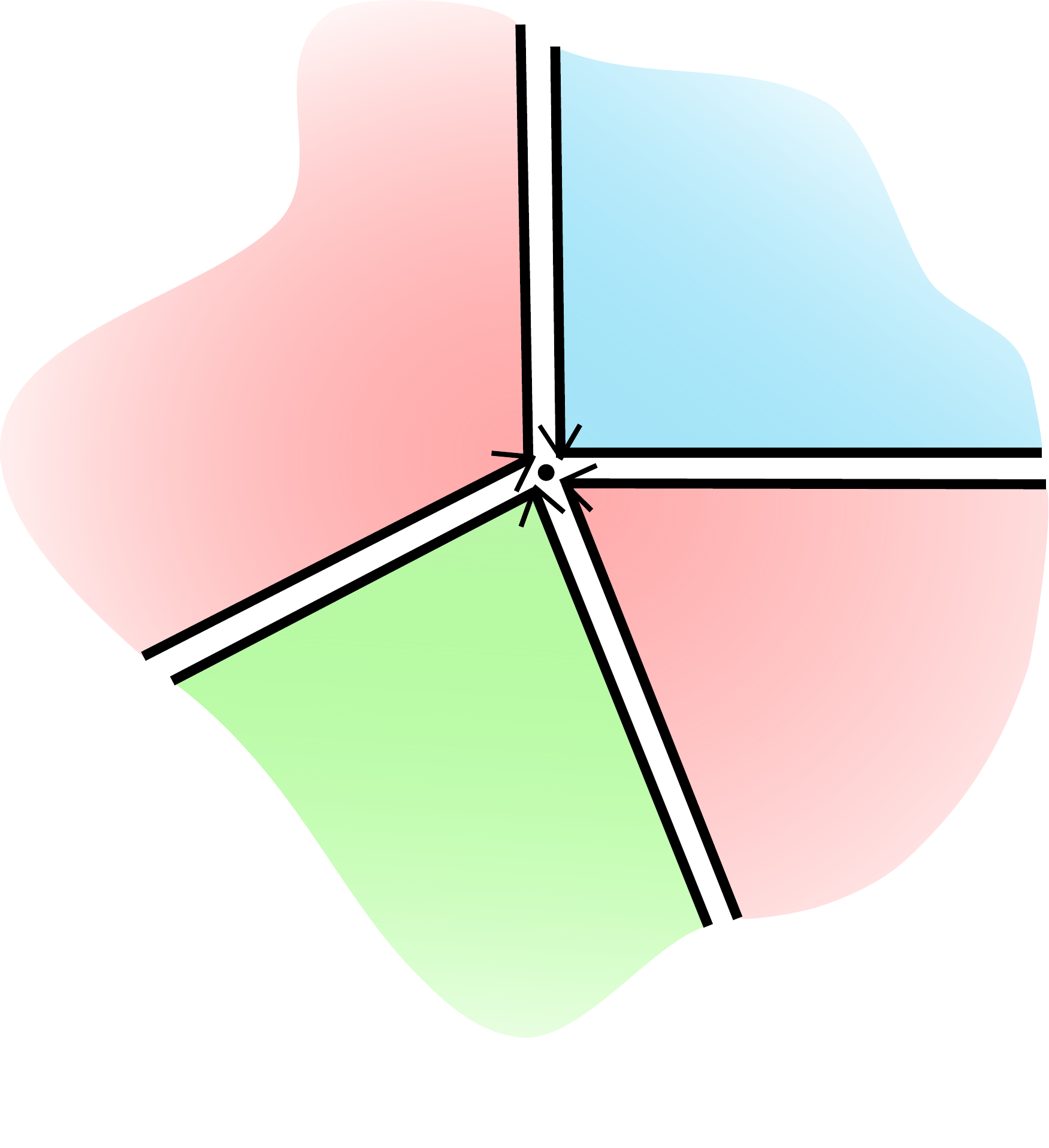
\end{subfigure}
\vspace{0.4cm}

\caption{(1) A non-simple map. (2) A non-fully simple map. These two non-fully maps give a zero contribution because they contain two half-edges belonging to boundaries $(i,l) \in H^{\partial}$ in the same cycle of $\sigma$, i.e. incident to the same vertex (and hence with equal indices $j_{i,l}$ since the indices are constant around a vertex, which is forbidden from the definition of the $\mathcal{P}$'s).}
\label{fig:test}
\end{figure}

Thus,
\bea
\Big\langle \prod_{i = 1}^n \mathcal{P}_{\gamma_i}(M)  \Big\rangle & = & \frac{1}{Z} \sum_{r \geq 0} \sum_{k_1,\ldots,k_r \geq 1} \frac{N^r}{r!} \prod_{m = 1}^r \frac{t_{k_m}}{k_m} N^{-\frac{|H|}{2}} \sum_{\alpha \in \mathfrak{I}_H} N^{|\mathcal{C}(\alpha\circ\varphi)|-L}. \nonumber \\
&= & \sum_{\substack{\partial \text{-connected fully}\\ \text{simple } \mathcal{M} = [(\sigma,\alpha)] \\ \text{with } \partial \text{ lengths}\,\,(L_i)_{i = 1}^n}} \frac{N^{\chi(\sigma,\alpha)- L}}{|{\rm Aut}(\sigma,\alpha)|}\,\prod_{f \in \mathcal{C}(\left.\varphi\right|_{H^u})}t_{\ell(f)}. \nonumber
\eea

The generalization to the measure \eqref{eq:mesS} is straightforward and gives rise to generating series for fully simple stuffed maps:
$$
\Big\langle \prod_{i = 1}^n \mathcal{P}_{\gamma_i}(M)  \Big\rangle_S = \sum_{\substack{\mathcal{M} \\ {\rm fully}\,\,{\rm simple}\,\,{\rm stuffed}\,\,{\rm map} \\ \partial\,{\rm perimeters}\,\,(L_i)_{i = 1}^n \\ \partial {\rm -connected}}} \frac{N^{\chi(\mathcal{M}) - L}}{|{\rm Aut}(\mathcal{M})|}\,\prod_{p = 1}^F t^{h_p}_{(\ell(c))_{c\in f_p}}.
$$

As before, the cumulant expectation values give the generating series of connected fully simple maps and stuffed maps for the more general measure:
\begin{proposition}\label{PPPPP}
\bea
\kappa_n(\mathcal{P}_{\gamma_1}(M) ,\ldots,\mathcal{P}_{\gamma_n}(M) ) & = & \sum_{g \geq 0} N^{2-2g-n-L}  H^{[g]}_{L_1,\ldots,L_n}, \nonumber \\ 
\kappa_n(\mathcal{P}_{\gamma_1}(M) ,\ldots,\mathcal{P}_{\gamma_n}(M) )_S & = & \sum_{g \geq 0} N^{2-2g-n-L} \widehat{H}^{[g]}_{L_1,\ldots,L_n}. \nonumber
\eea
\end{proposition}

\section{Towards a proof of the conjecture for usual maps}
\label{PRof}
\label{Section9}


%

In this section we give a sketch on the ideas towards a proof of the Conjecture~\ref{conj} for usual maps, indicating all the technicalities we skip. We manage to reduce the problem to a technical condition concerning a weaker version of symplectic invariance for the exchange transformation that we introduced in Section~\ref{sympinv}. We confirmed experimentally that this condition is satisfied for some particular case. However, we do not have a justification for this condition to be satisfied in general for the moment.

We believe the further analysis of our problem could help shed some clarity on this fundamental question of TR.

The starting point of our argument\footnote{The idea of this argument first appeared in the derivation of Bouchard-Mari\~{n}o conjecture proposed in \cite{BEMS}. In that article, generating series of simple Hurwitz numbers were represented in terms of a matrix model with external field for a complicated $V$, albeit it was later pointed out by D.~Zvonkine that this representation was ill-defined even in the realm of formal series. This issue is not relevant here as we start with a well-defined matrix model in formal series, and are careful to justify all steps by legal operations within formal series.} is the representation of the generating series of connected fully simple maps as the free energies $\ln \frac{\check{Z}(A)}{\check{Z}(0)}$ of the $1$-hermitian matrix model with external field \cite{Semenoff}. This model is considered here to be valued in formal series. The topological expansion of its correlators satisfies Eynard-Orantin topological recursion, for a well-characterized spectral curve $(\mathcal{S}_{A},x,y)$ \cite{EPf}. On the other hand, the generating series $X_{n}^{[g]}$ we are after are encoded into the $n$-th order Taylor expansion of $\ln \frac{\check{Z}(A)}{\check{Z}(0)}$ around $A = 0$. Using a milder version of symplectic invariance and the properties of topological recursion under deformations of the spectral curve \cite{EORev}, we relate these $n$-th order Taylor coefficients to TR amplitudes of the topological recursion applied to the curve $(\mathcal{S}_{0},y,x)$. As the matrix model $\check{Z}(0)$ generates usual maps, the spectral curve $\mathcal{S}_{0}$ must be the initial data mentioned in Theorem~\ref{TRRRR}.  Unfortunately, our idea of proof is not combinatorial and relies on the symplectic invariance itself.

Prior to applying the result of \cite{EPf}, we give the definition of the topological expansion of $\frac{\check{Z}(A)}{\check{Z}(0)}$ and sketch the computation of the spectral curve from Schwinger-Dyson equations. These aspects are well-known to physicists.

\subsection{The topological expansion}

Corollary~\ref{gudn} together with Proposition~\ref{PPPPP} to access the genus $g$ part yields 
\beq\label{gsHMMEF}
\hat{F}^{[g]}_A = \sum_{n \geq 1}\frac{1}{n!} \sum_{\ell_1,\ldots,\ell_n \geq 1} \frac{p_{\ell_1}(A)}{N\ell_1}\cdots \frac{p_{\ell_n}(A)}{N\ell_n} \kappa_n^{[g]}(\mathcal{P}^{(\ell_1)}(M),\ldots,\mathcal{P}^{(\ell_n)}(M)\big).
\eeq
Recall that we had identified $\kappa_n^{[g]}(\mathcal{P}^{(\ell_1)}(M),\ldots,\mathcal{P}^{(\ell_n)}(M)\big)$ with the generating series of fully simple maps $H^{[g]}_{\ell_1,\ldots,\ell_n}$. We will actually think of $\hat{F}^{[g]}_A$ in general as an element 
$$
\hat{F}^{[g]}_A(q_1,q_2,q_3,\ldots)\in \mathcal{R} :=\mathbb{Q}[[t_3,t_4,\ldots]][[q_1,q_2,\ldots]].
$$
In our concrete case, $\hat{F}^{[g]}_A=\hat{F}^{[g]}_A\left(\frac{{\rm Tr}\,A}{N},\frac{{\rm Tr}\,A^2}{N},\frac{{\rm Tr}\,A^3}{N},\ldots\right)$.

The same procedure gives the topological expansion of the correlators
\beq\label{HMMEFcorrelators}
\hat{W}_{n;A}(x_1,\ldots,x_n) = \kappa_{n}\Big({\rm Tr}\,\frac{1}{x_1 - M},\ldots,{\rm Tr}\,\frac{1}{x_n - M}\Big)_A = \sum_{g \geq 0} N^{2 - 2g - n}\,\hat{W}_{n;A}^{[g]}(x_1,\ldots,x_n),
\eeq
with $\hat{W}_{n;A}^{[g]} \in \mathcal{R}[[x_1^{-1},\ldots,x_n^{-1}]]$, where we think of $q_i$ as replacing ${\rm Tr}\,A^i/N$ as above.

We will need to handle more general observables, which involve expectation values of products of ${\rm Tr}\,[M^{k}]_{i,i}$ and of ${\rm Tr}\,M^{k'}$. Although we are skipping the details, their topological expansion can also be defined; the coefficients of this expansion will now belong to
$$
\tilde{\mathcal{R}} = \lim_{\infty \leftarrow \nu} \mathbb{Q}[[t_3,t_4,\ldots]][[a_1,\ldots,a_{\nu}]][[q_1,q_2,\ldots]],
$$
where $A$ is specialized to the matrix ${\rm diag}(a_1,\ldots,a_{\nu},0,\ldots,0)$. The restriction morphisms to define the projective limit consist in specializing some $a$'s to $0$. We can often work in a specialization of this ring $\tilde{\mathcal{R}}$, namely
$$
\mathcal{R}_{-} := \mathbb{Q}[[t_3,t_4,\ldots,]][[a_1,\ldots,a_{\nu}]][[N^{-1}]].
$$

\subsection{The spectral curve}

To be able to deduce the spectral curve for the hermitian matrix model with external field, one would need to generalize the notion of topological expansion here in order to extract the first term in the topological expansion of the first Schwinger-Dyson equation of this model, which is proved in \cite{EPf} in the more general context of the chain of matrices.
\begin{lemma}
We have the following identity in $\tilde{\mathcal{R}}[N,N^{-1}]][[x^{-1},y^{-1}]]$:
\bea
 0 & = & \kappa_{2}\Big({\rm Tr}\,\frac{1}{x - M},{\rm Tr}\,\frac{1}{x - M}\frac{1}{y - A}\Big)_A + (\hat{W}_{1;A}(x) - NV'(x) + Ny)\Big\langle {\rm Tr}\,\frac{1}{x - M}\,\frac{1}{y - A}\Big\rangle_{A} \nonumber \\
\label{SD1} && - N\hat{W}_{1;A}(x) + \Big\langle {\rm Tr}\,\frac{V'(x) - V'(M)}{x - M}\,\frac{1}{y - A} \Big\rangle_{A}\,. \nonumber
\eea
\end{lemma}
\noindent \textbf{Proof.} This is obtained from the relation
$$
0 = \sum_{i,j = 1}^N \sum\int \dd M\,\partial_{M_{i,j}}\bigg(\Big(\frac{1}{x - M}\,\frac{1}{y - A}\Big)_{j,i} \exp\big[N\,{\rm Tr}(MA - V(M))\big]\bigg)\,.
$$
\hfill $\Box$

Let us introduce simplified notations
$$
\hat{W}_{A}^{(i)}(x) = \bigg\langle \Big[\frac{1}{x - M}\Big]_{i,i}\Big\rangle_{A}^{[0]},\qquad \hat{W}_{A}(x) = \hat{W}_{1;A}^{[0]}(x) = \sum_{i = 1}^N \hat{W}_{A}^{(i)}(x)\,.
$$
We can write the planar limit of the first Schwinger-Dyson equation
\begin{proposition}
We have
\beq
\label{Sdd2}\hat{W}_{A}(x)^2 - [V'(x)\hat{W}_{A}(x)]_{-} + \sum_{i = 1}^N \tfrac{a_i}{N} \hat{W}_{A}^{(i)}(x) = 0\,,
\eeq
and for any $i \in \{1,\ldots,N\}$, we have
\beq
\label{Sdd3}\hat{W}_{A}^{(i)}(x)\big(\hat{W}_{A}(x) + a_i) - \tfrac{1}{N} [V'(x)\hat{W}_{A}^{(i)}(x)]_{-} = 0\,,
\eeq
where $[\ \cdot\ ]_{-}$ takes the negative part of the Laurent expansion when $x \rightarrow \infty$.
\end{proposition}
\noindent \textbf{Proof.} In the planar limit of \eqref{SD1}, $U_2$ disappears
$$
0 = (\hat{W}_{A}(x) - V'(x)  + y)\Big\langle {\rm Tr}\,\frac{1}{x - M}\frac{1}{y - A}\Big\rangle_{A} - \hat{W}_{A}(x) + \frac{1}{N}\,\Big\langle {\rm Tr}\,\frac{V'(x) - V'(M)}{x - M}\,\frac{1}{y - A}\Big\rangle_{A}^{[0]}\,.
$$
The right-hand side rational function of $y$, with simple poles at $y \rightarrow a_i$ and $y \rightarrow \infty$. Identifying the coefficient of these poles gives an equivalent set of equations. At $y \rightarrow \infty$ we get a trivial relation. At $y \rightarrow a_i$ we get
\beq
(\hat{W}_{A}(x) - V'(x) + \tfrac{a_i}{N})\hat{W}_{A}^{(i)}(x) + \Big\langle \Big[\frac{V'(x) - V'(M)}{x - M}\Big]_{i,i}\Big\rangle_{A}^{[0]} = 0\,,
\eeq
in which we recognize \eqref{Sdd3}. Summing this relation over $i \in \{1,\ldots,N\}$, we obtain \eqref{Sdd2}. \hfill $\Box$

For $A = 0$, we would obtain the planar limit of the first Schwinger-Dyson equation of the hermitian matrix model
\beq
\label{SD0} \hat{W}_{A=0}(x)^2 - [V'(x)\hat{W}_{A=0}(x)]_{-} = 0\,,\eeq
which is equivalent to Tutte's equation for the generating series of disks. Its solution is well-known, see \textit{e.g.} \cite{Eynardbook}. Observe that for $A=0$, $\hat{W}_{A=0}(x)=W(x)$.

\begin{lemma}
\label{Tutte000}The equation \eqref{SD0} determines $\hat{W}_{A=0}(x)$ completely. Let $\mathcal{R}^0 = \mathbb{Q}[[t_3,\ldots,t_d]]$. There exist unique $\alpha,\gamma \in \mathcal{R}^0$ such that
$$
x(\zeta;0) = \alpha + \gamma(\zeta + \zeta^{-1}),\qquad w(\zeta;0) = [V'(x(\zeta))]_{+}
$$
satisfy $\hat{W}_{1;A=0}^{[0]}(x(\zeta;0)) = V'(x(\zeta;0)) - w(\zeta;0)$. We have that $x(\zeta;0)=x(\zeta)$. Here, $[\ \cdot\ ]_+$ takes the polynomial part in $\zeta$. More precisely, $\alpha$ and $\gamma$ are determined by the conditions
$$
[\zeta^0]\,V'(x(\zeta)) = 0,\qquad [\zeta^{-1}] \,V'(x(\zeta)) = \gamma^{-1}\,.
$$
and we have $\gamma = 1 + O(\mathbf{t})$.
\hfill $\Box$
\end{lemma}

\begin{lemma}
\label{UniqueW}The equations \eqref{Sdd2}-\eqref{Sdd3} determine $\hat{W}_A^{(i)}(x)$ uniquely for all $i \in \{1,\ldots,p\}$.
\end{lemma}
\noindent \textbf{Proof.} By specializing $A$ to diagonal matrices of arbitrary size $\nu$ and $q_i$ to ${\rm Tr}\,A_i/N$ it is enough to work here in the ring $\mathcal{R}_- := \mathbb{C}[[t_3,\ldots,t_d,a_1,\ldots,a_{\nu},N^{-1}]]$. We introduce two gradings on $\mathcal{R}_{-}$: the first one denoted $\deg_{A}$ assigns a degree $1$ to the variables $a_i$, and $0$ to the other generators; the second one $\deg_{t}$ assigns a degree $1$ to the variables $t_{j}$ and $0$ to the other generators.  We denote $\hat{W}^{(i)}_{\alpha;A}$  and $\hat{W}_{\alpha;A}$ the homogeneous part of $\hat{W}_A^{(i)}$ and $\hat{W}_A$ with ${\rm deg}_{A} = \alpha$. We remark that $\hat{W}_{0;A}^{(i)}(x) = \hat{W}_{A=0}^{(i)}(x)$ is independent of $A$ and $i$ and thus
$$
\hat{W}_{0;A}^{(i)}(x) = \frac{\hat{W}_{A=0}(x)}{N}.
$$
Besides, we observe that
\beq 
\label{secondr} \forall \alpha \geq 1,\quad \forall i \in \{1,\ldots,N\},\qquad \hat{W}^{(i)}_{\alpha;A}(x) = O(x^{-2}).
\eeq
We proceed by induction on ${\rm deg}_{A}$. We already know that the $\deg_{A} = 0$ part of \eqref{Sdd2}-\eqref{Sdd3} has a unique solution given by Lemma~\ref{Tutte000}. Let $\alpha \geq 1$, and assume \eqref{Sdd2}-\eqref{Sdd3} determine uniquely $\hat{W}_{\alpha'}^{(i)}$ for $\alpha' < \alpha$. Decomposing \eqref{Sdd2} in homogeneous degree $\alpha \geq 1$, we find
\beq
\label{kalpha}\mathcal{K}[\hat{W}_{\alpha;A}(\cdot)](x) + \sum_{\substack{0 < \alpha_1,\alpha_2 < \alpha \\ \alpha_1 + \alpha_2 = \alpha}} \hat{W}_{\alpha_1;A}(x)\hat{W}_{\alpha_2;A}(x) + \sum_{i = 1}^N \frac{a_i}{N} \hat{W}^{(i)}_{\alpha - 1;A}(x) = 0,
\eeq
where $\mathcal{K}$ is the linear operator
$$
\mathcal{K}[f](x) =  2\hat{W}_{A=0}(x)f(x) - [V'(x)f(x)]_{-}.
$$
Let us write
$$
V(x) = \frac{x^2}{2} + \delta V(x),\qquad {\rm deg}_{t}( \delta V) = 1.
$$
When $f(x) = O(x^{-2})$, we have
$$
\mathcal{K}[f](x) = (2\hat{W}_{A=0}(x) - x)f(x) - [(\delta V)'(x)f(x)]_{-}.
$$
Under this assumption, the equation
\beq
\label{eqk}\mathcal{K}[f](x) = g(x)
\eeq
in $\mathcal{R}_{-}[x^{-1}]$ has a unique solution, which we denote $f(x) = \mathcal{K}^{-1}[g](x)$. Indeed, $(2\hat{W}_{A=0}(x) - x)$ is invertible and \eqref{eqk} determine recursively the $\deg_{t}$ homogeneous components of $f$ recursively in terms of those of $g$. Taking into account \eqref{secondr}, we can apply this remark to \eqref{kalpha} and find that
$$
\hat{W}_{\alpha;A}(x) = -\mathcal{K}^{-1}\bigg[\sum_{\substack{0 < \alpha_1,\alpha_2 < \alpha \\ \alpha_1 + \alpha_2 = \alpha}} \hat{W}_{\alpha_1;A}(\cdot)\hat{W}_{\alpha_2;A}(\cdot) + \sum_{i = 1}^N \frac{a_i}{N} \hat{W}^{(i)}_{\alpha - 1;A}(\cdot)\bigg](x)
$$
is determined. We turn to the $\deg_{A} = \alpha$ part of \eqref{Sdd3}
$$
\mathcal{K}[\hat{W}^{(i)}_{\alpha;A}(\cdot)](x) + \tfrac{1}{N}\big(\hat{W}_{A=0}(x)\hat{W}_{\alpha;A}(x) + \frac{a_i}{N} \hat{W}^{(i)}_{\alpha - 1;A}(x)\big) + \sum_{\substack{0 < \alpha_1,\alpha_2 < \alpha \\ \alpha_1 + \alpha_2 = \alpha}} \hat{W}^{(i)}_{\alpha_1;A}(x)\hat{W}^{(i)}_{\alpha_2;A}(x) = 0.
$$
Hence
$$
\hat{W}_A^{(i)}(x) = -\mathcal{K}^{-1}\bigg[\tfrac{1}{N}\big(\hat{W}_{A=0}(-)\hat{W}_{\alpha;A}(\cdot) + \frac{a_i}{N} \hat{W}^{(i)}_{\alpha - 1;A}(\cdot)\big) + \sum_{\substack{0 < \alpha_1,\alpha_2 < \alpha \\ \alpha_1 + \alpha_2 = \alpha}} \hat{W}^{(i)}_{\alpha_1;A}(\cdot)\hat{W}^{(i)}_{\alpha_2;A}(\cdot) \bigg](x)
$$
is determined as well. We conclude by induction.
\hfill $\Box$

\begin{lemma}
There exists a unique polynomial $w(z;A) \in \mathcal{R}_{-}[z]$ and $b_{k} = b_{k}(A) \in \mathcal{R}_{-}$ for $k \in \{1,\ldots,N\}$, such that
\beq
\label{xzA} x(z;A) = z + \frac{1}{N} \sum_{i = 1}^N \frac{1}{w'(b_i;A)(z - b_i)}\,,
\eeq
together with
\beq
\label{wVW} w(z;A) = V'(x(z)) + O(z^{-1}),\qquad w(b_k;A) = a_k\,.
\eeq
Besides, this unique data is such that
\beq
\label{911} w(z;A) = V'(x(z)) + z^{-1} + O(z^{-2}).
\eeq
\end{lemma} 
\noindent \textbf{Proof.} Let $\beta_i(A),b_i(A)$ be elements of $\mathcal{R}_{-}$, so far undetermined. We assume $\beta_i(A)$ invertible, and $\beta := \beta_i(0)$ and $b := b_i(0)$ independent of $i$. We define
\beq
\label{POlu}x(z;A) = z + \frac{1}{N}\,\sum_{i = 1}^N \frac{1}{\beta_i(A)(z - b_i(A))}
\eeq
and introduce the polynomial $w(z;A) \in \mathcal{R}_{-}[z]$ such that
\beq
\label{POlu2}V'(x(z;A)) = w(z;A) + O(z^{-2}),\qquad z \rightarrow \infty
\eeq
Equivalently
$$
w(z;A) = \oint \frac{\dd \tilde{z}}{2{\rm i}\pi}\,\frac{V'(x(\tilde{z};A))}{\tilde{z} - z},
$$
where the contour is close enough to $\tilde{z} = \infty$. We are going to prove that the system of equations
\beq
\label{sys}\forall i \in \{1,\ldots,N\},\qquad  \left\{\begin{array}{lll} w(b_{i}(A);A) & = & a_i \\ w'(b_i(A);A) & = & \beta_{i}(A)  \end{array}\right.
\eeq
has unique solutions $b_i(A)$ and $\beta_{i}(A)$ in $\mathcal{R}_{-}$, which we will adopt to define \eqref{POlu}-\eqref{POlu2}.

For $A = 0$, our definition gives
$$
x(z;0) = z + \frac{1}{\beta(z - b)}.
$$
Let $\alpha,\gamma \in \mathcal{R}^0$ be as in Lemma~\ref{Tutte000}. We recall that $\gamma = 1 + O(t)$, hence $\gamma$ is invertible in $\mathcal{R}^0$.  By making the change of variable $z = \gamma\zeta + \alpha$ and choosing
$$
\beta_{i}(A) = \gamma^{-2} + O(A),\qquad b_{i}(A) = \alpha + O(A),
$$
we find by comparing with Lemma~\ref{Tutte000} that
$$
\hat{W}_{A=0}(x(z)) = V'(x(z)) - w(z;0),\qquad w(b;0) = 0,\qquad w'(b;0) = \beta,
$$
in terms of the functions introduced in \eqref{POlu}-\eqref{POlu2}. 

Next, we introduce a grading in $\mathcal{R}_{-}$ by assigning degree $1$ to each $a_i$, and $0$ to all other generators. We write $x_{d}(z;A)$, $w_{d}(z;A)$, $\beta_{i,d}(A)$ and $b_{i,d}(A)$ for the degree $d$ component of the corresponding quantities. Fix $d \geq 1$, and assume we have already determined all these quantities in degree $d' < d$. Let us examine the degree $d$ part of the system \eqref{sys}, and isolate the pieces involving $\beta_{i,d}(A)$ and $b_{i,d}(A)$. We find for all $i \in \{1,\ldots,N\}$
\beq
\label{sys2} \left\{\begin{array}{lll} w'(\gamma^{-2};0) b_{i,d}(A) + \frac{c_2}{N}\Big( \sum_{i = 1}^N \frac{\beta_{i,d}(A)}{\gamma^{-4}}\Big)+ \frac{c_3}{N}\Big(\sum_{i = 1}^N \frac{b_{i,d}(A)}{\gamma^{-2}}\Big)  & = & Y_{i,d}(A), \\
w''(\gamma^{-2};0) b_{i,d}(A) + \frac{c_3}{N}\Big(\sum_{i = 1}^N \frac{\beta_{i,d}(A)}{\gamma^{-4}}\Big) + \frac{c_4}{N}\Big(\sum_{i = 1}^N \frac{b_{i,d}(A)}{\gamma^{-2}}\Big) & = & \tilde{Y}_{i,d}(A), \end{array}\right. 
\eeq
where
$$
c_{k} = \oint \frac{\dd\tilde{z}}{2{\rm i}\pi}\,\frac{V''(x(\tilde{z};0))}{(\tilde{z} - \alpha)^{k}}
$$
and $Y_{i,d}$ and $\tilde{Y}_{i,d}$ are polynomials in the $t_k$s, $a_i$s, and $\beta_{j,d'}(A)$ and $b_{j,d'}(A)$ with $d' < d$. In this formula we  used that $\beta_{j,0} = \gamma^{-2}$ and $b_{j,0} = \alpha$ for all $j$. For instance, in degree $1$
$$
Y_{i,1}(A) = a_i,\qquad \tilde{Y}_{i,1}(A) = 0.
$$
As $V''(x) = 1 + O(t)$, we deduce by moving the contour to surround $\tilde{z} = \alpha$ that
$$
\forall k \geq 2,\qquad c_{k} = O(t).
$$
Therefore, the system \eqref{sys2} takes the matrix form
$$
\Bigg[\left(\begin{array}{cc} \gamma^{-2}\,{\rm Id}_{N} & 0 \\ 0 & -{\rm Id}_{N}\end{array}\right) + N^{-1}O(t)\Bigg]\left(\begin{array}{c} b_{\bullet,d} \\ \beta_{\bullet,d} \end{array}\right) = \left(\begin{array}{l} Y_{\bullet,d}(A) \\ \tilde{Y}_{\bullet,d}(A) \end{array}\right).
$$
The matrix in the left-hand side is invertible in $\mathcal{R}_{-}$, hence $\beta_{i,d}(A)$ and $b_{i,d}(A)$ are uniquely determined. By induction, we conclude to the existence of unique $\beta_{i}(A)$ and $b_{i}(A)$ in $\mathcal{R}_{-}$ satisfying \eqref{sys}.  

Let $c_{\infty} \in \mathcal{R}_{-}$ be such that 
\beq
\label{Vexp} V'(x(z;A)) = w(z;A) + c_{\infty}\,z^{-1} + O(z^{-2}).
\eeq
We claim that $c_{\infty} = 1$. Indeed, the second set of equations in \eqref{sys} imply
$$
\sum_{i = 1}^N \Res_{z \rightarrow b_i} x(z;A)\dd w(z;A) = 1,
$$
while as $x(z;A) = O(z)$ when $z \rightarrow \infty$, we obtain by moving the contour to $\infty$ and using \eqref{Vexp}
$$
\sum_{i = 1}^N \Res_{z \rightarrow b_i} x(z;A)\dd w(z;A) = -\Res_{z \rightarrow \infty} x(z;A)\dd w(z;A) = c_{\infty} - \Res_{x \rightarrow \infty} x\,\dd V' = c_{\infty}.
$$
The last equality holds because $V'$ is a polynomial in $x$. \hfill $\Box$

\begin{lemma}
There exists a unique polynomial $P_i(\xi;A) \in \mathcal{R}_{-}[\xi]$ of degree $d - 2$ with leading coefficient $\tfrac{t_{d}}{N}$, such that
\beq
\label{WPs} V'(x(z;A)) - w(z;A) = \sum_{i = 1}^N \frac{P_i(x(z);A)}{w(z;A) - a_i}.
\eeq
\end{lemma}
\noindent \textbf{Proof.} We have
$$
x(z;A) = \frac{S_{N + 1}(z)}{T_{N}(z)},\qquad w(z;A) = U_{d - 1}(z),
$$
where $S,T,U$ are polynomials in $z$ with coefficients in $\mathcal{R}_{-}$, of degree indicated by the subscript, and $T_N$ is monic. Therefore, the resultant
$$
Q(X,Y) = {\rm res}_{z}\big[XT_N(z) - S_{N + 1}(z),T_N(z)(Y - U_{d - 1}(z))\big]
$$
is a polynomial with coefficients in $\mathcal{R}_{-}$ and of degree $N + d - 1$ in $X$, and $N + 1$ in $Y$, which gives a polynomial relation
$$ 
Q(x(z;A),w(z;A)) = 0.
$$ 

\begin{center}
\begin{figure}[h!]
\begin{center}
\includegraphics[width=0.65\textwidth]{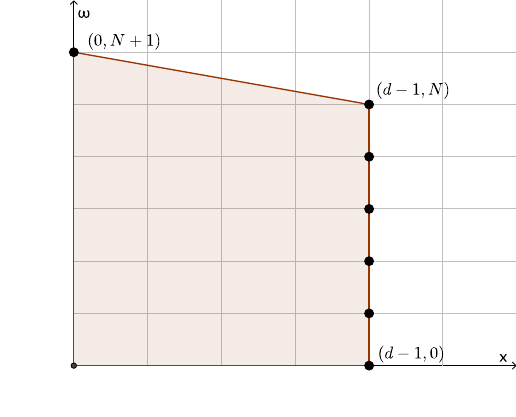}
\vspace{-1.3cm}
\caption{\label{Poly} The Newton polygon of $Q(x,w)$. The coefficients identified in \eqref{QXY} correspond to the nodes in the picture.}
\end{center}
\end{figure}
\end{center}

In fact, we can argue that the degree in $X$ smaller, as follows. We study the slopes of the Newton polygon of $Q$ associated to $x \rightarrow \infty$ or $w \rightarrow \infty$. We have $w \rightarrow \infty$ if and only if $z \rightarrow \infty$, and in this case $x \rightarrow \infty$ and we have
\beq
\label{wtd} w \sim -t_dx^{d - 1}.
\eeq
Besides, the only other situation where $x \rightarrow \infty$ is when $w \rightarrow a_i$ for some $i \in \{1,\ldots,N\}$, and in this case we read from \eqref{xzA} that
\beq
\label{wai} (w - a_i)x \  \sim \  1.
\eeq
This determines slopes which must be in the Newton polygon of $Q$.  A closer look to the determinant defining the resultant shows, using $T_N(0) = (-1)^{N} \prod_{i = 1}^N a_i$ and that $XT_N(z) - S_{N + 1}(z) = -z^{N + 1} + O(z^{N})$, that the top degree term of $Q(X,Y)$ in the variable $Y$ is $Y^{N + 1} \prod_{i = 1} a_i^{N}$. In particular, the coefficient of $X^iY^{N + 1}$ which could a priori exist, vanish for $i \in \{1,\ldots,d - 1 + N\}$. The existence of the previous slopes then forces the Newton polygon to be included in the shaded region of Figure~\ref{Poly}. In particular, $Q$ must be irreducible. And the precise behaviors \eqref{wtd}-\eqref{wai} leads to a decomposition
\beq
\label{QXY}Q(X,Y) = c \Big(Y^{N + 1} + t_{d}X^{d - 1} \prod_{i = 1}^N (Y - a_i)\Big) + \tilde{Q}(X,Y),\qquad c := \prod_{i = 1}^N a_i^{N},\eeq
for some polynomial $\tilde{Q}(X,Y)$ with coefficients in $\mathcal{R}_{-}$ such that
\beq
\label{degP}\deg_{X} \tilde{Q} \leq d - 2,\qquad \deg_{Y} \leq N.
\eeq
Now, let us examine
$$
L(z) = \big(V'(x(z;A)) - w(z;A)\big) \prod_{i = 1}^N (w(z;A) - a_i).
$$
It can also be written
$$
L(z) = -w(z;A)^{N + 1} - t_{d}x(z;A)^{d - 1} \prod_{i = 1}^N (w(z;A) - a_i) + \tilde{L}(x(z;A),w(z;A)),
$$
where $\tilde{L}(X,Y)$ is a polynomial satisfying the same degree bound as \eqref{degP}. Using the relation \eqref{QXY}, we can eliminate the first terms and get
$$
L(z) = \hat{L}(x(z;A),w(z;A)),
$$
for some polynomial $\hat{L}(X,Y)$ with coefficients in the localization $\mathcal{R}_{-}^{(c)}$ of $\mathcal{R}_{-}$ at $c$, and with $\deg_{X} \hat{L} \leq d - 2$ and $\deg_{Y} \hat{L} \leq N$. We deduce the existence of polynomials $(P_i(X))_{i = 1}^{N}$ and $P_{\infty}(X)$ of degree $\leq d - 2$ and with coefficients in $\mathcal{R}_{-}^{(c)}$, such that
$$ 
V'(x(z;A)) - w(z;A) = P_{\infty}(x(z;A)) + \sum_{i = 1}^N \frac{P_i(x(z;A))}{w(z;A) - a_i},
$$
after partial fraction decomposition. According to \eqref{wVW}, the left-hand side behaves is $O(z^{-1})$ when $z \rightarrow \infty$. As $w(z;A) = -t_{d}z^{d - 1}$ and $P_i(x(z;A)) = O(z^{d - 2})$, this is also true for the sum over $i$. Any non-zero monomial in $P_{\infty}$ would disagree with this behavior at $z \rightarrow \infty$. Hence $P_{\infty} = 0$, and we have prove the existence result for polynomials $P_i(x(z;A))$ with coefficients in $\mathcal{R}_{-}^{(c)}$.

Now, we prove uniqueness. By construction, $w(z;A) - a_i$ is a polynomial of degree $d - 1$ in $z$ with coefficients in the local ring $\mathcal{R}_{-}$, and $b_i(A)$ is a root. We remark that
$$
w(z;A) = -t_{d}z^{d - 1} + z + \tilde{w}(z;A),
$$
where $\tilde{w}(z;A) = O(A,t)$ is a polynomial of degree $\leq (d - 2)$. Therefore, $w(z;A) - a_i$ has $(d - 2)$-other roots, counted with multiplicity, which belong to the local ring $\widehat{\mathcal{R}}_{-}$ obtained from $\mathcal{R}_{-}$ by adjunction of a finite set $\mathfrak{b}$ consisting of $t_{d}^{-1/(d - 2)}$ and all its Galois conjugates. Besides, in $\widehat{\mathcal{R}}_{-}$ those $(d - 2)$ roots are pairwise distinct. 

By construction, $V'(x(z);A) - w(z;A)$ is a rational function of $z$, with poles at $z = b_i$ and $z = \infty$. As the denominator of the $i$-th term has $(d - 2)$ roots $\mathfrak{b}_{i}$ which are not poles of $\mathcal{W}(z;A)$ as set of root, we deduce that $P_i(x(z);A)$ must have roots at $\mathfrak{b}_i$, hence
$$
P_i(\xi;A) =  \tfrac{t_d}{N}\,\prod_{\rho(A) \in \mathfrak{b}_{i}(A)} (\xi - x(\rho;A)).
$$
Therefore, \eqref{WPs} determines uniquely the polynomial $P_i(\xi;A)$. Note that the coefficients of this polynomial belong to the localization of $\mathcal{R}_{-}^{(t_{d})}$ of $\mathcal{R}_{-}$ at $t_{d}$, and not only to the localization of $\widehat{\mathcal{R}}_{-}$ at $t_{d}$, as the product runs over Galois orbits. By comparison, the $P_i$ constructed in the existence part had coefficients in $\mathcal{R}_{-}^{(c)}$. We deduce that $P_i$ has coefficients in $\mathcal{R}_{-}^{(c)} \cap \mathcal{R}_{-}{(t_{d})} = \mathcal{R}_{-}$.
\hfill $\Box$

\begin{corollary}
We have
$$
\hat{W}_A(x(z)) = V'(x(z);A) - w(z;A),\qquad \hat{W}_A^{(i)}(x(z)) = \frac{P_i(x(z);A)}{w(z;A) - a_i}.
$$
\end{corollary}
\noindent \textbf{Proof.} As $x = z + O(z^{-1})$, we can perform a Lagrange inversion and define unique elements $\hat{\mathcal{W}}(x;A)$ and $\hat{\mathcal{W}}_A^{(i)}(x)$ in $ \mathcal{R}_{-}[[x^{-1}]]$ such that
$$
\hat{\mathcal{W}}_A(x(z;A)) = V'(x(z;A)) - w(z;A),\qquad \hat{\mathcal{W}}_A^{(i)}(x(z;A)) = \frac{P_i(x(z;A);A)}{w(z;A) - a_i}.
$$
By construction, we have
\beq
\label{WSum}\sum_{i = 1}^N \hat{\mathcal{W}}^{(i)}_A(x) = \hat{\mathcal{W}}_A(x),
\eeq
and
$$
\forall i \in \{1,\ldots,N\},\qquad \hat{\mathcal{W}}_A^{(i)}(x)(\hat{\mathcal{W}}_A(x) - V'(x) + a_i) + P_i(x) = 0.
$$
Therefore, $\hat{\mathcal{W}}_A^{(i)}$ and $\hat{\mathcal{W}}_A$ satisfy \eqref{Sdd2}-\eqref{Sdd3}. Note that \eqref{911} ensures that
$$
\hat{\mathcal{W}}_A(x) = \tfrac{1}{x} + O(\tfrac{1}{x^2}),\qquad x \rightarrow \infty,
$$
and that $V'(x) = -t_{d}x^{d - 1}$ also implies
$$
\hat{\mathcal{W}}_A^{(i)}(x) = \tfrac{1}{Nx} + O(\tfrac{1}{x^{2}}),\qquad x \rightarrow \infty.
$$
Since the solution of these equations is unique according to Lemma~\ref{UniqueW}, we conclude that $\hat{W}_A = \hat{\mathcal{W}}_A$ and $\hat{W}^{(i)}_A = \hat{\mathcal{W}}_A^{(i)}$. 
\hfill $\Box$

\subsection{Topological recursion}

Eynard and Prats-Ferrer analyzed in \cite{EPf} the (topological expansion) of tower of Schwinger-Dyson equations which results from variation of the potential $V$, and involve the $n$-point correlators. Their final result reads:

\begin{theorem}
Let $\omega_{g,n}^{A}$ be the TR amplitudes for the initial data
\beq
\label{rare}\left\{\begin{array}{l} \mathcal{C} = \mathbb{P}^1 \nonumber \\ p(z) = x(z;A) \\ \lambda(z) = -w(z;A) \\ B(z_1,z_2) = \frac{\dd z_1\dd z_2}{(z_1 - z_2)^2}\,.  \end{array}\right. 
\eeq
For any $g,n \geq 0$, the equality
$$
\hat{W}_{n;A}^{[g]}(p(z_1),\ldots,p(z_n))\prod_{i = 1}^n \dd p(z_i) = \omega_{g,n}^{A}(z_1,\ldots,z_n) - \delta_{g,0}\delta_{n,2}\frac{\dd p(z_1)\dd p(z_2)}{(p(z_1) - p(z_2))^2} + \delta_{g,0}\delta_{n,1}\,\dd V(p(z_1)) 
$$
holds in Laurent expansion near $z_i \rightarrow \infty$.
\end{theorem}

%

\subsubsection{Deformations of the spectral curve}

\begin{lemma}
We have, for any $i \in \{1,\ldots,N\}$,
\beq
\label{omedef}\Omega_{i}(z;A) = \partial_{a_i} x(z;A) w'(z;A) - \partial_{a_i} w(z;A) x'(z;A) = \frac{1}{w(b_{i};A)}\,\frac{1}{(z - b_i)^2},\qquad
\eeq
where the derivative with respect to $a_i$ is taken at $z$ fixed, and $\,\,'$ denotes the derivative with respect to the variable $z$.
\end{lemma} 
\noindent \textbf{Proof.} From the form of $x(z;A)$ and $w(z;A)$, we know that $\Omega_{i}(z,A)$ is a rational function of $z$, with at most double poles at $z \rightarrow b_{k}$ for $k \in \{1,\ldots,N\}$, and maybe a pole at $\infty$. We are going to identify $\Omega_{i}(z,A)$ from its singular behavior at these poles. It is easier to start by computing
$$
\tilde{\Omega}_{j}(z;A) := \partial_{b_j} x(z;A) w'(z;A) - \partial_{b_j} w(z;A) x'(z;A)
$$
and then use the relation
\beq
\label{Ometil} \Omega_{i}(z;A) = \sum_{j = 1}^N \frac{\partial b_j}{\partial a_i}\,\tilde{\Omega}_{j}(z;A)\,.
\eeq
We start by examining $z \rightarrow \infty$. From the equation
$$
w(z;A) = V'(x(z;A)) + \tfrac{1}{z} + O(\tfrac{1}{z^2})\,,
$$
we deduce
\bea
w'(z;A) & = & x'(z;A)V''(x(z;A)) - \tfrac{1}{z^2} + O(\tfrac{1}{z^3})\,, \nonumber \\
\partial_{b_j} w(z;A) & = & \partial_{b_{j}}x(z;A) V''(x(z;A)) + O(\tfrac{1}{z^2})\,, \nonumber
\eea 
and from the form of $x$
$$
\partial_{b_j} x(z;A) = O(1),\qquad x'(z;A) = O(1)\,.
$$
This implies
\bea
\tilde{\Omega}_{i}(z;A) & = & \partial_{b_j} x(z;A)\big(x'(z;A)V''(x(z;A)) + O(\tfrac{1}{z^2})\big) \nonumber \\
&-& \big(\partial_{b_{j}} x(z;A) V''(x(z;A) + O(\tfrac{1}{z^2})\big)x'(z;A) \nonumber \\
& = & O(\tfrac{1}{z^2})\,, \nonumber
\eea
and therefore $\tilde{\Omega}_{i}(z;A)$ has no pole at $\infty$.

Next, we examine $z \rightarrow b_{k}$. We have
\bea
\partial_{b_{j}} x(z;A) & = & \frac{\delta_{j,k}}{N}\bigg(-\frac{w''(b_{k};A)}{(w'(b_{k};A))^2}\,\frac{1}{z - b_{k}} + \frac{1}{w'(b_{k};A)}\,\frac{1}{(z - b_{k})^2}\bigg) \nonumber \\
&- &\frac{1}{N}\,\frac{\partial_{b_{j}}w'(b_{k};A)}{(w'(b_{k};A))^2}\,\frac{1}{z - b_{k}} + O(1)\,, \nonumber \\
w'(z;A) & = & w'(b_{k};A) + w''(b_{k};A)(z - b_{k}) + O(z - b_{k})^2\,, \nonumber \\ 
\partial_{b_{j}} w(z;A) & = & \partial_{b_{j}} w(b_{k};A) + \partial_{b_{j}} w'(b_{k};A)(z - b_{k}) + O(z - b_{k})^2\,, \nonumber \\
x'(z) & = & -\frac{1}{w'(b_{k};A)}\,\frac{1}{(z - b_{k})^2} + O(1)\,. \nonumber
\eea
Hence, we obtain after simplification
$$
\tilde{\Omega}_{j}(z;A) = \frac{1}{N}\bigg(\delta_{j,k} + \frac{\partial_{b_{j}} w'(b_{k};A)}{w'(b_{k};A)}\bigg)\,\frac{1}{(z - b_{k})^2} + O(1) \,.
$$
So, the only singularities of $\tilde{\Omega}_{j}(z;A)$ are double pole without residues at $b_{k}$, and we get
$$
\tilde{\Omega}_{j}(z;A) = \frac{1}{N} \sum_{k = 1}^N \bigg(\delta_{j,k} + \frac{\partial_{b_{j}} w'(b_{k};A)}{w'(b_{k};A)}\bigg)\,.
$$

We finally return to $\Omega_j(z;A)$. Differentiating the relation $w(b_k;A) = a_k$ with respect to $a_i$ we get
$$
\partial_{a_i} w(b_{k};A) + \partial_{a_i}b_{k} w'(b_{k};A) = \delta_{i,k}\,.
$$
We insert it in \eqref{Ometil} and find a simplification
$$
\Omega_i(z;A) = \frac{1}{N} \frac{1}{w'(b_{i};A)}\,\frac{1}{(z - b_i)^2}\,.
$$

\begin{flushright}
$\Box$
\end{flushright}

A general property of the TR amplitudes, which we only state here for spectral curves of genus $0$, is the following:
\begin{theorem}\label{deformationsTR}\cite{EOFg}
Let $(\mathbb{P}^1,p_t,\lambda_t,B = \frac{\dd z_1\dd z_2}{(z_1 - z_2)^2})$ be a holomorphic family of initial data for TR, depending on a parameter $t \in \mathbb{C}$, and $\omega_{g,n}^{t}$ its TR amplitudes. Assume there exists a generalized cycle $\gamma$ in $\mathbb{P}^1$ whose support does not contain the zeroes of $\dd p$, such that 
$$
\partial_{t} \lambda_t(z) \dd p_t(z) - \partial_{t}p_t(z) \dd \lambda_t(z) = \int_{\gamma} B(z,\cdot).
$$
where the derivatives are taken with $z$ fixed.
Then, for any $g+n > 1$, $n,m > 0$,
$$
\partial_{t}^m \omega_{g,n}(z_1,\ldots,z_n) = -\int_{\gamma^m} \omega_{g,n + m}^{t}(z_1,\ldots,z_n,\zeta_1,\ldots, \zeta_m),
$$
where the derivatives are taken at $z_i$ fixed and the $\zeta_i$ are integrated over the cycles $\gamma$. 

For $n = 0$ and $g\geq 1$, or $g =0$ and $m\geq 3$, we have:
$$
\partial_{t}^m \mathfrak{F}_g= \int_{\gamma^m} \omega_{g, m}^{t}(\zeta_1,\ldots, \zeta_m).
$$ 
\end{theorem}

We use the notation $\mathfrak{F}_g[x,y]$ for the topological recursion $n=0$ invariants, with $(x,y)$ emphasizing the spectral curve they come from.
\begin{remark}
We remark that the Theorem \ref{deformationsTR} cannot be applied to our spectral curve 
$$
(x(z;A),-w(z;A)), \text{ when } A\rightarrow 0.
$$ 
It may seem there is no problem from a naive perspective, but the behavior of the branchpoints is pathological in the limit, since they coalesce to $\infty$. This gives a setting in which the usual topological recursion does not apply.

On the other hand, the exchanged spectral curve $(w(z;A),x(z;A))$ behaves like a regular spectral curve to which we can apply the usual topological recursion, and hence Theorem \ref{deformationsTR}. We denote by $\check{\omega}_{g,n}^A(z_1,\ldots,z_n)$ its corresponding TR amplitudes.
\end{remark}

If we specialize to $n = 0$ and the deformation \eqref{omedef}, we find
\begin{corollary}
\label{coniu}
We have for any $n \geq 1$
$$
\partial_{a_1}\cdots\partial_{a_n} \mathfrak{F}_{g}^A[w,x] = \frac{\check{\omega}_{g,n}^{A}(z_1,\ldots,z_n)}{\dd w(z_1;A)\cdots \dd w(z_n;A)},
$$
where the $z_i$ are points in $\mathcal{C}$ defined by $w(z_i;A) = a_i$.
\end{corollary}
\noindent \textbf{Proof.} $\Omega_i$ represents the infinitesimal deformation of the TR initial data $(\lambda = w(z;A),p = x(z;A))$, and an equivalent form of \eqref{omedef} is
$$
\Omega_{i}(z;A)\dd z = \Res_{\zeta \rightarrow b_i} \frac{B(z,\zeta)}{w(\zeta;A) - w(b_i;A)}.
$$
Therefore, from Theorem \ref{deformationsTR}, we obtain
\bea
\partial_{a_1}\cdots\partial_{a_n} \mathfrak{F}_{g}^A[w,x] & = & \Res_{\zeta_1 \rightarrow b_{1}} \cdots \Res_{\zeta_n \rightarrow b_n} \frac{\check{\omega}_{g,n}^{A}(\zeta_1,\ldots,\zeta_n)}{\prod_{i = 1}^n (w(\zeta_i;A) - w(b_i;A))} \nonumber \\
\label{reg} & = & \frac{\check{\omega}_{g,n}^{A}(b_1,\ldots,b_n)}{\dd w(b_1;A) \cdots \dd w(b_n;A)},
\eea 
where the differential acts on the first variable of $w$.
\hfill $\Box$

\subsection{Conclusion}

Now we give the technical condition under which our conjecture would be true for usual maps. For $n>0$:
\beq\label{technicalCond}
\partial_{a_1}\cdots\partial_{a_n}\hat{F}^{[g]}_A =\partial_{a_1}\cdots\partial_{a_n}\mathfrak{F}_{g}^A[x,-w]=\partial_{a_1}\cdots\partial_{a_n}\mathfrak{F}_{g}^A[w,x].
\eeq
Observe that we only need this equality for $n>0$, i.e. for the non-constant Taylor coefficients.
We believe the first equality is true since the free energy coming from the matrix model and the TR invariants should differ by a constant not depending on $A$.
We also think the second equality is true and can be proved by checking that the correction terms of symplectic invariance, which are still being analyzed in general, do not depend on $A$.

In any case, the computations we did for quadrangulations to support our conjecture indicate that if there existed a non-zero difference, it would also have a combinatorial interpretation, which may help understand the nature of the still mysterious property of symplectic invariance.


We sketch the final argument which would lead to a proof of the conjecture if we suppose \eqref{technicalCond} is true.

We view $\mathcal{R}$ as a graded ring by assigning degree $1$ to each generator $q_i$.
Recall from \eqref{gsHMMEF} that, as an element of $\mathcal{R}$, the free energy decomposes as
$$
\hat{F}^{[g]}_A = \sum_{n \geq 1}\frac{1}{n!} \sum_{\ell_1,\ldots,\ell_n} H_{\ell_1,\ldots,\ell_n}^{[g]}\,\prod_{i = 1}^n \frac{p_{\ell_i}(A)}{N\ell_i},\qquad H_{\ell_1,\ldots,\ell_n}^{[g]} \in \mathcal{R}^0.
$$

\begin{corollary}
For $2g - 2 + n > 0$, we have in $\tilde{R}[[N^{-1}]]$
$$
\frac{\check{\omega}_{g,n}^{A = 0}(b_1,\ldots,b_n)}{\prod_{i = 1}^n\dd w(b_i;0)} = \sum_{\ell_1,\ldots,\ell_n \geq 1} H_{\ell_1,\ldots,\ell_n}^{[g]}\prod_{i = 1}^n a_i^{\ell_i - 1},\qquad {\rm with}\quad w(b_i;0) = a_i.
$$
\end{corollary}
\noindent\textbf{Proof.} 
Let $n \geq 1$ be an integer. We denote $\pi_{\mathcal{R},n}$ the projection from $\mathcal{R}$ to its degree $n$ subspace. We introduce the ring $\mathcal{Q}^{(n)} = \mathcal{R}^0[[a_1,\ldots,a_n]][[q_1,q_2,\ldots]]$, where $q_i\equiv \tfrac{p_i(A)}{N}= \tfrac{{\rm Tr}\, A^i}{N}$. We make it a graded ring by assigning degree $1$ to each generator $q_{\ell}$.  We denote $\pi_{\mathcal{Q},0}$ the projection from $\mathcal{Q}^{(n)}$ to its degree~$0$ subspace. We can define a linear map $\Upsilon^{(n)}\,:\,\mathcal{R}^{\mathfrak{S}_{N}} \rightarrow \mathcal{Q}^{(n)}$ by
$$
\Upsilon^{(n)}(f) = a_1\partial_{a_1}\cdots a_n\partial_{a_n} f.
$$
The map $\Upsilon^{(n)}$ is homogeneous of degree $-n$, in particular it sends $p_{\lambda}$ with $\ell(\lambda) < n$ to zero. Besides, from the degree $n$ part to the degree $0$ part it induces an isomorphism, which is just the change of basis from power sums $\tfrac{p_i(A)}{N}$ to (unnormalized) symmetric monomials in the $a_i$'s, and we have
\beq
\label{comutd}\Upsilon^{(n)} \circ \pi_{\mathcal{R},n} = \pi_{\mathcal{Q},0}\circ \Upsilon^{(n)}.
\eeq

We would like to access
\beq
\label{dgfsg222}\Upsilon^{(n)} \circ \pi_{\mathcal{R},n}(\hat{F}^{[g]}_A) = \sum_{\ell_1,\ldots,\ell_n \geq 1} H_{\ell_1,\ldots,\ell_n}^{[g]}\,\prod_{i  = 1}^n a_i^{\ell_i}.
\eeq
Using the technical condition \eqref{technicalCond} and that $w(b_i;A) = a_i$, we obtain from Corollary~\ref{coniu} that
$$
\Upsilon^{(n)}(\hat{F}^{[g]}_A) = \check{\omega}_{g,n}^A(b_1,\ldots,b_n) \prod_{i = 1}^n \frac{ w(b_i;A)}{\dd w(b_i;A)}.
$$
According to \eqref{comutd} the degree $0$ part of the right-hand side computes the quantity in \eqref{dgfsg222}. The spectral curve $(x(z;A), w(z;A))$ is symmetric in $a_1,\ldots,a_{N}$. Therefore when we compute the right-hand side with TR, we obtain a formal series in $\tfrac{p_{\ell}(A)}{N}$, whose term of degree $0$ is
$$
\check{\omega}_{g,n}^{A = 0}(b_1,\ldots,b_n)\,\prod_{i = 1}^n \frac{w(b_i;0)}{\dd w(b_i;0)},\qquad {\rm with}\quad  w(b_i;0) = a_i.
$$ 
\hfill $\Box$
\chapter{Applications}
\label{chap:apps}


\section{Relation with free probability}
\label{SFree}
\label{Section10}

We explain how the results of Chapter \ref{chap:intro} fit in the context of free probability, and give a possible application of our Conjecture~\ref{conj} and more general conjectures of Section \ref{Section6}.

More concretely, we give a combinatorial interpretation of higher order free cumulants in terms of fully simple maps. We refer the reader to Section \ref{IntroFP} for a summary on free probability, and an introduction to higher order freeness and free cumulants.


\subsection{Enumerative interpretation of higher order free cumulants}\label{HOFCsFS}

Let $M=(M_N)_{N\in\N}$ be a unitarily invariant hermitian random matrix ensemble. If the following large $N$ limit of the scaled limits of classical cumulants of $n$ traces of powers of our matrices
\beq\label{correlationMom}
\varphi_{\ell_1,\ldots,\ell_n}^M\coloneqq \lim_{N\to \infty} N^{n-2}\kappa_n({\rm Tr}\, M_N^{\ell_1},\ldots,{\rm Tr}\, M_N^{\ell_n})
\eeq
exists for $n \geq 1$, we recall from \eqref{correlationMoments} that it defines a higher order non-commutative probability space generated by a single element $M$. These limits \eqref{correlationMom} were called the $n$-th order correlation moments and constitute the limiting distribution of all orders of $M$, turning the space of hermitian unitarily invariant random matrix ensembles into a higher order probability space.

Higher order free cumulants are defined in terms of the correlation moments (see Definition \ref{defFreeCumulants}) through complicated combinatorial objects called partitioned permutations (see \ref{partitionedPermutations}). However, in the important setting of random matrices, Theorem \ref{freeCum} expressed the $n$th order free cumulants also as scaled limits of classical cumulants, but this time of entries of the matrices $M_N=(M^{(N)}_{r,s})_{r,s=1}^{N}$:
\beq
k_{\ell_1,\ldots,\ell_n}^M = \lim_{N \rightarrow \infty} N^{n - 2 + \sum_{i} \ell_i} \kappa_{
\sum_{i = 1}^n \ell_i}\big((M^{(N)}_{i_{j,k},i_{j,k + 1\,\,{\rm mod}\,\,\ell_j}})\big),
\eeq
where $\gamma_j := (i_{j,1},\ldots,i_{j,\ell_j})_{j = 1}^n$ are pairwise disjoint cycles of respective lengths $\ell_j$. Note also that we can express the free cumulants in a more compact way in terms of the fully simple observables we introduced:


\begin{lemma}
\label{bgsfgK}We have
$$ 
\kappa_{\sum_{i = 1}^n \ell_i}\big(M_{i_{j,k},i_{j,k + 1\,\,{\rm mod}\,\,\ell_j}}\big) = \kappa_n\big(\mathcal{P}_{\gamma_1}(M),\ldots,\mathcal{P}_{\gamma_n}(M)\big).
$$
\end{lemma} 
\noindent \textbf{Proof.} Both sides can be expressed as a linear combination of terms of the form 
$$
\prod_{\alpha}\Big\langle \prod_{(j,k) \in I_{\alpha}} M_{i_{j,k},i_{j,k + 1\,\,{\rm mod}\,\,\ell_j}}\Big\rangle,
$$
where $(I_{\alpha})_{\alpha}$ is a partition of $\mathbf{I} = \big\{(j,k)\,\,\big|\,\,1 \leq j \leq n,\,\,1 \leq k \leq \ell_j\big\}$. The term corresponding to the partition consisting of a single set $I = \mathbf{I}$ appears on both sides with a coefficient $1$. We have in general
\bea
\label{Iaprof}&& \Big\langle \prod_{(j,k) \in I_{\alpha}} M_{i_{j,k},i_{j,k + 1\,\,{\rm mod}\,\,\ell_j}}\Big\rangle \\
& = & \sum_{b\,:\,I_{\alpha} \rightarrow \llbracket 1,N \rrbracket} \Big\langle \prod_{(j,k) \in I_{\alpha}} \lambda_{b_{j,k}} \Big\rangle  \int_{U(N)} \dd U\,\bigg[\prod_{(j,k) \in I_{\alpha}} U_{i_{j,k},b_{j,k}} U^{\dagger}_{b_{j,k},i_{j,k + 1\,\,{\rm mod}\,\,\ell_j}} \bigg], \nonumber 
\eea
where we have used the $U_N$-invariance of the distribution of $M$, and $(\lambda_{a})_{a = 1}^N$ are the (unordered) eigenvalues of $M$. The integral over $U_N$ is computed by Weingarten calculus, Theorem~\ref{UNmoment}. To match the notations of Theorem~\ref{UNmoment} we have
\bea
A_{\alpha} & := & \{a_{l}\,\,|\,\,1 \leq l \leq N\} = \{i_{j,k}\,\,|\,\,(j,k) \in I_{\alpha}\} \nonumber \\
A'_{\alpha} & := & \{a'_{l}\,\,|\,\,1 \leq l \leq N\} = \{i_{j,k + 1\,\,{\rm mod}\,\,\ell_j}\,\,|\,\,(j,k) \in I_{\alpha}\} \nonumber
\eea
and the product of Kronecker deltas in Theorem~\ref{UNmoment} tells us that if $A_{\alpha} \neq A'_{\alpha}$ as subsets of $\llbracket 1,N \rrbracket$, then \eqref{Iaprof} will be zero. This is indeed the case when $I_{\alpha}$ is strictly included in $\mathbf{I}$,  for the $i_{j,k}$ are pairwise disjoint. This claimed equality follows.
\hfill $\Box$

\vspace{0.1cm}

We have hence expressed the $n$th order correlation functions and the $n$th order free cumulants for a higher order probability space given by a hermitian unitarily invariant random matrix ensemble in terms of the connected ordinary and fully simple correlators that we defined (in their disconnected versions) in \eqref{Obsdisc}.

More concretely, for a general measure \eqref{eq:mesS}, the propositions ~\ref{propStuffed} (for ordinary stuffed) and \ref{PPPPP} (for fully simple stuffed) represent the $n$th order correlation functions and the $n$th order free cumulants as generating series of planar ordinary and fully simple stuffed maps, respectively:
\bea\label{FP-Maps1}
\varphi_{\ell_1,\ldots,\ell_n}^M & = & \widehat{F}_{\ell_1,\ldots,\ell_n},\\
k_{\ell_1,\ldots,\ell_n}^M & = & \widehat{H}_{\ell_1,\ldots,\ell_n}.\label{FP-Maps2}
\eea

For the particular case of matrix ensembles given by the measure \eqref{eq:mes}, we obtain generating series of planar ordinary and fully simple usual maps instead.


\subsection{$R$-transform machinery in terms of maps}

With the identifications \eqref{FP-Maps1}-\eqref{FP-Maps2} we just exposed in mind, we see that the relation from Propositions~\ref{0,1}-\ref{0,2} between fully simple and ordinary generating series in the case of disks and cylinders recover the important $R$-transform formulas from Theorem \ref{R-transform} that related the generating series of first and second order moments and free cumulants. The relation between the generating series of fully simple disks $X$ and the $R$-transform (generating series of free cumulants) $\mathcal{R}$ is as follows: $\mathcal{R}(w)=X(w)-w^{-1}$. In this thesis, we have proved the formulas relating $X_{1}$ with $W_1$  and $X_{2}$ with $W_2$ via combinatorics of maps -- instead of non-crossing partitions -- independently of \cite{Secondorderfreeness}. We remark that in the context of topological recursion it is clear that the second order formula \eqref{speicher} can be re-written in a symmetric way, which was not obvious from the free probability point of view. We recall here our formula for cylinders in this symmetric form:
\beq
W_2(x_1,x_2)\dd x_1 \dd x_2 + \frac{\dd x_1 \dd x_2}{(x_1-x_2)^2} = X_2(w_1,w_2)\dd w_1 \dd w_2 + \frac{\dd w_1 \dd w_2}{(w_1-w_2)^2}.
\eeq
Recovering these formulas was one of the first motivations for considering fully simple maps.


Although Weingarten calculus and the HCIZ integral is also used in \cite{Secondorderfreeness} to relate higher order free cumulants to moments, we have explained in Section~\ref{SecM} that the relation is naturally expressed in terms of monotone Hurwitz numbers. As Hurwitz theory develops rapidly, this fact may give insight into the structure of higher order cumulants generating series. 

Our Conjecture~\ref{conj} applies to matrix ensembles in which $\ln\big(\frac{\dd\mu(M)}{\dd M}\big)$ is linear in the trace of powers of $M$, \textit{i.e.} with $T_{h,k} = 0$ for $(h,k) \neq (0,1)$, and it turns into Conjecture~\ref{ConT02} if we turn on $T_{0,2}$ as well. These are the models governed by the topological recursion, and whose combinatorics is captured by usual maps, or maps carrying a loop model. These conjectures would therefore provide a computational tool for the free cumulants in these models, via the topological recursion restricted to genus $0$: going free amounts to performing the exchange transformation $x \mapsto y$. 

More general unitarily invariant ensembles are rather governed by the blobbed topological recursion of \cite{Bstuff} and related to stuffed maps. Concretely, the initial data for the blobbed topological recursion is a spectral curve as in \eqref{inidataordmap}, supplemented with blobs $(\phi_{g,n})_{2g - 2 + n > 0}$ which play the role of extra initial data intervening in topology $(g,n)$ and beyond. For matrix ensembles of the form \eqref{eq:mesS}, there exist specific values for the blobs, such that the blobbed topological recursion computes the large $N$ expansion of the correlators. It would be interesting to know if the $x \leftrightarrow y$ can be supplemented with a transformation of the blobs, in such a way that the blobbed topological recursion for the transformed initial data computes the free cumulants. Restricting to genus $0$, it would give a computational scheme to handle free cumulants of any order -- in full generality -- via the blobbed topological recursion. 

The higher genus theory should capture finite size corrections to freeness, and the universality of the topological recursion suggests that it may be possible to formulate a universal theory of approximate freeness, for which unitarily invariant matrix ensembles would provide examples.

Generalizing the $R$-transform machinery to higher order free cumulants proved to be a very complicated problem and in the most general article \cite{Secondorderfreeness} they only managed to do it for second order, and in a quite intricate way. If our conjecture for usual maps~\ref{conj} is true, topological recursion for the pair of pants, i.e. for the topology $(0,3)$, gives directly the $R$-transform formula of third order \ref{03formula} for the specific measure \eqref{eq:mes}, as illustrated in Section~\ref{RtransPairPants}. This is already interesting to the free probability community and gives a hint on how to generalize those complicated relations. Finding other generalized $R$-transform formulas is not easy, even if we had topological recursion. This constitutes one of the motivations of Section~\ref{virasoro}.

%


\section{ELSV formula for $2$-orbifold monotone Hurwitz numbers}
\label{Section11}


ELSV\footnote{The original ELSV formula \cite{ELSV} relates simple Hurwitz numbers to Hodge integrals. It was the main tool of proofs of Witten's conjecture that appeared after Kontsevich's proof, like the one by Okounkov-Pandharipande \cite{OkPand}. Later, Kazarian \cite{KazarianELSV} gave a unified way to deduce most known results in the intersection theory of $\overline{\mathcal{M}}_{g,n}$.}-type formulas relate connected Hurwitz numbers to the intersection theory of the moduli space of stable curves $\overline{\mathcal{M}}_{g,n}$, enabling the tranfer of results from one world to the other. In this section, we provide a new ELSV-like formula for $2$-orbifold monotone Hurwitz numbers.

\subsubsection{Connected $2$-orbifold monotone Hurwitz numbers}

We are going to consider double strictly monotone Hurwitz numbers, with ramification profiles $\mu$ arbitrary and $\lambda = (2,\ldots,2)$: $[E_{k}]_{\mu,(2,\ldots,2)}$, which are called {\it $2$-orbifold} strictly monotone Hurwitz numbers. Notice that
$$
|{\rm Aut}\,(2,\ldots,2)| = 2^{L/2}\,(L/2)!,\text{ with } L = |\mu|, \text{ since } \ell((2,\ldots,2))=\frac{|\mu|}{2},
$$ 
and remember that $2^{L/2}\,(L/2)!\,[E_k]_{\mu,(2,\ldots,2)}$ is the number of strictly monotone $k$-step paths from $C_{\mu}$ to some (arbitrary but) fixed permutation $\sigma \in C_{(2,\ldots,2)}$ in the Cayley graph of $\mathfrak{S}_{L}$. We say that such a path is connected when the group generated by all permutations met along the path acts transitively on $\llbracket 1,L \rrbracket$.  This matches the usual definition of connectedness in the language of branched coverings. We define $2^{L/2}(L/2)!\,[E^{\circ}_{k}]_{\mu,(2,\ldots,2)}$ to be the same weighted enumeration but restricted to connected paths. We can express these disconnected Hurwitz numbers in terms of the connected ones as follows. A disconnected path from $C_{\mu}$ to $\sigma$ can be broken in connected components, i.e in paths in $\prod_{i = 1}^{s} \mathfrak{S}_{J_i}$ where $(J_i)_{i = 1}^{s}$ is an unordered partition of $\llbracket 1,L \rrbracket$ into subsets such that each $2$-cycle in $\sigma$ is contained in some $J_i$. In particular these sets all have even cardinalities. Each connected component starts from a conjugacy class $C_{\mu^{(i)}}$, where $(\mu^{(i)})_{i = 1}^s$ are partitions whose concatenation is $\mu$, and ends at $\sigma|_{J_i}$ -- which has again type $(2,\ldots,2)$ with $\ell((2,\ldots,2))=\frac{|\mu^{(i)}|}{2}$. Therefore, we get
$$ 
2^{\frac{L}{2}}(L/2)!\,[E_k]_{\mu,(2,\ldots,2)} =\sum_{s \geq 1} \frac{1}{s!} \sum_{\substack{\mu^{(1)} \cup \cdots \cup \mu^{(s)} = \mu \\ k_1 + \cdots + k_s = k}} \frac{\frac{L}{2}!}{\frac{|\mu^{(1)}|}{2}!\cdots \frac{|\mu^{(s)}|}{2}!}\,\prod_{i = 1}^{s} 2^{|\mu^{(i)}|/2}\,\big(|\mu^{(i)}|/2\big)!\,[E^{\circ}_{k_i}]_{\mu^{(i)},(2,\ldots,2)}.
$$
Hence, the symmetry factors disappear and we get
$$ 
[E_k]_{\mu,(2,\ldots,2)} = \sum_{s \geq 1} \frac{1}{s!} \sum_{\substack{\mu^{(1)} \cup \cdots \cup \mu^{(s)} = \mu \\ k_1 + \cdots + k_s = k}} \prod_{i = 1}^s [E^{\circ}_{k_i}]_{\mu^{(i)},(2,\ldots,2)}.
$$
It is convenient to rename $[E^{\circ,g}]_{\mu,(2,\ldots,2)}$ the $2$-orbifold connected Hurwitz numbers of genus $g$, which is related to the above $k$ by Riemann-Hurwitz formula in the branched covering interpretation:
$$
k = 2g - 2 + \ell(\mu) + \frac{|\mu|}{2}.
$$

\subsection{GUE and monotone Hurwitz numbers}
The Gaussian Unitary Ensemble is the probability measure on the space of hermitian matrices of size $N$
\bea
\dd\mu(M) & = &  2^{-\frac{N}{2}}\pi^{-\frac{N^2}{2}}\dd M\,\exp(- N\,{\rm Tr}\,\tfrac{M^2}{2}) \nonumber \\
& = & 2^{-\frac{N}{2}}\pi^{-\frac{N^2}{2}}\prod_{i = 1}^N \dd M_{i,i}\,e^{-NM_{i,i}^2/2} \prod_{i < j} \dd {\rm Re}\,M_{i,j}\,e^{-N({\rm Re} M_{i,j})^2}  \dd {\rm Im}\,M_{i,j}\,e^{- N({\rm Im}\,M_{i,j})^2}. \nonumber 
\eea
In this section we consider expectation values and cumulants with respect to the GUE measure.

We recall that the cumulants of the GUE have a topological expansion
$$
\kappa_{n}({\rm Tr}\,M^{\mu_1},\ldots,{\rm Tr}\,M^{\mu_n}) = \sum_{g \geq 0} N^{2 - 2g - n}\,\kappa_{n}^{[g]}({\rm Tr}\,M^{\mu_1},\ldots,{\rm Tr}\,M^{\mu_n}).
$$
For any fixed positive integers $\mu_i$, this sum is finite. The coefficients $\kappa_{n}^{[g]}$ count genus $g$ maps whose only faces are the $n$ marked faces, and are sometimes called generalized Catalan numbers. They have been extensively studied by various methods \cite{LehmanWalsh,HarerZagier,GNica,ChapuyZ,AndersenGaussian}. We are able to relate them to $2$-orbifold strictly monotone Hurwitz numbers, observing that the only fully simple maps without internal faces are the degenerate ones:
\begin{proposition}
\label{propdh} For any $g \geq 0$, $n \geq 1$ and partition $\mu$,
$$
\frac{\kappa_{n}^{[g]}({\rm Tr}\,M^{\mu_1},\ldots,{\rm Tr}\,M^{\mu_n})}{|{\rm Aut}\,\mu|} = [E^{\circ,g}]_{\mu,(2,\ldots,2)},
$$
where $[E_{g}^{\circ}]_{\mu,\lambda}$ are the connected double weakly monotone Hurwitz numbers of genus $g$. 
\end{proposition}
\noindent \textbf{Proof.} As the entries of $M$ are independent and gaussian, we can easily evaluate
$$
\langle \mathcal{P}_{\lambda}(M) \rangle = \prod_{i = 1}^{\ell(\lambda)} \frac{\delta_{\lambda_i,2}}{N}.
$$
From Theorem~\ref{Transi} we deduce
$$
|{\rm Aut}\,\mu|^{-1}\Big\langle \prod_{i = 1}^n {\rm Tr}\,M^{\mu_i}\Big \rangle =N^{\frac{|\mu|}{2}}\sum_{k \geq 0} N^{-k}[E_{k}]_{\mu,(2,\ldots,2)}.
$$
For the purpose of this proof, we name $\tilde{p}_{\mu}$ the power sum basis of $\mathcal{B}$. We have
\bea
\sum_{\mu} \frac{\langle p_{\mu}(M) \rangle \tilde{p}_{\mu}}{|{\rm Aut}\,\mu|} & = &  \Big\langle \exp\Big(\sum_{m \geq 1}  \frac{{\rm Tr}\,M^m\,\tilde{p}_{m}}{m}\Big)\Big\rangle \\ 
& = & \exp\bigg(\sum_{n \geq 1} \frac{1}{n!} \sum_{m_1,\ldots,m_n \geq 1} \kappa_{n}({\rm Tr}\,M^{m_1},\cdots,{\rm Tr}\,M^{m_n}) \prod_{i = 1}^n \frac{\tilde{p}_{m_i}}{m_i}\bigg) \nonumber \\
& = & \exp\bigg(\sum_{\mu} \frac{\kappa(p_{\mu_1}(M),\ldots,p_{\mu_{\ell(\mu)}}(M))\,\tilde{p}_{\mu}}{|{\rm Aut}\,\mu|}\bigg) \nonumber \\
& = & \exp\bigg(\sum_{\mu} \sum_{g \geq 0} N^{2 - 2g - \ell(\mu)}\,\kappa_{n}^{[g]}({\rm Tr}\,M^{\mu_1},\ldots,{\rm Tr}\,M^{\mu_{\ell(\mu)}})\,\frac{\tilde{p}_{\mu}}{|{\rm Aut}\,\mu|} \bigg).
\eea 
But on the other hand we have
\bea
\sum_{\mu} \frac{\langle p_{\mu}(M) \rangle\,\tilde{p}_{\mu}}{|{\rm Aut}\,\mu|} & = & \sum_{\mu} N^{\frac{|\mu|}{2}} \Big(\sum_{k \geq 0} N^{-k}[E_{k}]_{\mu,(2,\ldots,2)}\Big)\tilde{p}_{\mu} \nonumber \\
& = & \sum_{s \geq 1} \frac{1}{s!} \sum_{\substack{\mu^{(1)},\ldots,\mu^{(s)} \\ k_1,\ldots,k_s \geq 0}} \prod_{i = 1}^s N^{\frac{|\mu^{(i)}|}{2} -k_i}\,[E_{k_i}^{\circ}]_{\mu^{(i)},(2,\ldots,2)}\tilde{p}_{\mu^{(i)}} \nonumber \\
& = & \exp\bigg(\sum_{\mu} \sum_{k \geq 0} N^{\frac{|\mu|}{2} - k} [E_{k}^{\circ}]_{\mu,(2,\ldots,2)} \tilde{p}_{\mu} \bigg) \nonumber \\
& = & \exp\bigg(\sum_{\mu}  \sum_{g \geq 0} N^{2 - 2g - \ell(\mu)}\,[E^{\circ,g}]_{\mu,(2,\ldots,2)} \tilde{p}_{\mu} \bigg). \nonumber
\eea
Comparing the two formulas yields the claim. \hfill $\Box$
\vspace{0.2cm}

This specialization of our result recovers a particular case of \cite[Prop. 4.8]{ALS} which says that the enumeration of hypermaps is equivalent to the strictly monotone orbiforld Hurwitz problem. This suggests it is natural to investigate if our results can be extended to the more general setting of hypermaps.

It is well-known that the GUE correlation functions are computed by the topological recursion for the spectral curve,
$$
\mathcal{C} = \mathbb{P}^1,\qquad p(z) = z + \frac{1}{z},\qquad \lambda(z) = \frac{1}{z},\qquad B(z_1,z_2) = \frac{\dd z_1\dd z_2}{(z_1 - z_2)^2}\,.
$$ 
Therefore, Proposition~\ref{propdh} gives a new proof that the $2$-orbifold strictly monotone Hurwitz numbers are computed by the topological recursion, a fact already known as a special case of more general results, see \textit{e.g.} \cite{DOPS14,EynardHarnad}.

\subsection{GUE and Hodge integrals}

Dubrovin, Liu, Yang and Zhang \cite{DiYang} recently discovered a relation between Hodge integrals and the even GUE moments. For $2g - 2 + n > 0$, the Hodge bundle $\mathbb{E}$ is the holomorphic vector bundle over Deligne-Mumford compactification of the moduli space of curves $\overline{\mathcal{M}}_{g,n}$ whose fiber above a curve with punctures $(\mathcal{C},p_1,\ldots,p_n)$ is the $g$-dimensional space of holomorphic $1$-forms on $\mathcal{C}$. We denote
$$
\Lambda(t) = \sum_{j = 0}^{g} {\rm c}_{j}(\mathbb{E})\,t^{j}
$$
its Chern polynomial. Let $\psi_i$ be the first Chern class of the line bundle $T^*_{p_i}\mathcal{C}$, and introduce the formal series
$$ 
Z_{{\rm Hodge}}(t;\hbar) = \exp\bigg(\sum_{2g - 2 + n > 0} \frac{\hbar^{2g - 2 + n}}{n!} \sum_{i_1,\ldots,i_n \geq 0} \int_{\overline{\mathcal{M}}_{g,n}} \Lambda(-1)\Lambda(-1)\Lambda(\tfrac{1}{2}) \prod_{i = 1}^n \psi_{i}^{d_i} t_{d_i}\bigg).
$$ 
On the GUE side, the cumulants have a topological expansion
$$
\kappa_{n}({\rm Tr}\,M^{\ell_1},\ldots,{\rm Tr}\,M^{\ell_n}) = \sum_{g \geq 0} N^{2 - 2g - n}\,\kappa_{n}^{[g]}({\rm Tr}\,M^{\ell_1},\ldots,{\rm Tr}\,M^{\ell_n}),
$$
where $\kappa_{n}^{[g]}$ are independent of $N$, and the sum is always finite. We introduce the formal series 
\beq 
\label{Zeven} Z_{{\rm even}}(s;N) = \frac{e^{-A(s;N)}\,\prod_{j = 1}^{N - 1}j!}{2^{N}\pi^{\frac{N(N + 1)}{2}}}\,\int_{\mathcal{H}_{N}} \dd M\,\exp\bigg[N\,{\rm Tr}\,\Big(-\tfrac{M^2}{2} + \sum_{j \geq 1} s_{j}\,{\rm Tr}\,M^{2j}\Big)\bigg],
\eeq
where we choose
\bea  
A(s;N) & =  & \frac{\ln N}{12} - \zeta'(-1) \nonumber \\
&& + N^2\bigg[-\frac{3}{4} + \sum_{j \geq 1} \frac{1}{j + 1}\,{2j \choose j}s_j + \frac{1}{2} \sum_{j_1,j_2 \geq 1} \frac{j_1j_2}{j_1 + j_2}{2j_1 \choose j_1}{2j_2 \choose j_2}s_{j_1}s_{j_2}\bigg]. \nonumber
\eea
The normalization factor in \eqref{Zeven} is related to the volume of $U_N$ and the factor $e^{-A(s;N)}$ cancels the non-decaying terms in its large $N$ asymptotics, as well as the contributions of $\kappa_{1}^{[0]}$ and $\kappa_{2}^{[0]}$. The large $N$ asymptotics of the outcome  reads
\bea
Z_{{\rm even}}(s;N) & = &  \exp\bigg( \sum_{g \geq 2} \frac{N^{2 - 2g}B_{2g}}{4g(g - 1)} \nonumber \\
&& + \sum_{2g - 2 + n > 0} \frac{N^{2g - 2 + n}}{n!}\,\sum_{\ell_1,\ldots,\ell_n \geq 0}  \kappa_n^{[g]}({\rm Tr}\,M^{2\ell_1},\ldots,{\rm Tr}\,M^{2\ell_n}) \prod_{i = 1}^n s_{\ell_i} \bigg),  \nonumber 
\eea 
and we consider it as an element of $N^2\mathbb{Q}[N^{-1}][[s_1,s_2,\ldots]]$.

\begin{theorem} \cite{DiYang}
\label{thDi}With the change of variable
$$
T_{i,\pm}(s;N) = \sum_{k \geq 1} k^{i + 1}{2k \choose k} s_{k} - \delta_{i \geq 2} \pm \frac{\delta_{i,0}}{2N},
$$
we have the identity of formal series
$$
Z_{{\rm even}}(s;N) = Z_{{\rm Hodge}}(T_+(s;N),\sqrt{2}N^{-1}) Z_{{\rm Hodge}}(T_{-}(s;N),\sqrt{2}N^{-1}).
$$
\end{theorem}

We can extract from this result an explicit formula for the even GUE moments, which gives an ELSV-like formula for the monotone Hurwitz numbers with even ramification above $\infty$, and $(2,\ldots,2)$ ramification above $0$.
\begin{corollary}
For $g \geq 0$ and $n \geq 1$ such that $2g - 2 + n > 0$ and $\mu=(2m_1,\ldots,2m_n)$, we have
\bea
|{\rm Aut}\,\mu|[E^{\circ}_{g}]_{\mu,(2,\ldots,2)}\!\! \!& = &\!\!\! \kappa_n^{[g]}({\rm Tr}\,M^{2m_1},\ldots,{\rm Tr}\,M^{2m_n}) \nonumber \\
\!\!\!& = &\!\!\! 2^{g} \int_{\overline{\mathcal{M}}_{g,n}} [\Delta]\,\cap\,\Lambda(-1)\Lambda(-1)\Lambda(\tfrac{1}{2})\exp\Big(-\sum_{d \geq 1} \frac{\kappa_{d}}{d}\Big) \prod_{i = 1}^n \frac{m_i {2m_i \choose m_i}}{1 - m_i\psi_i}. \nonumber
\eea
$\kappa_{d}$ are the pushforwards of $\psi_{n + 1}^{d + 1}$ via the morphism forgetting the last puncture. Denoting $[\Delta_h]$ the class of the boundary strata $\overline{\mathcal{M}}_{g - h,n + 2h} \subset \overline{\mathcal{M}}_{g,n}$ which comes from the pairwise gluing of the last $2h$ punctures, we have introduced
$$ 
[\Delta] = \sum_{h \geq 0} \frac{[\Delta_h]}{2^{3h}(2h)!}.
$$
\end{corollary}
\noindent \textbf{Proof.} Identifying the coefficient of $\frac{N^{2 - 2g - n}}{n!}\,s_{m_1}\cdots s_{m_n}$ in Theorem~\ref{thDi} yields
\bea
\label{kappaaaa}&& \kappa_{n}^{[g]}({\rm Tr}\,M^{2m_1},\ldots,{\rm Tr}\,M^{2m_n}) \nonumber \\
& = & \sum_{h = 0}^{\lfloor \frac{g}{2} \rfloor} \sum_{\ell \geq 0} \frac{2^{g - 3h}}{(2h)!\ell!} \int_{\overline{\mathcal{M}}_{g - h,n + \ell + 2h}} \Lambda(-1)\Lambda(-1)\Lambda(\tfrac{1}{2}) \prod_{i = 1}^n \frac{m_i {2m_i \choose m_i}}{1 - m_i\psi_i} \prod_{i = n + 1}^{n + \ell} \frac{-\psi_{i}^2}{1 - \psi_i}.
\eea
This sum is actually finite as the degree of the class to integrate goes beyond the dimension of the moduli space. We can get rid of the $\ell$ factors of $\psi$-classes by using the pushforward relation
\beq 
\label{relka} (\pi_{\ell})_*\Big(X \prod_{i = 1}^{\ell} \psi_i^{d_i + 1}\Big) = X\,\bigg(\sum_{\sigma \in \mathfrak{S}_{\ell}} \prod_{\gamma \in \mathcal{C}(\sigma)} \kappa_{\sum_{i \in \gamma} d_i}\bigg),
\eeq
where $\pi_{\ell}\,:\,\overline{\mathcal{M}}_{g',k + \ell} \rightarrow \overline{\mathcal{M}}_{g',k}$ is morphism forgetting the last $\ell$ punctures, and $X$ is the pullback via $\pi_{\ell}$ of an arbitrary class on $\overline{\mathcal{M}}_{g',k}$. In general, if we introduce formal variables $\hat{a}_{1},\hat{a}_{2},\ldots$, we deduce from \eqref{relka} the relation
$$
\sum_{\ell \geq 0} \frac{1}{\ell!} \sum_{d_1,\ldots,d_{\ell} \geq 1} (\pi_{\ell})_{*}\bigg(X\prod_{i = k + 1}^{k + \ell} \psi_{i}^{d_i + 1}\bigg) \prod_{i = 1}^{\ell} \hat{a}_{d_j} = X \exp\Big(\sum_{d \geq 1} a_{d}\kappa_{d}\Big),
$$
where
$$
\sum_{d \geq 1} a_{d}v^{d} = -\ln\Big(1 - \sum_{d \geq 1} \hat{a}_{d}v^{d}\Big).
$$
To simplify \eqref{kappaaaa} we should apply this relation with $\hat{a}_{d} = -1$ for all $d \geq 1$. Therefore $a_{d} = -\frac{1}{d}$. Consequently
\bea
&&  \kappa_{n}^{[g]}({\rm Tr}\,M^{2m_1},\cdots,{\rm Tr}\,M^{2m_n}) \nonumber \\
& = & \sum_{h = 0}^{\lfloor \frac{g}{2} \rfloor} \frac{2^{g - 3h}}{(2h)!} \int_{\overline{\mathcal{M}}_{g - h,n + 2h}} \Lambda(-1)\Lambda(-1)\Lambda(\tfrac{1}{2})\exp\Big(-\sum_{d \geq 1} \frac{\kappa_{d}}{d}\Big) \prod_{i = 1}^n \frac{m_i {2m_i \choose m_i}}{1 - m_i\psi_i}. \nonumber
\eea
The claim is a compact rewriting of this formula, using the pushforward via the inclusions 
$$
\iota_{h}\,:\,\overline{\mathcal{M}}_{g - h,n + 2h} \rightarrow \overline{\mathcal{M}}_{g,n}.
$$ \hfill $\Box$

\section{Virasoro constraints for fully simple maps}\label{virasoro}
This section is based on work in progress with G.~Borot and D.~Lewa\'nski. Using our interpretation of the matrix model with external field in terms of fully simple maps, and the fact that we can go from fully simple to ordinary observables through double weakly monotone Hurwitz numbers, we deduce a method to compute Virasoro constraints for fully simple maps, making use of the semi-infinite wedge formalism.

Our goal is to obtain some explicit results for low topologies in order to provide some insight for the $R$-transform machinery of free probability for $n\geq 3$ and to find Tutte-like recursions for fully simple maps, which otherwise seem too involved to be deduced combinatorially. Our motivation is also to gain some understanding on both combinatorial problems involved and how they relate to each other: fully simple maps and partitioned permutations (used to define higher order free cumulants).

\subsection{Virasoro constraints for ordinary maps}\label{themmm}

We give here the explicit Virasoro constraints for ordinary maps that we introduced in Section~\ref{VirasoroIntro}. 

Let $V_0$ be a function such that all the moments of $\dd\rho(x) = e^{-NV_0(x)}$ on $\mathbb{R}$ exist. We consider again the 1-hermitian matrix model, but here with fixed $\mathbf{p^{(0)}}=(p_1^{(0)},p_2^{(0)},\ldots)$ and set of formal parameters that we denote $\mathbf{p}=(p_0,p_1,p_2,\ldots)$:
\beq 
\label{ZT}Z(\mathbf{p}) = e^{p_0N} \int_{\mathcal{H}_{N}} \dd\rho(M),
\eeq 
with $\dd\rho(M) := \dd M\,\exp[-N\,{\rm Tr}\,V(M)]$ and  $V(x) = V_0(x) - \sum_{k > 0} \frac{p_{k}}{N}\frac{x^k}{k}$.

For $V_0(x) = \frac{x^2}{2}$, we saw that \eqref{ZT} governs the combinatorics of ordinary usual maps, in the sense that $n$th derivatives with respect to $p_{k_1},\ldots,p_{k_n}$ enumerate maps with $n$ ordinary boundaries of respective perimeters $k_1,\ldots,k_n$:
$$
[N^{2-2g-n}]\  \frac{\partial}{\partial_{k_1}}\cdots\frac{\partial}{\partial_{k_n}} Z(\mathbf{p}) \Bigg{|}_{\textbf{p}=0} = \frac{F_{k_1,\ldots,k_n}^{[g]}}{k_1\cdots k_n}\,.
$$


The Virasoro constraints for the 1-hermitian matrix model easily follows from integration by parts in the matrix integral, see e.g. \cite{MMM,Eynardbook}:
\begin{theorem}
Assume $V_0(x) = - \sum_{j \geq 1} \frac{p_{j}^{(0)}}{j}\,x^j$ and we use the convention $p_{-1} = 0$. The differential operators defined for $n \geq -1$ by
$$
L_{n} = \sum_{m = 0}^n \frac{m(n - m)}{N}\,\partial_{p_m}\partial_{p_{n - m}} + \sum_{m \geq 1} (n + m)\left(\frac{p_{m}}{N} + p_{m}^{(0)}\right)\partial_{p_{n + m}} 
+ \frac{p_0}{N} n \partial_{p_n} 
$$
have the commutation relations $[L_m,L_n] = (m - n)L_{m + n}$, and are such that
\beq
\label{Vrc}\forall n \geq -1,\qquad L_n \cdot Z(\mathbf{p}) = 0.
\eeq
\end{theorem}
For usual maps, \eqref{Vrc} is equivalent to the generalization of Tutte's equation in all topologies, as explained in \cite{Eynardbook}.



\subsection{Matrix model with external field and its Virasoro constraints}

We consider again the matrix model with a fixed matrix $A\in\mathcal{H}_{N}$ as an external field:
$$
\check{Z}(A) = e^{p_0N } \!\!\! \int_{\mathcal{H}_N} \dd\rho_{A}(M), \ \ \ \ \  \dd\rho_{A}(M) = \dd\rho(M)\,e^{N\,{\rm Tr}\,MA}.
$$


We proved in Corollary~\ref{gudn} that $\hat{\mathcal{Z}}(A) = \frac{\check{Z}(A)}{\check{Z}(0)}$ can be identified with the generating series of fully simple maps.
Let us define the \emph{charge} and \emph{energy} operators
$$
H_0 :=  \sum_{k > 0} p_{k}\partial_{p_k}, \qquad \qquad  E :=  \sum_{k > 0} k p_{k}\partial_{p_k}. \\
$$
We will say that an operator has energy (or charge) $c\in\mathbb{Z}$ if it is an eigenvector of $[E,\cdot]$ $([H_0,\cdot])$ with eigenvalue $c$.
We introduce Okounkov and Pandharipande's energy $0$ operator $\mathcal{E}_0(z)$ and the operator associated to double weakly monotone Hurwitz numbers $\mathcal{O}$ on the so-called semi-infinite wedge formalism:
\bea
\mathcal{E}_0(z) & := & \frac{1}{\varsigma(z)} \sum_{s \geq 1} \sum_{m,n \geq 1} \frac{1}{m!n!} \bigg(\sum_{\substack{k_1 + \cdots + k_m = s \\ k_i \geq 1}} \prod_{i = 1}^m \frac{\varsigma(k_iz)}{k_i} p_{k_i}\bigg)\bigg(\sum_{\substack{\ell_1 + \cdots + \ell_n = s \\ \ell_i \geq 1}} \prod_{i = 1}^n \varsigma(\ell_i z) \partial_{p_{\ell_i}}\bigg), \nonumber 
\\
\mathcal{O}(N) & := & \exp\bigg(- \bigg[ \frac{\mathcal{E}_{0}(\partial_{N})}{\varsigma(\partial_{N})} - E \bigg] \ln N\bigg), \nonumber
\eea
where $\varsigma(z) = e^{z/2} - e^{-z/2}$.  We are going to prove in Section~\ref{proofZAOZ}, after introducing the semi-infinite wedge formalism, the following result:

\begin{theorem} \label{thm:OZ} Let $A \in \mathcal{H}_N$. Then 
\begin{equation}
\hat{\mathcal{Z}}(A) = \mathcal{O}(N). Z(\textbf{p})\Big{|}_{p_k =\, \Tr A^k},
\end{equation}
where the operator $\mathcal{O}(N)$ acts on the Schur function expansion of $Z(\textbf{p})$.
\end{theorem}
This theorem provides a recipe to calculate Virasoro contraints for fully simple maps, as we deduce in the following corollary:
\begin{corollary}
The operators $\mathcal{O}L_{n}\mathcal{O}^{-1}$ annihilate $\hat{\mathcal{Z}}(A)\Big{|}_{\Tr A^k =\, p_k}$. 
\end{corollary}
 

It is therefore important to compute the conjugated operators $\mathcal{O}\partial_{p_k}\mathcal{O}^{-1}$ and $\mathcal{O}p_{k}\mathcal{O}^{-1}$ to get explicit Virasoro constraints for the 1-hermitian matrix model with external field. 

We remark that the same technique can be used to produce Virasoro constraints for fully simple stuffed maps, after conjugating the Virasoro constraints for stuffed maps with the same operator $\mathcal{O}$.

%

\subsection{Semi-infinite wedge formalism}

We briefly introduce here the so-called semi-infinite wedge formalism, whose pioneers were Okounkov and Pandharipande \cite{Okounkov,OkoPand1,OkoPand2} and which turned into a standard tool in Hurwitz theory. For more details, one can check for example \cite{Oko01,SemiInfWedge} and references therein.

\subsubsection{Clifford and Heisenberg algebras}

Let $\mathbb{Z}_{F} = \mathbb{Z} + \tfrac{1}{2}$.
 We consider the Clifford algebra $\mathfrak{cl}$, defined by a family of generators $\psi_{s}$ and $\psi_{s}^{\dagger}$ indexed by $s \in \mathbb{Z}_{F}$, with relations
$$
\{\psi_{r},\psi_{s}\} = \{\psi_{r}^{\dagger},\psi_{s}^{\dagger}\} = 0,\qquad \{\psi_{r},\psi_{s}^{\dagger}\} = \delta_{r,s}\,.
$$
The generators with $s < 0$ (resp. $s > 0$) are called creation operators (resp. annihilation operators). If~$P$ is a non-constant monomial in $\psi$'s and $\psi^{\dagger}$'s, we define its normal ordering $:P:$ to be the monomial where the annihilation operators are located to the right of creation operators, multiplied by the sign of the permutation that had to be applied to achieve this ordering. The normal ordering is well-defined, since the $\psi$'s and the $\psi^{\dagger}$ commute among themselves. We also take the convention $:1: = 0$ and the normal ordering can be extended by linearity to any polynomial in $\psi$ and $\psi^{\dagger}$. For instance,
$$
:\psi_{r}\psi_{s}^{\dagger}:= \left\{\begin{array}{lll} -\psi_{s}^{\dagger}\psi_{r}, & & \text{if } s < 0, \\ \psi_{r}\psi_{s}^{\dagger}, & & {\rm otherwise.} \end{array}\right.
$$
Using the anticommutation relations, we can also write
$$
:\psi_{r}\psi_{s}^{\dagger}:\,= \psi_{r}\psi_{s}^{\dagger} - \delta_{r,s}\delta_{s < 0}\,.
$$
It is convenient to collect them in generating series
$$
\psi(\zeta) = \sum_{s \in \mathbb{Z}_{F}} \psi_{s}\zeta^{s - \frac{1}{2}},\qquad \psi^{\dagger}(\zeta) = \sum_{s \in \mathbb{Z}_{F}} \psi_{s}^{\dagger}\zeta^{-s-\frac{1}{2}}\,.
$$
The following vector space
$$
\widehat{\mathfrak{gl}}_{\infty} = \Big\{c + \sum_{r,s} X_{r,s}\,:\psi_{r}\psi_{s}^{\dagger}: \ \ \Big|\ \ \ c,X_{r,s} \in \mathbb{Q},\quad X_{r,s} = 0,\text{ for } |r - s| \text{ large enough}\Big\}\,.
$$
is a Lie algebra. One defines a group
$$
\widehat{{\rm GL}}_{\infty} = \big\{\exp(X_1)\cdots \exp(X_n),\quad X_i \in \widehat{\mathfrak{gl}}_{\infty}\big\} \,.
$$

We introduce the hamiltonians
$$
H(\zeta) =\,: \psi(\zeta)\psi^{\dagger}(\zeta):\,\,\in \widehat{\mathfrak{gl}}_{\infty}((\zeta))\,,
$$
which can be decomposed as
$$
H(\zeta) = \sum_{n \in \mathbb{Z}} H_n\,\zeta^{-n - 1},\qquad H_n = \sum_{s \in \mathbb{Z}_{F}} :\psi_{s}\psi_{n + s}^{\dagger}:\,.
$$
From the relations in the Clifford algebra, we can deduce the commutation relation of these operators
$$
[H_{m},H_{n}] = m\delta_{m + n,0},\qquad [H_n,\psi(\zeta)] = \zeta^n\psi(\zeta),\qquad [H_n,\psi^{\dagger}(\zeta)] = -\zeta^n\psi^{\dagger}(\zeta)\,.
$$  
We call diagonal operators the $:\psi_{s}\psi_{s}^{\dagger}:$ for $s \in \mathbb{Z}_{F}$. They form a Lie subalgebra $\mathfrak{d}_{\infty} \subset \widehat{\mathfrak{gl}}_{\infty}$. We denote $\overline{\mathfrak{d}}_{\infty}$ its completion, whose elements are infinite formal series 
$$ 
\sum_{s \in \mathbb{Z}_{F}} a_{s}\,:\psi_{s}\psi_{s}^{\dagger}:\,,\qquad a_{s} \in \mathbb{Q}\,.
$$

\subsubsection{Semi-infinite wedge}

Let $\mathcal{V} = z^{\frac{1}{2}}\mathbb{Q}[[z^{-1},z]$ and $\mathcal{V}_- = z^{\frac{1}{2}}\mathbb{Q}[[z^{-1}]]$. We denote $e_{s} = z^{s - \frac{1}{2}}$ the basis elements for $s \in \mathbb{Z}_{F}$, and $(e_s^*)$ the dual basis. The semi-infinite wedge space $\mathcal{F}$ is the vector space spanned by elements $z^{n_1} \wedge z^{n_2} \wedge \cdots$ where $(n_i)_{i \geq 1}$ is a strictly decreasing sequence of half integers, such that $n_{i + 1} = n_i -1$, for $i$ large enough. We define a scalar product $\langle \cdot,\cdot \rangle$ on $\mathcal{F}$ by declaring this basis to be orthonormal. The space $\mathcal{F}$ is equipped with a representation of $\widehat{\mathfrak{gl}}_{\infty}$, defined as
$$
\psi_{s} = e_{s} \wedge \cdot,\qquad \psi_{s}^{\dagger} = \iota_{e_{s}^*}\,,
$$
where $\iota$ is the interior product. The action of $:\psi_{r}\psi_{s}^{\dagger}:$ amounts (up to a sign) to sending $z^{s - \frac{1}{2}}$ to $z^{r - \frac{1}{2}}$, and all other basis elements to $0$, in a semi-infinite wedge vector. The action of $H_n$ is induced on $\mathcal{F}$ by the endomorphism of $\mathcal{V}$ sending $f(z)$ to $z^{-n}f(z)$. The space $\mathcal{F}$ decomposes as a direct sum
$$
\mathcal{F} = \bigoplus_{N \in \mathbb{Z}} \mathcal{F}_{N}
$$
where $\mathcal{F}_{N}$ is the eigenspace of $H_0$ for the eigenvalue $N$. $H_0$ is called the charge operator. $E$ is called energy operator. We denote $\Omega_{\emptyset} \in \mathcal{F}_{0}$ the semi-infinite wedge vector $z^{-\frac{1}{2}} \wedge z^{-\frac{3}{2}} \wedge z^{-\frac{5}{2}} \wedge \cdots $, and $\Omega_{\emptyset}^{(\ell)} \in \mathcal{F}_{\ell}$ the semi-infinite wedge vector $z^{-\ell -\frac{1}{2}} \wedge z^{-\ell -\frac{3}{2}} \wedge \cdots$. In general, if $\lambda$ is a partition, we complete it by putting $\lambda_i = 0$ for $i > \ell(\lambda)$, and one can define a semi-infinite wedge vector $\Omega_{\lambda} \in \mathcal{F}_{0}$ as $z^{\lambda_1-\frac{1}{2}} \wedge z^{\lambda_2-\frac{3}{2}} \wedge z^{\lambda_3-\frac{5}{2}} \wedge \cdots $.
If $\mathbf{p} = (u_1,u_2,\ldots)$ are formal variables, let us define
$$
H_{\pm}(\mathbf{p}) \coloneqq \sum_{k > 0} \frac{p_{k}}{k}\,H_{\pm k}.
$$

\begin{theorem} (Boson-fermion correspondence)
The linear map
$$
\begin{array}{crcl} T: & \mathcal{F} & \longrightarrow & \mathbb{Q}[[p_1,p_2,\ldots]][[\zeta^{\pm 1}]] \\ & v & \mapsto & \sum_{\ell \in \mathbb{Z}} \langle\Omega_{\emptyset}^{(\ell)}, \exp(H_+(\mathbf{u})) v \rangle\,\zeta^{\ell} \end{array}
$$
is an isomorphism. Besides
\bea
T(H_{k}v) & = & \left\{\begin{array}{lll} k\partial_{p_k} T(v), && {\rm if}\,\,k > 0, \\ p_{-k}T(v), & & {\rm if}\,\,k < 0, \end{array}\right. \nonumber \\
T(\psi(z)v) & = & \exp\Big(\sum_{k > 0} \frac{p_{k}}{k}\,z^{k}\Big)\exp\Big(-\sum_{k > 0} z^{-k}\,\partial_{p_k}\Big) \zeta \exp(z\zeta\partial_{\zeta})\,T(v), \nonumber \\
T(\psi^{\dagger}(z)v) & = & \exp\Big(-\sum_{k > 0} \frac{p_{k}}{k}\,z^k\Big)\exp\Big(\sum_{k > 0} z^{-k}\partial_{p_k}\Big)\zeta^{-1}\exp(-z\zeta\partial_{\zeta})\,T(v)\,. \nonumber
\eea
We have in terms of the Schur polynomials:
$$
T(\Omega_{\lambda}) = s_{\lambda},
$$
where $p_k$ are interpreted as the $k$th power sum.
\end{theorem}
This last remark shows that $T$ restricts to a linear isomorphism $T^{(0)}$
\beq
\label{3iso}\mathcal{F}_{0} \mathop{\simeq}^{T^{(0)}} \mathbb{Q}[[u_1,u_2,\ldots]] \simeq \mathcal{B} \simeq \bigoplus_{L \geq 0} Z(\mathbb{Q}[\mathfrak{S}_{L}]),
\eeq
where $\mathcal{B}$ is the vector space of symmetric polynomials in infinitely many variables -- which we saw (in Section \ref{SymmetricFunctions}) is in fact a (graded) ring. 


\subsubsection{From the ordinary to the fully simple partition function}\label{proofZAOZ}

We are now armed to prove the theorem:

\vspace{0.1cm}

\noindent \textbf{Proof of Theorem \ref{thm:OZ}.}
Using Corollary \ref{gudn} and \eqref{contentfct}, one can show that
\begin{align}\label{eq:ZAschur}
\hat{\mathcal{Z}}(A) = \frac{\check{Z}(A)}{\check{Z}(0)} &= \sum_{\lambda} \frac{N^{|\lambda|} \chi_{\lambda}({\rm id})}{|\lambda|! s_{\lambda}({\rm 1}_N)} \langle s_{\lambda}(M) \rangle s_{\lambda}(A) \\
\nonumber
& = \sum_{\lambda}  \langle s_{\lambda}(M) \rangle \sum_{k \geq 1} (-N)^{-k} h_k({\rm cont}\,\lambda)  s_{\lambda}(A),
\end{align}
where the $\lambda$-sum is taken over all partitions $\lambda$ and the measure is unitary invariant. The generating series for the $h_k({\rm cont}\,\lambda)$ is by \cite{ALS} the eigenvalue of the operator $\mathcal{O}(N)$ applied to the basis vector $\Omega_{\lambda} $ in the charge zero sector $\mathcal{F}_0$ of $\mathcal{F}$. The operator $\mathcal{O}$ here is related to the operator in \cite{ALS} by $\mathcal{D}^{(h)}(-z^{-1}) = \mathcal{O}(z)$. 
The action of $\mathcal{O}(N)$ on the space of symmetric functions can be easily read from the boson-fermion correspondence $T^{(0)}$ that sends $\Omega_{\lambda}$ to the Schur function $s_\lambda$. Hence we obtain
$$
 \hat{\mathcal{Z}}(A) = \sum_{\lambda} \langle s_{\lambda}(M) \rangle \left[\mathcal{O}^{(h)}(N). s_{\lambda}\right](A) = \left[  \mathcal{O}^{(h)}(N). \sum_{\lambda}  \langle s_{\lambda}(M) \rangle s_{\lambda}\right](A).
 $$
On the other hand, we can express the partition funtion $Z$ as a function of the variables $\textbf{p}$ as follows:
\begin{align*}
Z(\textbf{p}) &= \int_{\mathcal{H}_N} \dd \rho_0 (M) e^{N \sum_{k=1}^{\infty} \frac{p_k \Tr(M^k)}{N \, k}}  = \left\langle \exp\left( \sum_{k=1}^{\infty} \frac{p_k \Tr(M^k)}{k} \right) \right\rangle_0
\end{align*}
where $\langle \cdot \rangle_0$ is the expectation value taken with respect to the measure $d \rho_0 ( M )= dM e^{-N \Tr V_0(M)}$. Evaluating the variables $p_k$ in $p_k = \Tr A^k$ for a matrix $A \in \mathcal{H}_N$ we obtain that
\begin{align*}
Z(\textbf{p})\Big{|}_{p_k =\, \Tr A^k}  &= \left\langle \exp\left( \sum_{k=1}^{\infty} \frac{\Tr(A^k) \Tr(M^k)}{k} \right) \right\rangle_0 = \left\langle \exp\left( \sum_{k=1}^{\infty} \frac{p_k(A) p_k(M)}{k} \right) \right\rangle_0 = \\
& =  \left\langle \sum_{\lambda} s_{\lambda}(M) s_{\lambda}(A) \right\rangle_0 =  \sum_{\lambda} \left\langle s_{\lambda}(M) \right\rangle_0 s_{\lambda}(A) 
\end{align*}
as functions on $\mathcal{H}_N$. Applying $\mathcal{O}(N)$ to this expression and chosing the expectation value $\langle \cdot \rangle_0$ in \eqref{eq:ZAschur} concludes the proof of the theorem.\hfill $\Box$

\subsection{Conjugation of $L_n$}

Following \cite{OkoPand2}, we introduce the following notation for functions related to the hyperbolic sine, which will be useful for us:
$$
\varsigma(z) = e^{z/2} - e^{-z/2},
$$
\begin{equation}
\mathcal{S}(z) = \frac{\varsigma(z)}{z}= \frac{\sinh(z/2)}{z/2} = \sum_{k\geq 0}\frac{z^{2k}}{2^{2k}(2k+1)!},
\end{equation}

\begin{equation}
\frac{1}{\mathcal{S}(z)} = \frac{z/2}{\sinh(z/2)} =  1 + \sum_{k\geq 1} \frac{(2^{2k-1} -1)}{2^{2k-1}}\frac{|B_{2k}|}{(2k)!} z^{2k-1}.
\end{equation}

For any $a \in \mathbb{Z}$, we define the Okounkov and Pandharipande's operator with energy $a$:
\bea 
\mathcal{E}_{a}(z) & = & \frac{1}{\varsigma(z)} \sum_{r \geq 0} \frac{1}{r!} \sum_{\substack{m_1,\ldots,m_r \in \mathbb{Z} \setminus \{0\} \\ \sum_{i} m_i = a}} \bigg[ \prod_{i = 1}^r \frac{\varsigma(m_i z)}{m_i}\bigg]\,: H_{m_1} \cdots H_{m_r} :   \\
& = & \frac{1}{\varsigma(z)} \sum_{s \geq 0} \sum_{n,m \geq 0} \frac{1}{m!n!} \bigg(\sum_{\substack{k_1 + \cdots + k_m = s \\ k_i \geq 1}} \prod_{i = 1}^m \frac{\varsigma(k_iz)}{k_i} p_{k_i}\bigg)\bigg(\sum_{\substack{\ell_1 + \cdots + \ell_{n} = s + a \\ \ell_i > 0}} \prod_{i = 1}^n \varsigma(\ell_i z)\partial_{p_{\ell_i}}\bigg), \nonumber
\eea 
with the following commutation relations:
\beq
\label{Ecomut} [\mathcal{E}_{a}(z),\mathcal{E}_{b}(w)] = \varsigma(aw - bz)\mathcal{E}_{a + b}(z + w).
\eeq

%
The operators $\mathcal{F}_{r}^{(a)}$ are defined by the expansion
$$
\mathcal{E}_{a}(z) = \sum_{r\geq 0} \mathcal{F}_{r}^{(a)} z^r.$$

\begin{lemma}\label{lem:conjbasicoper}
The operators $\mathcal{O} b \partial_{p_b} \mathcal{O}^{-1} $ and $\mathcal{O} p_b \mathcal{O}^{-1} $ are power series in $N^{-1}$, and polynomial in $N^{-1}$ respectively, whose coefficients are explicit polynomials in the $\mathcal{F}_{r}^{(a)} $:
\begin{align}
\mathcal{O} b \partial_{p_b} \mathcal{O}^{-1} &=  \sum_{k\geq 0} (-1)^k \left[ 
\frac{b!}{(b-k)!} \sum_{a=0}^{k} (-1)^a\mathcal{F}_a^{(b)} [z^{k-a}].\left( \mathcal{S}(z)^{-b-1}\right) 
 \right]
  N^{-k},
\\
\mathcal{O} p_b \mathcal{O}^{-1} &= \sum_{k=0}^{b} (-1)^k\left[ 
 \frac{(k + b - 1)!}{(b-1)!} \sum_{a=0}^{k} \mathcal{F}_{a}^{(-b)} [z^{k-a}]. \left(\mathcal{S}(z)^{b-1}\right)
  \right] N^{-k}\,.
\end{align}
Note that since $\mathcal{S}$ is an even function only the summands where $a$ and $k$ have the same parity contribute.
\end{lemma}
\noindent \textbf{Proof.}
From \cite[Corollary~4.3]{KLS}, we get
\begin{align}
\mathcal{O} b \partial_{p_b} \mathcal{O}^{-1} &=  \sum_{k\geq 0} \frac{b!}{(b-k)!} (-N^{-1})^k [z^k]. \mathcal{S}(z)^{-b-1} \mathcal{E}_b(-z),\\
\mathcal{O} p_b \mathcal{O}^{-1} &= \sum_{k=0}^{b}   \frac{(k + b - 1)!}{(b-1)!} (-N^{-1})^k [z^k]. \mathcal{S}(z)^{b-1}\mathcal{E}_{-b}(z)\,.
\end{align}\hfill $\Box$

 We want to compute the operator
 \begin{align*}
 \mathcal{O} L_n \mathcal{O}^{-1} = &\, \, \frac{1}{N} \!\sum_{\substack{b_1 + b_2 = n \\ b_1, b_2 \geq 1}} [\mathcal{O} (b_1 \partial_{p_{b_1}} ) \mathcal{O}^{-1}][\mathcal{O} (b_2 \partial_{p_{b_2}} ) \mathcal{O}^{-1}] 
 \, + \, 
 \frac{1}{N} \!\sum_{b\geq 0} [ \mathcal{O} p_b \mathcal{O}^{-1}][\mathcal{O} ((b+n) \partial_{p_{b + n}} ) \mathcal{O}^{-1}]\\
& \, + \sum_{b\geq 1}  p^{(0)}_b[\mathcal{O} ((b+n) \partial_{p_{b + n}} ) \mathcal{O}^{-1}]
 \end{align*}
 as an operator in the $p$'s. For $r,\, b\in\mathbb{Z}$, we introduce the notation: $\mathcal{S}_{r}^{(b)}  \coloneqq [z^r].\mathcal{S}(z)^{b}$. By Lemma \ref{lem:conjbasicoper}, we obtain that $ \mathcal{O} L_n \mathcal{O}^{-1}$ is equal to
  \begin{align}\label{eq:explicitOLO}
& \sum_{\substack{b_1 + b_2 = n \\ b_1, b_2 \geq 1}} 
\sum_{k\geq 0} (-1)^k\! \left[ 
\sum_{k' + k'' = k}\sum_{a'= 0}^{k'} \sum_{a''=0}^{k''}  \frac{ b_1! \, b_2!}{(b_1 - k')! (b_2 - k'')!} (-1)^{a' + a''} \mathcal{S}_{k' - a'}^{(-b_1 -1)} \mathcal{S}_{k'' - a''}^{(-b_2 -1)}
\mathcal{F}_{a'}^{(b_1)}\mathcal{F}_{a''}^{(b_2)}
\right]\! N^{-k-1} \nonumber
 \\ 
 &+\sum_{b\geq 0}
 \sum_{k \geq 0}(-1)^k\! \left[ 
\sum_{k' + k'' = k}\sum_{a'= 0}^{k'} \sum_{a''=0}^{k''}  \frac{(b + n)! (k'' + b - 1)!}{( b + n - k')! (b - 1)!} (-1)^{a'} \mathcal{S}_{k' - a'}^{(-b - n -1)} \mathcal{S}_{k'' - a''}^{( b -1)}
\mathcal{F}_{a''}^{(-b)}\mathcal{F}_{a'}^{(b + n)}
\right]\! N^{-k-1} \nonumber
 \\
 &+\sum_{b\geq 1} p^{(0)}_b \sum_{k\geq 0} (-1)^k\!\left[ \frac{(b + n)!}{(b + n - k)!}  \sum_{a=0}^{k} (-1)^a \mathcal{S}_{k-a}^{(-b-n -1)} \mathcal{F}_a^{(b)} \right]\! N^{-k}.
 \end{align}

\subsection{Tutte's recursion for fully simple disks}
In this section we test our operators in the case of disks, i.e., for the $(g,n) = (0,1)$ topology. We first compute Tutte's equation for fully simple disks from Tutte's equation for ordinary disks. Then we derive it from the Virasoro operators $\mathcal{O} L_n \mathcal{O}^{-1}$ by selecting the coefficient of $N^{2-2g-n}=N$ in $\mathcal{O} L_n \mathcal{O}^{-1}\big(\hat{\mathcal{Z}}(A)\big{|}_{\Tr(A)^k = p_k}\big)$ and show that the two equations match.
\subsubsection{Derivation from Tutte's recursion for ordinary disks}Tutte's recursion for ordinary disks, which can be easily deduced from bijective combinatorics, reads
\begin{equation}\label{eq:Tutteusual}
F_{l+1} = \sum_{j=1}^{l-1} F_{j} F_{l-1-j} + \sum_{j=3}^d p_j^{(0)} F_{l-1 + j}.
\end{equation}
We remark that here the fixed parameters $p_j^{(0)}$, $j\geq 3$, play the role of weight per internal $j$-gon that we denoted $t_j$ in the rest of the thesis.
After the identification established in Section \ref{HOFCsFS} between correlation moments and generating series of ordinary maps, and free cumulants and generating series fully simple maps, we can make use of the moment-(free) cumulant relations from free probability. The case of disks corresponds to first order (or classical) free probability \ref{1stOrderFCsSection}, hence the relation is just via non-crossing partitions \eqref{1stOrderFCs}:
\begin{equation}\label{}
F_{l} = \sum_{\pi \in NC(l)} \prod_{B\in \pi}H_{|B|},
\end{equation}
where $NC(l)$ is the finite set of non-crossing partitions of the set $\{1, 2, \dots, l\}$. We can restate the last equation summing over partitions of the number $l$ (we stress that the term partition is used to indicate two distinct combinatorial objects):
\begin{equation}\label{eq:FHc}
F_{l} = \sum_{\lambda \vdash l} H_{\lambda_1} \cdots H_{\lambda_{\ell(\lambda)}} c_{\lambda},
\end{equation}
where $c_{\lambda}$ is the number of non-crossing partitions of the set $\{1, 2, \dots, l \}$ with exactly $\ell(\lambda)$ blocks, and whose sizes are the parts of $\lambda$. Substituting \eqref{eq:FHc} in \eqref{eq:Tutteusual}, we get

\begin{align*}
\sum_{\lambda \vdash l + 1} H_{\lambda_1}\cdots H_{\lambda_{\ell(\lambda)}} c_{\lambda} = \sum_{\lambda \vdash l - 1} H_{\lambda_1}\cdots H_{\lambda_{\ell(\lambda)}} \Big( \sum_{\substack{\lambda = \mu^{(1)} \sqcup \mu^{(2)} \\ \mu^{(i)} \neq \emptyset }} c_{\mu^{(1)}} c_{\mu^{(2)}}\Big) 
+ \!\!\!\! \sum_{\substack{ j = 3 \\ \lambda \vdash l - 1 + j}} p_j^{(0)} H_{\lambda_1}\cdots H_{\lambda_{\ell(\lambda)}} c_{\lambda}.
\end{align*}
The constants $c_{\lambda}$ have been computed combinatorially, and they are given by the following formula:
\begin{theorem}\cite{Kreweras}\label{thm:Kreweras} Let $m_i(\lambda)$ denote the number of parts $\lambda_j$ that are equal to $i$. With the notation $M(\lambda)\coloneqq\prod_{i\geq 1} m_i(\lambda)!$, we have
\begin{equation}
c_{\lambda} = \frac{1}{M(\lambda)}\frac{|\lambda|!}{(|\lambda | - (\ell(\lambda)- 1 ))!}\,.
\end{equation}
\end{theorem}
Observe that we have the following relation between $M(\lambda)$ and what we denoted $|{\rm Aut}\, \lambda|=\frac{|\lambda|!}{|C_{\lambda}|}$ in this thesis:
$$
M(\lambda)=\frac{|{\rm Aut}\, \lambda|}{\prod_{i=1}^{\ell(\lambda)}\lambda_i}\,.
$$
Carrying every term on the same side and using the theorem above, we obtain the following explicit Tutte equation for the fully simple disks.
\begin{align}\label{eq:tuttefs}
&\sum_{k=1} \sum_{\substack{\lambda \vdash l - 1 \\ \ell(\lambda) = k}}  H_{\lambda_1} \cdots H_{\lambda_{k}} 
\Bigg( \!\!\!\!\!\!\!\!
\sum_{\substack{\lambda = \mu^{(1)} \sqcup \, \mu^{(2)}
\\
\ell(\mu^{(1)}) = k' \geq 1, \, \ell(\mu^{(2)}) = k'' \geq 1
\\
k' + k'' = k
}}
\!\!\!\!\!\!\!\!
 \frac{1}{M(\mu^{(1)})  M(\mu^{(2)})} \frac{|\mu^{(1)}|!}{(|\mu^{(1)}| - k' + 1)!}\frac{|\mu^{(2)}|!}{(|\mu^{(2)}| - k'' + 1)!}
\Bigg) \\
&+ \sum_{k=1}\sum_{\substack{ j = 1 \\ \lambda \vdash l - 1 + j \\ \ell(\lambda) = k }} p_j^{(0)} \frac{H_{\lambda_1}\cdots H_{\lambda_{k} }}{M(\lambda)} \frac{(l-1+j)!}{(l + j - k)!}\Big{|}_{p_1^{(0)} =\, 0,\,\,  p_2^{(0)} =\, -1} = 0\,. \nonumber
\end{align}
The specialization $p_1^{(0)} = 0$, $p_2^{(0)}=-1$ corresponds precisely to the convention we chose for the $t_j$'s in the matrix model for ordinary maps, which amounts combinatorially to count maps without internal $1$-gons and $2$-gons.

\subsubsection{Derivation from the conjugated Virasoro operators}

We compute the constraints on generating series of fully simple disks from the conjugated Virasoro operators $\mathcal{O} L_{n} \mathcal{O}^{-1}$. Concretely, we shall prove that equation \eqref{eq:tuttefs} coincides with the equation
\begin{equation}\label{eq:selN1}
[N^1] \ \ \mathcal{O} L_{n} \mathcal{O}^{-1}. \left(\hat{\mathcal{Z}}(A)\big{|}_{\Tr(A)^k = p_k} \right) \Bigg{|}_{\textbf{p} = 0} = 0,
\end{equation}
under the substitution $ p_1^{(0)} =\, 0,\,\,  p_2^{(0)} =\, -1$, where the expansion of the partition function in the $p_j$'s reads
$$
\hat{\mathcal{Z}}(A)\big{|}_{\Tr(A)^k = p_k } = \exp \left( \sum_{\substack{g \geq 0 \\n \geq 1}} N^{2 - 2g - n}\sum_{l_1,\ldots,l_n\geq 0} H_{l_1, \dots, l_n}^{[g]} \frac{p_{l_1}\cdots p_{l_n}}{l_1 \cdots l_n}\right).
$$

We are going to show that, when selecting the coefficient of $N^1$ in \eqref{eq:selN1}, many of the terms in its explicit expression -- given in \eqref{eq:explicitOLO} -- cannot contribute. Indeed, let us select an operator corresponding to the index $\bar{k}$ in the $k$-sum of one of the three lines of \eqref{eq:explicitOLO}. This operator multiplies $N^{-\bar{k}-1}$ if it is selected from the first or the second line, whereas it multiplies $N^{-\bar{k}}$ if it is selected from the third line. In order for the resulting term to be of order $N^1$, we need $\bar{k} + 2$ (or $\bar{k} + 1$ if it is selected from the third line) powers of $N$ from the application of the operator. Observe that applying a single derivation at the time to pick up a $(g,n) = (0,1)$ term from the partition function is the only way to obtain strictly positive powers of $N$ out of the application of each summand of the operator:
\begin{equation}\label{eq:pulldownN1}
[N^{>0}]\ \ \ \alpha \partial_{p_\alpha} \exp \left( \sum_{\substack{g \geq 0 \\n \geq 1}} N^{2 - 2g - n}\sum_{l_1,\ldots,l_n\geq 0} H_{l_1, \dots, l_n}^{[g]} \frac{p_{l_1}\cdots p_{l_n}}{l_1 \cdots l_n}\right) \Bigg{|}_{\textbf{p}=0} = H_{\alpha}^{[0]} N^{1}\,, \text{ for } \alpha\in\mathbb{Z}_{\geq 0}.
\end{equation}
Therefore we need at least $\bar{k} + 2$ (respectively,  $\bar{k} + 1$) derivatives. On the other hand, $\bar{k} + 2$ (respectively,  $\bar{k} + 1$) derivatives is actually the maximum amount of derivatives that we can extract from the operator. To see this, recall that $\mathcal{F}_{a}^{(b)}$ equals
\begin{equation}\label{eq:expansionFab}
[z^a]\frac{1}{\varsigma(z)} \sum_{s \geq 0} \sum_{m', m'' \geq 0} \frac{1}{m'!m''!} \bigg(\sum_{\substack{k_1 + \cdots + k_{m'} = s \\ k_i \geq 1}} \prod_{i = 1}^{m'} \frac{\varsigma(k_iz)}{k_i} p_{k_i}\bigg)\bigg(\sum_{\substack{\ell_1 + \cdots + \ell_{m''} = s + b \\ \ell_i > 0}} \prod_{i = 1}^{m''} \varsigma(\ell_i z)\partial_{p_{\ell_i}}\bigg)
\end{equation}
and that 
$$
\frac{1}{\varsigma(z)} = \frac{1}{z} + O(z), \qquad \varsigma(z) = z + O(z^3).
$$
Therefore collecting the maximum amount of derivatives corresponds to picking $z^{-1}$ from $\varsigma(z)^{-1}$, $m'=0$, $m'' = a+1, s=0$, and the leading terms from each of the $\varsigma(l_i z)$, obtaining the term
$$
\frac{1}{(a+1)!} \sum_{\substack{l_1 + \dots + l_{a+1} = b \\ l_j \geq 1}}\prod_{i=1}^{a+1} l_i \partial_{p_{l_i}}.
$$
Now, observe that the only possibility to obtain the desired amount of derivatives is to choose maximal indices in the $a$-sums (that is, pick the summands $a' = k'$ and $a'' = k''$ in the first and second line, and $a=k$ in the third line of \eqref{eq:explicitOLO}). This simplifies the formula considerably: the $\mathcal{S}$-contributions involved are all of the form 
$$
\mathcal{S}_0^{(r)} = [z^0]. (\mathcal{S}(z))^r = [z^0](1 + O(z^2))^r = 1,
$$
there are no subleading contributions from the $\varsigma(z)$ functions in the $\mathcal{F}$ operators, and the $(-1)^k$ terms cancel out with the minus signs coming from the $a$-sums.

We want to put in evidence the part of the conjugated Virasoro operator with trivial $\mathcal{S}$-function contributions, i.e. the summands obtained from \eqref{eq:explicitOLO} by selecting $a'= k', a'' = k'', a = k$.

\begin{definition}\label{def:simplifiedVirasoro}
We call \textit{simplified version of the conjugated Virasoro} \eqref{eq:explicitOLO} the following operator:
 \begin{align} \label{eq:olo1a=kfirst}
& \sum_{\substack{b_1 + b_2 = n \\ b_1, b_2 \geq 1}} 
\sum_{k\geq 0} 
\sum_{k' + k'' = k} \frac{ b_1! \, b_2!}{(b_1 - k')! (b_2 - k'')!}  
\mathcal{F}_{k'}^{(b_1)}\mathcal{F}_{k''}^{(b_2)}
 N^{-k-1} 
 \\
 +&\sum_{b\geq 0}
 \sum_{k\geq 0}
\sum_{k' + k'' = k} \frac{(b + n)! (k' + b - 1)!}{( b + n - k'')! (b - 1)!} (-1)^{k'} 
\mathcal{F}_{k'}^{(-b)}\mathcal{F}_{k''}^{(b + n)}N^{-k-1} \label{eq:olo1a=ksecond}
 \\
 +&\sum_{b\geq 1} p^{(0)}_b \sum_{k\geq 0} \frac{(b + n)!}{(b + n - k)!}  \mathcal{F}_k^{(b+n)} N^{-k}.
 \label{eq:olo1a=kthird}
 \end{align}
\end{definition}

We have shown that only the simplified version of the operator as in definition \ref{def:simplifiedVirasoro} can contribute. Let us compute the contribution of each term separately:

\vspace{0.2cm}

\noindent $\bullet\ $ \textit{Contribution of \eqref{eq:olo1a=kfirst}.}
We showed that this contribution must come from the coefficient of $N^1$ in 
\begin{equation*}
\sum_{\substack{b_1 + b_2 = n \\ b_1, b_2 \geq 1}} 
\sum_{k\geq 0}
\sum_{\substack{k' + k'' = k \\ k', k'' \geq 0}} \frac{b_1!\, b_2!}{(b_1 - k')! (b_2 - k'')!} 
\sum_{\substack{
l_1^{(1)} + \dots + l_{k'+1}^{(1)} = b_1 
\\
l_1^{(2)} + \dots + l_{k''+1}^{(2)} = b_2
\\
 l_j^{(1)} \geq 1, l_j^{(2)} \geq 1 
 }}
 \frac{\prod_{i=1}^{k'+1} l_i^{(1)} \partial_{p_{l_i^{(1)}}}}{(k' + 1)!}
  \frac{\prod_{i=1}^{k''+1} l_i^{(2)} \partial_{p_{l_i^{(2)}}}}{(k''+1)!} 
 N^{-k-1}
 \end{equation*}
 applied to the partition function. We know that every derivative must pull exactly one power of $N$ down, together with a factor $H_{\alpha}^{[0]}$, as in equation \eqref{eq:pulldownN1}, or otherwise the entire term will not have the desired power of $N$. By collecting the coefficient of $N^1$, setting $\textbf{p} = 0$, and shifting the indices of the $k$-sums by one we obtain the contribution
 \begin{equation*}
 \sum_{\substack{b_1 + b_2 = n \\ b_1, b_2 \geq 1}} 
\sum_{\substack{ k', k'' \geq 1}} \frac{b_1!\, b_2!}{(b_1 - k' + 1)! (b_2 - k'' + 1)!} 
\sum_{\substack{
l_1^{(1)} + \dots + l_{k'}^{(1)} = b_1 
\\
l_1^{(2)} + \dots + l_{k''}^{(2)} = b_2
\\
 l_j^{(1)} \geq 1, l_j^{(2)} \geq 1 
 }}
 \!\!\!\!
 \frac{\prod_{i=1}^{k'} H_{l_i^{(1)}}^{[0]}}{k'!}
 \frac{ \prod_{i=1}^{k''} H_{l_i^{(2)}}^{[0]}}{k''!}.
 \end{equation*}
 
\noindent $\bullet\ $ \textit{Contribution of \eqref{eq:olo1a=ksecond}.}
We show that this contribution is equal to zero by observing that each of its summands involves a nontrivial monomial in the $p_j$ multiplied on the left. This monomial then cannot be annihilated by any derivative, and therefore the summand vanishes when we set $\textbf{p}=0$. To see this, note that each $\bar{k} > 0$ summand involves the operator $\mathcal{F}^{-b < 0}_{a'' = \bar{k} > 0}$ with negative energy, which must have positive $m'$, and hence a monomial in the $p_j$ on the left. In case $\bar{k} = 0$, the second summand simply reduces to the second term in the non-conjugated case: $p_b (b+n)\partial_{p_{b+n}}$.

\vspace{0.1cm}

\noindent $\bullet\ $ \textit{Contribution of \eqref{eq:olo1a=kthird}.}
We showed that this contribution must come from the coefficient of $N^1$ in 
\begin{equation*}
\sum_{b\geq 1} p^{(0)}_b \sum_{k\geq 0} \frac{(b + n)!}{(b + n - k)!}
\frac{1}{(k+1)!} \sum_{\substack{l_1 + \dots + l_{k+1} = b+n \\ l_j \geq 1}}\prod_{i=1}^{k+1} l_i \partial_{p_{l_i}}
 N^{-k}
\end{equation*}
applied to the partition function. Again, every derivative must pull exactly one power of $N$ down, together with a factor $H_{\alpha}^{[0]}$. By collecting the coefficient of $N^1$, setting $\textbf{p} = 0$, and shifting the indices of the $k$-sums by one we obtain the contribution
\begin{equation*}
\sum_{b\geq 1} p^{(0)}_b \sum_{k\geq 1} \frac{(b + n)!}{(b + n - k + 1 )!}
\sum_{\substack{l_1 + \dots + l_{k} = b+n \\ l_j \geq 1}}
\!\!\!\!
 \frac{\prod_{i=1}^{k} H_{l_i}^{[0]}}{k!}.
\end{equation*}


\begin{notation}
In order to write the formula in a more compact way, let us introduce the notation
\begin{equation}
\mathcal{H}_b((0,1)^k) :=
\sum_{\substack{l_1 + \dots + l_{k} = b\\ l_j \geq 1}}
 \prod_{i=1}^{k} H_{l_i}^{[0]},
\end{equation}
for $b >0$. 
Moreover, every term involving a negative factorial is considered to be zero.
\label{not:CC}
\end{notation}

Writing the contributions according to Notation \ref{not:CC} and putting them together proves that relation \eqref{eq:selN1} explicitly reads
  \begin{align*}
&\sum_{\substack{b_1 + b_2 = n \\ b_1, b_2 \geq 1}} 
\sum_{ k', k'' \geq 1} \frac{b_1!\, b_2!}{(b_1 - k' + 1)! (b_2 - k'' + 1)!} 
 \frac{\mathcal{H}_{b_1}((0,1)^{k'})}{k'!}
 \frac{ \mathcal{H}_{b_2}((0,1)^{k''})}{k''!}
 \\
 &+\sum_{b\geq 1} p^{(0)}_b \sum_{k\geq 1} \frac{(b + n)!}{(b + n - k + 1 )!}
 \frac{\mathcal{H}_{b+n}((0,1)^k)}{k!}
 = 0.
 \end{align*}
 
After the specialization of the parameters $p^{(0)}_1 = 0$ and $p^{(0)}_2 = -1$, it is enough to pick $n = l-1$ (the $b$-sum corresponds to the $j$-sum) and to take care of the difference between summing over $k$-uples of integers and summing over (unordered) partitions of a number (multiplying by the corresponding factors $g_{\lambda}=\frac{\ell(\lambda)!}{M(\lambda)}$ that we gave in \eqref{g}) to obtain exactly the equation appearing in \eqref{eq:tuttefs}.

\begin{remark}
We can express $F_l$ directly in terms of the notation \ref{not:CC}. In fact, expanding \eqref{eq:FHc} by Theorem~\ref{thm:Kreweras} and changing the summation over partitions into summation over independent indices, we obtain
\begin{equation}\label{eq:FaCC}
F_l = \sum_{k\geq 1} \frac{l!}{(l - k+ 1)!} \frac{\mathcal{H}_l((0,1)^k)}{k!}, \qquad \qquad l \geq 1.
\end{equation}
\end{remark}

The explicit derivation for usual disks serves as a toy model for more complicated topologies and also as a check of our method, since in this case we were able to compare the outcome with the result coming from combinatorics. Our first next goal is to give explicit results at least for the topologies: $(0,2)$, $(0,3)$ and $(1,1)$, using  the same techniques. 

We have done the computations for those topologies from the conjugated Virasoro point of view. We are in the process of turning the results into more compact formulas summing over all lengths and collecting the terms into correlators. As for the case of disks, we also substitute in the ordinary Virasoro constraints for $(0,2)$ the formula from free probability expressing ordinary cylinders in terms of fully simple ones, and match the already known terms. We think we can already extract information about the complicated partitioned permutations (introduced in Section~\ref{partitionedPermutations}) from the unknown terms in this case. We plan to apply the same scheme for topology $(0,3)$ and investigate if there is a pattern to be generalized to arbitrary topology $(g,n)$.

We remark that an advantage of this approach is that in principle it can be generalized to stuffed maps, i.e., to general unitary invariant measures, which may eventually give a hint for our vague conjecture from Section~\ref{vagueConjStuffed}.

\part[Large random maps with small and big boundaries]{Large random maps \\ with small and big boundaries \\ \ \\ \large{Nesting statistics in the $O(\mathsf{n})$ loop model for arbitrary topologies}}


\chapter{Introduction}
\label{chap:intro}

This part of the thesis is based on joint work with G.~Borot \cite{BGF16}, which is submitted for publication.
Our main goal is to analyze the nesting statistics in the $O(\mathsf{n})$ loop model on random maps of arbitrary topologies. The nesting properties of the loops in a configuration was initiated for disks and cylinders in \cite{BBD}. For this purpose we rely on the topological recursion results of \cite{BEOn,BEO} for the enumeration of maps in the $O(\mathsf{n})$ model. We characterize the generating series of maps of genus $g$ with $k$ boundaries and $k'$ marked points which realize a fixed nesting graph. These generating series are amenable to explicit computations in the loop model with bending energy on triangulations, and we characterize their behavior at criticality in the dense and in the dilute phase. In this part of the thesis, we will often call maps endowed with a loop configurations just maps (or as in the introduction configurations), and again we will use the name usual to emphasize when our maps do not carry a loop configuration.


After a Riemann conformal mapping,  the critical loop configurations on a fixed disk is believed to be described in the continuous limit by the so-called \emph{conformal loop ensemble} \cite{sheffield2009,zbMATH06121652}, denoted by ${\rm CLE}_{\kappa}$  and depending on a continuous index $\kappa\in(8/3,8)$,  with the correspondence $\mathsf{n} = 2\cos \pi(1 - \frac{4}{\kappa})$ for $\mathsf{n}\in (0,2]$  \cite{MR1964687,kg2004guide_to_sle,MR2112128}.

In \cite{BBD}, Borot, Bouttier and Duplantier investigated the nesting properties of loops on disks and cylinders weighted by an $O(\mathsf{n})$ model, and showed that they are in perfect agreement with the known nesting properties of ${\rm CLE}_{\kappa}$ \cite{MWW} after taking into account a suitable version of the KPZ relations \cite{KPZDS}. In this part of the thesis, we push this analysis forward and investigate rigorously the nesting properties of maps of any topology weighted by an $O(\mathsf{n})$ model. This includes as a special case the description of the critical behavior of maps without loops (i.e. in the class of pure gravity) having possibly marked points, microscopic and macroscopic boundaries. This generalization is non-trivial as the combinatorics of maps with several boundaries, marked points, and arbitrary genus, is much more involved than in the cases of disks and cylinders. Our approach is based on analytic combinatorics, and relies on two main ingredients: (1) the \textbf{substitution approach} developed in \cite{BBG12a,BBG12b} for planar maps; and (2) the \textbf{topological recursion} of \cite{EOFg,BEO} to reduce by a universal algorithm the enumeration of maps -- possibly carrying an $O(\mathsf{n})$ loop model -- of any topology to the enumeration of disks and cylinders. Obtaining the desired asymptotics for generating series of maps subjected to various constraints is then a matter of careful analysis of singularities.





\section{Combinatorial decomposition of configurations}\label{CombDecompConf}


We begin by recalling the notion of separating loops that we already introduced in Definition \ref{separating}. We also define the generating series of configurations with only non-separating loops through the substitution approach. The nesting graphs encoding the nesting information of a loop configuration will become of great relevance in this part of the thesis. The reader can review their definitions in Section \ref{Markm}. We also introduce the refined generating series of configurations realizing a fixed nesting graph, which will allow us to study the nesting properties of loop configurations. Finally, we give a combinatorial decomposition of configurations in terms of their associated nesting graph, which permits to study the critical behavior of the whole configurations via the analysis of the critical behavior of every type of piece.

\subsection{Substitution approach}

\label{SSub} 
In maps with the topology of a disk, there is a notion of inside and outside a loop, from the point of view of the boundary. Then, the nested loop approach \cite{BBG12b} puts in bijection disks $\mathcal{M}$ with a loop configuration with triples $(\mathcal{D},\mathcal{R},\mathcal{M}')$, where:
\begin{itemize}
\item[$\bullet$] $\mathcal{D}$ is a usual disk, called the \emph{gasket} of $\mathcal{M}$. It is obtained as the connected component containing the boundary in the complement of all loops in $\mathcal{M}$, filling the interior of each outermost loop by a face.
\item[$\bullet$] $\mathcal{R}$ is a disjoint union of sequences of faces visited by a single loop so as to form an annulus, which is rooted on its outer boundary. It is obtained as the collection of faces crossed by the outermost loops in $\mathcal{M}$ -- from the point of view of the boundary -- and the root edge on the outer boundary of each ring (call it $B$) is conventionally defined to be the edge outgoing from the vertex in $B$ which is reached by the shortest leftmost geodesic between the origin of the root edge on the boundary of $\mathcal{M}$, and $B$.
\item[$\bullet$] $\mathcal{M}'$ is a disjoint union of disks carrying loop configurations. These are the inside of the outermost loops.
\end{itemize}

\begin{figure}[h!]
\begin{center}
\def\svgwidth{\columnwidth}
\scalebox{0.8}{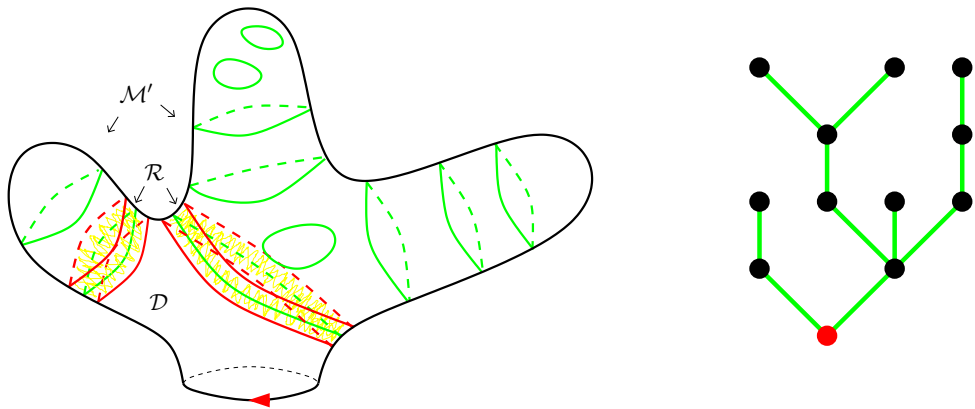}
  \caption{  \label{fig:nestingtree} Left: schematic representation of a loop configuration on a
    planar map with one boundary, illustrating the gasket $\mathcal{D}$, the outermost annuli $\mathcal{R}$ and the disjoint union of disks with loops $\mathcal{M}'$. Right: the associated primary nesting tree
    (the red vertex corresponds to the gasket).}
  \end{center}
\end{figure}


This translates into a functional relation for the generating series of disks: we can express the generating series of disks endowed with loop configurations $\mathcal{F}_{\ell}$ in terms of the generating series of usual disks with special face weights. 
\beq
\label{fixedpoint} \mathcal{F}_{\ell} = F_{\ell}(T_1,T_2,\ldots),
\eeq
where the weights $T_l$ of a face of degree $l$ must satisfy the following fixed point condition
\begin{equation}
  \label{eq:fixp}
  T_l = t_l + \sum_{\ell \geq 0} A_{l,\ell} F_{\ell}(T_1,T_2,\ldots) = t_l + \sum_{\ell \geq 0} A_{l,\ell}\,\mathcal{F}_{\ell}.
\end{equation}
We have denoted by $A_{l,\ell}$ the generating series of sequences of faces visited by a loop, which are glued together so as to form an annulus, in which the outer boundary is rooted and has length $l$, and the inner boundary is unrooted and has length $\ell$. Compared to the notations of \cite{BBG12b}, we decide to include in $A_{l,\ell}$ the weight $n$ for the loop crossing all faces of the annulus. We call $T_l$ the \emph{renormalized face weights}. Note that, although $t_l$ could be zero for $l = 1,2$ and for $l \geq l_0$, $T_1,T_2$ and $T_l$ for $l \geq l_0$ are a priori non-zero. For this reason, it was necessary to consider the model of usual maps with general face weights \eqref{usualmap}, while we could restrict to faces (visited or not) of perimeter larger or equal to $3$ in the definition of the Boltzmann weight for the general $O(\mathsf{n})$ loop model \eqref{Bweight}.

In general, the generating series of usual maps evaluated at renormalized face weights $(T_1,T_2,\ldots)$ will play an important role for us and we denote it by 
\beq\label{RenormalizedFs}
\mathsf{F}^{[g,\bullet k']}_{\ell_1,\ldots,\ell_k}\coloneqq F^{[g,\bullet k']}_{\ell_1,\ldots,\ell_k}(T_1,T_2,\ldots).
\eeq
We remark the following equalities for the cases with only one marked element, i.e. for $k+k'=1$:
\begin{itemize}
\item If there is only one boundary, $\mathcal{F}^{[g]}_{\ell}=\mathsf{F}^{[g]}_{\ell}$, as \eqref{fixedpoint} claimed for the case $g=0$.
\item If there is only one marked vertex, $\mathcal{F}^{[g,\bullet]}=\mathsf{F}^{[g,\bullet]}$.
\end{itemize}
According to the nested loop approach, $\mathsf{F}^{[g,\bullet k']}$ enumerates maps in the $O(\mathsf{n})$ model where the $k$ boundaries and the $k'$ marked points all belong to the same connected component after removal of all loops. In particular, $\mathsf{F}^{\bullet}$ is the generating series of disks pointed in the gasket and $\mathsf{F}_{\ell_1,\ell_2}$ the generating series of cylinders with the two boundaries belonging to the same vertex in the associated nesting graph.

Now we describe two marking procedures that behave in the same way for general configurations as for usual maps. The operation of turning an internal face into a boundary of length $\ell$ is realized by the operator $\ell\,\frac{\partial}{\partial t_{\ell}}$, while marking a vertex amounts to applying $u\,\frac{\partial}{\partial u}$:
\bea\label{marking1}
\mathcal{F}_{\ell_1,\ldots,\ell_{k+1}}^{[g,\bullet k']} & = & \ell_{k+1}\partial{t_{\ell_{k+1}}} \mathcal{F}_{\ell_1,\ldots,\ell_k}^{[g,\bullet k']}, \\
\label{marking2} \mathcal{F}_{\ell_1,\ldots,\ell_k}^{[g,\bullet k']} & = & (u\partial_{u})^{k'} \mathcal{F}_{\ell_1,\ldots,\ell_k}^{[g]}.
\eea

On the other hand, notice that due to the constraints on the relative position of the marked points and boundaries with respect to the loops for usual maps specialized to renormalized face weights, marking a face or a vertex for this kind maps is done before the evaluation: 
\bea
\mathsf{F}_{\ell_1,\ldots,\ell_{k+1}}^{[g,\bullet k']} & = & \left.(\ell_{k+1}\partial{t_{\ell_{k+1}}} F_{\ell_1,\ldots,\ell_k}^{[g,\bullet k']})\right|_{\{t_l=T_l\}_{l\geq 1}} \neq \ell_{k+1}\partial{t_{\ell_{k+1}}} (F_{\ell_1,\ldots,\ell_k}^{[g]}(T_1,T_2,\ldots)), \nonumber \\
\mathsf{F}_{\ell_1,\ldots,\ell_k}^{[g,\bullet k']} & = & \left.((u\partial_{u})^{k'} F_{\ell_1,\ldots,\ell_k}^{[g]})\right|_{\{t_l=T_l\}_{l\geq 1}}  \neq (u\partial_{u})^{k'} (F_{\ell_1,\ldots,\ell_k}^{[g]}(T_1,T_2,\ldots)). \nonumber
\eea
The difference comes from the order of differentiation and evaluation at renormalized face weights $(T_1,T_2,\ldots)$, which depend on $u$ and $t_{\ell_{k+1}}$.

Functional relations for more general planar maps can be deduced from the fixed point equations \eqref{fixedpoint}-\eqref{eq:fixp}, using these operations of marking a face or a vertex for general configurations \eqref{marking1}-\eqref{marking2}, here for pointed disks and cylinders:
\bea
\label{Cylfixed} \mathcal{F}^{[0,2]}_{\ell_1,\ell_2} & = & \mathsf{F}_{\ell_1,\ell_2}^{[0,2]} + \sum_{l,l' \geq 1} \mathsf{F}_{\ell_1,l}^{[0,2]}\,R_{l,l'}\,\mathcal{F}_{l',\ell_2}^{[0,2]}, \\
\label{Cyl2fixed} \mathcal{F}^{\bullet}_{\ell} & = & \mathsf{F}_{\ell}^{\bullet} + \sum_{l' \geq 1, l'' \geq 0} \mathsf{F}_{\ell,l'}^{[0,2]}\,R_{l',l''} \mathcal{F}_{l''}^{\bullet}, 
\eea
where $R_{l,\ell} = A_{l,\ell}/l$ is the generating series of annuli visited by a single loop, whose outer and inner boundaries are both unrooted.

\subsection{Separating loops and refined enumeration}
\label{Srefff} 

We recall now the definition of separating loop already given in \ref{separating}.
In a map $\mathcal{M}$ with a non empty set of marked elements $\mathcal{E}$, a loop is \emph{separating}\footnote{The intuitive idea of this definition is very clear for maps of genus $0$, where separating loops always ``separate'' two marked elements in different connected components of $\mathcal{M} \setminus \mathcal{E}$. However, with this definition for maps of arbitrary topology, non-contractible loops in $\mathcal{M}$ are separating, even though the name could be misleading in such a case. For example, a non-contractible loop in a torus with only one marked element is also called separating, even if it cannot ``separate'' the marked element from any other marked element. So separating loops also help keeping track of the more complicated structure of higher genus maps.} if it is not contractible in $\mathcal{M} \setminus \mathcal{E}$. The separating loops (or sequences of separating loops) were encoded in the edges of the nesting graph. If the map is planar, we can equivalently say that a loop is separating if it does not bound a disk (in the underlying surface) which contains no marked element.

\begin{center}
\begin{figure}[h!]
\begin{center}
\scalebox{0.2}{\includegraphics{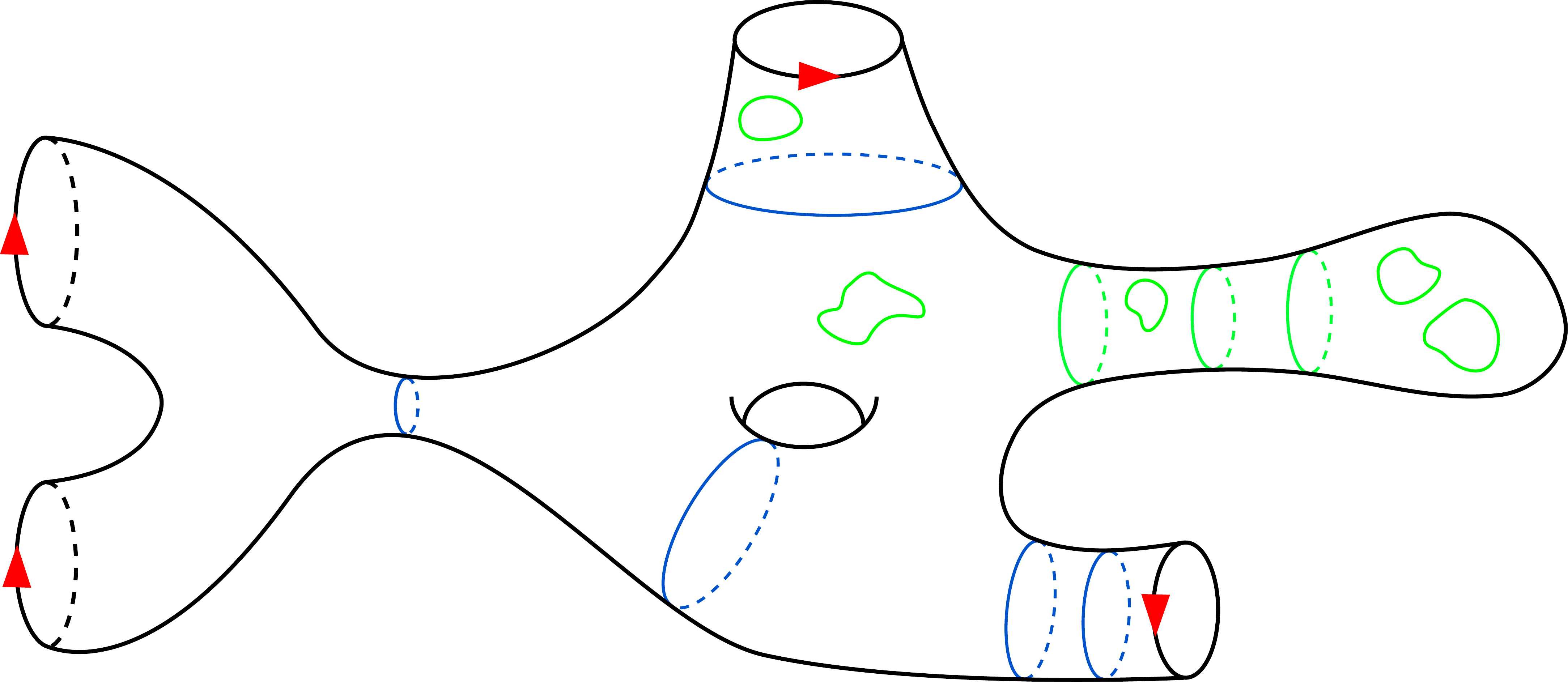}}
\caption{Blue: Separating loops. Green: Non-separating loops.}
\end{center}
\end{figure}
\end{center}

First of all, we remark that the usual maps with renormalized face weights we introduced in the previous section are exactly configurations with no separating loops, i.e. configurations which only carry non-separating loops. Therefore, $\mathsf{F}_{\ell_1,\ldots,\ell_{k}}^{[g,k,\bullet k']}$ is the generating series of configurations with no separating loops.

Let us examine now the separating loops in the simple case of two marked elements in a planar map. Then, either the two marked elements are not separated by any loop (the nesting graph consists of a single vertex carrying the two marks), or they are separated by $P \geq 1$ loops (the nesting graph consists of an edge of length $P$ between two vertices). We recall that $P$ was called the depth or arm length. To fix ideas, let us say that the first marked element is a boundary and the second one is a boundary (or a vertex). Then, we can put such a map $\mathcal{M}$ in bijection either with a cylinder (or a pointed disk) having no separating loop, or a triple consisting of a cylinder with no separating loops, an annulus of faces visited by a single loop, and another cylinder (or pointed disk) $\mathcal{M}'$ with $p - 1$ separating loops. This is the combinatorial meaning of \eqref{Cylfixed}-\eqref{Cyl2fixed}, and it allows an easy refinement. Namely, let $\mathcal{F}^{[0,2]}_{\ell_1,\ell_2}[s]$ (resp. $\mathcal{F}^{\bullet}_{\ell}[s]$) be the generating series of cylinders (resp. pointed disks) where the Boltzmann weight includes an extra factor $s^{P}$ with $P$ the depth. There generating series receive the name of \emph{refined} generating series. We obtain from the previous reasoning:
\bea
\label{Cylfixedref} \mathcal{F}_{\ell_1,\ell_2}[s] & = & \mathsf{F}_{\ell_1,\ell_2} + s\sum_{l,l' \geq 1} \mathsf{F}_{\ell_1,l}\,R_{l,l'}\,\mathcal{F}_{l',\ell_2}[s], \\
\label{Cyl2fixedref} \mathcal{F}^{\bullet}_{\ell}[s] & = & \mathsf{F}_{\ell}^{\bullet} + s \sum_{l',l'' \geq 1} \mathsf{F}_{\ell,l'}\,R_{l',l''} \mathcal{F}_{l''}^{\bullet}[s].
\eea

In full generality, we are interested in computing $\mathscr{F}^{[g,k,\bullet k']}_{\ell_1,\ldots,\ell_k}[\Gamma,\star;\mathbf{s}]$, the refined generating series of configurations of genus $g$ with $k$ boundaries and $k'$ marked points which achieve the fixed nesting graph $(\Gamma,\star)$, and for which the usual Boltzmann weight contains an extra factor:
$$
\prod_{\mathsf{e} \in E(\Gamma)} s(\mathsf{e})^{P(\mathsf{e})}.
$$

The construction of the nesting graph provides a combinatorial decomposition of maps, illustrated in Figure \ref{decomp}. Indeed, we can retrieve bijectively the original map from $(\Gamma,\star,\mathbf{P})$, by gluing together:
\begin{itemize}
\item[$\bullet$] for each vertex $\mathsf{v}$ of valency $d(\mathsf{v})$, a usual map (with renormalized weights) of genus $h(\mathsf{v})$ with $k(\mathsf{v})$ labeled boundaries and $d(\mathsf{v})$ other unlabeled boundaries, and $k'(\mathsf{v})$ marked points;
\item[$\bullet$] for each edge $\mathsf{e}$ of length $1$, an annulus visited by a single loop;
\item[$\bullet$] for each each $\mathsf{e}$ of length $P(\mathsf{e}) \geq 2$, two annuli visited by a single loop capping a cylinder with $P(\mathsf{e}) - 2$ separating loops. 
\end{itemize}

\begin{center}
\begin{figure}[h!]
\begin{center}
\def\svgwidth{\columnwidth}
\scalebox{1}{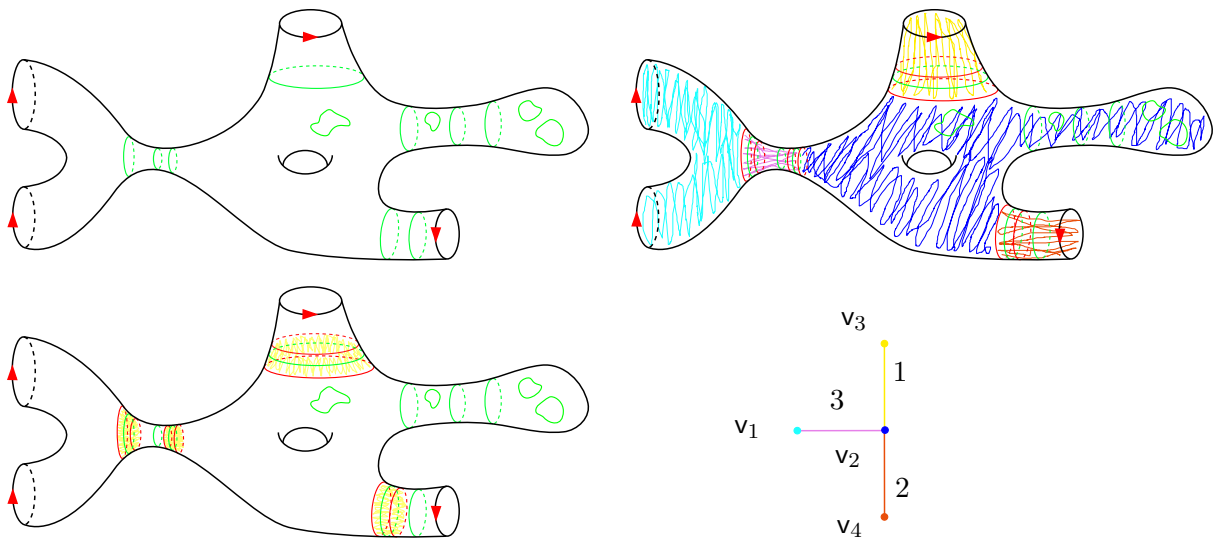}
\caption{The vertices $\mathsf{v}_1, \mathsf{v}_2 \in \tilde{V}(\Gamma)$. They carry the following information regarding the associated genus, number of boundaries and valency: $h(\mathsf{v}_1)=0, k(\mathsf{v}_1)=2, d(\mathsf{v}_1)=1, h(\mathsf{v}_2)=1, k(\mathsf{v}_2)=0, d(\mathsf{v}_2)=3.$ The other vertices correspond to connected components of topology $(0,2)$ with only one labeled boundary: $\mathsf{v}_3, \mathsf{v}_4\in V_{0,2}(\Gamma)$.}\label{decomp}
\end{center}
\end{figure}
\end{center}

Let us denote $E(\Gamma)$ the set of edges and $V(\Gamma)$ the set of vertices of the nesting graph $\Gamma$. At a given vertex $\mathsf{v}$, $\mathsf{e}(\mathsf{v})$ is the set of outcoming half-edges, and for a given edge $\mathsf{e}$, $\{\mathsf{e}_{+},\mathsf{e}_{-}\}$ is its set of half-edges. $\partial(\mathsf{v})$ the set of boundaries which are registered on marked elements on $\mathsf{v}$ -- if there are no marked elements on $\mathsf{v}$ or just $k(\mathsf{v}) = 0$, then $\partial(\mathsf{v}) = \emptyset$. Let $V_{0,2}(\Gamma)$ be the set of univalent vertices $\mathsf{v}$ of genus $0$ which carry exactly $1$ boundary; the outgoing half-edge (pointing towards the boundary) is then denoted $\mathsf{e}_+(\mathsf{v})$ and $\tilde{V}(\Gamma)=V(\Gamma) \setminus V_{0,2}(\Gamma)$. Let $E_{{\rm un}}(\Gamma)$ be the set of edges which are incident to vertices in $V_{0,2}(\Gamma)$, and $\tilde{E}(\Gamma) = E(\Gamma)\setminus E_{{\rm un}}(\Gamma)$. We define the set of glueing half-edges as follows: 
$$
E_{{\rm glue}}(\Gamma)= \bigcup_{\mathsf{e}\in \tilde{E}(\Gamma)}\{\mathsf{e}_{+},\mathsf{e}_{-}\} \cup \bigcup_{\mathsf{v}\in V_{0,2}(\Gamma)} \mathsf{e}_+(\mathsf{v}).
$$

Let us introduce the generating series of cylinders with one annulus (with unrooted outer boundary) glued to one of the two boundaries
\beq
\label{Fihat} \hat{\mathcal{F}}_{\ell_1,\ell_2}[s] = s \sum_{l \geq 0} R_{\ell_1,l} \mathcal{F}_{l,\ell_2}[s]
\eeq
and the generating series of cylinders cuffed with two annuli with unrooted outer boundaries
\beq
\label{Fiilde} \tilde{\mathcal{F}}_{\ell_1,\ell_2}[s] = s R_{\ell_1,\ell_2} + s^2 \sum_{l,l' \geq 0} R_{\ell_1,l} \mathcal{F}_{l,l'}[s]\,R_{l',\ell_2}\,.
\eeq
By convention, we included in the latter an extra term corresponding to a single annulus with its two boundaries unrooted.
We refer to these two generating series as the generating series of \emph{cuffed cylinders}.

We can determine the desired refined generating series of maps, whose corresponding nesting graph is fixed, using the decomposition of any such map  into the previously introduced pieces.
\begin{proposition}
\label{magi}
\bea
\mathscr{F}^{[g,k,\bullet k']}_{\ell_1,\ldots,\ell_k}[\Gamma,\star,\mathbf{s}] & = &  \sum_{l\,:\,E_{{\rm glue}}(\Gamma) \rightarrow \mathbb{N}}\,\,\prod_{\mathsf{v} \in \tilde{V}(\Gamma)} \frac{\mathsf{F}^{[h(\mathsf{v}),k(\mathsf{v}) + d(\mathsf{v}),\bullet k'(\mathsf{v})]}_{\ell(\partial(\mathsf{v})),l(\mathsf{e}(\mathsf{v}))}}{d(\mathsf{v})!}
\nonumber \\
& & \qquad  \prod_{\mathsf{e} \in \tilde{E}(\Gamma)} \tilde{\mathcal{F}}_{l(\mathsf{e}_-),l(\mathsf{e}_+)}^{[0,2]}[s(\mathsf{e})] \prod_{\mathsf{v} \in V_{0,2}(\Gamma)} \hat{\mathcal{F}}^{[0,2]}_{l(\mathsf{e}_+(\mathsf{v})),\ell(\partial(\mathsf{v}))}[s(\mathsf{e}_+(\mathsf{v}))],
\eea
where $\ell : \bigcup_{\mathsf{v}\in V(\Gamma)}\partial(\mathsf{v}) \rightarrow \mathbb{N}$ is given by $\ell_1,\ldots, \ell_k$.
\end{proposition}


\section{Analytic properties of generating series}
\label{Sanal} 
So far, all the parameters of the model were formal. We now would like to assign them real values, as we did in Section \ref{largeRandom}. In this section, we review the properties of generating series of maps obtained by recording all possible boundary perimeters at the same time.

\subsection{Usual maps}
\label{UsSection} 
In the context of usual maps (here not specialized to the renormalized face weigths), we called $u$ and a sequence $(t_l)_{l \geq 1}$ of nonnegative real numbers admissible if $F_{\ell}^{\bullet} < \infty$ for any $\ell$. By extension, we say that $u$ and a sequence $(t_l)_{l \geq 1}$ of real numbers are \emph{admissible} if $u$ and $(|t_l|)_{l \geq 1}$ are admissible. For admissible vertex and face weights, we can define
$$
W(x) = \sum_{\ell \geq 0} \frac{F_{\ell}}{x^{\ell + 1}} \in \mathbb{Q}[[x^{-1}]].
$$

Then, $W(x)$ satisfies the one-cut lemma and a functional relation coming from Tutte's combinatorial decomposition of rooted disks.\begin{theorem} \cite{BBG12b}
\label{ponecut}If $(t_l)_{l \geq 1}$ is admissible, then the formal series $W(x)$ is the Laurent series expansion at $x = \infty$ of a function, still denoted $W(x)$, which is holomorphic for $x \in \mathbb{C}\setminus \gamma$, where $\gamma = [\gamma_-,\gamma_+]$ is a segment of the real line depending on the vertex and face weights. Its endpoints are characterized so that $\gamma_{\pm} = \mathfrak{s} \pm 2\mathfrak{r}$, and $\mathfrak{r}$ and $\mathfrak{s}$ are the evaluation at the chosen weights of the unique formal series in the variables $u$ and $(t_l)_{l \geq 1}$ such that
\bea
\oint_{\gamma} \frac{\dd y}{2{\rm i}\pi}\,\frac{\big(y - \sum_{l \geq 1} t_l\,y^{l - 1}\big)}{\sigma(z)} & = & 0, \\
u + \oint_{\gamma} \frac{\dd y}{2{\rm i}\pi}\,\frac{y\big(y - \sum_{l \geq 1} t_l\,y^{l - 1}\big)}{\sigma(z)} & = & 0,
\eea
where $\sigma(x) = \sqrt{x^2 - 2\mathfrak{s}x + \mathfrak{s}^2 - 4\mathfrak{r}}$. Besides, the endpoints satisfy $|\gamma_-| \leq \gamma_+$, with equality iff $t_{l} = 0$ for all odd $l$'s.
\end{theorem}

\begin{theorem}\cite{BBG12b}
\label{prop2} $W(x)$ is uniformly bounded for $x \in \mathbb{C}\setminus \gamma$. Its boundary values on the cut satisfy the functional relation:
\beq
\label{eq:funcnu}\forall x \in \gamma,\qquad W(x + {\rm i}0) + W(x - {\rm i}0) = x - \sum_{l \geq 1} t_l\,x^{l - 1},
\eeq
and $W(x) = u/x + O(1/x^2)$ when $x \rightarrow \infty$. These properties uniquely determine $\gamma_-,\gamma_+$ and $W(x)$.
\end{theorem}

Although \eqref{eq:funcnu} arise as a consequence of Tutte's equation and analytic continuation, it has itself not received a combinatorial interpretation yet. 

With Theorem~\ref{ponecut} at hand, the analysis of Tutte's equation for generating series of maps with several rooted boundaries and their analytic continuation has been performed (in a more general setting) in \cite{BEO,Bstuff}. The first outcome is that, if $u$ and $(t_l)_{l \geq 1}$ are admissible, then $F_{\ell_1,\ldots,\ell_k}^{[g,k,\bullet k']} < \infty$, for all $g$, $k$ and $k'$, so that we can define
$$
W^{[g,k,\bullet k']}(x_1,\ldots,x_k) = \sum_{\ell_1,\ldots,\ell_k \geq 0} \frac{F^{[g,k,\bullet k']}_{\ell_1,\ldots,\ell_k}}{x_1^{\ell_1 + 1} \cdots x_{k}^{\ell_k + 1}} \in \mathbb{Q}[[x_1^{-1},\ldots,x_k^{-1}]].
$$
The second outcome is that these are also Laurent series expansions at $\infty$ of functions, still denoted $W^{[g,k,\bullet k']}(x_1,\ldots,x_k)$, which are holomorphic for $x_i \in \mathbb{C}\setminus\gamma$, with the same $\gamma$ as in Theorem~\ref{ponecut}, and which have upper/lower boundary values when $x_i$ approaches $\gamma$ while $(x_j)_{j \neq i} \in (\mathbb{C}\setminus \gamma)^{k - 1}$ are fixed. More specifically, for cylinders:
\begin{theorem}
We have that $\sigma(x_1)\sigma(x_2)W^{[0,2]}(x_1,x_2)$ remains uniformly bounded for $x_1,x_2 \in \mathbb{C}\setminus\gamma$ and the following functional relation, for $x_1 \in (\gamma_-,\gamma_+)$ and $x_2 \in \mathbb{C}\setminus\gamma$:
$$
W^{[0,2]}(x_1 + {\rm i}0,x_2) + W^{[0,2]}(x_1 - {\rm i}0,x_2) = -\frac{1}{(x_1 - x_2)^2}\,.
$$
Moreover, $W^{[0,2]}(x_1,x_2) \in O(x_1^{-2}x_2^{-2})$ when $x_1,x_2 \rightarrow \infty$ and these properties uniquely determine the generating series of usual cylinders $W^{[0,2]}(x_1,x_2)$.
\end{theorem}

Once $\gamma_{\pm}$ have been obtained, the following formula is well-known:
\beq
\label{F2solep} W^{[0,2]}(x_1,x_2) = \frac{1}{2(x_1 - x_2)^2}\bigg\{-1 + \frac{x_1x_2 - \frac{\gamma_- + \gamma_+}{2}(x_1 + x_2) + \gamma_-\gamma_+}{\sigma(x_1)\sigma(x_2)}\bigg\}.
\eeq

The generating series for usual pointed disks is also particularly simple (see \textit{e.g.} \cite{BBG12b}):
\beq
\label{diskpointed} W^{\bullet}(x) = \frac{1}{\sigma(x)}.\eeq

\begin{theorem} \cite{Eynardbook,BEO}
Let $2g - 2 + k > 0$. There exists $r(g,k) > 0$ such that
$$
\sigma(x_1)^{r(g,k)}W^{[g,k]}(x_1,\ldots,x_k)
$$
remains bounded when $x_1$ approaches $\gamma$, while $(x_i)_{i = 2}^k$ are kept fixed away from $\gamma$. \\
The series $W^{[g,k]}(x_1,\ldots,x_k)$ has upper/lower boundary values for $x_1 \in (\gamma_-,\gamma_+)$ and $x_I = (x_i)_{i = 2}^k$ fixed away from $\gamma$; it satisfies under the same conditions:
$$
W^{[g,k]}(x_1 + {\rm i}0,x_I) + W^{[g,k]}(x_1 - {\rm i}0,x_I) = 0,
$$
and $W^{[g,k]}(x_1,x_I) \in O(x_1^{-2})$ when $x_1 \rightarrow \infty$.
\end{theorem}

\subsection{In the $O(\mathsf{n})$ loop model}\label{ONsection}

In the context of the $O(\mathsf{n})$ model, we say that two sequences of real numbers $(t_{l})_{l \geq 3}$ and $(A_{l_1,l_2})_{l_1,l_2}$ are admissible if the corresponding sequence of renormalized face weights $(T_1,T_2,\ldots)$ computed by \eqref{eq:fixp} is admissible. For admissible face weights, we can define:
$$
\mathcal{W}(x) = \sum_{\ell \geq 0} \frac{\mathcal{F}_{\ell}}{x^{\ell + 1}} \in \mathbb{Q}[[x^{-1}]]\,.
$$
In the remaining of the article, we always assume admissible face weights.

As consequence of \eqref{fixedpoint}, $\mathcal{W}(x)$ satisfies the one-cut property (the analogue of Theorem~\ref{ponecut}), and we still denote $\gamma_{\pm}$ the endpoints of the cuts, which now depend on face weights $(t_l)_{l \geq 3}$ and annuli weights $(A_{l,l'})_{l,l' \geq 0}$. Admissibility also implies that the annuli generating series
\bea
\mathbf{R}(x,y) & = & \sum_{l + l' \geq 1} R_{l,l'}x^{l}y^{l'} \ \  \text{ and} \nonumber \\
\mathbf{A}(x,y) & = & \sum_{l \geq 1} \sum_{l' \geq 0} A_{l,l'}\,x^{l - 1}y^{l'} = \partial_{x}\mathbf{R}(x,y) \nonumber 
\eea
are holomorphic in a neighborhood of $\gamma\times\gamma$. And, $\mathcal{W}(x)$'s boundary values on the cut satisfy the following functional relation:
\begin{theorem}\label{onecutloop}\cite{BBG12b}
\label{funcF} We have that $\mathcal{W}(x)$ is uniformly bounded for $x \in \mathbb{C}\setminus\gamma$ and has upper/lower boundary values on $\gamma$. Moreover, for $x \in \gamma$,
\beq
\label{eq:funcF}\mathcal{W}(x + {\rm i}0) + \mathcal{W}(x - {\rm i}0) + \oint_{\gamma} \frac{\dd z}{2{\rm i}\pi}\,\mathbf{A}(x,z)\,\mathcal{W}(z) = x - \sum_{k \geq 1} t_k\,x^{k - 1}
\eeq
and $\mathcal{W}(x) = u/x + O(1/x^2)$ when $x \rightarrow \infty$. These properties uniquely determine $\mathcal{W}(x)$ and $\gamma_{\pm}$.
\end{theorem}

Now with Theorem~\ref{onecutloop} at hand, the analysis of Tutte's equation for the partition functions of maps having several boundaries in the loop model, and their analytic continuation, has also been performed in \cite{BEO,Bstuff}. The outcome is that
$$
\mathcal{W}^{[g,k,\bullet k']}(x_1,\ldots,x_k) = \sum_{\ell_1,\ldots,\ell_k \geq 0} \frac{\mathcal{F}^{[g,k,\bullet k']}_{\ell_1,\ldots,\ell_k}}{x_1^{\ell_1 + 1} \cdots x_{k}^{\ell_k + 1}} \in \mathbb{Q}[[x_1^{-1},\ldots,x_k^{-1}]]
$$
are also well-defined and Laurent series expansions at infinity of functions, still denoted 
$$
\mathcal{W}^{[g,k,\bullet k']}(x_1,\ldots,x_k),
$$ 
which are holomorphic for $x_i \in \mathbb{C}\setminus\gamma$, with the same $\gamma$ independently of $g$, $k$ and $k'$, and admit upper/lower boundary values for $x_i \in \gamma$ while $(x_j)_{j \neq i} \in (\mathbb{C}\setminus \gamma)^{k - 1}$ are kept fixed. Besides:
\begin{theorem}
\label{prop4loop} 
We have that $\sigma(x_1)\sigma(x_2)\mathcal{W}^{[0,2]}(x_1,x_2)$ remains uniformly bounded for $x_1,x_2 \in \mathbb{C}\setminus\gamma$. For $x_1 \in (\gamma_-,\gamma_+)$ and $x_2 \in \mathbb{C}\setminus\gamma$, we have the following functional relation:
$$
\mathcal{W}^{[0,2]}(x_1 + {\rm i}0,x_2) + \mathcal{W}^{[0,2]}(x_1 - {\rm i}0,x_2) + \oint_{\gamma} \frac{\dd y}{2{\rm i}\pi}\,\mathbf{A}(x_1,y) \mathcal{W}^{[0,2]}(y,x_2) = -\frac{1}{(x_1 - x_2)^2},
$$
and $\mathcal{W}^{[0,2]}(x_1,x_2) \in O(x_1^{-2}x_2^{-2})$, when $x_1,x_2 \rightarrow \infty$. These properties uniquely determine the generating series of cylinders $\mathcal{W}^{[0,2]}(x_1,x_2)$.
\end{theorem}
\begin{theorem}\label{2g2m} \cite{Eynardbook,BEO}
Let $2g - 2 + k > 0$. There exists $r(g,k) > 0$ such that
$$
\sigma(x_1)^{r(g,k)}\mathcal{W}^{[g,k]}(x_1,\ldots,x_k)
$$
remains bounded when $x_1$ approaches $\gamma$ while $(x_i)_{i = 2}^k$ are kept fixed away from $\gamma$. \\
Moreover, $\mathcal{W}^{[g,k]}(x_1,\ldots,x_k)$ has upper/lower boundary values for $x_1 \in (\gamma_-,\gamma_+)$ and $x_I = (x_i)_{i = 2}^k$ fixed away from $\gamma$, and it satisfies under the same conditions:
$$
\mathcal{W}^{[g,k]}(x_1 + {\rm i}0,x_I) + \mathcal{W}^{[g,k]}(x_1 - {\rm i}0,x_I) + \oint_{\gamma} \frac{\dd y}{2{\rm i}\pi}\,\mathbf{A}(x,y)\,\mathcal{W}^{[g,k]}(y,x_I) = 0
$$
and $\mathcal{W}^{[g,k]}(x_1,x_I) \in O(x_1^{-2})$ when $x_1 \rightarrow \infty$.
\end{theorem}

\subsubsection{Refined generating series}

We now recall the results of \cite{BBD} for the refined generating series of pointed disks and cylinders. First of all, for admissible weights and $s \in \mathbb{R}$ at least in a neighborhood of $[-1,1]$,
\bea
\mathcal{W}_{s}^{\bullet}(x_1) & = & \sum_{\ell \geq 0} \frac{\mathcal{F}_{\ell}^{\bullet}[s]}{x_1^{\ell_1 + 1}} \in \mathbb{Q}[[x_1^{-1}]], \nonumber \\
\mathcal{W}_{s}^{[0,2]}(x_1,x_2) & = & \sum_{\ell_1,\ell_2 \geq 1} \frac{\mathcal{F}_{\ell_1,\ell_2}^{[0,2]}[s]}{x_1^{\ell_1 + 1}x_2^{\ell_2 + 1}} \in \mathbb{Q}[[x_1^{-1},x_2^{-1}]]
\eea
are well-defined, and are Laurent series expansions at infinity of functions, still denoted in the same way, which are holomorphic of $x_i \in \mathbb{C}\setminus \gamma$, for the same $\gamma$ appearing in Section~\ref{ONsection}, independently of $s$. Besides, we have linear functional relations very similar to those satisfied by the unrefined generating series:
 
\begin{theorem}\cite{BBD}
\label{prop76} We have that $\sigma(x_1)\sigma(x_2)\mathcal{W}_{s}^{[0,2]}(x_1,x_2)$ is uniformly bounded for $x_1,x_2 \in \mathbb{C}\setminus\gamma$. For any $x_1 \in (\gamma_-,\gamma_+)$ and $x_2 \in \mathbb{C}\setminus\gamma$ fixed, we have:
$$
\mathcal{W}_{s}^{[0,2]}(x_1 + {\rm i}0,x_2) + \mathcal{W}^{[0,2]}_{s}(x_1 - {\rm i}0,x_2) +  s \oint_{\gamma} \frac{\dd y}{2{\rm i}\pi}\,\mathbf{A}(x_1,y)\,\mathcal{W}^{[0,2]}_{s}(y,x_2) = - \frac{1}{(x_1 - x_2)^2}
$$
and $\mathcal{W}_{s}^{[0,2]}(x_1,x_2) \in O(x_1^{-2}x_2^{-2})$, when $x_1,x_2 \rightarrow \infty$. These properties uniquely determine the refined generating series of cylinders $\mathcal{W}_{s}^{[0,2]}(x_1,x_2)$.
\end{theorem}

\begin{theorem}\cite{BBD}
\label{prop77} We have that $\sigma(x)\mathcal{W}_{s}^{\bullet}(x)$ is uniformly bounded, when $x \in \mathbb{C}\setminus\gamma$. For $x \in (\gamma_-,\gamma_+)$,
$$
\mathcal{W}^\bullet_{s}(x + {\rm i}0) + \mathcal{W}^\bullet_{s}(x - {\rm i}0) + s \oint_{\gamma} \frac{\dd y}{2{\rm i}\pi}\,\mathbf{A}(x,y)\,\mathcal{W}^\bullet_{s}(y) = 0
$$
and $\mathcal{W}_{s}^{\bullet}(x) = u/x + O(1/x^2)$, when $x \rightarrow \infty$. These properties uniquely determine $\mathcal{W}_{s}^{\bullet}(x)$.
\end{theorem}
From the analytic properties of $\mathcal{W}_{s}^{[0,2]}$ and $\mathbf{R}$, it follows that
\beq
\hat{\mathcal{W}}_{s}^{[0,2]}(x_1,x_2) = \sum_{\ell_1,\ell_2 \geq 0} \hat{\mathcal{F}}_{\ell_1,\ell_2}^{[0,2]}[s]\,\frac{x_1^{\ell_1}}{x_2^{\ell_2 + 1}} = s \oint_{\gamma} \frac{\dd y}{2{\rm i}\pi}\,\mathbf{R}(x_1,y) \mathcal{W}^{[0,2]}_{s}(y,x_2) \nonumber
\eeq
is the series expansion when $x_1 \rightarrow 0$ and $x_2 \rightarrow \infty$ of a function denoted likewise, which is holomorphic for $x_1$ in a neighborhood of $\gamma$ and $x_2$ in $\mathbb{C}\setminus\gamma$. And,
\bea
\label{Ftilded}\tilde{\mathcal{W}}_{s}^{[0,2]}(x_1,x_2) & = & \sum_{\ell_1,\ell_2 \geq 0} \tilde{\mathcal{F}}_{\ell_1,\ell_2}^{[0,2]}[s]\,x_1^{\ell_1}x_2^{\ell_2} \\
& = &  s\,\mathbf{R}(x_1,x_2) + s^2 \oint_{\gamma} \frac{\dd y_1}{2{\rm i}\pi}\,\frac{\dd y_2}{2{\rm i}\pi}\,\mathbf{R}(x_1,y_1) \mathcal{W}^{[0,2]}_{s}(y_1,y_2) \mathbf{R}(y_2,x_2) \nonumber
\eea
is the series expansion at $x_i \rightarrow 0$ of a function denoted likewise, which is holomorphic for $x_i$ in a neighborhood of $\gamma$. This fact and the analytic properties of $\mathsf{W}^{[g,k,\bullet k']}$ for any $g,k,k'$ described in Section~\ref{ONsection} imply, together with the formula of Proposition~\ref{magi}:

\begin{proposition}
\label{P212} Let $\Gamma$ be a fixed nesting graph. If $u$, $(t_{l})_{l \geq 3}$ and $(A_{l,\ell})_{l,\ell}$ are admissible, and $s(\mathsf{e}) \in \mathbb{R}$ in a neighborhood of $[-1,1]$ for each $\mathsf{e} \in E(\Gamma)$, then the generating series for fixed nesting graph $\Gamma$
$$
\mathscr{W}^{[g,k,\bullet k']}_{\Gamma,\star,\mathbf{s}}(x_1,\ldots,x_k) = \sum_{\ell_1,\ldots,\ell_k \geq 0} \frac{\mathscr{F}^{[g,k,\bullet k']}_{\ell_1,\ldots,\ell_k}[\Gamma,\star,s]}{x_1^{\ell_1 + 1}\cdots x_k^{\ell_k + 1}}
$$
are well-defined, and are the Laurent expansions at $\infty$ of functions, denoted with same symbol, which are holomorphic in $(x_1,\ldots,x_k) \in (\mathbb{C}\setminus\gamma)^k$ for the same segment $\gamma$ appearing in Section~\ref{ONsection}. If $I$ is a finite set, $(x_i)_{i \in I}$ a collection of variables and $J$ a subset of $I$, we denote $x_{J} = (x_j)_{j \in J}$. The formula of Proposition~\ref{magi} can be translated into
\bea
\label{FixedNGcorrelators}\mathscr{W}^{[g,k]}_{\Gamma,\star,\mathbf{s}}(x_1,\ldots,x_k) & = &  \oint_{\gamma^{E_{{\rm glue}}(\Gamma)}} \prod_{\mathsf{e} \in E_{{\rm glue}}(\Gamma)}  \frac{\dd y_{\mathsf{e}}}{2{\rm i}\pi} \prod_{\mathsf{v} \in \tilde{V}(\Gamma)} \frac{\mathsf{W}^{[h(\mathsf{v}),k(\mathsf{v}) + d(\mathsf{v}),\bullet k'(\mathsf{v})]}(x_{\partial(\mathsf{v})},y_{\mathsf{e}(\mathsf{v})})}{d(\mathsf{v})!} \nonumber\\
&& \qquad \prod_{\mathsf{e} \in \tilde{E}(\Gamma)} \tilde{\mathcal{W}}^{[0,2]}_{s(\mathsf{e})}(y_{\mathsf{e}_+},y_{\mathsf{e}_-}) \prod_{\mathsf{v} \in V_{0,2}(\Gamma)}\hat{\mathcal{W}}^{[0,2]}_{s(\mathsf{e}_+(\mathsf{v}))}(y_{\mathsf{e}_+(\mathsf{v})},x_{\partial(\mathsf{v})}). 
\eea
\end{proposition}

\subsubsection{Topological recursion and outline}\label{TROn}

\begin{theorem}\cite{Eynardbook,BEOn,BEO}
The generating series $\mathcal{W}^{[g,k]}$ for arbitrary topologies can be obtained from the generating series of disks $\mathcal{W}^{[0,1]} = \mathcal{W}$ and of cylinders $\mathcal{W}^{[0,2]} = \mathcal{W}^{[0,2]}$ by topological recursion. By specialization, the generating series of usual maps at renormalized face weights $\mathsf{W}^{[g,k]}$ is also given by topological recursion: the initial data of the recursion is then $\mathsf{W}(x) = \mathcal{W}(x)$ and $\mathsf{W}^{[0,2]}$ given by \eqref{F2solep}.
\end{theorem}



We remind the reader that the Section \ref{TRIntro} in Introduction is devoted to an overview of the general topological recursion method. 
We shall describe its somewhat simpler application to the bending energy model in the next Section.

For the general $O(\mathsf{n})$ model, we cannot go much further at present. Let us summarize the logic of computation of $\mathscr{W}_{\Gamma,\star,\mathbf{s}}^{[g,k]}$, which is the main quantity of interest in this article. Firstly, one tries to solve for $\mathcal{W}(x)$ the linear equation of Theorem~\ref{funcF}, as a function of $\gamma_{\pm}$, only exploiting that $\sigma(x)\mathcal{W}(x)$ remains uniformly bounded for $x \rightarrow \gamma_{\pm}$ -- for the moment, we do not use the stronger fact that $\mathcal{W}(x)$ is bounded. This problem is known a priori to have a unique solution for any choice of $\gamma_{\pm}$, but is hardly amenable to an explicit solution. Secondly, imposing that $\mathcal{W}(x)$ is actually uniformly bounded for $x \in \mathbb{C}\setminus \gamma$ gives two non-linear equations which determine $\gamma_{\pm}$. These equations may not have a unique solution, but we look for the unique solution such that $\gamma_{\pm}$ are evaluations at the desired weights of formal power series of $\sqrt{u}$, $t_l$, $n$ and $A_{k,l}$. Thirdly, now knowing $\gamma_{\pm}$ -- or assuming to know them -- one tries to solve for $\mathcal{W}^{[0,2]}_{s}(x_1,x_2)$ the linear equation of Theorem~\ref{prop76}, in a uniform way for any $s$. This problem is as difficult as the first step\footnote{As a matter of fact, there exists a general and explicit linear formula to extract $\mathcal{W}(x)$ (resp. $\mathcal{W}^{\bullet}_{s}$) from the knowledge of $\mathcal{W}^{[0,2]}_{s = 1}(x_1,x_2)$ (resp. $\mathcal{W}^{[0,2]}_{s}(x_1,x_2)$), which we will not need here.}. In a fourth step, if $\gamma_{\pm}$, $\mathcal{W}(x)$, and $\mathcal{W}^{[0,2]}(x_1,x_2)$ are known or assumed so, the topological recursion allows the explicit computation of $\mathcal{W}^{[g,k]}(x_1,\ldots,x_k)$ by induction on $2g - 2 + k$. We now have all the ingredients to compute in a fifth step the generating series $\mathscr{W}_{\Gamma,\star,\mathbf{s}}^{[g,k]}$ in absence of marked points. 

\subsection{Adding marked points}
\label{addinm}
The computation of generating series of maps with marked points is done a posteriori. For the generating series of maps with loops where the position of the marked points is not constrained, we simply have
$$
\mathcal{W}^{[g,k,\bullet k']} = (u\partial_u)^{k'} \mathcal{W}^{[g,k]}.
$$
To force marked points and boundaries to be all together, not separated by loops, \textit{i.e.} to compute $\mathsf{W}^{[g,k,\bullet k']}$, we proceed differently.

Consider a usual map of genus $g$ with $k$ boundaries of perimeters $(\ell_i)_{i = 1}^k$. Denote $V$ the number of vertices, $E$ the number of edges, and $(N_m)_{m \geq 1}$ the number of (non-marked) faces of degree $m$. We have the Euler relation
$$
2 - 2g - k = V - E + \sum_{m \geq 1} N_m,
$$
and counting half-edges gives
$$
2E = \sum_{m \geq 1} mN_{m} + \sum_{i = 1}^k \ell_i.
$$
Then, the number of vertices is
$$
V = 2 - 2g - k + \sum_{m \geq 1} (\tfrac{m}{2} - 1)N_m + \sum_{i = 1}^k \tfrac{1}{2} \ell_i.
$$

Therefore, the operation of marking a point is realized at the level of generating series by application of the operator
$$
2 - 2g - k + \sum_{m \geq 1} (\tfrac{1}{2} - \tfrac{1}{m}) mt_m \partial_{t_m} - \sum_{i = 1}^k \tfrac{1}{2}\partial_{x_i} x_i.
$$
In particular, if we denote $V(x) = \tfrac{1}{2}x^2 - \sum_{m \geq 1}\tfrac{t_m}{m}x^m$, the generating series of usual maps with marked points and (non-renormalized) face weights $\{t_m\}_{m \geq 1}$ satisfies, for all $k' \geq 1$
\bea
W^{[g,k,\bullet k']}(x_1,\ldots,x_k) & = &  \Big(2 - 2g - k - \sum_{i = 1}^k \tfrac{1}{2}\,\partial_{x_i} x_i\Big)W^{[g,k,\bullet (k' - 1)]}(x_1,\ldots,x_k) \nonumber \\
&& - \oint_{\gamma} \frac{\dd y}{2{\rm i}\pi}\,\big(\tfrac{y}{2}V'(y) - V(y)\big)W^{[g,k + 1,\bullet (k' - 1)]}(y,x_1,\ldots,x_k). \nonumber
\eea
For renormalized face weights, we have to take into account the shift \eqref{eq:fixp}, resulting in
\bea
\mathsf{W}^{[g,k,\bullet k']}(x_1,\ldots,x_k) & = & \Big(2 - 2g - k - \sum_{i = 1}^k \tfrac{1}{2}\,\partial_{x_i} x_i\Big)\mathsf{W}^{[g,k,\bullet (k' - 1)]}(x_1,\ldots,x_k) \nonumber \\
\label{renFmarked} &&- \oint_{\gamma} \frac{\dd y}{2{\rm i}\pi}\,\big(\tfrac{y}{2}\tilde{V}'(y) - \tilde{V}(y)\big)\mathsf{W}^{[g,k + 1,\bullet (k' - 1)]}(y,x_1,\ldots,x_k), 
\eea
where
$$
\tilde{V}(x) = V(x) - \oint_{\gamma} \frac{\dd z}{2{\rm i}\pi}\,\mathbf{R}(x,z)\,\mathcal{W}(z).
$$


\section{The bending energy model}\label{S4}

We shall focus on the class of loop models with bending energy on triangulations studied in \cite{BBG12b}, for which the computations can be explicitly carried out. On top of the loop fugacity $\mathsf{n}$ and the vertex weight $u$, it features a weight $\mathsf{g}$ per unvisited triangle, $\mathsf{h}$ per visited triangle, and $\alpha$ per consecutive pair of visited triangles pointing in the same direction. In this simplified model the annuli generating series, which can be directly computed combinatorially, take a particularly simple form:
\bea
\label{rin}\mathbf{R}(x,z) & = &  \mathsf{n}\ln\left(\frac{1}{1 - \alpha \mathsf{h}(x + z) - (1 - \alpha^2)\mathsf{h}^2xz}\right) \\
&  = &  \mathsf{n}\ln\left(\frac{1}{z - \varsigma(x)}\right) + \frac{\mathsf{n}}{2}\ln\left(\frac{\varsigma'(x)}{-\mathsf{h}^2}\right), \nonumber \\
\mathbf{A}(x,z) & = & \partial_{x}\mathbf{R}(x,z) \nonumber \\
& = & \mathsf{n}\left(\frac{\varsigma'(x)}{z - \varsigma(x)} + \frac{\varsigma''(x)}{2\varsigma'(x)}\right), \nonumber
\eea
where
\beq
\label{varsigma}\varsigma(x) = \frac{1 - \alpha \mathsf{h} x}{\alpha \mathsf{h} + (1 - \alpha^2)\mathsf{h}^2 x}
\eeq
is a rational involution. We assume that the weights are admissible, and thus all relevant generating series of maps with boundaries have a cut $[\gamma_-,\gamma_+]$.

Technically, the fact that $\mathbf{A}(x,y)$ is a rational function with a single pole allows for an explicit solution of the linear equation for $\mathcal{W}(x)$ and $\mathcal{W}_{s}^{[0,2]}(x_1,x_2)$, assuming $\gamma_{\pm}$ are known (see Section~\ref{slili}). Then, $\gamma_{\pm}$ are determined implicitly by two complicated equations -- cf. \eqref{determinab} below. This is nevertheless explicit enough to analyze the critical behavior of the model (see Section~\ref{critit}).

\subsection{Solving the linear equation}
\label{elparam}

If $f$ is a holomorphic function in $\mathbb{C}\setminus\gamma$ such that $f(x) \sim c_{f}/x$ when $x \rightarrow \infty$, we can evaluate the contour integral:
\beq
\label{contueq}\oint_{\gamma} \frac{\dd y}{2{\rm i}\pi}\,\mathbf{A}(x,y)\,f(y) = -\mathsf{n} \varsigma'(x)\,f(\varsigma(x)) + \mathsf{n} c_{f}\,\frac{\varsigma''(x)}{2\varsigma'(x)},
\eeq
where we notice that
$$
\frac{\varsigma''(x)}{2\varsigma'(x)} = -\frac{1}{x - \varsigma(\infty)}.$$
Therefore, a linear equation of the form
$$
f(x + {\rm i}0) + f(x - {\rm i}0) + s\oint_{\gamma} \frac{\dd y}{2{\rm i}\pi}\,\mathbf{A}(x,y)\,f(y) = \phi(x),\qquad \forall x \in (\gamma_-,\gamma_+),
$$
becomes
\beq
\label{fhomo} f(x + {\rm i}0) + f(x - {\rm i}0) - \mathsf{n}\,s\,\varsigma'(x)f(\varsigma(x)) = \tilde{\phi}(x) \coloneqq \phi(x)  -\mathsf{n}\,s\, c_{f}\,\frac{\varsigma''(x)}{2\varsigma'(x)},\ \  \forall x \in (\gamma_-,\gamma_+).
\eeq
When $\mathsf{n}s \neq \pm 2$, which is assumed here, 
\beq
\label{fhom} f^{{\rm hom}}(x) = f(x) - \frac{2\tilde{\phi}(x) + \mathsf{n} s \varsigma'(x)\tilde{\phi}(\varsigma(x))}{4 - \mathsf{n}^2s^2},
\eeq
with $f$ a solution of (\ref{fhomo}), satisfies the following homogeneous linear equation:
\beq
\label{fhomoh} \forall x \in (\gamma_-,\gamma_+),\qquad f^{{\rm hom}}(x + {\rm i}0) + f^{{\rm hom}}(x - {\rm i}0) - \mathsf{n}s\varsigma'(x)f^{{\rm hom}}(\varsigma(x)) = 0.
\eeq
If we assume that $\phi(x)$ is a given rational function with poles $q$ away from $\gamma$, $f^{{\rm hom}}(x)$ acquires poles at the same points, and we have:
$$
f^{{\rm hom}}(x) = \delta_{q,\infty}\,\frac{c_{f}}{x} - \frac{2\tilde{\phi}(x) + \mathsf{n} s \varsigma'(x)\tilde{\phi}(\varsigma(x))}{4 - \mathsf{n}^2s^2} + O(1),\qquad x \rightarrow q.
$$
So, we are left with the problem of solving \eqref{fhomoh} with vanishing right-hand side, but admitting rational singularity with prescribed divergent part at a finite set of points $q \in \mathbb{C}\setminus \gamma$.

\begin{figure}
\begin{center}
\includegraphics[width=0.7\textwidth]{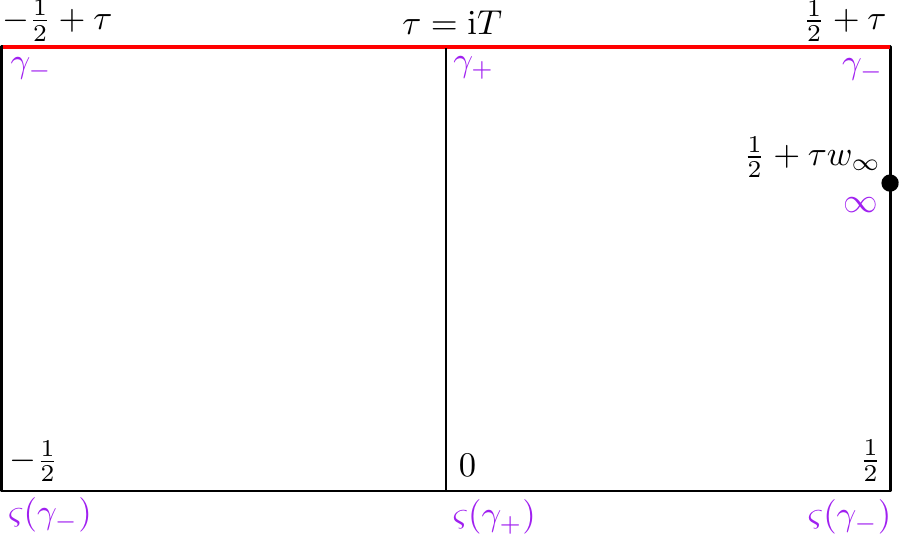}
\end{center}
\caption{\label{ParamF} Two copies of the fundamental rectangle in the $v$-plane. We indicate the image of special values of $x$ in purple, and the image of the cut $\gamma$ in red. The left (resp. right) panel is the image of ${\rm Im}\,x > 0$ (resp. ${\rm Im}\,x < 0$).}
\end{figure}

The key to the solution is the use of an elliptic parametrization $x = x(v)$.
Considering values of $\gamma_{\pm}$ and $\varsigma(\gamma_{\pm})$ such that
\beq
\label{orderg}\gamma_- < \gamma_+ < \varsigma(\gamma_+) < \varsigma(\gamma_-),
\eeq
we set
\beq
\label{io} v = {\rm i}C\,\int^{x}_{\varsigma(\gamma_+)} \frac{\dd y}{\sqrt{(y - \varsigma(\gamma_-))(y - \varsigma(\gamma_+))(y - \gamma_+)(y - \gamma_-)}}.
\eeq
The normalizing constant $C$ is chosen such that, for $x$ moving from the origin $\varsigma(\gamma_+)$ to $\varsigma(\gamma_-)$ with a small negative imaginary part, $v$ is moving from $0$ to $\tfrac{1}{2}$. When $x$ moves on the real axis from $\varsigma(\gamma_+)$ to $\gamma_+$, $v$ moves from $0$ to a purely imaginary value denoted $\tau = {\rm i}T$. We give some properties of the parametrization $x$, which follow from studying the analytic continuation of the functional inverse of $v$. For more information on this parametrization, we refer the reader to the Appendix \ref{App1}. 
The domain $\mathbb{C}\setminus\big(\gamma\cup\varsigma(\gamma)\big)$ is mapped to the fundamental rectangle (Figure~\ref{ParamF})
\beq
\big\{v \in \mathbb{C},\  0 < \mathrm{Re}\,v < 1/2,\  |\mathrm{Im}\,v| < T\big\},
\eeq
with values at the corners:
\beq
\begin{array}{lcl}
x(\tau) = x(-\tau) = \gamma_{+}, & \qquad &  x(\tau + 1/2) = x(-\tau + 1/2) = \gamma_{-}, \\ 
x(0) = \varsigma(\gamma_{+}), & \qquad & x(1/2) = \varsigma(\gamma_{-}). \end{array} 
\eeq
Besides, when $x$ is in the physical sheet, $$v(\varsigma(x)) = \tau - v(x).$$ Since the involution $\varsigma$ is decreasing, $\varsigma(\gamma_-)$ belongs to the union $(\varsigma(\gamma_+),+\infty) \sqcup (-\infty,\gamma_-)$, and therefore $x = \infty$ is mapped to $v_{\infty} = \frac{1}{2} + \tau w_{\infty}$ with $0 < w_{\infty} < 1/2$. When $\alpha = 1$, by symmetry we must have $w_{\infty} = 1/2$.

The function $v \mapsto x(v)$ is analytically continued for $v \in \mathbb{C}$ by the relations:
\beq
x(-v) = x(v + 1) = x(v + 2\tau) = x(v).
\eeq
This parametrization allows the conversion \cite{EKOn,BBG12b} of the functional equation
\beq
\forall x \in \mathring{\gamma},\qquad f(x + {\rm i}0) + f(x - {\rm i}0) - \mathsf{n}\,\varsigma'(x)\,f(\varsigma(x)) = 0
\eeq
for an analytic function $f(x)$ in $\mathbb{C}\setminus\gamma$, into the functional equation:
\beq
\label{reuh}\forall v \in \mathbb{C},\qquad \tilde{f}(v + 2\tau) + \tilde{f}(v) - \mathsf{n}\,\tilde{f}(v - \tau) = 0, \qquad \text{with } \tilde{f}(v) = \tilde{f}(v + 1) = -\tilde{f}(-v),
\eeq
for the analytic continuation of the function $\tilde{f}(v) = f(x(v))x'(v)$. The second condition in \eqref{reuh} enforces the continuity of $f(x)$ on $\mathbb{R}\setminus\gamma$. We set:
\beq
b = \frac{\mathrm{arccos}(\mathsf{n}/2)}{\pi}\,.
\eeq
The new parameter $b$ ranges from $\tfrac{1}{2}$ to $0$ when $\mathsf{n}$ ranges from $0$ to $2$. Solutions of the first equation of \eqref{reuh} with prescribed meromorphic singularities can be build from a fundamental solution $\Upsilon_b$, defined uniquely by the properties:
\beq
\label{propoup}\Upsilon_{b}(v + 1) = \Upsilon_{b}(v),\qquad \Upsilon_{b}(v + \tau) = e^{{\rm i}\pi b}\Upsilon_{b}(v),\qquad \Upsilon_{b}(v) \mathop{\sim}_{v \rightarrow 0} \frac{1}{v}\,.
\eeq
Its expression and main properties are reminded in Section~\ref{AppUp} of the Appendix.

\subsubsection{Elementary generating series}\label{slili} We present the solution for the generating series of disks, and of refined disks and cylinders.
\label{SectionG}
Let $\mathcal{G}(v)$ be the analytic continuation of 
\beq
\label{EquationG} x'(v)\mathcal{W}(x(v)) - \partial_{v}\Bigg(\frac{2V(x(v)) + \mathsf{n} V(\varsigma(x(v)))}{4 - \mathsf{n}^2} - \frac{\mathsf{n}\, u \ln \big[\varsigma'(x(v))\big]}{2(2 + \mathsf{n})}\Bigg),
\eeq
where $V(x) = \tfrac{1}{2}x^2 - \sum_{k \geq 1} \tfrac{t_l}{l}x^{l}$ collects the weights of empty faces. In the model we study, empty faces are triangles counted with weight $\mathsf{g}$ each, so $V(x) = \tfrac{1}{2}x^2 - \tfrac{\mathsf{g}}{3}x^3$. Let us introduce $(\tilde{\mathsf{g}}_l)_{l \geq 0}$ as the coefficients of expansion:
$$
\label{deftildeg} \frac{\partial}{\partial v}\Big(-\frac{2V(x(v))}{4 - \mathsf{n}^2} + \frac{2 \ln x(v)}{2 + \mathsf{n}}\Big) \mathop{=}_{v \rightarrow v_{\infty}} \sum_{k \geq 1} \frac{\tilde{\mathsf{g}}_{k - 1}}{(v - v_{\infty})^k} + O(1).
$$
Their expressions for the model where all faces are triangles are recorded in Section~\ref{Appgdeter} of the Appendix.

\begin{proposition}[Disks] \cite{BBG12b}
\label{theimdisk}We have that
$$
\mathcal{G}(v) = \sum_{l \geq 0} \frac{1}{2}\,\frac{(-1)^l\tilde{\mathsf{g}}_l}{l!}\,\partial_{v_{\infty}}^{l}\big[\Upsilon_{b}(v + v_{\infty}) + \Upsilon_{b}(v - v_{\infty}) - \Upsilon_{b}(-v + v_{\infty}) - \Upsilon_{b}(-v - v_{\infty})\big].
$$
The endpoints $\gamma_{\pm}$ are determined by the two conditions:
\beq
\label{determinab}\mathcal{G}(\tau + \varepsilon) = 0,\qquad \varepsilon = 0,\tfrac{1}{2},
\eeq
which follow from the fact that $\mathcal{W}(x)$ remains bounded when $x \rightarrow \gamma_{\pm}$.
\end{proposition}

For use in refined generating series, let us define
$$
b(s) = \frac{{\rm arccos}(\mathsf{n}s/2)}{\pi}.
$$

\begin{proposition} \label{their}\cite{BBD} Define $\mathcal{G}^\bullet_{s}(v)$ as the analytic continuation of
$$
\label{second}x'(v)\mathcal{W}_{s}^\bullet(x(v)) + \partial_{v}\Bigg(\frac{\mathsf{n}su\,\ln[\varsigma'(x(v))]}{2(2 + \mathsf{n}s)}\Bigg).
$$
We have:
$$
\mathcal{G}_{s}^\bullet(v) = \frac{u}{2 + \mathsf{n}s}\Big[-\Upsilon_{b(s)}(v + v_{\infty}) - \Upsilon_{b(s)}(v - v_{\infty}) + \Upsilon_{b(s)}(-v + v_{\infty}) + \Upsilon_{b(s)}(-v - v_{\infty})\Big].
$$
\end{proposition}

\begin{proposition} \label{p15}\cite{BEOn,BBD} Define $\mathcal{G}^{[0,2]}_{s}(v_1,v_2)$ as the analytic continuation of
$$
x'(v_1)x'(v_2)\mathcal{W}^{[0,2]}_{s}(x(v_1),x(v_2)) + \frac{\partial}{\partial v_1}\frac{\partial}{\partial v_2}\Bigg(\frac{2\ln\big[x(v_1) - x(v_2)\big] + \mathsf{n}s\ln\big[\varsigma(x(v_1)) - x(v_2)\big]}{4 - \mathsf{n}^2s^2}\Bigg).
$$
We have:
$$
\mathcal{G}^{[0,2]}_{s}(v_1,v_2) = \frac{1}{4 - \mathsf{n}^2s^2}\Big[\Upsilon_{b(s)}'(v_1 + v_2) - \Upsilon_{b(s)}'(v_1 - v_2) - \Upsilon_{b(s)}'(-v_1 + v_2) + \Upsilon_{b(s)}'(-v_1 - v_2)\Big].
$$
\end{proposition}

\begin{remark} When there is no bending energy, i.e. $\alpha = 1$, the $4$-terms expression of Propositions~\ref{theimdisk}-\ref{their} can be reduced to $2$ terms using $\tau - v_{\infty} = v_{\infty}\,\,{\rm mod}\,\,\mathbb{Z}$ and the pseudo-periodicity of the special function $\Upsilon_{b}$. 
\end{remark}


\section{Topological recursion}\label{TRtr}


Theorem~\ref{2g2m} in the special case of the bending energy model shows that $\mathcal{W}^{[g,k]}(x_1,x_2,\ldots,x_k)$ for $2g - 2 + k > 0$ satisfies the homogeneous linear equation with respect to $x_1$, for fixed $(x_i)_{i = 2}^k$. Following Section~\ref{elparam}, we can thus introduce a meromorphic function $\mathcal{G}^{[g,k]}(v_1,\ldots,v_k)$ as the analytic continuation of
\beq
\label{140}\mathcal{W}^{[g,k]}(x(v_1),\ldots,x(v_k))\,\prod_{i = 1}^k x'(v_i).
\eeq
It is also convenient to introduce a shift for the case of cylinders. We consider:
\bea
\overline{\mathcal{G}}^{[g,k]}(v_1,\ldots,v_k) & = & \mathcal{G}^{[g,k]}(v_1,\ldots,v_k) \nonumber \\
\label{H5s} & & + \delta_{g,0}\delta_{k,2}\bigg(\frac{2 - \mathsf{n}^2}{4 - \mathsf{n}^2}\,\frac{x'(v_1)x'(v_2)}{(x(v_1) - x(v_2))^2} - \frac{\mathsf{n}}{4 - \mathsf{n}^2}\,\frac{\varsigma'(x(v_1))x'(v_1)x'(v_2)}{(x(v_1 - \tau) - x(v_2))^2} \bigg). \nonumber
\eea
While $\mathcal{G}^{[0,2]}(v_1,v_2)$ satisfied the homogeneous linear equation, $\overline{\mathcal{G}}^{[0,2]}(v_1,v_2)$ satisfies, with respect to $v_1$, the inhomogeneous version of equation \eqref{reuh} with right-hand side $1/(x(v_1) - x(v_2))^2$.

Our starting point is the topological recursion residue formula proved in \cite{BEOn} or \cite[Section 5]{BEO}. Let us define the recursion kernel, for $\varepsilon \in \{0,1/2\}$:
\beq
K_{\varepsilon}(v_0,v) = -\frac{\dd v}{2}\,\frac{\int_{2\tau - v}^{v} \dd v'\,\overline{\mathcal{G}}^{[0,2]}(v',v_0)}{\mathcal{G}(v) + \mathcal{G}(2\tau - v)}.
\eeq
If $k \geq 2$, let $I = \{2,\ldots,k\}$, and if $k = 1$, $I = \emptyset$. If $J$ is a set, we denote $v_{J} = (v_{j})_{j \in J}$. 

\begin{theorem}
\label{them1o} For $2g - 2 + k > 0$, we have
\bea
\overline{\mathcal{G}}^{[g,k]}(v_1,v_I) & = & \sum_{\varepsilon \in \{0,1/2\}} \Res_{v \rightarrow \tau + \varepsilon} K_{\varepsilon}(v_1,v) \Bigg[\overline{\mathcal{G}}^{[g - 1,k + 1]}(v,2\tau - v,v_{I}) \nonumber \\
&& + \sum_{\substack{h + h' = g \\ J \sqcup J' = I}}^{{\rm no}\,\,{\rm disks}} \overline{\mathcal{G}}^{[h,1 + |J|]}(v,v_{J})\,\overline{\mathcal{G}}^{[h',1 + |J'|]}(2\tau - v,v_{J'})\Bigg], \nonumber
\eea
where ``no disks'' means that we exclude the terms containing disk generating series, that is $(h,J)$ or $(h',J')$ equal to $(0,\emptyset)$.
\end{theorem}

We are going to rewrite this recursion without involving residues. We first need to introduce some notations. Let us define the \emph{elementary blocks}:
\beq
\label{BB0}\varepsilon \in \{0,\tfrac{1}{2}\},\qquad \mathcal{B}_{\varepsilon,l}(v) = \frac{\partial^{2l}}{\partial v_2^{2l}} \overline{\mathcal{G}}^{[0,2]}(v,v_2)\Big|_{v_2 = \tau + \varepsilon}.
\eeq
Since $x(\tau + \varepsilon + w)$ is an even function of $w$, formula \eqref{BB0} is insensitive to replacing $\overline{\mathcal{G}}^{[0,2]}$ by $\mathcal{G}^{[0,2]}$. From the structure of $\mathcal{G}^{[0,2]} = \mathcal{G}^{[0,2]}_{s = 1}$ shown in Proposition~\ref{p15}, we see that
\beq
\label{BB1} \mathcal{B}_{\varepsilon,l}(v) = \partial_{v}^{2l} \mathcal{B}_{\varepsilon,0}(v).
\eeq

\begin{proposition}
\label{2g2mr}For $2g - 2 + k > 0$, we have a decomposition
$$
\overline{\mathcal{G}}^{[g,k]}(v_1,\ldots,v_k) = \mathcal{G}^{[g,k]}(v_1,\ldots,v_k) = \sum_{\substack{l_1,\ldots,l_k \geq 0 \\ \varepsilon_1,\ldots,\varepsilon_{k} \in \{0,\frac{1}{2}\}}} \mathcal{C}^{[g,k]}\bigl[{}^{l_1}_{\varepsilon_1} \cdots {}^{l_k}_{\varepsilon_k}\bigr]  \prod_{i = 1}^k \mathcal{B}_{\varepsilon_i,l_i}(v_i),
$$
where the sum contains only finitely many non-zero terms.
\end{proposition}

We outline now our procedure to arrive to the main result of Chapter~\ref{Chapter6} (Theorem~\ref{ouqusf}), which is the main tool necessary to deduce all the nesting properties of Chapter \ref{LoopNesting}:
\begin{itemize}
\item Recursion for the coefficients $\mathcal{C}^{[g,k]}[{}^{l_I}_{\varepsilon_I}\bigr]$ with two kinds of coefficients that we denote $\mathcal{K}$ and $\tilde{\mathcal{K}}$ given in Proposition~\ref{cocor1}, as a consequence of TR (Theorem~\ref{them1o}). 

\item Diagrammatic representation by trivalent vertices of $\mathcal{K}$, $\tilde{\mathcal{K}}$, and the initial cases $\mathcal{C}^{(0,3)}$ and $\mathcal{C}^{(1,1)}$, with different properties of their incident edges (Definition~\ref{trivVert}).

\item Expression for $\mathcal{C}^{[g,k]}$'s as a sum over graphs composed by the previous four kinds of pieces in Proposition~\ref{cosums}. 

\item Critical behavior of the four kinds of pieces and of the elementary blocks in Lemma~\ref{pieces}.

\item Fixed the coloring of the $k$ legs $\varepsilon_1,\ldots,\varepsilon_k$, determine which graph and coloring (of the graph) give the leading contribution to $\mathcal{C}^{[g,k]}$ in the critical regime. This is solved Lemma~\ref{Cbehavior}, and constitutes the most technical step.

\item Critical behavior of $\mathcal{W}^{[g,k]}$ and $\mathsf{W}^{[g,k]}$, obtained summing all these contributions over the possible colorings of the legs (Theorem~\ref{ouqusf}).
\end{itemize}

\subsubsection{Initial conditions}
\label{Seinia} We denote $y_{\varepsilon,1}$ and $y_{\varepsilon,2}$ the first two coefficients in the Taylor expansion at $w \rightarrow 0$:
\beq
\label{316} \Delta_{\varepsilon} \mathcal{G}(w) \coloneqq \mathcal{G}(w + \tau + \varepsilon) + \mathcal{G}(-w + \tau + \varepsilon) = y_{\varepsilon,1}\,w^2 + \frac{y_{\varepsilon,2}}{6}\,w^4 + O(w^6).
\eeq
We also need the constants
\beq
\label{ububub} \upsilon_{b,2m+1} = \lim_{w \rightarrow 0} \Big(\Upsilon_{b}^{(2m+1)}(w) + \frac{(2m+1)!}{w^{2m+2}}\Big),
\eeq
introduced in Appendix~\ref{AppUp}. The initial conditions for the recursion concern $(g,k) = (0,3)$ and $(1,1)$:
$$
\boxed{\mathcal{C}^{[0,3]}\big[{}^{l_1}_{\varepsilon_1}\,{}^{l_2}_{\varepsilon_2}\,{}^{l_3}_{\varepsilon_3}\bigr] = -\frac{2\,\delta_{l_1,l_2,l_3,0}\,\delta_{\varepsilon_1,\varepsilon_2,\varepsilon_3}}{y_{\varepsilon_1,1}},
\qquad \mathcal{C}^{[1,1]}\bigl[{}^{l}_{\varepsilon}\bigr] = \delta_{l,0}\Big(\frac{y_{\varepsilon,2}}{24y_{\varepsilon,1}^2} + \frac{\upsilon_{b,1}}{y_{\varepsilon,1}}\Big) -\frac{\delta_{l,1}}{24 y_{\varepsilon,1}}.}
$$

\subsubsection{The recursion coefficients}
We first define
\beq
\label{Khoubi} \boxed{\mathcal{K}\bigl[{}^{l}_{\varepsilon}\,{}^{m}_{\,\sigma}\,{}^{m'}_{\,\sigma'}\bigr] = \Res_{w \rightarrow 0} \frac{-w^{2l + 1}\dd w}{(2l + 1)!\,\Delta_{\varepsilon}\mathcal{G}(w)}\,\mathcal{B}_{\sigma,m}(w + \tau + \varepsilon)\,\mathcal{B}_{\sigma',m'}(-w + \tau + \varepsilon).}
\eeq
Since $\Delta_{\varepsilon}\mathcal{G}(w)$ is even, we have the symmetry
\beq
\mathcal{K}\bigl[{}^{l}_{\varepsilon}\,{}^{m}_{\,\sigma}\,{}^{m'}_{\,\sigma'}\bigr] = \mathcal{K}\bigl[{}^{l}_{\varepsilon}\,{}^{m'}_{\,\sigma'}\,{}^{m}_{\,\sigma}\bigr].
\eeq
By counting the degree of the integrand at $w = 0$, we find that there are finitely many indices for which $\mathcal{K}$ does not vanish:
\bea
\label{selvK} &&\,\,\, \big\{\varepsilon = \sigma = \sigma' \,\,\mathrm{and}\,\, l \leq m + m ' + 2\big\}, \nonumber \\
& \mathrm{or}& \,\,\,\big\{\varepsilon = \sigma \neq \sigma'\,\,\mathrm{and}\,\,l \leq m + 1\big\}, \nonumber \\
& \mathrm{or}& \,\,\,\big\{\varepsilon \neq \sigma = \sigma'\,\,\mathrm{and}\,\,l = 0\big\}. \nonumber
\eea
We also define
\beq
\label{Ktildeou}\boxed{\tilde{\mathcal{K}}\bigl[{}^{l}_{\varepsilon}\,{}^{l'}_{\varepsilon'}\,{}^{m}_{\,\sigma}\bigr] =\,\frac{-\delta_{\varepsilon,\varepsilon'}}{(2l + 1)!\,(2l')!}\Res_{w \rightarrow 0} \dd w\,\frac{w^{2(l + l')+1}}{\Delta_{\varepsilon}\mathcal{G}(w)}\,\mathcal{B}_{\sigma,m}(\tau + \varepsilon + w).}
\eeq
Again, there are finitely many values of the parameters for which $\tilde{\mathcal{K}}$ does not vanish:
\bea
\label{selvKt} &&\,\,\,\big\{\varepsilon = \varepsilon' = \sigma\,\,\mathrm{and}\,\,l + l' \leq m + 1\big\}, \nonumber \\
& \mathrm{or}& \,\,\,\big\{\varepsilon = \varepsilon' \neq \sigma\,\,\mathrm{and}\,\,(l,l') = (0,0)\big\}. \nonumber
\eea

\subsection{The recursion formula without residues}

\begin{proposition}\label{cocor1}
Assume $2g - 2 + k \geq 2$, and denote $L = \{2,\ldots,k\}$. The coefficients of the decomposition in Proposition~\ref{2g2mr} satisfy:
\bea
\mathcal{C}^{[g,k]}\bigr[{}^{l_1}_{\varepsilon_1} \cdots\,{}^{l_k}_{\varepsilon_k}\big] & = & \sum_{\substack{m,m' \geq 0 \\\sigma,\sigma' \in \{0,1/2\}}} \mathcal{K}\bigl[{}^{l_1}_{\varepsilon_1}\,{}^{m}_{\,\sigma}\,{}^{m'}_{\,\sigma'}\bigr]\,\mathcal{C}^{[g - 1,k+1]}\bigl[{}^{m}_{\,\sigma}\,{}^{m'}_{\,\sigma'}\,{}^{l_{L}}_{\varepsilon_{L}}\bigr] \nonumber \\
& & + \sum_{\substack{h + h' = g \\ J \sqcup J' = L \\ m,m' \geq 0,\,\,\sigma,\sigma' \in \{0,1/2\}}}^{{\rm stable}} \mathcal{K}\bigl[{}^{l_1}_{\varepsilon_1}\,{}^{m}_{\,\sigma}\,{}^{m'}_{\,\sigma'}\bigr]\,\mathcal{C}^{[h,|J| + 1]}\big[{}^{m}_{\,\sigma}\,{}^{l_{J}}_{\varepsilon_J}\bigr]\,\mathcal{C}^{[h',1 + |J'|]}\big[{}^{m'}_{\,\sigma'}\,{}^{l_{J'}}_{\varepsilon_{J'}}\bigr] \nonumber \\
\label{chack}& & + \sum_{\substack{i \in L,\,\,m \geq 0 \\ \sigma \in \{0,1/2\}}} 2\,\tilde{\mathcal{K}}\bigl[{}^{l_1}_{\varepsilon_1}\,{}^{l_i}_{\varepsilon_i}\,{}^{m}_{\,\sigma}\bigr]\,\mathcal{C}^{[g,k - 1]}\bigl[{}^{m}_{\,\sigma}\,{}^{l_{L\setminus\{i\}}}_{\varepsilon_{L\setminus\{i\}}}\bigr],
\eea
where ``stable'' means that we exclude the terms involving disk or cylinder generating series, i.e. for which $(h, |J|+1)$ or $(h', |J'|+1)$ belongs to $\{(0,1),(0,2)\}$.
\end{proposition}

Although this recursion gives a non symmetric role to the first boundary, the result ensuing from the initial conditions of \S~\ref{Seinia} is symmetric. This must be true by consistency, and this is in fact a general property of the topological recursion, cf. \cite[Theorem 4.6]{EOFg}.


\subsubsection{Properties of the elementary blocks}

We have called elementary blocks the following functions:
\beq
\label{Bled}\mathcal{B}_{\varepsilon,l}(v) = \frac{\partial^{2l}}{\partial v_2^{2l}} \mathcal{G}^{[0,2]}(v,v_2)\Big|_{v_2 = \tau + \varepsilon}.
\eeq
\begin{lemma}
\label{regood} $\mathcal{B}_{\varepsilon,l}(\tau + \varepsilon' + w)$ is regular at $w = 0$ if $\varepsilon \neq \varepsilon'$, and behaves like $(2l + 1)!w^{-(2l + 2)} + O(1)$ when $w \rightarrow 0$ if $\varepsilon = \varepsilon'$.
\end{lemma}
\noindent \textbf{Proof.} We compute using Proposition~\ref{p15} and the properties \eqref{propoup} of $\Upsilon_{b}$:
\beq
\label{blbelb}\mathcal{B}_{\varepsilon,l}(\tau + \varepsilon' + w) = \frac{(e^{2{{\rm i}\pi b}} - 1)\Upsilon_{b}^{(2l + 1)}(\varepsilon + \varepsilon' + w) + (e^{-2{\rm i}\pi b} - 1)\Upsilon_{b}^{(2l + 1)}(\varepsilon + \varepsilon' - w)}{4 - \mathsf{n}^2}.
\eeq
We deduce its behavior when $w \rightarrow 0$. Since $\Upsilon_{b}$ is regular at the value $\tfrac{1}{2}$, \eqref{blbelb} is regular at $w = 0$ when $\varepsilon \neq \varepsilon'$. If $\varepsilon = \varepsilon'$, the simple pole of $\Upsilon_b$ produces the divergent behavior:
$$
\mathcal{B}_{\varepsilon,l}(\tau + \varepsilon + w) = \frac{(2l + 1)!}{w^{2l + 2}} + O(1).
$$
\hfill $\Box$

We shall need later in the computation of $\mathcal{G}^{[1,1]}$:
\begin{lemma}
\label{Lupb}
\beq
\label{upb2}\overline{\mathcal{G}}^{[0,2]}(\tau + \varepsilon + w,\tau + \varepsilon - w) \,\mathop{=}_{w \rightarrow 0}\,\, \frac{1}{4w^2} - \upsilon_{b,1} + o(1),
\eeq
where $\upsilon_{b,1}$ is the constant computed in \eqref{upb1def}.
\end{lemma}
\noindent\textbf{Proof.} We compute with \eqref{H5s}:
\bea
\label{G2taue}  && \overline{\mathcal{G}}^{[0,2]}(\tau + \varepsilon + w,\tau + \varepsilon - w)\\
& = & -\frac{2 - \mathsf{n}^2}{4 - \mathsf{n}^2}\bigg(\upsilon_{b,1} + \frac{S_{x}(\tau + \varepsilon + w)}{6}\bigg) \nonumber \\
&& + \frac{\mathsf{n}}{4 - \mathsf{n}^2}\,\frac{x'(\varepsilon + w)x'(\tau + \varepsilon + w)}{(x(\varepsilon + w) - x(\tau + \varepsilon + w))^2} - \frac{\Upsilon_{b}'(2w) + \Upsilon_{b}'(-2w)}{4 - \mathsf{n}^2}, \nonumber
\eea
where we introduced the Schwarzian derivative:
\beq
S_{x}(v) = \frac{x'''(v)}{x'(v)} - \frac{3}{2}\Big(\frac{x''(v)}{x'(v)}\Big)^2.
\eeq
Since $x'(\tau + \varepsilon + w)$ is an odd function of $w$, the second term in \eqref{G2taue} is $o(1)$ when $w \rightarrow 0$. We also compute:\bea
\frac{1}{6}\,S_{x}(\tau + \varepsilon + w) & = & \frac{1}{6}\Bigg[\frac{x''''(\tau + \varepsilon)}{x''(\tau + \varepsilon)} - \frac{3}{2w^2}\Bigg(\frac{1 + \frac{x''''(\tau + \varepsilon)}{2x''(\tau + \varepsilon)}\,w^2}{1 + \frac{x''''(\tau + \varepsilon)}{6x''(\tau + \varepsilon)}\,w^2}\Bigg)^2 + O(w^2)\Bigg] \nonumber \\
& = & -\frac{1}{4w^2} + O(w^2) \nonumber
\eea
and
\beq
\Upsilon_b'(2w) + \Upsilon_{b}'(-2w) = 2\Big(-\frac{1}{4w^2} + \upsilon_b\Big) + o(w^2). \nonumber
\eeq
Collecting all terms in \eqref{G2taue} we find \eqref{upb2}.
\hfill $\Box$

\subsubsection{Computing the residues}

Now we are ready to examine the formula of Theorem~\ref{them1o}. In order to compute the residues at $v \rightarrow \tau + \varepsilon$, we should first compute the expansion of the recursion kernel near those points. If we set $v = (\tau + \varepsilon) + w$ and $\Delta_{\varepsilon}\mathcal{G}(w) = \mathcal{G}(\tau + \varepsilon + w) + \mathcal{G}(\tau + \varepsilon - w)$, we find around $w\rightarrow 0$:
\bea
K_{\varepsilon}(v_0,\tau + \varepsilon + w) & = & \frac{-1}{2\Delta_{\varepsilon}\mathcal{G}(w)}\,\int_{-w}^{w} \dd z\Bigg(\sum_{l \geq 0} \mathcal{B}_{\varepsilon,l}(v_0)\,\frac{z^{2l}}{(2l)!} + ({\rm odd}\,\,{\rm terms})\Bigg) \nonumber \\
\label{expKq}& = & - \sum_{l \geq 0} \frac{w^{2l + 1}}{(2l + 1)!\Delta_{\varepsilon}\mathcal{G}(w)}\,\mathcal{B}_{\varepsilon,l}(v_0),
\eea
in terms of the elementary blocks \eqref{Bled}. Since we consider a model with off-critical weights, $\Delta_{\varepsilon}\mathcal{G}(w)$ has exactly a double zero at $w \rightarrow 0$. Subsequently, $K_{\varepsilon}(v_0,\tau + \varepsilon + w)$ has a simple pole at $w = 0$, and the term indexed by $l$ in the sum has a simple pole if $l = 0$, and has a zero of order $(2l - 1)$ if $l \geq 1$.

We prove Propositions~\ref{2g2mr}-\ref{cocor1} by induction on $\chi = 2g - 2 + k > 0$. The first case to consider is $\chi = 1$, i.e.  $(g,k) = (0,3)$ or $(1,1)$. For $(g,k) = (0,3)$, Theorem~\ref{them1o} yields
\bea
&&  \mathcal{G}^{[0,3]}(v_1,v_2,v_3) \nonumber \\
& = & \sum_{\varepsilon \in \{0,1/2\}} \Res_{w \rightarrow 0} K_{\varepsilon}(v_1,\tau + \varepsilon + w)\Big[\overline{\mathcal{G}}^{[0,2]}(\tau + \varepsilon + w,v_2)\overline{\mathcal{G}}^{[0,2]}(\tau + \varepsilon - w,v_3) \nonumber \\
& & \qquad\qquad\qquad\qquad\qquad\qquad\qquad + \overline{\mathcal{G}}^{[0,2]}(\tau + \varepsilon + w,v_3)\overline{\mathcal{G}}^{[0,2]}(\tau + \varepsilon - w,v_2)\Big]. \nonumber
\eea
As one can check from Proposition~\ref{p15}, $\mathcal{G}^{[0,2]}(\tau + \varepsilon + w,v')$ is regular when $w \rightarrow 0$. Therefore, the residue picks up the term $l = 0$ in the expansion of the recursion kernel, and evaluates the function between brackets to $w = 0$. The result is thus of the form announced in Proposition~\ref{2g2mr}, with only non-zero coefficients:
\beq
\label{Coini}\mathcal{C}^{[0,3]}\bigl[{}^{0}_{\varepsilon}\,{}^{0}_{\varepsilon}\,{}^{0}_{\varepsilon}\bigr] = -\frac{2}{y_{\varepsilon,1}},\qquad \varepsilon \in \{0,\tfrac{1}{2}\},
\eeq
computed using also the expansion \eqref{316} of $\Delta_{\varepsilon}\mathcal{G}(w)$. \\
For $(g,k) = (1,1)$, Theorem~\ref{them1o} yields
$$
\mathcal{G}^{[1,1]}(v_1) = \sum_{\varepsilon \in \{0,1/2\}} \Res_{w \rightarrow 0} K_{\varepsilon}(v_1, \tau + \varepsilon + w)\cdot \overline{\mathcal{G}}^{[0,2]}(\tau + \varepsilon + w,\tau + \varepsilon - w).
$$
We have seen in Lemma~\ref{Lupb} that the last factor has a double pole when $w \rightarrow 0$, with no simple pole and constant term $-\upsilon_{b,1}$ defined in \eqref{upb1def}. Then, we have to expand the recursion kernel up to $O(w^2)$ in order to obtain the final answer for $\mathcal{G}^{[1,1]}$. In other words, we only need to include the terms $l = 0$ and $l = 1$, and use the expansion \eqref{316} of the denominator to perform the computation:
$$
K_{\varepsilon}(v_1,\tau + \varepsilon + w) = -\frac{\mathcal{B}_{\varepsilon,0}(v_1)}{y_{\varepsilon,1}}\,\frac{1}{w} + \Big(\frac{y_{\varepsilon,2}\mathcal{B}_{\varepsilon,0}(v_1)}{y_{\varepsilon,1}^2} - \frac{\mathcal{B}_{\varepsilon,1}(v_1)}{y_{\varepsilon,1}}\Big)\,\frac{w}{6} + o(w).
$$
We find eventually
$$
\mathcal{G}^{[1,1]}(v_1) = \sum_{\varepsilon \in \{0,1/2\}} \Big(\frac{y_{\varepsilon,2}}{24y_{\varepsilon,1}^2} + \frac{\upsilon_{b,1}}{y_{\varepsilon,1}}\Big)\mathcal{B}_{\varepsilon,0}(v_1)  - \frac{\mathcal{B}_{\varepsilon,1}(v_1)}{24y_{\varepsilon,1}}.
$$
The answer is of the form of Proposition~\ref{2g2mr}, with only non-zero coefficients:
\beq
\label{C1ini} \mathcal{C}^{[1,1]}\bigl[{}^{0}_{\varepsilon}\bigr] = \frac{y_{\varepsilon,2}}{24y_{\varepsilon,1}^2} + \frac{\upsilon_b}{y_{\varepsilon,1}},\qquad \mathcal{C}^{[1,1]}\bigl[{}^{1}_{\varepsilon}\bigr] = -\frac{1}{24y_{\varepsilon,1}}.
\eeq

Now, take $\chi \geq 2$, and assume the result is true for all $\overline{\mathcal{G}}^{[g',k']} = \mathcal{G}^{[g',k']}$, with $0 < 2g' - 2 + k' < \chi$. We would like to compute $\mathcal{G}^{[g,k]}$ for a topology such that $2g - 2 + k = \chi$. The residue formula of Theorem~\ref{them1o} involves $\mathcal{G}^{[g',k']}$ for $0 < 2g' - 2 + k' < \chi$, which we replace by the decomposition of Proposition~\ref{2g2mr}, as well as $\overline{\mathcal{G}}^{[0,2]}$.

The terms which do not contain $\overline{\mathcal{G}}^{[0,2]}$ give a contribution which is the sum over indices $(l_j,\varepsilon_j)_{j \in I}$ and indices $(m,\sigma),(m',\sigma')$ of terms containing the factor:
\bea
& & \Big[\prod_{j \in I} \mathcal{B}_{\varepsilon_j,l_j}(v_j)\Big]\,\Res_{w \rightarrow 0} K_{\varepsilon}(v_0,\tau + \varepsilon + w)\,\mathcal{B}_{\sigma,m}(\tau + \varepsilon + w)\,\mathcal{B}_{\sigma',m'}(\tau + \varepsilon - w) \nonumber \\
\label{233} & = & \sum_{l \geq 0} \Big[\prod_{j \in I} \mathcal{B}_{\varepsilon_j,l_j}(v_j)\cdot\mathcal{B}_{\varepsilon,l}(v_0)\Big]\,\mathcal{K}\bigl[{}^{l}_{\varepsilon}\,{}^{m}_{\sigma}\,{}^{m'}_{\sigma'}\bigr]. \nonumber
\eea
We computed the residue thanks to the expansion of $K_{\varepsilon}$ given in \eqref{expKq}, and we introduced the coefficient \eqref{Khoubi}:
\beq
\mathcal{K}\bigl[{}^{l}_{\varepsilon}\,{}^{m}_{\,\sigma}\,{}^{m'}_{\,\sigma'}\bigr] = \Res_{w \rightarrow 0} \frac{-\dd w\,w^{2l + 1}}{(2l + 1)!\Delta_{\varepsilon}\mathcal{G}(w)}\,\mathcal{B}_{\sigma,m}(\tau + \varepsilon + w)\,\mathcal{B}_{\sigma',m'}(\tau + \varepsilon - w) \nonumber. 
\eeq
These terms thus form a linear combination of products of elementary blocks in the variables $v_0,(v_j)_{j \in I}$, which contribute to $\mathcal{C}^{[g,k]}\big[{}^{l}_{\varepsilon}\,{}^{l_{I}}_{\varepsilon_{I}}\bigr]$ by the two first lines in \eqref{chack}.

Since $2g - 2 + k \geq 2$, the contribution to $\mathcal{G}^{[g,k]}(v_0,v_I)$ containing $\overline{\mathcal{G}}^{[0,2]}$ is precisely the sum over $\varepsilon \in \{0,1/2\}$ and $i \in I = \{2,\ldots,k\}$ of
\beq
\label{resiJ}\Res_{w \rightarrow 0} K_{\varepsilon}(v_0,\tau + \varepsilon + w)\Big[\overline{\mathcal{G}}^{[0,2]}(\tau + \varepsilon + w,v_i)\,\mathcal{G}^{[g,k - 1]}(\tau + \varepsilon - w,v_{I\setminus \{i\}}) + (w \rightarrow -w)\Big].
\eeq
The quantity in brackets can be decomposed using odd and even parts:
\bea
& & 2\Big[\overline{\mathcal{G}}^{[0,2]}_{{\rm even}}(\tau + \varepsilon + w,v_i)\,\mathcal{G}^{[g,k - 1]}_{{\rm even}}(\tau + \varepsilon + w,v_{I\setminus\{i\}})\nonumber \\
& &  - \overline{\mathcal{G}}^{[0,2]}_{{\rm odd}}(\tau + \varepsilon + w,v_i)\,\mathcal{G}^{[g,k - 1]}_{{\rm odd}}(\tau + \varepsilon + w,v_{I\setminus\{i\}})\Big]. \nonumber
\eea
When we insert in this expression the decomposition of Proposition~\ref{2g2mr} for $\mathcal{G}^{[g,k - 1]}$, we have to deal with the sum over indices $(l_{j},\varepsilon_{j})_{j \neq i}$ and $(m,\sigma)$ of terms of the form
\bea
& & 2\mathcal{C}^{[g,k - 1]}\bigl[{}^{m}_{\,\sigma}\,{}^{l_{I\setminus\{i\}}}_{\varepsilon_{I\setminus\{i\}}}\bigr]\prod_{j \in I\setminus\{i\}} \mathcal{B}_{\varepsilon_j,l_j}(v_j)\cdot
\Big[\overline{\mathcal{G}}^{[0,2]}_{{\rm even}}(\tau + \varepsilon + w,v_i)\,\mathcal{B}_{\sigma,m}^{{\rm even}}(\tau + \varepsilon + w) \nonumber \\ \label{235}& & \phantom{ 2\,\mathcal{C}_{(g,k-1)}\bigl[{}^{m}_{\sigma}\,{}^{l_{I\setminus\{i\}}}_{\varepsilon_{I\setminus\{i\}}}\bigr]\prod_{j \in I\setminus\{i\}} \mathcal{B}_{\varepsilon_j,l_j}(v_j)\cdot}\, - \overline{\mathcal{G}}^{[0,2]}_{{\rm odd}}(\tau + \varepsilon + w,v_i)\,\mathcal{B}_{\sigma,m}^{{\rm odd}}(\tau + \varepsilon + w)\Big]. 
\eea
According to Lemma~\ref{regood}, $\mathcal{B}_{\sigma,m}^{{\rm odd}}(\tau + \varepsilon + w) \in O(w)$ when $w \rightarrow 0$. Since $\overline{\mathcal{G}}^{[0,2]}(\tau + \varepsilon + w,v_i)$ is regular when $w \rightarrow 0$, this implies that the product of the odd parts does not contribute to the residue \eqref{resiJ}. Besides, the expansion at $w \rightarrow 0$ of the product of even parts in \eqref{235} can be expressed in terms of the elementary blocks. We thus obtain a contribution
\beq
\label{238}\sum_{l,l_i \geq 0} 2\,\mathcal{C}^{[g,k - 1]}\bigl[{}^{m}_{\,\varepsilon}\,{}^{l_{I\setminus\{i\}}}_{\varepsilon_{I\setminus\{i\}}}\bigr] \Bigg[\prod_{j \in I\setminus\{i\}} \mathcal{B}_{\varepsilon_j,l_j}(v_j) \cdot \mathcal{B}_{\varepsilon,l}(v_0)\mathcal{B}_{\varepsilon,l_i}(v_i) \Bigg]\,\tilde{\mathcal{K}}\bigl[{}^{l}_{\varepsilon}\,{}^{l'}_{\varepsilon'}\,{}^{m}_{\,\sigma}\bigr]
\eeq
and we have defined
\beq
\tilde{\mathcal{K}}\bigl[{}^{l}_{\varepsilon}\,{}^{l'}_{\varepsilon'}\,{}^{m}_{\,\sigma}\bigr] = \delta_{\varepsilon,\varepsilon'}\,\Res_{w \rightarrow 0} \frac{-\dd w\,w^{2l + 1}}{(2l + 1)!\Delta_{\varepsilon}\mathcal{G}(w)}\cdot \frac{w^{2l'}}{(2l')!} \cdot \mathcal{B}_{\sigma,m}(w + \tau + \varepsilon), \nonumber
\eeq
which is the coefficient announced in \eqref{Ktildeou}. Since the prefactor of $\mathcal{B}$ in the residue is an odd $1$-form in $w$, the residue picks up the even part of $\mathcal{B}$, so it did not change the result to replace $\mathcal{B}^{{\rm even}}$ by $\mathcal{B}$. Let us examine the cases for which $\tilde{K}$ does not vanish. If $\varepsilon = \sigma$, we take into account the behavior at $w \rightarrow 0$ of $\mathcal{B}_{\varepsilon,m}(\tau + \varepsilon + w)$ given by Lemma~\ref{regood}, and find
\beq
\label{fikii}\tilde{\mathcal{K}}\bigl[{}^{l}_{\varepsilon}\,{}^{l'}_{\varepsilon'}\,{}^{m}_{\,\sigma}\bigr] = \delta_{\varepsilon,\varepsilon'}\,\frac{1}{(2l + 1)!(2l')!}\,\Res_{w \rightarrow 0} \frac{-\dd w\, w^{2(l + l')}}{\Delta_{\varepsilon}\mathcal{G}(w)}\left(\frac{ (2m + 1)! w^{-2m}}{w} + w\, \upsilon_{b,2m+1} \right),
\eeq
where $\upsilon_{b,2m+1}$ are the constants introduced in \eqref{expord1Upsilon}.
Since $\Delta_{\varepsilon}\mathcal{G}(w)$ has a double zero at $w = 0$, \eqref{fikii} vanishes if $l + l' \geq m + 2$. If $\varepsilon \neq \sigma$, Lemma~\ref{regood} tells us that $\mathcal{B}_{\sigma,m}(\tau + \varepsilon + w)$ is regular at $w = 0$, hence $\tilde{K}$ vanish unless $(l,l') = (0,0)$, and we have
$$
\tilde{\mathcal{K}}\bigl[{}^{0}_{\varepsilon}\,{}^{0}_{\varepsilon}\,{}^{m}_{\,\sigma}\bigr] = - \frac{\mathcal{B}_{\sigma,m}(\tau + \varepsilon)
}{y_{\varepsilon,1}} = \frac{\Upsilon_{b}^{(2m + 1)}(\tfrac{1}{2})}{y_{\varepsilon,1}}.
$$
The last equality follows from \eqref{blbelb} and the properties of $\Upsilon_{b}$ described in Appendix~\ref{AppA}. We can study in a similar way the cases for which $K$ does not vanish.

Collecting all the terms from \eqref{233} and \eqref{238}, we arrive to Formula ~\eqref{chack} and conclude the recursive proof. \hfill $\Box$

\subsection{Diagrammatic representation}
\label{diagTR}
Unfolding the recursion yields a formula for $\mathcal{G}^{[g,k]}$ with $2g - 2 + k > 0$ as a sum over the set $\mathcal{S}^{[g,k]}$ of graphs $\mathscr{G}$ with first Betti number $g$, trivalent vertices equipped with a cyclic order of their incident edges, and legs (univalent vertices) labeled $\{1,\ldots,k\}$. With this definition, if there is an edge from a trivalent vertex to itself (a loop), the cyclic order is just the transposition of the two distinct incident edges. The weight given to a graph actually depends on the choice of an initial leg $i_0$, but the sum over graphs is independent of those choices \cite{EORev}.

Before stating the formula, we need a preliminary construction. If $\mathscr{G} \in \mathcal{S}^{[g,k]}$, we denote $V(\mathscr{G})$ the set of trivalent vertices and a $E(\mathscr{G})$ the set of edges. We also denote $V_o(\mathscr{G})$ the set of trivalent vertices with a loop. If $\mathsf{v}$ is a vertex, we denote $\mathsf{e}[\mathsf{v}]$ its set of incident edges. A simple counting gives:
\beq
|E(\mathscr{G})| = 3g - 3 + 2k,\qquad |V(\mathscr{G})| = 2g - 2 + k.
\eeq

\subsubsection{Exploration of a cyclically ordered graph}
\label{explodocus}

The choice of an initial leg and the data of the cyclic order determines a way to explore $\mathscr{G}$, i.e. two bijections
$$\varphi\,:\,\{1,\ldots,|E(\mathscr{G})|\} \rightarrow E(\mathscr{G}),\qquad \eta\,:\,\{1,\ldots,|V(\mathscr{G})| + k\} \rightarrow V(\mathscr{G}) \cup \{1,\ldots,k\}$$
which record in which order the edges, and the vertices or legs, are visited. Let us describe how $\varphi$ and $\eta$ are constructed.

We declare that $\eta(1)$ is the initial leg, and $\phi(1)$ is the edge incident to the initial leg $i_0$. Since $2g - 2 + k > 0$, $\mathcal{G}$ must have at least a trivalent vertex, so $\phi(1)$ is also incident to a trivalent vertex that we declare to be $\eta(2)$. We define a seed with initial value $(\phi(1),\eta(2))$. Then, we apply the following algorithm. Let $(\mathsf{e} = \phi(j_1),\mathsf{v} = \eta(j_2))$ be the seed. If $\mathsf{v}$ is not a leg, let $\mathsf{e}^+$ (resp. $\mathsf{e}^-$) be the edge following (resp. preceding) $\mathsf{e}$ in the cyclic order around $\mathsf{v}$.

\smallskip

\noindent $\bullet$ \textbf{First cases:} either $\mathsf{v}$ is a leg or, otherwise, $\mathsf{e}^+$ and $\mathsf{e}^-$ have already been explored (i.e. are equal to $\phi(i^+)$ and $\phi(i^-)$ for some $i^{\pm} < j_1$). If actually all vertices have already been explored (i.e. $j_2 = |V(\mathscr{G})| + k$), the algorithm terminates; otherwise, we consider the maximal $j_2' < j_2$ such that $\eta(j_2')$ is not a leg, and the maximal $j_1' \leq j_1$ such that $\phi(j_1')$ is incident to $\eta(j_2')$, and reset the seed to $(\phi(j_1'),\eta(j_2'))$.\smallskip

\noindent $\bullet$ \textbf{Second case:} $\mathsf{e}^+$ has not been explored. We define $\phi(j_1 + 1) = \mathsf{e}^+ = \{\mathsf{v},\mathsf{v}^+\}$ and $\eta(j_2 + 1) = \mathsf{v}^+$, and reset the seed to $(\mathsf{e}^+,\mathsf{v}^+)$.
\smallskip

\noindent $\bullet$ \textbf{Third case:} $\mathsf{e}^+$ has already been explored, but not $\mathsf{e}^-$. We define $\phi(j_1 + 1) = \mathsf{e}^- = \{\mathsf{v},\mathsf{v}^-\}$ and $\eta(j' + 1) = \mathsf{v}^-$, and reset the seed to $(\mathsf{e}^-,\mathsf{v}^-)$.

\smallskip

Now, at any trivalent vertex $\mathsf{v}$ which does not have a loop, we can label the incident edges $\mathsf{e}^0_{\mathsf{v}},\mathsf{e}^1_{\mathsf{v}},\mathsf{e}^2_{\mathsf{v}}$, starting from the edge such that $\phi^{-1}(\mathsf{e}_{\mathsf{v}}^0)$ is minimal among $\mathsf{e}[\mathsf{v}]$, and following the cyclic order. If a trivalent vertex $\mathsf{v}$ has a loop, we can just label $\mathsf{e}^0_{\mathsf{v}}$ the incident edge which is not a loop, and $\mathsf{e}_{\mathsf{v}}^1$ the other one; this definition also agrees with the order of exploration at $\mathsf{v}$.
\begin{definition}\label{trivVert}
A trivalent vertex $\mathsf{v}$ is \emph{bi-terminal} if $\mathsf{e}_{\mathsf{v}}^{1}$ and $\mathsf{e}_{\mathsf{v}}^{2}$ are incident to legs. It is \emph{terminal} if $\mathsf{e}_{\mathsf{v}}^{1}$ xor $\mathsf{e}_{\mathsf{v}}^{2}$ is incident to a leg. We denote $V_{t}(\mathscr{G})$ (resp. $V_{tt}(\mathscr{G})$) the set of (bi-)terminal vertices, and $V'(\mathscr{G})$ the set of trivalent vertices which are neither terminal, neither bi-terminal, nor have a loop.
\end{definition}
We stress that, for a given graph, all these notions depend on the choice of an initial leg.

\subsubsection{The unfolded formula}

Let ${\rm Col}(\mathscr{G};(\bs{l},\bs{\varepsilon}))$ be the set of colorings of edges by labels in $\mathbb{N}\times\{0,\tfrac{1}{2}\}$ such that
\begin{itemize}
\item[$\bullet$] the coloring of edges incident to legs agrees with the fixed coloring $(\bs{l},\bs{\varepsilon})$ of the legs;
\item[$\bullet$] the color of a loop is identical to the color of the other edge incident to the vertex where the loop is attached.
\end{itemize}
If $(\bs{m},\bs{\sigma})$ is such a coloring, and $\mathsf{v}$ is a trivalent vertex which does not have a loop, we define $\bs{m}[\mathsf{v}]$ to be the sequence $(m(\mathsf{e}_{\mathsf{v}}^0),m(\mathsf{e}_{\mathsf{v}}^{1}),m(\mathsf{e}_{\mathsf{v}}^{2}))$, and similarly for the sequence $\bs{\sigma}[\mathsf{v}]$. One proves by induction:

\begin{proposition}
\label{cosums}
For $2g - 2 + k > 0$, we have
\bea
 && \mathcal{C}^{[g,k]}\bigl[{}^{l_1}_{\varepsilon_1}\,\cdots\,{}^{l_k}_{\varepsilon_k}\bigr] \nonumber \\
& = & \sum_{\substack{\mathscr{G} \in \mathcal{S}^{[g,k]} \\ (\bs{m},\bs{\sigma}) \\ \in {\rm Col}(\mathscr{G};(\bs{l},\bs{\varepsilon}))}} \prod_{\mathsf{v} \in V'(\mathscr{G})} \mathcal{K}\bigl[{}^{\bs{m}[\mathsf{v}]}_{\,\bs{\sigma}[\mathsf{v}]}\bigr] \prod_{\mathsf{v} \in V_{t}(\mathscr{G})} \tilde{\mathcal{K}}\bigl[{}^{\bs{m}[\mathsf{v}]}_{\,\bs{\sigma}[\mathsf{v}]}\bigr]\prod_{\mathsf{v} \in V_{tt}(\mathscr{G})} \mathcal{C}^{[0,3]}\bigl[{}^{\bs{m}[\mathsf{v}]}_{\,\bs{\sigma}[\mathsf{v}]}\bigr] \prod_{\mathsf{v} \in V_{o}(\mathscr{G})} \mathcal{C}^{[1,1]}\bigr[{}^{m(\mathsf{e}_{\mathsf{v}}^0)}_{\,\sigma(\mathsf{e}_{\mathsf{v}}^{0})}\bigr]. \nonumber
\eea
 \hfill $\Box$
\end{proposition}

\subsection{Usual maps with renormalized face weights}

$\mathcal{W}^{[g,k]}|_{\mathsf{n} = 0}$ is the generating series of usual triangulations, with weight $t_3$ per triangle. This is different from $\mathsf{W}^{[g,k]}$, which is by definition the generating series of usual maps with renormalized face weights \eqref{eq:fixp}, i.e. the generating series of configurations with only non-separating loops, and still depends on $\mathsf{n}$. 

Recall that $\mathcal{W}^{[g,k]}$ depends on $\mathsf{n}$ in two ways. Firstly, $\mathsf{n}$ appears as a proportionality coefficient in $\mathbf{A}(x,y)$ -- see \eqref{rin} -- in the linear functional relation of Theorem~\eqref{2g2m}. Secondly, the linear equation for $\mathcal{W}(x)$ gives two equations determining $\gamma_{\pm}$ as functions of $\mathsf{n}$, and this data gives the interval $x \in (\gamma_{-},\gamma_+)$ on which the linear equation for $\mathcal{W}^{[g,k]}$ holds. For $(g,k) \neq (0,1)$, we can disentangle the two dependences in $\mathsf{n}$: let us call $\mathsf{n}_1$ the variable appearing linearly in the linear equation, and $\mathsf{n}_2$ the variable on which $\gamma_{\pm}$ depends. We denote momentarily $\mathcal{W}^{[g,k]}_{\mathsf{n}_1,\mathsf{n}_2}$ the corresponding generating series. Note that the parametrization $x(v)$ only depends on $\mathsf{n}_2$.

The previous remarks show that the generating series of maps in the $O(\mathsf{n})$ model is
$$
\mathcal{W}^{[g,k]} = \mathcal{W}^{[g,k]}_{\mathsf{n}_1 = \mathsf{n},\mathsf{n}_2 = \mathsf{n}},
$$
while the generating series of usual maps with renormalized face weights is
$$
\mathsf{W}^{[g,k]} = \mathcal{W}^{[g,k]}_{\mathsf{n}_1 = 0,\mathsf{n}_2 = \mathsf{n}}.
$$
Note however that $\mathsf{W}(x) = \mathcal{W}(x)$, since disks do not contain separating loops.

Let us use non-curly letters to denote the analogue, in the context of usual maps with renormalized face weights, of all quantities defined in the context of maps of the $O(\mathsf{n})$ model. We have
\bea
\overline{\mathsf{G}}^{[0,2]}(v_1,v_2) & = & \frac{1}{4}\bigg[\Upsilon'_{1/2}(v_1 + v_2) - \Upsilon_{1/2}'(v_1 - v_2) - \Upsilon'_{1/2}(-v_1 + v_2) + \Upsilon'_{1/2}(-v_1 - v_2)\bigg] \nonumber \\
&& + \frac{x'(v_1)x'(v_2)}{2(x(v_1) - x(v_2))^2}, \nonumber
\eea
where $\Upsilon_{1/2}$ is a function of the elliptic modulus $\tau$, thus a function of $\mathsf{n}_2$. The modified building block is defined as:
\bea
\mathsf{B}_{\varepsilon,l}(v) = \partial_{v}^{2l}\mathsf{B}_{\varepsilon,0}(v),\qquad \mathsf{B}_{\varepsilon,0}(v) = \overline{\mathsf{G}}^{[0,2]}(v,\tau + \varepsilon).
\eea
As the generating series of disks are $\mathcal{W}(x) = \mathsf{W}(x)$ and the parametrization $x(v)$ only depends on $\mathsf{n}_2$, we have
$$
\Delta_{\varepsilon}\mathsf{G}(v) = \Delta_{\varepsilon}\mathcal{G}(v).
$$
The modified recursion coefficients (compare with \eqref{Khoubi}-\eqref{Ktildeou}) are
\bea
\mathsf{K}\bigl[{}^{l}_{\varepsilon}\,{}^{m}_{\sigma}\,{}^{m'}_{\sigma'}\bigr] & = & \Res_{w \rightarrow 0} \frac{-w^{2l + 1}\dd w}{(2l + 1)!\Delta_{\varepsilon}\mathcal{G}(w)}\,\mathsf{B}_{\sigma,m}(w + \tau + \varepsilon) \mathsf{B}_{\sigma',m'}(-w + \tau + \varepsilon), \nonumber \\
\tilde{\mathsf{K}}\bigl[{}^{l}_{\varepsilon}\,{}^{l'}_{\varepsilon'}\,{}^{m}_{\sigma}\bigr] & = & \frac{-\delta_{\varepsilon,\varepsilon'}}{(2l - 1)!\,(2l')!} \Res_{w \rightarrow 0} \frac{\dd w}{w}\,\frac{w^{2(l + l')}}{\Delta_{\varepsilon}\mathcal{G}(w)}\,\mathsf{B}_{m,\sigma}(\tau + \varepsilon + w). \nonumber
\eea
Following the proof of Proposition~\ref{cocor1}, the non-zero modified initial data read:
$$
\boxed{\mathsf{C}^{[0,3]}\big[{}^{l_1}_{\varepsilon_1}\,{}^{l_2}_{\varepsilon_2}\,{}^{l_3}_{\varepsilon_3}\bigr] = -\frac{2\,\delta_{l_1,l_2,l_3,0}\,\delta_{\varepsilon_1,\varepsilon_2,\varepsilon_3}}{y_{\varepsilon_1,1}},
\qquad \mathsf{C}^{[1,1]}\bigl[{}^{l}_{\varepsilon}\bigr] = \delta_{l,0}\Big(\frac{y_{\varepsilon,2}}{24y_{\varepsilon,1}^2} + \frac{\upsilon_{1/2,1}}{y_{\varepsilon,1}}\Big) -\frac{\delta_{l,1}}{24 y_{\varepsilon,1}}.}$$
Compared to the initial conditions for $\mathcal{C}$'s, the only difference is the replacement of $\upsilon_{b,1}$ by $\upsilon_{1/2,1}$ (see \eqref{ububub} for their definition) in $\mathsf{C}^{[1,1]}\bigl[{}^{0}_{\varepsilon}\bigr]$. Then, the analogue of Propositions \ref{2g2mr}-\ref{cosums} is:
\begin{proposition}
\label{cosums2} For $2g - 2 + k > 0$, we have a decomposition into a finite sum:
$$
\mathsf{G}^{[g,k]}(v_1,\ldots,v_k) = \sum_{\substack{l_1,\ldots,l_k \geq 0 \\ \varepsilon_1,\ldots,\varepsilon_k \in \{0,\frac{1}{2}\}}} \mathsf{C}^{[g,k]}\bigl[{}^{l_1}_{\varepsilon_1}\,\cdots\,{}^{l_k}_{\varepsilon_k}\bigr]\prod_{i = 1}^k \mathsf{B}_{\varepsilon_i,l_i}(v_i)\,.
$$
The coefficients are given by the unfolded formula:
\bea
&& \mathsf{C}^{[g,k]}\bigl[{}^{l_1}_{\varepsilon_1}\,\cdots\,{}^{l_k}_{\varepsilon_k}\bigr] \nonumber \\
& = & \sum_{\substack{\mathscr{G} \in \mathcal{S}^{[g,k]} \\ \substack{(\bs{m},\bs{\sigma}) \\ \in {\rm Col}(\mathscr{G};(\bs{l},\bs{\varepsilon}))}}} \prod_{\mathsf{v} \in V'(\mathscr{G})} \mathsf{K}\bigl[{}^{\bs{m}[\mathsf{v}]}_{\,\bs{\sigma}[\mathsf{v}]}\bigr] \prod_{\mathsf{v} \in V_{t}(\mathscr{G})} \tilde{\mathsf{K}}\bigl[{}^{\bs{m}[\mathsf{v}]}_{\,\bs{\sigma}[\mathsf{v}]}\bigr]\prod_{\mathsf{v} \in V_{tt}(\mathscr{G})} \mathsf{C}^{[0,3]}\bigl[{}^{\bs{m}[\mathsf{v}]}_{\,\bs{\sigma}[\mathsf{v}]}\bigr] \prod_{\mathsf{v} \in V_{o}(\mathscr{G})} \mathsf{C}^{[1,1]}\bigr[{}^{m(\mathsf{e}_{\mathsf{v}}^0)}_{\,\sigma(\mathsf{e}_{\mathsf{v}}^{0})}\bigr]\,. \nonumber
\eea
\hfill $\Box$
\end{proposition}
\chapter{Critical behavior for large maps}\label{Chapter6}


\section{In the bending energy model (disregarding nesting)}
\label{critit}
\label{S5}

The definition of criticality, the different universality classes of the $O(\mathsf{n})$ loop model and the phase diagram for the bending energy model can be reviewed in Section \ref{largeRandom}. We start by recalling the most important definitions.
For fixed values $(\mathsf{n},\alpha,\mathsf{g},\mathsf{h})$, we introduced
$$
u_{c} = \sup\{u \geq 0\,:\, \mathcal{F}_{\ell}^{\bullet} < \infty\}
$$
in terms of the generating series of pointed disks defined in \eqref{Cyl2fixed}. If $u_{c} = 1$ (resp. $u_{c} < 1$, $u_{c} > 1$), we say that the model is at a critical (resp. subcritical, supercritical) point. At a critical point, the generating series $\mathsf{W}(x) = \mathcal{W}(x)$ has a singularity when $u \rightarrow 1^{-}$, and the nature (universality class) of this singularity is characterized by some critical exponents. The phase diagram of the model with bending energy was rigorously determined in \cite{BBG12b,BBD}, and is plotted qualitatively in Figure~\ref{Qualiphas}. We now review the precise results obtained in \cite{BBG12b,BBD}.


In the model with bending energy, we find the same three universality classes characteristic of the general $O(\mathsf{n})$ model: \emph{generic}, non-generic \emph{dilute} and non-generic \emph{dense}. For $\mathsf{n} > 0$, we find a dense critical line, which ends with a dilute critical point, and continues as a generic critical line. For $\mathsf{n} = 0$, only the generic critical line remains. The generic universality class, called pure gravity, is already present in maps without loops. On the contrary, the non-generic universality class is specific to the loop model, and it corresponds to a regime where macroscopic loops continue to exist in maps of volume $V \rightarrow \infty$ \cite{KOn,BEThese}. In order to explore the nesting statistics of the $O(\mathsf{n})$ loop model, we will describe our various generating series on the non-generic critical line.

A non-generic critical point occurs when $\gamma_+$ approaches the fixed point of $\varsigma$:
$$
\gamma_+^* = \varsigma(\gamma_+^*) = \frac{1}{h(\alpha + 1)}.
$$
In this limit, the two cuts $\gamma$ and $\varsigma(\gamma)$ merge at $\gamma_+^*$, and one can justify on the basis of combinatorial arguments \cite[Section 6]{BBG12b} that $\gamma_- \rightarrow \gamma_-^*$ with
$$
|\gamma_-^*| < |\gamma_+^*|\qquad {\rm and}\qquad \varsigma(\gamma_-^*) \neq \gamma_-^*.
$$
In terms of the parametrization $x(v)$, it amounts to letting $T \rightarrow 0$, and this is conveniently measured in terms of the parameter
$$
q = e^{-\frac{\pi}{T}} \rightarrow 0.
$$
After establishing the behavior of $x(v)$ and the special function $\Upsilon_{b}(v)$ in this regime (see Appendix~\ref{AppA} for a summary), one can prove:

\begin{theorem}\cite{BBG12b}
\label{thphase}
Assume $\alpha = 1$, and introduce the parameter
$$
\rho = 1 - 2\mathsf{h}\gamma_-^* = 1 - \frac{\gamma_-^*}{\gamma_+^*}.
$$
There is a non-generic critical line, parametrized by $\rho \in (\rho_{\min},\rho_{\max}]$:
\bea
\frac{\mathsf{g}}{\mathsf{h}} & = & \frac{4(\rho b\sqrt{2 + \mathsf{n}} - \sqrt{2 - \mathsf{n}})}{\rho^2(b^2 - 1)\sqrt{2 - \mathsf{n}} + 4\rho b\sqrt{2 + \mathsf{n}} - 2\sqrt{2 - \mathsf{n}}}, \nonumber \\
\mathsf{h}^2 & = & \frac{\rho^2 b}{24\sqrt{4 - \mathsf{n}^2}}\,\frac{\rho^2\,b(1 - b^2)\sqrt{2 + \mathsf{n}}  - 4\rho\sqrt{2 - \mathsf{n}} + 6b\sqrt{2 + \mathsf{n}}}{-\rho^2(1 - b^2)\sqrt{2 - \mathsf{n}} + 4\rho b\sqrt{2 + \mathsf{n}} - 2\sqrt{2 - \mathsf{n}}}.
\nonumber 
\eea
It realizes the dense phase of the model. The endpoint 
$$
\rho_{\max} = \frac{1}{b}\,\sqrt{\frac{2 - \mathsf{n}}{2 + \mathsf{n}}}
$$
corresponds to the fully packed model $\mathsf{g} = 0$, with the critical value $\mathsf{h} = \frac{1}{2\sqrt{2}\sqrt{2 + \mathsf{n}}}$. The endpoint
$$
\rho_{\min} = \frac{\sqrt{6 + \mathsf{n}} - \sqrt{2 - \mathsf{n}}}{(1 - b)\sqrt{2 + \mathsf{n}}}
$$
is a non-generic critical point realizing the dilute phase, and it has coordinates:
\bea
\frac{\mathsf{g}}{\mathsf{h}} & = & 1 + \sqrt{\frac{2 - \mathsf{n}}{6 + \mathsf{n}}}, \nonumber \\
\mathsf{h}^2 & = &  \frac{b(2 - b)}{3(1-  b^2)(2 + \mathsf{n})}\Big(1 - \frac{1}{4\sqrt{(2 - \mathsf{n})(6 + \mathsf{n})}}\Big). \nonumber
\eea
\end{theorem}
The fact that the non-generic critical line ends at $\rho_{\max} < 2$ is in agreement with $|\gamma_-^*| < |\gamma_+^*|$.

\begin{theorem}\cite{BBD}
\label{alphanotlarge} There exists $\alpha_c(\mathsf{n}) > 1$ such that, in the model with bending energy $\alpha < \alpha_c(\mathsf{n})$, the qualitatitive conclusions of the previous theorem still hold. For $\alpha = \alpha_c(\mathsf{n})$, only a non-generic critical point in the dilute phase exist, and for $\alpha > \alpha_c(\mathsf{n})$, non-generic critical points do not exist.
\end{theorem}

\begin{theorem}\cite{BBD}
\label{th38} Assume $(g,h)$ are chosen such that the model has a non-generic critical point for vertex weight $u = 1$. When $u < 1$ tends to $1$, we have
$$
q \sim \Big(\frac{1 - u}{q_*}\Big)^{c},
$$
with the universal exponent
$$
c = \left\{\begin{array}{lll} \frac{1}{1 - b} & & {\rm dense}, \\ 1 & & {\rm dilute}. \end{array}\right.
$$
The non-universal constant reads, for $\alpha = 1$:
$$
q_* = \left\{\begin{array}{lll} \frac{6(\mathsf{n} + 2)}{b}\,\frac{\rho^2(1 - b)^2\sqrt{2 + \mathsf{n}} + 2\rho(1 - b)\sqrt{2 - \mathsf{n}} - 2\sqrt{2 + \mathsf{n}}}{\rho^2b(1 - b^2)\sqrt{2 + \mathsf{n}} - 4\rho(1 - b^2)\sqrt{2 - \mathsf{n}} + 6b\sqrt{2 + \mathsf{n}}} & & {\rm dense}, \\ \frac{24}{b(1 - b)(2 - b)} & & {\rm dilute}. \end{array}\right. $$
For $\alpha \neq 1$, its expression is much more involved, see \cite[Appendix E]{BBD}.
\end{theorem}


\subsubsection{Small and large boundaries}

The generating series of connected maps of genus $g$ in the $O(\mathsf{n})$ model with fixed volume $V$ and fixed boundary lengths $\ell_1,\ldots,\ell_k$ reads
$$
[u^{V}] \mathcal{F}_{\ell_1,\ldots,\ell_k}^{[g,k]} = \oint \frac{\dd u}{2{\rm i}\pi\,u^{V + 1}} \oint \prod_{i = 1}^k \frac{\dd x_i\,x_i^{\ell_i}}{2{\rm i}\pi} \mathcal{W}^{[g,k]}(x_1,\ldots,x_k).
$$
The contour for integration of $x_i$ is originally around $\infty$ with negative orientation, but we can move it to surround $\gamma$. At a critical point, the asymptotics when $V \rightarrow \infty$ are dominated by the behavior of the generating series at $u = 1$. If we want to keep $\ell_i$ finite, we can leave the contour integral over $x_i$ in a neighborhood of $\infty$, and by setting $x_i = x(\tfrac{1}{2} + \tau w_i)$ we trade it for a contour surrounding $w_i = w_{\infty}^*$. If we want to let $\ell_i \rightarrow \infty$ at a rate controlled by $V \rightarrow \infty$, the asymptotics will be dominated by the behavior of the generating series for $x_i$ near the singularity $\gamma_+ \rightarrow \gamma_+^*$, i.e. for $x_i = x(\tau w_i)$ with $w_i$ of order $1$. The same principle holds for any of the unrefined generating series $\mathsf{W}$ and $\mathscr{W}_{\Gamma}$.

If $\mathbf{H}_s(x_1\ldots,x_k)$ is a refined generating series of maps with $k$ boundaries (with $s$ a Boltzmann weight for certain separating loops), we can compute the number of such maps having fixed volume $V$, fixed number $P$ of such separating loops, and fixed boundary perimeters, by$$
\Big[s^Pu^{V} \prod_{i} x_i^{-(\ell_i + 1)}\Big] \mathbf{H}_s(x_1,\ldots,x_k) = \oint \frac{\dd s}{2{\rm i}\pi}\,\oint \frac{\dd u}{2{\rm i}\pi} \oint \Big[\prod_{i = 1}^k \frac{x_i^{\ell_i}\dd x_i}{2{\rm i}\pi}\Big] \frac{\mathbf{H}_s(x_1,\ldots,x_k)}{s^{P + 1}u^{V + 1}}.$$
In the regime $P,V \rightarrow \infty$, the contour integral over $s$ will be determined by the behavior of the generating series near the dominant singularity in the variable $s$, and $u \rightarrow 1$.

To summarize, we need to study the behavior of generating series approaching criticality, i.e. $q = e^{-\frac{\pi}{T}}$ with $\tau = {\rm i}T \rightarrow 0$, while $x = x(v)$ with $v = \varepsilon + \tau w$ and $w$ is in a fixed compact. With $\varepsilon = \frac{1}{2}$ we have access to the regime of finite (also called ``small'') boundaries, and with $\varepsilon = 0$ to the regime of large boundaries.

\subsubsection{Organization of the computations}

In the present Section~\ref{critit}, we will study maps without marked points. The modifications arising to include a number $k' > 0$ of marked points will be discussed in Section~\ref{crititmarked}. We will find, as can be expected, that marked points behave -- as far as critical exponents are concerned -- as small boundaries.

Our first goal is to determine the behavior of the generating series of maps $\mathcal{W}^{[g,k]}$ and of usual maps with renormalized face weights $\mathsf{W}^{[g,k]}$. To obtain it, we first determine the behavior of the building blocks of Propositions ~\ref{cosums}-\ref{cosums2} in the next paragraph, and then study the behavior of the sum over colorings and graphs to derive the behavior of $\mathcal{C}^{[g,k]}$ and $\mathsf{C}^{[g,k]}$ (Lemma~\ref{Cbehavior}). This step is rather technical, and the result for the critical exponent for $\mathcal{C}$'s and $\mathsf{C}$'s is not particularly simple. Yet, the final result for the critical behavior of the generating series of maps themselves turns out to be much simpler (Theorem~\ref{ouqusf}). We recall that the $\mathcal{C}$'s do not have a combinatorial interpretation in terms of maps, so this technical part should only be seen as a (necessary) intermediate step to arrive to the $\mathcal{W}$'s and $\mathsf{W}$'s.


\subsection{Critical behavior of the building blocks}
\label{Sbuildi}
We first examine the behavior at criticality, i.e. $q = e^{-\frac{\pi}{T}} \rightarrow 0$, of the various bricks appearing in Proposition~\ref{cosums}.  Let us define
$$
\varepsilon,\varepsilon' \in \{0,\tfrac{1}{2}\},\qquad \varepsilon \oplus \varepsilon' \coloneqq \left\{\begin{array}{rcl} 0 & & {\rm if}\,\,\varepsilon = \varepsilon', \\ \tfrac{1}{2} & & {\rm if}\,\,\varepsilon \neq \varepsilon', \end{array}\right.
$$
and for $\varepsilon,\sigma,\sigma'\in \{0,\tfrac{1}{2}\}$
\beq\label{fCrit}
f(\varepsilon,\sigma,\sigma'\vert B) \coloneqq  B\big[(\varepsilon \oplus \sigma) + (\varepsilon \oplus \sigma')\big] + \big(\mathfrak{d}\tfrac{b}{2} - 1\big)(1 - 2\varepsilon),
\eeq
with $\mathfrak{d} = 1$ in the dense phase, and $\mathfrak{d} = -1$ in the dilute phase. We give its table of values (dense on the left, dilute on the right) for $B=b$:
\begin{center}
\begin{tabular}{|c||c|c|c|}
\hline $\bs{\sigma + \sigma'}$ & $\bs{0}$ & $\bs{\tfrac{1}{2}}$ & $\bs{1}$ \\
\hline\hline
$\bs{\varepsilon = 0}$ & $\tfrac{b}{2} - 1$ & $b - 1$ & $\tfrac{3b}{2} - 1$ \\
\hline
$\bs{\varepsilon = \tfrac{1}{2}}$ & $b$ & $\tfrac{b}{2}$ & $0$ \\
\hline
\end{tabular}
\quad \begin{tabular}{|c||c|c|c|}
\hline $\bs{\sigma + \sigma'}$ & $\bs{0}$ & $\bs{\tfrac{1}{2}}$ & $\bs{1}$ \\
\hline\hline
$\bs{\varepsilon = 0}$ & $-\tfrac{b}{2} - 1$ & $-1$ & $\tfrac{b}{2} - 1$ \\
\hline
$\bs{\varepsilon = \tfrac{1}{2}}$ & $b$ & $\tfrac{b}{2}$ & $0$ \\
\hline
\end{tabular}
\end{center}

\begin{lemma}
\label{pieces}
In the critical regime $\tau = {\rm i}T$ with $T \rightarrow 0^+$, we have for the building blocks of the generating series of maps in the bending energy model
\bea
\mathcal{K}\bigl[{}^{l}_{\varepsilon}\,{}^{m}_{\sigma}\,{}^{m'}_{\sigma'}\bigr] & = & \Big(\frac{\pi}{T}\Big)^{2(m + m' - l) + 1}\,q^{f(\varepsilon,\sigma,\sigma'\vert b)}\Big\{K^*\bigl[{}^{l}_{\varepsilon}\,{}^{m}_{\sigma}\,{}^{m'}_{\sigma'}\bigr] + O(q^{b})\Big\}, \nonumber \\
\tilde{\mathcal{K}}\bigl[{}^{l}_{\varepsilon}\,{}^{l'}_{\varepsilon}\,{}^{m}_{\sigma}\bigr] & = &  \Big(\frac{\pi}{T}\Big)^{2(m - l - l') - 1}\,q^{f(\varepsilon,\varepsilon,\sigma\vert b)}\Big\{\tilde{K}^*\bigl[{}^{l}_{\varepsilon}\,{}^{m}_{\sigma}\,{}^{m'}_{\sigma'}\bigr] + O(q^{b})\Big\}, \nonumber \\
\mathcal{C}^{[0,3]}\bigl[{}^{0}_{\varepsilon}\,{}^{0}_{\varepsilon}\,{}^{0}_{\varepsilon}\bigr] & = &  \Big(\frac{\pi}{T}\Big)^{-3}\,q^{f(\varepsilon,\varepsilon,\varepsilon\vert b)}\Big\{\mathcal{C}^{[0,3]}_{*}\bigl[{}^{0}_{\varepsilon}\,{}^{0}_{\varepsilon}\,{}^{0}_{\varepsilon}\bigr] + O(q^{b})\Big\}, \nonumber \\
\mathcal{C}^{[1,1]}\bigl[{}^{l}_{\varepsilon}\bigr] & = &  \Big(\frac{\pi}{T}\Big)^{-(2l + 1)}\,q^{f(\varepsilon,\varepsilon,\varepsilon\vert b)}\Big\{\mathcal{C}^{[1,1]}_*\bigl[{}^{l}_{\varepsilon}\bigr] + O(q^{b})\Big\}, \nonumber \\
\mathcal{B}_{\varepsilon,l}(\tau\phi + \varepsilon') & = &  \Big(\frac{\pi}{T}\Big)^{2l + 2}\,q^{b(\varepsilon \oplus \varepsilon')}\Big\{\mathcal{B}_{\varepsilon\oplus \varepsilon',l}^{*,(2l + 1)}(\pi \phi) + O(q^{b})\Big\}. \nonumber 
\eea
And, for the building blocks of the generating series of usual maps with renormalized face weights
\bea
\mathsf{K}\bigl[{}^{l}_{\varepsilon}\,{}^{m}_{\sigma}\,{}^{m'}_{\sigma^{\prime}}\bigr] & = &  \Big(\frac{\pi}{T}\Big)^{2(m + m' - l) + 1}\,q^{f\left(\varepsilon,\sigma,\sigma'\vert \frac{1}{2}\right)}\Big\{\mathsf{K}^*\bigl[{}^{l}_{\varepsilon}\,{}^{m}_{\sigma}\,{}^{m'}_{\sigma'}\bigr] + O(q^{b})\Big\}, \nonumber \\
\tilde{\mathsf{K}}\bigl[{}^{l}_{\varepsilon}\,{}^{l'}_{\varepsilon}\,{}^{m}_{\sigma}\bigr] & = &  \Big(\frac{\pi}{T}\Big)^{2(m - l - l') - 1}\,q^{f\left(\varepsilon,\varepsilon,\sigma\vert \frac{1}{2}\right)}\Big\{\tilde{\mathsf{K}}^*\bigl[{}^{l}_{\varepsilon}\,{}^{m}_{\sigma}\,{}^{m'}_{\sigma'}\bigr] + O(q^{b})\Big\}, \nonumber \\
\mathsf{C}^{[0,3]}\bigl[{}^{0}_{\varepsilon}\,{}^{0}_{\varepsilon}\,{}^{0}_{\varepsilon}\bigr] & = &  \Big(\frac{\pi}{T}\Big)^{-3}\,q^{f\left(\varepsilon,\varepsilon,\varepsilon \vert \frac{1}{2}\right)}\Big\{\mathsf{C}^{[0,3]}_*\bigl[{}^{0}_{\varepsilon}\,{}^{0}_{\varepsilon}\,{}^{0}_{\varepsilon}\bigr] + O(q^{b})\Big\}, \nonumber \\
\mathsf{C}^{[1,1]}\bigl[{}^{l}_{\varepsilon}\bigr] & = &  \Big(\frac{\pi}{T}\Big)^{-(2l + 1)}\,q^{f\left(\varepsilon,\varepsilon,\varepsilon\vert \frac{1}{2}\right)}\Big\{\mathsf{C}^{[1,1]}_*\bigl[{}^{l}_{\varepsilon}\bigr] + O(q^{b})\Big\}, \nonumber \\
\mathsf{B}_{\varepsilon,l}(\tau\phi + \varepsilon') & = &  \Big(\frac{\pi}{T}\Big)^{2l + 2}\,q^{\frac{1}{2}(\varepsilon \oplus \varepsilon')}\Big\{\mathsf{B}_{\varepsilon\oplus \varepsilon',l}^{*,(2l + 1)}(\pi \phi) + O(q^{\frac{1}{2}})\Big\}. \nonumber 
\eea 
\end{lemma}
We will do many computations just for the $\mathcal{C}^{[g,k]}$'s, but they will work analogously for the $\mathsf{C}^{[g,k]}$'s specifying the exponent of $\mathsf{B}$ to $b = \tfrac{1}{2}$ and $B=\frac{1}{2}$ in the rest of the exponents given by $f$.

The expressions for the leading order coefficients -- here denoted with $*$ -- are provided in Appendix~\ref{AppBB}, where we provide a proof of the lemma. They are non-zero and satisfy the same selection rules as the unstarred quantities on the left-hand side. An interesting feature of the result is that, in the formula of Proposition~\ref{cosums} (resp. Proposition~\ref{cosums2}), the contribution to $\mathcal{C}^{[g,k]}$ (resp. $\mathsf{C}^{[g,k]}$) of a colored graph $(\mathscr{G};\bs{\sigma})$ has order of magnitude $q^{f(\mathscr{G};\bs{\sigma})}$ with
$$
f(\mathscr{G};\bs{\sigma}) = \sum_{\mathsf{v} \in V(\mathscr{G})} f(\bs{\sigma}[\mathsf{v}]\,\vert\, B), \text{ with } B=b \text{ (resp. } B=1/2 \text{)}.
$$
We remark that $f(\bs{\sigma}[\mathsf{v}]\,\vert\, B)$ does not depend on the vertex being terminal, bi-terminal, having a loop or not. Since $q = e^{-\frac{\pi}{T}} \rightarrow 0$ when $T \rightarrow 0^+$, the leading term in $\mathcal{C}^{[g,k]}$ and $\mathsf{C}^{[g,k]}$ are given by the colored graphs minimizing $f(\mathscr{G};\bs{\sigma})$. We will study the minimizing graphs and their exponent in Section~\ref{NextS}.

\subsubsection{Minimization over colorings}

\begin{lemma}
\label{lemanun}
For a given graph $\mathscr{G}$ of genus $g$ with $k$ legs, the coloring assigning $0$ to each edge realizes the minimum of $f(\mathscr{G};\bs{\sigma})$, which is 
$$
(2g - 2 + k)\left(\mathfrak{d}\tfrac{b}{2}-1\right).
$$
\end{lemma}
\noindent \textbf{Proof.} Every $f(\varepsilon,\sigma,\sigma')$ realizes its minimum $\left(\mathfrak{d}\frac{b}{2}-1\right)$ at $(\varepsilon,\sigma,\sigma')=(0,0,0)$, and the coloring with $\bs{\sigma}[\mathsf{v}] = (0,0,0)$ for all $\mathsf{v} \in V(\mathscr{G})$ receives a non-zero contribution at this order.  \hfill $\Box$

\subsection{Study of the critical exponents of the coefficients $\mathcal{C}$ and $\mathsf{C}$}
\label{NextS}

Let $\lfloor x \rfloor$ denote the unique integer such that $\lfloor x \rfloor \leq x < \lfloor x \rfloor + 1$. Let us define
\bea
\beta_1(i_{1/2}) & \coloneqq & \lfloor \tfrac{i_{1/2}}{2}\rfloor+2\delta_{i_{1/2},1}, \nonumber \\
\beta_2(g,k,i_0) & \coloneqq & 2 g -2 +\lfloor\tfrac{k}{2}\rfloor+\lfloor\tfrac{i_0 + (k \,{\rm mod}\, 2)}{2}\rfloor  .\nonumber\eea
We then define a function of three integers $g,i_0,i_{1/2}$ such that $2g - 2 + i_0 + i_{1/2} \geq 1$:
\beq
\label{betadef} \beta(g,i_{0},i_{1/2}\vert B) = \left\{\begin{array}{lll} \beta_1(i_{1/2})\tfrac{B}{2}+ \beta_2(g,i_{0}+i_{1/2},i_0)(\mathfrak{d} \tfrac{b}{2}-1) && {\rm if}\,\,\beta_2( g,i_0 + i_{1/2},i_0) > 0, \\ 0 & & {\rm otherwise}. \end{array}\right.
\eeq
It will be useful later to know what happens when we decrement $i_0$ and increment $i_{1/2}$.
\begin{lemma}
\label{betadecret} For $i_0>0$, we have $\beta(g,i_0,i_{1/2}\vert B)+\Delta=\beta(g,i_0-1,i_{1/2}+1\vert B)$, where 
\beq
\Delta =  \left\{\begin{array}{rl}
2 \frac{B}{2} -\left( \mathfrak{d}\frac{b}{2} -1 \right), &\text{ if } i_{1/2}=0, \\
-\frac{B}{2}, &\text{ if } i_{1/2}=1, \\
-\left( \mathfrak{d}\frac{b}{2} -1 \right), &\text{ if } i_{1/2}>0 \text{ even}, \\
\frac{B}{2}, &\text{ if } i_{1/2}>1 \text{ odd},
\end{array} \right.  \nonumber
\eeq
except for the exceptional cases $(g,k)=(0,3)$, $(0,4)$ and $(g,k,i_0)=(1,1,1), (0,5,1)$. In the last cases, we obtain
\beq
\Delta =  \left\{\begin{array}{rl}
 -\left( \mathfrak{d}\frac{b}{2} -1 \right), &\text{ if } (g,k)=(1,1), \, i_0=1, \\
-2\frac{B}{2}-\left( \mathfrak{d}\frac{b}{2} -1 \right), &\text{ if } (g,k)=(0,5), \, i_0=1,
\end{array} \right.  \nonumber
\eeq
and, in the other exceptional cases, where some configurations $(i_0, i_{1/2})$ give $\mathcal{C}^{[g,k]}=0$, we only record the variations between configurations giving non-zero $\mathcal{C}$'s:
\begin{itemize}
\item $\beta(0,3,0) -\left(\mathfrak{d}\frac{b}{2}-1\right)=\beta(0,0,3)=0$,
\item $\beta(0,4,0) +\frac{B}{2}-\left(\mathfrak{d}\frac{b}{2}-1\right)=\beta(0,2,2)$,
\item $\beta(0,2,2)-\frac{B}{2} -\left(\mathfrak{d}\frac{b}{2}-1\right)=\beta(0,0,4)=0$.
\end{itemize}
\end{lemma}
\noindent \textbf{Proof.} The exceptional cases can be easily checked with the expression for $\beta$. For the general situation, we separate cases according to the parity of $i_0$ and $k$, and we check first how $\beta_2(g,i_0+i_{1/2},i_0)$ varies depending on the parity of $i_{1/2}$:
\begin{itemize}
\item If $i_{1/2}$ is even, then $\beta_2(g,k,i_0-1)=\beta_2(g,k,i_0)-1$.
\item If $i_{1/2}$ is odd, then $\beta_2(g,k,i_0-1)=\beta_2(g,k,i_0)$.
\end{itemize}
For the variation of $\beta_1(i_{1/2})$, we distinguish four cases:
\begin{itemize}
\item $\beta_1(1)=\beta_1(0)+2=2$.
\item $\beta_1(2)=\beta_1(1)-1=1$.
\item If $i_{1/2}>0$ is even, then $\beta_1(i_{1/2}+1)=\beta_1(i_{1/2})$.
\item If $i_{1/2}>1$ is odd, then $\beta_1(i_{1/2}+1)=\beta_1(i_{1/2})+1$.
\end{itemize}
\hfill $\Box$


\begin{lemma}
\label{Cbehavior}
Let $g \geq 0$ and $k \geq 1$ such that $2g - 2 + k > 0$. Let $\varepsilon_1, \ldots, \varepsilon_k \in \{0,\tfrac{1}{2}\}$ be fixed, and denote $i_0$ (resp. $i_{1/2}$) be the number of $\varepsilon_i = 0$ (resp. $=\tfrac{1}{2}$). Then, in the critical regime $\tau = {\rm i}T$ with $T \rightarrow 0^+$ we obtain
\bea
\mathcal{C}^{[g,k]}\bigl[{}^{l_1}_{\varepsilon_1}\,\cdots\,{}^{l_k}_{\varepsilon_k}\bigr] & = &  \Big(\frac{\pi}{T}\Big)^{- \sum_{i = 1}^k (2l_i + 1)}\,q^{\beta(g,i_{0},i_{1/2}\vert b)}\Big(\mathcal{C}^{[g,k]}_*\bigl[{}^{l_1}_{\varepsilon_1}\,\cdots\,{}^{l_k}_{\varepsilon_k}\bigr] + O(q^{\frac{b}{2}})\Big),\nonumber \\
\mathsf{C}^{[g,k]}\bigl[{}^{l_1}_{\varepsilon_1}\,\cdots\,{}^{l_k}_{\varepsilon_k}\bigr] & = &  \Big(\frac{\pi}{T}\Big)^{-\sum_{i = 1}^k (2l_i + 1)}\,q^{\beta\left(g,i_{0},i_{1/2}\vert \frac{1}{2}\right)}\Big(\mathsf{C}^{[g,k]}_*\bigl[{}^{l_1}_{\varepsilon_1}\,\cdots\,{}^{l_k}_{\varepsilon_k}\bigr] + O(q^{\frac{b}{2}})\Big), \nonumber
\eea
where the leading coefficients indicated with $*$ are non-zero.
\end{lemma}
 
\noindent\textbf{Proof.} We shall do the reasoning for $\mathcal{C}^{[g,k]}$, i.e. for $B=b$, but all the comparisons we do will work also for the special case of $B=\frac{1}{2}$, so the final scaling exponent will be the same for $\mathsf{C}^{[g,k]}$ specifying $B=\frac{1}{2}$, instead of $B=b$. For simplicity, we will write $\beta(g,i_0,i_{1/2})\equiv\beta(g,i_0,i_{1/2}\vert b)$ in this proof. The determination of the exponent of $\tfrac{\pi}{T}$ will be addressed in the third part of the proof. For the moment, we only focus on the powers of $q$. Since we know $\mathcal{C}^{[g,k]}\bigl[{}^{l_1}_{\varepsilon_1}\,\cdots\,{}^{l_k}_{\varepsilon_k}\bigr]$ is invariant by permutation of the pairs $(l_i,\varepsilon_i)_{i = 1}^k$, the scaling exponent will only depend on $g$, $i_0$ and $i_{1/2}$. In the case $i_{1/2} = 0$, we have:
$$
\beta(g,k,0)=\left(2 g -2 +\lfloor\tfrac{k}{2}\rfloor+\lfloor\tfrac{k + (k \,{\rm mod}\, 2)}{2}\rfloor\right)\left(\mathfrak{d} \tfrac{b}{2}-1\right) = (2 g -2 +k)\left(\mathfrak{d} \tfrac{b}{2}-1\right),
$$ 
so the claim is correct according to Lemma \ref{lemanun}.

We prove all the other cases by induction on $2 g -2 +k$, starting by the two base cases with $2 g -2 +k=1$. In both base cases there is only one graph with a single vertex.

\smallskip

\noindent $\bullet$ $(g,k)=(0,3)$. Remember $\mathcal{C}^{[0,3]}\big[{}^{l_1}_{\varepsilon_1}\,{}^{l_2}_{\varepsilon_2}\,{}^{l_3}_{\varepsilon_3}\bigr] = 0$ in case we do not have $\varepsilon_1=\varepsilon_2=\varepsilon_3$. So the only case to consider is $i_{1/2}=3$ and we have $f(\frac{1}{2},\frac{1}{2},\frac{1}{2})=0$, which is equal to $\beta(0,0,3)$ since $\beta_2(0,3,0)=-1 < 0$.

\smallskip

\noindent $\bullet$ $(g,k)=(1,1)$. Remember the color of a loop should be identical to the color of the other edge. So in the case $i_{1/2}=1$, we get $f(\frac{1}{2},\frac{1}{2},\frac{1}{2})=0$, which is equal to $\beta(1,0,1)$ since $\beta_2(1,1,0)=0$.

\smallskip

Now we will prove the result for cases with $2 g -2 +k$, supposing it is true for all cases $( \overline{g},\overline{k})$ with $2 \overline{g} -2 +\overline{k} <2 g -2 +k$. We can decompose graphs $\mathscr{G}\in\mathcal{S}^{(g,k)}$ in terms of a graph $\mathcal{P}$ which consists of only one trivalent vertex $\mathsf{v}_0$ without loops, and either one graph $\tilde{\mathscr{G}}\in\mathcal{S}^{(g - 1,k+1)}$, or two graphs $\mathscr{G}'\in \mathcal{S}^{(g',k'+1)}$ and $\mathscr{G}''\in\mathcal{S}^{(g'',k''+1)}$, with $g' + g'' = g$ and $k'+k''=k-1$, excluding the cases $(g',k')=(0,0)$ and $(g'',k'')=(0,0)$.

The two last legs of $\mathcal{P}$ are shared either with two legs of $\tilde{\mathscr{G}}$, or with one in $\mathscr{G}'$ and one in $\mathscr{G}''$. Consider the following decompositions $\tilde{k}=\tilde{i}_0+\tilde{i}_{1/2}$, $k'=i_0'+i_{1/2}'$ and $k^{\prime\prime}=i_0''+i_{1/2}''$, with $\tilde{k}+2=k+1$ and $(k'+1)+(k''+1)=k+1$, where $\tilde{k},k'$ and $k''$ correspond to the number of legs which are not shared with $\mathcal{P}$ in the respective subgraphs $\tilde{\mathscr{G}}$, $\mathscr{G}'$ and $\mathcal{G''}$.

In order to extend a coloring for the corresponding subgraph $\tilde{\mathscr{G}}$, or $\mathscr{G}'$ and $\mathscr{G}''$ to a coloring of the whole $\mathscr{G}$, we will pick $\bs{\sigma}[\mathsf{v}_0]=(\sigma_0,\sigma_1,\sigma_2)$ in a compatible way, i.e. the colorings $\sigma_1$ and $\sigma_2$ of the two legs of $\mathcal{P}$ which are shared with the corresponding subgraphs will coincide with the given ones for these legs on the subgraphs. We will make these choices to minimize $f(\mathscr{G};\bs{\sigma})$, which will be $f(\bs{\sigma}[\mathsf{v}_0])+\sum_{\mathsf{v}\in V(\tilde{\mathscr{G}})}f(\tilde{\bs{\sigma}}[\mathsf{v}])$ or $f(\bs{\sigma}[\mathsf{v}_0])+\sum_{\mathsf{v}\in V(\mathscr{G}')}f(\bs{\sigma}'[\mathsf{v}])+\sum_{\mathsf{v}\in V(\mathscr{G}'')}f(\bs{\sigma}''[\mathsf{v}])$.

For every configuration $i_0+i_{1/2}=k$, we will first build a graph $\mathscr{G}\in \mathcal{S}^{(g,k)}$ with a coloring which is compatible with the fixed colorings of the legs $\bs{\sigma}\in \mathsf{Col}(\mathscr{G};(\bs{l},\bs{\varepsilon}))$ from the ones from previous induction steps such that $f(\mathscr{G};\bs{\sigma})=\beta(g, i_0,i_{1/2})$, i.e. a graph realizing the desired value. Secondly, we will have to prove that for every other graph $\overline{\mathscr{G}}\in \mathcal{S}^{(g,k)}$ there is no other coloring $\overline{\bs{\sigma}}\in \mathsf{Col}(\mathscr{G};(\bs{l},\bs{\varepsilon}))$ such that $f(\overline{\mathscr{G}};\overline{\bs{\sigma}})<\beta(g, i_0,i_{1/2})$, i.e. that $\beta(g, i_0,i_{1/2})$ is actually the minimum.

Remember that the cases with $i_0=k$ and $i_{1/2}=0$ were already checked, so we do not consider them in the following.

\bigskip

\begin{center}
\textit{First part: special cases}
\end{center}

\bigskip

We will deal first with the two special cases $(g,k)=(0,4), (0,5)$.

\medskip

\noindent $\bullet$ $(g,k)=(0,4)$. 
The graphs in $\mathcal{S}^{(0,4)}$ have only two vertices, one terminal and one bi-terminal. This implies that the only options with $\mathcal{C}^{[0,4]}\big[{}^{l_1}_{\varepsilon_1}\,\cdots\,{}^{l_4}_{\varepsilon_4}\bigr]\neq 0$ are $i_0=0, 2, 4$.
We show in Figure~\ref{fig:graphs(0,4)} the graphs with a suitable coloring which realize the desired value in every remaining case. 
\begin{figure}[h!]
\begin{center}
\includegraphics[width=.6\textwidth]{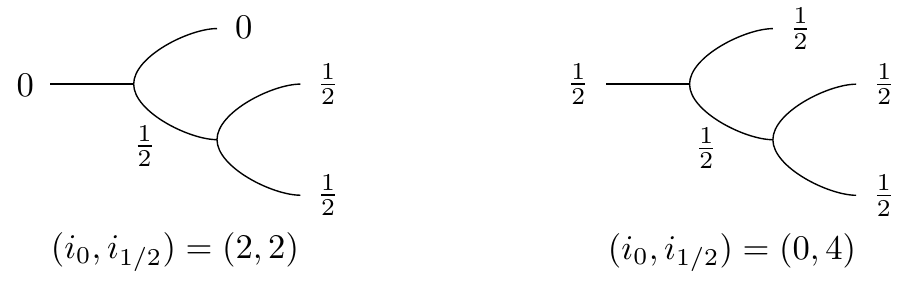}
  \caption{\label{fig:graphs(0,4)} $(g,k)=(0,4)$} 
  \end{center}
\end{figure}
Observe that $\mathscr{G}''$ is the only graph in $\mathcal{S}^{(0,3)}$. Since $i_{1/2}''=2$ and the only vertex is biterminal, we have to set $\sigma_2=\tfrac{1}{2}$ here. We already checked that $f(\mathscr{G}'';\bs{\sigma}'')=\beta(0,0,3)=0$. Therefore, for $(i_0,i_{1/2})=(2,2)$ we obtain $f(\mathscr{G};\bs{\sigma})= f(0,0,\frac{1}{2}) = \mathfrak{d}\tfrac{b}{2} -1 + \tfrac{b}{2}=\beta(0,2,2)$ and for $(i_0,i_{1/2})=(0,4)$, $f(\mathscr{G};\bs{\sigma})=f(\frac{1}{2},\frac{1}{2},\frac{1}{2})=0= \beta(0,0,4)$, as we wanted.

\medskip

\noindent $\bullet$ $(g,k)=(0,5)$. For every possible $(i_0,i_{1/2})$ we choose the graph with the corresponding coloring shown in Figure~\ref{fig:graphs(0,5)}:
\begin{figure}[h!]
\begin{center}
\includegraphics[width=.9\textwidth]{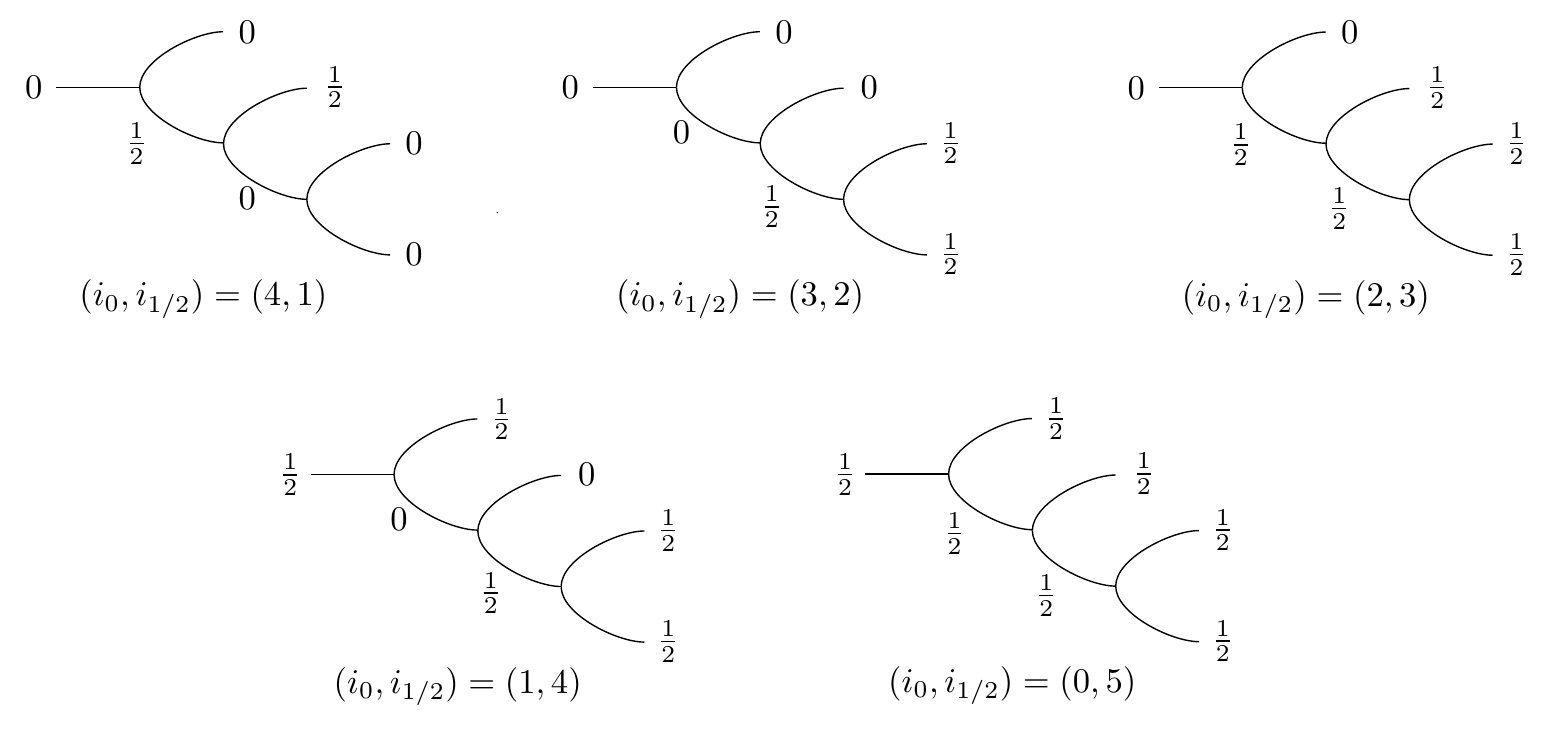}
  \caption{\label{fig:graphs(0,5)} $(g,k)=(0,5)$} 
  \end{center}
\end{figure}
Observe that $\mathscr{G}'' \in \mathcal{S}^{(0,4)}$, which also makes the choice of $\sigma_2$ special. 

\smallskip 

\noindent $\diamond$ If $i_0=4$ and $i_{1/2}=1$, with the chosen graph we can only set $\sigma_2= \tfrac{1}{2}$. By induction hypothesis, we have $f(\mathscr{G}'';\bs{\sigma}'')=\beta(0,2,2)=\frac{b}{2}+\mathfrak{d}\frac{b}{2}-1$. Therefore
$$
f(\mathscr{G};\bs{\sigma})= f(0,0,\tfrac{1}{2}) + \beta(0,2,2) = (\mathfrak{d} + 1)\tfrac{b}{2}-1+(\mathfrak{d} + 1)\tfrac{b}{2} -1 = 2\tfrac{b}{2} + 2\left(\mathfrak{d}\tfrac{b}{2}-1\right)= \beta(0,4,1).
$$

\smallskip

\noindent $\diamond$ If $i_0=3$ and $i_{1/2}=2$, we can choose $\sigma_2=0$. By induction hypothesis, we have $f(\mathscr{G}'';\bs{\sigma}'')=\beta(0,2,2)=\mathfrak{d}\tfrac{b}{2} -1 + \tfrac{b}{2}$. Therefore$$
f(\mathscr{G};\bs{\sigma})= f(0,0,0) + \beta(0,2,2) = \mathfrak{d}\tfrac{b}{2} -1+ \tfrac{b}{2} +\mathfrak{d}\tfrac{b}{2}-1 = \beta(0,3,2).
$$

\smallskip

\noindent $\diamond$ If $i_0=2$ and $i_{1/2}=3$, we can only choose $\sigma_2=\frac{1}{2}$. By induction hypothesis, we have $f(\mathscr{G}'';\bs{\sigma}'')=\beta(0,0,4)=0$. Therefore
$$
f(\mathscr{G};\bs{\sigma})= f(0,0,\tfrac{1}{2}) = (\mathfrak{d} + 1)\tfrac{b}{2}-1= \beta(0,2,3).
$$

\smallskip

\noindent $\diamond$ If $i_0=1$ and $i_{1/2}=4$, we can only choose $\sigma_2=0$. Therefore $f(\mathscr{G};\bs{\sigma})= f(\frac{1}{2},\frac{1}{2},0)+ \beta(0,2,2)= \frac{b}{2}+\frac{b}{2}+\mathfrak{d}\frac{b}{2}-1= \beta(0,1,4)$.

\smallskip

\noindent $\diamond$ If $i_0=0$ and $i_{1/2}=5$, we can only choose $\sigma_2= \frac{1}{2}$. Thus $f(\mathscr{G};\bs{\sigma})= f(\frac{1}{2},\frac{1}{2},\frac{1}{2})+ \beta(0,0,4)= 0= \beta(0,0,5)$.

\bigskip 

\begin{center}
\textit{First part: general cases}
\end{center}

\bigskip

For the general cases with $(g,k)\neq (0,4), (0,5)$ we will consider four cases. If $k>2$, we will be automatically in one of the first two cases.

\noindent \textbf{Case I: $\mathbf{i_0 \geq 2}$.} We will choose the graph $\mathscr{G}$ constructed from $\mathscr{G}'$ and $\mathscr{G}''$, with $k'=i_0'=1$ and $g=0$. Observe that $\mathscr{G}'$ has $2$ legs, both with coloring $0$.
In this case, $\mathsf{v}_0$ is a terminal vertex, so for the contribution to be non-zero, we have $\sigma_0=\sigma_1$ and we know that $\sigma_1=0$ because the leg is shared with $\mathscr{G}'$. Note that $i_0=i_0''+2$ and $i_{1/2}=i_{1/2}''$.
\begin{figure}[h!]
\begin{center}
\includegraphics[width=.4\textwidth]{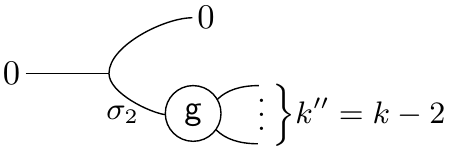} 
\caption{\label{fig:11} $i_0 \geq 2$} 
\end{center}
\end{figure}
In the general case, we can always choose $\sigma_2=0$. By the induction hypothesis, we can choose $(\mathscr{G}'',\bs{\sigma}'')$ such that $f(\mathcal{G}'';\bs{\sigma}'')= \beta(g,i_0''+1,i_{1/2}'')$. Therefore 
\bea
f(\mathscr{G};\bs{\sigma})
 & =& \mathfrak{d}\tfrac{b}{2}-1 + \beta_1(i_{1/2}'')\tfrac{b}{2} + \beta_2(g, k''+1, i_0''+1)\left(\mathfrak{d}\tfrac{b}{2}-1\right) \nonumber\\
 &=& \beta_1(i_{1/2})\tfrac{b}{2} + (\beta_2(g, k-1, i_0-1)+1)\left(\mathfrak{d}\tfrac{b}{2}-1\right)=\beta(g,i_0,i_{1/2}). \nonumber
\eea
The last step is a simple computation separating the cases where $k-1$ is even and odd.

\medskip

\noindent  \textbf{Case II: $\mathbf{i_{1/2}\geq 2}$.} Again we choose the graph $\mathscr{G}$ constructed from $\mathscr{G}'$ and $\mathscr{G}''$, with $k'=1$ and $g=0$, but with $i'_{1/2}=1$ because in this case we have no assumption on $i_0$. And again $\mathsf{v}_0$ is a terminal vertex, so for the contribution to be non-zero, we have $\sigma_0=\sigma_1$, but here $\sigma_1= \frac{1}{2}$. Note that $i_0=i_0''$ and $i_{1/2}=i_{1/2}''+2$. 
\begin{figure}[h!]
\begin{center}
\includegraphics[width=.38\textwidth]{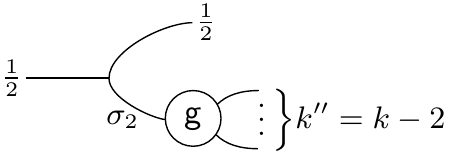}  
\caption{\label{fig:12} $i_{1/2}\geq 2$} 
\end{center}
\end{figure}
It will minimize to choose $\sigma_2=\frac{1}{2}$, if $i_{1/2}''=1$, and $\sigma_2=0$, otherwise. If $i_{1/2}''=1$, we have
\bea
f(\mathscr{G};\bs{\sigma})
 & =& 0 + \beta_1(2)\tfrac{b}{2} + \beta_2(g, k''+1, i_0'')\left(\mathfrak{d}\tfrac{b}{2}-1\right)\nonumber\\
 &=& \beta_1(i_{1/2})\tfrac{b}{2} + \beta_2(g, k-1, i_0)\left(\mathfrak{d}\tfrac{b}{2}-1\right)=\beta(g,i_0,i_{1/2}). \nonumber
\eea
In the last step we separate the cases where $k-1$ is even and odd, and we use $i_0+3=k$ to deduce the parity of $i_0$ in every case. If $i_{1/2}''\neq 1$, we have:
\bea
f(\mathscr{G};\bs{\sigma}) & =& \tfrac{b}{2} + \beta_1(i_{1/2}'')\tfrac{b}{2} + \beta_2(g, k''+1, i_0''+1)\left(\mathfrak{d}\tfrac{b}{2}-1\right) \nonumber\\
 &=& \beta_1(i_{1/2}''+2)\tfrac{b}{2} + \beta_2(g, k-1, i_0+1)\left(\mathfrak{d}\tfrac{b}{2}-1\right)=\beta(g,i_0,i_{1/2}). \nonumber
\eea
The last step is again a simple computation separating cases according to the parity of $k-1$.

\medskip

\noindent \textbf{Case III: $\mathbf{i_0=1, i_{1/2}=1}$.}  This is the remaining case of $k=2$. Observe that here $g>0$ so that $2g-2+2>0$. We distinguish two cases:

\smallskip

\noindent $\bullet$ $g = 1$. We choose the graph $\mathscr{G}$ constructed from $\tilde{\mathscr{G}}$ with $\tilde{i}=\tilde{i}_{1/2}=1$. Since the vertex of $\tilde{\mathscr{G}}$ is terminal in $\mathscr{G}$, at least $\sigma_1$ or $\sigma_2$ should be $\frac{1}{2}$ for the contribution of the graph to be non-zero. Actually if we set $\sigma_1=\sigma_2=\frac{1}{2}$, we get $f(\mathscr{G};\bs{\sigma})=f(0,\frac{1}{2},\frac{1}{2})+f\left(\frac{1}{2},\frac{1}{2},\frac{1}{2}\right)= \mathfrak{d}\tfrac{b}{2} - 1 +  b = \beta(1,1,1)$.

\begin{figure}[h!]
\begin{center}
\includegraphics[width=.3\textwidth]{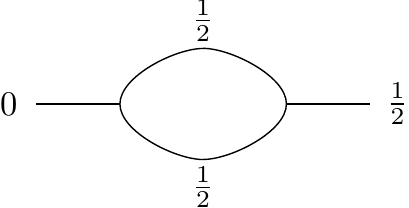}
\caption{\label{fig:(1,2)131} $(g,i_0,i_{1/2}) = (1,1,1)$}\end{center}    \end{figure}

\smallskip

\noindent $\bullet$ $g > 1$. We build $\mathscr{G}$ from $\mathscr{G}'$ and $\mathscr{G}''$ with $k'=i_{1/2}'=1$, $g'>0$, $k''=0$ and $g''> 0$, which we can choose because $g>1$. Observe that if $\overline{g}>0$, then $2 \overline{g} -2+1 >0$ and hence in our cases we will have $\beta_2 >0$. Since $i_0', i_0'' =0$ and $i_0=1$, $\sigma_0=0$ and we can choose $\sigma_1=\sigma_2=0$.
\begin{figure}[h!]
\begin{center}
\includegraphics[width=.25\textwidth]{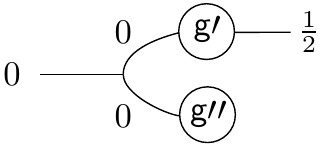}
\caption{\label{fig:132} $g>1, (i_0,i_{1/2}) = (1,1)$ } 
\end{center}
\end{figure}
Therefore 
\bea
f(\mathscr{G};\bs{\sigma}) &=& \mathfrak{d}\tfrac{b}{2}-1 + \beta_1(1)\tfrac{b}{2}+\beta_2(g',2,1)\left(\mathfrak{d}\tfrac{b}{2}-1\right) + \beta_1(0)\tfrac{b}{2}+\beta_2(g'',1,1)\left(\mathfrak{d}\tfrac{b}{2}-1\right) \nonumber \\
&=& b + (2g-2+1)\left(\mathfrak{d}\tfrac{b}{2}-1\right) = \beta(g,1,1). \nonumber
\eea

\medskip

\noindent \textbf{Case IV: $\mathbf{k=1}$.} The case $(1,1)$ was already a base one, so here we suppose $g>1$. We consider the case $k=i_{1/2}=1$. So $\sigma_0=\varepsilon_1=\frac{1}{2}$ and we construct $\mathscr{G}$ from $\mathscr{G}'$ and $\mathscr{G}''$ with $k'=k''=0$ and $g', g'' >0$. We can choose $\sigma_1=\sigma_2=0$.
\begin{figure}[h!]
\begin{center}
\includegraphics[width=.2\textwidth]{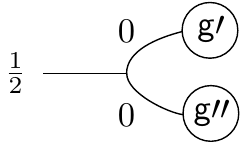}
\caption{\label{fig:14} $k=1$} 
\end{center}
\end{figure}
Therefore 
\bea
f(\mathscr{G};\bs{\sigma}) &=& b +\beta_1(0)\tfrac{b}{2}+\beta_2(g',1,1)\left(\mathfrak{d}\tfrac{b}{2}-1\right) + \beta_1(0)\tfrac{b}{2}+\beta_2(g'',1,1)\left(\mathfrak{d}\tfrac{b}{2}-1\right) \nonumber \\
&=& b + (2g-2)\left(\mathfrak{d}\tfrac{b}{2}-1\right) = \beta(g,0,1). \nonumber
\eea

\medskip

\begin{center}
\textit{Second part: disconnected cases}
\end{center}

\medskip

For the second part of the proof, we will check that all other possible graphs and colorings for every case do not give a smaller exponent. We first discuss the disconnected case, i.e. the case where $\mathscr{G}$ is constructed from $\mathscr{G}'$ and $\mathscr{G}''$  so that $\mathscr{G}', \mathscr{G}''\not\in\mathcal{S}^{(0,2)}$. The cases with $\mathscr{G}'$ or $\mathscr{G}''$ in $\mathcal{S}^{(0,3)}$ or $\mathcal{S}^{(0,4)}$ will be considered apart because they will have some extra restrictions to choose $\sigma_1$ and $\sigma_2$ and will be called the exceptional cases in this part.
\begin{figure}[h!]
\begin{center}
\includegraphics[width=.26\textwidth]{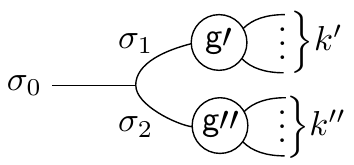}
\caption{\label{fig:21} $\mathscr{G}$ from $\mathscr{G}'$ and $\mathscr{G}''$} 
\end{center}
\end{figure}

Moreover, remember that a graph in $\mathcal{S}^{(0,5)}$ and with $i_0=0, i_{1/2}=5$ was also giving the special value of $\beta_2(0,5,0)=0$ and hence $\beta(0,0,5)=0$ automatically. However, observe that when one of the pieces $\mathscr{G}'$ or $\mathscr{G}''$ is in $\mathcal{S}^{(0,5)}$, we will not have any exceptional situation here because a graph in $\mathcal{S}^{(0,5)}$ with $i_{1/2}'=4$ and $\sigma_1=\tfrac{1}{2}$, or $i_{1/2}''=4$ and $\sigma_2=\tfrac{1}{2}$ will never be chosen to minimize; it will always be better to choose $\sigma_1$ or $\sigma_2$ to be $0$, which in this case is possible.

\smallskip

\noindent \textbf{Case} $\mathbf{\sigma_0=0}$. Let us check that choosing $\sigma_1=0$, if possible, always minimizes. Observe that $f(0,0,\sigma_2)=f\left(0,\frac{1}{2},\sigma_2\right)-\frac{b}{2}$. Then, making use of Lemma~\ref{betadecret}, we get\beq
f(0,0,\sigma_2)  +\beta(g',i_0'+1,i_{1/2}') \leq f(0,\tfrac{1}{2},\sigma_2)+\beta(g',i_0',i_{1/2}'+1). \nonumber
\eeq
Indeed, it is clear if $i_{1/2}'\neq 1$, and if $i_{1/2}'= 1$, we always have an equality because
\beq
f(0,0,\sigma_2)  +\beta(g',i_0'+1,1) = f(0,\tfrac{1}{2},\sigma_2)-\tfrac{b}{2}+\beta(g',i_0',2)+\tfrac{b}{2}. \nonumber
\eeq
The same argument works for $\sigma_2$. Now, we should check that the exceptional cases, where we cannot choose $\sigma_1=\sigma_2=0$, do not minimize further.

\smallskip

\noindent $\bullet$ If $\sigma_1=\tfrac{1}{2}, \sigma_2=0, (g',k'+1)=(0,3), i_{1/2}'=2$, we have
$$
f(\mathscr{G};\bs{\sigma}) = \big((\mathfrak{d} + 1)\tfrac{b}{2} - 1\big) + 0 + \beta(g,i_0,i_{1/2} - 1) = \delta_{i_{1/2},3}\,b + \beta(g,i_0,i_{1/2}),
$$
where we have used that here $i_{1/2}\geq 2$.

\smallskip

\noindent $\bullet$ If $\sigma_1=\frac{1}{2}, \sigma_2=0, (g',k'+1)=(0,4), i_{1/2}'=1$, we have
\bea
f(\mathscr{G};\bs{\sigma})
& = &(\mathfrak{d} + 1)\tfrac{b}{2}-1 + \tfrac{b}{2}+\left(\mathfrak{d}\tfrac{b}{2}-1\right) + \beta_1(i_{1/2}-1)\tfrac{b}{2}+\beta_2(g,k-4,i_0-2)\left(\mathfrak{d}\tfrac{b}{2}-1\right) \nonumber \\
&=& (\beta_1(i_{1/2}-1)+2)\tfrac{b}{2}+(\beta_2(g,k-4,i_0-2)+2)\left(\mathfrak{d}\tfrac{b}{2}-1\right) \nonumber \\
&=& (\beta_1(i_{1/2}-1)+2)\tfrac{b}{2}+(\beta_2(g,k,i_0)-1)\left(\mathfrak{d}\tfrac{b}{2}-1\right) \geq \beta(g,i_0,i_{1/2}). \nonumber
\eea
In the last step we have used that $\beta_1(i_{1/2}-1)+2>\beta(i_{1/2})$.\smallskip

\noindent $\bullet$ If $\sigma_1=\frac{1}{2}, \sigma_2=0, (g',k'+1)=(0,4), i_{1/2}'=3$, we have
\bea
f(\mathscr{G};\bs{\sigma})
&=&\big(\mathfrak{d}\tfrac{b}{2}-1 + \tfrac{b}{2}\big)+ 0 + \beta_1(i_{1/2}-3)\tfrac{b}{2}+\beta_2(g,k-4,i_0)\left(\mathfrak{d}\tfrac{b}{2}-1\right) \nonumber \\
&=& (\beta_1(i_{1/2}-3)+1)\tfrac{b}{2}+(\beta_2(g,k-4,i_0)+1)\left(\mathfrak{d}\tfrac{b}{2}-1\right) \nonumber \\
&=& (\beta_1(i_{1/2}-3)+1)\tfrac{b}{2}+(\beta_2(g,k,i_0)-1)\left(\mathfrak{d}\tfrac{b}{2}-1\right)  \nonumber \\
&\geq& (\beta_1(i_{1/2})-1)\tfrac{b}{2}+(\beta_2(g,k,i_0)-1)\left(\mathfrak{d}\tfrac{b}{2}-1\right) \nonumber \\
&=& \beta(g,i_0,i_{1/2}) -\big(\mathfrak{d}\tfrac{b}{2}-1 + \tfrac{b}{2}\big) \geq \beta(g,i_0,i_{1/2}). \nonumber
\eea

\smallskip

\noindent $\bullet$ The remaining cases with $\sigma_1=\frac{1}{2}, \sigma_2=\frac{1}{2}$ consist of $\mathscr{G}', \mathscr{G}'' \in\mathcal{S}^{(0,4)}$; $\mathscr{G}'\in\mathcal{S}^{(0,3)}$ and $\mathscr{G}'\in\mathcal{S}^{(0,4)}$, and symmetric ones by exchanging the role of $\sigma_1$ and $\sigma_2$. They can be checked easily from the results for the base cases $(0,3)$ and $(0,4)$. 

\medskip
Choosing $\sigma_1=\sigma_2=0$ for the non-exceptional cases, we obtain:
\bea
f(\mathscr{G};\bs{\sigma})&=& \big(\mathfrak{d}\tfrac{b}{2} - 1\big) + \beta(g',i_0'+1,i_{1/2}')+\beta(g-g',i_0''+1,i_{1/2}'')\nonumber \\ 
&=&(\beta_1(i_{1/2}')+\beta_1(i_{1/2}''))\tfrac{b}{2} \nonumber \\
& & + \big(\beta_2(g',k'+1,i_{0}'+1)+\beta_2(g'',k''+1,i_{0}''+1)+1\big)\left(\mathfrak{d}\tfrac{b}{2}-1\right).\nonumber
\eea
On the one hand, separating cases according to the parity of $k'$ and $k''$, and the parity of $i_0'$ and $i_0''$, we check that
\bea
\beta_2(g',k'+1,i_{0}'+1)+\beta_2(g'',k''+1,i_{0}''+1)+1 \nonumber \\
\label{ineq} \leq \beta_2(g'+g'',k'+k''+1,i_0'+i_0''+1)=\beta_2(g,k,i_0) 
\eea
and hence with this part we cannot minimize further.

On the other hand, distinguishing cases according to the parity of $i'_{1/2}$, $i_{1/2}''$ and $i_{1/2}=i'_{1/2}+i_{1/2}''$, and considering the special cases with some of them equal to $1$, we see that $\beta_1(i_{1/2}')+\beta_1(i_{1/2}'')=\beta_1(i_{1/2})-1$, if both $i_{1/2}'$ and $i_{1/2}''$ are odd and $>1$, and $\beta_1(i_{1/2}')+\beta_1(i_{1/2}'')\geq \beta_1(i_{1/2})$, otherwise.

Finally, we check easily that in the case of odd $i_{1/2}',i_{1/2}''>1$, where we have minimized $\beta_1$ by $1$, we lie in the cases with 
$\beta_2(g',k'+1,k_{0}'+1)+\beta_2(g'',k''+1,k_{0}''+1)+1 <\beta_2(g,k,i_0)$. Therefore, we also do not minimize globally, i.e.
\beq
f(\mathscr{G};\bs{\sigma}) \geq \beta(g,i_0,i_{1/2}), \nonumber
\eeq
because $-\frac{b}{2} >\mathfrak{d}\frac{b}{2}-1$ and thus, with a minimizing purpose, we prefer $\beta_2+1$ to $\beta_1-1$.

\medskip

\noindent \textbf{Case $\mathbf{\sigma_0=\frac{1}{2}}$.} As in the previous cases, it can be checked first that the exceptional cases do not minimize further. Let us check now which $\sigma_1$ we should choose to minimize, making use of Lemma~\ref{betadecret}.
\beq
f(\tfrac{1}{2},0,\sigma_2)+\beta_1(g',i_0'+1,i_{1/2}')=f(\tfrac{1}{2},\tfrac{1}{2},\sigma_2)+ \tfrac{b}{2}+\beta_1(g',i_0',i_{1/2}'+1)-\Delta. \nonumber
\eeq
If $i_{1/2}'\neq 1$, $\Delta \geq \frac{b}{2}$ and hence $\sigma_1=0$ minimizes. But, if $i_{1/2}'= 1$, $\Delta = -\frac{b}{2}$ and hence $\sigma_1=\frac{1}{2}$ is the minimizing choice. By the symmetry of the situation, the same argument works for the choice of $\sigma_2$ depending on $i_{1/2}''$.

\smallskip

\noindent $\bullet$ $i_{1/2}'=i_{1/2}''=1$. Using the inequality \eqref{ineq}, we have
\bea
f(\mathscr{G};\bs{\sigma})&=& 0 +\beta(g',i_0',i_{1/2}'+1)+\beta(g'',i_0',i_{1/2}'+1)\nonumber \\ 
&=& \beta_1(2) b +(\beta_2(g',k'+1,i_{0}')+\beta_2(g'',k''+1,i_{0}''))\left(\mathfrak{d}\tfrac{b}{2}-1\right) \nonumber \\
& \geq & \beta_1(3)\tfrac{b}{2}+(\beta_2(g,k,i_0'+i_0''-1)-1)\left(\mathfrak{d}\tfrac{b}{2}-1\right)  \nonumber\\
& \geq & \beta_1(3)\tfrac{b}{2}+\beta_2(g,k,i_0'+i_0'')\left(\mathfrak{d}\tfrac{b}{2}-1\right) = \beta(g,i_0,i_{1/2}). \nonumber
\eea

\smallskip

\noindent $\bullet$ $i_{1/2}'=1, i_{1/2}''\neq 1$ (and the analogous case $i_{1/2}'\neq1, i_{1/2}''=1$). Again using \eqref{ineq}, we obtain
\bea
f(\mathscr{G};\bs{\sigma})&=& \tfrac{b}{2} +\beta(g',i_0',i_{1/2}'+1)+\beta(g'',i_0''+1,i_{1/2}'')\nonumber \\ 
&=&\tfrac{b}{2}+(\beta_1(2)+\beta_1(i_{1/2}''))\tfrac{b}{2}+(\beta_2(g',k'+1,i_{0}')+\beta_2(g'',k''+1,i_{0}''+1))\left(\mathfrak{d}\tfrac{b}{2}-1\right) \nonumber \\
& \geq &\tfrac{b}{2}+ \beta_1(i_{1/2}''+2)\tfrac{b}{2}+(\beta_2(g,k,i_0'+i_0'')-1)\left(\mathfrak{d}\tfrac{b}{2}-1\right) \nonumber\\
& \geq & \beta_1(i_{1/2})\tfrac{b}{2}+\beta_2(g,k,i_0'+i_0'')\left(\mathfrak{d}\tfrac{b}{2}-1\right) = \beta(g,i_0,i_{1/2}). \nonumber
\eea

\smallskip

\noindent $\bullet$ $i_{1/2}'\neq 1, i_{1/2}''\neq 1$.
\bea
f(\mathscr{G};\bs{\sigma})&=& b +\beta(g',i_0'+1,i_{1/2}')+\beta(g'',i_0''+1,i_{1/2}'') \nonumber \\
&=& b+(\beta_1(i_{1/2}')+\beta_1(i_{1/2}''))\tfrac{b}{2} \nonumber \\
&& + (\beta_2(g',k'+1,i_{0}'+1)+\beta_2(g'',k''+1,i_{0}''+1))\left(\mathfrak{d}\tfrac{b}{2}-1\right) \nonumber \\
& \geq &(2+\beta_1(i_{1/2}')+\beta_1(i_{1/2}''))\tfrac{b}{2}+(\beta_2(g,k,i_0'+i_0''+1)-1)\left(\mathfrak{d}\tfrac{b}{2}-1\right) \nonumber \\
& \geq & \beta_1(i_{1/2}'+i_{1/2}''+1)\tfrac{b}{2}+(\beta_2(g,k,i_0'+i_0'')+1-1)\left(\mathfrak{d}\tfrac{b}{2}-1\right) = \beta(g,i_0,i_{1/2}). \nonumber
\eea
	
\medskip

\begin{center}
\textit{Second part: connected case}
\end{center}

\medskip

Now let us examine the case in which $\mathscr{G}$ is constructed from $\tilde{\mathscr{G}}$. Firstly it can be easily checked apart that special cases with $\tilde{\mathscr{G}}\in \mathcal{S}^{(0,3)}, \mathcal{S}^{(0,4)}$ do not minimize further.
\begin{figure}[h!]
\begin{center}
\includegraphics[width=.26\textwidth]{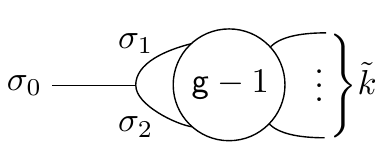}
\caption{\label{fig:22}  $\mathscr{G}$ from $\tilde{\mathscr{G}}$} 
\end{center}
\end{figure}

\noindent \textbf{Case $\mathbf{\sigma_0=0}$.} When we are not in the exceptional cases, we can always choose $\sigma_1=\sigma_2=0$ to minimize.
\bea
f(\mathscr{G};\bs{\sigma})&=& \big(\mathfrak{d}\tfrac{b}{2} - 1\big) +\beta_1(\tilde{i}_{1/2})\tfrac{b}{2}+\beta_2(\tilde{g},\tilde{k}+2,\tilde{i}_0+2)\left(\mathfrak{d}\tfrac{b}{2}-1\right)\nonumber \\ 
&=&\left(\mathfrak{d}\tfrac{b}{2}-1\right)+\beta_1(i_{1/2})\tfrac{b}{2}+\beta_2(g-1,k+1,i_0+1)\left(\mathfrak{d}\tfrac{b}{2}-1\right) \nonumber \\
&=&  \beta_1(i_{1/2})\tfrac{b}{2}+(\beta_2(g,k+1,i_0+1)-1)\left(\mathfrak{d}\tfrac{b}{2}-1\right)= \beta(g,i_0,i_{1/2}), \nonumber
\eea
where for the last computation we distinguish cases according to the parity of $k$.

\medskip

\noindent \textbf{Case $\mathbf{\sigma_0=\frac{1}{2}}$.} By the symmetry argument at the beginning of the proof, the only case remaining to be checked is the one corresponding to $i_0=0$ and $i_{1/2}=k$. By a computation similar to the one in the previous case with $\sigma_0=\frac{1}{2}$, we get that if $\tilde{i}_{1/2}=1$, $(\sigma_1,\sigma_2)=(\frac{1}{2},0)$ or $(\sigma_1,\sigma_2)=(0,\frac{1}{2})$ are the minimizing choices and if if $\tilde{i}_{1/2}\neq 1$, then we choose $(\sigma_1,\sigma_2)=(0,0)$ to minimize.

\smallskip

\noindent $\bullet$ $\tilde{i}_{1/2}=1\,\, (k=i_{1/2}=2)$.
\bea
f(\mathscr{G};\bs{\sigma})&=& \tfrac{b}{2} +\beta_1(2)\tfrac{b}{2}+\beta_2(g-1,3,1)\left(\mathfrak{d}\tfrac{b}{2}-1\right)\nonumber \\ &=&\tfrac{b}{2}+\beta_1(2)\tfrac{b}{2}+(\beta_2(g,2,0)-1)\left(\mathfrak{d}\tfrac{b}{2}-1\right) \nonumber \\
&\geq&\beta_1(2)\tfrac{b}{2}+\beta_2(g,2,0)\left(\mathfrak{d}\tfrac{b}{2}-1\right)= \beta(g,0,2). \nonumber \nonumber
\eea

\smallskip

\noindent $\bullet$ $\tilde{i}_{1/2}\neq 1$.
\bea
f(\mathscr{G};\bs{\sigma})&=& b +\beta_1(\tilde{i}_{1/2})\tfrac{b}{2}+\beta_2(g-1,k+1,2)\left(\mathfrak{d}\tfrac{b}{2}-1\right)\nonumber \\ 
&=&b+\beta_1(i_{1/2}-1)\tfrac{b}{2}+\beta_2(g-1,k+1,2)\left(\mathfrak{d}\tfrac{b}{2}-1\right)\left(\mathfrak{d}\tfrac{b}{2}-1\right) \nonumber \\
&\geq&\beta_1(i_{1/2})\tfrac{b}{2}+\beta_2(g,k,0)\left(\mathfrak{d}\tfrac{b}{2}-1\right)= \beta(g,0,i_{1/2}), \nonumber \nonumber
\eea
where the last inequality is simple to check distinguishing the usual cases.

This exhausts all possible graphs and shows that the colored graphs constructed in the \textit{first part} for each $(g,i_0,i_{1/2})$ achieve the minimal value for the exponent, and this value is given by $\beta(g,i_0,i_{1/2})$ of \eqref{betadef}.

\medskip

\begin{center}
\textit{Third part}
\end{center}

\medskip

We show by induction on $\chi = 2g - 2 + k \geq 1$ that $\mathcal{C}^{[g,k]}\bigl[{}^{l_1}_{\varepsilon_1}\,\cdots\,{}^{l_k}_{\varepsilon_k}\bigr]$ receives a power $(\frac{\pi}{T})^{- \sum_{i = 1}^k (2\ell_i + 1)}$ as prefactor. It is already correct for $(g,k) = (0,3)$ and $(1,1)$ according to Lemma~\ref{pieces}. If it is true for all $(\overline{g},\overline{k})$ such that $2\overline{g} - 2 + \overline{k} < \chi$,  then one easily checks with the recursive formula of Proposition~\ref{cocor1}, the behavior of $K$ and $\tilde{K}$, and the induction hypothesis that it continues to holds for all $(g,k)$ such that $2g - 2 + k = \chi$.\medskip

Together with the identification of the leading power of $q$ in the previous steps, this concludes the proof for the critical behavior of $\mathcal{C}^{[g,k]}$. The arguments are identical for $\mathsf{C}^{[g,k]}$.

\hfill $\Box$


\subsection{Generating series of configurations}
\label{SGNAP}

We arrive to the final result for the generating series of maps in the $O(\mathsf{n})$ model.
\begin{theorem}
\label{ouqusf} Let $k= k_0+k_{1/2} \geq 1$ and $g \geq 0$ such that $2g - 2 + k > 0$. Let $x_j = x(\tfrac{1}{2} + \tau \phi_j)$ for $j \in \{1,\ldots,k_{1/2}\}$, i.e. $x_j$ remains away from $[\gamma_-^*,\gamma_+^*]$. Let $y_j = x(\tau \psi_j)$
 for $j \in \{1,\ldots,k_0\}$, i.e. $y_j$ scales with $q \rightarrow 0$ such that $y_j - \gamma_+ \in O(q^{\frac{1}{2}})$. Then, we have in the critical regime $q \rightarrow 0$:
\bea
&& \mathcal{W}^{[g,k]}(x_1,\ldots,x_{k_{1/2}},y_1,\ldots,y_{k_0}) \nonumber \\
& = & \Big(\frac{\pi}{T}\Big)^{k} q^{(2g - 2 + k)(\mathfrak{d}\frac{b}{2}-1)-\frac{k}{2} + \frac{b + 1}{2}k_{1/2}}\Big( \mathcal{W}^{[g,k]}_{*}(\phi_1,\ldots,\phi_{k_{1/2}},\psi_1,\ldots,\psi_{k_0}) + O(q^{\frac{b}{2}})\Big), \nonumber
\eea
and for the generating series of usual maps with renormalized face weights:
\bea
&& \mathsf{W}^{[g,k]}(x_1,\ldots,x_{k_{1/2}},y_1,\ldots,y_{k_0}) \nonumber \\
& = & \Big(\frac{\pi}{T}\Big)^{k} q^{\widetilde{\beta}(g,k,k_{1/2})}\Big(\mathsf{W}^{[g,k]}_{*}(\phi_1,\ldots,\phi_{k_{1/2}},\psi_1,\ldots,\psi_{k_0}) + O(q^{\frac{b}{2}})\Big), \nonumber 
\eea
with $\widetilde{\beta}(g,k,k_{1/2})=(2g - 2 + k)(\mathfrak{d}\frac{b}{2}-1)-\frac{k}{2} + \frac{3}{4}k_{1/2}$. Recall that $\mathfrak{d} = 1$ in the dense phase and $\mathfrak{d} = -1$ in the dilute phase. In both cases, the errors are uniform for $\phi_j,\psi_j$ in any compact.
\end{theorem}
The result for $(g,k) = (0,2)$ is much easier to derive: this is done in Corollary~\ref{dgfsgg} below, and the outcome is that Theorem~\ref{ouqusf} is still valid for $(g,k) = (0,2)$. Remark that in this case, the first term in the critical exponent vanishes so the result is the same in the dense and dilute phase -- only the relation between $u$ and $q$ differ, according to Theorem~\ref{th38}.

\vspace{0.2cm}

\noindent \textbf{Proof.} First, we study the critical behavior of $\mathcal{G}^{[g,k]}(v_1,\ldots,v_k)$. From the decomposition of Proposition~\ref{2g2mr}, the critical behavior for its coefficients $\mathcal{C}^{[g,k]}$  from Lemma~\ref{Cbehavior} and the asymptotic behavior for $\mathcal{B}_{\varepsilon,l}(v)$ in the two regimes $v = \varepsilon' + \tau\phi$ with $\varepsilon'=0,\tfrac{1}{2}$ given in Lemma \ref{pieces}, it follows that each summand with $\varepsilon_1,\ldots,\varepsilon_k$ fixed behaves like $q^{\bar{\beta}(g,i_0,i_{1/2},j_0,j_{1/2}\vert b)}$, with
\beq
\label{bbar} \bar{\beta}(g,i_0,i_{1/2},j_0,j_{1/2}\vert b)=\beta(g,i_0,i_{1/2}\vert b) + (j_0+j_{1/2})\tfrac{b}{2}, 
\eeq
where $j_0+j_{1/2}=|\{j\in \{1,\ldots,k\} \,\,:\,\, \varepsilon_j \neq \varepsilon'_j\}|$ and, more concretely, 
\bea
j_{1/2} &\coloneqq &|\{ j\in \{1,\ldots,k_{1/2}\} \,\,:\,\, 0=\varepsilon_j \neq \varepsilon'_j=\tfrac{1}{2}\}|, \nonumber \\
j_0 &\coloneqq &|\{ j\in \{k_{1/2}+1,\ldots,k_{1/2}+k_0\} \,\,:\,\, \tfrac{1}{2}=\varepsilon_j \neq \varepsilon'_j=0\}|. \nonumber
\eea
Since we are interested in the dominant behavior of $\mathcal{G}^{[g,k]}\big((v_i)_{i = 1}^k\big)$ with fixed $k_0$ and $k_{1/2}$, we need to decide which $0\leq j_0\leq k_0$ and $0\leq j_{1/2}\leq k_{1/2}$ minimize $\bar{\beta}(g,i_0,i_{1/2},j_0,j_{1/2}\vert b)$. Observe that $i_0=k_0-j_0+j_{1/2}$ and $i_{1/2}=k_{1/2}-j_{1/2}+j_0$. We have to take into account the already known behavior of $\beta(g,i_0,i_{1/2}\vert b)$ varying $i_0$ and $i_{1/2}$ to find in the end the quadruple $(i_0,i_{1/2},j_0,j_{1/2})$ minimizing $\bar{\beta}(g,i_0,i_{1/2},j_0,j_{1/2}\vert b)$ for fixed $g, k_0$ and $k_{1/2}$.
We will consider first the special base cases, where some configurations $(i_0, i_{1/2})$ give vanishing~$\mathcal{C}$'s. 

\smallskip

\noindent $\bullet$ $(g,k) = (0,3)$.  Since the only configurations with $\mathcal{C}^{[g,k]}\neq 0$ are $(i_0, i_{1/2})=(3,0)$ and $(i_0, i_{1/2})=(0,3)$, we automatically have $(j_0, j_{1/2})=(0,k_{1/2})$ and $(j_0, j_{1/2})=(k_{0},0)$, respectively.
\beq
\bar{\beta}(0,3,0,0,k_{1/2}\vert b)=\beta(0,3,0\vert b)+k_{1/2}\tfrac{b}{2} \leq 0 \leq \beta(0,0,3\vert b)+k_{0}\tfrac{b}{2} = \bar{\beta}(0,0,3,k_0,0\vert b). \nonumber\eeq

\smallskip

\noindent $\bullet$ $(g,k) = (0,4)$. First, similarly to the previous case, we have
\beq
\bar{\beta}(0,4,0,0,k_{1/2}\vert b)=\beta(0,4,0\vert b)+k_{1/2}\tfrac{b}{2}  \leq 0 \leq \beta(0,0,4\vert b)+k_{0}\tfrac{b}{2} = \bar{\beta}(0,0,4,k_0,0\vert b). \nonumber
\eeq
The only possibility remaining to compare is $(i_0, i_{1/2})=(2,2)$. Here we use $\beta(0,2,2\vert b)=\beta(0,4,0\vert b)+1+(1-\mathfrak{d})\frac{b}{2}$ from Lemma ~\ref{betadecret}.
\begin{itemize}
\item[$\diamond$] $\bar{\beta}(0,2,2,2,k_{1/2}\vert b)=\beta(0,4,0\vert b)+1+(1-\mathfrak{d})\frac{b}{2}+(2+k_{1/2})\tfrac{b}{2} \geq \bar{\beta}(0,4,0,0,k_{1/2}\vert b)$.
\item[$\diamond$] $\bar{\beta}(0,2,2,1,k_{1/2}-1\vert b)=\beta(0,4,0\vert b)+1+(1-\mathfrak{d})\frac{b}{2}+k_{1/2}\tfrac{b}{2}\geq \bar{\beta}(0,4,0,0,k_{1/2}\vert b)$.
\item[$\diamond$] $\bar{\beta}(0,2,2,0,k_{1/2}-2\vert b)=\beta(0,4,0\vert b)+1+(1-\mathfrak{d})\frac{b}{2}+(k_{1/2}-2)\tfrac{b}{2}\geq \bar{\beta}(0,4,0,0,k_{1/2}\vert b)$.
\end{itemize}

\smallskip

Using again Lemma~\ref{betadecret}, observe that in all the remaining cases, for $i_0>0$, we have
$$
\beta(g,i_0,i_{1/2}\vert b)+\Delta=\beta(g,i_0-1,i_{1/2}+1\vert b),$$
with $\Delta \geq \frac{b}{2}$, except for $i_{1/2}=1$.

\smallskip

\noindent $\bullet$ We now justify that it is always better to decrease $j_0$. If $i_{1/2}\neq 1$,
\bea
\bar{\beta}(g,i_0-1,i_{1/2}+1,j_0,j_{1/2}\vert b)& = & \beta(g,i_0-1,i_{1/2}+1\vert b) + (j_0+j_{1/2})\tfrac{b}{2}\nonumber \\
& = & \beta(g,i_0,i_{1/2}\vert b)+\Delta + (j_0+j_{1/2})\tfrac{b}{2} \nonumber \\ 
& \geq &\beta(g,i_0,i_{1/2}\vert b) + (j_0+j_{1/2}-1)\tfrac{b}{2}\nonumber \\
&=& \bar{\beta}(g,i_0,i_{1/2},j_0-1,j_{1/2}\vert b). \nonumber
\eea
The equality still holds for $i_{1/2}=1$, more concretely
\beq
\bar{\beta}(g,i_0-1,2,j_0,j_{1/2}\vert b) =  \beta(g,i_0,1\vert b)-\tfrac{b}{2} + (j_0+j_{1/2})\tfrac{b}{2} = \bar{\beta}(g,i_0,1,j_0-1,j_{1/2}\vert b). \nonumber
\eeq

\smallskip

\noindent $\bullet$ Now, if $i_{1/2}\neq 1$, it is also better to increase $j_{1/2}$:
\bea
\bar{\beta}(g,i_0-1,i_{1/2}+1,j_0,j_{1/2}\vert b)& = & \beta(g,i_0,i_{1/2}\vert b)+\Delta + (j_0+j_{1/2})\tfrac{b}{2} \nonumber \\ 
& \geq &\beta(g,i_0,i_{1/2}) + (j_0+j_{1/2}+1\vert b)\tfrac{b}{2}\nonumber \\
&=& \bar{\beta}(g,i_0,i_{1/2},j_0,j_{1/2}+1\vert b). \nonumber
\eea
And if $i_{1/2}=1$, it is better to increase $j_{1/2}$ by $2$:
\bea
\bar{\beta}(g,k-2,2,j_0,j_{1/2}\vert b)& = & \beta(g,k,0\vert b)-\tfrac{b}{2}-\left(\mathfrak{d}\tfrac{b}{2}-1\right) + (j_0+j_{1/2})\tfrac{b}{2} \nonumber \\ 
& \geq &\beta(g,k,0) + b + (j_0+j_{1/2}\vert b)\tfrac{b}{2}\nonumber \\&=& \bar{\beta}(g,k,0,j_0,j_{1/2}+2\vert b). \nonumber
\eea
Observe that in the key case $i_{1/2}=1$, if we have $j_{1/2}=k_{1/2}-1$, i.e. not the maximum but with no possibility of being increased by $2$, we will always have $j_0=1$ $(i_{1/2}+1=2)$ and if we decrease that before we will not lie in the case $i_{1/2}=1$ anymore. So this pathological case is not a real problem.

\smallskip

Therefore, the minimal exponent corresponds to the minimum $j_0$ and the maximum $j_{1/2}$, i.e. $j_0=0$ and $j_{1/2}=k_{1/2}$, and $i_0=k$ and $i_{1/2}=0$:
\beq
\bar{\beta}(g,k,0,0,k_{1/2}\vert b)=\beta(g,k,0\vert b) + k_{1/2}\tfrac{b}{2}=(2g - 2 + k)\left(\mathfrak{d}\tfrac{b}{2}-1\right)+ k_{1/2}\tfrac{b}{2}. \nonumber
\eeq
The final result follows from
$$
\mathcal{W}^{[g,k]}(x(v_1),\ldots,x(v_k)) \Big[\prod_{i  = 1}^k x'(v_i)\Big]= \mathcal{G}^{[g,k]}(v_1,\ldots,v_k),
$$
the critical behavior we just found for $\mathcal{G}^{[g,k]}(v_1,\ldots,v_k)$ and the asymptotic behavior for $x(v)$ in the two regimes $v = \tau\phi$ and $v = \tfrac{1}{2} + \tau\phi$ given in Appendix~\ref{App1}. The resulting power of $q$ is
\bea
&& (2g - 2 + k)\left(\mathfrak{d}\tfrac{b}{2}-1\right)+k_{1/2}\tfrac{b}{2}-\tfrac{1}{2} k_0 \nonumber \\
& = & (2g - 2 + k_0 + k_{1/2})\left(\mathfrak{d}\tfrac{b}{2}-1\right)-\tfrac{k_0 + k_{1/2}}{2} + k_{1/2}\,\tfrac{b + 1}{2}. \nonumber
\eea

For the $\mathsf{W}$'s, the only differences compared to \eqref{bbar} are the factor $\frac{1}{4}$ instead of $\frac{b}{2}$ in the total exponent for fixed $j_0,j_{1/2}$ and $B=\frac{1}{2}$ instead of $b$ in $\beta$:
$$
\bar{\beta}(g,i_0,i_{1/2},j_0,j_{1/2}\vert \tfrac{1}{2})=\beta(g,i_0,i_{1/2}\vert\tfrac{1}{2}) + (j_0+j_{1/2})\tfrac{1}{4}.
$$
This is a particular case of the previous analysis, so the minimum of this exponent is again reached when $j_0 = 0$ and $j_{1/2} = k_{1/2}$, and $i_0=k$ and $i_{1/2}=0$, and this entails the claim. Since in this case, $\beta(g,k,0\vert b)=\beta(g,k,0\vert\frac{1}{2})$, in the end only the first difference matters.
\hfill $\Box$


\subsection{Generating series of configurations with marked points}

\label{crititmarked}

We now generalize Theorem~\ref{ouqusf} to allow marked points. 
\begin{lemma}
Let $k = k_0 + k_{1/2} \geq 1$ and $g \geq 0$ such that $(g,k) \neq (0,1)$. Let $x_j = x(\tfrac{1}{2} + \tau \phi_j)$ for $j \in \{1,\ldots,k_{1/2}\}$, i.e. $x_j$ remains finite and away from $[\gamma_-^*,\gamma_+^*]$. Let $y_j = x(\tau \psi_j)$
 for $j \in \{1,\ldots,k_0\}$, i.e. $y_j$ scales with $q \rightarrow 0$ such that $y_j - \gamma_+ \in O(q^{\frac{1}{2}})$. We have in the critical regime $q \rightarrow 0$
\bea
&& \mathsf{W}^{[g,k,\bullet k']}(x_1,\ldots,x_{k_{1/2}},y_{1},\ldots,y_{k_0}) \nonumber \\
& = & \Big(\frac{\pi}{T}\Big)^{k+k'} q^{\widetilde{\beta}(g,k + k',k_{1/2} + k')}\big\{\mathsf{W}^{[g,k,\bullet k']}_*(\phi_1,\ldots,\phi_{k_{1/2}},\psi_{1},\ldots,\psi_{k_0}) + O(q^{\frac{b}{2}})\big\}. \nonumber
\eea
This is also true for $(g,k) = (0,1)$.
\end{lemma} 
\vspace{0.2cm}
The outcome is that marked points behave as small boundaries. Subsequently, the asymptotics of the generating series $\mathscr{W}_{\Gamma,\star,\mathbf{s}}^{[g,k]}$ given by Proposition~\ref{P212} in presence of $k'$ marked points are the same as obtained in Theorem~\ref{CoscrF}, provided one replaces $k_{1/2}$ with $k_{1/2} + k'$, and likewise for Theorem~\ref{LAPA} concerning fixed volume asymptotics, and Theorem~\ref{igfsgb} concerning fixed volume and fixed arm lengths asymptotics.

\vspace{0.2cm}

\noindent \textbf{Proof.} First assume $(g,k) \neq (0,1)$. We proceed by recursion, starting from the base case $k' = 0$ obtained in Theorem~\ref{ouqusf}:
$$
\mathsf{W}^{[g,k]}(\mathbf{x},\mathbf{y}) = q^{\widetilde{\beta}(g,k,k_{1/2})}\,\Phi\Big[u;(x_i)_{i = 1}^{k_{1/2}};\Big(\frac{y_i - \gamma_+^*}{q^{\frac{1}{2}}}\Big)_{i = 1}^{k_0}\Big],
$$
where $\Phi$ is a function which has a uniform limit when $u \rightarrow 1$ and its other variables remain in a compact, and
\beq
\widetilde{\beta}(g,k,k_{1/2}) = (2g - 2 + k)(\mathfrak{d}\tfrac{b}{2} - 1) - \tfrac{k}{2} + \tfrac{3}{4}k_{1/2}.
\eeq
We shall use \eqref{renFmarked} to decrease the value of $k'$. Assume the claim holds for $k''$ marked points with $k'' < k'$. Equation \eqref{renFmarked} gives us
\bea
\mathsf{W}^{[g,k,\bullet k']}(\mathbf{x},\mathbf{y}) & = & \Big(2 - 2g - k - \sum_{i = 1}^{k_{1/2}} \tfrac{1}{2}\,\partial_{x_i} x_i - \sum_{i = 1}^{k_{0}} \tfrac{1}{2}\,\partial_{y_i}y_i\Big)\mathsf{W}^{[g,k,\bullet (k' - 1)]}(\mathbf{x},\mathbf{y}) \nonumber \\
\label{tehrer} & & - \oint_{\gamma} \frac{\dd z}{2{\rm i}\pi}\,\big(\tfrac{z}{2}\tilde{V}'(z) - \tilde{V}(z)\big)\mathsf{W}^{[g,k + 1,\bullet (k' - 1)]}(z,\mathbf{x},\mathbf{y}), 
\eea
with
\bea 
\tilde{V}'(x) & = & V'(x) - \oint_{\gamma} \mathbf{A}(x,z)\mathcal{W}(z) \nonumber \\
& = & V'(x) + \mathsf{n}\varsigma'(x)\mathcal{W}(\varsigma(x)) - \frac{\mathsf{n} u\varsigma''(x)}{2\varsigma'(x)}. \nonumber
\eea
We can substitute in this expression the function $\mathcal{G}$ introduced in \eqref{EquationG}:
\beq
\tilde{V}'(x) = V'(x) -\mathsf{n}\frac{\mathcal{G}(\tau - v)}{x'(v)} + \frac{\mathsf{n}\big(2\varsigma'(x)V'(\varsigma(x)) + \mathsf{n} V'(x)\big)}{4 - \mathsf{n}^2} - \frac{\mathsf{n}u\varsigma''(x)}{\varsigma'(x)}. \nonumber
\eeq
The critical behavior of $\mathcal{G}(v)$ when $v = \varepsilon + \tau w$ with $\varepsilon \in \{0,\tfrac{1}{2}\}$, and $q = e^{{\rm i}\pi \tau} \rightarrow 0$ is obtained from substituting its expression from Proposition~\ref{theimdisk}, using the asymptotics of the function $\Upsilon_b$ in \ref{lemUp}, and the identities \eqref{D1}-\eqref{D2}. The result takes the form
$$
\mathcal{G}(\tau - v) = q^{(1 - 2\varepsilon)(1 - \mathfrak{d}\frac{b}{2})}\big\{\tilde{\mathcal{G}}^*_{\varepsilon}(\phi) + O(q^{\frac{b}{2}})\big\}.
$$ 
Besides, the induction hypothesis tells us that the order of magnitude of$$
\mathsf{W}^{[g,k + 1,\bullet (k' - 1)]}(x(v),\mathbf{x},\mathbf{y})
$$
receives an extra factor of $q^{\frac{3}{4}}$ when $v = \frac{1}{2} + \tau\phi$ with $\phi$ in a compact. As $b \in (0,\tfrac{1}{2})$, in any case we have $\tfrac{3}{4} < 1 - \mathfrak{d}\tfrac{b}{2}$ and therefore the contribution of the vicinity (at scale $q^{\frac{1}{2}}$) of $\gamma_+^*$ in the contour integral over $\gamma$ in the second line of  \eqref{tehrer} remains negligible compared to the contribution of the bulk of the contour (given by the regime $\varepsilon = \tfrac{1}{2}$). And, by induction hypothesis, this contribution is of order $q^{\widetilde{\beta}(g,(k + k' - 1) + 1,(k_{1/2} + k' - 1) + 1)}$, where the $+1$ come from the variable $z \in \gamma$. On the other hand, the first line in \eqref{tehrer} has a contribution of order $q^{\widetilde{\beta}(g,k + k' - 1,k_{1/2} + k' - 1)}$. As 
$$
\widetilde{\beta}(g,k + k' - 1,k_{1/2} + k' - 1) - \widetilde{\beta}(g,k + k',k_{1/2} + k') = \tfrac{1}{2} - \mathfrak{d}\tfrac{b}{2}  > 0
$$
the first line is always negligible compared to the second line, and this gives the claim for $k'$ marked points. We conclude for all $(g,k) \neq (0,1)$ by induction.
 
 \vspace{0.2cm}
 
Now consider $(g,k) = (0,1)$. For $k' = 1$, we have from \eqref{diskpointed}:
$$
\mathsf{W}^{\bullet}(x) = \frac{1}{\sqrt{(x - \gamma_+)(x - \gamma_-)}},
$$
Therefore with $x = x(\tau \phi) = \gamma_+^* + q^{\frac{1}{2}}x_0^*(\phi)$ in the critical regime
$$
\mathsf{W}^{\bullet}(x) \sim q^{-\frac{1}{4}}\,\mathsf{W}^{\bullet}_{*}(\phi),
$$
whose exponent agrees with  $\widetilde{\beta}(g,k + k' = 2,k_{1/2} + k' = 1)$. On the other hand, for $x = x(\tfrac{1}{2} + \tau\phi)$ in the critical regime, we have
$$
\mathsf{W}^{\bullet}(x) = \frac{1}{\sqrt{(x - \gamma_+^*)(x - \gamma_-^*)}} + q^{\frac{1}{2}}\,\tilde{\mathsf{W}}^{\bullet}_{*}(\phi) + O(q)
$$
coming from the behavior of $\gamma_+$ when $q \rightarrow 0$ as given by Corollary~\ref{CoB5}. This exponent $\frac{1}{2}$ agrees with $\widetilde{\beta}(g = 0,k + k' = 2,k_{1/2} + k' = 2)$. With these two cases as initial conditions and the previous results, we can repeat the previous steps to show from \eqref{tehrer} that the claim holds for $(g,k) = (0,1)$ for any $k' > 0$. \hfill $\Box$

\section{Critical behavior of the TR invariants}\label{GeneralizationCriticality}

Our analysis of the critical behavior of the topological recursion amplitudes for the bending energy model is in fact more general than the $O(\mathsf{n})$ loop model and it may be used to study the critical behavior of other problems in enumerative geometry. Here we summarize the initial conditions we need and give the result in general. This generalization is not present in the article in which this part of the thesis is based on \cite{BGF16}.

Let us consider a spectral curve that has two types of branchpoints: dominant singularities and non singular ones. For simplicity, we can assume it has one branchpoint of each type: $a_1$ will be the dominant singularity and $a_2$ the regular branchpoint. If the branchpoints are only of one type, the analysis also holds, but it is much simpler. The really complicated case which will use the technical analysis we performed in the particular case of the bending energy model corresponds only to the case with both kinds of branchpoints. Moreover, having more than one dominant singularity or non singular branchpoint only modifies the constant prefactor. In any case, here we focus on the critical exponent, which will be the same.

The setting is a family of regular spectral curves parametrized by a complex parameter $u$, which remains always regular around $a_2$, in the sense we explained in Section \ref{TRIntro}. We remind the reader that a regular spectral curve looks locally like a square root around the critical points.  Our family of spectral curves will become singular around the dominant singularity $a_1$, when the parameters are tuned to the critical values.

In order to study large size asymptotics we always have to study the generating series around a singularity.
We assume here the system is at a critical point, i.e. a certain parameter $u_c$ above which the spectral curve $y(x)$ is singular is equal to $1$. So the spectral curve $y(x)$ has a singularity at $u=1$. We will consider a parameter $\delta$ on the spectral curve that will control the distance to criticality (when it becomes small) in order to study the critical behavior of the TR amplitudes $\omega_n^{[g]}$, where every variable will be either in a small region of size $\delta$ around the singularity $a_1$ of $\omega_1^{[0]}$ or away from $a_1$.
Let $\hat{x}_1\coloneqq x(a_1)$ and $\hat{y}_1\coloneqq y(a_1)$. We expand $x(z)$ and $y(z)$ around $a_1$, considering $z=a_1+\delta\phi$, and find:
$$
\begin{cases}
x(z) = \hat{x}_1+  \delta^{\mathsf{q}} x_*(\phi) + O(\delta^{\mathsf{q}+1}), \text{ with } \mathsf{q}\geq 2 \\
y(z)= \hat{y}_1+  \delta^{\mathsf{p}} y_*(\phi) + O(\delta^{\mathsf{p}+1}).
\end{cases}
$$
The curve $(x_*,y_*)$ is called the blow up of the curve $(x,y)$. Since we assumed that $\hat{x}_1$ was a dominant singularity of $y(x)$, we have
$$
\left.y(x)\right|_{\rm sing} \stackrel{\bigcdot}{\sim} (x-\hat{x}_1)^{\frac{\mathsf{p}}{\mathsf{q}}},
$$
with $\frac{\mathsf{p}}{\mathsf{q}}\notin\mathbb{Z}$ and $\frac{\mathsf{p}}{\mathsf{q}}>\frac{1}{2}$. We remark that $\frac{\mathsf{p}}{\mathsf{q}}$ is not necessarily a rational number. On the other hand, for the non singular branchpoint, we would have the behavior $\left.y(x)\right|_{\rm sing}\stackrel{\bigcdot}{\sim} (x-\hat{x}_1)^{\frac{1}{2}}$. We control the distance to the singularity by finding the critical behavior of delta near the singularity: $\delta\stackrel{\bigcdot}{\sim}(1-u)^{\theta}$ with $\theta > 0$ for delta the zooming variable around the singularity and 
$\delta\stackrel{\bigcdot}{\sim}1$ for delta the zooming size around any other point. We remark that adding a holomorphic function of $x$ to $y$ does not change the result of TR, and that is the reason why we present the general setting in this way\footnote{Sometimes for the enumerative problem to have the correct meaning in lower topologies, such a function may be necessary.}.

We consider two possible regimes for the variable $z$:
\begin{enumerate}
\item It scales with $(1-u)\rightarrow 0$ such that $z-a_1\in O((1-u)^{\theta})$, i.e. it approaches the singularity. 
\item It remains away from the singularity $a_1$. 
\end{enumerate}

Our goal is to find the critical exponent of the TR amplitudes $\omega_n^{[g]}(z_1,\ldots,z_k)$ with $k_L=k_L(z_1,\ldots,z_k)$ being the number of $z$'s in the first regime and $k_S=k_S(z_1,\ldots,z_k)$ the number of $z$'s in the second regime. We always have $k_L+k_S=k$.

We need the critical behavior of the following initial data:
\bea
\dd x (z) & \stackrel{\bigcdot}{\sim} & (1-u)^{\alpha_x^{[k_L(z)]}}, \nonumber \\
\omega_1^{[0]}(z) = y(z) \dd x(z) & \stackrel{\bigcdot}{\sim} & (1-u)^{\alpha^{[0,1,k_L(z)]}}, \nonumber \\
B(z_1,z_2) & \stackrel{\bigcdot}{\sim} &
(1-u)^{\alpha^{[0,2,k_L(z_1,z_2)]}}, \nonumber
\eea
where $k_L$ is here the number of $z$'s in the first regime for every initial piece.

With the scaling exponents we introduced, the initial critical exponents for this general TR problem are, when $u\rightarrow 1$:
\bea
\alpha_x^{[k_L(z)]} &=& \begin{cases} \mathsf{q}\,\theta &, \text{ if } k_L=1,  \\
0 &, \text{ if } k_L=0, \nonumber
\end{cases} \\
\alpha^{[0,1,k_L(z)]} &=& \begin{cases} (\mathsf{p}+\mathsf{q})\theta &, \text{ if } k_L=1,  \\
0 &, \text{ if } k_L=0, \nonumber
\end{cases} \\
\alpha^{[0,2,k_L(z_1,z_2)]} &=& \begin{cases} \theta &, \text{ if } k_L=1,  \\
0 &, \text{ if } k_L=0,2. \nonumber
\end{cases} \\
\eea

Now we can generalize the function $f$ we introduced in \eqref{fCrit}, which will determine the critical behavior of all the building pieces of the recursion formula we found, which is equivalent to the classical recursion formula of TR using residue computations, as in Lemma \ref{pieces}:
$$
f(z_1,z_2,z_3)\coloneqq \alpha^{[0,2,k_L(z_1,z_2)]}+ \alpha^{[0,2,k_L(z_1,z_3)]}- \alpha^{[0,1,k_L(z_1)]}.
$$
We illustrate its table of values:
\begin{center}
\begin{tabular}{|c||c|c|c|}
\hline $\bs{k_L(z_2)+k_L(z_3)}$ & $\bs{2}$ & $\bs{1}$ & $\bs{0}$ \\
\hline\hline
$\bs{k_L(z_1) = 1}$ & $-(\mathsf{p}+\mathsf{q})\theta$ & $(1-(\mathsf{p}+\mathsf{q}))\theta$ & $(2-(\mathsf{p}+\mathsf{q}))\theta$ \\
\hline
$\bs{k_L(z_1) = 0}$ & $2\theta$ & $\theta$ & $0$ \\
\hline
\end{tabular}
\end{center}

Observe that we always have $2-(\mathsf{p}+\mathsf{q})\leq 0$ and $\alpha^{[0,2,1]}=\theta >0$. So the $6$ possible values compare in the same way as for the bending energy model and all our arguments in the proofs of the technical Lemma \ref{Cbehavior} and the final Theorem \ref{ouqusf} work for this general case as well.

Therefore, we are ready to give the critical exponent of the TR amplitudes $\omega_n^{[g]}(z_1,\ldots,z_k)$ with $k_L$ $z$'s in the first regime and $k_S$ $z$'s with the second type of behavior. Let $k=k_L+k_S\geq 1$ and $g\geq 0$ such that $2g-2+k>0$. Then, we have in the critical regime $u\rightarrow 1$:
\beq\label{TRbehavior}
\frac{\omega_n^{[g]}(z_1,\ldots,z_k)}{\dd x(z_1)\cdots \dd x(z_k)} \stackrel{\bigcdot}{\sim} (1-u)^{\beta^{[g,k,k_L]}},
\eeq
with $\beta^{[g,k,k_L]}=(2-2g-k)\alpha^{[0,1,1]}-k_L\alpha_x^{[1]}+k_S\alpha^{[0,2,1]}= (2-2g-k)(\mathsf{p}+\mathsf{q})\theta-k_L\mathsf{q}\,\theta+k_S\theta$.
We remark that this general critical exponent still shows an affine dependence on the Euler characteristic associated to the corresponding correlator: $\chi= 2-2g-k$.

\subsection{Ordinary, usual maps (pure gravity)}

Now we are going to apply this general result to the case of usual maps, and more concretely to the universality class of pure gravity. Here we refer the reader to \cite[Chapter 5]{Eynardbook}, where the critical behavior for usual maps with $k=k_L$, i.e., with all the variables close to the singularity, which in this setting we know corresponds to the lengths of the boundaries going to infinity, is already given and one can find the initial conditions we will need now. Here we generalize the result from \cite[Chapter 5]{Eynardbook} that gives the critical behavior for maps with all boundaries in the large regime $(k_L,k_S)=(k,0)$ to the critical behavior of maps with large and small boundaries, i.e. to all possible intermediate configurations $(k_L,k_S)$. We remark that in this setting the spectral curve is rational and comes with a natural uniformizing coordinate $z$.

In the case of pure gravity, considering for example triangulations, one has $\mathsf{q}=2, \mathsf{p}=3$ and $\theta=\frac{1}{4}$. Therefore, for this case we obtain the following critical behavior for the generating series of usual maps when $u\rightarrow 1$:
\beq
W_k^{[g]}(x(z_1),\ldots,x(z_k)) \stackrel{\bigcdot}{\sim} (1-u)^{\beta^{[g,k,k_L]}},
\eeq
with
\beq
\beta^{[g,k,k_L]}=(2-2g-k)\frac{5}{4}-\frac{k_L}{2}+\frac{k_S}{4}.
\eeq
Note that this critical exponent coincides with the one we found for the bending energy model specialized to $b=\frac{1}{2}$ (which corresponds to $\mathsf{n}=0$) in the dilute phase ($\mathfrak{d}=-1$), i.e. when we consider the generic phase of the $O(\mathsf{n})$ loop model, which coincides with pure gravity.

The case of general usual maps with large boundaries is treated in \cite[Chapter 5]{Eynardbook} in a context a bit different from ours in which the Boltzmann weights are allowed to take negative values. In that context, one cannot define a probability measure to rigorously define random maps, as we did in \eqref{probMeasure}, and more pairs $(\mathsf{p},\mathsf{q})$ are possible. Actually, in the space of the parameters $t_l$'s, there exist critical submanifolds which contain the various singular behaviors of the spectral curves $(x,y)$ such that $y\stackrel{\bigcdot}{\sim}(x-\hat{x}_1)^{\frac{\mathsf{p}}{\mathsf{q}}}$, with $\mathsf{q}=2$ and $\mathsf{p}=2m+1$ for $m\in\mathbb{Z}_{\geq 1}$. The case $\frac{\mathsf{p}}{\mathsf{q}}>\frac{3}{2}$ corresponds to multicritical points for which more than one $t_l$ is set to criticality and more derivatives of $y$ vanish (in the critical non-multicritical case, we had just $y'(a_1)=0$ and $y''(a_1)\rightarrow\infty$).  Moreover, it is computed that
$$
\theta = \frac{1}{\mathsf{p}+\mathsf{q}-1}.
$$ 
In this case, the critical exponent when $u\rightarrow 1$ reads:
\beq
\beta^{[g,k,k_L]}=(2-2g-k)\frac{\mathsf{p}+\mathsf{q}}{\mathsf{p}+\mathsf{q}-1}-\frac{k_L\mathsf{q}}{\mathsf{p}+\mathsf{q}-1}+\frac{k_S}{\mathsf{p}+\mathsf{q}-1}.
\eeq

In the setting we considered in Section \ref{largeRandom}, maps decorated by an Ising model allow to reach any rational $\frac{\mathsf{p}}{\mathsf{q}}$ singularity.

\subsection{Fully simple maps}

Let $(\mathcal{S},x,y)$ be the spectral curve of a TR problem. Now we briefly study the critical behavior of the TR amplitudes for the exchanged spectral curve $(\mathcal{S},y,x)$, assuming that dominant singularities for the exchanged curve approach the ones of the original curve. We know this happens in the case of usual maps, which is the setting we are interested in illustrating here. For a general argument on any family of exchanged curves, one would certainly need a further analysis which may require further assumptions on the family. More concretely, if our Conjecture \ref{conj} is true, we find the critical behavior for fully simple maps with large and small boundaries.

The spectral curve had the following general behavior around a singularity $\hat{x}_1$:
$$
y(x)=\hat{y}_1+K_1(x-\hat{x}_1)+K_2(x-\hat{x}_1)^{\frac{\mathsf{p}}{\mathsf{q}}}+O((x-\hat{x}_1)^{\frac{\mathsf{p}}{\mathsf{q}}}).
$$
Therefore, the exchanged spectral curve will behave as 
$$
\left.x(y)\right|_{\rm sing} \stackrel{\bigcdot}{\sim} (y-\hat{y}_1)^{\mathsf{r}}, \quad \text{with }\quad \mathsf{r}=\begin{cases}
\frac{\mathsf{q}}{\mathsf{p}}, & \text{ if }\; 0< \frac{\mathsf{p}}{\mathsf{q}} < 1, \\
\frac{\mathsf{p}}{\mathsf{q}}, & \text{ if }\; 1< \frac{\mathsf{p}}{\mathsf{q}} < 2.
\end{cases}
$$

Then, the critical exponent of TR amplitudes for the exchanged spectral curve $(\mathcal{S},y,x)$ when $u\rightarrow 1$ remains as before \eqref{TRbehavior} for $1< \frac{\mathsf{p}}{\mathsf{q}} < 2$ and reads:
\beq
\beta^{[g,k,k_L]}=(2-2g-k)(\mathsf{p}+\mathsf{q})\theta-k_L\mathsf{p}\,\theta+k_S\theta, \text{ for } 0< \frac{\mathsf{p}}{\mathsf{q}} < 1.
\eeq

Thus, assuming the Conjecture \ref{conj} where $y\equiv w$, we obtain the critical exponent for fully simple maps when $u\rightarrow 1$:
\beq
X_k^{[g]}(w(z_1),\ldots,w(z_k)) \stackrel{\bigcdot}{\sim} (1-u)^{\beta^{[g,k,k_L]}},
\eeq
with
\beq
\beta^{[g,k,k_L]}=(2-2g-k)\frac{\mathsf{p}+\mathsf{q}}{\mathsf{p}+\mathsf{q}-1}-\frac{k_L\mathsf{q}}{\mathsf{p}+\mathsf{q}-1}+\frac{k_S}{\mathsf{p}+\mathsf{q}-1}.
\eeq

And more concretely, for the ``exchanged pure gravity'', for example for triangulations, we obtain that the critical behavior of the generating series of fully simple disks $X(w)$ (identifying $y\equiv w$ and $x(y)\equiv X(w)$) remarkably remains unchanged in comparison to ordinary disks:
$$
x(y)=\hat{x}_1+\left(\frac{y-\hat{y}_1}{K_1}\right)^{\frac{3}{2}}+O((y-\hat{y}_1)^{\frac{3}{2}}),
$$
and, equally, for higher topologies:
\beq
\beta^{[g,k,k_L]}=(2-2g-k)\frac{5}{4}-\frac{k_L}{2}+\frac{k_S}{4}.
\eeq

It would be interesting to perform a more refined study of the critical behavior of fully simple maps in the future.

%
%
%
%
\chapter{Application: Loop nesting}
\label{LoopNesting}




\section{Critical behavior of nestings in the bending energy model}
\label{S6}


Our first main goal here is to determine the behavior of the generating series $\mathscr{W}$ of maps realizing a given nesting graph $\Gamma$, without remembering  the arm lengths -- i.e. setting $s(\mathsf{e}) = 1$ -- and in absence of marked points. For this purpose, we perform a saddle point analysis of the expression of Proposition~\ref{P212} using the previous results on the behavior of $\mathsf{W}$, and of the generating series of cuffed cylinders $\hat{\mathcal{W}}_s^{[0,2]}$ and $\tilde{\mathcal{W}}_{s}^{[0,2]}$ in Section~\ref{armsS}. The final result is Theorem~\ref{CoscrF} below. The second goal is to extend these computations to the refined generating series $\mathscr{W}_{\Gamma,\star,\mathbf{s}}^{[g,k]}$ of maps realizing a given nesting graph. Here, we just need to repeat the computations of our first goal in presence of the variable $s$, which roughly amounts to replacing $b$ by $b(s)$ when necessary. The only important difference is that we wish to extract the leading contribution containing the dominant singularity in the variable $s$, and this sometimes brings some modification to the hierarchy of dominant terms. The result is described in Section~\ref{CoscrFs}.

In Section~\ref{Fixedas} we convert the critical behavior of all those generating series into asymptotics for fixed large volume $V$ and fixed boundary perimeters $(L_i)_i$ in the regime of small or large boundaries. In Section~\ref{Fixedasarms}, we also examine the critical behavior in this setting of the probability of having fixed arm lengths $P(\mathsf{e})$ tending to $\infty$ at rate $\ln V$ -- which naturally appears from the analysis. In particular, we compute the large deviation function for the arm lengths.

Finally, in Section~\ref{crititmarked}, we show that all these results continue to be valid in presence of marked points, provided one treats each marked point as a small boundary.

\subsection{Cylinders and cuffed cylinders}

\label{armsS}

In order to derive the critical behavior of $\mathscr{W}_{\Gamma,\star,\mathbf{s}}^{[g,k]}$, we need one more ingredient, namely the critical behaviors of $\tilde{\mathcal{W}}_{s}^{[0,2]}(x_1,x_2)$ and $\hat{\mathcal{W}}_{s}^{[0,2]}(x_1,x_2)$.

For this purpose, we first derive the critical behavior of $\mathcal{G}_{s}^{[0,2]}$ in the various regimes, which can be straightforwardly obtained using the expression in Proposition~\ref{p15} together with the asymptotic behavior of the special function $\Upsilon_b$ in Lemma~\ref{lemUp} in Appendix.

\begin{lemma}\label{behaviorG2}
Set $v_i = \varepsilon_i + \tau w_i$ for $\varepsilon_i \in \{0,\tfrac{1}{2}\}$. In the limit $q \rightarrow 0$, we have
\bea
\mathcal{G}_{s}^{[0,2]}(v_1,v_2) & = & \frac{(\frac{\pi}{T})^2}{4 - \mathsf{n}^2s^2}\,\frac{q^{(\varepsilon_1 \oplus \varepsilon_2)b(s)}}{1 - q^{b(s)}} \nonumber \\
&& \times \left\{\begin{array}{lll} H_{b(s),0}(w_1,w_2) - q^{b(s)}H_{b(s) + 2,0}(w_1,w_2) + O(q^{2 - b(s)}) & & {\rm if}\,\,\varepsilon_1 = \varepsilon_2, \\ H_{b(s),\frac{1}{2}}(w_1,w_2) - q^{1 - b(s)}H_{b(s) - 2,\frac{1}{2}}(w_1,w_2) + O(q) && {\rm if}\,\,\varepsilon_1 \neq \varepsilon_2, \end{array}\right. \nonumber
\eea
where
\bea
H_{b,0}(w_1,w_2) & = & (b - 1)\Big(\frac{\sin \pi(b - 1)(w_1 + w_2)}{\sin \pi(w_1 + w_2)} - \frac{\sin \pi(b - 1)(w_1 - w_2)}{\sin \pi(w_1 - w_2)}\Big) \nonumber \\
&& + \frac{\cos \pi(w_1 + w_2)\cos\pi (b - 1)(w_1 + w_2)}{\sin^2 \pi(w_1 + w_2)} \nonumber \\
&& - \frac{\cos\pi (w_1 - w_2) \cos \pi(b - 1)(w_1 - w_2)}{\sin^2\pi(w_1 - w_2)}, \nonumber \\
H_{b,\frac{1}{2}}(w_1,w_2) & = & 8b\,\sin\pi b w_1\,\sin\pi b w_2. \nonumber
\eea
The errors are uniform for $w_1,w_2$ in any compact and stable under differentiation.
\end{lemma}

The first consequence of this Lemma is the critical behavior of the ``singular part'' of $\mathcal{W}_{s}^{[0,2]}(x_1,x_2)$ with respect to the variables $u$ and $(x_1,x_2)$, which will be used in Theorem~\ref{LAPB} to obtain the asymptotics of the cylinder generating series for fixed large volumes and fixed boundary perimeters. We warn the reader about two subtleties in this analysis regarding what we mean by this ``singular part''. $\mathcal{W}_{s}^{[0,2]}$ is directly expressed in terms of $\mathcal{G}_{s}^{[0,2]}$ in Proposition~\ref{p15} up to a shift term. This shift term can actually be dropped as far as fixing boundary perimeter is concerned, as it gives a zero contribution when performing contour integrations of the form $\oint \frac{\dd x_1\,x_1^{L_1}}{2{\rm i}\pi}\,\frac{\dd x_2\,x_2^{L_2}}{2{\rm i}\pi}\,\mathcal{W}_{s}^{[0,2]}(x_1,x_2)$. Powers $q^0$ should also be dropped from this ``singular term'' as they disappear in contour integrals $\oint \frac{\dd u}{2{\rm i}\pi u^{V + 1}}\mathcal{W}_{s}^{[0,2]}$ used to fix the volume; in such a case, the next-to-leading order will play the leading role in the computations for fixed volume. Taking these subtleties into account, the result for this ``singular part'' of $\mathcal{W}_{s}^{[0,2]}$ straightforwardly follows from Lemma~\ref{behaviorG2} and the behavior of $x(v)$ given in Lemma~\ref{LemB3} from the Appendix:

\begin{corollary}
\label{dgfsgg} Set $x_i = x(\varepsilon_i + \tau\phi_i)$ for $\varepsilon_i \in \{0,\tfrac{1}{2}\}$. In the limit $q \rightarrow 0$, the singular parts (for this we use the sign $\equiv$) with respect to the variables $u$, $x_1,x_2$ of the cylinder generating series are
\bea 
\mathcal{W}^{[0,2]}_{s}(x_1,x_2) & \equiv & \frac{(\frac{\pi}{T})^2}{4 - \mathsf{n}^2s^2} \frac{q^{\widetilde{\beta}^{(0,2)}(s,\varepsilon_1,\varepsilon_2)}}{1 - q^{b(s)}}\big\{\mathcal{W}^{[0,2]}_{s\,*}(\phi_1,\phi_2) + q^{b(s)}\mathcal{W}^{[0,2]}_{s\,**}(\phi_1,\phi_2) + O(q^{2b(s)})\big\}, \nonumber \\
\mathsf{W}^{[0,2]}(x_1,x_2) & \equiv & \left(\frac{\pi}{2T}\right)^2 \frac{q^{\widetilde{\beta}^{(0,2)}(1,\varepsilon_1,\varepsilon_2)}}{1 - q^{\frac{1}{2}}}\big\{\mathcal{W}^{[0,2]}_{s = 0\,*}(\phi_1,\phi_2) + q^{\frac{1}{2}}\,\mathcal{W}^{[0,2]}_{s = 0\,**}(\phi_1,\phi_2) + O(q)\big\}, \nonumber 
\eea
where 
\beq
\label{betti} \widetilde{\beta}^{(0,2)}(s,\varepsilon_1,\varepsilon_2) = \left\{\begin{array}{lll} -1 & & {\rm if}\,\,\varepsilon_1 =  \varepsilon_2 = 0, \\ \tfrac{b(s) - 1}{2} && {\rm if}\,\,\varepsilon_1 \neq \varepsilon_2, \\ b(s) & & {\rm if}\,\,\varepsilon_1 = \varepsilon_2 = \tfrac{1}{2},\end{array}\right.
\eeq
$$
\mathcal{W}_{s\,*}^{[0,2]} = \frac{H_{b(s),\varepsilon_1 \oplus \varepsilon_2}(\phi_1,\phi_2)}{(x_{\varepsilon_1}^*)'(\phi_1)(x_{\varepsilon_2}^*)'(\phi_2)},
$$
and for $\varepsilon_1 = \varepsilon_2$:
$$
\mathcal{W}_{s\,**}^{[0,2]}(\phi_1,\phi_2) = \frac{ - H_{b(s) + 2,0}(\phi_1,\phi_2)}{(x_{\varepsilon_1}^*)'(\phi_1)(x_{\varepsilon_2}^*)'(\phi_2)}.
$$
The value of $\mathcal{W}_{s\,**}^{[0,2]}$ for $\varepsilon_1 \neq \varepsilon_2$ will be irrelevant.
\end{corollary}

The second consequence of Lemma~\ref{behaviorG2} is the critical behavior of the generating series of cuffed cylinders $\hat{\mathcal{W}}_{s}^{[0,2]}$ and $\tilde{\mathcal{W}}_{s}^{[0,2]}$ which appear in the evaluation of $\mathscr{W}$ via Proposition~\ref{P212}.

\begin{lemma}\label{arms} Let $x_j=x(\varepsilon_j + \tau \phi_j)$ for $j=1,2$, and consider the critical regime $q \rightarrow 0$. Let $\bs{\mathcal{H}}(x)$ be a generating series which is holomorphic for $x \in \mathbb{C}\setminus[\gamma_-,\gamma_+]$ such that $\bs{\mathcal{H}}(x) \in O(1/x^2)$ when $x \rightarrow \infty$, and when $x = x(\varepsilon + \tau\phi)$ admits the critical behavior
$$
\bs{\mathcal{H}}(x) =  \Big(\frac{\pi}{T}\Big)^{C}\,q^{\frac{3}{2}\,\varepsilon}\big\{\bs{\mathcal{H}}_{\varepsilon,*}(\phi) + O(q^{b})\big\},
$$
where $C$ stands for an arbitrary real number. When computing the integral
\beq
\label{contH1} \oint_{\gamma} \frac{\dd x_1}{2{\rm i}\pi}\,\bs{\mathcal{H}}(x_1)\hat{\mathcal{W}}_{s}^{[0,2]}(x_1,x_2), 
\eeq
the relevant singular part (for this we use the sign $\equiv$) of $\hat{\mathcal{W}}_{s}^{[0,2]}\!\!$ is 
\beq
\label{FEF0} \hat{\mathcal{W}}_{s}^{[0,2]}(x_1,x_2) \equiv q^{\hat{\varkappa}(\varepsilon_2)}\big\{\hat{\mathcal{W}}_{s;\varepsilon_2,*}^{[0,2]}(\phi_1,\phi_2) + O(q^{b(s)})\big\},
\eeq
with $\varepsilon_1 = 0$ and the exponent
$$
\hat{\varkappa}(\varepsilon_2) = \left\{\begin{array}{lll} -\tfrac{1}{2} & & {\rm if}\,\,\varepsilon_2 = 0, \\ \tfrac{b(s)}{2} & & {\rm if}\,\,\varepsilon_2 = \tfrac{1}{2}. \end{array}\right.\,
$$
Likewise, let $\tilde{\bs{\mathcal{H}}}(x_1,x_2)$ be a generating series which is holomorphic for $(x_1,x_2) \in (\mathbb{C}\setminus[\gamma_-,\gamma_+])^2$ and such that $\tilde{\bs{\mathcal{H}}}(x_1,x_2) \in O(x_1^{-2}x_2^{-2})$ when $x_i \rightarrow \infty$, and admitting the following critical behavior when $x_j = x(\varepsilon_j + \tau\phi_j)$:
$$
\tilde{\bs{\mathcal{H}}}(x_1,x_2) =  \Big(\frac{\pi}{T}\Big)^{C} q^{\frac{3}{2}(\varepsilon_1 + \varepsilon_2)}\big\{\tilde{\bs{\mathcal{H}}}_{\varepsilon_1,\varepsilon_2,*}(\phi_1,\phi_2) + O(q^{b})\big\},
$$
where $C$ is an arbitrary number. When computing the contour integral
\beq 
\label{contH2} \oint_{\gamma} \frac{\dd x_1}{2{\rm i}\pi}\oint_{\gamma} \frac{\dd x_2}{2{\rm i}\pi}\,\tilde{\bs{\mathcal{H}}}(x_1,x_2)\,\tilde{\mathcal{W}}_{s}^{[0,2]}(x_1,x_2),
\eeq
the singular part of $\tilde{\mathcal{W}}_{s}^{[0,2]}$ is
\beq
\label{FEFE0}\tilde{\mathcal{W}}_{s}^{[0,2]}(x_1,x_2) \equiv \tilde{\mathcal{W}}_{s\,*}^{[0,2]}(\phi_1,\phi_2) + O(q^{b(s)}),
\eeq
with $\varepsilon_1 = \varepsilon_2 = 0$. 
The non-zero constant prefactors are given in \eqref{FFFFFF0}-\eqref{FFFFFF} in the course of the proof. 
\end{lemma}



\noindent\textbf{Proof.} We shall estimate the contour integrals \eqref{contH1} and \eqref{contH2} in the regime $q \rightarrow 0$ by the steepest descent method. In particular, we will have to determine which region of the complex plane gives the dominant contribution of the integral, and the proof will show that it is always the vicinity of $\gamma_+^*$. It is however convenient to first transform the expressions of $\hat{\mathcal{W}}_{s}^{[0,2]}$ and $\tilde{\mathcal{W}}_{s}^{[0,2]}$.

Using $\partial_{x} \mathbf{R}(x,y) = \mathbf{A}(x,y)$, the evaluation \eqref{contueq} of the contour integral of a function against $\mathbf{A}(x,y)$ and the definition of $\mathcal{G}_{s}^{[0,2]}(v_1,v_2)$ in Proposition~\ref{p15}, setting $x_i = x(v_i)$ and analytically continuing in $(v_1,v_2)$, we obtain
\bea
\hat{\mathcal{W}}_{s}^{[0,2]}(x_1,x_2) & = & s\oint_{\gamma}\frac{\dd y}{2{\rm i}\pi}  \mathbf{R}(x_1,y)\mathcal{W}_{s}^{[0,2]}(y,x_2) \nonumber \\
& = &  s\int^{x_1} \dd \tilde{x}_1 \oint_{\gamma} \frac{\dd y}{2{\rm i}\pi}\,\mathbf{A}(\tilde{x}_1,y)\mathcal{W}_{s}^{[0,2]}(y,x_2) + C(x_2) \nonumber\\
& = & -\int^{x_1} \dd \tilde{x}_1\,\mathsf{n}s\, \varsigma'(\tilde{x}_1)\mathcal{W}_{s}^{[0,2]}(\varsigma(\tilde{x}_1),x_2) + C(x_2) \nonumber\\
& = & \mathsf{n}s \int^{v(x_1)} \dd\tilde{v}_1\,\frac{\mathcal{G}_{s}^{[0,2]}(\tau - \tilde{v}_1, v_2)}{x'(v_2)}  \nonumber\\
\label{Fhat222} & + & \frac{\mathsf{n}s}{4-\mathsf{n}^2s^2}\left(\frac{2}{x_2-\varsigma(x_1)}+\frac{\mathsf{n}s\,\varsigma'(x_2)}{\varsigma(x_2)-\varsigma(x_1)}\right) + C(x_2),
\eea
where we stress that $C(x_2)$ does not depend on $x_1$, and for this reason will disappear when performing contour integration against $\bs{\mathcal{H}}(x_1)$ as $\bs{\mathcal{H}}(x_1) \in O(x_1^{-2})$. We can then do a partial fraction expansion with respect to $x_1$:
\bea
\frac{1}{x_2-\varsigma(x_1)} & = & \frac{-\varsigma'(x_2)}{x_1 - \varsigma(x_2)} + \frac{1}{x_2 - \varsigma(\infty)}, \nonumber \\
\frac{\varsigma'(x_2)}{\varsigma(x_2) - \varsigma(x_1)} & = & -\frac{1}{x_1 - x_2} + \frac{\varsigma'(x_2)}{\varsigma(x_2) - \varsigma(\infty)}.\nonumber
\eea
Therefore:
\bea
\oint_{\gamma} \frac{\dd x_1}{2{\rm i}\pi}\,\bs{\mathcal{H}}(x_1)\,\hat{\mathcal{W}}_{s}^{[0,2]}(x_1,x_2) & = & ns \oint_{\gamma} \frac{\dd x_1}{2{\rm i}\pi}\, \bs{\mathcal{H}}(x_1) \int^{v(x_1)} \frac{\dd\tilde{v}_1\,\mathcal{G}^{[0,2]}_{s}(\tau - \tilde{v}_1,v_2)}{x'(v_2)} \nonumber \\
\label{contqq} && + \frac{\mathsf{n}s}{4 - \mathsf{n}^2s^2}\big(2\varsigma'(x_2)\bs{\mathcal{H}}(\varsigma(x_2)) + \mathsf{n}s\bs{\mathcal{H}}(x_2)\big).
\eea

The second term is of order of magnitude $q^{\frac{3}{2}\,\varepsilon_2}$. To examine the behavior of the first term, we fix the value of $\varepsilon_2 \in \{0,\tfrac{1}{2}\}$. When the variable $x_1$ is in the regime $x_1 = x(\varepsilon_1 + \tau\phi_1)$, the integrand (including $\dd x_1$) is of order of magnitude
\beq
\label{qqpower} q^{\frac{3}{2}\,\varepsilon_1 + (\frac{1}{2} - \varepsilon_1) - (\frac{1}{2} - \varepsilon_2) + b(s) (\varepsilon_1 \oplus \varepsilon_2)}.
\eeq
If $\varepsilon_2 = 0$, this is for $\varepsilon_1 = 0$ equal to $q^{0}$, while for $\varepsilon_1 = \tfrac{1}{2}$ it is equal to $q^{\frac{1}{4} + \frac{b(s)}{2}}$ -- which is negligible compared to the former. If $\varepsilon_2 = \tfrac{1}{2}$, \eqref{qqpower} is equal for $\varepsilon_1 = 0$ to $q^{\frac{b(s) + 1}{2}}$, while for $\varepsilon_1 = \tfrac{1}{2}$ it is equal to $q^{\frac{3}{4}}$ -- which is negligible compared to the former. So, independently of the value of $\varepsilon_2$, we move the contour for $x_1$ to pass close to $\gamma_+^*$ and the integral will be dominated by the regime $x_1 = x(\varepsilon_1 + \tau\phi_1)$ with $\varepsilon_1 = 0$. And, the first term in \eqref{contqq} is of order $q^{0}$ when $\varepsilon_2 = 0$, and of order $q^{\frac{b(s) + 1}{2}}$ when $\varepsilon_2 = \tfrac{1}{2}$. Since $b(s) \in (0,\tfrac{1}{2})$, we deduce that \eqref{contqq} is of order $q^{0}$ if $\varepsilon_2 = 0$, and of order $q^{\frac{b(s) + 1}{2}}$ if $\varepsilon_2 = \tfrac{1}{2}$.

Combining everything, the singular part of $\hat{\mathcal{W}}_{s}^{[0,2]}$ which is relevant to extract the leading term in \eqref{contqq} is
$$
\hat{\mathcal{W}}_{s}^{[0,2]}(x_1,x_2) \equiv q^{\hat{\varkappa}(\varepsilon_2)}\big\{\hat{\mathcal{W}}_{s;\varepsilon_2,*}^{[0,2]}(\phi_1,\phi_2) + O(q^{b(s)})\big\},
$$
with $\varepsilon_1 = 0$, the exponent
$$
\hat{\varkappa}(\varepsilon_2) = \left\{\begin{array}{lll} -\tfrac{1}{2} & & {\rm if}\,\,\varepsilon_2 = 0, \\ \tfrac{b(s)}{2}  & & {\rm if}\,\,\varepsilon_2 = \tfrac{1}{2}, \end{array}\right.
$$
and the prefactors:
\bea
\label{FFFFFF0}  \hat{\mathcal{W}}_{s;0,*}^{[0,2]}(\phi_1,\phi_2) & = & \frac{\mathsf{n}s}{4 - \mathsf{n}^2s^2}\bigg\{ \int^{\phi_1} \dd\tilde{\phi}_1\,\frac{H_{b(s),0}(1 - \tilde{\phi}_1,\phi_2)}{(x^*_0)'(\phi_2)}  \\
&&  + \frac{(x_0^*)'(1 - \phi_2)}{(x_0^*)'(\phi_2)}\,\frac{2}{x_0^*(\phi_1) - x_0^*(1 - \phi_2)} - \frac{\mathsf{n}s}{x_0^*(\phi_1) - x^*_0(\phi_2)}\bigg\}, \nonumber  \\
\label{FFFFFF} \hat{\mathcal{W}}_{s;\frac{1}{2},*}^{[0,2]}(\phi_1,\phi_2) & = & -\frac{8 b(s) \mathsf{n} s\,\cos \pi b(s) \phi_1\,\sin \pi b(s) \phi_2}{4 - \mathsf{n}^2s^2}\,.
\eea

Let us now turn to $\tilde{\mathcal{W}}^{[0,2]}_{s}$. We compute from the definition \eqref{Ftilded}:
\bea
&&  \tilde{\mathcal{W}}_{s}^{[0,2]}(x_1,x_2) \nonumber \\
& = & s\mathbf{R}(x_1,x_2) + s^2\oint_{\gamma^2}\frac{\dd y_1}{2{\rm i}\pi}\,\frac{\dd y_2}{2{\rm i}\pi} \mathbf{R}(x_1,y_1)\mathbf{R}(x_2,y_2)\mathcal{W}_{s}^{[0,2]}(y_1,y_2) \nonumber\\
& = & s\mathbf{R}(x_1,x_2) + s^2\int^{x_1} \dd\tilde{x}_1 \int^{x_2} \dd\tilde{x}_2 \oint_{\gamma} \frac{\dd y_1}{2{\rm i}\pi}\,\frac{\dd y_2}{2{\rm i}\pi}\,\mathbf{A}(\tilde{x}_1,y_1)\mathbf{A}(\tilde{x}_2,y_2)\mathcal{W}_{s}^{[0,2]}(y_1,y_2) \nonumber \\
&& + C_1(x_1) + C_2(x_2) \nonumber \\
& = & s\mathbf{R}(x_1,x_2) + \mathsf{n}^2s^2\int^{x_1} \dd\tilde{x}_1 \int^{x_2}\dd\tilde{x}_2\, \varsigma'(\tilde{x}_1)\varsigma'(\tilde{x}_2) \mathcal{W}_{s}^{[0,2]}(\varsigma(\tilde{x}_1),\varsigma(\tilde{x}_2))
 \dd \tilde{x}_1 \dd \tilde{x}_2 \nonumber \\
 && + C_1(x_1) + C_2(x_2) \nonumber \\
 & = & s\mathbf{R}(x_1,x_2) + \mathsf{n}^2s^2 \bigg(\int^{v(x_1)} \dd\tilde{v}_1\,\int^{v(x_2)}\dd\tilde{v}_2\,\mathcal{G}_{s}^{[0,2]}(\tau - \tilde{v}_1,\tau - \tilde{v}_2)  \nonumber \\
\label{Ftilde222} & &-  \frac{2\,\ln\big[\varsigma(x_1) - \varsigma(x_2)\big] + \mathsf{n}s\,\ln\big[x_1 - \varsigma(x_2)\big]}{4 - \mathsf{n}^2s^2} + \tilde{C}_1(x_1) + \tilde{C}_2(x_2)\bigg).
\eea
The functions $C_1(x_1)$, $\tilde{C}_1(x_1)$, $C_2(x_2)$ and $\tilde{C}_2(x_2)$ do not depend simultaneously on $x_1$ and $x_2$ and will thus disappear when we perform the contour integral against $\tilde{\bs{\mathcal{H}}}(x_1,x_2)$ as it behaves like $O(x_1^{-2}x_2^{-2})$ when $x_i \rightarrow \infty$. Given the expression \eqref{rin} for $\mathbf{R}$, the term $s\mathbf{R}(x_1,x_2)$ in the first line combines with the ratio in the second line, up to an extra term which only depends on $x_2$ and will also disappear:
\bea 
\tilde{\mathcal{W}}_{s}^{[0,2]}(x_1,x_2) & = & \mathsf{n}^2s^2\int^{v(x_1)}\int^{v(x_2)}  \dd\tilde{v}_1\,\dd\tilde{v}_2\,\mathcal{G}_{s}^{[0,2]}(\tau - \tilde{v}_1,\tau - \tilde{v}_2) \nonumber \\
&& - \frac{2\mathsf{n}s(\mathsf{n}s\,\ln[x_1 - x_2] + 2\,\ln[x_1 - \varsigma(x_2)])}{4 - \mathsf{n}^2s^2}+ \hat{C}_1(x_1) + \hat{C}_2(x_2),\nonumber 
\eea
where again $\hat{C}_i(x_i)$ does not depend simultaneously on both $x_1$ and $x_2$ so that they will disappear in the next step. Now, we can compute
\bea
&& \oint_{\gamma} \frac{\dd x_1}{2{\rm i}\pi}\,\oint_{\gamma} \frac{\dd x_2}{2{\rm i}\pi}\,\tilde{\bs{\mathcal{H}}}(x_1,x_2)\,\tilde{\mathcal{W}}_{s}^{[0,2]}(x_1,x_2)  \nonumber \\
&=& \mathsf{n}^2s^2 \oint_{\gamma} \frac{\dd x_1}{2{\rm i}\pi} \oint_{\gamma} \frac{\dd x_2}{2{\rm i}\pi} \tilde{\bs{\mathcal{H}}}(x_1,x_2) \int^{v(x_1)} \bigg\{\dd\tilde{v}_1\,\int^{v(x_2)} \dd \tilde{v}_2\,\mathcal{G}_{s}^{[0,2]}(\tau - \tilde{v}_1,\tau - \tilde{v}_2) \nonumber \\&& \qquad \qquad - \frac{2 ns \, \dd \tilde{x}_1}{4 - \mathsf{n}^2s^2}\Big(\frac{2}{\tilde{x}_1 - \varsigma(x_2)} + \frac{ns\varsigma'(\tilde{x}_1)}{\varsigma(\tilde{x}_1) - \varsigma(x_2)}\Big)\bigg\} \nonumber \\
&= &  \mathsf{n}^2s^2 \bigg\{ \oint_{\gamma} \frac{\dd x_1}{2{\rm i}\pi} \oint_{\gamma} \frac{\dd x_2}{2{\rm i}\pi}\,\tilde{\bs{\mathcal{H}}}(x_1,x_2)  \int^{v(x_1)} \int^{v(x_2)} \dd\tilde{v}_1 \dd\tilde{v}_2\,\mathcal{G}_{s}^{[0,2]}(\tau - \tilde{v}_1,\tau - \tilde{v}_2)  \nonumber \\
\label{lassa}&& \qquad\qquad - \frac{2 \mathsf{n}s}{4 - \mathsf{n}^2s^2}\oint_{\gamma} \frac{\dd x_1}{2{\rm i}\pi}\bigg(2\int^{x_1} \dd\tilde{x}_1\,\tilde{\bs{\mathcal{H}}}(\tilde{x}_1,\varsigma(x_1))\,\varsigma'(x_1)  + \mathsf{n}s \int^{x_1}  \dd\tilde{x}_1\,\tilde{\bs{\mathcal{H}}}(\tilde{x}_1,x_1)\bigg)\bigg\}. \nonumber
\eea
The same arguments we used for $\hat{\mathcal{W}}_{s}^{[0,2]}$ show that the dominant contribution to the integrals always comes from the part of the integration where $x_j = x(\tau\phi_j)$ with $\phi_j$ of order $1$. So, the singular part of $\tilde{\mathcal{W}}_{s}^{[0,2]}$ which allows us to extract the dominant contribution of \eqref{lassa} is:
$$
\tilde{\mathcal{W}}_{s}^{[0,2]}(x_1,x_2) \equiv \tilde{\mathcal{W}}_{s\,*}^{[0,2]}(\phi_1,\phi_2) + O(q^{b(s)}),
$$
where $x_j = x(\varepsilon_j + \tau\phi_j)$ with $\varepsilon_1 = \varepsilon_2 = 0$ and:
\bea
\label{fnun0} \tilde{\mathcal{W}}_{s\,*}^{[0,2]} & = & \frac{\mathsf{n}^2s^2}{4 - \mathsf{n}^2s^2} \int^{\phi_1}\int^{\phi_2} \dd\tilde{\phi}_1\dd\tilde{\phi}_2\,H_{b(s),0}(1 - \tilde{\phi}_1,1 - \tilde{\phi}_2)  \\
&&  - \frac{2\mathsf{n}s}{4 - \mathsf{n}^2s^2}\Big(2\ln[x_0^*(\phi_1) - x_0^*(1 - \phi_2)] + \mathsf{n}s\ln[x_0^*(1 - \phi_1) - x_0^*(1 - \phi_2)]\Big). \nonumber
\eea
\hfill $\Box$
\vspace{0.2cm}

\subsection{Fixed nesting graph}

\label{CoscrFs}

Now we can deduce the critical behavior of the  generating series of maps with a fixed nesting graph $\Gamma$. Recall that we denoted $V_{0,2}(\Gamma)$ the set of univalent vertices of genus $0$ carrying exactly one boundary. Let us introduce the notations $V_{0,2}^0(\Gamma)$ (resp. $V_{0,2}^{1/2}(\Gamma)$) for the vertices for which we keep the boundary large (resp. small). Let $k^{(0,2)}$, $k_0^{(0,2)}$ and $k_{1/2}^{(0,2)}$ denote the cardinalities of $V_{0,2}(\Gamma)$, $V_{0,2}^0(\Gamma)$ and $V_{0,2}^{1/2}(\Gamma)$, respectively.
\begin{theorem}
\label{CoscrF} Let $x_j=x(\varepsilon_j + \tau \phi_j)$ for $j=1, \ldots, k$, and $k_0$ and $k_{1/2}$ denote the number of $\varepsilon_j =0$ (large boundaries) and of $\varepsilon_j =1/2$ (small boundaries). When $q \rightarrow 0$, we have for the singular part with respect to $u$ and $x_i$'s:
$$ 
\mathscr{W}_{\Gamma,\star,\mathbf{s} = \mathbf{1}}^{[g,k]}(x_1,\ldots,x_k) = \Big(\frac{\pi}{T}\Big)^{k} q^{\varkappa(g,k,k_{1/2},k_{1/2}^{(0,2)})} \big\{[\mathscr{W}_{\Gamma,\star,\mathbf{s} = \mathbf{1}}^{[g,k]}]_{*}(\phi_1,\ldots, \phi_k) + O(q^{\frac{b}{2}})\big\},
$$
where $\varkappa(g,k,k_{1/2},k_{1/2}^{(0,2)})=(2g - 2 + k)(\mathfrak{d}\tfrac{b}{2} - 1) - \tfrac{k}{2} + \tfrac{3}{4}\,k_{1/2} + (\tfrac{b}{2}- \tfrac{1}{4})k_{1/2}^{(0,2)}.$
The errors are uniform for $\phi_j$ in any compact.
\end{theorem}
Remarkably, the result does not depend on the details of $\Gamma$, but only on its genus $g$, and number of boundaries of different types. For a fixed topology $(g,k)$, the graphs minimizing the number of small boundaries have the biggest contribution and if we also fix a configuration $(k_0,k_{1/2})$, the graphs maximizing $k_{1/2}^{(0,2)}$ contribute the most.

\vspace{0.2cm}

\noindent\textbf{Proof.}
We want to estimate the expression of Proposition~\ref{P212} for $\mathscr{W}_{\Gamma,\star,\mathbf{s}}^{[g,k]}$ in the regime $q\rightarrow 0$. Given that the vertex weights are $\mathsf{W}$'s whose leading term according to Theorem~\ref{ouqusf} has the property to receive an extra factor $q^{\frac{3}{4}}$ whenever a boundary variable $x_i$ is not close to $\gamma_+^*$ at scale $q^{\frac{1}{2}}$, we are in the conditions of Lemma~\ref{arms}. We can apply the steepest descent method to approximate the integral, and we have argued in the proof of Lemma~\ref{arms} that the contour should be moved to pass close to $\gamma_+^*$ because the dominant contribution comes from the regime where each $y_{\mathsf{e}} - \gamma_+^* \in O(q^{\frac{1}{2}})$, i.e. $y_{\mathsf{e}} = x(\tau\phi_{\mathsf{e}})$ for $\phi_{\mathsf{e}}$ of order $1$. Therefore, combining Theorem~\ref{ouqusf} for $\mathsf{W}$'s and Lemma~\ref{arms} for $\hat{\mathcal{W}}_{s}^{[0,2]}$ and $\tilde{\mathcal{W}}_{s}^{[0,2]}$, we arrive to: 
\bea
&& \mathscr{W}^{[g,k]}_{\Gamma,\star,\mathbf{1}}(x_1,\ldots,x_k) \nonumber \\
& = & \oint_{\gamma^{E_{{\rm glue}}(\Gamma)}} \prod_{\mathsf{e} \in E_{{\rm glue}}(\Gamma)}  \frac{\dd y_{\mathsf{e}}}{2{\rm i}\pi} \prod_{\mathsf{v} \in \tilde{V}(\Gamma)} \frac{\mathsf{W}^{[h(\mathsf{v}),k(\mathsf{v}) + d(\mathsf{v})]}(x_{\partial(\mathsf{v})},y_{\mathsf{e}(\mathsf{v})})}{d(\mathsf{v})!} \nonumber \\
&& \qquad \qquad \times \prod_{\mathsf{e} \in \tilde{E}(\Gamma)} \tilde{\mathcal{W}}^{[0,2]}_{s = 1}(y_{\mathsf{e}_+},y_{\mathsf{e}_-}) \prod_{\mathsf{v} \in V_{0,2}(\Gamma)}\hat{\mathcal{W}}^{[0,2]}_{s = 1}(y_{\mathsf{e}_+(\mathsf{v})},x_{\partial(\mathsf{v})}) \nonumber\\
& = & \prod_{\mathsf{e} \in E_{{\rm glue}}(\Gamma)}q^{\frac{1}{2}}\prod_{\mathsf{v} \in \tilde{V}(\Gamma)}  q^{[2h(\mathsf{v}) -2 +k(\mathsf{v})+d(\mathsf{v})](\mathfrak{d}\frac{b}{2} - 1)-\frac{k(\mathsf{v})+d(\mathsf{v})}{2}+\frac{3}{4}k_{1/2}(\mathsf{v})} \prod_{\mathsf{v} \in V_{0,2}^0}q^{-\frac{1}{2}} \prod_{\mathsf{v} \in V_{0,2}^{1/2}} q^{\frac{b}{2}} \nonumber\\
\label{laisft}&&\times \big\{[\mathscr{W}_{\Gamma,\star,\mathbf{1}}^{[g,k]}]_{*}(\phi_1,\ldots, \phi_k) + O(q^{\frac{b}{2}})\big\},
\eea
with
\bea
[\mathscr{W}_{\Gamma,\star,\mathbf{1}}^{[g,k]}]_{*}(\phi_1,\ldots, \phi_k) & = & \oint_{\overline{\mathcal{C}}^{E_{{\rm glue}}(\Gamma)}} \prod_{\mathsf{e} \in E_{{\rm glue}}(\Gamma)} \frac{\dd x_0^*(\tilde{\phi}_{\mathsf{e}})}{2{\rm i}\pi}\,\prod_{\mathsf{v} \in \tilde{V}(\Gamma)} \frac{\mathsf{W}_{*}^{[h(\mathsf{v}),k(\mathsf{v}) + d(\mathsf{v})]}(\phi_{\partial(\mathsf{v})},\tilde{\phi}_{\mathsf{e}(\mathsf{v})})}{d(\mathsf{v})!} \nonumber \\
&& \times \prod_{\mathsf{e} \in \tilde{E}(\Gamma)} \tilde{\mathcal{W}}_{s = 1\,*}^{[0,2]}(\tilde{\phi}_{\mathsf{e}_+},\tilde{\phi}_{\mathsf{e}_{-}}) \prod_{\mathsf{v} \in V_{0,2}(\Gamma)} \hat{\mathcal{W}}^{[0,2]}_{s = 1\,*}(\tilde{\phi}_{\mathsf{e}_+(\mathsf{v})},\phi_{\partial(\mathsf{v})}). \nonumber
\eea

Since we refer all the time to a fixed  nesting graph $\Gamma$, we omit it in the notations for simplicity. Let us now simplify the total exponent. The first Betti number of the graph is 
$$
1-|V|+|E| = g-\sum_{\mathsf{v}\in\tilde{V}}h(\mathsf{v}),
$$
and we recall that $|V| = |\tilde{V}| + k^{(0,2)}$ and $|E| = |\tilde{E}| + k^{(0,2)}$. Then, we observe that
$$ 
\sum_{\mathsf{v} \in \tilde{V}} k(\mathsf{v}) = k - k^{(0,2)}\ \  \text{ and }\ \  \sum_{\mathsf{v} \in \tilde{V}} k_{1/2}(\mathsf{v}) = k_{1/2} - k_{1/2}^{(0,2)}.
$$
By counting inner half-edges we also find
$$
\sum_{\mathsf{v} \in \tilde{V}} d(\mathsf{v}) = 2|E| - |E_{{\rm un}}| = 2|\tilde{E}| + k^{(0,2)} = |E_{{\rm glue}}|\,.
$$
Moreover, we obviously have $k^{(0,2)} = k^{(0,2)}_{0} + k^{(0,2)}_{1/2}$. Substituting these relations in \eqref{laisft} gives a total exponent\bea
\varkappa & = & \tfrac{1}{2}\,|E_{{\rm glue}}| +\big(2(g-|E|+|V|-1)-2|\tilde{V}| + k - k^{(0,2)} + 2|E|-k^{(0,2)}\big) \left(\mathfrak{d}\tfrac{b}{2}-1\right) \nonumber \\
&& -\tfrac{1}{2}(k-k^{(0,2)})-\tfrac{1}{2}\,|E_{{\rm glue}}|+\tfrac{3}{4}(k_{1/2}-k_{1/2}^{(0,2)})-\tfrac{1}{2}\,k_0^{(0,2)} + \tfrac{b}{2}\,k_{1/2}^{(0,2)} \nonumber \\
&= & (2g-2+k)\left(\mathfrak{d}\tfrac{b}{2}-1\right)-\tfrac{k}{2}+\tfrac{3}{4}k_{1/2}+ (\tfrac{b}{2} - \tfrac{1}{4})k^{(0,2)}_{1/2}. \nonumber
\eea
\hfill $\Box$


\section{Large volume asymptotics}
\label{FixedV}

Recall from Theorem~\ref{th38} the scaling of $q$ with respect to the variable $u$ coupled to the volume
$$
q \sim \Big(\frac{1 - u}{q_*}\Big)^{c},\qquad c = \frac{1}{1 - \frac{b}{2} - \mathfrak{d}\frac{b}{2}}, 
$$
with $\mathfrak{d} = 1$ in dense phase, $\mathfrak{d} = -1$ in dilute phase.

\subsection{Relative amplitude of nesting graphs}
\label{Fixedas}

We now extract from Theorem~\ref{CoscrF} the leading asymptotics of the generating series of maps of given volume $V$, given boundary perimeters, and given nesting graph $\Gamma$, not keeping track of the number of separating loops on each arm -- \textit{i.e.} for $s(\mathsf{e}) = 1$, in the regime $V \rightarrow \infty$, while we impose either small or large boundaries.

\begin{theorem}
\label{LAPA} Take $(\mathsf{g},\mathsf{h})$ on the non-generic critical line. Assume $2g - 2 + k > 0$. The generating series of connected maps of volume $V$, of genus $g$, with $k_{1/2}$ boundaries of finite perimeter $L_i=\ell_i$, among which $k_{1/2}^{(0,2)}$ are carried by a genus $0$ leaf as only mark, and $k_{0}$ boundaries of perimeters $L_i=\ell_i V^{c/2}$ -- for fixed positive $\bs{\ell} = (\ell_i)_{i = 1}^k$ -- and realizing the nesting graph $(\Gamma,\star)$, behaves when $V \rightarrow \infty$ as
\beq
\label{lneq} \Big[u^{V}\prod_{i = 1}^k x_i^{-(L_i+1)}\Big]\mathscr{W}_{\Gamma,\star,\mathbf{1}}^{[g,k]} \sim \pmb{\mathscr{A}}_{\Gamma,\star,\mathbf{1}}^{[g,k]}(\bs{\ell})\,V^{[-1 + c((2g - 2 + k)(1 - \mathfrak{d}\frac{b}{2}) - \frac{1}{4}k_{1/2} + (\frac{1}{4} - \frac{b}{2})k_{1/2}^{(0,2)})]},
\eeq
where $k = k_0 + k_{1/2}$ is the total number of boundaries, and an expression for the non-zero prefactor is given in \eqref{Agks1}.
\end{theorem}

Several remarkable conclusions can be drawn from this result. Firstly, if we keep all boundaries large, we have
$$ 
\mathscr{W}_{\Gamma,\star,\mathbf{1}}^{[g,k]} \stackrel{\bigcdot}{\sim} V^{-1 + c(2g - 2 + k)(1 - \mathfrak{d}\frac{b}{2})} 
$$
and the order of magnitude only depends on the global topology of $\Gamma$, i.e. on the genus $g$ and the number of boundaries $k$. In other words, for given $g$ and $k$, all nesting graphs have comparable probabilities to be realized. 

\begin{figure}[h!]
\begin{center}
\includegraphics[width=0.8\textwidth]{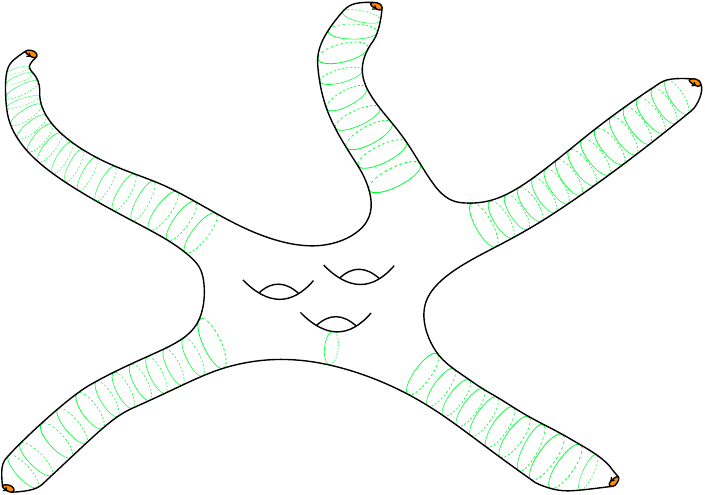}
\caption{\label{starfig} A typical map of the $O(\mathsf{n})$ model with small boundaries. These are most likely to be incident to distinct long arms (containing $O(\ln V)$ separating loops). We have only drawn in green the loops which are separating.}
\end{center}
\end{figure}

Secondly, if we keep a certain number $k_{1/2} > 0$ of small boundaries, the nesting graphs most likely to be realized when $V \rightarrow \infty$ at criticality are the ones with $k_{1/2}^{(0,2)} = k_{1/2}$, \textit{i.e.}, where each small boundary belongs as the only marked element to a connected component with the topology of a cylinder on the complement of all loops (see Figure~\ref{starfig}). And, all nesting graphs with this property have comparable probabilities.
For completeness, we also study the case of cylinders $(g,k) = (0,2)$ -- for which the computations already appeared in \cite{BBD}. There are only two possible nesting graphs:
\beq
\label{Gcyl} (\Gamma_1,\star) = \bullet^{1,2}\qquad\qquad (\Gamma_2,\star) = {}^{1}\!\bullet\!\!\!-\!\!\!-\!\!\bullet^{2}
\eeq
Before conditioning on the volume and the boundary perimeters, the generating series for $(\Gamma_1,\star)$ is $\mathsf{W}^{[0,2]}(x_1,x_2)$, while the generating series for the $(\Gamma_2,\star)$ is $(\mathcal{W}_{s = 1}^{[0,2]} - \mathsf{W}^{[0,2]})(x_1,x_2)$. We derive from Corollary~\ref{dgfsgg}:

\begin{theorem}
\label{LAPB} Take $(\mathsf{g},\mathsf{h})$ on the non-generic critical line. Fix $\ell_i$ positive independent of $V$, and $\varepsilon_i \in \{0,\tfrac{1}{2}\}$. If $\varepsilon_i = 0$, we choose $L_i = \ell_i V^{c/2}$, and if $\varepsilon_i = \tfrac{1}{2}$, we rather choose $L_i = \ell_i$. We have when $V \rightarrow \infty$:
\bea   
\label{qdoguns}\left[u^{V}x_1^{-(L_1+1)}x_2^{-(L_2+1)}\right]\,\mathscr{W}_{\Gamma_1,\star,s = 0}^{[0,2]}(x_1,x_2) & \sim & \mathscr{W}_{\Gamma_1,\star,s = 0}^{[0,2]}(\ell_1,\ell_2)\,V^{-1 - \frac{c}{2}(1-(\varepsilon_1 \oplus \varepsilon_2))}, \\
\label{qdoguns2} \left[u^{V}x_1^{-(L_1+1)}x_2^{-(L_2+1)}\right]\,\mathscr{W}_{\Gamma_2,\star,s = 1}^{[0,2]}(x_1,x_2) & \sim & \mathscr{W}_{\Gamma_2,\star,s = 1}^{[0,2]}(\ell_1,\ell_2)\,V^{-1 - cb(1-(\varepsilon_1 \oplus \varepsilon_2))},  
\eea
with a non-zero prefactor.
\end{theorem}

The constant prefactors $\pmb{\mathscr{A}}$ are computed in the course of the proofs.  Although their structure is combinatorially clear -- we essentially have to replace in the formula of Proposition~\ref{P212} all the factors by their effective leading asymptotics derived throughout the previous Section, and perform the extra contour integrations in $\tilde{u}$ and $\tilde{x}$ whose effect is simply displayed in \eqref{Agks1} -- it is however a formidable task to obtain explicit formulas (as functions of $\ell_i$) for a given nesting graph $\Gamma$. For us, the formula serves as showing that this prefactor is non trivial.

We remark that the formula for the exponent in Theorem \ref{LAPB} does not agree with the one in Theorem \ref{LAPA} taking $(g,k) =(0,2)$.

\vspace{0.2cm}

\noindent \textbf{Proofs.} We briefly sketch the proof as the details of the saddle point analysis are essentially the same as in \cite[Section 6.4 and 6.5]{BBD}. Let $\partial_0(\Gamma)$ denote the set of boundaries for which we want to impose perimeter $L_i = \ell_i\,V^{c/2}$ (\textit{i.e.} we declare $\varepsilon_i = 0$), and $\partial_{1/2}(\Gamma)$ the set of boundaries for which we rather impose $L_i = \ell_i$ (\textit{i.e.} we declare $\varepsilon_i = \tfrac{1}{2}$). The analysis reveals that this scaling $V^{c/2}$ for large boundaries is the one for which a non-trivial behavior will be obtained.

\begin{center}
\textit{Conditioning on boundary perimeters}
\end{center}

\vspace{0.1cm}

\begin{figure}
\begin{center}
\includegraphics[width=0.5\textwidth]{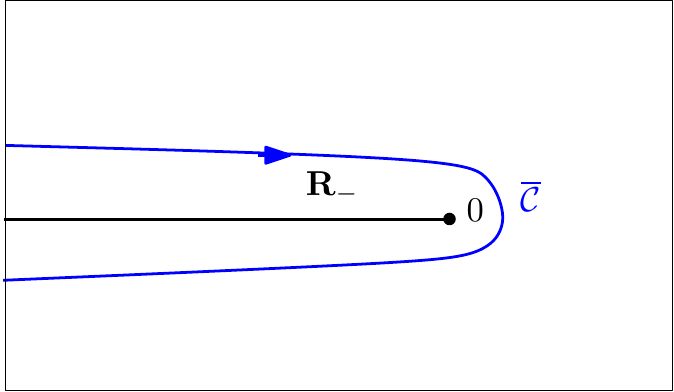}
\caption{\label{ContCbar} The contour $\overline{\mathcal{C}}$.}
\end{center}
\end{figure}

We first study integrals of the form
\beq
\label{Iform} \mathcal{I}(u) =   \prod_{i \in \partial_{0}(\Gamma)} \oint_{\gamma} \frac{x_i^{\ell_i V^{c/2}}\,\dd x_i}{2{\rm i}\pi}\ \prod_{i \in \partial_{1/2}(\Gamma)} \oint_{\gamma} \frac{x_i^{\ell_i}\,\dd x_i}{2{\rm i}\pi}\,\Phi\bigg[u;(x_i)_{i \in \partial_{1/2}(\Gamma)};\Big(\frac{x_i - \gamma_+}{q^{\frac{1}{2}}}\Big)_{i \in \partial_{0}(\Gamma)}\bigg],
\eeq
where $\Phi$ is a function which has a non-zero limit when $u \rightarrow 1$, and the convergence is uniform when its variables belong to any compact. We also take from Corollary~\ref{CoB7} in Appendix that
$$
\gamma_+^* - \gamma_+ = O(q).
$$
We use the change of variables
$$ 
x_i = \left\{\begin{array}{lll} \gamma_+^* + q^{\frac{1}{2}}\,x_0^*(\phi_i) && {\rm if}\,\,i \in \partial_{0}(\Gamma), \\ \gamma_+^* + x_{\frac{1}{2}}^*(\phi_i) && {\rm if}\,\,i \in \partial_{1/2}(\Gamma), \end{array}\right.
$$ 
and deform the contour in $\big(\tilde{x}_i = q^{-\frac{1}{2}}(x_i - \gamma_+^*)\big)_{i \in \partial_{0}(\Gamma)}$ so that it passes close to the cut (see Figure~\ref{ContCbar}). In the limit $u \rightarrow 1$, the properties of the integrand on those steepest descent contours ensure that we can use the monotone convergence theorem to find
\bea 
\mathcal{I}(u) & \sim & q^{\frac{1}{2}k_0} \prod_{i \in \partial_{0}(\Gamma)} \oint \frac{\dd x_0^*(\phi_i)\,e^{x_0^*(\phi_i)\ell_i/\gamma_+^*}}{2{\rm i}\pi}\prod_{i \in \partial_{1/2}(\Gamma)} \oint \frac{(x_{\frac{1}{2}}^*(\phi_i))^{\ell_i}\,\dd x_{\frac{1}{2}}^*(\phi_i)}{2{\rm i}\pi} \nonumber \\
 && \times\Phi\Big[1;\big(x_{\frac{1}{2}}^*(\phi_i)\big)_{i \in \partial_{1/2}(\Gamma)};\big(x_0^*(\phi_i)\big)_{i \in \partial_{0}(\Gamma)}\Big].  \nonumber
\eea

\vspace{0.2cm}

\begin{center}
\textit{Conditioning on volume $V$}
\end{center}

\vspace{0.1cm}

\begin{figure}[h!]
\begin{center}
\includegraphics[width=0.5\textwidth]{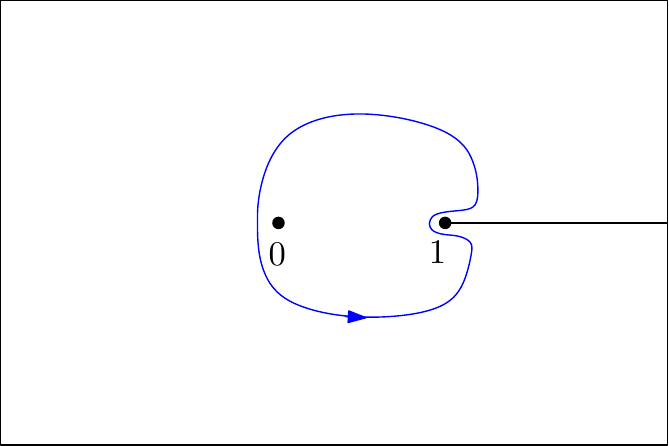}
\caption{\label{Cont1} The contour of integration for $\tilde{u}$.}
\end{center}
\end{figure}

Next, we would like to estimate integrals of the form
$$
\overline{\mathcal{I}} = \oint \frac{\dd u}{2{\rm i}\pi\,u^{V + 1}}\,\mathcal{I}(u) q^{\eta}
$$
for some exponent $\eta$. We recall $q$ is a function of $u$ for which Theorem~\ref{th38} gives
$$ 
q \sim \Big(\frac{1 - u}{q_*}\Big)^{c},\qquad u \rightarrow 1.
$$
We perform the change of variables
$$
u = 1 - \frac{\tilde{u}}{V}
$$
and deform the contour in $u$ to the one shown in Figure~\ref{Cont1}. Now assume
$$  
\eta + \tfrac{1}{2}k_0 \neq 0.
$$ 
In the limit $V \rightarrow \infty$, by the properties of the integrand on this steepest descent contour, we can complete the integral to a contour which is again $\overline{\mathcal{C}}$ shown in Figure~\ref{ContCbar} and find
\bea  
\overline{\mathcal{I}} & \sim & V^{-1 - c(\eta + \frac{1}{2}k_0)}\,\oint_{\overline{\mathcal{C}}} \frac{-\dd\tilde{u}\,e^{\tilde{u}}}{2{\rm i}\pi} \Big(\frac{\tilde{u}}{q_*}\Big)^{c(\eta + \frac{1}{2}k_0)} \prod_{i \in \partial_{0}(\Gamma)} \oint_{(x_0^*)^{-1}(\overline{\mathcal{C}})} \frac{\dd x_0^*(\phi_i)\,e^{x_0^*(\phi_i)\ell_i/\gamma_+^*}}{2{\rm i}\pi} \nonumber \\
&& \prod_{i \in \partial_{1/2}(\Gamma)} \oint_{(x_{\frac{1}{2}}^*)^{-1}(\gamma)} \frac{(x_{\frac{1}{2}}^*(\phi_i))^{\ell_i}\,\dd x_{\frac{1}{2}}^*(\phi_i)}{2{\rm i}\pi}\,\Phi\Big[1;(x_{\frac{1}{2}}^*(\phi_i))_{i \in \partial_{1/2}(\Gamma)};\big(x_0^*(\phi_i)\big)_{i \in \partial_{0}(\Gamma)}\Big].  \nonumber
\eea
The integral over $\tilde{u}$ factors out and yields a Gamma function
\bea 
\overline{\mathcal{I}} & = & \frac{V^{-1 - c(\eta + \frac{1}{2}k_0)}}{-\Gamma\big[-c\big(\eta + \frac{1}{2}k_0\big)\big]} \prod_{i \in \partial_{0}(\Gamma)} \oint_{(x_0^*)^{-1}(\overline{\mathcal{C}})} \frac{\dd x_0^*(\phi_i)\,e^{x_0^*(\phi_i)\ell_i/\gamma_+^*}}{2{\rm i}\pi}\nonumber \\
\label{Ibar} &&  \prod_{i \in \partial_{1/2}(\Gamma)} \oint_{(x_{\frac{1}{2}}^*)^{-1}(\gamma)} \frac{(x_{\frac{1}{2}}^*(\phi_i))^{\ell_i}\,\dd x_{\frac{1}{2}}^*(\phi_i)}{2{\rm i}\pi}\, \Phi\Big[1;(x_{\frac{1}{2}}^*(\phi_i))_{i \in \partial_{1/2}(\Gamma)};\big(x_0^*(\phi_i)\big)_{i \in \partial_{0}(\Gamma)}\Big] .
\eea

\vspace{0.2cm}

\begin{center}
\textit{Specialization to Theorem~\ref{LAPA}}
\end{center}
\vspace{0.1cm}
We obtain Theorem~\ref{LAPA} for $\mathscr{W}_{\Gamma,\star,\mathbf{1}}^{[g,k]}$ with $2g - 2 + k > 0$ by taking from the proof of Theorem~\ref{CoscrF} the exponent
\beq
\label{etadenom} \eta \coloneqq (2g - 2 + k)(\mathfrak{d}\tfrac{b}{2} - 1) - \tfrac{k}{2} + \tfrac{3}{4}\,k_{1/2} + (\tfrac{b}{2}- \tfrac{1}{4})k_{1/2}^{(0,2)}
\eeq
and
\beq
\Phi\Big[1;(x_{\frac{1}{2}}^*(\phi_i))_{i \in \partial_{1/2}(\Gamma)};\big(x_0^*(\phi_i)\big)_{i \in \partial_{0}(\Gamma)}\Big] = [\mathscr{W}_{\Gamma,\star,\mathbf{1}}^{[g,k]}]_{*}(\phi_1,\ldots,\phi_k).
\eeq
Since $k = k_0 + k_{1/2}$, we remark that
$$
\eta + \tfrac{1}{2}k_0 = (2g - 2 + k)(\mathfrak{d}\tfrac{b}{2} - 1) + \tfrac{1}{4}\,k_{1/2} + (\tfrac{b}{2}- \tfrac{1}{4})k_{1/2}^{(0,2)}
$$
is non-zero. The constant prefactor is thus
\bea
\pmb{\mathscr{A}}^{[g,k]}_{\Gamma,\star,\mathbf{1}}(\bs{\ell}) & = & -\Gamma^{-1}\big[-c\big(\eta + \frac{1}{2}k_0\big)\big] \prod_{i \in \partial_{0}} \oint_{(x_0^*)^{-1}(\overline{\mathcal{C}})} \frac{\dd x_0^*(\phi_i)\,e^{x_0^*(\phi_i)\ell_i/\gamma_+^*}}{2{\rm i}\pi}\nonumber \\
\label{Agks1} &&  \prod_{i \in \partial_{1/2}} \oint_{(x_{\frac{1}{2}}^*)^{-1}(\gamma)} \frac{(x_{\frac{1}{2}}^*(\phi_i))^{\ell_i}\,\dd x_{\frac{1}{2}}^*(\phi_i)}{2{\rm i}\pi}\, [\mathscr{W}_{\Gamma,\star,\mathbf{1}}^{[g,k]}]_{*}(\phi_1,\ldots,\phi_k).
\eea

\vspace{0.2cm}

\begin{center}
\textit{Specialization to Theorem~\ref{LAPB}}
\end{center}

\vspace{0.1cm}

We first consider $\mathscr{W}^{[0,2]}_{\Gamma,\star,s = 1}$. From Corollary~\ref{dgfsgg}, the first term leads us to the previous setting with
\beq
\label{exonde}\eta + \tfrac{1}{2}k_0 = \widetilde{\beta}^{(0,2)}(s,\varepsilon_1,\varepsilon_2) + \tfrac{1}{2}k_0 = \left\{\begin{array}{lll} 0 & & {\rm if}\,\,\varepsilon_1 =  \varepsilon_2 = 0, \\ \tfrac{b(s)}{2} && {\rm if}\,\,\varepsilon_1 \neq \varepsilon_2, \\ b(s) & & {\rm if}\,\,\varepsilon_1 = \varepsilon_2 = \tfrac{1}{2}, \end{array}\right.
\eeq
with $s = 0$ for $(\Gamma_1,\star)$ and $s = 1$ for $(\Gamma_2,\star)$. However, in the case of two large boundaries ($\varepsilon_1 = \varepsilon_2 = 0$), we see that this first term contains no power of $q$, so is regular in $u$. The leading contribution in this case comes from the second term, hence corresponds to an exponent
\beq
\label{exonde2} \eta + \tfrac{1}{2}k_0 = b(s), \quad {\rm if}\,\,\varepsilon_1 = \varepsilon_2 = 0.
\eeq
So, we obtain the desired result by specializing \eqref{Ibar} to the exponent \eqref{exonde} corrected by \eqref{exonde2} and
\beq
\label{Phoih} \Phi\Big[1;(x_{\frac{1}{2}}^*(\phi_i))_{i \in \partial_{1/2}(\Gamma)};\big(x_0^*(\phi_i)\big)_{i \in \partial_{0}(\Gamma)}\Big] = \left\{\begin{array}{lll} \mathcal{W}_{s=\mathbf{1}\,**}^{[0,2]}(\phi_1,\phi_2) & & {\rm if}\,\,\varepsilon_1 = \varepsilon_2 = 0, \\ \mathcal{W}_{s=\mathbf{1}\,*}^{[0,2]}(\phi_1,\phi_2) && {\rm otherwise}, \end{array}\right. 
\eeq
with again $s = 0$ for $(\Gamma_1,\star)$ and $s = 1$ for $(\Gamma_2,\star)$. If we define $[\mathscr{W}^{[0,2]}_{\Gamma,\star,\mathbf{1}}]_*(\phi_1,\phi_2)$ to be the right-hand side of \eqref{Phoih}, the prefactors in \eqref{qdoguns}-\eqref{qdoguns2} are then also given by \eqref{Agks1} with the exponents $\eta$ we just saw.

\hfill $\Box$

\subsection{Large deviation for arm lengths in a fixed nesting graph}\label{Fixedasarms}

Next, we also determine the asymptotics of the probability
\beq
\label{Pgk} \mathbb{P}^{[g,k]}\big[\mathbf{P}|\Gamma,\star,V,\mathbf{L}\big] \coloneqq \frac{\Big[u^{V} \prod_{\mathsf{e} \in E(\Gamma)} s(\mathsf{e})^{P(\mathsf{e})} \prod_{i = 1}^k x_i^{-(L_i+1)}\Big]\,\mathscr{W}^{[g,k]}_{\Gamma,\star,\mathbf{s}}(x_1,\ldots,x_k)}{\Big[ u^{V} \prod_{i = 1}^k x_i^{-(L_i+1)}\Big]\,\mathscr{W}^{[g,k]}_{\Gamma,\star,\mathbf{1}}(x_1,\ldots,x_k)}
\eeq
that a connected map of genus $g$, of fixed volume $V$, with $k$ boundaries of fixed perimeters $\mathbf{L} = (L_i)_{i = 1}^k$, fixed nesting graph $(\Gamma,\star)$, has a number $P(\mathsf{e})$ of separating loops on every arm $\mathsf{e} \in E(\Gamma)$. We assume that $\Gamma$ has at least one arm for this to make sense. 

Consider configurations which are of topology higher than $(0,1)$ or $(0,2)$.
We argued in the last section that large lengths for gluing annuli carrying loops give effectively dominant contributions.
Therefore, given a nesting graph $\Gamma$, we define the following subsets of edges, which will be the only relevant ones in this section:
\begin{itemize}
\item $E_{0,2}^S(\Gamma)\coloneqq \{\mathsf{e}_+(\mathsf{v})\mid \mathsf{v}\in V_{0,2}^{1/2}(\Gamma)\}$ the edges incident to a genus $0$ univalent vertex $\mathsf{v}$ carrying as only mark one microscopic boundary. This type of edges correspond to arms whose ends are a small boundary and an annulus (carrying a loop). 

\item $E_{0,2}^{L,L}(\Gamma)\coloneqq E(\Gamma)\setminus E_{0,2}^S(\Gamma) = \tilde{E}(\Gamma) \cup \{\mathsf{e}_+(\mathsf{v})\mid \mathsf{v}\in V_{0,2}^{0}(\Gamma)\}$ the rest of the edges, which correspond to inner arms (whose ends are two annuli) and an arm whose end is a large boundary and an annulus.
\end{itemize}

\begin{center}
\begin{figure}
\begin{center}
\includegraphics[width=0.6\textwidth]{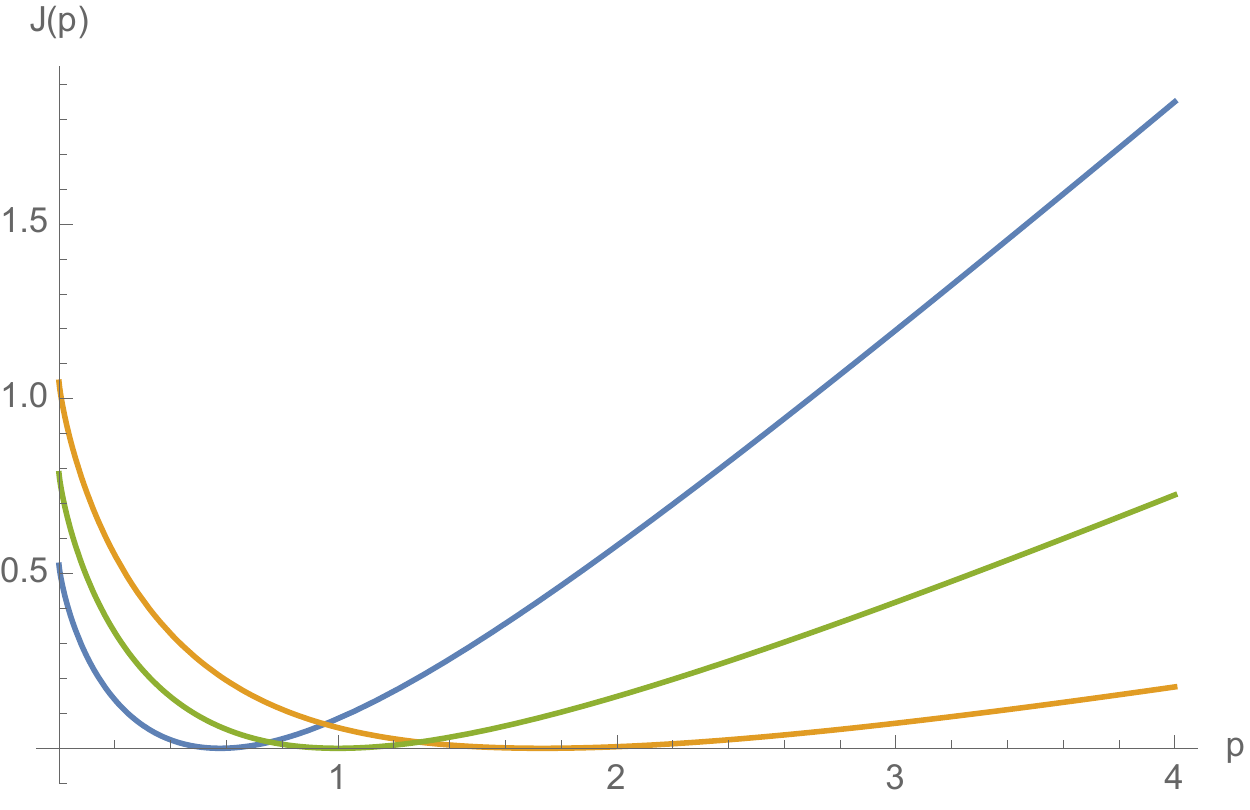}
\caption{\label{Jplot} The function $J(p)$: blue for $\mathsf{n} = 1$, green for $\mathsf{n} = \sqrt{2}$ (Ising), and orange for $\mathsf{n} = \sqrt{3}$ ($3$-Potts).}
\end{center}
\end{figure}
\end{center}

We introduce
\bea
J(p) & = & \sup_{s \in [0,2/\mathsf{n}]} \big\{p\ln(s) + {\arccos}(\mathsf{n}s/2) - {\rm arccos}(\mathsf{n}/2)\big\} \nonumber \\
\label{Jdef} & = & p\ln\Big(\frac{2}{\mathsf{n}}\,\frac{p}{\sqrt{1 + p^2}}\Big) + {\rm arccot}(p) - {\rm arccos}(\mathsf{n}/2).
\eea
This function is plotted in Figure~\ref{Jplot}.

\begin{theorem}
\label{igfsgb}Take $(\mathsf{g},\mathsf{h})$ on the non-generic critical line and fix $\Gamma$ a nesting graph. Assume $2g - 2 + k > 0$, fix positive variables $\bs{\ell} = (\ell_i)_{i = 1}^k$ independent of $\,V\!$, and positive $\,\mathbf{p} = \big(p(\mathsf{e})\big)_{\mathsf{e} \in E(\Gamma)}$ such that $\,p(\mathsf{e}) \ll \ln V\!$. We consider the regime where $k_0$ boundaries have perimeter $L_i = \ell_i V^{c/2}$, $k_{1/2}$ boundaries have perimeter $L_i = \ell_i$, and 
\beq
\label{factor2} P(\mathsf{e}) = \frac{c\ln V\,p(\mathsf{e})}{\jmath(\mathsf{e})\pi}, \; \text{ with } \jmath(\mathsf{e}) = \left\{\begin{array}{lll} 2, & &  \text{ if }\,\mathsf{e}\in E_{0,2}^S(\Gamma), \\ 1, & & \text{ otherwise}. \end{array}\right.
\eeq
In the limit $V \rightarrow \infty$, we have
\beq
\label{Omun} \mathbb{P}^{[g,k]}\big[\mathbf{P}|\Gamma,\star,V,\mathbf{L}\big]
 \sim  \mathscr{P}^{[g,k]}\big[\mathbf{p}|\Gamma,\star,\bs{\ell}\big]\,\prod_{\mathsf{e} \in E_{0,2}^{L,L}(\Gamma)} V^{-\frac{c}{\pi}\,p(\mathsf{e})\ln \frac{2}{\mathsf{n}}}\, \prod_{\mathsf{e} \in E_{0,2}^S(\Gamma)} \frac{V^{-\frac{c}{2\pi}\,J[p(\mathsf{e})]}}{\sqrt{\ln V}}, 
\eeq
where $\mathscr{P}^{[g,k]}\big[\mathbf{p}|\Gamma,\star,\bs{\ell}\big]$ is a non-zero prefactor.
\end{theorem}

For completeness we recall the result for $(g,k) = (0,2)$ from \cite[Theorem 7.1]{BBD}. 

\begin{theorem}
\label{CylGG}Take $(\mathsf{g},\mathsf{h})$ on the non-generic critical line. Fix positive variables $(\ell_1,\ell_2)$ independent of $\,V\!$, and $p$ positive such that $p \ll \ln V\!$. We have when $V \rightarrow \infty$
\bea
\label{O92}\mathbb{P}^{[0,2]}\Big[P = \frac{c\ln V}{\pi}\,p\,\Big|\,V\,,\,L_1 = \ell_1\,,\,L_2 = \ell_2\Big] & \sim & \mathscr{P}^{[0,2]}_{{\rm 1}}(p,\ell_1,\ell_2)\,\frac{V^{-\frac{c}{\pi} J(p)}}{\sqrt{\ln V}},  \\
\mathbb{P}^{[0,2]}\Big[P = \frac{c\ln V}{2\pi}\,p\,\Big|\,V\,,\, L_1 = \ell_1\,,\,L_2 = \ell_2 V^{c/2}\Big] & \sim & \mathscr{P}^{[0,2]}_{2}(p,\ell_1,\ell_2)\,\frac{V^{-\frac{c}{2\pi}\,J(p)}}{\sqrt{\ln V}},  \nonumber \\
\mathbb{P}^{[0,2]}\Big[P = \frac{c\ln V}{\pi}\,p\,\Big|\,V\,,\,L_1 = \ell_1 V^{c/2}\,,\,L_2 = \ell_2 V^{c/2} \Big]  & \sim & \mathscr{P}^{[0,2]}_{3}(p,\ell_1,\ell_2)\,V^{-\frac{c}{\pi}\,p(\mathsf{e})\ln \frac{2}{\mathsf{n}}}. \nonumber 
\eea
\end{theorem}


From Theorem~\ref{igfsgb} one concludes that, for a given nesting graph $\Gamma$, the arm lengths for configurations of higher topology behave like independent random variables. Recall that the analysis of the generating series of configurations with a fixed nesting graph showed that large lengths for the gluing annuli, which contain the inner boundaries of the arms, give effectively dominant contributions. 

The arms that correspond to an edge in $E_{0,2}^{L,L}(\Gamma)$, i.e., the arms with both boundaries large (either both interior, or one interior and one exterior), will typically contain finitely many separating loops, with exactly the same behavior of separate cylinders with both boundaries large from Theorem  \ref{CylGG}. On the other hand, the arms that correspond to an edge $\mathsf{e}$ in $E_{0,2}^S(\Gamma)$ have a depth of order $\ln V$, with large deviation function proportional to $J(p)$. This is exactly the same behavior as for separate arms conditioned to have infinite volume with at least one small boundary. 

Focusing around the point
$$
p_{\rm opt} = \frac{\mathsf{n}}{\sqrt{4 - \mathsf{\mathsf{n}}^2}},
$$
where $J$ reaches its minimum value $0$, we obtain:
\begin{corollary}
\label{Gaussflu} Consider the ensemble of connected maps of genus $g$ with $k$ boundaries of perimeters~$\mathbf{L}$, with volume $V\!$, realizing a fixed nesting graph $(\Gamma,\star)$. Under the assumptions and the regime described in Theorem~\ref{igfsgb}, the vector of random variables
$$
\left(\frac{P(\mathsf{e}) - \frac{c\,p_{{\rm opt}}\ln V}{2\pi}}{\sqrt{\ln V}}\right)_{\mathsf{e}\in E_{0,2}^S(\Gamma)}
$$
converges in law when $V \rightarrow \infty$ to the random Gaussian vector $\big(\mathscr{N}(0,\sigma^2)\big)_{\mathsf{e}\in E_{0,2}^S(\Gamma)}$ with variances
$$
\sigma^2 = \frac{2 \, \mathsf{n} \, c}{\pi(4 - \mathsf{n}^2)^{\frac{3}{2}}}.$$
\end{corollary}

\noindent \textbf{Proof of Theorem~\ref{igfsgb}.}  
The asymptotic behavior for the denominator in \eqref{Pgk} when $V\rightarrow\infty$ was already obtained in Theorem~\ref{LAPA}:
$$
\overline{\mathcal{I}}\stackrel{\bigcdot}{\sim} V^{-1-c\left(\varkappa(g,k,k_{1/2},k_{1/2}^{(0,2)})+\frac{1}{2}k_0\right)}.
$$
In order to study the asymptotic behavior of the numerator we need to determine the behavior of the the singular part of the generating series of configurations with a fixed nesting graph $\Gamma$, this time with respect to $s,u$ and $x_i$'s:
\bea
\mathscr{W}_{\Gamma,\star,\mathbf{s}}^{(\mathsf{g},k)}(x_1,\ldots,x_k) & = & \Big[\prod_{\mathsf{e} \in E} \frac{1}{4 - \mathsf{n}^2s^2}\Big]\,q^{\varkappa(\mathsf{g},k,k_{1/2},k_{1/2}^{(0,2)}) + \sum_{\mathsf{e} \in E_{0,2}^{S}} \frac{1}{2}(b[s(\mathsf{e})] - b)} \nonumber \\
&& \bigg\{\sum_{\tau\,:\,E_{0,2}^{L,L} \rightarrow \{0,1\}} \Big[\prod_{\mathsf{e} \in E_{0,2}^{L,L}} q^{\tau(e)\,b[s(\mathsf{e})]} \Big]\cdot \bigg([\mathscr{W}_{\Gamma,\star,\mathsf{s}}^{(\mathsf{g},k)}]_{*,\tau} + O\Big(\sum_{\mathsf{e} \in E(\Gamma)} q^{\frac{b[s(\mathsf{e})]}{2}}\Big)\bigg) \bigg\}. \nonumber \\
&& \label{718e} 
\eea 
We denote $\overline{\mathcal{I}}^{(\tau)}_{\mathbf{s}}$ the contribution attached to $\tau\,:\,E_{0,2}^{L,L} \rightarrow \{0,1\}$ in the above sum and get an expression of the form
\bea
\mathcal{J}^{(\tau)} & = & \oint \prod_{\mathsf{e} \in E} \frac{\dd s(\mathsf{e})}{2\pi {\rm i}\,s(\mathsf{e})^{P(\mathsf{e}) + 1}}\,\prod_{i = 1}^{k} \frac{\dd x_i}{2\pi {\rm i}}\,x_i^{L_i}\,\frac{\dd u}{2 \pi {\rm i}\,u^{V + 1}}\,\overline{\mathcal{I}}^{(\tau)}_{\mathsf{s}} \nonumber \\
& \sim & V^{-1 - c(\varkappa + \tfrac{1}{2}k_0)} \oint \prod_{\mathsf{e} \in E} \frac{1}{4 - \mathsf{n}^2s^2} \prod_{\mathsf{e} \in E} \frac{\dd s(\mathsf{e})}{2\pi {\rm i}\,s(\mathsf{e})^{P(\mathsf{e}) + 1}} \nonumber \\
& & \cdot\, \frac{V^{-c\eta_{\tau}(\mathbf{s})}}{-\Gamma\big[-c(\varkappa + \frac{1}{2}k_0 + \eta_{\tau}(\mathbf{s}))\big]} \nonumber \\
&& \cdot\prod_{i \in \partial_{0}(\Gamma)} \oint_{(x_0^*)^{-1}(\overline{\mathcal{C}})} \frac{\dd x_0^*(\phi_i)\,e^{x_0^*(\phi_i)\ell_i/\gamma_+^*}}{2\pi {\rm i}} \prod_{i \in \partial_{1/2}(\Gamma)} \oint_{(x_{1/2}^*)^{-1}(\gamma)} \frac{(x_{1/2}^*(\phi_i))^{\ell_i}\,\dd x_{1/2}^*(\phi_i)}{2\pi {\rm i}} \nonumber \\
&& \cdot\, \Phi_{\mathbf{s}}^{(\tau)}\Big[1\,;\,\big(x_{1/2}(\phi_i)\big)_{i \in \partial_{1/2}(\Gamma)}\,;\,\big(x_0^*(\phi_i)\big)_{i \in \partial_0(\Gamma)}\Big], \nonumber 
\eea 
with
$$
\eta_{\tau}(\mathbf{s}) = \sum_{\mathsf{e} \in E_{0,2}^{S}(\Gamma)} \tfrac{1}{2}(b[s(\mathsf{e})] - b) + \sum_{\mathsf{e} \in E_{0,2}^{L,L}(\Gamma) } \tau(e)\,b[s(\mathsf{e})].
$$

We study the regime
$$
P(\mathsf{e}) = \frac{cp(\mathsf{e})\ln V}{\jmath(\mathsf{e})\pi}
$$
for $p(\mathsf{e}) > 0$ independent of $V$. If we extend the map $\tau\,:\,E_{0,2}^{L,L}  \rightarrow \{0,1\}$ to a map $\tau\,:\,E(\Gamma) \rightarrow \{0,1\}$ by declaring $\tau(\mathsf{e}) = 1$ for $\mathsf{e} \in E_{0,2}^{S}(\Gamma)$, the singular part of the integrand is of the form
\bea
\prod_{\mathsf{e}\in E} s(\mathsf{e})^{-P(\mathsf{e})} \prod_{\mathsf{e}\in\tau^{-1}(1)} V^{-c \frac{b(s(\mathsf{e}))}{\jmath(\mathsf{e})}} \prod_{\mathsf{e}\in\tau^{-1}(0)}\frac{1}{4-\mathsf{n}^2s(\mathsf{e})^2}=\\ \prod_{\mathsf{e} \in \tau^{-1}(1)} \exp\Big(\jmath(\mathsf{e})^{-1}\,\ln V\,\mathcal{S}_{p(\mathsf{e})}[s(\mathsf{e})]\Big) \prod_{\mathsf{e} \in \tau^{-1}(0)} \frac{s(\mathsf{e})^{-\frac{cp}{\pi}\,\ln V}}{4 - \mathsf{n}^2s(\mathsf{e})^2},
\eea
with
$$
\mathcal{S}_{p}(s) = -\frac{cp\ln s}{\pi} - cb(s).
$$
We first compute the saddle point $\mathfrak{s}(p)$ of $\mathcal{S}_{p}$, i.e., the point such that $\mathcal{S}'_{p}(\mathfrak{s}(p)) = 0$. We find
$$
\mathfrak{s}(p) = \frac{2}{\mathsf{n}}\,\frac{p}{\sqrt{1 + p^2}}.
$$
We also compute in terms of the function $J$ introduced in \eqref{Jdef}
$$
\mathcal{S}_{p}(\mathfrak{s}(p)) = \frac{-c (\pi b + J(p))}{\pi}.
$$
For $\mathsf{e} \in \tau^{-1}(1)$ we perform the change of variables
$$
s(\mathsf{e}) = \mathfrak{s}[p(\mathsf{e})] + \frac{\tilde{s}(\mathsf{e})}{\sqrt{\ln V}}
$$
and find by Taylor expansion of $\mathcal{S}_{p}$ at order $2$ around $s = \mathfrak{s}(p)$:
$$ 
\frac{\dd s(\mathsf{e})}{2\pi {\rm i}\,s(\mathsf{e})^{P + 1}}\,V^{-cb(s)/\jmath(\mathsf{e})} \sim \frac{\dd \tilde{s}(\mathsf{e})}{2\pi {\rm i}\,\mathfrak{s}(p(\mathsf{e}))}\, (\ln V)^{-\frac{1}{2}}\,V^{\frac{\mathcal{S}_{p(\mathsf{e})}(\mathfrak{s}(p(\mathsf{e})))}{\jmath(\mathsf{e})}}\,\exp\Big(\frac{c\mathsf{n}^2(p(\mathsf{e})^2 + 1)^2}{8\jmath(\mathsf{e})\pi p(\mathsf{e})}\,\tilde{s}(\mathsf{e})^2\Big),
$$ 
which remains valid when $p(\mathsf{e})$ is allowed to depend on $V$ such that $p(\mathsf{e}) \ll \ln V$ and $P(\mathsf{e}) \gg 1$. We then deform the contour in $\tilde{s}(\mathsf{e})$ to a steepest descent contour ${\rm i}\mathbb{R}$, and the properties of the integrand imply we can apply the monotone convergence theorem and computation of the Gaussian integral in $\tilde{s}$ yields when $V \rightarrow \infty$:
\bea 
\mathcal{J}^{(\tau)} & \dot{\sim} & \frac{V^{-1 - c(\varkappa + \frac{1}{2}k_0)}}{-\Gamma\big[-c\big(\varkappa + \frac{1}{2}k_0 +\eta_{\tau}(\underline{\mathfrak{s}}_{\tau}(\mathbf{p}))\big)\big]}\,\prod_{\mathsf{e} \in \tau^{-1}(1)} \frac{V^{-cb/\jmath(\mathsf{e}) -cJ(p(\mathsf{e}))/\jmath(\mathsf{e})\pi}}{\sqrt{2j(\mathsf{e})^{-1}cp(\mathsf{e})(p^2(\mathsf{e}) + 1)\ln V}} \prod_{\mathsf{e} \in E_{0,2}^{{\rm S}}} V^{-\frac{b}{2}} \nonumber \\
\label{Seconffu}&& \cdot \prod_{\mathsf{e} \in \tau^{-1}(0)} V^{-\frac{c}{\pi}\,p(\mathsf{e})\,\ln(2/\mathsf{n})},
\eea
where
$$
\underline{\mathfrak{s}}_{\tau}(\mathbf{p}) = \Big(\big(\mathfrak{s}[p(\mathsf{e})]\big)_{\mathsf{e} \in \tau^{-1}(1)}\,,\,\big(\tfrac{2}{\mathsf{n}}\big)_{\mathsf{e} \in \tau^{-1}(0)}\Big)\,.
$$
The contour integral over $s(\mathsf{e})$ for $\tau(\mathsf{e}) = 0$ was easy to calculate and just produces $(\mathsf{n}/2)^{P(\mathsf{e})}$ and appears in an equivalent form in the second line of \eqref{Seconffu}. Here we had to separate cases for $P(\mathsf{e})$ even or odd, and check they both give the same contribution, taking into account the prefactors $\hat{\mathcal{W}}^{[0,2]}_{s;0,*}$ \eqref{FFFFFF0} and $\tilde{\mathcal{W}}^{[0,2]}_{s;*}$ \eqref{fnun0}.

As we need to sum over $\tau$ as in \eqref{718e}, we have to compare for $\mathsf{e} \in E_{0,2}^{L,L} (\Gamma)$ the factor coming from $\tau(\mathsf{e}) = 0$
$$
V^{-\frac{c}{\pi}\,p(\mathsf{e})\,\ln(2/\mathsf{n})}
$$ 
to the factor coming from $\tau(\mathsf{e}) = 1$
$$
V^{-c(\frac{J(p(\mathsf{e})}{\pi} + b)}.
$$
Since
$$
c\left(\frac{J(p)}{\pi}+b\right) > \ln\left(\frac{2}{\mathsf{n}}\right)\frac{cp}{\pi}, \; \text{ for all } p \geq 0,
$$
the term with $\tau(\mathsf{e}) = 0$, for all $\mathsf{e} \in E_{0,2}^{L,L}(\Gamma)$, dominates. We conclude by dividing by the asymptotic exponent of the numerator which has been previously obtained.
\hfill $\Box$

\appendix

\chapter{Appendix}
\label{chap:appendix1}

\section{The special function $\Upsilon_b$}

\label{AppA}
\label{AppUp}
Let $\tau$ be a complex number in the upper-half plane. The Jacobi theta function is the entire function of~$v \in \mathbb{C}$ defined by
\beq
\vartheta_{1}(v|\tau) = -\sum_{m \in \mathbb{Z}} e^{{\rm i}\pi \tau (m + \frac{1}{2})^2 + {\rm i}\pi (w  + \frac{1}{2})(2m + 1)}.
\eeq
Its main properties are
\beq
\vartheta_{1}(-v|\tau) = \vartheta_{1}(v + 1|\tau) = -\vartheta_{1}(v|\tau),\qquad \vartheta_{1}(v + \tau|\tau) = -e^{-2{\rm i}\pi (v  + \frac{\tau}{2})}\,\vartheta_{1}(v|\tau)
\eeq
and the effect of the modular transformation:
\beq
\label{modularS}\vartheta_{1}(v|\tau) = \frac{e^{-\frac{{\rm i}\pi v^2}{\tau}}}{\sqrt{-{\rm i}\tau}}\,\vartheta_{1}(\tfrac{v}{\tau}|-\tfrac{1}{\tau}).
\eeq

\begin{definition}
We define $\Upsilon_{b}(v)$ as the unique meromorphic function with a simple pole at $v = 0$ with residue $1$, and the pseudo-periodicity properties:
$$
\Upsilon_{b}(v + 1) = \Upsilon_{b}(v),\qquad \Upsilon_{b}(v + \tau) = e^{{\rm i}\pi b}\Upsilon_{b}(v).
$$
We have several expressions:
\bea
\Upsilon_{b}(v) & = & \sum_{m \in \mathbb{Z}} e^{-{\rm i}\pi b m}\,\mathrm{cotan}\,\pi(v + m\tau) \nonumber \\
& = & \frac{\vartheta_{1}'(0|\tau)}{\vartheta_{1}(-\frac{b}{2}|\tau)}\,\frac{\vartheta_{1}(v - \frac{b}{2}|\tau)}{\vartheta_{1}(v|\tau)} \nonumber \\
\label{210eq}& = &  \frac{e^{\frac{{\rm i}\pi b v}{\tau}}}{{\rm i}T}\,\frac{\vartheta_1'\big(0|-\frac{1}{\tau}\big)}{\vartheta_1\big(-\frac{b}{2\tau}\big|-\frac{1}{\tau}\big)}\,\frac{\vartheta_{1}\big(\frac{v - \frac{b}{2}}{\tau}\big|-\frac{1}{\tau}\big)}{\vartheta_{1}\big(\frac{v}{\tau}|-\frac{1}{\tau}\big)}.
\eea
\end{definition}

We have the expansion:
\beq
\label{expord1Upsilon}\Upsilon_{b}(w) = \frac{1}{w} + \sum_{j\geq 0}  \upsilon_{b,j} w^j,\qquad w \rightarrow 0,
\eeq
with
\bea
\label{upb1def} \upsilon_{b,1} & = &  \frac{1}{2}\,\frac{\vartheta_{1}''(\frac{b}{2}|\tau)}{\vartheta_{1}(\frac{b}{2}|\tau)} - \frac{1}{6}\,\frac{\vartheta_{1}'''(0|\tau)}{\vartheta_{1}'(0|\tau)}  \\
& = & \frac{1}{({\rm i}T)^2}\Big(\frac{1}{2}\,\frac{\vartheta_{1}''(\frac{b\tilde{\tau}}{2}|\tilde{\tau})}{\vartheta_{1}(\frac{b\tilde{\tau}}{2}|\tilde{\tau})} -\frac{1}{6}\,\frac{\vartheta_{1}'''(0|\tilde{\tau})}{\vartheta_{1}'(0|\tilde{\tau})} + {\rm i}\pi b \frac{\vartheta_{1}'(\frac{b\tilde{\tau}}{2}|\tilde{\tau})}{\vartheta_{1}(\frac{b\tilde{\tau}}{2}|\tilde{\tau})} - \frac{\pi^2b^2}{2}\Big), \nonumber
\eea 
where $\tilde{\tau} = -1/\tau$ and $\tau = {\rm i}T$. The value of the constant term $v_{b,0}$ is irrelevant for our purposes. The expressions involving $\tilde{\tau}$ or
$$
q = e^{{\rm i}\pi\tilde{\tau}} = e^{-\frac{\pi}{T}}
$$
are convenient to study the regime $T \rightarrow 0$, i.e. $q \rightarrow 0$.
\begin{lemma}
\label{lemUp}Let $v = \varepsilon + \tau w$. We have, for $b \in (0,1)$:
{\small $$
\Upsilon_{b}(v) = \frac{2\pi q^{\varepsilon b}}{T(1 - q^b)} \cdot \left\{\begin{array}{lll} \Upsilon_{b,0}^*(w)  - q^b\Upsilon_{b + 2,0}^*(w) + O(q^{2 - b}) & & {\rm if}\,\,\varepsilon = 0, \\ \Upsilon_{b,\frac{1}{2}}^*(w) - (q^{1 - b} - q)\Upsilon_{b - 2,\frac{1}{2}}^*(w) + q\Upsilon_{b + 2,\frac{1}{2}}^*(w) + O(q^{1 + b}) & & {\rm if}\,\,\varepsilon = \tfrac{1}{2}.\end{array}\right.
$$}
The errors are uniform for $w$ in any compact independent of $\tau \rightarrow 0$, stable under differentiation, and the expressions for the limit functions are
\bea
\label{theone}\Upsilon_{b,0}^*(w) & = & \frac{e^{{\rm i}\pi (b -1)w}}{2{\rm i}\sin(\pi w)}, \\
\label{thehalf}\Upsilon_{b,\frac{1}{2}}^*(w) & = & - e^{{\rm i}\pi b w}.\eea
We also have
\beq
\upsilon_{b,1} = \Big(\frac{\pi}{T}\Big)^2\Big\{\frac{1}{3} + b + \frac{b^2}{2} + O(q^{b})\Big\}.
\eeq
\hfill $\Box$
\end{lemma}

\section{The parametrization $x \leftrightarrow v$}
\label{App1}
\label{xbeh}

Consider given values of $\gamma_{\pm}$ and $\varsigma(\gamma_{\pm})$ such that
\beq
\label{orderg}\gamma_- < \gamma_+ < \varsigma(\gamma_+) < \varsigma(\gamma_-).
\eeq
We set
\beq
\label{io} v = {\rm i}C\,\int^{x}_{\varsigma(\gamma_+)} \frac{\dd y}{\sqrt{(y - \varsigma(\gamma_-))(y - \varsigma(\gamma_+))(y - \gamma_+)(y - \gamma_-)}}.
\eeq
The normalizing constant is chosen such that, for $x$ moving from the origin $\varsigma(\gamma_+)$ to $\varsigma(\gamma_-)$ with a small negative imaginary part, $v$ is moving from $0$ to $\tfrac{1}{2}$. When $x$ moves on the real axis from $\varsigma(\gamma_+)$ to~$\gamma_+$, $v$ moves from $0$ to a purely imaginary value denoted $\tau = {\rm i}T$. Then, the function $v \mapsto x(v)$ has the properties:
$$
x(v + 2\tau) = x(v + 1) = x(-v) = x(v),\qquad \varsigma(x(v)) = x(v - \tau),  
$$
and is depicted in Figure~\ref{ParamF}. $x'(v)$ has zeroes when $v \in \tfrac{1}{2}\mathbb{Z} + \tau\mathbb{Z}$, and double poles at $v = v_{\infty} + \mathbb{Z} + 2\tau\mathbb{Z}$. From \eqref{io}, paying attention to the determination of the squareroot at infinity obtained by analytic continuation, we can read in particular:
\beq
x'(v) \sim \frac{{\rm i}C}{(v - v_{\infty})^2},\qquad v \rightarrow v_{\infty}.
\eeq
From \eqref{orderg}, we know that $v_{\infty} = \frac{1}{2} + \tau w_{\infty}$, where $w_{\infty} \in (0,1)$ is determined as a function of $\gamma_{\pm}$ and~$\varsigma(\gamma_{\pm})$. 

There is an alternative expression for \eqref{io} in terms of Jacobi functions:
$$
v = \frac{2{\rm i}C\,{\rm arcsn}^{-1}\Big[\sqrt{\frac{\varsigma(\gamma_+) - \gamma_-}{\varsigma(\gamma_-) - \gamma_-}\,\frac{x - \varsigma(\gamma_+)}{x - \varsigma(\gamma_-)}}\,;\,k\Big]}{\sqrt{(\varsigma(\gamma_+) - \gamma_-)(\varsigma(\gamma_-) - \gamma_+)}},
$$
with
$$
k = \sqrt{\frac{(\varsigma(\gamma_-) - \gamma_-)(\varsigma(\gamma_+) - \gamma_+)}{(\varsigma(\gamma_-) - \gamma_+)(\varsigma(\gamma_+) - \gamma_-)}}.
$$
By specialization at $x = \gamma_-$ and $x = \varsigma(\gamma_-)$, we deduce the expressions:
\bea
\label{XCeq} C & = & \frac{\sqrt{(\varsigma(\gamma_+) - \gamma_-)(\varsigma(\gamma_-) - \gamma_+)}}{4K'(k)}, \\
\label{TCeq} T & = & \frac{K(k)}{2K'(k)},
\eea
in terms of the complete elliptic integrals. By matching poles and zeroes, we can infer an expression for $x(v) - \gamma_+$ in terms of Jacobi theta functions:
\beq
\label{Xtheta} x(v) - \gamma_+ = -{\rm i}C\,\frac{\vartheta_{1}'(0|2\tau)\vartheta_{1}(2v_{\infty}|2\tau)}{\vartheta_{1}(v_{\infty} - \tau|2\tau)\vartheta_{1}(v_{\infty} + \tau|2\tau)}\,\frac{\vartheta_{1}(v - \tau|2\tau)\vartheta_{1}(v + \tau|2\tau)}{\vartheta_{1}(v - v_{\infty}|2\tau)\vartheta_{1}(v + v_{\infty}|2\tau)}.
\eeq
From \eqref{io}, one can derive the expansion of $x(v)$ when $v \rightarrow v_{\infty}$. 
\begin{lemma}
\label{Xinfexp} When $v \rightarrow v_{\infty}$, we have the expansion
$$
x(v) = \frac{-{\rm i}C}{v - v_{\infty}} + \frac{E_{1}}{4} + \frac{{\rm i}}{C}\,\frac{3E_{1}^2 - 8E_{2}}{48}\,(v - v_{\infty}) + \frac{-E_1^3 + 4E_1E_2 - 8E_3}{64C^2}\,(v - v_{\infty})^2 + O(v - v_{\infty})^3,
$$
where we introduced the symmetric polynomials in the endpoints:
\bea
 E_{1} & = & \gamma_- + \gamma_+ + \varsigma(\gamma_+) + \varsigma(\gamma_-), \label{DefE} \\
 E_{2} & = & \gamma_-\big\{\gamma_+ + \varsigma(\gamma_+) + \varsigma(\gamma_-)\big\} + \gamma_+\big\{\varsigma(\gamma_+) + \varsigma(\gamma_-)\big\} + \varsigma(\gamma_+)\varsigma(\gamma_-),  \label{DefE2} \\
 E_{3} & = & \gamma_-\gamma_+\varsigma(\gamma_+) + \gamma_-\gamma_+\varsigma(\gamma_-)  + \gamma_-\varsigma(\gamma_-)\varsigma(\gamma_+) + \gamma_+\varsigma(\gamma_+)\varsigma(\gamma_-). \label{DefE3}
\eea
More generally, the coefficient of $(v - v_{\infty})^k$ in this expansion is a homogeneous symmetric polynomial of degree $(k + 1)$ with respect to the endpoints, with rational coefficients up to an overall factor $({\rm i}C)^{-k}$.\hfill $\Box$
\end{lemma}

In the study of non-generic critical points, we want to take the limit where $\gamma_+$ and $\varsigma(\gamma_+)$ collide to the fixed point of the involution:
$$
\gamma_+^* = \frac{1}{(\alpha + 1)\mathsf{h}},
$$
while $\gamma_- \rightarrow \gamma_-^*$ remains distinct from $\varsigma(\gamma_-^*)$. This implies $T\rightarrow 0$, or equivalently $k \rightarrow 0$. This limit is easily studied using the modular transformation \eqref{modularS} in \eqref{Xtheta}, or the properties of the elliptic integrals. If we set
$$
q = e^{-\frac{\pi}{T}},
$$
we arrive to:
\begin{lemma}
\bea
q & = & \Big(\frac{k}{4}\Big)^{4}\Big\{1 + O(k^2)\Big\}, \nonumber \\
w_{\infty} & = & w_{\infty}^*\big\{1 + O(q^{\frac{1}{2}})\big\}. \nonumber 
\eea
\hfill $\Box$
\end{lemma}

We can then derive the critical behavior of the parametrization $x(v)$ in the two regimes of interest:
\begin{lemma}
\label{LemB3}Let $v = \varepsilon + \tau w$ for $\varepsilon \in \{0,\frac{1}{2}\}$. We have
$$
x(v) - \gamma_+ = q^{\frac{1}{2} - \varepsilon}\,\big\{x_{\varepsilon}^*(w) + O(q^{\frac{1}{2}})\big\}.
$$
The error is uniform for $w$ in any compact independent of $\tau \rightarrow 0$, and this is stable under differentiation with respect to $v$. It is actually an asymptotic series in $q^{\frac{1}{2}}$. The limit functions are
\bea
x_{0}^*(w) & = & 8\sqrt{(\varsigma(\gamma_-^*) - \gamma_+^*)(\gamma_+^* - \gamma_-^*)}\,\sin(\pi w_{\infty}^*)\,\cos^2\Big(\frac{\pi w}{2}\Big), \nonumber \\
x_{\frac{1}{2}}^*(w) & = & \sqrt{(\varsigma(\gamma_-^*) - \gamma_+^*)(\gamma_+^* - \gamma_-^*)}\,\frac{\sin(\pi w_{\infty}^*)}{\cos(\pi w) - \cos (\pi w_{\infty}^*)}. \nonumber
\eea
\hfill $\Box$
\end{lemma}
If we specialize the second equation to $v = \frac{1}{2} + \tau$, use the expression \eqref{varsigma} of $\varsigma(x)$ and perform elementary trigonometric manipulations, we find:
\begin{corollary}
\label{CoB4}
$$
{\rm cos}(\pi w_{\infty}^*) = \frac{1 - \alpha}{1 + \alpha}\cdot \frac{1 - \mathsf{h}(1 + \alpha)\gamma_-^*}{1 + \mathsf{h}(1 - \alpha)\gamma_-^*}.
$$ 
\hfill $\Box$ 
\end{corollary}
We may consider $w_{\infty}^*$ as a parameter for the non-generic critical line.  Specializing again Lemma~\ref{LemB3} to $v = \varepsilon + \tau$ and using Corollary~\ref{CoB4} yields:
\begin{corollary}
\label{CoB5} There exists a constant $\rho_1$ such that:
\bea
2\mathsf{h}(\gamma_+^* - \gamma_+) & = & \frac{16\cos(\pi w_{\infty}^*)}{(1 - \alpha^2)}\,q^{\frac{1}{2}} + O(q), \nonumber \\
2\mathsf{h}(\gamma_-^* - \gamma_-) & = & \rho_{1} q^{\frac{1}{2}} + O(q),  \nonumber
\eea
and
\bea
E_1 & = & \frac{1 - \alpha\sin^2(\pi w_{\infty}^*)}{(1 - \alpha^2)\mathsf{h} \sin^2(\pi w_{\infty}^*)}+ \frac{2\rho_1\cos(\pi w_{\infty}^*)}{\mathsf{h}(1 - \cos(\pi w_{\infty}^*))^2}\,q^{\frac{1}{2}} + O(q), \nonumber \\
E_2 & = & \frac{2\big((3\alpha^2 - 1)\sin^2(\pi w_{\infty}^*) - 2(3\alpha - 2)\big)}{(\alpha^2 - 1)^2\mathsf{h}^2\sin^2(\pi w_{\infty}^*)} + \frac{2\rho_1(3\alpha - 2)}{\mathsf{h}^2(1 - \alpha^2)(1 - \cos(\pi w_{\infty}^*))^2}\,q^{\frac{1}{2}} + O(q), \nonumber \\
E_3 & = & \frac{4\big(\alpha^2\sin^2(\pi w_{\infty}^*) - \alpha(2 + \cos^2(\pi w_{\infty}^*) + 1)\big)}{(1 - \alpha)^2(1 + \alpha)^3\sin^2(\pi w_{\infty}^*)h^3} + O(q^{\frac{1}{2}}). \nonumber
\eea
\end{corollary}
The first four lines are used in \cite{BBD} to describe the phase diagram and the critical exponents of the model. Straightforward computations with \eqref{XCeq}-\eqref{TCeq} yield:
\begin{corollary}
\label{CBCB6} \bea
\frac{\pi C}{T} & = & \sqrt{(\varsigma(\gamma_-) - \gamma_+^*)(\gamma_+^* - \gamma_-)} + O(q), \nonumber \\
& = & \frac{2\,{\rm cot}(\pi w_{\infty}^*)}{(1 - \alpha^2)\mathsf{h}} + \frac{(1 + \cos(\pi w_{\infty}^*))\rho_1}{2(1 - \cos(\pi w_{\infty}^*))\sin(\pi w_{\infty}^*)}\,q^{\frac{1}{2}} + O(q). \nonumber
\eea
\hfill $\Box$
\end{corollary}

There are some simplifications in absence of bending energy, \textit{i.e.}, $\alpha = 1$. We then have $w_{\infty}^* = \frac{1}{2}$ which is in agreement with Corollary~\ref{CoB4}. The non-generic critical line is then parametrized by $\rho = 1 - 2\mathsf{h}\gamma_-^*$, which is related to the former parametrization by letting $\alpha \rightarrow 1$ and $w_{\infty}^* \rightarrow \frac{1}{2}$ in such a way that
\beq
\label{conflua}\Big(\frac{1}{2} - w_{\infty}^*\Big) \sim \frac{(1 - \alpha)}{2\pi}\,\rho.
\eeq
Corollary~\ref{CoB5} specializes to:
\begin{corollary}
\label{CoB7} For $\alpha  = 1$,  we have:
\bea
2\mathsf{h}(\gamma_+^* - \gamma_+) & = & O(q), \nonumber \\
E_1 & = & \frac{2}{\mathsf{h}} + O(q), \nonumber \\
E_2 & = & \frac{6 - \rho^2}{4\mathsf{h}^2} - \frac{\rho\rho_1}{2\mathsf{h}^2}\,q^{\frac{1}{2}} + O(q), \nonumber \\
E_3 & = & \frac{2 - \rho^2}{4\mathsf{h}^3} + O(q^{\frac{1}{2}}), \nonumber \\ 
\frac{\pi C}{T} & = & \frac{\rho}{2\mathsf{h}} + \frac{\rho_1}{2\mathsf{h}}\,q^{\frac{1}{2}} + O(q). \nonumber
\eea
\hfill $\Box$
\end{corollary}
The fact that $\varsigma(x) = \frac{1}{\mathsf{h}} - x$ and $\gamma_+^* = \frac{1}{2\mathsf{h}}$ gives the exact relation $E_1 = \frac{2}{\mathsf{h}}$, in agreement with the second line.

\section{The coefficients $\tilde{\mathsf{g}}_k$}
\label{Appgdeter}
In the loop model with bending energy where all faces are triangles, the parameters are: $\mathsf{g}$ (resp. $\mathsf{h}$) the weight per face not visited (resp. visited) by a loop, $\alpha$ the bending energy, and $\mathsf{n}$ the weight per loop. We can compute $\tilde{\mathsf{g}}_{k}$ from their definition \eqref{deftildeg} if we insert the expansion of Lemma~\ref{Xinfexp}. We recall that $C$ is the constant in \eqref{io}, and $E$'s are symmetric polynomials in the endpoints defined in Lemma~\eqref{Xinfexp}. If we introduce
$$
\tilde{\mathsf{g}}_k = ({\rm i}C)^k\,\widehat{\mathsf{g}}_{k},
$$
we find
\bea
\widehat{\mathsf{g}}_{3} & = & \frac{2\mathsf{g}}{4 - \mathsf{n}^2}, \nonumber \\
\widehat{\mathsf{g}}_{2} & = & \frac{2 - \mathsf{g}E_1}{4 - \mathsf{n}^2}, \nonumber \\
\widehat{\mathsf{g}}_{1} & = & \frac{\mathsf{g}(3E_{1}^2 - 4E_2) - 6E_1}{12(4 - \mathsf{n}^2)}, \nonumber \\
\widehat{\mathsf{g}}_{0} & = & -\frac{2u}{2 + \mathsf{n}}. \nonumber
\eea
We remark that $\widehat{\mathsf{g}}_{3}$ and $\widehat{\mathsf{g}}_{0}$ depend on the parameters of the model in a very simple way, whereas $\widehat{\mathsf{g}}_{1}$ and $\widehat{\mathsf{g}}_{2}$ have a non-trivial behavior in the non-generic critical regime, which can be deduced up to $O(q)$ from Corollary~\ref{CoB5}, either in terms of the parameter $w_{\infty}^*$, or the parameter $\rho$ if $\alpha = 1$.

\begin{corollary}
We have:
\bea 
\widehat{\mathsf{g}}_{2} & = & \frac{1}{4 - \mathsf{n}^2}\Big[1 + \frac{2\mathsf{g}}{\mathsf{h}}\Big(\alpha - \frac{1}{\sin^2(\pi w_{\infty}^*)}\Big)\Big]  \nonumber \\
&& - \frac{\mathsf{g}}{\mathsf{h}}\,\frac{\rho_1\cos(\pi w_{\infty}^*)}{(1 - \cos(\pi w_{\infty}^*))^2(4 - \mathsf{n}^2)}\,q^{\frac{1}{2}} + O(q), \nonumber \\
\widehat{\mathsf{g}}_{1} & = & \frac{2g\big[(3\alpha^2 + 1)\sin^4(\pi w_{\infty}^*) + 2(3\alpha - 2)\sin^2(\pi w_{\infty}^*) + 6\big]}{3(1 - \alpha^2)^2\mathsf{h}^2(4 - \mathsf{n}^2)\sin^4(\pi w_{\infty}^*)} \nonumber \\  
&& + \frac{3\mathsf{h}\sin^2(\pi w_{\infty}^*)(1 - \alpha^2)(\alpha\sin^2(\pi w_{\infty}^*) + 1)}{3(1 - \alpha^2)^2\mathsf{h}^2(4 - \mathsf{n}^2)\sin^4(\pi w_{\infty}^*)} \nonumber \\
&& + \frac{\rho_1\cos(w)\big\{2\mathsf{g}\big[4 -3\alpha\sin^2(\pi w_{\infty}^*) + 2\cos^2(\pi w_{\infty}^*)\big] -  3\mathsf{h}\sin^2(\pi w_{\infty}^*)(1 - \alpha^2)\big\}}{(1 - \cos(\pi w_{\infty}^*))^2\sin^2(\pi w_{\infty}^*)(1 - \alpha^2)\mathsf{h}^2(4 - \mathsf{n}^2)}\,q^{\frac{1}{2}} \nonumber \\
&& + O(q). \nonumber
\eea 

\hfill $\Box$
\end{corollary}

There are some simplifications for $\alpha = 1$. Owing to the exact relation $E_1 = \frac{2}{\mathsf{h}}$, only $\widehat{\mathsf{g}}_{1}$ has a non-trivial dependence in the non-critical regime:
\begin{corollary}
For $\alpha = 1$, we have:
\bea
\widehat{\mathsf{g}}_{2} & = & \frac{2}{4 - \mathsf{n}^2}\Big(1 - \frac{\mathsf{g}}{\mathsf{h}}\Big), \nonumber \\
\widehat{\mathsf{g}}_{1} & = & \frac{1}{\mathsf{h}(4 - \mathsf{n}^2)}\Big(- 1 + \frac{\mathsf{g}}{\mathsf{h}}(\rho^2 + 6)\Big) + \frac{\mathsf{g}\rho\rho_1}{\mathsf{h}^2(4 - \mathsf{n}^2)}\,q^{\frac{1}{2}} + O(q). \nonumber 
\eea
\hfill $\Box$
\end{corollary}

\section{Determination of the endpoints and phase diagram}
\label{proofbeh}
\label{AppD}
In this section, we recall the elements leading to the proofs of the theorems of the beginning of Section~\ref{critit}, see \cite{BBD} for more details. The equations $\Delta_{\varepsilon}\mathcal{G}^{\bullet}(0) = 0$ for $\varepsilon \in \{0,\frac{1}{2}\}$ determine $\gamma_{\pm}$ in terms of the weights of the model. We compute from Proposition~\ref{theimdisk} and the behavior of $\Upsilon_{b}(\tau\phi + 1/2)$ given in Lemma~\ref{lemUp}:
\bea
\label{D1} \mathcal{D}Y_{b,0}(\pi w_{\infty}) - q^{1 - b}\mathcal{D}Y_{b - 2,0}(\pi w_{\infty}) + O(q) & = & 0, \label{eq1} \\
\label{D2} \mathcal{D}Y_{b,\frac{1}{2}}(\pi w_{\infty}) - q^{b} Y_{b + 2,\frac{1}{2}}^{(k)}(\pi w_{\infty}) + O(q) & = & 0, \label{eq2}
\eea
where
\beq
\label{DYB} Y_{b,0}(w) = \cos(b w),\qquad Y_{b,\frac{1}{2}}(w) = \frac{\sin[(1 - b)w]}{\sin w},\qquad \mathcal{D} = \sum_{l = 0}^3 \frac{(-1)^l \widehat{\mathsf{g}}_{l}}{l!}\,\Big(\frac{\pi C}{T}\Big)^{l} \partial_{\pi w_{\infty}}^l.
\eeq

Exactly at criticality, we must have $u = 1$ and $q = 0$, thus using Corollary~\ref{CBCB6}:
$$
-\frac{2}{2 + \mathsf{n}} + \sum_{k = 1}^{3} \frac{(-1)^k \widehat{\mathsf{g}}_k^*}{k!}\,\Big(\frac{2{\rm cot}(\pi w_{\infty}^*)}{(1 - \alpha^2)\mathsf{h}}\Big)^k\,\frac{Y_{b,\varepsilon}^{(k)}(\pi w_{\infty}^*)}{Y_{b,\varepsilon}(\pi w_{\infty}^*)} = 0,\qquad \varepsilon \in \{0,\tfrac{1}{2}\}.
$$
We note that the critical values $\widehat{\mathsf{g}}_k^*$ obtained in Section~\ref{Appgdeter} are such that \eqref{eq1}-\eqref{eq2} give a system of two linear equations determining $\frac{\mathsf{g}}{\mathsf{h}}$ and $\mathsf{h}^2$ in terms of the parameter $w_{\infty}^*$. For $\alpha = 1$, we rather use $\rho$ as parameter, and the solution is
\bea
\label{gsurh} \frac{\mathsf{g}}{\mathsf{h}} & = & \frac{4(\rho b\sqrt{2 + \mathsf{n}} - \sqrt{2 - \mathsf{n}})}{\rho^2(b^2 - 1)\sqrt{2 - \mathsf{n}} + 4\rho b\sqrt{2 + \mathsf{n}} - 2\sqrt{2 - \mathsf{n}}}, \\
\label{h2eq} \mathsf{h}^2 & = & \frac{\rho^2 b}{24\sqrt{4 - \mathsf{n}^2}}\,\frac{\rho^2\,b(1 - b^2)\sqrt{2 + \mathsf{n}}  - 4\rho\sqrt{2 - \mathsf{n}} + 6b\sqrt{2 + \mathsf{n}}}{-\rho^2(1 - b^2)\sqrt{2 - \mathsf{n}} + 4\rho b\sqrt{2 + \mathsf{n}} - 2\sqrt{2 - \mathsf{n}}}.
\eea
Since $\tfrac{\mathsf{g}}{\mathsf{h}}$ and $\mathsf{h}^2$ must be nonnegative, we must have $\rho \in [\rho_{\min}',\rho_{\max}]$ with
\bea
\label{E55} \rho'_{\min}  & = & \frac{2\sqrt{1 - b^2}\sqrt{2 - \mathsf{n}} - \sqrt{2}\sqrt{(10 + \mathsf{n})b^2 - 4 + 2\mathsf{n}}}{b\sqrt{1 - b^2}\sqrt{2 - \mathsf{n}}},  \\
\label{E56} \rho_{\max}  & = & \frac{1}{b}\,\sqrt{\frac{2 - \mathsf{n}}{2 + \mathsf{n}}}.\eea
However, we will see later that the non-generic critical line only exists until some value $\rho_{\min} > \rho'_{\min}$, so \eqref{E55} will become irrelevant.
For $\alpha \neq 1$, see \cite[Appendix D]{BBD}.

Now, let us examine the approach of criticality. We fix $(\mathsf{g},\mathsf{h})$ on the non-generic critical line for $u = 1$, and we now study the behavior when $u \neq 1$ but $u \rightarrow 1$ of the endpoints $\gamma_{\pm}$. In particular, since the behavior of the elliptic functions is conveniently expressed in this regime in terms of $q = e^{-\frac{\pi}{T}}$, our first task is to relate $(1 - u)$ to $q \rightarrow 0$. For this purpose, we look at \eqref{eq1}, and note that $u$ only appears in $\widehat{\mathsf{g}}_{0}$. There could be a term of order $q^{\frac{1}{2}}$ stemming from near-criticality corrections to $w_{\infty}$, $\widehat{\mathsf{g}}_{k}$ and $\frac{\pi C}{T}$, but computation reveals that it is absent. Therefore, we obtain:
$$
1 - u = \frac{\mathsf{n} + 2}{2}\bigg(\sum_{l = 0}^3 \frac{(-1)^{l} \widehat{\mathsf{g}}_{l}^*}{l!}\Big(\frac{2\,{\rm cot}(\pi w_{\infty}^*)}{(1 - \alpha^2)\mathsf{h}}\Big)^l\,\frac{Y_{b - 2,0}^{(l)}(\pi w_{\infty}^*)}{Y_{b,0}(\pi w_{\infty}^*)}\bigg) q^{1 - b} + O(q),
$$
where $\widehat{\mathsf{g}}^*_{0} = -\frac{2}{2 + \mathsf{n}}$ and $(\widehat{\mathsf{g}}^*_k)_{k \geq 1}$ should be replaced by their values in terms of $(\mathsf{g},\mathsf{h},w_{\infty}^*)$ from Section~\ref{Appgdeter}, and $(\mathcal{\mathsf{g}},\mathcal{\mathsf{h}})$ by their parametrization \eqref{gsurh}-\eqref{h2eq} on the critical line.

We examine the case $\alpha = 1$. Using the parametrization \eqref{gsurh}-\eqref{h2eq}, the resulting formula is:
\beq
\label{Deltastar}1 - u = q_*\,q^{1 - b} + (q_{*,1} + c'\rho_1)q + o(q).
\eeq
with:
\bea
q_* & = & \frac{12}{b}\,\frac{\rho^2(1 - b)^2\sqrt{2 + \mathsf{n}} + 2\rho(1 - b)\sqrt{2 - \mathsf{n}} - 2\sqrt{2 + \mathsf{n}}}{-\rho^2 b(1 - b^2)\sqrt{2 + \mathsf{n}} + 4\rho(1 - b^2)\sqrt{2 - \mathsf{n}} - 6b\sqrt{2 + \mathsf{n}}}, \nonumber \\
q_{*,1} & = & \frac{24}{b}\,\frac{-\rho^2(b^2 + 1)\sqrt{2 + \mathsf{n}} + 2\rho b\sqrt{2 - \mathsf{n}} + 2\sqrt{2 + \mathsf{n}}}{-\rho^2b(1 - b^2)\sqrt{2 + \mathsf{n}} + 4\rho(1 - b^2)\sqrt{2 - \mathsf{n}} - 6b\sqrt{2 + \mathsf{n}}}. \nonumber
\eea
The value of $c'$ is irrelevant because we will soon show that $\rho_1 = 0$. As $(1 - u)$ should be nonnegative for $q > 0$, we must have $q_* \geq 0$, which demands $\rho \in [\rho_{\min},\rho_{\max}]$ with:
\beq
\label{rhominf}\rho_{\min} = \frac{\sqrt{6 + \mathsf{n}} - \sqrt{2 - \mathsf{n}}}{(1 - b)\sqrt{2 + \mathsf{n}}}.
\eeq
We observe that this lower bound is larger than $\rho_{\min}'$ given by \eqref{E55} for any $n \in [0,2]$, therefore the non-generic critical line can only exist in the range $\rho \in [\rho_{\min},\rho_{\max}]$ provided by \eqref{rhominf}-\eqref{E56}. These necessary conditions were also obtained in \cite{BBG12b} -- where the lower bound arose from the constraint of positivity of the spectral density associated with the generating series of disks $\mathcal{W}(x)$ -- and it was checked that these conditions are sufficient.

We now turn to the second equation \eqref{eq2}. We have checked that the term in $q^{b}$ vanishes, as we expect by consistency with \eqref{Deltastar}. Then, the term of order $q^{\frac{1}{2}}$ is proportional to $\rho_{1}$, therefore we must have, in both dense and dilute phase:
$$
\rho_{1} = 0,
$$
which means that $\gamma_{-} - \gamma_{-}^* \in O(q)$.

\vspace{0.2cm}

We see that for $\rho \in (\rho_{\min},\rho_{\max}]$:
$$
q \sim q_*\,(1 - u)^{\frac{1}{1 - b}}.
$$
for some constant $q_* > 0$. This corresponds, by definition, to the dense phase. For $\rho = \rho_{\min}$ (\textit{i.e.} the dilute phase) we have $q_* = 0$, and \eqref{Deltastar} specializes to:
$$
1 - u = \frac{24}{b(1 - b)(2 - b)}\,q + o(q).
$$

For general $\alpha$ not too large (see the statement of Theorem~\ref{alphanotlarge}), the result is qualitatively the same, only the non-zero constant prefactor differs -- see \cite[Appendix D]{BBD}.

\section{Proof of Lemma~\ref{pieces}}
\label{AppBB}

The goal in this appendix is to obtain the critical behavior of the building blocks. We give expressions valid for both universality classes using
$$
\mathfrak{d} = \left\{\begin{array}{lll} 1 & & {\rm dense}, \\ -1 & & {\rm dilute}. \end{array} \right.
$$
 From the expression \eqref{BB0}-\eqref{BB1} and the behavior of the special function $\Upsilon_{b}$ from Lemma~\ref{lemUp} we find:
\label{aabas}
\begin{lemma}
\label{lmH5} We have in the regime $T \rightarrow 0$:
\bea
\mathcal{B}_{\varepsilon,l}(\varepsilon' + \tau\phi) & = & \frac{2(-1)^{l + 1}}{\sqrt{4 - \mathsf{n}^2}}\,\Big(\frac{\pi}{T}\Big)^{2l + 2}\,q^{b(\varepsilon \oplus \varepsilon')}\big\{B_{\varepsilon\oplus \varepsilon',b}^{*,(2l + 1)}(\pi\phi) + O(q^{b})\big\}, \nonumber \\
\mathsf{B}_{\varepsilon,l}(\varepsilon' + \tau\phi) & = & (-1)^{l + 1} \Big(\frac{\pi}{T}\Big)^{2l + 2} q^{\frac{1}{2}(\varepsilon \oplus \varepsilon')}\big\{B_{\varepsilon \oplus \varepsilon',\frac{1}{2}}^{*,(2l + 1)}(\pi\phi) + O(q^{\frac{1}{2}})\big\} ,\nonumber
\eea
where
$$
B_{0,b}^*(\phi) = \frac{\sin (1 - b)\phi}{\sin \phi},\qquad B_{\frac{1}{2},b}^*(\phi) = 2\cos b\phi.
$$
The error is uniform for $\phi$ in any compact, and stable by differentiation.
\end{lemma}

We next focus on the denominator of the recursion kernel.

\begin{lemma}
\label{lmH6} We have in the regime $T \rightarrow 0$
\beq
(\Delta_{\varepsilon}\mathcal{G})(\tau\phi) = \Big(\frac{\pi}{T}\Big)\,q^{(1 - \mathfrak{d}\frac{b}{2})(1 - 2\varepsilon)}\,\big\{G^*_{\varepsilon}(\phi) + O(q^b)\big\}. \nonumber
\eeq
We have
$$
G^*_{0}(\phi) = \left\{\begin{array}{lll} -D^*_{b - 2}\,\sin \pi\phi\,\sin \pi(1 - b)\phi & & {\rm in}\,\,{\rm dense}\,\,{\rm phase}, \\ D^*_{b + 2}\,\sin\pi\phi\,\sin\pi(1 + b)\phi & & {\rm in}\,\,{\rm dilute}\,\,{\rm phase}, \end{array}\right.
$$
with $D^*_b$ given in \eqref{Dstarb} below. In each phase, $T_0^*(\phi) \neq 0$. We have
$$
T_{1/2}^*(\phi) = {\rm i}\sqrt{4 - \mathsf{n}^2}\,\mathcal{D}^*\bigg(2\frac{\sin\pi(1 - b)w_{\infty}^*}{\sin\pi w_{\infty}^*} - \frac{\sin\pi(1 - b)(w_{\infty}^* - \phi)}{\sin\pi(w_{\infty}^* - \phi)} - \frac{\sin\pi(1 - b)(w_{\infty}^* + \phi)}{\sin\pi(w_{\infty}^* + \phi)}\bigg),
$$
where $\mathcal{D}^*$ is a differential operator given in \eqref{Diffopsa} below.
\end{lemma}
\noindent\textbf{Proof.} From Proposition~\ref{theimdisk} and the behavior of $\Upsilon_{b}(\tau\phi + \tfrac{1}{2})$ given in Lemma~\ref{lemUp}, we repeat in a finer way the computation of the beginning of Section~\ref{critit}.:
\bea
\label{D0G}  && (\Delta_{0}\mathcal{G})(\tau\phi) \\
& = & \sqrt{4 - \mathsf{n}^2}\,\frac{8{\rm i}\pi}{T}\,\frac{q^{b/2}}{1 - q^b}\Big\{ -\cos \pi b \phi\,\mathcal{D}Y_{b,0}(\pi w_{\infty}) + q^{1 - b}\,\cos \pi(b - 2)\phi\,\mathcal{D}Y_{b - 2,0}(\pi w_{\infty}) \nonumber \\
&& - q\big(\cos\pi(b - 2)\phi\,\mathcal{D}Y_{b - 2,0}(\pi w_{\infty}) + \cos\pi(b + 2)\phi\,\mathcal{D} Y_{b + 2,0}(\pi w_{\infty})\big) + O(q^{1 + b})\Big\}, \nonumber
\eea
where $Y_{b,0}$ are $\mathcal{D}$ were introduced in \eqref{DYB}. One of the exact condition determining the endpoint was $\Delta_0\mathcal{G}(0) = 0$, \textit{i.e.}
$$
\mathcal{D}Y_{b,0}(\pi w_{\infty}) = q^{1 - b}\mathcal{D} Y_{b - 2,0}(\pi w_{\infty}) - q \mathcal{D}(Y_{b - 2,0} + Y_{b + 2,0})(\pi w_{\infty}) + O(q^{1 + b}) = 0,
$$
which we can substitute in \eqref{D0G} to obtain
\bea
&& (\Delta_{0}\mathcal{G})(\tau\phi) \nonumber \\
& = & \sqrt{4 - \mathsf{n}^2}\,\frac{16{\rm i}\pi}{T}\,\frac{q^{b/2}}{1 - q^b}\bigg\{q^{1 - b}\sin\pi\phi \sin\pi(1 - b)\phi \mathcal{D}Y_{b - 2,0}(\pi w_{\infty}) \nonumber \\
&& + q \sin\pi\phi \Big(\sin\pi(1 - b)\phi\,\mathcal{D} Y_{b - 2,0}(\pi w_{\infty}) + \sin\pi(1 + b)\phi\,\mathcal{D}Y_{b + 2,0}(\pi w_{\infty})\Big) + O(q^{1 + b})\bigg\}. \nonumber
\eea
The dense phase was characterized by
\beq
\label{Diffopsa} \mathcal{D}^*Y_{b - 2,0}(\pi w_{\infty}^*) \neq 0,\qquad \mathcal{D}^* = \sum_{l \geq 0} \frac{(-1)^l\widehat{\mathsf{g}}_{l}^*}{l!}\,\partial_{\pi w_{\infty}^*}^{l}.
\eeq
Therefore, the first term, of order $q^{1 - b}$, is indeed the leading term. The dilute phase is characterized by $\mathcal{D}^*Y_{b - 2}(\pi w_{\infty}^*) = 0$ and then one can check that $\mathcal{D}^*Y_{b + 2}(\pi w_{\infty}^*) \neq 0$. So, in the dilute phase the leading term is of order $O(q)$. This gives the announced results with
\beq
\label{Dstarb} D^*_{b} = 16{\rm i}\sqrt{4 - \mathsf{n}^2}\,\mathcal{D}^*Y_{b}(\pi w_{\infty}^*).
\eeq
For $(\Delta_{\frac{1}{2}}\mathcal{G})(\tau\phi)$, we easily arrive to the result using the beh\-avior of $\Upsilon_{b}(\tau\phi)$ from Lem\-ma~\ref{lemUp}, and exploiting the freedom to subtract $\Delta_{\frac{1}{2}}\mathcal{G}(0) = 0$. \hfill $\Box$

\begin{corollary}
\label{coE3} We have when $T \rightarrow 0$, for $r = 1,2$ and $\varepsilon \in \{0,\tfrac{1}{2}\}$:
$$
y_{\varepsilon,r} = \Big(\frac{\pi}{T}\Big)^{2r + 1}q^{(1 - 2\varepsilon)(1 - \mathfrak{d}b/2)}\big\{y_{\varepsilon,r}^* + O(q^{b})\big\}, 
$$
with $y_{\varepsilon,r}^* \neq 0$ computable from Lemma~\ref{lmH6}, and $y_{0,2}^*/y_{0,1}^* = - 2 + 2b - b^2$.
\end{corollary}

Inserting the previous results into the expressions \eqref{Coini}-\eqref{C1ini} for the initial conditions, we find:
\begin{corollary}
When $T \rightarrow 0$, we have
\bea
\mathcal{C}^{[0,3]}\bigl[{}^{0}_{\varepsilon}\,{}^{0}_{\varepsilon}\,{}^{0}_{\varepsilon}\bigr] & = & \Big(\frac{\pi}{T}\Big)^{-3}\,q^{(1 - 2\varepsilon)(\mathfrak{d}b/2 - 1)}\Big\{\mathcal{C}^{[0,3]}_{*}\bigl[{}^{0}_{\varepsilon}\,{}^{0}_{\varepsilon}\,{}^{0}_{\varepsilon}\bigr] + O(q^{b})\Big\}, \nonumber \\
\mathcal{C}^{[1,1]}\bigl[{}^{0}_{\varepsilon}\bigr] & = & \Big(\frac{\pi}{T}\Big)^{-1}\,q^{(1 - 2\varepsilon)(\mathfrak{d}b/2 - 1)}\Big\{\mathcal{C}^{[1,1]}_{*}\bigl[{}^{0}_{\varepsilon}\bigr] + O(q^{b})\Big\}, \nonumber \\
\mathcal{C}^{[1,1]}\bigl[{}^{1}_{\varepsilon}\bigr] & = & \Big(\frac{\pi}{T}\Big)^{-3}\,q^{(1 - 2\varepsilon)(\mathfrak{d}b/2 - 1)}\Big\{\mathcal{C}^{[1,1]}_{*}\bigl[{}^{1}_{\varepsilon}\bigr] + O(q^{b})\Big\}, \nonumber \\
\eea
and likewise for the $\mathsf{C}$'s, with: 
$$
\begin{aligned}
\mathcal{C}^{[0,3]}_{*}\bigl[{}^{0}_{\varepsilon}\,{}^{0}_{\varepsilon}\,{}^{0}_{\varepsilon}\bigr] \ & = \  -\frac{2}{y_{\varepsilon,1}^{*}}, & \ \ \ \  \mathsf{C}^{[0,3]}_{*}\bigl[{}^{0}_{\varepsilon}\,{}^{0}_{\varepsilon}\,{}^{0}_{\varepsilon}\bigr] \ & = \  -\frac{2}{y_{\varepsilon,1}^{*}}, \\
\mathcal{C}^{[1,1]}_{*}\bigl[{}^{0}_{0}\bigr] \ & = \  \frac{6 + 26b + 11b^2}{24y_{0,1}^*}, & \ \ \ \  \mathsf{C}^{[1,1]}_{*}\bigl[{}^{0}_{0}\bigr] \ & = \  \frac{29 + 26b + 11b^2}{24y_{0,1}^*},   \\
\mathcal{C}^{[1,1]}_{*}\bigl[{}^{\,0}_{\frac{1}{2}}\bigr] \ & = \ \frac{y_{\frac{1}{2},2}^*}{24(y_{\frac{1}{2},1}^*)^2} + \frac{2 + 6b + 3b^2}{6y_{\frac{1}{2},1}^*}, & \ \ \ \ \mathsf{C}^{[1,1]}_{*}\bigl[{}^{\,0}_{\frac{1}{2}}\bigr] \ & = \  \frac{y_{\frac{1}{2},2}^*}{24(y_{\frac{1}{2},1}^*)^2} + \frac{23}{24y_{\frac{1}{2},1}^*}, \\
\mathcal{C}^{[1,1]}_*\bigl[{}^{1}_{\varepsilon}\bigr] \ & = \  -\frac{1}{24y_{\varepsilon,1}^{*}}, & \ \ \ \  \mathsf{C}^{[1,1]}_*\bigl[{}^{1}_{\varepsilon}\bigr] \ & = \   -\frac{1}{24y_{\varepsilon,1}^{*}}.
\end{aligned}
$$
\end{corollary}

From Corollary~\ref{coE3} we can then deduce the critical behavior of $K$'s and $\tilde{K}$'s.

\begin{corollary}
For $\varepsilon,\sigma,\sigma' \in \{0,\frac{1}{2}\}$, we denote:
$$
f(\varepsilon,\sigma,\sigma'\vert B) \coloneqq B\big[(\varepsilon \oplus \sigma) + (\varepsilon \oplus \sigma')\big] + (\mathfrak{d}\tfrac{b}{2} - 1)(1 - 2\varepsilon).
$$
When $T \rightarrow 0$, we have
\bea
\mathcal{K}\bigl[{}^{l}_{\varepsilon}\,{}^{m}_{\sigma}\,{}^{m'}_{\sigma'}\bigr] & = & \Big(\frac{\pi}{T}\Big)^{2(m + m' - l) + 1}\,q^{f(\varepsilon,\sigma,\sigma'\vert b)}\Big\{K^*\bigl[{}^{l}_{\varepsilon}\,{}^{m}_{\sigma}\,{}^{m'}_{\sigma'}\bigr] + O(q^{b})\Big\}, \nonumber \\
\tilde{\mathcal{K}}\bigl[{}^{l}_{\varepsilon}\,{}^{l'}_{\varepsilon}\,{}^{m}_{\sigma}\bigr] & = & \Big(\frac{\pi}{T}\Big)^{2(m - l - l') - 1}\,q^{f(\varepsilon,\varepsilon,\sigma\vert b)}\Big\{\tilde{K}^*\bigl[{}^{l}_{\varepsilon}\,{}^{l'}_{\varepsilon}\,{}^{m}_{\sigma}\bigr] + O(q^{b})\Big\}, \nonumber
\eea
and 
\bea
\mathsf{K}\bigl[{}^{l}_{\varepsilon}\,{}^{m}_{\sigma}\,{}^{m'}_{\sigma'}\bigr] & = & \Big(\frac{\pi}{T}\Big)^{2(m + m' - l) + 1}\,q^{f\left(\varepsilon,\sigma,\sigma'\vert \frac{1}{2}\right)}\Big\{K^*\bigl[{}^{l}_{\varepsilon}\,{}^{m}_{\sigma}\,{}^{m'}_{\sigma'}\bigr] + O(q^{b})\Big\}, \nonumber \\
\tilde{\mathsf{K}}\bigl[{}^{l}_{\varepsilon}\,{}^{l'}_{\varepsilon}\,{}^{m}_{\sigma}\bigr] & = & \Big(\frac{\pi}{T}\Big)^{2(m - l - l') - 1}\,q^{f\left(\varepsilon,\varepsilon,\sigma\vert \frac{1}{2}\right)}\Big\{\tilde{K}^*\bigl[{}^{l}_{\varepsilon}\,{}^{l'}_{\varepsilon}\,{}^{m}_{\sigma}\bigr] + O(q^{b})\Big\}, \nonumber
\eea
with
\bea
\mathcal{K}^*\bigl[{}^{l}_{\varepsilon}\,{}^{m}_{\sigma}\,{}^{m'}_{\sigma'}\bigr] & = & \frac{4(-1)^{l + m + m'}}{4 - \mathsf{n}^2} \Res_{\phi \rightarrow 0} \frac{\dd \phi\,\phi^{2l + 1}}{(2l + 1)!\,G^*_{\varepsilon}(\phi)}\,B_{\varepsilon \oplus \sigma}^{*,(2m + 1)}[\pi(\phi + 1)]\,B_{\varepsilon \oplus \sigma'}^{*,(2m' + 1)}[\pi(\phi - 1)], \nonumber \\
\tilde{\mathcal{K}}^*\bigl[{}^{l}_{\varepsilon}\,{}^{l'}_{\varepsilon}\,{}^{m}_{\sigma}\bigr] & = & \frac{2(-1)^{m + l + l' + 1}}{\sqrt{4 - \mathsf{n}^2}}\,\Res_{\phi \rightarrow 0} \frac{\dd \phi\,\phi^{2(l + l') + 1}}{(2l + 1)!(2l')!\,G^*_{\varepsilon}(\phi)}\,B_{\varepsilon \oplus \sigma}^{*,(2m + 1)}[\pi(\phi + 1)], \nonumber
\eea
and likewise for the $\mathsf{K}^*$ and $\tilde{\mathsf{K}}^*$.
\end{corollary}
\noindent \textbf{Proof.} This is a direct computation from Lemma~\ref{lmH5}-\ref{lmH6}. We note that for $\tilde{\mathcal{K}}$ (and resp. $\tilde{\mathsf{K}}$), we find an exponent $q^{\tilde{f}(\varepsilon,\sigma\vert B)}$, with $\tilde{f}(\varepsilon,\sigma\vert B) = B(\varepsilon \oplus \sigma) + (\mathfrak{d}\tfrac{b}{2} - 1)(1 - 2\varepsilon)$, with $B=b$ (and resp. $B=\frac{1}{2}$). But since $\varepsilon \oplus \varepsilon = 0$, this is also equal to $f(\varepsilon,\varepsilon,\sigma\vert B)$. \hfill $\Box$

We also remark that the order of magnitude of $\mathcal{C}^{[0,3]}\bigl[{}^{0}_{\varepsilon}\,{}^{0}_{\varepsilon}\,{}^{0}_{\varepsilon}\bigr]$ and $\mathcal{C}^{[1,1]}\bigl[{}^{l}_{\varepsilon}]$, and of $\mathsf{C}^{[0,3]}\bigl[{}^{0}_{\varepsilon}\,{}^{0}_{\varepsilon}\,{}^{0}_{\varepsilon}\bigr]$ and $\mathsf{C}^{[1,1]}\bigl[{}^{l}_{\varepsilon}]$, is $q^{f(\varepsilon,\varepsilon,\varepsilon\vert b)}=q^{f\left(\varepsilon,\varepsilon,\varepsilon\vert \frac{1}{2}\right)}$. Therefore, for a given graph $\mathcal{G}$ and coloring $\bs{\sigma}$ of its edges appearing in the sum of Proposition~\ref{cosums}, and any vertex $\mathsf{v} \in V(\mathscr{G})$, the factor associated to $\mathsf{v}$ -- either $\mathcal{K}$, $\tilde{\mathcal{K}}$, $\mathcal{C}^{[0,3]}$ or $\mathcal{C}^{[1,1]}$ -- is of order of magnitude $q^{f_{\mathsf{v}}(b)}$ with
$$
f_{\mathsf{v}}(b) = f\big(\sigma(\mathsf{e}_{\mathsf{v}}^{0}),\sigma(\mathsf{e}_{\mathsf{v}}^{1}),\sigma(\mathsf{e}_{\mathsf{v}}^{2})\vert b\big).
$$
Similarly, any factor $\mathsf{K}$, $\tilde{\mathsf{K}}$, $\mathsf{\mathsf{C}}^{[0,3]}$ or $\mathsf{\mathsf{C}}^{[1,1]}$ associated to a vertex $\mathsf{v}\in V(\mathscr{G})$ scales like $q^{f_{\mathsf{v}}\left(\frac{1}{2}\right)}$.

\bibliographystyle{amsalpha}
\bibliography{Thesis}


%
%

\end{document}